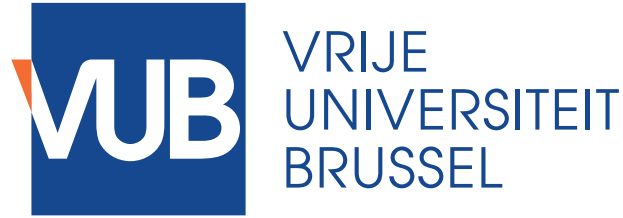

# Aspects of astrophysical particle production and beyond the Standard Model phenomenology

Matthias Vereecken

*Promotors*
Prof. Dr. Alexander Sevrin
Prof. Dr. Nick van Eijndhoven
Prof. Dr. Jorgen D'Hondt

*Co-promotor*
Prof. Dr. Alberto Mariotti

10th October 2019

## Doctoral examination commission

**Chair**: Prof. Dr. Stijn Buitink
*Astronomy and Astrophysics – Vrije Universiteit Brussel*

**Secretary**: Prof. Dr. Krijn De Vries
*Interuniversity Institute for High Energies (IIHE) – Vrije Universiteit Brussel*

**Promotor:** Prof. Dr. Alexander Sevrin
*Theoretische Natuurkunde (TENA) – Vrije Universiteit Brussel*

**Promotor:** Prof. Dr. Nick van Eijndhoven
*Interuniversity Institute for High Energies (IIHE) – Vrije Universiteit Brussel*

**Promotor:** Prof. Dr. Jorgen D'Hondt
*Interuniversity Institute for High Energies (IIHE) – Vrije Universiteit Brussel*

**Co-promotor:** Prof. Dr. Alberto Mariotti
*Interuniversity Institute for High Energies (IIHE) – Vrije Universiteit Brussel*

Prof. Dr. Dominique Maes
*Structural Biology – Vrije Universiteit Brussel*

Prof. Dr. Laura Lopez Honorez
*Theoretische Natuurkunde (TENA) – Vrije Universiteit Brussel*

Prof. Dr. Kumiko Kotera
*Institut d'Astrophysique de Paris (IAP)*

Prof. Dr. Benjamin Fuks
*Laboratoire de Physique Théorique et Hautes Énergies (LPTHE)*

# List of publications

The work presented in this thesis has lead to several publications. The first part of the thesis, which is devoted to Beyond the Standard Model physics, is based on the following publication

- Seng Pei Liew et al. 'Z-peaked excess in goldstini scenarios'. In: *Phys. Lett.* B750 (2015), pp. 539–546. DOI: 10.1016/j.physletb.2015.09.035. arXiv: 1506.08803 [hep-ph]

The second part focuses on high-energy neutrino astronomy. Some of the material there has been published in

- G. Maggi et al. 'Obscured flat spectrum radio active galactic nuclei as sources of high-energy neutrinos'. In: *Phys. Rev.* D94.10 (2016), p. 103007. DOI: 10.1103/PhysRevD.94.103007. arXiv: 1608.00028 [astro-ph.HE]

- Krijn D. de Vries et al. 'Constraints and prospects on gravitational-wave and neutrino emissions using GW150914'. In: *Phys. Rev.* D96.8 (2017), p. 083003. DOI: 10.1103/PhysRevD.96.083003. arXiv: 1612.02648 [astro-ph.HE]

and in the conference proceedings

- Krijn D. de Vries et al. 'Implications of GW related searches for IceCube'. In: *Proceedings, 52nd Rencontres de Moriond on Electroweak Interactions and Unified Theories: La Thuile, Italy, March 18-25, 2017.* 2017, pp. 477–480. arXiv: 1709.07430 [astro-ph.HE]

- Gwenhaël de Wasseige et al. 'Constraints and prospects on gravitational wave and neutrino emission using GW150914'. In: *PoS* ICRC2017 (2018), p. 959. DOI: 10.22323/1.301.0959. arXiv: 1709.04880 [astro-ph.HE]

# Acknowledgements

The PhD experience does not rely on hard science and many repeated (sometimes futile) attempts at solving seemingly simple problems alone. It involves a whole network of people who help guide the PhD in the right direction, give advice when you did not know you needed it or provide support when things get tough.

First of all, there are my (co)promotors, Alex, Nick, Jorgen, and Alberto, who accepted me when I applied for a position and helped me write the project proposal, as well as the FWO Vlaanderen, which provided me with funding. Thank you for providing me with the opportunity to pursue a PhD! I want to specifically thank Alberto for much of the support and advice, both scientific and personal, as well as the many learning opportunities. Without you, I would probably never have been able to bring this PhD or the thesis to a successful end. I'm also grateful to Krijn for the enriching discussions we had on both our projects. Last, I want to explicitly thank the members of the doctoral examination commission for the interesting discussions we had during the private defense as well as for the constructive comments to improve the manuscript.

Next, there are the colleagues who helped shape my experience at the VUB. I want to especially thank Gwen, whose enthusiasm and drive is infecting. Our many discussions at the office and conferences were some of the most fun I have had these years. I am also grateful to my office mates Karen and Dries, who were able to turn some of the less interesting days into absurd moments. Last, but not least, is Marleen, who is responsible for much of the positivity around the office.

Of course, also friends and family played a major role these past few years. Céline, you gave me useful advice, particularly during the first year when I really needed it. Véronique: your ability to accidentally write award-winning papers and presentations is truly inspiring! Our chats and brunches were absolutely hilarious and a welcome diversion. Naturally, there is also my family, in particular my parents and grandparents. Your support, which went above and beyond, was absolutely essential in making all of this possible. Also, your continued and sincere interest in what I was doing, even when the subject must have been almost incomprehensible, meant more than you could ever know. Finally, I want to thank Bert. Whether you were here or on the other side of the world, you were with me in every step of the way, experienced together with me the best and the worst moments and helped me put things back in perspective when I needed it. Your unwavering support during these years, especially this past year, meant the world to me.

# Contents

# List of Figures

# List of Tables

# Introduction

The building blocks of the universe are described by the particles and interactions contained within the Standard Model of particle physics. Together with gravity, successfully described by general relativity, this theory can explain almost all phenomena we observe, from the smallest scales of fundamental particles to the largest scales of galaxies and galaxy clusters, within a single framework. Likewise, our continuous journey to improve our understanding of these interactions and potentially uncover new physics leads us to study all of these phenomena using the toolset of particle physics; both to understand and to observe these phenomena. In this thesis, we focus on two aspects of this. Using the high precision of particle accelerators at Earth, we seek to unearth new physics which might be looming around the corner. On the other hand, we also investigate the interactions occurring in extreme environments provided by the universe, which accelerate particles to energies which are orders of magnitude higher than those we can currently reach on Earth.

Despite its many successes, we know that the Standard Model itself can not be complete. The mass of the Higgs boson is not protected by any symmetries of the theory. Therefore, its value is sensitive to the highest energy scales up to which our theory is valid. As a result, unless nature if fine-tuned, its small value suggests that there is new physics which might loom around the corner. Moreover, while the Standard Model is very successful, it provides no explanation for the existence of dark matter, the nature of the neutrino masses or the predominance of matter compared to antimatter.

One possibility to extend the Standard Model is to include supersymmetry (and its breaking at low energy), which doubles the particle content of the Standard Model whilst maintaining a predictive framework for their possible interactions. It and other new physics can be sought for in a myriad of ways. At the forefront of these are particle accelerators. By colliding together particles at high energy, new and heavy particles can appear. However, searches for new physics in proton-proton collisions at the Large Hadron Collider are all compatible with the Standard Model prediction. Therefore, we are prompted to investigate the smallest deviations in the data and explore less conventional signatures of new physics.

On the other hand, the Standard Model's success in describing all known interactions, except gravity, can also be used to study the most extreme objects in the universe. Such objects emit high-energy cosmic rays, gamma rays, and neutrinos, whose production can be described with particle physics, as well as gravitational waves. By combining information from the different types of emission in a multimessenger approach, we can increase our understanding of their sources.

While the existence of a diffuse flux of high-energy neutrinos is firmly established, the exact sources of these neutrinos are not yet identified. Recently, one particular source has been detected: TXS 0506+056, a flaring blazar, the centre of an active galaxy spewing out high-energy radiation in Earth's direction. However, sources of this type can not account for the full diffuse flux of high-energy neutrinos detected by IceCube. One hint on the properties of possible sources comes from the gamma-ray background observed by Fermi. The processes which produce neutrinos must also produce gamma rays in similar quantities. However, the relative brightness of the observed diffuse neutrino and gamma-ray fluxes suggests that the neutrino sources must be obscured in gamma rays.

Recently, gravitational waves have become observable as well. Whereas neutrino and gamma-ray emission depends on an interplay of the source properties and its interaction with the environment, gravitational waves are uniquely capable of probing the inner engine. The first ever gravitational wave event detected, GW150904, was a merger of two black holes. This detection triggered a large follow-up campaign searching for associated emission of electromagnetic radiation or neutrinos. Typically, no such emission is expected. However, in light of a candidate gamma-ray signal, several models have been developed where such emission can occur.

The first part of this thesis concerns the exploration of beyond the Standard Model physics, introduced in Chapter 1. In particular, we study in Chapter 2 a model of supersymmetry with an extended supersymmetry-breaking sector. In such a model, the particle spectrum is enriched, while its predictiveness is maintained. We apply this model to an excess seen at the LHC in its first run and also update our analysis using the newest data from the second run. In the second part of this thesis, we turn our attention to neutrino astronomy, introduced in Chapter 3. We start by developing a model where the astrophysical neutrino flux is produced by objects which are obscured by gas in Chapter 4. Our model is applied to a set of obscured objects and we investigate the sensitivity to our model and constraints on the parameter space. We also investigate whether our model can explain the unresolved neutrino flux without violating the constraints from the gamma-ray background. Finally, in Chapter 5, we test the hypothesis that binary black hole mergers do not emit neutrinos, using information from direct searches for such neutrinos and exploiting the information gained from the diffuse astrophysical neutrino flux.

# Prologue:
# the Standard Model of particle physics

The Standard Model of particle physics is the current theory for fundamental particles and their interactions, which can account for almost all our observations with high accuracy. It is the culmination of many years of theoretical and experimental work in high-energy particle physics.

On the theoretical side, the Standard Model was built starting with the Yang-Mills theory for non-Abelian gauge theories [1], through the Glashow model for unifying electromagnetic and weak interactions [2] to the current Weinberg-Salam theory for electroweak interactions [3, 4], which incorporates the Brout-Englert-Higgs mechanism [5–7]. Simultaneously, the theory for strong interactions was built, starting from the quark model developed by Gell-Mann [8, 9], together with the work on asymptotic freedom of non-Abelian gauge theories for the strong interaction of Gross and Wilczek [10–12] and Politzer [13, 14], laying the basis for quantum chromodynamics (QCD). Experimentally, the first discoveries occurred at the end of the 19$^{th}$ century all the way up to 2012, with the discovery of the electron by Thomson [15], the atomic nucleus and the proton by Rutherford [16, 17], the positron by Anderson [18], the muon by Neddermeyer and Anderson [19] the first neutrino by Reines and Cowan [20], partons in deep inelastic scattering at SLAC [21, 22], the $W^{\pm}$ and $Z$ bosons by Carlo Rubbia, Simon van der Meer and the UA1 collaboration [23, 24], the top quark at Fermilab [25, 26] and finally the discovery of the Higgs boson at the LHC [27, 28], as well as many others not mentioned here.

The Standard Model contains all known interactions, except gravity, and all particles in a predictive but relatively simple and elegant mathematical framework. Its structure is completely determined by the requirement that the Lagrangian is Lorentz invariant and possesses certain internal, local symmetries called gauge symmetries. The strong and electroweak forces are then associated to the symmetry group $SU(3)_c \times SU(2)_L \times U(1)_Y$ (i.e. colour, weak isospin and hypercharge) and the fermions and Higgs boson live in various representations of these symmetry groups, as indicated in Table 1. The interactions between these particles, mediated by gauge bosons associated to the gauge symmetries, are a direct consequence of requiring that the theory is invariant under these symmetries.

The Standard Model follows then automatically as the most general renormalisable Lagrangian compatible with $SU(3)_c \times SU(2)_L \times U(1)_Y$ symmetry. In its most compact

Table 1: Standard Model fermions and scalar and their representation under the different gauge groups.

| Field | Spin | $SU(3)_c$ | $SU(2)_L$ | $Y$ | $Q = T^3 + Y$ |
|---|---|---|---|---|---|
| $q = \begin{pmatrix} u_L \\ d_L \end{pmatrix}$ | 1/2 | **3** | **2** | 1/6 | $\begin{pmatrix} +2/3 \\ -1/3 \end{pmatrix}$ |
| $u_R$ | 1/2 | **3** | **1** | 2/3 | $+2/3$ |
| $d_R$ | 1/2 | **3** | **1** | $-1/3$ | $-1/3$ |
| $l = \begin{pmatrix} \nu_L \\ e_L \end{pmatrix}$ | 1/2 | **1** | **2** | $-1/2$ | $\begin{pmatrix} 0 \\ -1 \end{pmatrix}$ |
| $e_R$ | 1/2 | **1** | **1** | $-1$ | $-1$ |
| $\phi = \begin{pmatrix} \phi^+ \\ \phi^0 \end{pmatrix}$ | 0 | **1** | **2** | 1/2 | $\begin{pmatrix} +1 \\ 0 \end{pmatrix}$ |

form, it is written as[1]

$$
\begin{aligned}
\mathcal{L}_{SM} = & -\frac{1}{4} G^a_{\mu\nu} G^{a\mu\nu} - \frac{1}{4} W^i_{\mu\nu} W^{i\mu\nu} - \frac{1}{4} B_{\mu\nu} B^{\mu\nu} \\
& + (D_\mu \phi)^\dagger (D^\mu \phi) + \mu^2 \phi^\dagger \phi - \frac{\lambda^2}{2} (\phi^\dagger \phi)^2 \\
& + i \bar{\psi}_i \slashed{D} \psi_i - \left( \bar{q}^i \lambda^u_{ij} u^j_R \tilde{\phi} + \bar{q}^i \lambda^d_{ij} d^j_R \phi + \bar{l}^i \lambda^l_{ij} e^j_R + \text{h.c.} \right) .
\end{aligned}
\tag{1}
$$

Here, we defined $\slashed{D} = D_\mu \gamma^\mu$ and $\bar{\psi} = \psi^\dagger \gamma^0$, with $\psi = q_L,\ u_R,\ d_R,\ l,\ e_R$ a dirac spinor. The covariant derivatives are given by

$$
D_\mu = \partial_\mu + i g_s G^a_\mu T^a + i g W^i_\mu T^i + i g' B_\mu Y,
\tag{2}
$$

with $G^a_\mu$, $W^i_\mu$ and $B_\mu$ the gauge bosons of the strong and electroweak symmetry groups, where the indices $a$ and $i$ run over the adjoint representation. The hypercharge is given by $Y$, the generators for the strong interaction are $T^a = \lambda^a/2$ with $\lambda^a$ the Gell-Mann matrices (defined in Appendix A.4) and for weak isospin $T^i = \sigma^i/2$, with $\sigma^i$ the Pauli matrices (defined in Appendix A.3). If a field upon which the covariant derivative operates is a singlet under a symmetry group, its corresponding generator is zero and drops out. The field strengths are defined as $B_{\mu\nu} = \partial_\mu B_\nu - \partial_\nu B_\mu$ for $U(1)_Y$, $W^i_{\mu\nu} = \partial_\mu W^i_\nu - \partial_\nu W^i_\mu - g \epsilon^{ijk} W^j_\mu W^k_\nu$ for $SU(2)_L$, and $G^a_{\mu\nu} = \partial_\mu G^a_\nu - \partial_\nu G^a_\mu - g_s f^{abc} G^b_\mu G^c_\nu$ for $SU(3)_c$, with $\epsilon^{ijk}$ and $f^{abc}$ the structure constants of $SU(2)$ and $SU(3)$.

While the strong interaction, mediated by gluons, remains intact, the same is not true for the electroweak interactions. Due to the Mexican-hat potential for the Higgs field[2] and the Brout-Englert-Higgs mechanism, the electroweak symmetry $SU(2)_L \times U(1)_Y$ is spontaneously broken to its subgroup $U(1)_{\text{em}}$, i.e. electromagnetism. The would-be Goldstone bosons which appear as a consequence of this breaking are "eaten" by the gauge bosons of the broken symmetries to give mass to the $W^\pm$ and $Z$ bosons, with masses $m_{W^\pm} = 80$ GeV and $m_Z = 91$ GeV respectively (the "weak scale"), while the photon $\gamma$ associated to the unbroken $U(1)_{\text{em}}$ remains massless. The weakness of the weak interactions is then explained by the mass of its mediators, the $W^\pm$ and $Z$ bosons. The remaining degree of freedom of the Higgs field appears as a Higgs boson, with a mass $m_h = 125$ GeV. Electroweak symmetry breaking also gives mass to the leptons and the quarks through their Yukawa couplings with the Higgs boson. The lepton masses are given by $m_e = 512$ keV, $m_\mu = 106$ MeV, and $m_\tau = 1.777$ GeV, while the neutrinos are massless in the Standard Model. The quark masses are given by $m_u = 2.16$ MeV, $m_d = 4.67$ MeV, $m_c = 1.27$ GeV, $m_s = 93$ MeV, $m_b \sim 4.18$ GeV and $m_t \approx 172$ GeV, with the top quark as an outlier. All these masses are shown in Figure 1. The Yukawa couplings $\lambda_{ij}^f$, which are $3 \times 3$ matrices in flavour space, are not flavour-diagonal. As a consequence, when writing the electroweak interactions in the mass eigenbasis instead of the gauge eigenbasis, the charged weak interaction no longer conserves flavour. This is expressed through the Cabibbo-Kobayashi-Maskawa (CKM) matrix [30, 31], or quark mixing matrix, and the Pontecorvo-Maki-Nakagawa-Sakata (PMNS) matrix [32, 33], or lepton mixing matrix. Finally, lepton and baryon number conservation appear for free as accidental symmetries in this Lagrangian.

---

[1] Here, we ignore the possibility of a $\theta_{\text{QCD}}$-term, which should anyway vanish or almost vanish.

[2] There is no generally accepted convention for the parameter in front of the quartic coupling in the Higgs potential. Here, we chose the one of [29], Chapter 10. The same form for the coupling constants also appears in supersymmetry. A more common notation for the prefactor in the context of Standard Model studies is $\lambda$ instead of $\lambda^2/2$, as appears in Chapter 11 of the same reference.

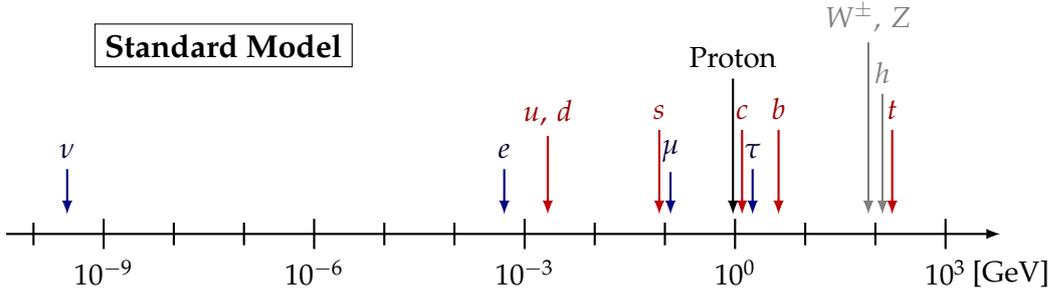

Figure 1: Masses of the Standard Model particles. Note that neutrinos, while massless in the Standard Model, have been observed to possess mass. The proton, which obtains its mass not from the Higgs mechanism, but dynamically from QCD, is shown as a reference. The gluon $g$ and photon $\gamma$, which mediate the strong and electromagnetic interactions, are massless and thus not shown. Figure modified from [34].

This mathematical description of the Standard Model allows for precise predictions of any observable currently accessible to particle physics experiments. These predictions, despite the simplicity of the Standard Model Lagrangian as it is written above, require very complex calculations and must often be performed using Monte Carlo generators. With these numerical codes, it is possible to describe interactions in many different contexts using a single framework. Indeed, in this thesis, we discuss two topics which are quite different in scope. However, many of the methodologies, both for the description of the physics from a theory point of view and for the detection of the associated phenomena from an experimental point of view are shared between these seemingly unrelated fields. In particular, here we use in both parts the calculation of interactions between (fundamental) particles using Monte Carlo generators. Likewise, the subsequent analysis of the results is performed in a similar way. In the first part of this thesis, we deal with phenomena at the smallest scales, namely the search for extensions of the Standard Model. Such extensions, typically describing particles and interactions at and above the weak scale (but sufficiently below the Planck scale, see Chapter 1) must be described by a theory within the same framework as the Standard Model. As we will see, the numerical tools to describe interactions in the Standard Model have been extended to allow the investigation of any beyond the Standard Model theory. In part two of this thesis, we use the fundamental particles and their interactions already present in the Standard model to gain insight into objects at the largest scale in our Universe, such as active galactic nuclei and binary black holes.

Before moving on to the main text, it is worth mentioning some of the tools that were utilised to produce the material in this thesis. In addition to the specialised scientific packages, which will be highlighted in the relevant sections, most of the work here is built upon a set of broad-purpose programming tools. Most importantly, this thesis was made using the `Python` ecosystem, more specifically its packages for scientific computing [35, 36]. These packages are `Scipy` [37], `Numpy` [38, 39], `Matplotlib` [40], and `Pandas` [41]. In addition to these, some features of `ROOT` [42] were also used. Finally, Feynman diagrams were drawn using `Jaxodraw` [43].

**Part I**

# Beyond the Standard Model physics at colliders

# Physics beyond the Standard Model

In the first part of this thesis, we study physics beyond the Standard Mode, in particular supersymmetry, and the search for it at the Large Hadron Collider (LHC). Supersymmetry is a well-motivated theory, which can naturally solve some of the issues from which the Standard Model suffers. As such, it has been one of the main candidates for beyond the Standard Model physics in the past decades. In this chapter, we lay the groundwork for testing supersymmetry at the LHC. We start by motivating the need for physics beyond the Standard Model. Next, we review how to build supersymmetric Lagrangians and, more importantly, how to achieve supersymmetry breaking. We see how, as a result of supersymmetry breaking, there appears a goldstone mode, the goldstino. In the case of gauge-mediated supersymmetry breaking, this goldstino essentially defines the collider phenomenology. Finally, we review the current status of supersymmetry at the LHC.

## 1.1   Problems with the Standard Model

The Standard Model (SM) as a theory has been successful[1] at describing almost all our current experiments with incredible precision. Why then, should one look for physics beyond the Standard Model (BSM)? We know that the Standard Model can not be complete. It can not account for several observational facts. Moreover, there are also theoretical grounds for why the Standard Model appears to be inconsistent. Here, we will briefly discuss these arguments pointing towards new physics.

First, we list some of the observations for which there is no explanation within the Standard Model. From measurement of the cosmic microwave background (CMB), which is background radiation which decoupled from all other particles when the universe was a mere $380\,000$ years old, we know that the universe contains a large component of dark matter[2], matter which has no or very weak interactions with baryonic

---

[1]Some might even say "too successful".

[2]In addition, measurements of the expansion of the universe indicate that this expansion is accelerat-

matter [46] except for gravity. While not necessary, dark matter could be made out of new, possibly fundamental, particles with potential interactions with Standard Model particles. One argument for this is the WIMP[3] miracle [46, 47], which is the observation that if dark matter is made out of a massive particle with interactions on the order of the weak scale which is initially in thermal equilibrium with the rest of the universe, then it obtains automatically the correct relic abundance. Even if dark matter is not such a particle, its production or appearance is still likely to require new physics. Also from cosmology, and cosmic ray experiments, we know that the universe is made predominantly out of matter and not anti-matter, despite the fact that the Standard Model is (almost) completely symmetric under the exchange of these two (*CP*-symmetry). This matter-antimatter asymmetry and the three Sakharov conditions [48] needed to produce it require new physics.

There are also particle physics observations which can not be explained. While propagating, neutrinos oscillate between the three flavours, as has been observed in neutrino fluxes coming from the sun [49], the atmosphere [50] and from reactors [51]. This oscillating behaviour can only be explained if the neutrinos have masses [52, 53]. However, in the Standard Model, neutrinos are exactly massless. In order to give mass to the neutrinos, a new particle (such as a right-handed neutrino, which is completely sterile) is necessary. Finally, there are also some anomalies which have shown up in particle physics experiments. The muon anomalous magnetic moment $g_\mu - 2$ [54] can be calculated to great accuracy in the Standard Model. While the measurement of the electron anomalous magnetic moment agrees with the prediction[4], that of the muon deviates significantly: $3.5\sigma$ [57] to $4\sigma$ [58]. There are also several anomalies in *B*-physics experiments [59]: when comparing the decays into different leptons of certain mesons containing a *b*-quark, the observed ratio does not agree with the Standard Model prediction[5]. It is interesting that both of these anomalies distinguish between different flavours.

Next, we discuss some of the more theoretical problems with and shortcomings of the Standard Model. The first of these is the most aesthetic one: what is the origin of the difference between flavours? In particular, the structure of the CKM matrix [30, 31], or quark mixing matrix, and the PMNS matrix [32, 53], or lepton mixing matrix, has no explanation in the Standard Model. Similarly, there is no explanation for the value of the quark and lepton masses, in particular the large hierarchy between the neutrino masses[6] and all other particles. These issues are related, since the fermions in the Standard Model receive their mass after electroweak symmetry breaking through their Yukawa couplings with the Higgs. The CKM and PMNS matrices are then the consequence of diagonalising

---

ing [44, 45]. This requires the presence of a cosmological constant or dark energy. However, this component is even farther away from being understood than dark matter.

[3]Weakly Interacting Massive Particle.

[4]Recently, a new measurement with improved precision of the fine structure constant $\alpha$, which is the main source of uncertainty on the prediction for $g_e$, has resulted in a $2.5\sigma$ deviation also for this $g_e$ [55], see also [56].

[5]Although there are still some hadronic uncertainties in this prediction.

[6]Only an upper bound on the sum of the neutrino masses is known.

these mass matrices. Therefore, the question is how the values of the Yukawa matrices are determined. Although this is not necessarily an issue, since one can simply assume their values, such a solution is unsatisfactory.

Another issue is the unification of the electroweak and strong forces [60]. Although the electroweak forces themselves are already described together in a single theory, they are not strictly unified in a single description, since they are still composed out of different gauge groups. On top of this, the strong force is treated completely separately. There are several arguments to suspect that these forces should be unified in a single theory. The first of these is anomaly cancellations [61]: the $U(1)_Y$-charges of the different Standard Model fermions are such that gauge-gravity anomalies are cancelled, and the classical gauge symmetries are retained in the quantum theory. Unless this structure is somehow enforced, there is no reason why the $U(1)_Y$-charges should be distributed like this. Another reason is the observation that the coupling constants of the different gauge forces, which depend on the energy scale where interactions are probed, approximately (but not exactly) meet around an energy of $10^{16}$ GeV. This suggests that at this energy, there is a grand unified theory (GUT) [60] with a single gauge group, which contains the Standard Model gauge groups. The Standard Model fermions can then be fit in different representations of this group and anomaly cancellation is immediately ensured.

One step further, one would also like a theory which unifies the gauge forces with gravity to form a Theory of Everything [62]. Currently, these theories are incompatible, since the gravitational constant has negative mass dimension. In a quantum field theory, this leads to a non-renormalisable theory, for which a UV-completion is necessary. There is thus need for an encompassing theory (such as string theory). The energy at which this unified theory needs to appear is determined by the Planck scale, purely on the basis of dimensional analysis. The reduced[7] Planck mass is given by $m_P = \sqrt{\frac{\hbar c}{4\pi G}} \sim 10^{18}$ GeV$/c^2$. Currently, this scale, as well as the GUT scale (both shown in Figure 1.1), are far outside the reach of experimental tests.

Finally, there is also the issue of the Higgs mass. It is the only dimensionful parameter in the Standard Model, the value of which must be set by hand at the weak scale. More importantly, however, is the issue of protecting its value when taking into account quantum corrections. Naively put, since the Higgs is a scalar, its interactions are not very constrained. Therefore, in a quantum theory, it will receive corrections to its mass from all other particles. As a result, the Higgs mass is pulled up to the highest scale up to which the Standard Model theory is valid. In the current model, this means that the Higgs mass is pulled up to the GUT scale or even the Planck scale. Only by miraculous cancellations between the bare mass and the quantum corrections can the Higgs mass be kept low. While this is in principle possible, it requires extreme levels of fine-tuning which are deemed undesirable. This is known as the naturalness problem[8] or, more specifically, the hierarchy problem [63–66].

---

[7]It is called "reduced" because of the inclusion of a factor $4\pi$.

[8]The question of whether naturalness can be considered a good guide for model building or not is a very good one and its answer depends on who this question is asked to. The same can be said about the amount of fine-tuning allowed for a model to be considered natural or how to define such fine-tuning in the first place.

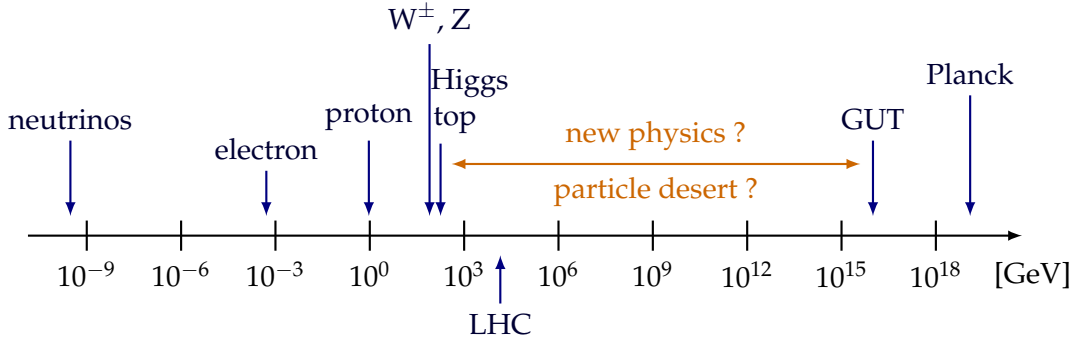

Figure 1.1: Energy scales in the Standard Model and beyond. Beyond the top mass, the only scales at which we know something must happen are the scale of a grand unified theory (GUT), where the gauge couplings of the Standard Model and its extensions have a similar value, and the Planck scale where gravity and quantum mechanics are both important. Figure modified from [34].

More concretely, in a quantum field theory, the Higgs squared-mass parameter $m_h^2$ receives quantum corrections from every particle to which it couples. In particular, if a theory contains a coupling between the Higgs and a fermion pair $hff$ with coupling strength $\lambda_f$ and is valid up to a scale $\Lambda_{\mathrm{UV}}$, the loop corrections to the Higgs mass, shown in Figure 1.2a, are of the form

$$\Delta m_h^2 = -\frac{|\lambda_f|^2}{8\pi^2} \Lambda_{\mathrm{UV}}^2 + \dots ,\qquad(1.1)$$

where the ellipsis denotes terms at most logarithmic in $\Lambda_{\mathrm{UV}}$ and proportional to $m_f^2$. The Higgs mass is thus pulled up to the cut-off scale of the theory, with no way to protect it. In the Standard model, this is mainly caused by the top quark, since its coupling to the Higgs is by far the strongest. Switching from a hard cut-off to dimensional regularisation does not solve the issue, but merely hides it: while now the $\Lambda_{\mathrm{UV}}^2$ piece is gone, there remains a part proportional to the scalar mass squared (see also Appendix C.1). The core of the problem is that the Higgs mass has a quadratic sensitivity to high mass scales [67]. This issue can only be solved by introducing new particles which cut off the loop at some energy above the electroweak scale.

Consider the case where also a scalar $\phi$ couples to the Higgs as $hh\phi\phi$. The corrections to the Higgs squared-mass parameter due to the loop in Figure 1.2b are then

$$\Delta m_h^2 = \frac{\lambda_S}{16\pi^2} \left[ \Lambda_{\mathrm{UV}}^2 - 2m_S^2 \ln(\Lambda_{\mathrm{UV}}/m_S) + \dots \right].\qquad(1.2)$$

This suggests a way out: if there is a symmetry which relates both corrections, then they can conspire to cancel [68–73]. Supersymmetry does exactly this: it introduces for every fermion in the theory two scalars, where their coupling constants are related as $\lambda_S = |\lambda_f|^2$. In this way, the quadratic sensitivity to high mass scales of the Higgs mass is "cured" (this is shown more explicitly in Appendix C.1). Note that of all the issues mentioned here, this is the only argument which points to new physics around the weak scale[9] (with the possible exception of dark matter).

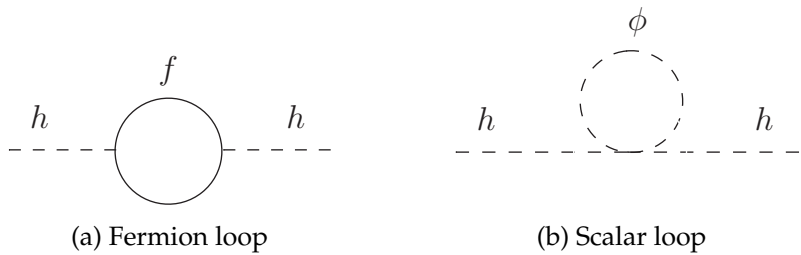

(a) Fermion loop                    (b) Scalar loop

Figure 1.2: Corrections to the Higgs mass due to fermions and scalars.

Finally, it is important to mention that there is a wealth of alternative theories which can also solve the hierarchy problem or predict new physics at the weak scale or slightly above. In composite Higgs models [74, 75], the Higgs boson is bound state of a new strong interaction. One of the possible explanations for the lightness of the Higgs is then to identify it as a pseudo-Goldstone boson due to the breaking of a flavour symmetry, similar to the pion (for a review, see e.g. [76]). Another alternative is neutral naturalness, where there appear non-coloured partners to the top quark, which can solve the issues with the Higgs mass while evading strong constraints on new coloured states. One example of this is the Twin Higgs scenario [77]. In this case, there is a Mirror of the Standard model and its interactions. The Higgs boson then appears again as a pseudo-goldstone boson of this extended symmetry. There are also Kaluza-Klein theories of extra dimensions [78], such as the Arkani-Hamed, Dimopoulos and Dvali (ADD) model [79], which can explain the weakness of gravity and predicts new physics to appear at the TeV scale, and the Randall-Sundrum model [80], which can explain why the weak scale is so much below the Planck scale and also predicts new states around the TeV scale.

Finally, more exotic models include the Relaxion [81], where the Higgs mass is set dynamically during inflation without requiring new dynamics at the weak scale, or clockwork theories [82–84], which achieve small coupling constants by linking together a large number of fields, each with couplings of $\mathcal{O}(1)$.

In the following, we will focus on supersymmetry, how to build supersymmetric extensions of the Standard Model and its constraints at LHC.

## 1.2 Supersymmetry basics

In this section, we introduce the basics of supersymmetry (SUSY) and how to build a supersymmetric Lagrangian. We do not attempt to provide the reader with a comprehensive review of the full theory of supersymmetry. For this, we refer to standard texts such as [85–87]. Rather, our main goal is to set up all of the basic ingredients necessary for the next chapter, where we attempt to explain a particular event signature at the LHC using a model with multiple supersymmetry-breaking sectors. However, since the discussion here is still quite detailed, a cheat sheet is available in Appendix B. This appendix shows only the essential supersymmetry ingredients from this section and can be used either as a summary, a reference, or a shortcut[10].

---

[9]Although there are models which can solve the hierarchy problem at higher scales. For example, composite models (see below) predict new physics in the multi-TeV range.

In this chapter and the next, we mostly[11] follow the conventions of [67] (see also [88]). In particular, we use two-component Weyl spinors to represent the Standard Model and supersymmetry fermions. This is a more natural convention for the Standard Model and its extensions compared to four-component Dirac or Majorana spinors, since it treats left-handed and right-handed particles, which transform differently under the Standard Model gauge groups, separately from the start. Moreover, Weyl spinors will appear as components of chiral superfields. Hermitian conjugation is indicated by a dagger, as opposed to the common notation using barred spinors, popularised by [86]. The equivalence between these notations is given by $\overline{\psi}_{\dot{\alpha}} \equiv \psi^{\dagger}_{\dot{\alpha}} \equiv (\psi_{\alpha})^{\dagger}$. Our conventions are defined in more detail in Appendix A, which also includes a discussion on spinors and some useful identities.

The following discussion is heavily based on the review [67] and also borrows elements from [89–91].

### 1.2.1 Motivation

The strongest motivation for investigating supersymmetric theories[12] comes from the Coleman-Mandula theorem [92] which singles out supersymmetry as the unique extension of Poincaré invariance in a quantum field theory (in more than two spacetime dimensions). It says (following the discussion in [93] and [89]):

> In a theory with non-trivial scattering in more than $1 + 1$ dimensions, the only possible conserved quantities that transform as tensors under the Lorentz group (i.e. without spinor indices) are the usual energy-momentum vector $P_{\mu}$, the generators of Lorentz transformations $J_{\mu\nu}$, as well as possible scalar "internal" symmetry charges $Z_i$ which commute with $P_{\mu}$ and $J_{\mu\nu}$.

The basic idea of this theorem is that the conserved charges $P_{\mu}$ and $J_{\mu\nu}$ leave only the scattering angle undetermined in e.g. a 2-body collision. If a theory would admit additional conservation laws, these would determine this scattering angle and leave only a discrete set of possibilities. Analyticity of the scattering amplitude then requires that it vanishes at all angles[13]. In other words, it is not possible for an interacting theory[14] to exhibit more conserved charges[15] with Lorentz indices, since this would be incompatible with non-trivial scattering.

---

[10]While the shortcut is definitely faster, it also skips all the physical motivation behind the structure of supersymmetry.

[11]The only significant difference is that we use the mostly-negative convention for the metric, see Appendix A.

[12]This is of course entirely subjective.

However, spinor charges can circumvent this theorem, as expressed by the Haag-Lopuszanski-Sohnius extension of the Coleman-Mandula theorem [94]. To see this, consider a fermionic charge $Q_\alpha$. Its anti-commutator with its conjugate $\{Q_\alpha, Q_{\dot\alpha}^\dagger\}$ is a spin 1 object. Since a conserved spin 1 charge exists, namely $P_\mu$, conservation of $Q_\alpha$ is allowed.

We can thus construct a theory with conserved fermionic charges $Q_\alpha$ and its conjugate $Q_{\dot\alpha}^\dagger$ defined by the following commutation relations

$$\{Q_\alpha, Q_{\dot\beta}^\dagger\} = 2i\sigma_{\alpha\dot\beta}^\mu \partial_\mu = 2\sigma_{\alpha\dot\beta}^\mu P_\mu, \tag{1.3}$$

$$\{Q_\alpha, Q_\beta\} = \{Q_{\dot\alpha}^\dagger, Q_{\dot\beta}^\dagger\} = 0, \tag{1.4}$$

$$[P^\mu, Q_\alpha] = [P^\mu, Q_{\dot\alpha}^\dagger] = 0, \tag{1.5}$$

$$[Q_\alpha, M_{\mu\nu}] = -\frac{1}{2}(\sigma_{\mu\nu})_\alpha{}^\beta Q_\beta, \tag{1.6}$$

$$\left[Q_{\dot\alpha}^\dagger, M_{\mu\nu}\right] = \frac{1}{2}(\overline\sigma_{\mu\nu})^{\dot\alpha}{}_{\dot\beta} Q_{\dot\beta}^\dagger. \tag{1.7}$$

Here, $P_\mu = i\partial_\mu$ is the usual generator of spacetime translations, $M_{\mu\nu}$ the generator of Lorentz transformations, and $Q_\alpha$ is, similarly, the generator of supersymmetry transformations[16]. In addition to these, the usual Poincaré algebra, given in Appendix A.2, still holds.

Single-particle states in a $3 + 1$ dimensional theory with Poincaré invariance must transform in unitary representations of the Poincaré algebra[17] (i.e. particles are scalars, tensors or spinors). Since supersymmetry is an extension of the Poincaré algebra, it relates the different particle representations to each other, forming supermultiplets. Since the supersymmetry generators are fermionic, it relates fermionic and bosonic states to each other, schematically

$$Q|\text{Boson}\rangle = |\text{Fermion}\rangle, \qquad Q|\text{Fermion}\rangle = |\text{Boson}\rangle. \tag{1.8}$$

The operator $P^\mu$ commutes with the $Q_\alpha$, $Q_{\dot\alpha}^\dagger$, so particles in same supermultiplet have same eigenvalues of $P^2$ and thus the same mass. While the structure of the algebra, as an extension of the usual Poincaré algebra, is strongly constrained, the supersymmetry generators commute with any internal symmetries, in particular gauge symmetries. Therefore, the particles related by supersymmetry fall in the same representation of gauge group.

---

[13]There are two assumptions in this reasoning. The first is the analyticity of the scattering amplitude (which is the reason that theories in $1 + 1$ dimension evade the theorem: there is only forward and backward scattering). The second is that the added conserved quantity is not too non-local (e.g. $p$-branes in string theory are extended objects and thus the theorem does not apply to them).

[14]In a free theory one can define an infinite number of charges.

[15]In a theory with only massless particles, the generators of conformal transformations are also allowed.

[16]In a way, one can then think of the supersymmetry generators as the square root of the spacetime transformations.

[17]Classified using Wigner's little groups [95].

In a theory with a single supersymmetry, two irreducible representations are the chiral and vector supermultiplets, each containing two fields[18]. The chiral supermultiplet contains a scalar and a two-component Weyl fermion, precisely the content needed to contain the Standard Model fermions and the Higgs. Consequently, the additional states introduced by supersymmetry are the scalar sfermions and the fermionic higgsinos. The vector multiplet contains a fermion and a vector and can thus be used to represent the Standard Model gauge fields. The additional fermions introduced by supersymmetry are the gauginos. In all cases, the number of degrees of freedom for the fermions and bosons are equal[19]. In renormalisable theories, all other possible representations can be formed from these two (e.g. a massles vector superfield eating a chiral superfield to become a massive vector superfield when gauge symmetry is broken).

It is well-known that supersymmetry *can* provide a solution to the hierarchy problem. Applied to the Standard Model, supersymmetry adds two complex scalar fields for each Dirac fermion with the same mass. Moreover, their couplings to the Higgs are related as $\lambda_S = |\lambda_f|^2$. Supersymmetry thus guarantees that quadratic divergences vanish to all orders in perturbation theory. An intuitive way to see that this must always be true is from the observation that supersymmetry relates the Higgs mass to the mass of its corresponding fermion. The latter has radiative corrections at most logarithmic in perturbation theory, so this must be true for the scalar as well.

Originally, supersymmetric theories were not designed to be applied to the Standard Model [96–101], so it is interesting that they can solve certain problems associated with it[20]. However, since we do not observe the supersymmetric partners, we know that supersymmetry must be broken at currently accessible energies, lifting up the superpartner masses. This reintroduces corrections to the Higgs squared-mass parameter. Although these are not quadratically divergent (see Section 1.2.5), if the superpartner masses are too high (i.e. above TeV scale[21]) these corrections would still increase $m_H^2$ too much for the model to be considered natural. Current constraints (see Section 1.5.2) seem to disfavour "conventional" models of supersymmetry which can solve the hierarchy problem[22].

---

[18]Since the supersymmetry charges anti-commute with themselves, one can only relate particles differing with $1/2$ in spin from one another. In theories with multiple supersymmetries, this is no longer the case. However, since then other issues pop up, we will restrict ourselves to the case of $\mathcal{N} = 1$ supersymmetry.

[19]At least, this is the case on-shell. Off-shell, i.e. before imposing the equations of motion, this is only the case after introducing auxiliary fields, see Section 1.2.2. This is also an example of why we restrict to $\mathcal{N} = 1$: in the case of $\mathcal{N} \geq 2$ SUSY, the issue of the off-shell degrees of freedom has not been completely solved yet.

[20]Also, the introduction of supersymmetry modifies the running of the gauge couplings with energy, such that the meet to better precision at the GUT scale [102–105].

[21]This estimate is somewhat dependent on the energy scale accessible by current colliders.

[22]At the UV scale, the argument for supersymmetry are still as strong as before. If one believes in Gell-Mann's totalitarian principle "Everything not forbidden is compulsory" (see footnote on page 859 of [106]), then supersymmetry must be there.

### 1.2.2 Superfields

It is possible to construct a supersymmetric theory term by term by imposing the correct supersymmetry transformations and following commutation relations on a theory with fermions with gauge interactions. However, such an approach is cumbersome, technical and not very illuminating. Instead, it is more intuitive to construct supersymmetric Lagrangians using the formalism of superspace, which is a geometrisation of supersymmetry. In the Poincaré algebra, the $P_\mu$ generate translations in spacetime and the $M_{\mu\nu}$ generate rotations and boosts. Since supersymmetry is an extension of this, we want a similar interpretation for the supersymmetry generators. By generalising spacetime to a space with Grassmann variables, the $Q_\alpha$ and $Q_{\dot\beta}^\dagger$ now also generate translations in those variables. In this section, we define superspace and find the possible superfields which can harbour the particle content of the Standard Model.

**Superspace**

Superspace is defined as an extension of ordinary spacetime, by appending the usual coordinates $x^\mu$ with the anti-commuting two-component spinors $\theta^\alpha$ and $\theta_{\dot\alpha}^\dagger$ (with mass dimension $-1/2$). Consider then a function of $x^\mu$, $\theta^\alpha$, and $\theta_{\dot\alpha}^\dagger$ in superspace, a superfield [107, 108]. Since the $\theta^\alpha$ and $\theta_{\dot\alpha}^\dagger$ are anti-commuting and each of these coordinates has two independent components, the expansion of any superfield in the fermionic coordinates terminates at the $\theta\theta\theta^\dagger\theta^\dagger$-component. The general superfield[23] expansion is then given by

$$S(x, \theta, \theta^\dagger) = a + \theta\xi + \theta^\dagger\chi^\dagger + \theta\theta b + \theta^\dagger\theta^\dagger c + \theta^\dagger\overline{\sigma}^\mu\theta v_\mu + \theta^\dagger\theta^\dagger\theta\eta + \theta\theta\theta^\dagger\zeta^\dagger + \theta\theta\theta^\dagger\theta^\dagger d. \quad (1.9)$$

This superfield contains both bosonic and fermionic fields. In this way, a superfield forms an object which groups together the fermionic and bosonic fields which are related to each other. This general superfield forms a reducible representation of supersymmetry. Restricting its form, we can construct chiral and vector superfields. Using superfields, one can make the invariance under supersymmetry transformations manifest.

Next, we need to define integration over superspace, in particular over the coordinates $\theta^\alpha$ and $\theta_{\dot\alpha}^\dagger$. Integration of such anti-commuting variables is defined in analogy with Grassman variables $\eta$ where

$$\int d\eta = 0, \qquad \int d\eta\,\eta = 1. \quad (1.10)$$

Defining the measure as

$$\int d^2\theta = -\frac{1}{4}\,d\theta^\alpha\,d\theta^\beta\epsilon_{\alpha\beta}, \qquad \int d^2\theta^\dagger = -\frac{1}{4}\,d\theta_{\dot\alpha}^\dagger\,d\theta_{\dot\beta}^\dagger\epsilon^{\dot\alpha\dot\beta}, \quad (1.11)$$

we find

$$\int d^2\theta\,\theta\theta = 1, \qquad \int d^2\theta^\dagger\theta^\dagger\theta^\dagger = 1, \quad (1.12)$$

---

[23]Note that the superfield can also carry additional indices (see e.g. Eq. (1.54)).

while integration with a different number of $\theta$ gives zero[24]. A useful relation for working with superfields and their integrals is[25]

$$(\theta\xi)(\theta\chi) = -\frac{1}{2}(\theta\theta)(\xi\chi),\tag{1.13}$$

since $\theta_\alpha\theta_\beta = \frac{1}{2}\epsilon_{\alpha\beta}\theta\theta$ and $\epsilon_{\alpha\beta}\epsilon^{\beta\gamma} = \epsilon^{\gamma\beta}\epsilon_{\beta\alpha} = \delta_\alpha^\gamma$. An analogous equation holds for dotted quantities. That and other relations are also given in Appendix A.6.

Infinitesimal translations in superspace are equivalent to global supersymmetry transformations. To see this, define the differential operators

$$\hat{Q}_\alpha = i\frac{\partial}{\partial\theta^\alpha} - (\sigma^\mu\theta^\dagger)_\alpha\partial_\mu, \qquad \hat{Q}^\alpha = -i\frac{\partial}{\partial\theta_\alpha} + (\theta^\dagger\overline{\sigma}^\mu)^\alpha\partial_\mu,\tag{1.14}$$

$$\hat{Q}^{\dagger\dot{\alpha}} = i\frac{\partial}{\partial\theta^\dagger_{\dot{\alpha}}} - (\overline{\sigma}^\mu\theta)^{\dot{\alpha}}\partial_\mu, \qquad \hat{Q}^\dagger_{\dot{\alpha}} = -i\frac{\partial}{\partial\theta^{\dagger\dot{\alpha}}} + (\theta\sigma^\mu)_{\dot{\alpha}}\partial_\mu,\tag{1.15}$$

which satisfy the same anti-commutation relations as Eqs. (1.3)-(1.7). The variation of a superfield $S$ under a supersymmetry transformation is then given by[26]

$$\sqrt{2}\delta_\epsilon S = -i(\epsilon\hat{Q} + \epsilon^\dagger\hat{Q}^\dagger)S = S(x^\mu + i\epsilon\sigma^\mu\theta^\dagger + i\epsilon^\dagger\overline{\sigma}^\mu\theta, \theta + \epsilon, \theta^\dagger + \epsilon^\dagger) - S(x^\mu, \theta, \theta^\dagger),\tag{1.16}$$

which is indeed equivalent to a translation in superspace. Writing out the superfield in components, we can find the supersymmetry transformations on the component fields. However, since we are only interested in the supersymmetry transformations of irreducible representations, we do not give them here. Instead, they can be found in Appendix C.2.1, Eqs. (C.4)-(C.12). The main point is that these supersymmetry transformations automatically give the correct relation between fermions and bosons in a supersymmetric theory.

Finally, the relation between the action of the differential operators $\hat{Q}_\alpha$ and $\hat{Q}^\dagger_{\dot{\alpha}}$ on functions in superspace and the corresponding action of the quantum mechanical operators $Q_\alpha$ and $Q^\dagger_{\dot{\alpha}}$ acting on Hilbert space is given by

$$\left[X, \epsilon Q + \epsilon^\dagger Q^\dagger\right] = (\epsilon\hat{Q} + \epsilon^\dagger\hat{Q}^\dagger)X,\tag{1.17}$$

$$\left[X, P_\mu\right] = \hat{P}_\mu X.\tag{1.18}$$

---

[24]As a consequence of this, the integral of a total derivative w.r.t. the fermionic coordinates vanishes

$$\int \mathrm{d}^2\theta \frac{\partial}{\partial\theta^\alpha}(\text{anything}) = 0, \qquad \int \mathrm{d}^2\theta^\dagger \frac{\partial}{\partial\theta^\dagger_{\dot{\alpha}}}(\text{anything}) = 0,$$

because any objects before derivation has at most a $\theta\theta$ or $\theta^\dagger\theta^\dagger$ component.

[25]The product $\theta\theta$ is not zero, since $\theta^\alpha$ do anti-commute but the $\epsilon_{\alpha\beta}$ which appears in the product is anti-symmetric in its indices.

[26]The factor $\sqrt{2}$ is a convention which differs by author, here we follow the choice of [67].

**Chiral superfields**

We first describe chiral superfields, which will describe fermions and their superpartner scalars. In order to obtain chiral superfields, we must first define chiral covariant derivatives. They are given by

$$D_\alpha = \frac{\partial}{\partial \theta^\alpha} - i(\sigma^\mu \theta^\dagger)_\alpha \partial_\mu, \qquad D^\alpha = -\frac{\partial}{\partial \theta_\alpha} + i(\theta^\dagger \overline{\sigma}^\mu)^\alpha \partial_\mu, \qquad (1.19)$$

$$\overline{D}^{\dot\alpha} = \frac{\partial}{\partial \theta^\dagger_{\dot\alpha}} - i(\overline{\sigma}^\mu \theta)^{\dot\alpha} \partial_\mu, \qquad \overline{D}_{\dot\alpha} = -\frac{\partial}{\partial \theta^{\dagger\dot\alpha}} + i(\theta \sigma^\mu)_{\dot\alpha} \partial_\mu. \qquad (1.20)$$

By construction, they anti-commute with the generators of supersymmetry transformations

$$\left\{ \hat{Q}_\alpha, D_\beta \right\} = \left\{ \hat{Q}^\dagger_{\dot\alpha}, D_\beta \right\} = \left\{ \hat{Q}_\alpha, \overline{D}_{\dot\beta} \right\} = \left\{ \hat{Q}^\dagger_{\dot\alpha}, \overline{D}_{\dot\beta} \right\} = 0. \qquad (1.21)$$

Therefore, acting with a chiral covariant derivative on a superfield, one obtains again a superfield. These operators are similar to the supersymmetry generators, but have an opposite relative sign (and an additional $-i$). Therefore, they satisfy again[27] Eqs. (1.3)-(1.7), but do not represent a second supersymmetry. One can check that the application of three $D_\alpha$ or $\overline{D}_{\dot\alpha}$ on anything gives zero.

A chiral superfield can then be obtained from a general superfield by imposing the condition

$$\overline{D}_{\dot\alpha} \Phi = 0. \qquad (1.22)$$

Similarly, an anti-chiral superfield is obtained by imposing

$$D_\alpha \Phi^* = 0. \qquad (1.23)$$

These conditions can be solved most easily by performing a change of variables from $x^\mu$ to $y^\mu$, defined as

$$y^\mu \equiv x^\mu - i\theta\sigma^\mu\theta^\dagger, \qquad (1.24)$$

while leaving $\theta_\alpha, \theta^\dagger_{\dot\alpha}$ unchanged. In these coordinates, the chiral covariant derivatives become

$$D_\alpha = \frac{\partial}{\partial \theta^\alpha} - 2i(\sigma^\mu \theta^\dagger)_\alpha \frac{\partial}{\partial y^\mu}, \qquad D^\alpha = -\frac{\partial}{\partial \theta_\alpha} + 2i(\theta^\dagger \overline{\sigma}^\mu)^\alpha \frac{\partial}{\partial y^\mu}, \qquad (1.25)$$

$$\overline{D}^{\dot\alpha} = \frac{\partial}{\partial \theta^\dagger_{\dot\alpha}}, \qquad \overline{D}_{\dot\alpha} = -\frac{\partial}{\partial \theta^{\dagger\dot\alpha}}. \qquad (1.26)$$

In other words, the constraint can be solved with a superfield that only depends on $y^\mu$ and $\theta_\alpha$, but not $\theta^\dagger_{\dot\alpha}$. The chiral superfield then becomes (with an analogous reasoning for the anti-chiral superfield using $y^{\mu*} = x^\mu + i\theta\sigma^\mu\theta^\dagger$)

$$\Phi = \phi(y) + \sqrt{2}\theta\psi(y) + \theta\theta F(y), \qquad (1.27)$$

$$\Phi^* = \phi^*(y^*) + \sqrt{2}\theta^\dagger\psi^\dagger(y^*) + \theta^\dagger\theta^\dagger F^*(y^*). \qquad (1.28)$$

---

[27]Sometime with an additional sign difference, depending on the convention

The full expansion in terms of the coordinates $x^\mu, \theta, \theta^\dagger$ can be found in Appendix C.2.2, Eqs. (C.13)-(C.14). However, for most calculations, the expansion in terms of $y^\mu$ is sufficient (i.e. to compute $F$-terms, see the following).

The chiral superfield contains a complex scalar $\phi$, a two-component left-handed Weyl fermion $\psi_\alpha$ and an auxiliary complex scalar field $F$. These fields have the canonical mass dimensions components if the chiral superfield has dimension [mass]$^1$. The supersymmetry transformations of these fields can be obtained from those of the general superfield (Eqs. (C.4)-(C.12)) or by applying the generator of supersymmetry transformations on the chiral superfield in terms of $x^\mu$. They are given by[28]

$$\delta\phi_i = \epsilon\psi_i, \qquad\qquad \delta\phi^{*i} = \epsilon^\dagger\psi^{\dagger i}, \tag{1.29}$$

$$\delta(\psi_i)_\alpha = -i(\sigma^\mu\epsilon^\dagger)_\alpha\partial_\mu\phi_i + \epsilon_\alpha F_i, \qquad \delta(\psi^{\dagger i})_{\dot\alpha} = i(\epsilon\sigma^\mu)_{\dot\alpha}\partial_\mu\phi^{*i} + \epsilon^\dagger_{\dot\alpha}F^{*i}, \tag{1.30}$$

$$\delta F_i = -i\epsilon^\dagger\overline{\sigma}^\mu\partial_\mu\psi_i, \qquad\qquad \delta F^{*i} = i\partial_\mu\psi^{\dagger i}\overline{\sigma}^\mu\epsilon. \tag{1.31}$$

The auxiliary field $F$ (with dimension [mass]$^2$) which appears as the highest component is necessary to make the supersymmetry algebra close off-shell. This can be seen by calculating $(\delta_{\epsilon_2}\delta_{\epsilon_1} - \delta_{\epsilon_1}\delta_{\epsilon_2})X$, which only vanishes on-shell (i.e. by using the equations of motion) if the field $F$ is not included. Including $F$ fixes this, since its variation is proportional to the equation of motion of the fermion. Alternatively, one can see that $F$ is needed to balance the degrees of freedom between the scalars and fermion off-shell, since an off-shell Weyl fermion has four degrees of freedom (reduced to two on-shell), while a single complex scalar only has two.

Finally, due to the way chiral superfields are defined, one can build a new chiral superfield from an arbitrary function $W(\Phi_i)$ which is holomorphic in the chiral superfields treated as complex variables.

**Real or vector superfield**

A vector or real superfield, which are the supersymmetric extension of gauge bosons, can be obtained from the general superfield by imposing the condition[29] $V = V^*$ and redefining

$$\eta_\alpha = \lambda_\alpha - \frac{i}{2}(\sigma^\mu\partial_\mu\xi^\dagger)_\alpha, \qquad v_\mu = A_\mu, \qquad d = \frac{1}{2}D - \frac{1}{4}\partial_\mu\partial^\mu a. \tag{1.32}$$

The vector superfield then has components

$$V(x, \theta, \theta^\dagger) = a + \theta\xi + \theta^\dagger\xi^\dagger + \theta\theta b + \theta^\dagger\theta^\dagger b^* + \theta\sigma^\mu\theta^\dagger A_\mu + \theta^\dagger\theta^\dagger\theta(\lambda - \frac{i}{2}\sigma^\mu\partial_\mu\xi^\dagger)$$
$$+ \theta\theta\theta^\dagger(\lambda^\dagger - \frac{i}{2}\overline{\sigma}^\mu\partial_\mu\xi) + \theta\theta\theta^\dagger\theta^\dagger(\frac{1}{2}D - \frac{1}{4}\partial_\mu\partial^\mu a). \tag{1.33}$$

---

[28]In short, each $\hat{Q}_\alpha$ and $\hat{Q}^\dagger_{\dot\alpha}$ has a part which removes a $\theta^{(\dagger)}$ and a part which adds a $\theta^{(\dagger)}$ and a derivative, shifting the components around.

[29]A non-trivial superfield can then not be both chiral and real. On the other hand, it is possible to obtain a real superfield from a chiral one with the combinations $\Phi + \Phi^*$, $i(\Phi - \Phi^*)$ and $\Phi\Phi^*$, the latter of which we will encounter in the next section.

As can be seen, it contains a gauge field $A_\mu^a$ and a two-component Weyl fermion $\lambda^a$. In order for these components to have the canonical mass dimensions, $V$ needs to be dimensionless. Just as with the chiral superfield, an auxiliary field $D$ of mass dimension 2 appears as the highest component[30]. As mentioned before, all components need to be in the same representation of the gauge group, in this case the adjoint representation. The supersymmetry transformations of the vector superfield are given in Appendix C.2.3, Eqs. (C.15)-(C.20).

In addition to the physical fields $A^\mu$, $\lambda$ and $D$, a real superfield contains also the fields $a$, $\xi_\alpha$ and $b$. However, these fields can be "supergauged" away. The proper extension of ordinary gauge transformation can be found as follows. Consider a $U(1)$ gauge symmetry (non-Abelian symmetries will be mentioned below). The supergauge transformation is defined as

$$V \to V + i(\Omega^* - \Omega), \tag{1.34}$$

with $\Omega$ a chiral superfield gauge parameter. The supergauge transformation of the components is given in Eqs. (C.21)-(C.26). Here, we are only interested in the vector field transformation, which is

$$A_\mu \to A_\mu - \partial_\mu(\phi + \phi^*). \tag{1.35}$$

This is indeed the correct gauge transformation of a gauge field, so that the supergauge transformation is the proper extension of gauge transformations in superspace. From the transformations in Eqs. (C.21)-(C.26), we see that the additional fields $a$, $\xi_\alpha$ and $b$ can be redefined arbitrarily, and thus one can gauge them away (as anticipated), with an appropriate choice of the fields $\text{Im}(\phi)$, $\psi_\alpha$ and $F$ of the supergauge parameter $\Omega$. Removing these fields brings the vector superfield in the Wess-Zumino gauge [109]

$$V_{\text{WZgauge}} = \theta\sigma^\mu\theta^\dagger A_\mu + \theta^\dagger\theta^\dagger\theta\lambda + \theta\theta\theta^\dagger\lambda^\dagger + \frac{1}{2}\theta\theta\theta^\dagger\theta^\dagger D. \tag{1.36}$$

Supersymmetry transformations do not leave $V$ in the Wess-Zumino gauge, but it is always possible to bring the new field back in Wess-Zumino gauge using supergauge transformations. Appropriate supersymmetry transformations in Wess-Zumino gauge then become non-linear and can be found in Eqs. (C.27)-(C.29).

For the non-Abelian case, we impose the transformation (with foresight of what is needed to build a gauge-invariant kinetic term for chiral superfields charged under a non-Abelian gauge symmetry, see Section 1.2.3)

$$e^V \to e^{i\Omega^\dagger} e^V e^{-i\Omega}. \tag{1.37}$$

with

$$V_i{}^j = 2g_a T_i^{aj} V^a, \tag{1.38}$$

$$\Omega_i{}^j = 2g_a T_i^{aj} \Omega^a, \tag{1.39}$$

---

[30] In this case, the counting of number of degrees of freedom goes as follows. On-shell, both the vector boson and the fermion have two helicity states. Off-shell, $\lambda_\alpha^a$ contains two complex components, while $A_\mu^a$ has three components (one is removed by gauge transformations). The difference is compensated by the auxiliary real bosonic field $D^a$

so that $V^a$ and $\Omega^a$ are in the adjoint representation, while $V$ and $\Omega$ are in a representation $R$. The gauge transformation equation above can then be expanded (see [67] for details), giving

$$V^a \to V^a + i(\Omega^{a*} - \Omega^a) + g_a f^{abc} V^b(\Omega^{c*} + \Omega^c) - \frac{i}{3} g_a^2 f^{abc} f^{cde} V^b V^d(\Omega^{e*} - \Omega^c) + \dots \tag{1.40}$$

which is indeed a generalisation of an Abelian gauge transformation. Once again, as a special case if $\Omega^{a*} = \Omega^a$, this supergauge transformation includes ordinary gauge transformations. The second term can be chosen to bring the gauge field in the Wess-Zumino gauge (which still leaves the possibility to perform ordinary gauge transformations), so that we have again

$$(V^a)_{\text{WZgauge}} = \theta\sigma^\mu\theta^\dagger A_\mu^a + \theta^\dagger\theta^\dagger\theta\lambda^a + \theta\theta\theta^\dagger\lambda^{\dagger a} + \frac{1}{2}\theta\theta\theta^\dagger\theta^\dagger D^a, \tag{1.41}$$

where each field now carries an index $a$.

### 1.2.3 A supersymmetric Lagrangian

In this section, we review the method to build actions or Lagrangians which are invariant under supersymmetry transformations.

#### Building the action

We can make use of the properties of superfields found in the previous sections to build supersymmetry-invariant Lagrangians. The first way is to look at the highest component of a superfield, whose variation (Eq. (C.12)) is a total derivative. Alternatively, since $\hat{Q}_\alpha$ and $\hat{Q}_{\dot\alpha}^\dagger$ only contain total derivatives in $x^\mu$ and $\theta, \theta^\dagger$, an integral over full superspace automatically gives zero. From both points of view, a good candidate for the action $\mathcal{S}$ of a supersymmetric theory is

$$\delta_\epsilon\mathcal{S} = 0, \quad \text{for } \mathcal{S} = \int \mathrm{d}^4x \int \mathrm{d}^2\theta\, \mathrm{d}^2\theta^\dagger\, S(x, \theta, \theta^\dagger). \tag{1.42}$$

Since the action needs to be real, $S$ should be a real superfield.

Performing only the integration over the fermionic coordinates, we obtain the Lagrangian $\mathcal{L}(x)$. Since this effectively selects the $\theta\theta\theta^\dagger\theta^\dagger$-component, the Lagrangian is a $D$-term

$$[V]_D \equiv \int \mathrm{d}^2\theta\, \mathrm{d}^2\theta^\dagger\, V(x, \theta, \theta^\dagger) = V(x, \theta, \theta^\dagger)\Big|_{\theta\theta\theta^\dagger\theta^\dagger} = \frac{1}{2}D - \frac{1}{4}\partial_\mu\partial^\mu a \tag{1.43}$$

Another way to construct a Lagrangian which is invariant under supersymmetry variations (up to a total derivative), is by selecting the $F$-term of a chiral superfield[31] $\Phi$,

---

[31]One could also try using the $D$-term of chiral superfield to build a Lagrangian, but this is a total derivative (see Eqs. (C.13)-(C.14)), so this is a futile attempt.

which also transforms into a total derivative

$$[\Phi]_F \equiv \Phi|_{\theta\theta} = \int \mathrm{d}^2\theta \; \Phi|_{\theta^\dagger=0} = \int \mathrm{d}^2\theta \, \mathrm{d}^2\theta^\dagger \, \delta^{(2)}(\theta^\dagger)\Phi = F. \tag{1.44}$$

In order to construct an action which is real, we need to add its complex conjugate. Since selecting the $\theta\theta$-component of a chiral superfield means dropping the parts containing $\theta^\dagger_{\dot\alpha}$, we can use the expansion for chiral superfields in terms of $y^\mu$ given in Eq. (1.27)-(1.28).

Therefore, one can build supersymmetry invariant Lagrangians using real and chiral superfields and combinations of them and selecting their $F$- or $D$-terms, as long as these are compatible with the gauge symmetries.

**Chiral superfield Lagrangian**

We construct now a Lagrangian for chiral superfields including only renormalisable interactions. The composite superfield $\Phi^{*i}\Phi_j$ is real. We can thus use its $D$-term[32] to construct a Lagrangian invariant under supersymmetry[33]

$$\mathcal{L}_{\text{free}} = \left[\Phi^{*i}\Phi_i\right]_D = \int \mathrm{d}^2\theta \, \mathrm{d}^2\theta^\dagger \Phi^{*i}\Phi_i = \partial^\mu\phi^{*i}\partial_\mu\phi_i + i\psi^{\dagger i}\overline{\sigma}^\mu\partial_\mu\psi_i + F^{*i}F_i + \dots, \tag{1.45}$$

where the ellipsis indicates a total derivative part. This is the massless free-field Lagrangian for chiral superfields.

In order to add mass terms and interactions, we consider the superpotential $W(\Phi_i)$, a holomorphic function of chiral superfields treated as complex variables, which is again a chiral superfield. Selecting its $F$-term[34], we find

$$[W(\Phi_i)]_F = W^i F_i - \frac{1}{2}W^{ij}\psi_i\psi_j, \tag{1.46}$$

where

$$W^i = \frac{\delta W}{\delta \Phi_i}\bigg|_{\Phi_i\to\phi_i}, \qquad W^{ij} = \frac{\delta^2 W}{\delta\Phi_i\delta\Phi_j}\bigg|_{\Phi_i\to\phi_i}. \tag{1.47}$$

The total Lagrangian for chiral superfields is then

$$\mathcal{L}_{\text{chiral}}(x) = \left[\Phi^{*i}\Phi_i\right]_D + \left([W(\Phi_i)]_F + \text{c.c.}\right). \tag{1.48}$$

The auxiliary field appears in the Lagrangian only as $F_iF^{*i} + W^iF_i + W_i^*F^{*i}$, i.e. at most quadratically and without derivatives, so there is no dynamics associated with this field. Their equations of motion

$$F_i = -W_i^*, \qquad F^{*i} = -W^i, \tag{1.49}$$

---

[32] The full expansion for the composite superfield $\Phi^{*i}\Phi_j$ is given in Appendix C.3.1 Eq. (C.30).

[33] This comes down to selecting all combinations with $\theta\theta\theta^\dagger\theta^\dagger$, for which we need the full expansion of $\Phi_i$ in $x^\mu$.

[34] We need to select combinations with $\theta\theta$, so we can use the $\Phi$-expansion in terms of $y^\mu$. Terms with this combination can come from selecting the $\psi_i$-component of two $\Phi_i$ or from the $F_i$-component of one $\Phi_i$, together with a function of scalar fields.

are algebraic and can thus be substituted back in the Lagrangian[35]. Performing this substitution, we find

$$\mathcal{L} = \partial^\mu \phi^{*i} \partial_\mu \phi_i + i\psi^{\dagger i} \overline{\sigma}^\mu \partial_\mu \psi_i - \frac{1}{2} \left( W^{ij} \psi_i \psi_j + W^*_{ij} \psi^{\dagger i} \psi^{\dagger j} \right) - W^i W^*_i. \tag{1.50}$$

The most general form of the superpotential, including only renormalisable terms (i.e. with mass dimension smaller than four), is[36,37]

$$W(\Phi_i) = L^i \Phi_i + \frac{1}{2} M^{ij} \Phi_i \Phi_j + \frac{1}{6} y^{ijk} \Phi_i \Phi_j \Phi_k, \tag{1.51}$$

which includes mass terms and yukawa interactions and is symmetric under the exchange of $i, j, k$. In practice, only the term compatible with gauge interactions appear[38,39]. Finally, we can plug in this superpotential in the Lagrangian, resulting in

$$\begin{aligned} \mathcal{L} = {} & \partial^\mu \phi^{*i} \partial_\mu \phi_i - V(\phi, \phi^*) + i\psi^{\dagger i} \overline{\sigma}^\mu \partial_\mu \psi_i - \frac{1}{2} M^{ij} \psi_i \psi_j - \frac{1}{2} M^*_{ij} \psi^{\dagger i} \psi^{\dagger j} \\ & - \frac{1}{2} y^{ijk} \phi_i \phi_j \psi_k - \frac{1}{2} y^*_{ijk} \phi^{*i} \psi^{\dagger j} \psi^{\dagger k}, \end{aligned} \tag{1.52}$$

with

$$\begin{aligned} V(\phi, \phi^*) = {} & W^k W^*_k = F^{*k} F_k \\ = {} & M^*_{ik} M^{kj} \phi^{*i} \phi_j + \frac{1}{2} M^{in} y^*_{jkn} \phi_i \phi^{*j} \phi^{*k} + \frac{1}{2} M^*_{in} y^{jkn} \phi^{*i} \phi_j \phi_k \\ & + \frac{1}{4} y^{ijn} y^*_{kln} \phi_i \phi_j \phi^{*k} \phi^{*l}. \end{aligned} \tag{1.53}$$

This Lagrangian for chiral superfields with only renormalisable interactions is called the Wess-Zumino model [100]. Also, this makes explicit the original claim that quartic scalar coupling (e.g. between two Higgs and two sfermions) is equal to the square of the Yukawa coupling between a scalar (e.g. the Higgs) and two fermions.

### Real superfield Lagrangian

Next, we build the Lagrangian for real superfields, starting with the case of Abelian symmetries. Analogous to ordinary Abelian gauge theories, we define a gauge-invariant field-strength superfield

$$\mathcal{W}_\alpha = -\frac{1}{4} \overline{DD} D_\alpha V, \qquad \mathcal{W}^\dagger_{\dot{\alpha}} = -\frac{1}{4} DD \overline{D}_{\dot{\alpha}} V, \tag{1.54}$$

---

[35]Since $F_i$ appears at most quadratically, this is also true at the quantum mechanical level and equivalent to solving the path integral.

[36]For completeness, the full expansion of the composite superfields appearing in $W(\Phi_i)$ is given in Eqs. (C.31)(C.32), although in practice selecting only the $F$-term is easier directly.

[37]An interesting property of the superpotential is its non-renormalisation: in a full quantum theory, its components only undergo wave-function renormalisation. Therefore, the masses and couplings appearing here are protected from quadratic divergences. This property is due to the structure of supersymmetry, in particular the fact that only holomorphic terms can appear.

[38]In the MSSM, only one mass term passes this criterion.

[39]The first term is only allowed in the presence of gauge singlets and affects only the scalar potential.

which are chiral and anti-chiral superfields (with a spinor index) by construction. They possess mass dimension [mass]$^{3/2}$. These superfields are indeed gauge invariant, which can be seen by plugging in the Abelian supergauge transformation Eq. (1.34), using the anti-commutation relations of the covariant derivatives and the defining property of (anti-)chiral superfields Eqs. (1.22) and (1.23) and definition of (anti-)chiral superfields.

We can find the expansion of $\mathcal{W}_\alpha$ and its conjugate by substituting $V$ in Wess-Zumino gauge. Expressed in coordinates $y^\mu, \theta, \theta^\dagger$ (the expansion of $V$ for which is given in Appendix C.3.2 Eq. (C.35)), we have

$$\mathcal{W}_\alpha(y, \theta, \theta^\dagger) = \lambda_\alpha + \theta_\alpha D - \frac{i}{2}(\sigma^\mu \overline{\sigma}^\nu \theta)_\alpha F_{\mu\nu} + i\theta\theta(\sigma^\mu \partial_\mu \lambda^\dagger)_\alpha, \quad (1.55)$$

$$\mathcal{W}^{\dagger\dot{\alpha}}(y^*, \theta, \theta^\dagger) = \lambda^{\dagger\dot{\alpha}} + \theta^{\dagger\dot{\alpha}} D + \frac{i}{2}(\overline{\sigma}^\mu \sigma^\nu \theta^\dagger)^{\dot{\alpha}} F_{\mu\nu} + i\theta^\dagger\theta^\dagger(\overline{\sigma}^\mu \partial_\mu \lambda)^{\dot{\alpha}}. \quad (1.56)$$

Because this object is gauge invariant, this expansion is valid for any gauge.

Since the field strength is a chiral superfield with a spinor index, we can build a gauge-invariant Lagrangian[40] from the following $F$-term

$$[\mathcal{W}^\alpha \mathcal{W}_\alpha]_F = D^2 + 2i\lambda\sigma^\mu\partial_\mu\lambda^\dagger - \frac{1}{2}F^{\mu\nu}F_{\mu\nu} + \frac{i}{4}\epsilon^{\mu\nu\rho\sigma}F_{\mu\nu}F_{\rho\sigma}, \quad (1.57)$$

all function of $x^\mu$. The action is then (eliminating a total derivative)

$$\int \mathrm{d}^4x\, \mathcal{L} = \int \mathrm{d}^4x\, \frac{1}{4}\left[\mathcal{W}^\alpha \mathcal{W}_\alpha\right]_F + \text{c.c.} = \int \mathrm{d}^4x\, \left[\frac{1}{2}D^2 + i\lambda^\dagger\overline{\sigma}^\mu\partial_\mu\lambda - \frac{1}{4}F^{\mu\nu}F_{\mu\nu}\right], \quad (1.58)$$

where we recognise the kinetic term for the gauge boson and for the gaugino.

In the case of non-Abelian superfields, we define instead the field-strength chiral superfield

$$\mathcal{W}_\alpha = -\frac{1}{4}\overline{DD}\left(e^{-V}D_\alpha e^V\right) \quad (1.59)$$

which transforms as (from Eq. (1.37))

$$\mathcal{W}_\alpha \to e^{i\Omega}\mathcal{W}_\alpha e^{-i\Omega}, \quad (1.60)$$

reminiscent of the field strength in Non-Abelian Yang-Mills theory, which transforms as $F_{\mu\nu} \to F'_{\mu\nu} = UF_{\mu\nu}U^{-1}$. Inspired by the kinetic term for the Abelian field-strength superfield and for the ordinary non-Abelian field strenth, we can build the supergauge-invariant chiral superfield $\text{Tr}\left[\mathcal{W}^\alpha \mathcal{W}_\alpha\right]$, of which the $F$-term will be a supersymmetry-invariant Lagrangian.

To obtain the expression for the non-Abelian field-strength superfield in components, we expand the factor appearing in brackets in Eq. (1.59) (for details, see [67])

$$e^{-V}D_\alpha e^V = D_\alpha V - \frac{1}{2}[V, D_\alpha V] + \frac{1}{6}[V, [V, D_\alpha V]] + \dots. \quad (1.61)$$

---

[40] We ignore here the possibility of a Fayet-Iliopoulos term [110] $\mathcal{L}_{\text{FI}} = -2\kappa\,[V]_D = -\kappa D$, allowed for an Abelian gauge symmetry since the $D$-term does not transform under a gauge transformation (Eq. (C.29)). This term can play a role in supersymmetry breaking, but will not be further discussed.

and write

$$\mathcal{W}_\alpha = 2g_a T^a \mathcal{W}_\alpha^a, \tag{1.62}$$

so that $\mathcal{W}_\alpha^a$ is in the adjoint representation, while $\mathcal{W}_\alpha$ is in a representation $R$. We find

$$\mathcal{W}_\alpha^a = -\frac{1}{4}\overline{DD}\left(D_\alpha V^a - ig_a f^{abc} V^b D_\alpha V^c + \dots\right). \tag{1.63}$$

Finally, in Wess-Zumino gauge, this is

$$(\mathcal{W}_\alpha^a)_{\text{WZ gauge}} = \lambda_\alpha^a + \theta_\alpha D^a - \frac{i}{2}(\sigma^\mu\overline{\sigma}^\nu\theta)_\alpha F_{\mu\nu}^a + i\theta\theta(\sigma^\mu D_\mu\lambda^{\dagger a})_\alpha. \tag{1.64}$$

with the usual non-Abelian field strength

$$F_{\mu\nu}^a = \partial_\mu A_\nu^a - \partial_\nu A_\mu^a - gf^{abc}A_\mu^b A_\nu^c \tag{1.65}$$

and the covariant derivative of the gaugino given by

$$D_\mu\lambda^a = \partial_\mu\lambda^a - gf^{abc}A_\mu^b\lambda^c \tag{1.66}$$

We then obtain the kinetic term and self-interactions from (using $\text{Tr}[T^a T^b] = k_a\delta_{ab}$, where $k_a = 1/2$ for the defining representation by convention)

$$\frac{1}{4k_a g_a^2}\text{Tr}\left[\mathcal{W}^\alpha\mathcal{W}_\alpha\right]_F = [\mathsf{W}^{a\alpha}\mathcal{W}_\alpha^a]_F, \tag{1.67}$$

which is invariant under supergauge and supersymmetry transformations. Evaluated in Wess-Zumino gauge, the $F$-term is

$$[\mathcal{W}^{a\alpha}\mathcal{W}_\alpha^a]_F = D^a D^a + 2i\lambda^a\sigma^\mu D_\mu\lambda^{\dagger a} - \frac{1}{2}F^{a\mu\nu}F_{\mu\nu}^a + \frac{i}{4}\epsilon^{\mu\nu\rho\sigma}F_{\mu\nu}^a F_{\rho\sigma}^a, \tag{1.68}$$

which is valid in any gauge. Adding the complex conjugate of this term, we obtain the final Lagrangian for the real superfield

$$\mathcal{L}_{\text{gauge}} = \frac{1}{4}\left[\mathcal{W}^{a\alpha}\mathcal{W}_\alpha^a\right]_F + \text{c.c.} = -\frac{1}{4}F_{\mu\nu}^a F^{\mu\nu a} + i\lambda^{\dagger a}\overline{\sigma}^\mu D_\mu\lambda^a + \frac{1}{2}D^a D^a, \tag{1.69}$$

**Adding gauge interactions to chiral superfields**

In this section, we add gauge interactions to the chiral superfields. A general gauge symmetry with generators $T^a$ on the chiral superfields $\Phi_i$ in a representation $R$ is realised by

$$\Phi_i \rightarrow (e^{i\Omega})_i^j\Phi_j,$$
$$\Phi^{*i} \rightarrow \Phi^{*j}(e^{-\Omega^\dagger})_j^i, \tag{1.70}$$

where $\Omega_i^j = 2ig_a\Omega^a T_i^{aj}$, with the chiral superfields $\Omega^a$ acting again as the supergauge transformation parameters. Each symmetry (or more precisely, each Lie algebra generator) is associated to a vector superfield $V^a$. As before, we define $V_i^j = 2g_a T_i^{aj} V^a$. We can then extend the kinetic term for chiral superfields in Eq. (1.45) to include gauge transformations as

$$\mathcal{L} = \left[ \Phi^{*i}(e^V)_i^j \Phi_j \right]_D,\tag{1.71}$$

which is indeed gauge invariant, given the supergauge transformations for the real superfield $V$ which we defined in Eq. (1.34) for Abelian gauge fields and Eq. (1.37) for non-Abelian gauge fields.

In the Wess-Zumino gauge, we can easily find the components of the Lagrangian in Eq. (1.71), since the power series from the exponential terminates

$$V^2 = -\frac{1}{2}\theta\theta\theta^\dagger\theta^\dagger A_\mu^a A^{a\mu},\tag{1.72}$$

$$V^n = 0 \quad (n \geq 3).\tag{1.73}$$

Therefore,

$$e^{2g_a T^a V^a} = 1 + 2g_a T^a \left( \theta\sigma^\mu\theta^\dagger A_\mu^a + \theta^\dagger\theta^\dagger\theta\lambda^a + \theta\theta\theta^\dagger\lambda^{\dagger a} + \frac{1}{2}\theta\theta\theta^\dagger\theta^\dagger D^a \right)$$
$$+ g_a^2 T^a T^a \theta\theta\theta^\dagger\theta^\dagger A_\mu^a A^{a\mu}.\tag{1.74}$$

The kinetic term for chiral superfields charged under a (non-)Abelian gauge symmetry is then

$$\left[ \Phi^{*i}(e^V)_i^j \Phi_j \right]_D = F^{*i}F_i + D_\mu\phi^{*i}D^\mu\phi_i + i\psi^{\dagger i}\overline{\sigma}^\mu D_\mu\psi_i$$
$$- \sqrt{2}g_a(\phi^* T^a \psi)\lambda^a - \sqrt{2}g_a\lambda^\dagger(\psi^\dagger T^a \phi) + g_a(\phi^* T^a \phi)D^a.\tag{1.75}$$

with covariant derivatives defined as

$$D_\mu\phi_i = \partial_\mu\phi_i + igA_\mu^a(T^a\phi)i,\tag{1.76}$$

$$D_\mu\phi^{*i} = \partial_\mu\phi^{*i} - igA_\mu^a(\phi^* T^a)^i,\tag{1.77}$$

$$D_\mu\psi_i = \partial_\mu\psi_i + igA_\mu^a(T^a\psi)_i.\tag{1.78}$$

As a result of the coupling to a gauge field, the susy transformations of chiral superfields are altered,

$$\delta\phi_i = \epsilon\psi_i,\tag{1.79}$$

$$\delta(\psi_i)_\alpha = -i(\sigma^\mu\epsilon^\dagger)_\alpha D_\mu\phi^i + \epsilon_\alpha F_i,\tag{1.80}$$

$$\delta F_i = -i\epsilon^\dagger\overline{\sigma}^\mu D_\mu\psi_i + \sqrt{2}g(T^a\phi)_i\epsilon^\dagger\lambda^{\dagger a},\tag{1.81}$$

where derivatives have been turned in covariant derivatives and the variation of $F$ now has an additional term containing the gaugino.

The same final result for the chiral superfield terms could have been found in a more hands-on way, in two steps. First, we convert all the derivatives appearing in the chiral superfield Lagrangian Eq. (1.45) into covariant derivatives. This provides the necessary couplings of all the chiral superfield components to the gauge field $A_\mu^a$. However, the other superfield components still need to couple to the chiral superfields as well. The extra allowed renormalisable interaction terms are given by

$$(\phi^* T^a \psi)\lambda^a, \quad \lambda^{\dagger a}(\psi^{\dagger} T^a \phi), \quad (\phi^* T^a \phi)D^a, \tag{1.82}$$

which indeed appear in Eq. (1.75) with a coupling which makes the Lagrangian supersymmetry invariant. The first two terms are the "supersymmetrised" version of difermion-gauge couplings for gauginos, while the third one includes the auxiliary field $D^a$ and will therefore add a (scalar)$^4$-term to the scalar potential when solving the $D^a$ equation of motion.

Summarising, the general renormalisable Lagrangian for supersymmetric gauge theory is

$$\mathcal{L} = \left(\frac{1}{4} - i\frac{g_a^2 \Theta_a}{32\pi^2}\right) [\mathcal{W}^{a\alpha} \mathcal{W}_\alpha^a]_F + \text{c.c.} + \left[\Phi^{*i}(e^{2g_a T^a V^a})_i{}^j \Phi_j\right]_D + ([W(\Phi_i)]_F + \text{c.c.}), \tag{1.83}$$

with the first term given by Eq. (1.69), the second term by Eq. (1.75) and the final term by Eq. (1.46) and where we allowed for the $\Theta$-term for completeness.

Finally, the part of the Lagrangian containing gauge fields is often rewritten by defining

$$\tau_a = \frac{1}{g_a^2} - i\frac{\Theta_a}{8\pi^2}, \tag{1.84}$$

and absorbing the coupling constants in the fields

$$\widehat{V}^a \equiv g_a V^a, \tag{1.85}$$

$$\widehat{\mathcal{W}}_\alpha^a \equiv g_a \mathcal{W}_\alpha^a = -\frac{1}{4}\overline{DD}\left(D_\alpha \widehat{V}^a - if^{abc}\widehat{V}^b D_\alpha \widehat{V}^c + \dots\right), \tag{1.86}$$

and similar for their component expressions. The Lagrangian is then

$$\mathcal{L} = \frac{1}{4}\left[\tau_a \widehat{\mathcal{W}}^{a\alpha} \widehat{\mathcal{W}}_\alpha^a\right]_F + \text{c.c.} + \left[\Phi^{*i}(e^{2T^a \widehat{V}^a})_i{}^j \Phi_j\right]_D + ([W(\Phi_i)]_F + \text{c.c.}). \tag{1.87}$$

In this way, the gauge coupling only appears in the prefactor $\tau_a$ (e.g. no longer in the covariant derivative, since also there $g_a$ is absorbed). From now on, we will always use this form, but drop the hat (the small changes due to this can be seen explicitly in Appendix B). This coupling constant can be seen as the scalar component of a chiral superfield ($\tau_a$ is complex and it appears holomorphic in the gauge kinetic term). Treating $\tau_a$ as an actual chiral superfield, a spurion, is useful for calculating loop diagrams. Moreover, and more importantly for us, a vacuum expectation value (VEV) for the scalar part of $\tau_a$ gives the coupling above, while an $F$-term VEV (which breaks supersymmetry, see Sections 1.2.5 and 1.4) can give mass to the gauginos.

Note that, in general, it is also possible to include terms with $D_\alpha \Phi_i$ and multiple chiral covariant derivatives, but these are typically suppressed with additional mass dimensions.

**Scalar potential**

In the Lagrangian derived above, there appeared terms quadratic in the auxiliary fields, which are scalars, giving rise to a scalar potential. This potential is important for the phenomenology of supersymmetric theory, since scalar fields are the only fields that do not carry Lorentz indices and thus the only fields which can obtain a vacuum expectation value. As we will see in Section 1.4, if the vacuum solution of a supersymmetric theory has a non-zero energy, then supersymmetry is spontaneously broken in the vacuum state.

As discussed above, the auxiliary fields appear at most quadratically and without derivatives in the Lagrangian, so we can solve their equations of motion algebraically and substitute the solution back in the Lagrangian[41]. In this way, the terms linear in the auxiliary fields will appear squared in the scalar potential. Concretely, we saw, when building the chiral superfield Lagrangian, that the equations of motion for the auxiliary field $F_i$ and its conjugate are

$$F_i = -W_i^*, \qquad F^{*i} = -W^i. \tag{1.88}$$

Similarly, for the real superfield auxiliary field $D^a$, the last term in Eq. (1.82) combines with $D^a D^a / 2$ in the gauge kinetic term Eq. (1.69) to give the equation of motion

$$D^a = -g(\phi^* T^a \phi). \tag{1.89}$$

Substituting this back in the Lagrangian, the scalar potential $V(\phi, \phi^*)$ becomes[42]

$$V(\phi, \phi^*) = F^{*i} F_i + \frac{1}{2} \sum_a D^a D^a = W_i^* W^i + \frac{1}{2} \sum_a g_a^2 (\phi^* T^a \phi)^2. \tag{1.90}$$

These two parts are called the *F*-term and *D*-term contributions to the scalar potential. For reference, the full expression for the scalar potential, substituting the most general superpotential containing only renormalisable terms is given in Eq. (C.36).

**Supercurrent**

From Noether's theorem [111], we know that every continuous symmetry is associated with a conserved current. Therefore, supersymmetry needs to be associated with a conserved supercurrent [112, 113], obtained from

$$\epsilon J^\mu + \epsilon^\dagger J^{\dagger \mu} \equiv \sum_X \delta X \frac{\delta \mathcal{L}}{\delta(\partial_\mu X)} - K^\mu, \tag{1.91}$$

with $\delta \mathcal{L} = \partial_\mu K^\mu$. The zero components of this current are conserved charges which generate the supersymmetry transformations. Since supersymmetry is a fermionic symmetry, the associated supercurrent carries a spinor index. For the general supersymmetric Lagrangian in Eq. (1.83), one finds the supercurrent

$$J^\mu_\alpha = (\sigma^\nu \overline{\sigma}^\mu \psi_i)_\alpha \, D_\nu \phi^{*i} + i(\sigma^\mu \psi^{\dagger i})_\alpha \, W_i^*$$
$$+ \frac{1}{2\sqrt{2}} (\sigma^\nu \overline{\sigma}^\rho \sigma^\mu \lambda^{\dagger a})_\alpha \, F^a_{\nu\rho} + \frac{i}{\sqrt{2}} g_a \phi^* T^a \phi (\sigma^\mu \lambda^{\dagger a})_\alpha. \tag{1.92}$$

---

[41]In addition, since these fields appear at most quadratically, this procedure remains correct quantum mechanically.

[42]The scalar potential appears in the Lagrangian with a minus sign $\mathcal{L} \supset -V(\phi, \phi^*)$.

This supercurrent will be important in the next chapter.

### 1.2.4 Non-renormalisable interactions

The result of Section 1.2.3 can be generalised to a Lagrangian including non-renormalisable interactions, which is

$$\mathcal{L} = \left[ K(\Phi_i, (\Phi^* e^V)^j) \right]_D + \left( \left[ \frac{1}{4} f_{ab}(\Phi_i) \hat{\mathcal{W}}^{a\alpha} \hat{\mathcal{W}}^b_\alpha + W(\Phi_i) \right]_F + \text{c.c.} \right). \tag{1.93}$$

The superpotential $W$ is now an arbitrary holomorphic function of the chiral superfields treated as complex variables (compatible with gauge symmetries and of mass dimension 3), i.e. terms with more fields than Eq. (1.51) with prefactors of negative mass dimension can appear. The Kähler potential $K$ is a real, gauge invariant function of both $\Phi_i$ and $\Phi^{*j}$ and of mass dimension 2 (in the previous section, we had $K = \Phi^{*i} e^V \Phi_i$). The gauge kinetic funcion $f_{ab}$ is chiral superfield and holomorphic function of $\Phi_i$, symmetric in $a, b$. Typically, $f_{ab} \propto \delta_{ab}$, except for kinetic mixing between multiple Abelian groups.

### 1.2.5 Soft supersymmetry-breaking masses

Since supersymmetry predicts mass-degenerate pairs of fermions and bosons, we know that supersymmetry must be broken at energies currently accessible by colliders. Therefore, we must add supersymmetry-breaking masses to the superpartners introduced by supersymmetry. However, in order to retain the protection of the Higgs mass from high mass scales, this breaking needs to be soft, i.e. not reintroduce quadratic divergences in the quantum correction to the Higgs squared-mass parameter in the Lagrangian.

While the full analysis is very technical [114], there is a simple argument for the form that allowed quantum corrections $\Delta m_H^2$ can take. The supersymmetry-breaking terms should not change the relation between the dimensionless couplings imposed by supersymmetry, since these guarantee that the problematic quadratic divergences cancel (see Section 1.1). Therefore, the non-supersymmetric corrections can only be due to the addition of soft masses and they must vanish in the limit $m_{\text{soft}} \to 0$. By dimensional analysis, this implies that the corrections to the Higgs squared-mass parameter can not be proportional to[43] $\Lambda_{\text{UV}}^2$. Thus, the introduced corrections by allowed soft mass terms must have the structure (this is shown explicitly in Appendix C.1)

$$\Delta m_H^2 = m_{\text{soft}}^2 \left[ \frac{\lambda}{16\pi^2} \ln(\Lambda_{\text{UV}}/m_{\text{soft}}) + \dots \right]. \tag{1.94}$$

---

[43] Corrections of the form $\Delta m_H^2 \sim m_{\text{soft}} \Lambda_{\text{UV}}$ are also not possible because loop integrals always diverge quadratically or logarithmically with $\Lambda_{\text{UV}}$.

If we do not want to reintroduce the hierarchy problem, these corrections, although no longer exhibiting a quadratic sensitivity to high mass scales, can not be too high. For $\Lambda_{\mathrm{UV}} \sim M_P$ and $\lambda \sim 1$, one obtains that $m_{\mathrm{soft}}$ can typically not be much higher than the TeV scale.

All the allowed soft supersymmetry-breaking terms are then[44,45] [114]

$$\mathcal{L}_{\mathrm{soft}} = -\left(\frac{1}{2}M_a\lambda^a\lambda^a + \frac{1}{6}a^{ijk}\phi_i\phi_j\phi_k + \frac{1}{2}b^{ij}\phi_i\phi_j + t^i\phi_i\right) + \mathrm{c.c.}$$
$$- (m^2)^i_j\phi^{j*}\phi_i \tag{1.95}$$

$$\mathcal{L}_{\mathrm{maybe\ soft}} = -\frac{1}{2}c_i^{jk}\phi^{*i}\phi_j\phi_k + \mathrm{c.c.} \tag{1.96}$$

These terms indeed break supersymmetry, since they only contain the scalars and gauginos. They form the gaugino masses $M_a$, scalar mass terms $(m^2)^j_i$ and $b^{ij}$, the latter of which splits the scalar and pseudoscalar masses, (scalar)$^3$ couplings[46] $a^{ijk}$ and $c_i^{jk}$ and tadpole couplings $t^i$ (which we ignore from now on). The gaugino and scalar mass terms (at least for $i = j$) are always allowed, while the other three couplings, which mirror the superpotential couplings, are only allowed if the corresponding terms in the superpotential are also allowed.

These soft-breaking terms in the Lagrangian can be either introduced by hand (explicit breaking), or be obtained from the spontaneous breaking of a supersymmetric theory. The requirement of having only soft-breaking terms seems arbitrary and difficult to obtain in a generic supersymmetric theory. However, models with spontaneous supersymmetry breaking automatically produce terms of this form[47]. Essentially, soft-breaking terms in such a spontaneously broken theory must come from some of the fields obtaining vacuum expectation values. Since these appeared originally in a supersymmetric Lagrangian with couplings which were "behaving nicely", the resulting theory which is broken at low energy must still possess this quality. Moreover, when restoring supersymmetry, i.e. letting the VEVs→ 0, we must reobtain the original protected Higgs squared-mass parameter, so we return to the argument above Eq. (1.94).

It happens that the allowed soft-breaking terms explain why we have not seen the superpartners to the Standard Model particles yet (if they exist). All currently known particles, except the Higgs, can not have a mass term in the Lagrangian (due to gauge invariance and, for the fermions, the fact that they are chiral). Therefore, they must be massless in the absence of electroweak symmetry breaking. The Standard Model masses due to electroweak symmetry breaking are then proportional to the Higgs VEV times a dimensionless coupling constant. On the other hand, all the other particles in the minimal supersymmetric extension of the Standard Model *can* have a mass term in the Lagrangian: the squarks, sleptons and Higgs scalar are complex scalars, so a term $m^2|\phi|^2$ is always allowed by gauge symmetries. The gauginos and neutral higgsinos after mixing (since they initally carry hypercharge) are fermions in the real representation of the gauge group, so their mass terms are not forbidden as for the Standard Model fermions.

---

[44]A soft mass term $\mathcal{L} = -\frac{1}{2}m^{ij}\psi_i\psi_j + \mathrm{c.c.}$ could also appear, but can be absorbed back by redefining the superpotential and the soft parameters $m^2$ and $c$.

[45]An additional possibility is a soft supersymmetry-breaking Dirac mass term between chiral superfield fermions and the gauginos $\mathcal{L} = -M^a_{\mathrm{Dirac}}\lambda^a\psi_a + \mathrm{c.c.}$, but for this one needs chiral fermions in the adjoint representation, which does not occur in the minimal extension to the Standard Model. See [67] and references therein.

[46]Terms proportional to $c_i^{jk}$ are only soft in the absence of gauge singlets. But this term is usually neglected since it turns out to be difficult to generate with non-negligible values in supersymmetry-breaking models.

[47]Similarly, it turns out that dynamical supersymmetry-breaking models do as well. The same is true for spontaneously broken supergravity (string theory), in a unitary gauge.

While supersymmetry itself is very restrictive on the structure and size of the couplings[48], the supersymmetry-breaking terms are (a priori) almost unrestricted, introducing a lot of new free parameters.

## 1.3 The Minimal Supersymmetric Standard Model

In order to find a supersymmetric theory which describes reality, we need to define a supersymmetric extension of the Standard Model [115–118]. We discuss here the Minimal Supersymmetric Standard Model (MSSM), which is the minimal supersymmetric model which contains all of the Standard Model particles and interactions. However, we focus only on the parts which are needed to understand the next chapter. In particular, we ignore the squark sector and the accompanying flavour structure. As with the previous section, most of the discussion here is based on [67].

### 1.3.1 MSSM particle content

The particles of the MSSM are given by all the Standard Model fermions and their complex scalar partners (the sfermions) in chiral superfields, shown in Table 1.1, the Standard Model gauge bosons and their fermionic partners (the gauginos) in real superfields, shown in Table 1.2, as well as the Higgs sector. However, in order to construct a consistent theory, the Higgs sector must be extended to include two scalar Higgs fields. This can be easily seen in two ways. First, consider the hypercharges. In the Standard model, the right-handed fields of the up and down quarks have different hypercharge. Therefore, in order to build correct Yukawa interaction terms which give mass to both, one must use the Higgs scalar field $\Phi$ and its conjugate $\tilde{\Phi} \equiv i\sigma^2\Phi^*$, which have opposite hypercharge. Indeed, the Yukawa terms then have the form $-(y_u\overline{u}_R\tilde{\Phi}^\dagger Q_L + y_d\overline{d}_R\Phi^\dagger Q_L + \text{c.c.})$. However, in order to build a correct supersymmetric theory, the superpotential (which is where Yukawa couplings show up) must be holomorphic in the chiral superfields. Therefore, a second Higgs doublet with opposite hypercharge needs to be introduced. Second, since supersymmetry introduces a fermionic partner to the Higgs boson, the cancellation of the gauge-gravity anomalies [61] will be spoiled, unless a second fermion with the opposite contribution is introduced. Consequently, the MSSM includes two

---

[48]Just as the Standard Model: without the scalar sector, there are only three gauge couplings and $\theta_{\text{QCD}}$. Including scalars typically opens Pandora's box.

Table 1.1: Chiral supermultiplets of the Minimal Supersymmetric Standard Model and their representation/charge under the gauge groups. The bar is part of the name, indicating that it is the conjugate of the right-handed part of a Dirac spinor, and does *not* represent conjugation of the unbarred field.

| Supermultiplet | spin 0 | spin 1/2 | $(SU(3)_C, SU(2)_L, U(1)_Y)$ |
|---|---|---|---|
| $Q$ | $\begin{pmatrix} \tilde{u}_L \\ \tilde{d}_L \end{pmatrix}$ | $\begin{pmatrix} u_L \\ d_L \end{pmatrix}$ | $(\mathbf{3}, \mathbf{2}, +\frac{1}{6})$ |
| $\overline{u}$ | $\tilde{u}_R^*$ | $u_R^\dagger$ | $(\overline{\mathbf{3}}, \mathbf{1}, -\frac{2}{3})$ |
| $\overline{d}$ | $\tilde{d}_R^*$ | $d_R^\dagger$ | $(\overline{\mathbf{3}}, \mathbf{1}, +\frac{1}{3})$ |
| $L$ | $\begin{pmatrix} \tilde{\nu} \\ \tilde{e}_L \end{pmatrix}$ | $\begin{pmatrix} \nu \\ e_L \end{pmatrix}$ | $(\mathbf{1}, \mathbf{2}, -\frac{1}{2})$ |
| $\overline{e}$ | $\tilde{e}_R^*$ | $e_R^\dagger$ | $(\mathbf{1}, \mathbf{1}, +1)$ |
| $H_u$ | $\begin{pmatrix} H_u^+ \\ H_u^0 \end{pmatrix}$ | $\begin{pmatrix} \tilde{H}_u^+ \\ \tilde{H}_u^0 \end{pmatrix}$ | $(\mathbf{1}, \mathbf{2}, +\frac{1}{2})$ |
| $H_d$ | $\begin{pmatrix} H_d^0 \\ H_d^- \end{pmatrix}$ | $\begin{pmatrix} \tilde{H}_d^0 \\ \tilde{H}_d^- \end{pmatrix}$ | $(\mathbf{1}, \mathbf{2}, -\frac{1}{2})$ |

Table 1.2: Gauge supermultiplets of the Minimal Supersymmetric Standard Model (gauge eigenstates) and their representation/charge under the gauge groups.

| Supermultiplet | spin 1/2 | spin 1 | $(SU(3)_C, SU(2)_L, U(1)_Y)$ |
|---|---|---|---|
| $SU(3)_C$ | $\tilde{g}$ | $g$ | $(\mathbf{8}, \mathbf{1}, 0)$ |
| $SU(2)_L$ | $\tilde{W}^\pm \ \tilde{W}^0$ | $W^\pm \ W^0$ | $(\mathbf{1}, \mathbf{3}, 0)$ |
| $U(1)_Y$ | $\tilde{B}$ | $B$ | $(\mathbf{1}, \mathbf{1}, 0)$ |

Higgs doublets with opposite hypercharge together with their fermionic partners (the higgsinos) in chiral superfields, shown in Table 1.1. As usual, the electromagnetic charge is defined by $Q_{EM} = T_3 + Y$.

By convention, all chiral supermultiplets are defined in terms of left-handed two-component Weyl–spinors. Thus, in the defining superfields in Table 1.1, the conjugates of right-handed quarks and leptons appear, indicated by a bar. A Dirac spinor is then formed as

$$\begin{pmatrix} e \\ \overline{e}^\dagger \end{pmatrix} \equiv \begin{pmatrix} e_L \\ e_R \end{pmatrix}. \tag{1.97}$$

In general, the superpartners shown in Tables 1.1 and 1.2 will mix with others with the same quantum numbers. In particular[49], this will be the case for the neutral gauginos and the neutral higgsinos, which we discuss in Section 1.3.5. The gluinos are the exception, due to their unique quantum numbers

---

[49]Another example is the stop squarks. The two scalar partners $\tilde{t}_L$ and $\tilde{t}_R$ (which are not left-or right-handed themselves, but merely the partners of the corresponding left-handed and right-handed fermions) will mix to form $\tilde{t}_1$ and $\tilde{t}_2$. One of these will be pushed down in mass, the other up.

## 1.3.2 MSSM superpotential

The MSSM superpotential, which is a holomorphic function of chiral superfields, is given by

$$W_{\text{MSSM}} = \overline{u}\mathbf{y_u}QH_u - \overline{d}\mathbf{y_d}QH_d - \overline{e}\mathbf{y_e}LH_d + \mu H_u H_d, \tag{1.98}$$

where the Yukawa couplings are matrices in family space. The $\mu$-term is the only allowed supersymmetric mass term in the MSSM. Note that in the above, contractions of $SU(2)$ doublets are performed with an $\epsilon^{\alpha\beta}$ to make gauge-invariant combinations. In particular, using Eqs. (1.46) and (1.49), this superpotential gives rise to supersymmetric higgsino masses

$$-\mathcal{L}_{\text{higgsino}} \supset \mu(\tilde{H}_u^+ \tilde{H}_d^- - \tilde{H}_u^0 \tilde{H}_d^0) + \text{c.c.}, \tag{1.99}$$

and Higgs squared-mass terms[50]

$$-\mathcal{L}_{\text{Higgs}} \supset |\mu|^2(|H_u^0|^2 + |H_u^+|^2 + |H_d^0|^2 + |H_d^-|^2). \tag{1.100}$$

Note that the scale of $\mu$ itself still needs to be put in by hand. So, while this supersymmetric version of the Higgs squared-mass parameter is now protected from quantum corrections, there is no a priori reason for it to be close to the soft masses and weak scale instead of e.g. $M_P$. This is known as the "$\mu$-problem" and is typically solved by relating its generation also to supersymmetry breaking terms.

As is, there are additional allowed terms which violate lepton number (proportional to $\lambda^{ijk}L_i L_j \overline{e}_k$, $\lambda'^{ijk}L_i Q_j \overline{d}_k$ and $\mu'^i L_i H_u$) or baryon number ($\lambda''^{ijk}\overline{u}_i \overline{d}_j \overline{d}_k$) conservation. These terms are problematic, since they can induce proton decay, the limits on which are very strong. The easiest way to avoid this problem is by forbidding these terms. Postulating $B$ or $L$ symmetry, while possible, seems to go against the precedent set by the Standard Model, where their conservation[51] is accidental and follows from the structure of the theory. Instead, one can impose $R$-parity [118] or matter parity [63, 71, 120, 121]. $R$-parity[52] is defined by[53]

$$P_R = (-1)^{3(B-L)+2s}. \tag{1.101}$$

Only terms with even $R$-parity are then allowed in the Lagrangian. This indeed forbids the terms which give rise to proton decay, while allowing the ones in Eq. (1.98).

Components of different spin in a superfield also have different $R$-parity[54]. All Standard Model particles have $P_R = +1$, while all supersymmetric partners have $P_R = -1$. If $R$-parity is exactly conserved, the lightest supersymmetric particle (LSP) must then be stable[55] such that all superpartner decay chains eventually end in LSP + Standard Model particles. In addition, supersymmetric partners must be produced in pairs at colliders. Finally, while they will not be discussed here, it is also possible to consider R-parity violating theories [118, 122–127], which are being actively searched for at the LHC.

---

[50]So in order to have the Higgs break electroweak symmetry, one needs an additional negative supersymmetry-breaking squared-mass soft term.

[51]Except for non-perturbative electroweak effects [119].

[52]Related, but not the same as, $R$-symmetry. See Section 1.4.3.

[53]Matter parity is defined by a similar relation, where the exponent $2s$ is omitted.

[54]But they have the same matter parity.

### 1.3.3 MSSM soft masses

Next, we show the soft masses which break supersymmetry and push up the sfermion and gaugino masses. The MSSM soft masses are (remember that these are component fields, not full superfields)

$$
\begin{aligned}
\mathcal{L}_{\text{soft}}^{\text{MSSM}} = &-\frac{1}{2}\left(M_3 \tilde{g}\tilde{g} + M_2 \tilde{W}\tilde{W} + M_1 \tilde{B}\tilde{B} + \text{c.c.}\right) \\
&-\left(\tilde{\bar{u}}\mathbf{a_u}\tilde{Q}H_u - \tilde{\bar{d}}\mathbf{a_d}\tilde{Q}H_d - \tilde{\bar{e}}\mathbf{a_e}\tilde{L}H_d + \text{c.c.}\right) \\
&-\tilde{Q}^\dagger \mathbf{m_Q^2}\tilde{Q} - \tilde{L}^\dagger \mathbf{m_L^2}\tilde{L} - \tilde{\bar{u}}\mathbf{m_{\bar{u}}^2}\tilde{\bar{u}}^\dagger - \tilde{\bar{d}}\mathbf{m_{\bar{d}}^2}\tilde{\bar{d}}^\dagger - \tilde{\bar{e}}\mathbf{m_{\bar{e}}^2}\tilde{\bar{e}}^\dagger \\
&- m_{H_u}^2 H_u^* H_u - m_{H_d}^2 H_d^* H_d - (bH_u H_d + \text{c.c.}).
\end{aligned} \tag{1.102}
$$

These terms give rise to gaugino masses, (scalar)$^3$ couplings (mirroring the Yukawa couplings, but a priori independent from them), squark and slepton masses as well as supersymmetry-breaking Higgs potential terms: scalar squared-mass terms and a $b$-term. Factors in bold are matrices in family space. In general, one expects from dimensional analysis

$$M_1, M_2, M_3, \mathbf{a_u}, \mathbf{a_d}, \mathbf{a_e} \sim m_{\text{soft}} \tag{1.103}$$

$$\mathbf{m_Q^2}, \mathbf{m_L^2}, \mathbf{m_{\bar{u}}^2}, \mathbf{m_{\bar{d}}^2}, \mathbf{m_{\bar{e}}^2}, m_{H_u}^2, m_{H_d}^2, b \sim m_{\text{soft}}^2. \tag{1.104}$$

The values of these masses are not fixed by the theory, but can be estimated. As we already saw, supersymmetry, if it exists, must be broken at energies currently probed by colliders. On the other hand, if supersymmetry is to provide a solution to the hierarchy problem, the superpartners can not be too heavy. Therefore, typically, one expects all of these to be around the TeV scale (if supersymmetry is realised in nature slightly above the electroweak scale).

While supersymmetry itself is very predictive and economic (i.e. a few couplings determine many interactions), supersymmetry breaking introduces many new parameters, mostly related to flavour [128]: the MSSM introduces 105 masses, phases and mixing angles extra compared to Standard Model[56] which can not be rotated away by redefinition of the fields. However, the non-observation of flavour-violating processes severely restricts the soft terms. One can get rid of flavour-violating effects by choosing the squared masses to be flavour-blind, the (scalar)$^3$-couplings to be proportional to the Yukawa couplings and making all parameters real so there is no additional source of CP-violation (although there are other possibilities as well, see [67]). While these choices seem arbitrary, there are large classes of supersymmetry-breaking models which guarantee that such simplifications occur. However, these simplifications are then only valid at a certain mass scale. Going down in energy, quantum corrections need to be taken into account using the renormalisation-group (RG) equations. As a result, the flavour terms will be more complicated, but their flavour- and CP-conserving behaviour remains, since the running due to the gauge groups respects flavour and the influence of the Yukawa couplings is small[57].

---

[55] Such a particle, if neutral, is then also a good dark matter candidate.

[56] So, in total, the MSSM contains the 18 Standard Model parameters (including $\theta_{\text{QCD}}$), one Higgs sector which is the analogue of the Standard Model HIggs mass and 105 new parameters; see e.g. [29].

### 1.3.4 Higgs section

Since we will discuss higgsinos in the next chapter, we gather here the most important elements of the MSSM Higgs sector.

The classical scalar potential for the Higgs sector is given by

$$
\begin{aligned}
V &= (|\mu|^2 + m_{H_u}^2)(|H_u^0|^2 + |H_u^+|^2) + (|\mu|^2 + m_{H_d}^2)(|H_d^0|^2 + |H_d^-|^2) \\
&\quad + [b(H_u^+ H_d^- - H_u^0 H_d^0) + \text{c.c.}] \\
&\quad + \frac{1}{8}(g^2 + g'^2)(|H_u^0|^2 + |H_u^+|^2 - |H_d^0|^2 - |H_d^-|^2)^2 \\
&\quad + \frac{1}{2}g^2|H_u^+ H_d^{0*} + H_u^0 H_d^{-*}|^2,
\end{aligned}
\tag{1.105}
$$

where the first two lines come from the *F*-terms and supersymmetry-breaking terms and the final two lines come from the *D*-terms (after rearranging). Since the minimum needs to break $SU(2)_L \times U(1)_Y \to U(1)_{EM}$ like in the Standard Model, we can make an $SU(2)_L$ gauge transformation such that one Higgs VEV vanishes. Taking $H_u^+ = 0$, one finds that in the minimum also $H_d^- = 0$, so electromagnetism indeed stays unbroken. The Higgs potential is then

$$
\begin{aligned}
V &= (|\mu|^2 + m_{H_u}^2)|H_u^0|^2 + (|\mu|^2 + m_{H_d}^2)|H_d^0|^2 - (b H_u^0 H_d^0 + \text{c.c.}) \\
&\quad + \frac{1}{8}(g^2 + g'^2)(|H_u^0|^2 - |H_d^0|^2)^2.
\end{aligned}
\tag{1.106}
$$

As in the Standard Model, the Higgs fields will obtain a VEV, which we denote by

$$
v_u = \langle H_u^0 \rangle, \qquad v_d = \langle H_d^0 \rangle.
\tag{1.107}
$$

These VEVs, again as in the Standard Model with a single Higgs fields, must be related to the mass of the *Z*-boson,

$$
v_u^2 + v_d^2 = v^2 = 2m_Z^2/(g^2 + g'^2) \approx (174 \text{ GeV})^2.
\tag{1.108}
$$

---

[57]To be completely flavour-blind after running, this small effect should be compensated somehow.

The Weinberg angle is then given by $\sin\theta_w = g'/\sqrt{g^2 + g'^2}$ and the $W$-boson and $Z$-boson masses are related as $\cos\theta_W = m_W/m_Z$. This also defines an angle $\beta$ which is given by the ratio of $v_u$ and $v_d$,

$$\tan\beta \equiv v_u/v_d, \tag{1.109}$$

such that $v_d = v\cos\beta$ and $v_u = v\sin\beta$. The size of $\tan\beta$ determines the relative coupling strength of the Higgs to the up- and down components of the quark doublets. Since the mass of the quarks is generated through Yukawa couplings, a large $\tan\beta$ implies that $v_u$ is large while $v_d$ is small. To obtain the same quark mass, in particular the top and bottom mass, a large $\tan\beta$ implies a smaller coupling of the Higgs to the top and a larger coupling to the bottom quark.

Minimising the Higgs potential ($\partial V/\partial H_i^0 = 0$) and requiring that the result is compatible with the experimentally observed electroweak symmetry breaking (i.e. the previous two equations) implies the relations[58]

$$m_{H_u}^2 + |\mu|^2 - b\cot\beta - (m_Z^2/2)\cos(2\beta) = 0, \tag{1.110}$$

$$m_{H_d}^2 + |\mu|^2 - b\tan\beta + (m_Z^2/2)\cos(2\beta) = 0. \tag{1.111}$$

The Higgs scalar fields in the MSSM consist of two complex $SU(2)_L$-doublets, so eight real scalar degrees of freedom. Three of them become the would-be Nambu-Goldstone bosons ($G^0$ and $G^\pm$), eaten by the $W^\pm$- and $Z$-bosons. Of the remaining degrees of freedom, there are two CP-even neutral scalars $h^0$ (the lightest) and $H^0$, one CP-odd neutral scalar $A^0$ and two charged scalars $H^+$ and $H^-$, which are each others conjugate. These particles are related to one another through rotations

$$\begin{pmatrix} H_u^0 \\ H_d^0 \end{pmatrix} = \begin{pmatrix} v_u \\ v_d \end{pmatrix} + \frac{1}{\sqrt{2}} R_\alpha \begin{pmatrix} h^0 \\ H^0 \end{pmatrix} + \frac{i}{\sqrt{2}} R_{\beta_0} \begin{pmatrix} G^0 \\ A^0 \end{pmatrix}, \tag{1.112}$$

and

$$\begin{pmatrix} H_u^+ \\ H_d^{-*} \end{pmatrix} = R_{\beta_\pm} \begin{pmatrix} G^+ \\ H^+ \end{pmatrix}, \tag{1.113}$$

with

$$R_\alpha = \begin{pmatrix} \cos\alpha & \sin\alpha \\ -\sin\alpha & \cos\alpha \end{pmatrix}, \qquad R_{\beta_i} = \begin{pmatrix} \sin\beta_i & \cos\beta_i \\ -\cos\beta_i & \sin\beta_i \end{pmatrix}. \tag{1.114}$$

These rotation angles are chosen such that the quadratic part of the potential is diagonal in the squared-masses,

$$\begin{aligned} V = {} & \frac{1}{2} m_{h^0}^2 (h^0)^2 + \frac{1}{2} m_{H^0}^2 (H^0)^2 + \frac{1}{2} m_{G^0}^2 (G^0)^2 + \frac{1}{2} m_{A^0}^2 (A^0)^2 \\ & + m_{G^\pm}^2 |G^+|^2 + m_{H^\pm}^2 |H^+|^2 + \dots \end{aligned} \tag{1.115}$$

When $v_u$ and $v_d$ minimise the tree-level potential, then $\beta_0 = \beta_\pm = \beta$ and the would-be Nambu-Goldstone bosons have zero mass. The mixing angle $\alpha$ is a function of $\beta$, the neutral Higgs masses and $m_Z$.

---

[58]At loop level, these relations are modified.

Finally, one often considers the so-called "decoupling limit", as we will in the next chapter, where $m_{A^0} \gg m_Z$. Then one finds that $A^0$, $H^0$ and $H^\pm$ are all much heavier than $h^0$ and are nearly degenerate. The angle $\alpha$ is then given by $\alpha \approx \beta - \pi/2$ and $h^0$ has nearly the same couplings to the quarks, leptons and gauge bosons as the Standard Model Higgs boson.

Note that the Higgs is unique in the MSSM, since it is the only particle where a supersymmetric mass is allowed. This is then the reason why, contrary to the other chiral supermultiplets, we see the scalar component as the lowest mass component instead of the fermionic: both of these have a mass term (either supersymmetric for both or soft SUSY-breaking for the scaler). On the other hand, in order to have successful electroweak symmetry breaking, these two masses and the $b$-term must be of comparable size, since otherwise this would screw up the Higgs potential. This immediately leads to another issue already mentioned before: the $\mu$-problem. While the Higgs mass parameter is now protected from quadratic divergences, there is no reason why it should be at such a low scale. More specifically, it needs to be of the same size as the supersymmetry-breaking scale. But, since it is a supersymmetric mass term, there is no reason for these to be related. This could be solved if the supersymmetric Higgs mass term is forbidden by some symmetry and instead generated together with SUSY breaking. This can be done either through the mediation mechanisms (discussed in Section 1.4.4), or in another way. For example, in the Next-to-Minimal Supersymmetric Standard Model (NMSSM), the supersymmetric Higgs mass parameter is instead generated by the vacuum expectation value of a singlet superfield $S$. In this way, the Higgs mass parameter is automatically of the order of the SUSY-breaking scale.

### 1.3.5 Neutralinos

The neutral gauginos and neutral higgsinos are all neutral and can thus mix, such that in general the mass eigenstates, called neutralinos, are different from the gauge eigenstates. Denote the gauge eigenstates by $\psi^0 = (\tilde{B}, \tilde{W}^{(3)}, \tilde{H}_d^0, \tilde{H}_u^0)$. Then the neutralino mass matrix is defined by writing the relevant quadratic term in the Lagrangian as

$$\mathcal{L}_{\text{neutralino mass}} = -\frac{1}{2}(\psi^0)^T \mathbf{M}_{\tilde{\chi}} \psi^0 + \text{c.c.}, \tag{1.116}$$

with

$$\mathbf{M}_{\tilde{\chi}} = \begin{pmatrix} M_B & 0 & -g'v_d/\sqrt{2} & g'v_u/\sqrt{2} \\ 0 & M_W & gv_d/\sqrt{2} & -gv_u/\sqrt{2} \\ -g'v_d/\sqrt{2} & gv_d/\sqrt{2} & 0 & -\mu \\ g'v_u/\sqrt{2} & -gv_u/\sqrt{2} & -\mu & 0 \end{pmatrix}, \tag{1.117}$$

which arise from the supersymmetry-breaking gaugino masses, the supersymmetric Higgs masses and the Higgs-higgsino-gaugino couplings (for the off-diagonal parts). This can equivalently be written as

$$\mathbf{M}_{\tilde{\chi}} = \begin{pmatrix} M_B & 0 & -m_Z s\theta_w c\beta & m_Z s\theta_w s\beta \\ 0 & M_W & m_Z c\theta_w c\beta & -m_Z c\theta_w s\beta \\ -m_Z s\theta_w c\beta & m_Z c\theta_w c\beta & 0 & -\mu \\ m_Z s\theta_w s\beta & -m_Z c\theta_w s\beta & -\mu & 0 \end{pmatrix}. \tag{1.118}$$

The mass eigenstate neutralinos[59] $\tilde{\chi}_i$ are then related to the gauge eigenstates by

$$\tilde{\chi}_i = \mathbf{N}_{ij}\psi_j^0 \tag{1.119}$$

where the mixing matrix $\mathbf{N}_{ij}$ is defined by

$$\mathbf{N}^*\mathbf{M}_{\tilde{\chi}}\mathbf{N}^{-1} = \mathbf{diag}(m_{\chi_1}, m_{\chi_2}, m_{\chi_3}, m_{\chi_4}). \tag{1.120}$$

The neutralino with the lowest mass, if it is the LSP, is a possible dark matter candidate.

Note that in a spontaneously broken supersymmetric theory, there is an additional neutral fermion appearing at low energies in the mass spectrum, the goldstino[60]. Thus, in general, is also mixes with the neutralinos. However, as we will see, its interactions with the Standard Model are suppressed by the scale of supersymmetry breaking, such that its contribution to the neutralino mass eigen states can usually be neglected. However, when considering goldstino phenomenology, as we will do in the next chapter, this mixing *is* important.

## 1.4 Origin of supersymmetry breaking

In the previous section, we saw that supersymmetry, if it exists, must be broken at some scale above the electroweak scale, by the appearance of soft-breaking terms. While it is possible to insert these terms by hand (and indeed, for much of the phenomenology, this is a good approach), in a true supersymmetric theory, these terms must appear as a result of spontaneous supersymmetry breaking. Spontaneous symmetry breaking occurs when a theory is invariant under a continuous global symmetry (i.e. the Lagrangian possesses this symmetry and the associated Noether current is conserved), but the ground state is not invariant under application of the symmetry transformation. Goldstone's theorem [129–131] then teaches us that there must appear a massless particle in the theory for each broken generator, with the same quantum numbers as that generator. In its original form, with a bosonic symmetry, this implies the existence of a massless boson, the so-called Nambu-Goldstone boson[61]. Since supersymmetry is a fermionic symmetry, with generators $Q_\alpha$ and $Q_{\dot{\alpha}}^\dagger$, there must appear a Goldstone fermion, the goldstino. In this section, we briefly review the appearance[62] of the goldstino as well as the mechanisms with which supersymmetry can be broken. This will be important for the next chapter, which deals with goldstino phenomenology. Most of this section is based on the reviews [67, 89, 135, 136].

---

[59] Here, we deviate from the convention of [67], where the neutralinos are denoted by $\tilde{N}_i$.

[60] Its degrees of freedom must come from a different sector, see Section 1.4.4

[61] If the symmetry is gauged, i.e. a local symmetry, then the Goldstone boson is "eaten" by the gauge bosons, which acquires a mass and receives a longitudinal polarisation. In addition, the Goldstone boson equivalence theorem [132] then also tells us that, at high energy, scattering amplitudes involving longitudinally polarised gauge bosons equal those calculated using the goldstone bosons instead. See also the discussion in [133]. Something similar happens for the goldstino, see below.

[62] For a more general discussion on the appearance of the goldstino, see [134].

### 1.4.1   Supersymmetry breaking and the goldstino

In this section, we will identify the goldstino in a spontaneously broken supersymmetric theory. Since the broken symmetry generator in this case is $Q_\alpha$, we expect that there is a massless Nambu-Goldstone mode, a Weyl fermion, which we call the goldstino. Consider for now a theory with only chiral superfields, which is spontaneously broken. Therefore, the ground state or vacuum is not invariant under supersymmetry transformations. More concretely, when we apply the generator of supersymmetry transformations on the massless goldstino, its variation is not zero in the vacuum, i.e. $\langle \delta\psi \rangle = \langle \{Q, \psi\} \rangle \neq 0$. Since the variation of a fermion is schematically given by $\delta\psi \sim F + \partial\phi$ (see Eqs. (1.29)-(1.31)), this means that we must have $\langle F \rangle \neq 0$ (since $\partial_\mu \phi$ can not get a Lorentz invariant expectation value). Therefore, the goldstino transforms with a shift, as expected for a spontaneously broken vacuum[63], the size of which is given by $F$. In other words, $F$ is the order parameter of supersymmetry breaking and the fermion which becomes the goldstino is the superpartner of the auxiliary field $F$ which obtains a non-vanishing vacuum expectation value[64].

Now, we can identify the goldstino in a general supersymmetric theory using a simple argument. Since the anti-commutator of the supersymmetry generators $Q_\alpha$ and $Q_{\dot\alpha}^\dagger$ is proportional to the linear momentum $P^\mu$, the Hamiltonian can be written as

$$H = P^0 = \frac{1}{4}\left(Q_1 Q_1^\dagger + Q_1^\dagger Q_1 + Q_2 Q_2^\dagger + Q_2^\dagger Q_2\right). \qquad (1.121)$$

Therefore, the energy of the vacuum state is given by

$$\langle 0|H|0\rangle = \frac{1}{4}\left(||Q_1^\dagger|0\rangle||^2 + ||Q_1|0\rangle||^2 + ||Q_2^\dagger|0\rangle||^2 + ||Q_2|0\rangle||^2\right) \geq 0, \qquad (1.122)$$

and is always positive. If the ground state has zero energy, $\langle 0|H|0\rangle = 0$, then from the above equation, we find that the ground state must also be annihilated by the supersymmetry generators $Q_\alpha|0\rangle = 0$ (and its conjugate). In other words, a zero-energy vacuum is invariant under supersymmetry. If, however, $\langle 0|H|0\rangle \neq 0$, then also $Q_\alpha|0\rangle \neq 0$ and the vacuum transforms under supersymmetry. Therefore, if the vacuum has non-zero energy, supersymmetry must be broken. Since we have $\langle 0|H|0\rangle = \langle 0|V(\phi, \phi^*)|0\rangle$ (ignoring fermion condensates[65]), with $V(\phi, \phi^*)$ the scalar potential given by Eq. (1.90), supersymmetry is broken if the VEV of one of the $F_i$- or $D^a$-terms is non-zero. The goldstino is then some linear combination of the fermionic partners of these auxiliary fields which obtain a VEV.

---

[63]See the analogy for the Higgs mechanism, where the goldstone boson in the ungauged theory is identified by the component which shifts along the Mexican hat potential under the symmetry transformation.

[64]Note that a VEV for the scalar component does not break supersymmetry. Indeed, since $Q_\alpha = i\partial/\partial\theta^\alpha - \sigma^\mu_{\alpha\dot\alpha}\theta^{\dagger\dot\alpha}\partial_\mu$, when we apply this operator on a superfield $\langle S \rangle$ with a constant expectation value for its scalar component, it gives zero. Such a ground state is therefore invariant under supersymmetry.

[65]These can appear in models of dynamical SUSY breaking, see [89].

To find the exact expression for the goldstino, we can consider the fermion mass matrix for a general supersymmetric Lagrangian. In the $(\lambda^a, \psi_i)$-basis, it is

$$\mathbf{m_F} = \begin{pmatrix} 0 & \sqrt{2} g_b \left( \langle \phi^* \rangle T^b \right)^i \\ \sqrt{2} g_a \left( \langle \phi^* \rangle T^a \right)^j & \langle W^{ji} \rangle \end{pmatrix}, \tag{1.123}$$

where the gauginos have no mass (this would break supersymmetry explicitly, since the gauge bosons are massless), the chiral fermion mass term is given by the superpotential (Eq. (1.46)) and the off-diagonal elements come from the supersymmetrised version of the gauge-difermion couplings (Eq. (1.75)). This mass matrix is annihilated by a state

$$\tilde{G} = \begin{pmatrix} \langle D^a \rangle / \sqrt{2} \\ \langle F_i \rangle \end{pmatrix}. \tag{1.124}$$

The first row is annihilated by the condition that the superpotential is gauge-invariant, given by $\delta_\epsilon W = \frac{\partial W}{\partial \phi_i} \delta_\epsilon \phi_i = \epsilon W^i (T^a \phi)_i = 0$. The second row is annihilated because $\langle \frac{\partial V}{\partial \phi_i} \rangle = 0$ in a local minimum, so with the scalar potential from Eq. (1.90), this becomes (see also the derivation in e.g. [90])

$$\begin{aligned} \frac{\partial V}{\partial \phi_i} &= W^{ij} W_j + \sum_a g_a^2 (\phi^* T^a \phi)(\phi^* T^a)^i \\ &= -W^{ij} F_j - \sum_a g_a (\phi^* T^a)^i D^a = 0. \end{aligned} \tag{1.125}$$

Therefore, we have identified a state with zero mass, the goldstino, which appears when the auxiliary fields obtain a VEV. Moreover, the weight of each fermion in the goldstino is determined by the size of these VEVs.

If we include gravity, where Poincaré transformations are only well-defined locally, supersymmetry (which is an extension of the Poincaré symmetry group) must be promoted to a local symmetry (i.e. $\epsilon^\alpha \to \epsilon^\alpha(x^\mu)$). The resulting theory, which unifies supersymmety (and thus particle physics[66]) with general relativity, is called supergravity [137–144]. The graviton (a spin-2 particle) receives a superpartner fermion (with spin 3/2) called the gravitino $\tilde{\Psi}_\mu^\alpha$ with both a tensor and a spinor index (and has $P_R = -1$).

When a local symmetry is broken, the goldstone mode is eaten by the gauge field. Since the goldstone mode in this case is a fermion and the gauge field is associated to gravity, it is the gravitino[67] which must eat the goldstino. The goldstino then becomes the longitudinal component of the gravitino, with spin $\pm 1/2$. This is known as the super-Higgs mechanism [142, 145–148]. As a result, the gravitino gets a mass $m_{3/2}$, which can be estimated as [146, 149]

$$m_{3/2} \sim \frac{\langle F \rangle}{M_P} \tag{1.126}$$

---

[66] Possibly even particle physics describing our world.

[67] Another way to see this, is as follows. The gravitino transforms as $\delta \tilde{\Psi}_\mu^\alpha = \partial_\mu \epsilon^\alpha + \ldots$ (which one can find by inspecting its structure). Since this is similar to the gauge transformation of $A_\mu$, the gravitino is like the gauge field of local supersymmetry.

for $F$-term breaking, from dimensional analysis, since its value must disappear for $\langle F \rangle \to 0$ (no SUSY breaking) and $M_P \to \infty$ (no gravitational interactions).

For collider phenomenology, only the goldstino is important. Indeed, analogous to the Goldstone boson equivalence theorem, there is now a goldstino equivalence theorem [148], such that we can think of the longitudinal components of the massive gravitino as just the goldstino. Since the massless gravitino (with spin 3/2) is part of the supergravity multiplet, its interactions with matter are suppressed by powers of $M_P$, while the spin-1/2 goldstino couples to matter component with a coupling proportional to $1/F$ (as we will see in the next chapter, Section 2.2.2). Since typically $F \ll M_P^2$, the supersymmetric Standard Model fields couple dominantly to the goldstino component.

Since the goldstino appears as the consequence of the breaking of a continuous symmetry, its couplings are dictated by supersymmetry. We will derive these exact couplings in detail in the next chapter, Section 2.2.2.

### 1.4.2 Supertrace theorem

If supersymmetry breaking occurs at tree level, including only renormalisable couplings, then the mass splitting between ordinary particles and their superpartners is strongly constrained by the structure of the theory. This is expressed by the supertrace theorem [150], which we discuss here.

Let us consider the masses of the particles and their superpartners in a general supersymmetric theory. In the previous section, we already showed the fermion mass matrix in such a theory. Now, we can also look at the squared scalar mass matrix, defined by

$$V = \frac{1}{2} \begin{pmatrix} \phi^{*j} & \phi_j \end{pmatrix} \mathbf{m}_S^2 \begin{pmatrix} \phi_i & \phi^{*i} \end{pmatrix}, \tag{1.127}$$

the eigenvalues of which give the scalar masses. This mass matrix $\mathbf{m}_S^2$ is given by

$$\begin{pmatrix} W_{jk}^* W^{ik} + g_a^2 (T^a \phi)_j (\phi^* T^a)^i - g_a T_j^{ai} D^a & W_{ijk}^* W^k + g_a^2 (T^a \phi)_i (T^a \phi)_j \\ W^{ijk} W_k^* + g_a^2 (\phi^* T^a)^i (\phi^* T^a)^j & W_{ik}^* W^{jk} + g_a^2 (T^a \phi)_i (\phi^* T^a)^j - g_a T_i^{aj} D^a \end{pmatrix}, \tag{1.128}$$

where the scalars in this matrix are replaced by their VEVs and $W^{ijk}$ is the superpotential derived three times w.r.t. the $\phi_i$. The elements of this matrix can be found by considering the scalar potential in Eq. (1.90). The terms from the chiral part are found by expanding $W^i$ to find its quadratic part in the scalars. The terms from the real superfield part are found by selecting the appropriate scalars in the second term, while replacing the other scalars by their VEVs.

Finally, also the vector masses must be taken into account, which can obtain masses through the Higgs mechanism, from the covariant derivatives on the scalar fields in Eq. (1.75) if these get a VEV. Their squared mass matrix is given by $\mathbf{m}_V^2 = g_a^2 (\phi^* \{T^a, T^b\} \phi)$.

The supertrace is then defined as the sum of the squared masses over particles of spin $j$, weighted by their spin multiplicities. It is given by

$$
\begin{aligned}
\text{STr}(m^2) &\equiv \sum_j (-1)^{2j}(2j+1)\text{Tr}(m_j^2) \\
&= \text{Tr}(\mathbf{m_S^2}) - 2\text{Tr}(\mathbf{m_F^\dagger m_F}) + 3\text{Tr}(\mathbf{m_V^2}) \\
&= -2g_a\text{Tr}(T^a)D^a = 0.
\end{aligned}
\tag{1.129}
$$

The last equality holds in the absence of gravitational anomalies [61] (i.e. the traces of $U(1)$ charges over chiral superfields are zero).

Consider now for simplicity a supersymmetric theory of chiral superfields only, including only renormalisable terms, which is spontaneously broken at tree-level. The supertrace theorem then implies that the masses of the two scalars in a chiral superfield must be shifted equally up and down compared to the fermions. Therefore, if reality is described by a supersymmetric extension of the Standard Model in which supersymmetry is broken at tree level, we should have observed superpartners of the Standard Model fermions at masses below these fermions. Since no such particles have been observed, such a model can not be an appropriate description of our reality. We will discuss in Section 1.4.4 how we can build models for supersymmetry breaking which can solve this issue.

### 1.4.3  *F*- and *D*-term breaking

Before we move on to how one can circumvent the supertrace theorem, we first discuss how one can construct a supersymmetry-breaking theory. As seen in Section 1.4.1, SUSY breaking occurs when the vacuum has non-zero energy, which occurs when the *F*- or *D*-terms acquire a VEV. Therefore, determining whether a theory has a vacuum which conserves or breaks supersymmetry is equivalent to finding solutions to the equation

$$
V(\phi, \phi^*) = F^{*i}F_i + \frac{1}{2}\sum_a D^a D^a = W_i^* W^i + \frac{1}{2}\sum_a g_a^2(\phi^* T^a \phi)^2 = 0.
\tag{1.130}
$$

Depending on which of the two terms appearing here is responsible for supersymmetry breaking, one speaks of *F*-term of *D*-term breaking.

Consider first a theory with only chiral superfields. Finding supersymmetric ground states is then equivalent to finding solutions to the equation $\partial W(\phi^i)/\partial\phi^i = 0$ for the superpotential. Since this is a set of $N$ complex equations in $N$ complex variables, there will generically[68] be a solution to these equations, i.e. supersymmetry is unbroken. However, if the superpotential is somehow constrained, this might no longer be true. It turns out (see the discussion in e.g. [89]), that making the theory invariant under a gauge symmetry does not change this conclusion. However, if one imposes a symmetry under which the components of the multiplets transform differently, supersymmetry can be generically broken. Such a symmetry is called an *R*-symmetry.

---

[68]Generic means that with arbitrary small change in the parameters of the theory, any point with no solution to this equation, will turn into a theory which does admit a solution [89].

More concretely, an *R*-symmetry generator is one that does not commute with the supersymmetry generators $Q$ and $Q^\dagger$. While the Coleman-Mandula theorem (see Section 1.2.1) requires that any global symmetry commutes with the Poincaré group, the same is not true for the supersymmetry algebra. It turns out that there can be at most one $U(1)_R$-symmetry. It is typically normalised such that $Q$ has $R$-charge $-1$ and $Q^\dagger$ has $+1$, such that the *R*-symmetry generator satisfies $[R, Q] = -Q$ and $[R, Q^\dagger] = Q^\dagger$. If the lowest component of a chiral superfield has $R$-charge $R(\phi) = r$, then the other components have $R(\psi) = r - 1$ and $R(F) = r - 2$, so that $\theta$ and $\theta^\dagger$ transform with charges $+1$ and $-1$ respectively. A superfield has the same charge as its lowest component[69]. In order to have a Lagrangian which is invariant under $U(1)_R$, the superpotential (which appears under $\int d\theta^2$) must have $R(W) = +2$. Since this is not the case for any arbitrary superpotential, some terms in Eq. (1.51) will be forbidden if this *R*-symmetry exists. It turns out (see e.g. [89]), that if this *R*-symmetry is spontaneously broken[70,71], then generic superpotentials will no longer have a solution to Eq. (1.130), such that supersymmetry must be broken. This general result is known as the Nelson-Seiberg theorem [152]:

> The existence of an *R*-symmetry is a necessary condition for SUSY breaking and a spontaneously broken *R*-symmetry is a sufficient condition provided two conditions are satisfied. These conditions are: *genericity*, i.e. the effective Lagrangian is a generic Lagrangian consistent with the symmetries of the theory (no fine tuning), and *calculability*, i.e. the low energy theory can be described by a supersymmetric Wess-Zumino effective Lagrangian without gauge fields.

A model for *F*-term supersymmetry breaking, or O'Raifeartaigh model [153], is given by the superpotential

$$W = hX(\Phi_1^2 - \mu^2) + m\Phi_1\Phi_2, \tag{1.131}$$

which is invariant under an $U(1)_R$-symmetry with charge assignments $r_{\phi_1} = r_{\phi_2} = 2$ and $r_X = 0$. Using Eq. (1.130), we find that for this superpotential, supersymmetry is indeed broken. Explicitly solving for the spectrum, there appear two massless particles. The scalar $\phi_1$ is massless because its value is undetermined, which corresponds to a flat direction in the potential[72]. However, this degeneracy is typically lifted by quantum corrections (thus giving the scalar a mass) since it is not protected by any symmetry (see e.g. [89]). On the other hand, there also appears a goldstino, which must always be massless. All other scalars and fermions are non-degenerate in mass, as required by broken supersymmetry. Note that the field $X$ must be a gauge singlet to give an invariant Lagrangian and the linear term is always necessary[73] in a normalisable models, otherwise $\{X = 0, \Phi_i = 0\}$ is always a solution. The scale of SUSY breaking in this model is determined by $h\mu^2$, which is arbitrarily imposed here. Therefore, there is no a-priori reason for it to be smaller than $M_P$. In models of dynamical supersymmetry breaking, i.e. in a strongly coupled theory with effective dynamics at lower energy, such small scales appear through dimensional transmutation (similar to generation of $\Lambda_{\mathrm{QCD}}$ in QCD). For a review of dynamical supersymmetry breaking, which is outside the scope of this work, see [154, 155].

---

[69]For the vector superfield $V$, the $R$-charge must vanish, since it is real. Therefore, the gauginos $\lambda$ must have $R$-charge 1. In other words, the Majorana gaugino mass $\frac{1}{2}M_\lambda\lambda\lambda$ necessarily breaks $U(1)_R$.

[70]This implies the appearance of an associated goldstone boson, which is not observed. Gravitational effects may make it massive enough to not be ruled out [151]. Otherwise, it turns out that quantum corrections can give rise to "non-generic" superpotentials which break supersymmetry without requiring an *R*-symmetry (see e.g. [89]).

[71]*R*-parity can survive as a remnant of broken *R*-symmetry. Note that *R*-parity secretly *does* commute with supersymmetry, since it differs from matter parity only by spin.

[72]Such flat directions are known as moduli.

Finally, it is also possible to have $D$-term or Fayet-Iliopoulos breaking [110, 156]. Such a breaking can occur if the Lagrangian contains a term $\mathcal{L}_{\mathrm{FI}} = -\kappa D$ (only possible for an Abelian symmetry), which can induce a VEV for the $D$ auxiliary field, since there is an extra $\kappa$ in Eq. (1.89). If the scalars charged under this symmetry have non-zero masses (i.e. there appears a term $\sum_i |m_i|^2 |\phi_i|^2$ in the scalar potential), then there is no solution to Eq. (1.130) and supersymmetry is broken.

More generically, when one has a theory with both chiral and vector superfields, one finds that if the $F$-term vacuum equations have a solution, then in the absence of a Fayet-Iliopoulos terms, there exists always a simultaneous solution to the $D$-term vacuum equations, i.e. there is a supersymmetric minimum. On the other hand, if there is a spontaneously broken $U(1)_R$ symmetry, then SUSY is generically broken, and the $D$-term vacuum equations just add additional constraint on the $F$-terms (see the review [89]).

It turns out that models with $F$-term SUSY breaking are often more phenomenologically interesting, so from now on we always assume that SUSY breaking occurs in this way, neglecting the possibility for $D$-term breaking.

### 1.4.4 Supersymmetry-breaking mechanisms

From the discussion in the previous sections, we know that it is not possible for supersymmetric extensions of the Standard Model to break supersymmetry at tree level. Generating gaugino masses is difficult, since there is no scalar-gaugino-gaugino coupling that can turn into a mass term when the scalars acquire a VEV. Moreover, using $U(1)_Y$ for $D$-term SUSY breaking does not give an acceptable spectrum and there is no MSSM gauge singlet which can induce $F$-term breaking. Most importantly, however, the supertrace theorem requires some scalars to be lighter than the fermions, which is not observed.

However, there is a way to "save" supersymmetry. If SUSY breaking occurs instead in a so-called hidden sector, with no renormalisable, tree-level couplings with the observable sector (i.e. some supersymmetric extension of the Standard Model), as shown in Figure 1.3, the supertrace theorem can be violated. Indeed, the supertrace theorem implicitly requires renormalisability, since renormalisability imposes the minimal (canonical) form of the kinetic terms. In a non-renormalisable theory, the kinetic terms for the matter and gauge multiplets can take a non-canonical form, which can induce scalar and gaugino masses. Therefore, the goal is to build an effective low-energy theory (which includes the goldstino), where high-mass fields have been integrated out, which has a non-vanishing supertrace (see the excellent discussion in [136]).

---

[73]Non-polynomial superpotentials from non-perturbative effects can circumvent this, see [67].

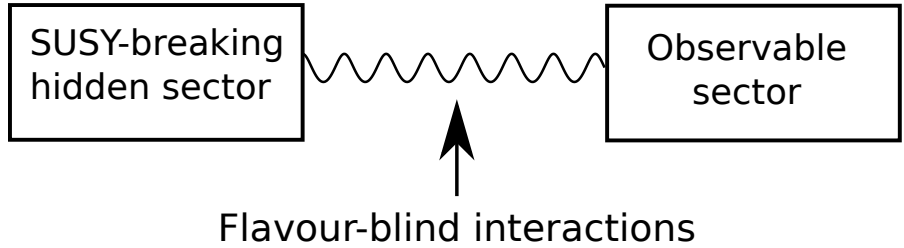

Figure 1.3: Supersymmetry breaking in a hidden sector, communicated to the observable sector (e.g. the MSSM) through some interactions.

There are two avenues[74] which can be pursued in order to achieve this. The first is to consider a theory which is altogether non-renormalisable, so the whole spectrum has a non-vanishing supertrace. Such a theory is given by supergravity [157–162], since gravitational interactions are non-renormalisable (see also the reviews [163–165]). The second way is to consider a theory which is renormalisable at tree level, but is described at low energy by an effective theory with non-renormalisable kinetic terms, induced by quantum corrections due to gauge interactions. This scenario is known as gauge-mediated supersymmetry breaking. In this section, we will briefly sketch how these two mediation scenarios work. While each of these also predicts their own typical mass spectra, a review of this is outside the scope of this thesis. For discussions, see [67, 166].

**Gravity-mediated supersymmetry breaking**

First, we discuss gravity-mediated or Planck-scale mediated supersymmetry breaking (PMSB) [157, 159, 160, 167–170]. If supersymmetry is indeed broken in a hidden sector, this way of communicating the breaking must always be present, though not necessarily dominant, since gravity couples to everything. Moreover, if observed, it is the first time that gravity plays a role in particle physics. For these two reasons, this is the standard scenario considered for SUSY breaking in the MSSM.

We can immediately estimate the size of the soft masses induced by gravity mediation. Since gravitational interactions are suppressed by the Planck scale $M_P$, dimensional analysis gives

$$m_{\text{soft}} \sim \frac{\langle F \rangle}{M_P},\tag{1.132}$$

so that the theory is unbroken for $\langle F \rangle \to 0$ and $M_P \to \infty$ (i.e. when gravitational interactions are turned off). In order to get soft masses slightly above the weak scale, $\mathcal{O}(100 \text{ GeV})$, we therefore approximately need SUSY breaking of size $\sqrt{\langle F \rangle} \sim 10^{10} - 10^{11}$ GeV.

---

[74]There are additional possibilities. However, we will discuss here only the two most important ones. For alternatives, see the review [67].

Now we make the procedure of how to obtain the soft masses more concrete. As we saw in Section 1.2.4, non-renormalisable interactions, in this case suppressed by powers of $1/M_P$, can be included by considering more general functions for the superpotential, Kähler potential and gauge kinetic function, which we can expand in powers of $1/M_P$. We can then obtain the supersymmetry-breaking effects by considering the so-obtained higher-dimensional operators involving a superfield $X$, of which the $F$-term obtains a VEV to break SUSY, which connect the hidden sector with the observable sector. This $X$ appears as $\sim \frac{1}{M_P} X$ (and similar for $X^*$ in the Kähler potential), in combination with the MSSM superfields. As mentioned before in Section 1.2.3, we can generalise the coupling constants appearing in the Lagrangian to spurion superfields, where the scalar components are the usual coupling constants which respect SUSY. Treating now these chiral superfields $X$ as spurions, we can obtain the SUSY-breaking terms by performing the substitutions

$$X \to \theta\theta F, \qquad X^* \to \theta^\dagger\theta^\dagger F^*, \tag{1.133}$$

with $F = \langle F_X \rangle$. In this way, for example, the gaugino masses are automatically generated from $f_{ab} = \frac{\delta_{ab}}{g_a^2}(1 - \frac{2}{M_P} f_a X + \dots)$ to give

$$\mathcal{L}_{\text{soft}}^{\text{gaugino}} = -\int \mathrm{d}\theta^2 \frac{X}{2M_P} f_a \mathcal{W}^a \mathcal{W}^b = -\frac{F}{2M_P} f_a \lambda^a \lambda^a. \tag{1.134}$$

Similarly, we obtain soft masses for the squarks from

$$\int \mathrm{d}\theta^4 \frac{z_Q}{M_P^2} X^\dagger X Q^\dagger Q \tag{1.135}$$

(and similar for the Higgs), as well as the $b$-term

$$\int \mathrm{d}\theta^2 \frac{b}{M_P} X H_u H_d. \tag{1.136}$$

In this way, it is also clear that indeed the soft masses have a size which can be estimated as $M_{\text{soft}} \sim \frac{\langle F_X \rangle}{M_P}$.

The estimate for the soft masses in Eq. (1.132) is the same as that of the gravitino (Eq. (1.126)). Therefore, the gravitino must be at the same energy scale as the other superpartners. However, because all its interactions with other particles are gravitational, it is irrelevant for collider phenomenology[75].

However, there is one important issue with gravity-mediated scenarios, related to flavour. While, in principle, interactions in general relativity are flavour-blind (except for the already-present Yukawa couplings), general relativity is only an effective low-energy theory. There must be a UV-completed theory and, from what is known from string theory, such a theory is unlikely to respect global symmetries such as the $U(3)^5$ flavour symmetry[76]. Moreover, suppose that the flavour-symmetry breaking already present in the Standard Model, i.e. the Yukawa-couplings, is generated at some scale $\Lambda_F$, which must be below or at the Planck scale, i.e. $\Lambda_F \lesssim M_P$. Above $\Lambda_F$ lies some dynamics which is responsible for flavour-symmetry breaking, while below $\Lambda_F$ this dynamics is frozen and only visible through the Yukawa couplings. There is then no reason why the soft terms in gravity mediation, which are generated at a scale above $\Lambda_F$, must be flavour invariant.

---

[75]It can still be important in cosmology [171, 172].

Such flavour-breaking contributions from the soft terms are dangerous[77] [71, 174, 175], since there exist strong bounds on the splittings between sparticles of different generations [176, 177]. Although there are models of gravity-mediated SUSY breaking where there these interactions are flavour-invariant (see references in [136]), typically these models are at best still at the edge of being allowed [178, 179].

**Gauge-mediated supersymmetry breaking**

Next, we discuss the case of gauge-mediated supersymmetry breaking (GMSB) [180–185] (for a review, see e.g. [135, 136]), where the interactions communicating SUSY breaking to the observable sector are the Standard Model gauge interactions. In this type of models, supersymmetry is broken in a hidden sector, which has tree-level interaction with some "messenger" fields. Messengers are chiral superfields which are charged under $SU(3)_C \times SU(2)_L \times U(1)_Y$ and due to their coupling to a supersymmetry-breaking VEV $\langle F \rangle$, their mass spectrum violates supersymmetry. The MSSM soft terms are then generated through loop diagrams involving messenger particles. From dimensional analysis, we can already estimate the induced soft masses as

$$m_{\text{soft}} \sim \frac{g_a^2}{16\pi^2} \frac{\langle F \rangle}{M_{\text{mess}}}. \tag{1.137}$$

So if $\langle F \rangle$ and $M_{\text{mess}}$ have a similar scale, then the scale of supersymmetry breaking can be down to $\sqrt{\langle F \rangle} \sim 10^4$ GeV with the lowest possible mass scale only slightly above this (see below). Therefore, supersymmetry in this type of models occurs at much lower scales than in models of gravity mediation.

Generally, there are many possibilities in which the messenger can obtain a broken mass spectrum, either in O'Raifeartaigh models [183–185] (where the scale is set by hand) or in dynamical SUSY breaking [183–185] (where it is not). However, the exact way in which the messengers obtain their SUSY-breaking masses is not important. In minimal gauge mediation, sketched in Figure 1.4, we parameterise SUSY breaking with a field $X$ which obtains a VEV for its $F$-component (e.g. with the model in Eq. (1.131), but we do not need to specify this). If this field $X$ is coupled to some chiral messenger fields $\Phi$ and $\tilde{\Phi}$ as $(M + \lambda X)\Phi\tilde{\Phi}$, it will give a SUSY-breaking mass to the components of these messenger fields. More concretely, the messenger Lagrangian has the form

$$\mathcal{L}_{\text{messenger}} \supset \int \mathrm{d}\theta^2 \, \mathcal{M}\tilde{\Phi}\Phi + \text{c.c.}, \tag{1.138}$$

with

$$\mathcal{M} = M + \theta\theta F. \tag{1.139}$$

---

[76]$U(3)$ for the three generations, five times for $Q$, $\bar{u}$, $\bar{d}$, $L$, and $\bar{e}$.

[77]For phenomenological analyses of the MSSM, one often considered models based on gravity mediation, with the additional assumption that there are no flavour-violating effects from the soft masses, that all gaugino masses are equal and the sfermion masses are equal. In this way, the huge MSSM parameter space is reduced to 4 parameters only. This is called minimal supergravity (MSUGRA) or Constrained Minimal Supersymmetric Standard Model (CMSSM) [173] (see also the reviews[67, 165]). While extremely predictive, this model has now been hopelessly excluded by the LHC, which is why this previously very important model has now been reduced to a footnote.

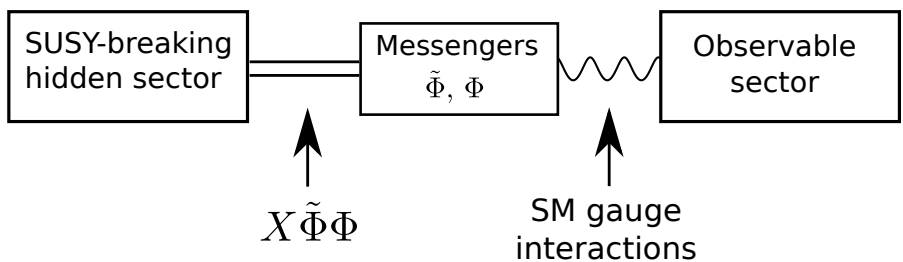

Figure 1.4: Minimal gauge mediation model, where supersymmetry breaking in some hidden sector, parameterised by a field $X$ which obtains a VEV $\langle X \rangle = \theta^2 F$, is communicated to the observable sector through some messenger fields $\tilde{\Phi}, \Phi$.

In order to ensure that gauge coupling unification remains, these messenger fields can be taken in $SU(5)$ multiplets, e.g. in $\mathbf{5} \oplus \overline{\mathbf{5}}$. Integrating out the auxiliary fields $F_\phi$ and $F_{\tilde{\phi}}$ (i.e. solving their equations of motion), we find

$$\mathcal{L}_{\text{messenger}} \supset -|M|^2(\phi^\dagger\phi + \tilde{\phi}^\dagger\tilde{\phi}) + (F\tilde{\phi}\phi + \text{c.c.}). \tag{1.140}$$

Making $F$ real with appropriate rotation of the scalar fields, we find the mass eigenstates $(\phi \pm \tilde{\phi})/\sqrt{2}$ with squared masses $|M|^2 \pm F$ (which also means that for stability of the vacuum, we must have $F \leq |M|^2$; this explains why the lowest messenger mass scale must be slightly above $\sqrt{F}$), while the fermions still have the masses $|M|$. This shows explicitly that the messenger mass spectrum breaks supersymmetry (but respects the supertrace theorem!).

The MSSM fields then receive their masses through loops involving the messengers, shown in Figure 1.5. For the gauginos, the leading contribution comes from 1-loop diagrams, set by the gauge couplings, with the result [183–185]

$$M_a \sim \frac{g^2}{16\pi^2}\frac{F}{M}. \tag{1.141}$$

The gauge boson masses receive no such contributions, since they are protected by gauge invariance. The scalars get mass through two-loop diagrams (since they do not couple directly to the messenger fields, but need to go through gauge interactions). The result is

$$m_{\tilde{f}}^2 \sim \left(\frac{g^2}{16\pi^2}\right)^2 \left(\frac{F}{M}\right)^2. \tag{1.142}$$

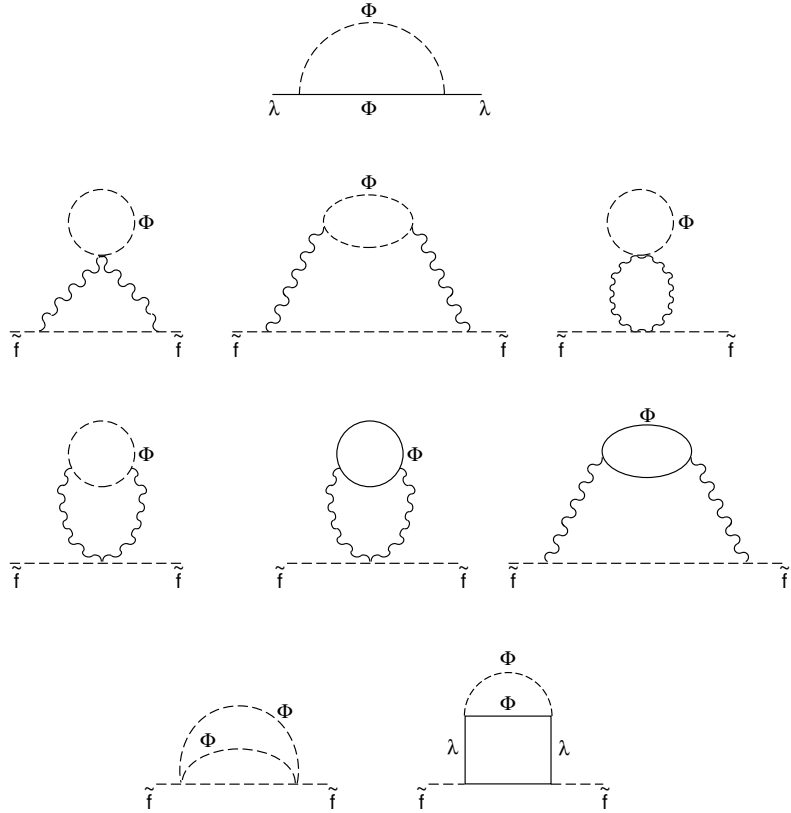

Figure 1.5: Feynman diagrams generating the SUSY-breaking gaugino mass ($\lambda$) and sfermion mass ($\tilde{f}$) at one- and two-loop respectively. Figure from [136].

These values agree with our estimate in Eq. (1.137). Moreover, the soft masses for the scalars and gauginos are at a similar order of magnitude (since the scalar masses have a square on both sides). The exact values of the soft masses (see [136]) can be found by explicit loop computation. Note that these expressions are valid at the messenger scale *M*. In order to obtain the values of the soft masses at a lower energy scale, their running due to quantum corrections needs to be included. In other words, the values of these masses need to be evolved down using the renormalisation-group equations. One can also use a simpler and more systematic method [186] to obtain the masses and the running. This makes use of the RG equations from the beginning, by replacing the coupling constants appearing there by superfields, the *F*-component of which contains the soft masses. A discussion of the typical mass spectra generated by gauge mediation models is outside the scope of this thesis. For a nice review, see [166].

There are different variations of gauge-mediation models (see [136, 187]). The discussion above corresponds to minimal gauge mediation, where a field *X* in some hidden sector obtains a VEV, which is communicated through a number of messenger fields *N* to the observable sector. Another variation is direct gauge mediation, where the supersymmetry-breaking sector is explicitly included in the model. Finally, there is also general gauge mediation [187], where instead of using messengers, the soft masses are computed from the contribution of a set of correlation functions of real superfields representing the hidden sector to the MSSM gauge currents. These contributions are determined completely by the gauge structure and make no reference to the details of the hidden sector or how the breaking is communicated. However, this does not imply that this way necessarily captures more possibilities than minimal models, only that it gets it for "free".

In gauge-mediation models, the flavour problem is solved automatically, which is one of its most attractive features. The squark and slepton masses depend only on their gauge quantum numbers and so the different generations are degenerate in mass (at the scale $M$), suppressing flavour-changing effects. More generically, there can be no flavour-violation from the hidden sector, since gauge mediation happens at a low scale, i.e. $M \ll \Lambda_F$, where the dynamics of the flavour sector is frozen and its remaining effect residing only in the Yukawa couplings. Any contribution of additional flavour-violating effects must thus be suppressed by powers of $M/\Lambda_F$. Therefore, there is a super-GIM mechanism at work, involving ordinary particles and their supersymmetric partners, and flavour-violating effects are expected to be negligible.

Note that gravity-mediation still occurs, from the same $F$-term VEV as the one which is responsible for the gauge-mediated soft SUSY breaking, since gravitational interactions are always present. If these contributions are sizeable, the flavour problem returns. However, gravity-mediation effects can be neglected if their contribution to the soft masses is negligible. This occurs approximately when[78]

$$\frac{g^2}{16\pi^2}\frac{\langle F \rangle}{M_{\text{messenger}}} \geq 10^{3/2}\frac{\langle F \rangle}{M_P},$$        (1.143)

or

$$M \leq \frac{g^2}{16\pi^2}10^{-3/2}M_P \sim 10^{15}\,\text{GeV}.$$        (1.144)

In other words, the upper bound on $M$ is up to the GUT scale, meaning that for any reasonable model of gauge mediation we can ignore gravity-mediation effects.

In models of gauge mediation, the gravitino has a much lower mass than in gravity mediation, because its mass is still estimated as in Eq. (1.126), but now $\langle F \rangle$ is much smaller. As a result, the gravitino is typically nearly massless and thus almost certainly the lightest supersymmetric particle (LSP). As mentioned in Section 1.4.1, the gravitino inherits the goldstino interactions. These are suppressed by $F$ (see Section 2.2.2 in the next chapter), which is however much smaller than in the gravity-mediated case. Therefore, the goldstino is relevant for the collider phenomenology and the main signature of gauge-mediated models. Indeed, since the goldstino is always the LSP, but has relatively small interactions with all other particles in the observable sector, all other particles decay first to the next-to-lightest supersymmetric particle (NLSP), which subsequently decays into the goldstino. The signature of gauge-mediated models of supersymmetry is then mainly determined by the nature of the NLSP, with varying phenomenology when it is, for example, the stau or one of the neutralinos. In the end, the NLSP decays into its Standard Model partner along with the neutral goldstino, which is only "detected" as missing energy. The decay width of a sparticle into the goldstino is given by (see e.g. [67, 88])

$$\Gamma(\tilde{X} \to X\tilde{G}) = \frac{m_{\tilde{X}}^5}{16\pi\langle F\rangle^2}\left(1 - \frac{m_X^2}{m_{\tilde{X}}^2}\right)^4,\tag{1.145}$$

where the main part[79] can be estimated from dimensional analysis, since the coupling of the goldstino is $\propto 1/\langle F\rangle$. The collider phenomenology of the goldstino will be important in the next chapter.

---

[78]Note that we are being sloppy here. For the SUSY-breaking mass communicated through the messengers and the gauge fields, there is an extra coupling $\lambda$ in front of $F$, from $\lambda X \bar{\Phi} \Phi$. For the gravity-mediation part, this coupling is not there, such that the breaking is induced by the "pure" $F$-term. Therefore, this $\lambda$ can somewhat change the estimate.

## 1.5  Searching for BSM physics

Beyond the Standard Model (BSM) physics scenarios, supersymmetry or otherwise, which are built to solve theoretical or observational channels at the electroweak scale, give, at least in general, testable predictions which can be falsified at current experimental facilities. These experiments cover a wide range of disciplines, including collider physics, heavy-flavour physics, precision experiments, neutrino physics, astrophysics, cosmology and astroparticle physics.

However, here, we focus on collider searches only. In these searches, particles (protons, electrons,...) are accelerated to high energy and made to collide. In these collisions, a lot of energy is pumped into the interactions of fundamental particles and new, thus-far unknown, particles are expected to appear if they are kinematically accessible and have sizeable interactions with the Standard Model.

### 1.5.1  Collider physics at the LHC

The Large Hadron Collider (LHC) [188], shown in Figure 1.6, is a circular particle accelerator which accelerates and collides protons[80] at centre of mass energies $\sqrt{s}$ of 7, 8, 13 and 14 TeV. It is the successor to the Large Electron-Positron Collider (LEP) [189–192], which collided electron-positron pairs at a centre of mass energies from 90 GeV up to 209 GeV, and the Tevatron [193, 194], which collided protons with anti-protons at a centre of mass energy of 2 TeV.

The timeline of LHC operations is shown in Figure 1.7. Run 1, which started in 2011 ($\sqrt{s} = 7$ TeV) and proceeded in 2012 ($\sqrt{s} = 8$ TeV), gathered in total $\mathcal{L} \sim 30$ fb$^{-1}$ of data (the "integrated luminosity"). This run saw the discovery of the Higgs boson [27, 28], completing the particle content of the Standard Model[81]. Run 2 started in 2015 and lasted until 2018, operating at $\sqrt{s} = 13$ TeV and gathering about $\mathcal{L} \sim 160$ fb$^{-1}$ of data (more than the projected amount of 150 fb$^{-1}$) [196]. In the near future, with Run 3, the LHC will run with increased luminosity, gathering twice as much data as in Run 2, while the planned High-luminosity LHC promises a tenfold increase in the amount of data.

---

[79]Half of the second factor is from derivatives in the interaction term evaluated on-shell, while half of it is from phase space integration.

[80]As well as heavy ions for some part of its run.

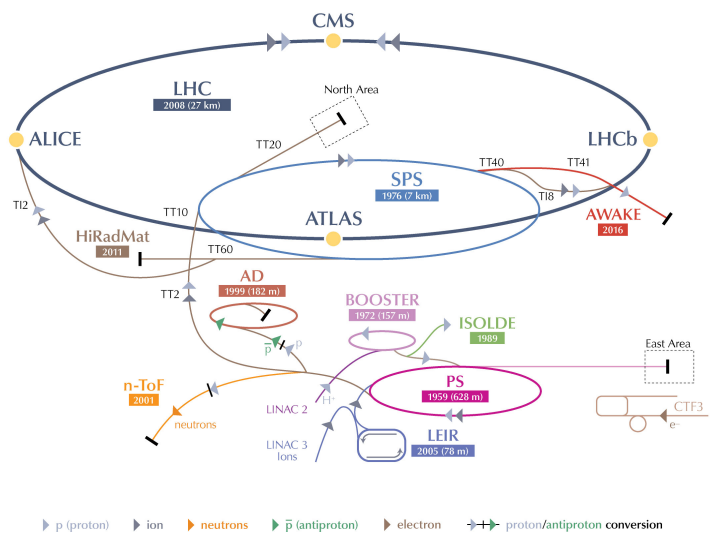

Figure 1.6: The CERN accelerator complex, featuring the Large Hadron Collider. Figure from [195].

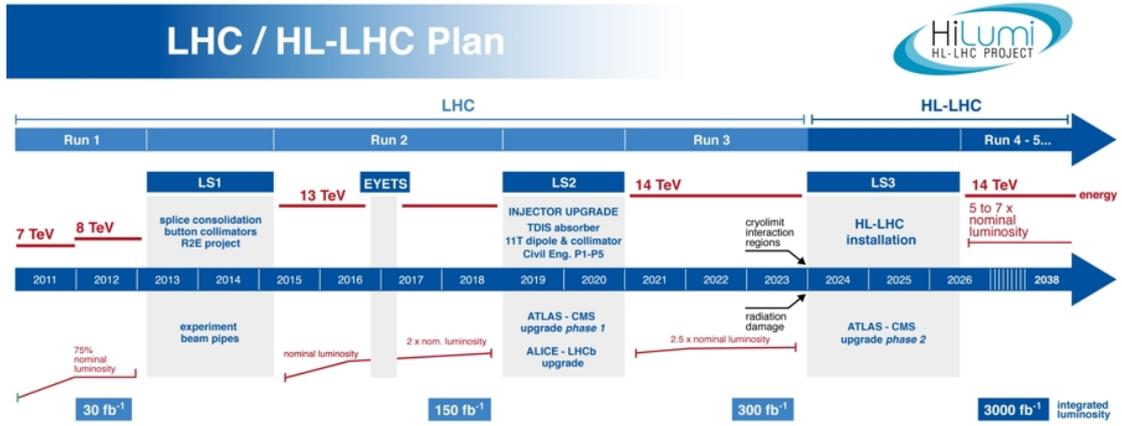

Figure 1.7: Timeline of LHC operations, showing the runs which occurred up to now and the planned future operations, as well as the associated gathered luminosities. Figure from [197].

The advantage of accelerating protons, compared to electrons/positron at LEP (which was built in the same tunnel), is that protons can be accelerated to higher energy[82] given a fixed radius of the accelerator. However, this also comes at a cost: since protons are hadrons, their collisions are inherently "messy", i.e. the collision events contain a lot of hadronic activity. This is clear in Figure 1.8, where a typical proton-proton collision is shown. The main scattering process involves quarks or gluons in the initial state. After the collision, all of the quarks and gluons in the "final state" hadronise, creating showers of strongly interacting particles such as pions, protons and neutrons which collectively form jets. In addition, there is also initial state radiation and final state radiation which also undergo hadronisation. Moreover, protons are accelerated and collided in bunches of about $10^{11}$ particles [198], such that multiple collisions occur at the same time (called pile-up[83]), most of which are pure-QCD events[84]. Finally, since it is the partons inside the proton which partake in the main scattering process, their exact energy is unknown. Therefore, for analyses one only knows that the transverse momentum of the total event is zero to a good approximation. Consequently, hadron colliders are typically considered "discovery machines", while lepton colliders are typically considered "precision machines", although the analysis techniques and computing power have advanced enough to partially compensate for this.

---

[81]The full Standard Model is not yet complete, since the parameters of the Higgs potential and light-flavour Yukawa couplings are not yet (accurately and independently) measured.

[82]Due to their higher mass, protons emit less synchrotron radiation as they are bent in the beam pipe, leading to less energy losses than for electrons.

[83]In 2017, the mean number of interactions per bunch crossing increased to 37, see e.g. [199].

[84]I.e. they are not the primary target when performing new physics searches, but still contain QCD physics which is not always well known. This is important to understand the underlying event in other physics analyses. These properties are studied in minimum-bias events which do not perform any selection on the event properties which are recorded, but are randomly saved from the collision events which occur

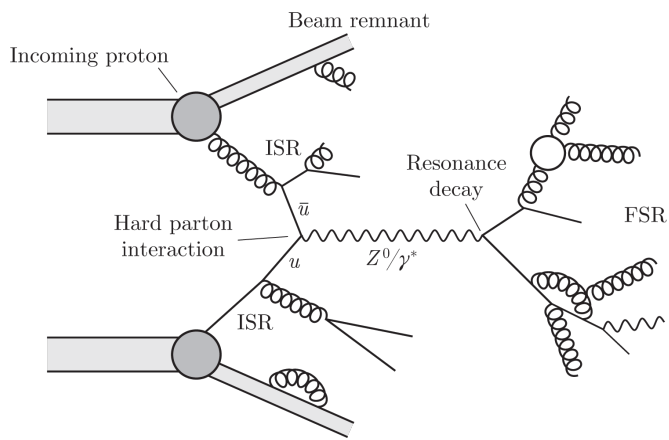

Figure 1.8: Schematic of a typical proton-proton collision at the LHC, showing the hard interaction producing a $Z/\gamma^*$, its decay into leptons or quarks and initial and final state radiation. Figure from [200].

The cross sections[85] for various Standard Model cross sections, such as $pp \to Z$, $pp \to t\bar{t}$ and large transverse momentum jets $pp \to gg$, $pp \to qg$, $pp \to qq$, etc. are shown in Figure 1.9. The total proton-proton cross section is around $\sigma_{pp} \sim 80$ mb at LHC energies. It is immediately clear that even the inelastic collisions are still dominated by strong-interaction processes, while the more interesting Standard Model processes, for example Higgs production, are relatively rare. From these cross sections, one can estimate that e.g. the number of top quark pairs produced during Run 2 $N_{t\bar{t}} = \mathcal{L}\sigma_{t\bar{t}}$ is about 128 million.

Along the beam line at specific collision points, the LHC has 4 large particle detectors: CMS [203] and ATLAS [204], focused mainly on the search for new particles at high masses and the focus in this thesis, LHCb [205] which investigates $b$-physics ($CP$ violation visible in $b$-hadron decays) and ALICE [206] which is built to study the the quark-gluon plasma. In addition, there are also three smaller experiments: TOTEM [207], which seeks to measure the total $pp$ cross section and operates in the forward regime[86] $3.1 \leq |\eta| \leq 6.5$, MoEDAL [208] which searches for magnetic monopoles or other highly-ionising (pseudo-)stable particles and LHCf [209] which measures neutral particles in

---

(see also below).

[85] As a reminder, we have the following conversion between units: 1 nb $= 10^3$ pb $= 10^6$ fb $= 10^{-33}$ cm$^2$.

[86] The pseudorapidity is defined as $\eta = -\ln\tan(\theta/2)$, where $\theta$ is the polar angle relative to the beam direction [203].

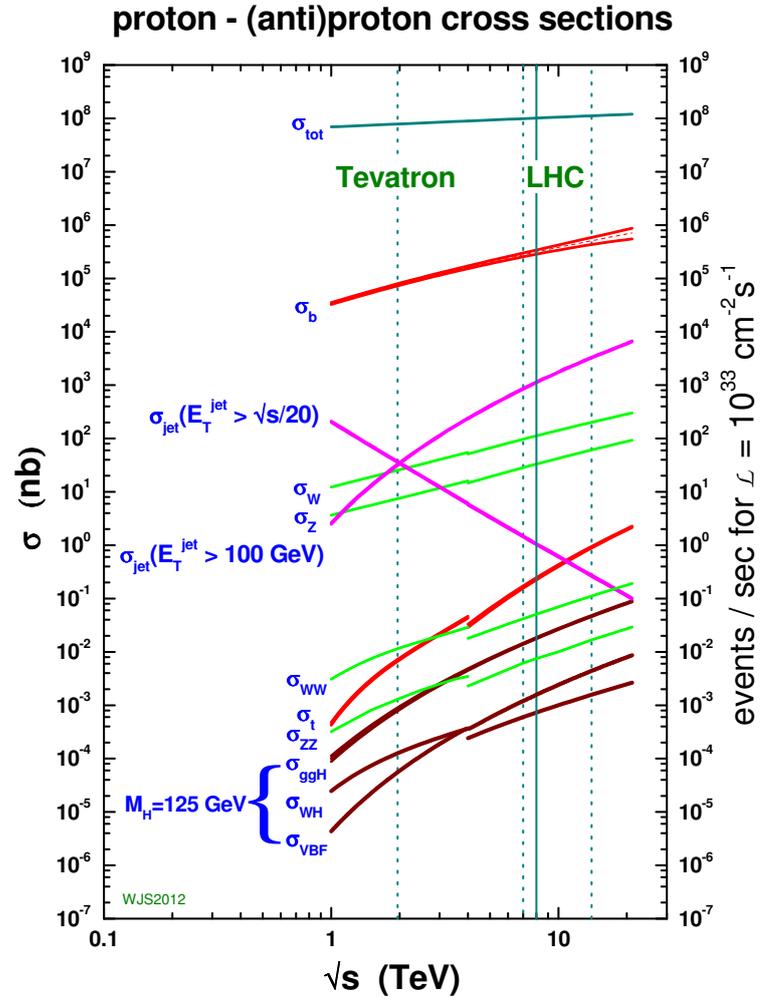

Figure 1.9: Cross sections of various Standard Model processes as a function of centre of mass energy $\sqrt{s}$. The discontinuities at $\sqrt{s} = 4$ TeV are due to the switch from $p - \overline{p}$ collisions to $p - p$ collisions. For reference, the peak instantaneous luminosity in 2017 was of the order $\mathcal{L} \sim 10^{34}$ cm$^{-2}$ s$^{-1}$ [198]. Figure from [201]; see also the review in [202].

the forward regime ($|\eta| > 8.4$).

The CMS (Compact Muon Solenoid) and ATLAS (A Toroidal LHC ApparatuS) detectors are shown in Figure 1.10. They are cylindrical detectors, almost completely hermetically closed[87], surrounding the collision point. A slice of the CMS detector[88] is shown in Figure 1.11. Using layers of different materials and detection techniques as well as a magnetic field, it is able to identify the different particles and measure or reconstruct their energy and momentum. The inner layer is the tracker, which can track the path of charged particles (such as electrons and muons). Subsequently, electrons and photons deposit all their energy in the electromagnetic calorimeter, allowing their energy to be measured. Similarly, hadronic particles such as proton, neutrons, pions and kaons deposit all of their energy in the hadronic calorimeter. Finally, muons, which due to their mass undergo few ionising interactions while passing through matter, are detected in the outermost muon chambers. As a consequence of the high event rate, it is impossible to save all events. Instead the detector "triggers" on certain pre-defined interesting signatures (e.g. high hadronic activity, the presence of photons, large transverse momentum imbalance,...), saving only those events. Using the various detector elements, it is possible to detect and identify all the Standard Model particles, except for neutrinos which interact only weakly. The final reconstructed objects, of which the energy and momentum are known, are electrons, muons, photons, and jets. Finally, also the missing transverse momentum $\not{E}_T = -|\sum_i \mathbf{p}_i|$ can then be determined. Together, these observables can be used to select interesting events (in addition to the minimum-bias events mentioned before) to either measure Standard Model parameters in increasing detail or search for BSM physics.

### 1.5.2 Current status of SUSY and BSM physics

Besides searching for the Higgs boson and measuring in more detail the top-quark properties (such as $\sigma_{t\bar{t}}$ and $m_t$ [213–215], single top quark production [214, 216] and the search for production of four top quarks [217]), one of the main goals of the LHC is to search for physics beyond the Standard Model. While some fluctuations were seen in the data[89], such as an excess in events with a lepton pair from $Z$-boson decay, jets and missing energy (discussed in the next chapter) and a 750 GeV diphoton excess [219, 220] which has since disappeared [221–224], no BSM physics has currently been detected. As such, limits have been put on the parameter space of many models, far outclassing previous limits from direct searches at colliders.

Seeing as supersymmetry is[90] one of the main candidates for new physics around the electroweak scale, a lot of effort has gone into detecting its signatures, covering as many corners of the parameter space as possible. Figure 1.12 shows the cross section[91]

---

[87]The CMS detector has detector elements up until $|\eta| \leq 5$ (where $\eta = 0$ is the central region) and the transition between the barrel and endcap region, visible in Figure 1.10a, is between $1.4 < |\eta| < 1.6$.

[88]The detection principle of the ATLAS detector is the same.

[89]Each of which was followed by a flood of interpretation papers [218].

[90]Or was, depending on your preference.

[91]Notice that this scale has different units that those of Figure 1.9.

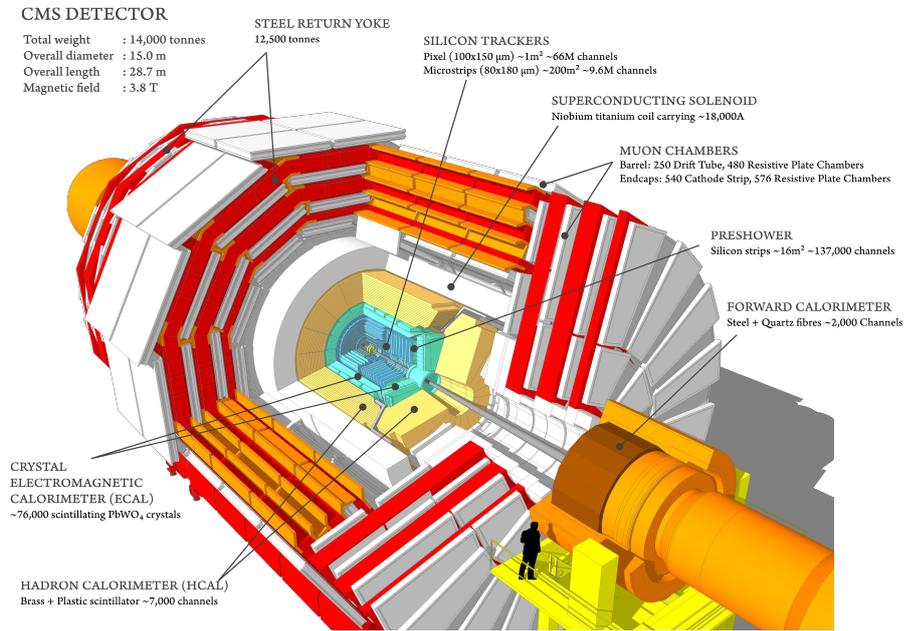

(a) CMS

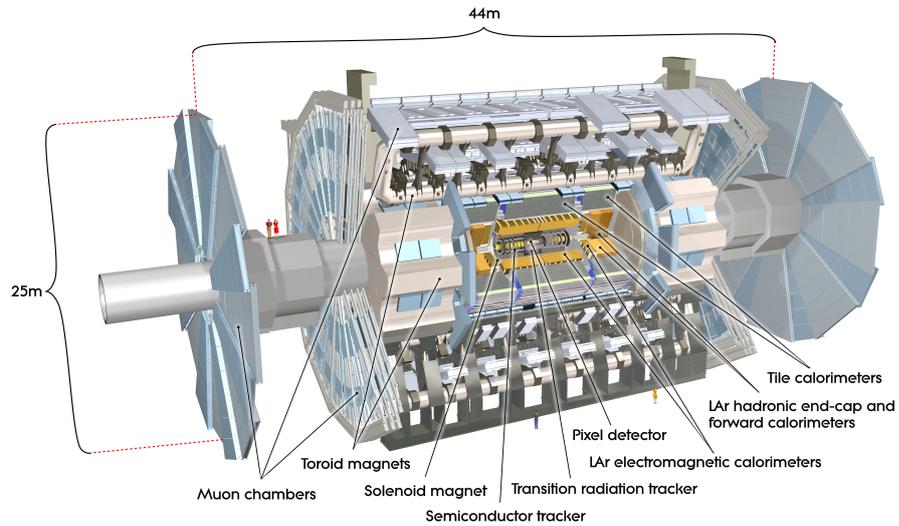

(b) ATLAS

Figure 1.10: CMS and ATLAS detectors. Both detectors operate on a similar principle, with a detector which is almost hermetically closed around the interaction point and consist of different detection layers. Figures from [210, 211].

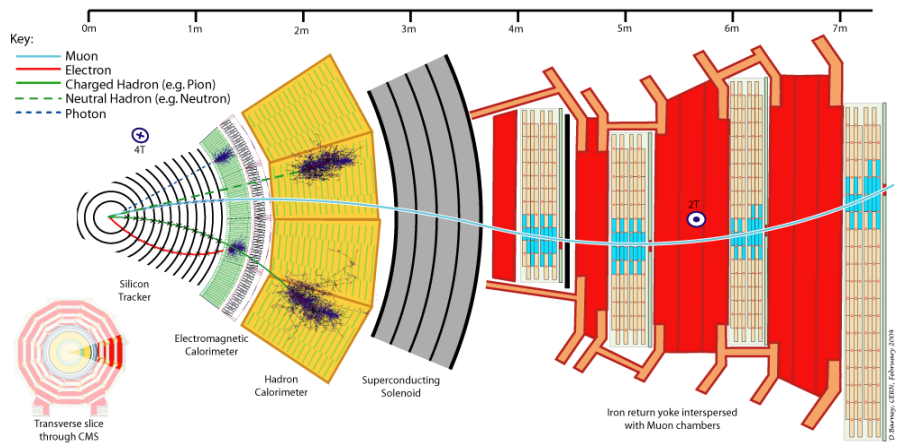

Figure 1.11: Slice of the CMS detector, showing the detector elements and the particles which they can detect. Figure from [212].

for pair production of the different superpartners as a function of their mass, at $\sqrt{s} = 13$ TeV. In typical scenarios, such particles are pair produced (the "production mode"), after which a decay chain is initiated to produce a final state with stable particles (e.g. $pp \rightarrow \tilde{g}\tilde{g} \rightarrow jj\tilde{\chi}_1^0\tilde{\chi}_1^0$, with $j$ a jet). It is clear that the search for supersymmetry is challenging: particles with the highest cross section need to compete with a strong QCD background from Standard Model processes, while the electroweak particles, which have more unique signature, have lower cross sections. Still, the most promising channels consist of squark and gluino pair production. Their high mass (taking into account earlier exclusion limits at lower mass from LEP) implies particularly hard hadronic activity compared to Standard Model background. In addition, particles produced in their decay can possibly lead to electroweak signatures. In addition, often the lightest supersymmetric particle (LSP) is neutral[92], leading to events with large missing transverse momentum if *R*-parity is conserved, which has a lower background from Standard Model processes[93].

Given this extensive search program, strong limits have been put on the parameter space of many SUSY models (see also the review in [29]; a more in-depth discussion (after Run 1), taking into account experimental constraints from theoretical point of view can be found in [235]). Before the LHC, exclusion limits on supersymmetry were mostly based on global fits in the very restrictive CMSSM model [67, 165, 173], taking into account many indirect constraints (e.g. flavour physics, low-energy experiments,...). Even taking into account the LEP results, many constraints were derived not directly, but from the non-observation of other particles (since the restriction to only five parameters[94]

---

[92]This is motivated by cosmology: any model where a charged stable particle is predicted is automatically ruled out. In gauge mediation models this issue does not appear, since the lightest particle is always the neutral goldstino.

[93]The main background coming from neutrinos from *Z* or *W*-boson decay or jet mismeasurement.

[94]The scalar and gaugino masses $M_0$ and $M_{1/2}$, the trilinear couplings $A_0$, $\tan\beta$, and sgn($\mu$).

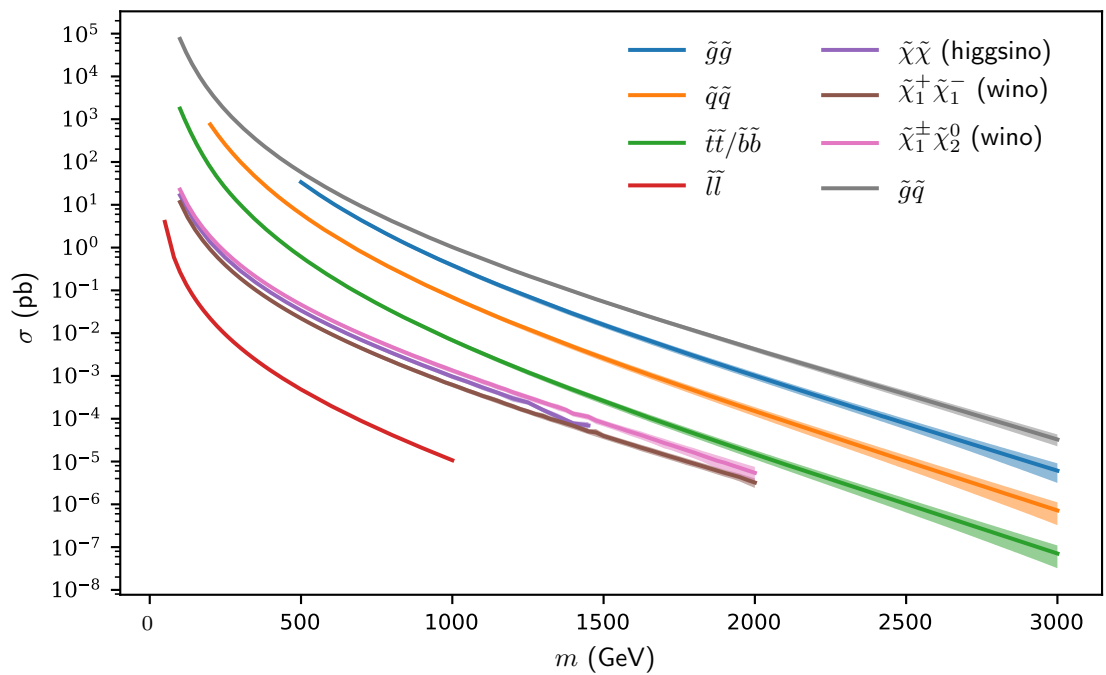

Figure 1.12: SUSY cross sections at $\sqrt{s} = 13$ TeV. The strong production channel $\tilde{g}\tilde{g}$ ($\tilde{q}\tilde{q}$) assumes the $\tilde{q}$ ($\tilde{g}$) are decoupled, while the $\tilde{g}\tilde{q}$ production channel assumes degenerate squarks and gluinos. The higgsino cross section includes both the neutral and charged components. The chargino-neutralino cross section assumes a wino-like NLSP and charginos and bino-like LSP $\tilde{\chi}_1^0$, in which case the process $\tilde{\chi}_2^0\tilde{\chi}_1^{\pm}$ dominates. Data from [225]; for cross sections in the coloured section see [226] and `NNLL-fast` [227], for slepton pair production [228–231], and for gaugino pair production [228, 232–234]. Notice the different scale compared to Figure 1.9.

means that there are strong correlations between the different particles).

However, the LHC is powerful enough to directly probe many of the superpartners, meaning it can quickly sweep through large parts of the parameter space. As a result, global fits became very constrained. After Run 1, the CMSSM became essentially excluded when combining constraints from direct searches at the LHC, observed Higgs boson properties, astrophysics and precision experiments [236].

Consequently, there was a shift towards the framework of simplified models [237, 238]. In such models, one assumes only the presence of a limited set of particles and possible decay chains (often with 100% or 50% branching ratios). In this way, one can constrain specific signatures, free from correlations with other, independent final states. Therefore, these results can be interpreted in many models. On the other hand, in this approach one can be less sensitive to specific decay chains for which a dedicated analysis does not exist (in principle, every final state defined by $n$ jets, $m$ leptons, etc. is covered by some analysis). Moreover, the interpretation of these results is dependent on the assumptions in the simplified model, such that one needs to be careful when applying these limits to other models.

A summary of different SUSY searches from ATLAS is shown in Figure 1.13, with similar figures available from CMS for SUSY (2016) [239] and Exotics (2019) [240]. In general, these searches probe gluinos around 2 TeV, first and second generation quarks up to 1–1.6 TeV, third generation squarks up to 300–800 GeV, and sleptons around 500 GeV [29]. As such, many parameter points have already been excluded. On the other hand, certain corners of the parameter space might still be unexplored. For example, in the last few years, significant attention has been directed towards compressed spectra. In these spectra, the mass difference between e.g. the gluino and the LSP is small, such that the Standard Model particles produced in gluino decay have low energy, and the event might not stick out of the Standard Model background or might not even trigger the detector.

Since simplified models might not capture all details of a full MSSM model (more complicated production modes, complex decay chains,...), but the CMSSM is too restrictive, another alternative has been developed. The phenomenological MSSM [242, 243], or pMSSM, aims to cover the MSSM parameter space, while restricting the number of parameters using reasonable arguments without imposing relations among the soft-breaking terms[95]. Since most of the MSSM parameters are due to flavour, many of the couplings have been chosen flavour-diagonal. In addition, $R$-parity conservation is assumed. In this way, the number of parameters in the pMSSM has been reduced to 19[96]. In addition, the allowed points in the parameter space are constrained by theory: there should appear no tachyons (particles with negative physical mass) in the spectrum, no colour- or charge-breaking minima and there must be successful electroweak symmetry

---

[95]Which is the case if one chooses a specific model for SUSY breaking, e.g. gauge mediation.

[96]The three gaugino masses $M_1$, $M_2$ and $M_3$, $\tan\beta$, $\mu$, the pseudoscalar Higgs mass $m_A$, 10 fermion masses (diagonal in flavour, first and second generation degenerate) and the trilinear couplings of third generation $A_t$, $A_b$ and $A_\tau$. All of these quantities are defined at a scale $M_{\text{SUSY}} = \sqrt{m_{\tilde{t}_1} m_{\tilde{t}_2}}$ and ran-down using RG running.

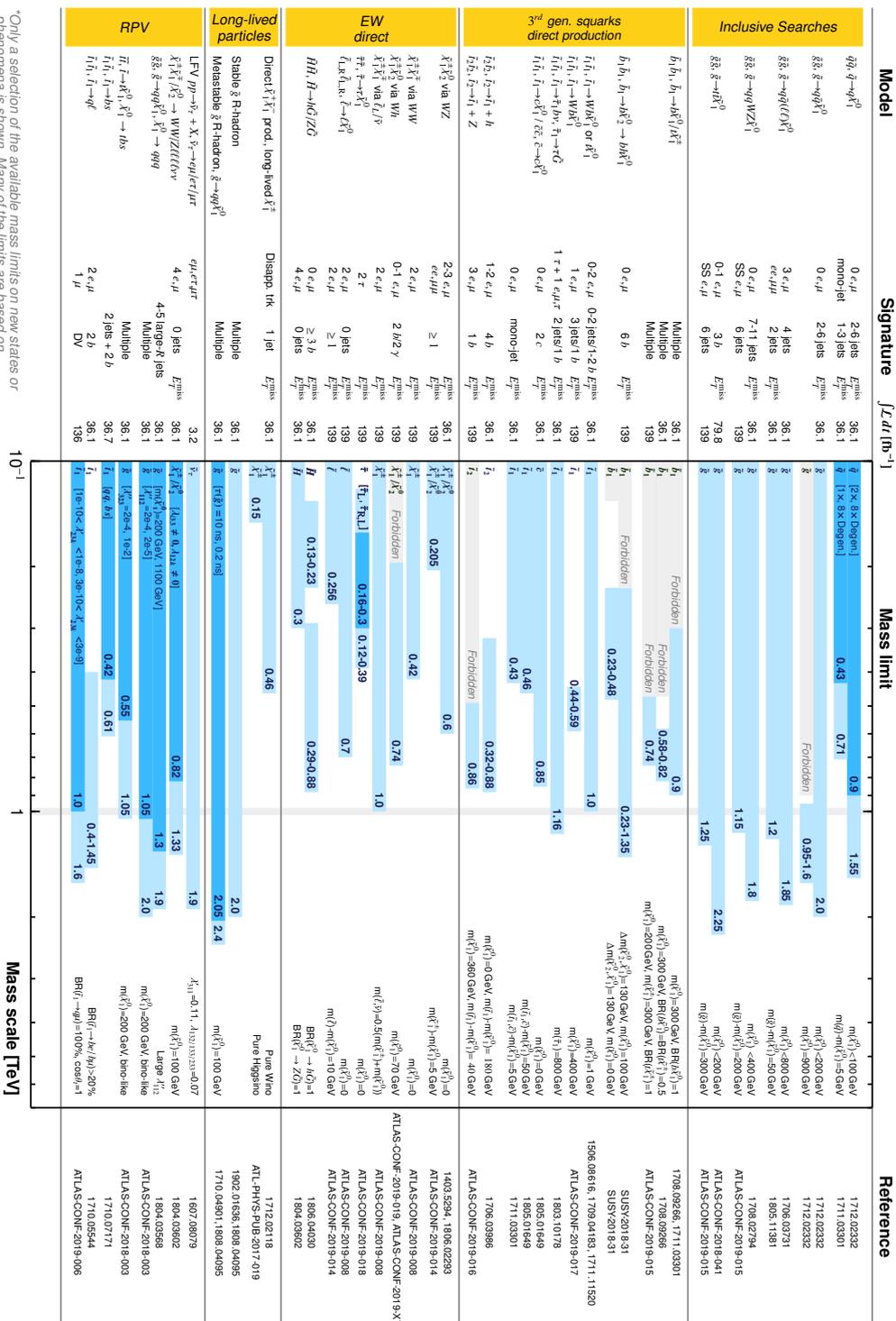

Figure 1.13: ATLAS SUSY summary plot. Note that limits can depend on choice of branching ratio or mass intermediate particles. Shows both maximal mass reach and alternative models. Figure from [241].

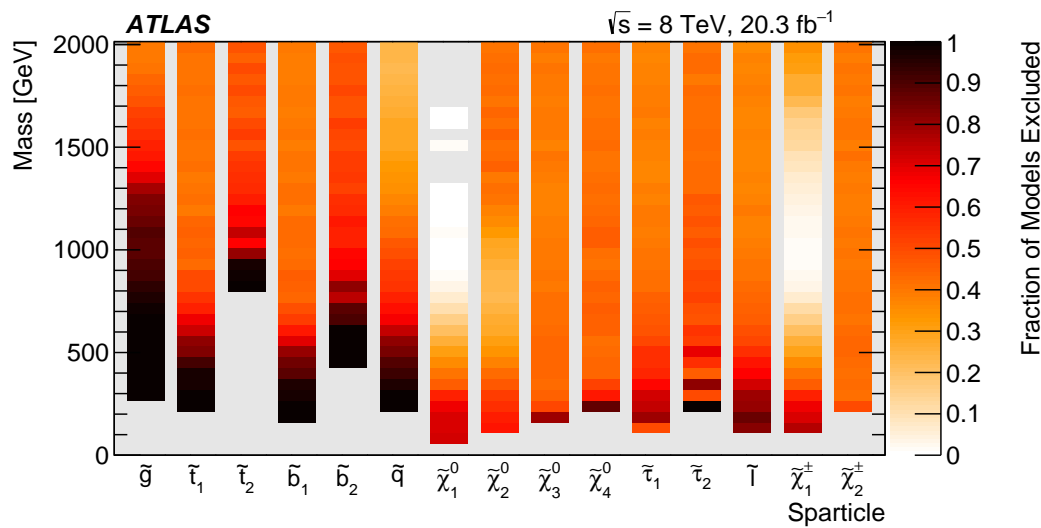

Figure 1.14: Constraints on PMSSM model points after Run 1 at $\sqrt{s} = 8$ TeV, by ATLAS. The colour coding represents the fraction of model points excluded for each sparticle. Figure from [241], see also [244, 245].

breaking.

A global study of the pMSSM model has been performed using the results from Run 1. A summary of its results are shown in Figure 1.14. For this study, it was additionally imposed that the LSP is a neutralino. These result illustrate that while the exclusion limit is typically very high, there is still a subset of model points that can escape the searches and evade current limits. While for gluinos the exclusion is rather tight, electroweak superpartners still have many allowed points in the parameter space below the typical exclusion limits.

The non-observation of supersymmetry at current energies implies that supersymmetry is unlikely to solve the hierarchy problem completely: the superpartner masses are too high to give the Higgs mass a completely natural value. This is known as the little hierarchy problem [246]. Therefore, new searches have gone into several new directions, all looking for signatures which might be a-typical for supersymmetry, for varying reasons. One direction is to look in corners of the parameter space that can evade current searches, for example compressed spectra (e.g. [247]), leading to soft final state particles which might not pass the trigger, or displaced tracks from longer living particles (e.g. [248]), which are not reconstructed well in standard searches. Another direction is to find other well-motivated models. One of the main scenarios in this direction is to look for particle dark matter (e.g. [249]), which shows up as large missing transverse energy in the detector. Summaries for these different signatures are shown in Appendix D. For now, these searches have not found evidence for new physics.

Further improvements are expected in the future, with Run 3 having higher luminosity, energy and improved detectors. However, the big jump in sensitivity between Run 1 and Run 2 was due to the highly increased centre of mass energy, in addition to the increased luminosity $\mathcal{L}$. While Run 3 will increase the amount of data and thus improve limits, its centre of mass energy remains essentially unchanged and the amount of data will "only" be doubled, such that a similar fast improvement can not be expected. On the other hand, new analysis techniques might probe more exotic signatures. In the far future, high-luminosity LHC will represent a tenfold increase in the amount of data, which might uncover particles produced in rare processes.

### 1.5.3 Phenomenological analyses

While there is a vast number of analyses looking at many different final states, each analysis can, due to finite manpower, only test a limited number of models, in which two parameters are typically allowed to vary (e.g. the mass of the pair-produced particle and the mass of the LSP). While these models are chosen to be representative for that specific final state, the resulting limits can not be directly used to constrain alternative models with a similar final state. Indeed, most experimental analyses are essentially[97] counting experiments[98], where the predicted number of signal and background events is compared to the observed number of events, given certain requirements (or "cuts") on the kinematic variables. Even if a model different from that used in the experimental analysis produces the same final state, it can still differ in the production mode, the decay chain and, as a result, the kinematic distributions. Therefore, imposing the cuts on kinematic variables used in a specific experimental analysis will lead to a different predicted number of events and, therefore, a different exclusion limit on the parameter space of the model. Therefore, a whole chain of phenomenological tools has been developed, which allows theorists to easily develop new BSM models and predict the observed number of events in experimental analyses.

First, the creation of model files needed to generate Feynman diagrams has been automated. Using `FeynRules 2.0` [250], one can simply write down a Lagrangian, from which all possible vertices are then extracted and written down in UFO format (Universal FeynRules Output) [251]. In order to obtain realistic mass spectra, determined by certain UV-parameters defined at a high energy scale and RG-evolving the particle masses and couplings down, as well as the corresponding decay widths to high precision (i.e. beyond tree-level), one can use the spectra generators `SARAH` [252], `SPHeno` [253, 254], `SOFTSUSY` [255], `SuSpect` [256] or `SUSY-HIT` [257].

Using the models thus obtained, events at the LHC can be produced with event generators such as `MadGraph5_aMC@NLO` [258], `CalcHEP` [259], `PYTHIA 8.2` [260], and `POWHEG` [261–263]. Of these, `MadGraph5_aMC@NLO` generates all possible Feynman diagrams contributing to a final state, given a specific model (i.e. the vertices extracted from an appropriate Lagrangian using the tools above) and generates Monte carlo events corresponding to this process, in addition to calculating the total cross section, while the latter two implement specific processes individually. Events produced in this way are parton-level events (i.e. the final state contains free quarks and gluons), which are parton showered (perturbative QCD, which is well-known) and hadronised (non-perturbative QCD, for which only phenomenological models exist) using `PYTHIA 8.2` or immediately inside `POWHEG`. Finally, the detection of these final state particles can be simulated approximately, but fast, using a Monte Carlo detector simulation[99] such as `DELPHES 3` [265][100].

---

[97]This is a gross simplification of the many different and very complex analysis techniques used in particle physics experiments. But, for our purposes, this simplification will suffice.
[98]Limits produced using kinematic fits might not strictly belong in this category.

The sample of "detected events" can then be analysed, for which specialised analysis packages have also been developed. `MadAnalysis 5` [266–268] allows one to easily implement different cuts on kinematic distributions and obtain in this way an analysis emulating the one used in experimental analysis. In addition, it contains an analysis library implementing some of the existing analyses, where each analysis has been verified in great detail. There also exist tools which perform several of the steps above down to the analysis level automatically, possibly interfaced to the tools above, such as `Rivet` [269] and `CheckMATE 2` [270] (which also builds upon [265, 271–274]).

Finally, other tools have been developed that make use of these same models and Monte Carlo codes to compute cross sections to probe dark matter observables, such as `MicrOMEGAs 2.0` [275] and `MadDM` [276]. Therefore, a single model implementation and a common tool chain can be used to probe BSM physics in completely different scenarios.

Using the extensive and mostly automated tool chain described above, it is relatively easy to investigate the exclusion potential of the LHC analyses on various models. One example of such a phenomenological analysis can be found in the next chapter.

## 1.6 Conclusion

In this chapter, we saw that the Standard Model of particle physics has shortcomings, which require that there is physics Beyond the Standard Model. While some of these issues do not point to a specific scale, others (mainly the hierarchy problem) suggest that new physics must be present around the weak scale. We saw how supersymmetry is one well-motivated theory which could, but does not necessarily, appear slightly above the weak scale. If it does, it can solve the hierarchy problem.

Next, we reviewed how to build a supersymmetric theory and the minimal super-symmetric extension to the Standard Model. In addition, we saw that supersymmetry, if it occurs in nature, must be broken. This breaking must occur in a hidden sector and the way of communicating this to the observable sector leaves its imprints on the allowed particle spectra. The most common way of communicating supersymmetry breaking are gravity mediation and gauge mediation. Moreover, if supersymmetry breaking is the result of spontaneous symmetry breaking, there must appear in the spectrum a goldstino. In the case of gauge mediation, this goldstino is always the lightest supersymmetric particle, such that all decay chains of superpartners must eventually end in a goldstino. Therefore, in gauge mediation, collider phenomenology is mainly determined by the nature of the next-to-lightest supersymmetric particle.

---

[99]The most detailed detector simulation used by experimental collaborations also uses a Monte Carlo method, using `Geant 4` [264]. However, this method simulates each interaction of each particle in each of the individual detector elements, which is therefore very slow. On the other hand, `DELPHES 3` simplifies the detection process to a coin-flip on whether each particle is detected or not and whether there are possible mismeasurements.

[100]Alternatively, this last step can be replaced by using the efficiency tables reported in a specific analysis, although this is less accurate.

Finally, we reviewed the current status of supersymmetry at the LHC. A large part of the parameter space has already been covered; all observations are compatible with Standard Model background. Therefore, standard supersymmetric scenarios are under pressure and supersymmetry might no longer provide a solution to the hierarchy problem (in fact, the requirement of naturalness might have to be reconsidered entirely). As such, supersymmetry has lost popularity and has become less of a focus in collider searches. Instead, searches for dark matter have received increased attention. On the other hand, supersymmetry (or another theory) might be hiding in a thus far unexplored corner of the parameter space. Therefore, also more exotic signatures, such as compressed spectra or displaced tracks, are now being considered.

# $Z$-peaked excess in goldstini scenarios

In this chapter, we study a possible explanation of a 3.0 $\sigma$ excess reported in 2015 by the ATLAS Collaboration at a centre-of-mass energy of $\sqrt{s} = 8$ TeV in events with a $Z$-peaked same-flavour opposite-sign lepton pair, jets and large missing transverse momentum. We consider a model in the context of gauge-mediated supersymmetry breaking with more than one hidden sector, the so-called goldstini scenario. In our model, with two supersymmetry-breaking sectors, the excess can be explained by gluino pair production, followed the decay $\tilde{g} \to g\tilde{\chi}_{1,2}^0 \to gZ\tilde{G}'$ to a higgsino-like neutralino and a pseudo-goldstino, where each step is a two-body decay. Due to the interplay between two separate supersymmetry-breaking sectors, the mass of the pseudo-goldstino, which is no longer protected by supersymmetry, can be appreciable. We find that a mass spectrum such as $m_{\tilde{g}} \sim 1000$ GeV, $m_{\tilde{\chi}_{1,2}^0} \sim 800$ GeV and $m_{\tilde{G}'} \sim 600$ GeV can fit the rate and the kinematic distributions of the excess, without conflicting with the stringent constraints from jets plus missing energy analyses and with the CMS constraint on the identical final state. This analysis, which was published in [277], will be discussed in Sections 2.1– 2.4, taking into account only 8 TeV data. Since the publication of this work, updated analyses for the same signature have been performed by both CMS and ATLAS at higher centre-of-mass energy $\sqrt{s} = 13$ TeV and with more data, finding that all data is compatible with the Standard Model background. We discuss the implications of these null results using the new data in Section 2.5.

## 2.1 Motivation

While the discovery of the Higgs boson was a significant achievement of Run 1 of the LHC, most of the searches for new physics have obtained only null-results. However, we should still keep our eyes open for the faintest possible imprint. In 2015, the ATLAS Collaboration reported on a search for supersymmetry (SUSY) with the final state containing a pair of same-flavour opposite-sign (SFOS) leptons, jets and large missing transverse momentum ($\not{E}_T$) at a centre-of-mass energy of 8 TeV [278]. The analysis, performed on an integrated luminosity of 20.3 fb$^{-1}$, observed an intriguing excess of 29 lepton pairs peaked at the invariant mass of the $Z$ boson ("on-$Z$"), while $10.6 \pm 3.2$ pairs are expected from the Standard Model (SM) prediction. This excess corresponds to a local significance of 3.0 $\sigma$.

### 2.1.1 ATLAS search for SFOS leptons, jets and missing energy

The excess was found in a search for supersymmetry in events containing a same-flavour opposite-sign dilepton pair, jets, and large missing transverse momentum in $\sqrt{s} = 8$ TeV $pp$ collisions using 20.3 fb$^{-1}$ gathered with the ATLAS detector in 2012 [278], which is summarised here.

The analysis targets two leptonic production mechanisms, which always give same-flavour opposite-sign (SFOS) pairs. The first one features the decay of neutralinos as $\tilde{\chi}_2^0 \to l^+ l^- \tilde{\chi}_1^0$ in a simplified model, either through the intermediate step $Z^* \tilde{\chi}_1^0$ or through $\tilde{l}^{\pm(*)} l^{\mp}$. In both cases, this results in a rising dilepton invariant mass distribution, terminating in a kinematic endpoint [279]. Therefore, a search is performed in the $m_{ll}$-distribution[1] off the $Z$-peak. For this simplified model, both squark[2] and gluino pair production are considered. The neutralino $\tilde{\chi}_2^0$ which decays into leptons is then produced in squark decay ($\tilde{q} \to q \tilde{\chi}_2^0$) or gluino decay ($\tilde{g} \to q \bar{q} \tilde{\chi}_2^0$).

The second production mechanism, shown in Figure 2.1, features on-shell $Z$-bosons in the final state, of which one then decays leptonically[3] into either an $e^+ e^-$-pair or a $\mu^+ \mu^-$-pair, each with a branching ratio of 3.36%, resulting in a peak in the $m_{ll}$-distribution at $m_Z$. This search targets a particular scenario of generalised gauge mediation (GGM), studied in [281], where the $Z$-boson is produced via the decay of the lightest neutralino to a (nearly) massless goldstino ($\tilde{\chi}_1^0 \to Z\tilde{G}$), which always appears in gauge mediation. Since the goldstino — which is neutral — escapes detection, it gives rise to missing momentum. Alternatively, the $Z$-boson could also be produced in decays of neutralinos like $\tilde{\chi}_2^0 \to Z\tilde{\chi}_1^0$, although the analysis focuses on, and is motivated by, the GGM scenario. In the particular scenario studied here, the production mode is the strong production of gluino pairs, which subsequently decay to produce the neutralinos (although squark production, which are likewise produced in strong interactions, could also be considered instead, which has little impact on the analysis [281]). As before, the search is performed in the $m_{ll}$-distribution, but now on the $Z$-peak. The on-$Z$ analysis shows an excess of events above the expected Standard Model background, while the results of the off-$Z$ search for an edge are well described by background. Therefore, in the following, we focus on the on-$Z$ scenario.

---

[1] The (Lorentz) invariant mass of the lepton pair is defined by $m_{ll} = \sqrt{(E_{l,1} + E_{l,2})^2 - (\mathbf{p}_{l,1} + \mathbf{p}_{l,2})^2}$.

[2] The simplified model includes the left-handed partners of the $u$, $d$, $c$ and $s$ quarks, while the partners of the right-handed quarks and $b$ and $t$ quarks are decoupled.

[3] The hadronic decay of the $Z$-boson, with a higher branching ratio of 69.9%, gives rise to a purely hadronic final state, which can give complementary information (see Section 2.1.2). The advantage of looking at the leptonic decay is that the dilepton invariant mass $m_{ll}$ is unaffected by uncertainties in the jet physics [280].

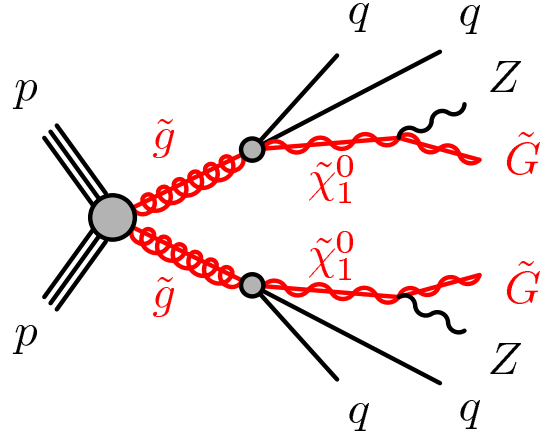

Figure 2.1: Feynman diagram for the general gauge mediation scenario considered by the ATLAS search for events with a same-flavour opposite-sign dilepton pair, jets, and large missing transverse momentum at $\sqrt{s} = 8$ TeV [278].

The analysis provides an interpretation in a generalised gauge mediation model, with a goldstino (or gravitino) LSP and a higgsino-like neutralino as the NLSP. The higgsino mass parameter $\mu$ and gluino mass $m_{\tilde{g}}$ are free parameters, while the $U(1)$ and $SU(2)$ gaugino mass parameters $M_1$ and $M_2$ are fixed to 1 TeV and all other sparticle masses are taken at at $\sim 1.5$ TeV. Parameters are chosen such that $\tilde{\chi}_1^0 \to Z\tilde{G}$ is the dominant NLSP decay, although its exact branching fraction varies with $\tan\beta$. As benchmark points, the analysis considered $\tan\beta = 1.5$ with 97% branching ratio [282] and $\tan\beta = 30$ with up to 40% contribution from $\tilde{\chi}_1^0 \to h\tilde{G}$, where $h$ is the lightest CP-even SUSY Higgs boson with $m_h = 126$ GeV and branching fractions similar to the Standard Model Higgs. The dominant production mode for this signature is gluino pair production, with gluino decay $\tilde{g} \to q\bar{q}\tilde{\chi}_1^0$, where $q = u, d, c, s$ with equal branching fraction. The gravitino mass is sufficiently small, such that NLSP decays are prompt. The decay length of the NLSP $c\tau_{\mathrm{NLSP}}$ varies with $\mu$, is longest at $\mu = 120$ GeV ($c\tau_{\mathrm{NLSP}} = 2$ mm) and decreases to $c\tau_{\mathrm{NLSP}} < 0.1$ mm for $\mu \geq 150$ GeV. This finite lifetime of the NLSP is taken into account by the analysis.

The event selection criteria of this search are the presence of at least two leptons, of which the two with the highest $p_T$ are used for the analysis. These must form a same-flavour opposite-sign pair and one of these leptons must have triggered the detector for this event. In addition, at least two jets must be identified. Moreover, for the search targeting the GGM scenario, the lepton pair must satisfy 81 GeV $< m_{ll} < 101$ GeV, the missing energy $\not{E}_T = |-\sum_{\mathrm{all}} \mathbf{p}_T^i|$ must satisfy $\not{E}_T > 225$ GeV and the hadronic activity[4] $H_T = \sum p_T^{jet} + p_T^{l,1} + p_T^{l,2}$ (a scalar sum) must satisfy $H_T > 600$ GeV. In this way, the analysis is optimised for high gluino mass and jet activity. No requirement is put on the presenece of *b*-jets. The final signal region acceptance × efficiency is 2–4%, including the *Z* branching ratio into leptons (for GGM models with $\mu > 400$ GeV).

---

[4]Technically, the hadronic activity should not include the $p_T$ of the leptons, although the definition used here often appears in the literature. Instead, the quantity used here should be more appropriately called visible activy, as can be found in the literature as well, typically denoted by $S_T$. Still, for consistency, we use the symbols and nomenclature as defined in this specific analysis.

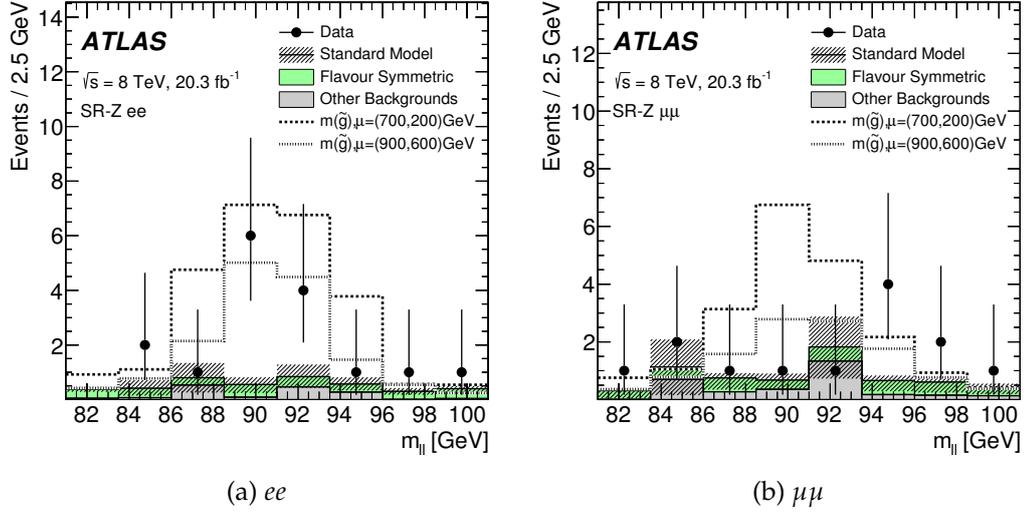

(a) *ee*            (b) *μμ*

Figure 2.2: Measured $m_{ll}$-distribution in the on-*Z* search (where an excess is seen) in the ATLAS search for events with a same-flavour opposite-sign dilepton pair, jets, and large missing transverse momentum [278].

The background for this event signature is formed by several processes. Most of the background is flavour-symmetric, leading to the *ee*, *μμ* and *eμ* to occur in a 1:1:2 ratio. Therefore, this background can be estimated from *eμ* data. It is dominated by $t\bar{t}$-production and also has contributions from *WW*, single top (*Wt*) and $Z \to \tau\tau$. Together, these make up about 60% of the on-*Z* background. Diboson background with on-shell *Z*-production can contribute up to 25% in the on-*Z* region and is estimated from Monte Carlo simulation. In addition, rare top backgrounds ($t\bar{t} + W$, $t\bar{t} + WW$ and $t\bar{t} + Z$) are also estimated from Monte Carlo simulation. Finally, mismeasured jets entering as leptons can contribute up to 10% to the background. The $Z/\gamma^*$+jets final state contributes due to artificially high $\not{E}_T$ from jet energy mismeasurements. This final state mimics the signal and is therefore important for the on-*Z* search, although its contribution to the total background is small.

The on-*Z* search found 16 events in the *ee* final state, 13 in *μμ*, or 29 combined, compared to an expected background of $4.2 \pm 1.6$, $6.4 \pm 2.2$ and $10.6 \pm 3.2$ respectively. This implies an excess with $1.7\sigma$ significance[5] in the *μμ*-channel, $3.0\sigma$ in *ee* and $3.0\sigma$ combined. This is also visible in Figure 2.2, which shows the observed $m_{ll}$-distribution of the on-*Z* search.

As a result of the excess, the exclusion limits at 95% confidence level (CL) are weaker than expected. The 95% CL limits on the cross section × acceptance × reconstruction efficiency are shown in Table 2.1. The interpretation within the GGM model is shown in Figure 2.3. For $\tan\beta = 1.5$, the analysis excludes gluino masses up to $m_{\tilde{g}} = 850$ GeV for $\mu > 450$ GeV, while for $\tan\beta = 30$, it excludes up to $m_{\tilde{g}} = 820$ GeV for $\mu > 600$ GeV. The slightly lower reach for the second benchmark is due to the smaller branching ratio into *Z*, which means the model itself predicts less events.

---

[5]This is a local significance, since it is not corrected for the look-elsewhere effect.

Table 2.1: Results from the ATLAS search for supersymmetry in events with a same-flavour opposite-sign dilepton pair on the *Z*-peak, jets and large missing transverse momentum at $\sqrt{s} = 8\,\text{TeV}$ [278]. The columns show the expected number of background events, the observed total number of events, the observed 95% CL limit on the number of signal events, the expected 95% CL limit on the number of signal events and the 95% CL limit on the visible cross section.

| Channel | Expected background | Observed events | $S^{95}_{\text{obs}}$ | $S^{95}_{\text{exp}}$ | $\langle \epsilon\sigma \rangle^{95}_{\text{obs}}$ (fb) |
|---|---|---|---|---|---|
| *ee* | $4.2 \pm 1.6$ | 16 | 20.2 | $8^{+4}_{-2}$ | 1.00 |
| $\mu\mu$ | $6.4 \pm 2.2$ | 13 | 14.7 | $9^{+4}_{-2}$ | 0.72 |
| Combined | $10.6 \pm 3.2$ | 29 | 29.6 | $12^{+5}_{-2}$ | 1.46 |

  Therefore, for the GGM model, parameter points outside the exclusion region are still allowed by this analysis. However, additional investigation is required to know whether there are also points in this parameter space, or in an alternative model, which can fit the observed number of events and the kinematic distribution. For example, while the excess number of events can be interpreted in the GGM model, the observed jet multiplicity is not well reproduced by this model. Possible model interpretations of the excess will be discussed in more detail in Section 2.1.3.

### 2.1.2  Constraints from other searches

Models that can explain the excess seen by ATLAS in the 8 TeV data are also constrained by different searches, the two most imporant ones of which we will discuss here (for the others, see the discussion in Section 2.1.3). The first of these is a CMS search for beyond the Standard Model physics in events with two leptons, jets and missing transverse momenum, with 19.4 $\text{fb}^{-1}$ of data at $\sqrt{s} = 8\,\text{TeV}$ [280] (i.e. the same event signature as the ATLAS search), summarised here. Similar to the ATLAS analysis, this search targets both scenarios in which the dilepton invariant mass distribution of a same-flavour opposite-sign lepton pair[6] exhibits an edge and in which it exhibits a peak in the window 81 GeV $< m_{ll} < 101$ GeV, compatible with the decay of an on-shell *Z*-boson. The edge search is performed over all values of the $m_{ll}$-distribution (i.e. also inside the *Z*-window), while a dedicated counting experiment is performed in the on-*Z* region searching for a peak. In both cases, the signal region is split in a central region, where both leptons have pseudorapidity[7] $|\eta_{\text{lepton}}| < 1.4$ and a forward region, where at least one lepton has rapidity $1.6 < |\eta_{\text{lepton}}| < 2.4$. Heavy resonance decays are expected to appear in the central region. The edge search is divided in three $m_{ll}$-windows: low mass, on-$Z$ and high mass. The largest deviation from Standard Model prediction found in the analysis appears in the central low-mass ($m_{ll}$ =20–70 GeV) region with a local significance of $2.6\sigma$. The excess is observed predominantly in events with at least one identified $b$-jet and diminishes if a veto on the presence of a $b$-jet is applied. The on-$Z$ search, where the ATLAS excess appeared, is compatible with the Standard Model prediction, both for $N_{\text{jets}} \geq 2$ and $N_{\text{jets}} \geq 3$ and for different regions of $\not{E}_T$ (100–200 GeV, 200–300 GeV and > 300 GeV).

---

[6] Contrary to the ATLAS analysis, the CMS analysis requires the presence of exactly two leptons.

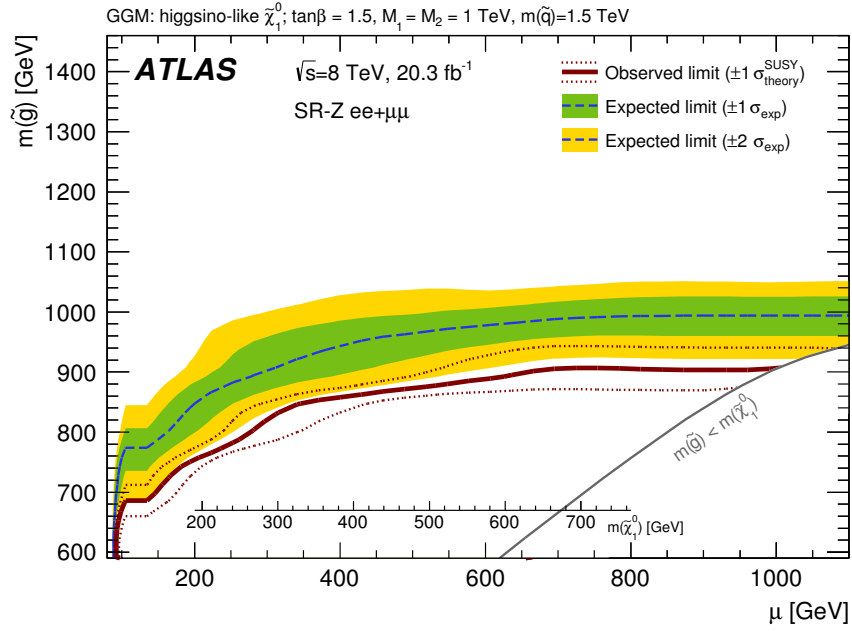

(a) $\tan \beta = 1.5$

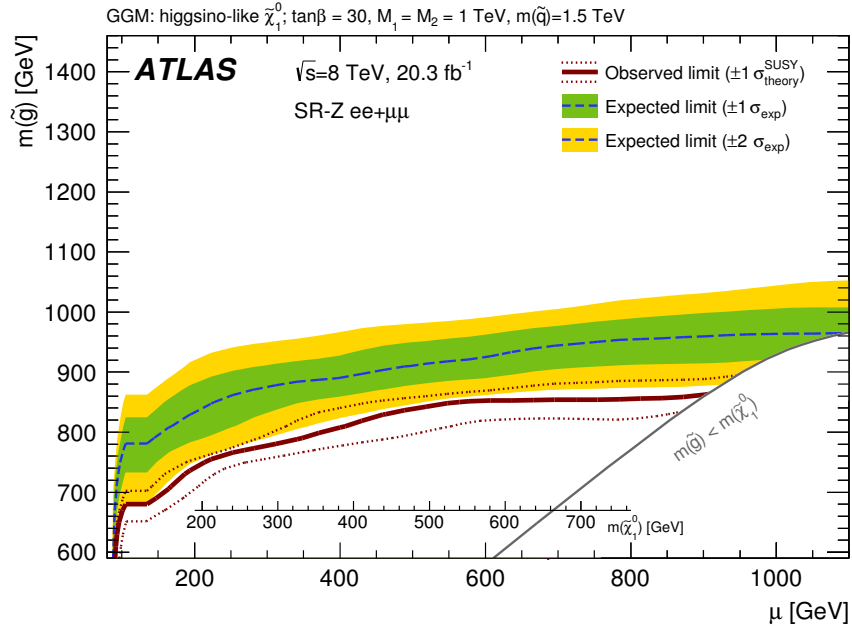

(b) $\tan \beta = 30$

Figure 2.3: Exclusion limits at 95% CL on the gluino mass and higgsino mass parameter in the GGM model (where the excess of events is seen), for two values of $\tan \beta$ in a search for supersymmetry in events with a same-flavour opposite-sign dilepton pair, jets, and large missing transverse momentum [278].

The second important constraint comes from the ATLAS search for squarks and gluinos in final states with jets and missing transverse momentum at $\sqrt{s} = 8$ TeV with 20.3 fb$^{-1}$ of data [283], which we describe here[8]. Such an event signature is always predicted in models which can explain the ATLAS $Z$-peak excess. The on-shell $Z$-boson producing the leptons in such models can also decay hadronically (with a branching ratio much larger than into leptons). When both $Z$-bosons in the event decay hadronically, the final state contains just jets and missing transverse momentum. The ATLAS analysis we consider now selects for high $p_T$-jets, missing transverse momentum and no electrons or muons, which are vetoed in order to avoid overlap with other searches. It targets the production of squarks and gluinos ($\tilde{g}\tilde{g}$, $\tilde{q}\tilde{q}$, and $\tilde{g}\tilde{q}$), which subsequently decay directly as $\tilde{q} \rightarrow q\tilde{\chi}_1^0$ and $\tilde{g} \rightarrow q\bar{q}\tilde{\chi}_1^0$ or produce charginos as $\tilde{q} \rightarrow q\tilde{\chi}^{\pm}$ or $\tilde{g} \rightarrow q\bar{q}\tilde{\chi}^{\pm}$, which subsequently decay as $\tilde{\chi}^{\pm} \rightarrow W^{\pm}\tilde{\chi}_1^0$ to produce more jets. The former scenario is of interest to the excess we wish to investigate.

Events are collected using a trigger[9] requiring a jet with $p_T > 80$ GeV and $\not{E}_T >$ 100 GeV. Signal regions are defined using the effective mass $m_{\text{eff}}(N_j)$ in $\geq N_j$ events, which is the scalar sum of the $p_T$ of the leading $N_j$ jets and $\not{E}_T$. The final selection puts requirements on $m_{\text{eff}}$(incl.) (sum over all jets with $p_T > 40$ GeV and $\not{E}_T$) and $\not{E}_T$, which suppresses multi-jet background where $\not{E}_T$ is generated by jet energy mismeasurements. Fifteen signal regions are then defined, with jet multiplicity (inclusive) ranging from 2 to 6 (to target squark versus gluino pair production, since their decays give different numbers of quarks) and with increasing background rejection (very loose "l-" to very tight "t+"), achieved with varying the requirements on $\not{E}_T/m_{\text{eff}}$ or $\not{E}_T/\sqrt{H_T}$.

The dominant background for this search is due to $Z$+jets (mostly from $Z \rightarrow \nu\nu$ to generate $\not{E}_T$), $W$+jets (most through $W \rightarrow \tau\nu$ with $\tau \rightarrow$ hadrons, some through $e\nu$ or $\mu\nu$ if the lepton is not reconstructed), $t\bar{t}$ (the semi-leptonic decay $t\bar{t} \rightarrow bb\tau\nu qq'$, with $\tau \rightarrow$ hadrons), mono-$t$ and multiple jets (from jet energy mismeasurement and semi-leptonic decays of heavy-flavour quarks). Diboson pair production with at least one boson decay to charged leptons is a small component.

None of the signal regions finds a significant excess, with the most significant devia-tion in a signal region with three jets having a p-value of 0.24. Therefore, 95% CL limits are placed in the parameter space of different models, including the CMSSM, a pMSSM model as well as simplified models. The simplified models, which we are interested in here, consider three cases: gluino pair production ($\tilde{g}\tilde{g}$) with the squarks decoupled, squark pair production ($\tilde{q}\tilde{q}$) with the gluinos decoupled or nearly degenerate squarks and gluinos ($m_{\tilde{q}} = 0.96 m_{\tilde{g}}$), in which case $\tilde{g}\tilde{q}$-production dominates. The exclusion limits for gluino or squark pair production followed by direct decay to a massive neutralino, relevant to the excess, are shown in Figure 2.4. In the case of a massless neutralino, the limit on the gluino mass is $m_{\tilde{g}} \geq 1330$ GeV, the limit on 8 degenerate squarks (a single squark flavour) is $m_{\tilde{q}} \geq 850$ GeV ($m_{\tilde{q}} \geq 440$ GeV) and the limit for squark-gluino production is $m_{\tilde{g}} \geq 1500$ GeV with $m_{\tilde{q}} = 0.96 m_{\tilde{g}}$ (not shown in the figure).

---

[7]See Section 1.5.1.

[8]There is an analogous search by CMS [284], which we will not explicitly discuss here, with similar results.

[9]Full trigger efficiency is reached above $p_T^{\text{jet}} > 130$ GeV and $\not{E}_T > 160$ GeV.

### 2.1.3  Interpreting the excess

While the ATLAS analysis provides (weaker than expected) exclusion limits on the parameter space of a GGM model, it does not attempt to explain the observed number of events and kinematic distributions. Interpretation of the excess in models of beyond the Standard Model physics is not a straightforward task, as null results from the other searches discussed above are placing considerably stringent constraints on the viable model parameter space. While relatively light gluinos with $m_{\tilde{g}} < 1.2$ TeV are required in order to produce the observed number of events [285], studies presented in [286, 287] have shown that explaining the excess within this scenario is difficult. Here, we will discuss some of the elements required to explain the ATLAS excess.

#### The production mode: gluinos, squarks,...

First, we discuss the production mode. In their analysis, ATLAS considers gluino pair production within a GGM model. Such a scenario had been previously (i.e. long before the appearance of an excess) investigated in [281], which explored the possibility to probe general neutralino NLSP at the LHC in the context of gauge mediation. In this case, the most promising channels to discover supersymmetry are those where the neutralinos appear through strong production (as opposed to direct electroweak production of the neutralinos and associated charginos), i.e. pair production of gluinos and/or squarks. While this earlier study considered gluino production, it was checked that using squarks instead does not significantly alter the results.

Therefore, it can be expected that both squark and gluino production are able to explain the excess. Naively, for an excess appearing in a search designed to be sensitive to strong production modes, both squark and gluino production should be able to supply the required cross section. Indeed, the observed signal rate typically requires a strong production cross section. We can estimate this easily. The observed excess is about 19 events, which corresponds to a visible cross section $\epsilon\sigma \approx 1$ fb (given the integrated luminosity of 20.3 fb$^{-1}$). The analysis quotes an efficiency 2–4%, although this depends strongly on model under study (i.e. this will be much lower for electroweak production since the selection favours strong hadronic activity). Retaining for now this estimate of the efficiency, the total cross section required is $\sigma \approx$ 25–50 fb$^{-1}$. From Figure 2.5, which shows the cross section of various SUSY production modes at $\sqrt{s} = 8$ TeV, we see that this can easily be supplied by the strong particles, while the electroweak sector requires low masses even *before* taking into account the analysis cuts which select for strong jet activity (i.e. for this channel, the acceptance $\times$ efficiency will certainly be lower than for squarks/gluinos). Therefore, squark or gluino pair production seems preferred[10].

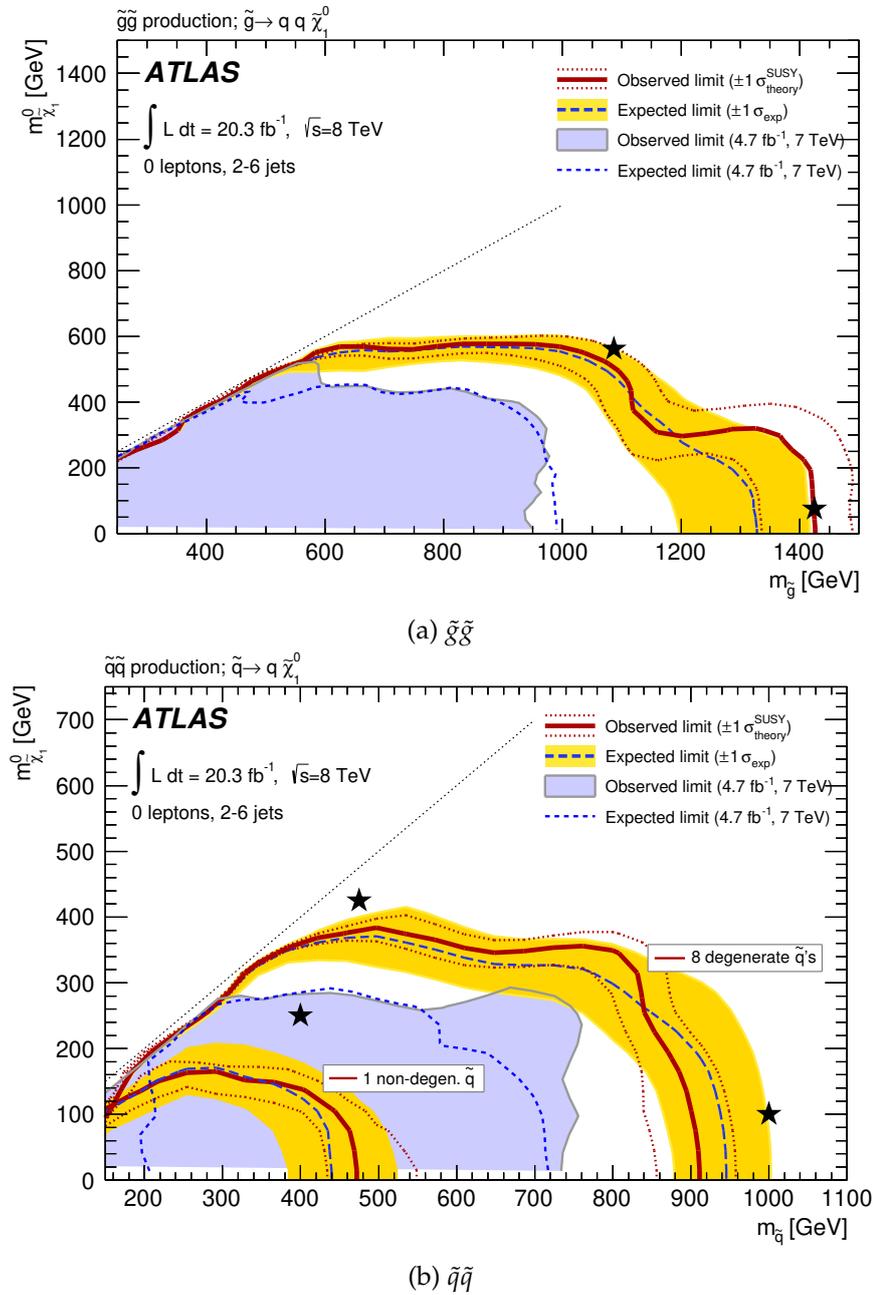

Figure 2.4: Exclusion limits at 95% CL on the gluino (squark) and neutralino masses, assuming the squarks (gluinos) are decoupled, in the $m_{\tilde{g}}$ ($m_{\tilde{q}}$)–$m_{\tilde{\chi}_1^0}$ plane for direct decay from the ATLAS search for squarks and gluinos in events with jets and missing transverse momentum [283]. Similar (stronger) exclusion limits for nearly degenerate squarks and gluinos from $\tilde{q}\tilde{g}$-production are not shown here.

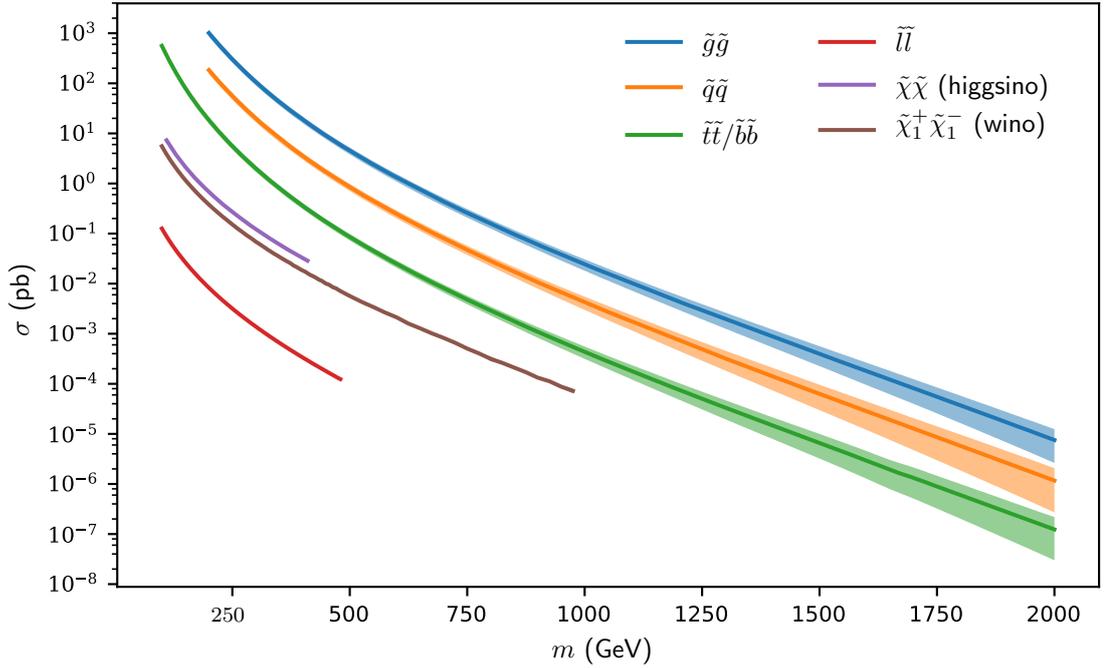

Figure 2.5: Cross sections for the various SUSY production modes at $\sqrt{s} = 8$ TeV. The gluino (squark) production cross section assumes decoupled squarks (gluinos). The higgsino cross section includes both the neutral and charged higgsinos. Data from [225], see also [288] and `NLL-fast` [289–296].

However, one could wonder whether other production modes are capable of supplying the required cross section, taking into account the analysis cuts, on general grounds. These possibilities were investigated in [285]. Stop pair production can not fit the excess: while the stop mass is only limited to $m_{\tilde{t}} \gtrsim 640$ GeV [297–299], only a small number of events is able to survive the cut on $H_T$, making this possibility not viable. Another possibility is direct weak production of neutralinos and charginos (which is not strongly constrained), the largest cross section of which is due to $\tilde{W}^0 \tilde{W}^{\pm}$ and $\tilde{W}^+ \tilde{W}^-$, i.e. $\tilde{\chi}_2^0 \tilde{\chi}_1^{\pm}$ and $\tilde{\chi}_1^+ \tilde{\chi}_1^-$. While light charginos could in principle supply the required cross section, too few events survive the cut on $H_T$, so that also this possibility is ruled out.

---

[10]In [285], it is claimed that only gluino (and not squark) pair production can supply the required cross section, while satisfying the existing bounds on squarks and gluinos. However, naively, one expects this to not be the case, since squarks and gluinos are constrained by the same search for jets+$\not{E}_T$ (although a slightly different signal region might drive the exclusion limit and squarks and gluinos exhibit different radiation patterns). More concretely, one can compare in Figure 2.5 the required cross section for the signal $\sigma \approx$ 25–50 fb$^{-1}$, or the upper limit on the cross section from Table 2.1 $\sigma \approx$ 36.5–73 fb$^{-1}$, with the strong production cross section at the bound on the gluino mass $m_{\tilde{g}} \geq 1330$ GeV and squark mass $m_{\tilde{q}} \geq 850$ GeV from the jets+$\not{E}_T$ analysis (Section 2.1.2). If anything, this naive estimate favours squark pair production.

Therefore, only gluino or squark pair production scenarios are potentially able to explain the excess. While most models consider gluino pair production, there is no fundamental reason to prefer it over squark pair production. Indeed, some models also use squark pair production [300–302][11]. However, introducing both squarks and gluinos at low energy is not allowed, since constraints from jets+$\slashed{E}_T$ searches become much stronger [285, 287].

**Gluino decay**

In this section, we focus on the possibility of gluino pair production and, more importantly, the imprints of its decay modes on the event signature. Gluino decay has been extensively discussed in [287]. The ATLAS analysis targets a GGM model where the gluino decays as[12] $\tilde{g} \to qq\tilde{\chi}_1^0$, with $q = u, d, c, s$ and where the neutralino $\tilde{\chi}_1^0$ is mostly higgsino. This decay is induced by an effective vertex due to the decay chain $\tilde{g} \to q\tilde{q}^* \to qq\tilde{\chi}_1^0$, mediated by a squark (of the first or second generation), as shown in Figure 2.6a. However, generically, in gauge mediation models, it is difficult to decouple only the first generation squarks while keeping the other squark flavours light, due to sum rules on the sfermion masses [187]. As a result, decays mediated by the third generation squarks, shown in Figure 2.6b, must also appear and, since the sbottom and stop squarks masses are typically below the other squark masses due to the usually negative contribution from the Yukawa coupling (which is largest for the third generation) to the RG running [67], they are usually less suppressed by the squark propagator. Moreover, while the first vertex in this decay chain is due to a gauge coupling, and therefore equal for the different squark flavours, the second vertex is proportional to a Yukawa coupling if the neutralino is a higgsino[13]. Therefore, generically, the decay channel into third generation quarks, in particular top quarks, dominates.

On the other hand, if the mass splitting between the gluino and the neutralino is not sufficient to produce two top quarks in the gluino decay, this decay channel is kinematically forbidden. In this case, however, it is possible to close the loop and emit a gluon (or a photon, although in this case the coupling is much lower), as shown in Figure 2.6c. While this channel is suppressed by one loop factor, the Yukawa coupling is still much stronger than for the other quarks. Moreover, in this case, the gluino undergoes a two-body decay, instead of a three-body decay, which is kinematically favoured.

---

[11]For another model which invokes squark pair production to explain a mild deviation in the analogous search by CMS, targeted at cascade decays, see [303].

[12]For simplicity, we denote both quarks and anti-quarks with $q$.

[13]If the neutralino is a bino or a wino, the coupling to the different quark generations is the same. In this case, the dominant channel (third or first/second generation) is determined by the allowed kinematics. In this case, the $\tilde{g} \to qq\tilde{\chi}_1^0$ channel can dominate.

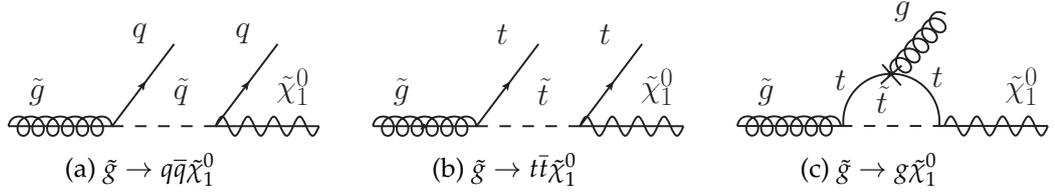

Figure 2.6: Feynman diagrams of the dominant gluino decays, where $q$ stands for first and second generation quarks and we do not distinguish between quark and anti-quark for simplicity.

These arguments have been verified in [287] by calculating the gluino branching ratio into the different channels as a function of the gluino-neutralino mass splitting, shown in Figure 2.7. Indeed, it can be seen that at $\Delta m \approx 2m_t$, there is a large qualitative change from the three-body decay into top quarks to the two-body decay into a gluon. In addition to the channels discussed above, also a chargino channel appears with a top and bottom quark, due to the charged higgsino components which are typically close in mass to the neutral higgsinos. As discussed in Section 1.3.4, the relative strength of the Higgs coupling to up-type/down-type quarks is determined by $\tan\beta$, with high $\tan\beta$ favouring down-type quarks.

**Nature of the neutralino**

Another important element in the analysis concerns the nature of the heavy neutralino which is created in the gluino decay. Depending on its nature (i.e. mostly bino, wino or higgsino), different decay channels are important. The dominant channels are into $\gamma$, $Z$ or $h$, plus the goldstino/gravitino[14]. The decay widths of a neutralino NLSP are given by [282, 304]

$$\Gamma(\tilde{\chi}_1^0 \rightarrow \tilde{G} + \gamma) = \frac{m_{\tilde{\chi}_1^0}^5}{16\pi F^2} a_\gamma, \tag{2.1}$$

$$\Gamma(\tilde{\chi}_1^0 \rightarrow \tilde{G} + Z) = \frac{m_{\tilde{\chi}_1^0}^5}{16\pi F^2} \left( a_{Z_T} + \frac{1}{2} a_{Z_L} \right) \left( 1 - \frac{m_Z^2}{m_{\tilde{\chi}_1^0}^2} \right)^4, \tag{2.2}$$

$$\Gamma(\tilde{\chi}_1^0 \rightarrow \tilde{G} + h) = \frac{m_{\tilde{\chi}_1^0}^5}{16\pi F^2} \frac{1}{2} a_h \left( 1 - \frac{m_h^2}{m_{\tilde{\chi}_1^0}^2} \right)^4, \tag{2.3}$$

---

[14]Decays into pairs of Standard Model fermions are also possible, but generally negligible [304].

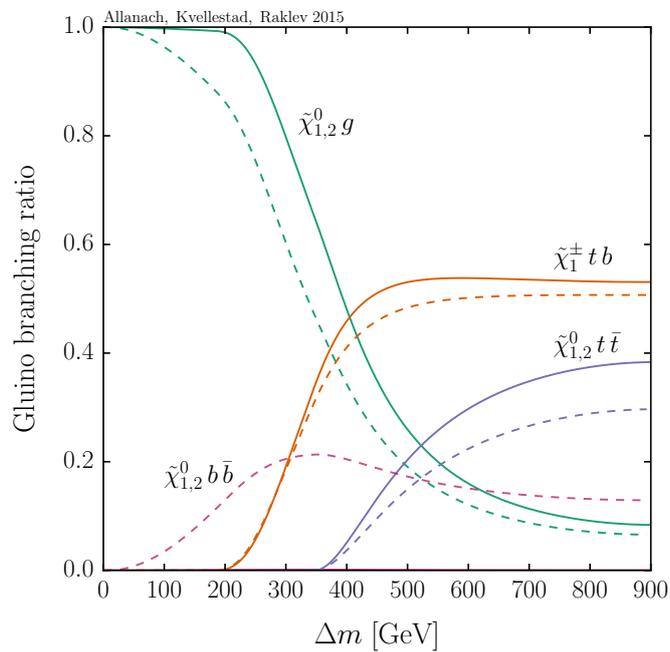

Figure 2.7: Gluino branching ratio as function of the gluino-neutralino mass splitting (gluino mass fixed at $m_{\tilde{g}} = 900$ GeV). The solid line is for $\tan \beta = 1.5$, while the dashed line is for $\tan \beta = 30$. Figure from [287].

with

$$a_\gamma = |N_{11} \cos\theta_W + N_{12} \sin\theta_W|^2, \tag{2.4}$$

$$a_{Z_T} = |N_{12} \cos\theta_W - N_{11} \sin\theta_W|^2, \tag{2.5}$$

$$a_{Z_L} = |N_{13} \cos\beta - N_{14} \sin\beta|^2, \tag{2.6}$$

$$a_h = |N_{13} \sin\alpha - N_{14} \cos\alpha|^2. \tag{2.7}$$

Here, $\sqrt{F}$ is the SUSY-breaking scale and $N_{ij}$ are elements of the neutralino mixing matrix. In the decoupling limit where the Higgs with the lowest mass behaves as the Standard Model Higgs, the coefficient in front of the decay with to $h + \tilde{G}$ becomes $|N_{13} \cos\beta + N_{14} \sin\beta|^2$. We will derive the coupling constants appearing in these formulas in a more general context in Section 2.2.

From these formulas, we find that for a pure bino NLSP, the possible decay channels are into a photon or into a $Z$-boson, plus the goldstino. Since the branching ratio is fixed by the Weinberg angle, in this case 77% of the neutralinos decay unavoidably into photons, which are strongly constrained from $\gamma + \not{E}_T$ searches, while only 23% of the neutralinos decay into $Z$-bosons. In the case of the wino, the situation is similar but reversed. Therefore, bino- and wino-like neutralinos are not suitable to explain the excess[15].

On the other hand, a higgsino NLSP can decay predominantly into $Z$-bosons. In the decoupling limit, this occurs for low $\tan\beta$. Therefore, since the excess is seen in a channel with the presence of a $Z$ boson, considering that decays into photons would be easy to rule out and that reduced branching fraction into $Z$ requires a higher production cross section (i.e. more easily constrained by other searches), a mostly-higgsino neutralino is the most natural explanation.

**Constraining models explaining the excess**

The consistency of the ATLAS excess in a simplified GGM model, testing also the constraints from other, complementary, searches, has been tested in [287]. Here, we summarise its main points. The simplified model under study is similar to the one in the ATLAS analysis: gluino production followed by decay into a mostly-higgsino neutralino, which subsequently decays into a goldstino. However, it fixes some of the inconsistencies in the ATLAS benchmark model. In the ATLAS analysis, the sfermion soft masses were fixed at $m_{\tilde{f}} = 1.5$ TeV. In this case, squark-gluino pair production dominates over gluino pair production if the gluino mass is larger than $m_{\tilde{g}} \sim 1$ TeV ($\tilde{q}\tilde{q}$ dominates for sufficiently high gluino masses). Moreover, the gaugino soft masses $M_1$ and $M_2$ were fixed at 1 TeV, in which case more complicated gluino decay chains open up and the NLSP is no longer mostly-higgsino for $\mu$ close to 1 TeV [287]. Therefore, for this analysis, these issues were avoided by fixing instead $M_1 = M_2 = 1.5$ TeV and $m_{\tilde{f}} = 4.5$ TeV, defined at a scale $\sqrt{m_{\tilde{t}_1} m_{\tilde{t}_2}} \sim 4.5$ TeV. In addition, this analysis takes into account properly the possible decay channels of the gluino, in particular the high branching ratio into top quarks for high gluino-neutralino mass splittings. While for the dilepton+jets+$\not{E}_T$ analysis this has relatively little impact compared to using the decay into first and second generation squarks, it is important when considering constraints from other searches.

---

[15]However, if $N_{12} \cos\theta_W - N_{11} \sin\theta_W \approx 1$, a cancellation occurs and the neutralino decay is mostly into $Z$. Such a situation is possible in general gauge mediation and has been considered as an explanation for the ATLAS excess in [285].

Additional leptons from top quark decays[16] (where the *Z*-boson can then decay hadronically, which has a high branching ratio) are constrained by the ATLAS stop search [297] and the CMS multi-lepton search [305]. In the compressed region, on the other hand, no top quarks are produced, but there are strong constraints from the jets+$\not{E}_T$ search discussed in the previous section due to the hadronic decay of the *Z*. Finally[17], also the CMS search for a similar final state (also discussed in the previous section) provides strong constraints on the largest part of the parameter space. These constraints are shown for two values of $\tan\beta$ in the gluino-neutralino mass plane in Figure 2.8, superimposed on the allowed region of the parameter space which can explain the ATLAS excess. We see that most of the parameter space is already ruled out by the CMS search for a similar final state, with the other small regions ruled out by the ATLAS jets+$\not{E}_T$ search and the CMS multilepton search. Therefore, it is difficult to explain the ATLAS excess within a GGM model (in particular variants similar to the original benchmark model considered in the ATLAS analysis). Moreover, since the exclusion is due to very general arguments and qualitatively different regions of the parameter space are covered by different searches, the same conclusion is likely true for most of the MSSM parameter space.

**Possible interpretations**

Given the strong constraints on possible models which can provide an explanation to the ATLAS excess, several solutions have been proposed in different models. Here, we discuss some of these proposed explanations, before moving on to our own interpretation in the next section. The study in [286] found that when both the intermediate and final state neutralinos are massive, the fit is significantly improved. In this case, some of the energy from gluino decay is taken by the massive neutralinos, such that the jets are softer, reducing the constraints from jets+$\not{E}_T$ searches.

Such a scenario can be realised in the NMSSM, with a bino-like NLSP and a singlino (fermionic partner of the singlet which generates the Higgs $\mu$-term) LSP with strong production through gluinos [286, 306] or through squarks [300, 301]; or with a higgsino-like NLSP through gluino pair production [307]. Alternatively, in [285], several benchmark points within the GGM framework have been considered (with a mixed bino-wino NLSP tuned to suppress decay into $\gamma$), which are on the border of exclusion from jets+$\not{E}_T$ searches. However, in [287], it has been found that these points are either excluded by other searches or do not predict a sufficient number of events to explain the ATLAS excess. Pure MSSM explanations, which do not suffer from the problems induced by an almost-massless LSP in gauge mediation, have also been considered. A scenario with a light sbottom or stop with a higssino-like NLSP and a bino-like LSP, was studied in [308], which can only marginally explain the excess due existing limits on the sbottom mass, while the stop dominant decay channel $\tilde{t}_1 \to b\tilde{\chi}_1^+$ does not produce $Z$-bosons. A scenario with mixed stops was discussed in [309], where the $Z$-boson is produced in $\tilde{t}_2 \to \tilde{t}_1 Z$ and the lightest stop is close in mass to the LSP so that the decay into tops in kinematically forbidden (decaying instead into $W^{(*)}b\tilde{\chi}_1^0$). Another model considers the pMSSM parameter space and finds a successful fit with light-flavour squark production in the 500–750 GeV range (so typically smaller than gluino models, since the gluino cross section is higher for the same mass), a bino-like NSLP and higssino-like LSP [302]. In this case, the jet+$\not{E}_T$ constraints are ameliorated both by the LSP mass and by the fact that squarks prefer to decay into the heavy bino-like NLSP instead of into the higssino-like LSP due to its small coupling to the latter (due to relatively small Yukawa couplings). Also, a split SUSY scenario, where sfermions are much heavier than the gauginos and higgsinos, was studied in [310], with gluinos decaying into a higgsino-like NLSP which subsequently decays into a bino-like LSP. Finally, also non-SUSY models have been studied. Some consider pair production of vector-like quarks, which subsequently decay into first generation quarks and a $Z$-boson [311] or heavy gluon production into pairs of bottom partner vector-like quarks which subsequently decay into $b$-quarks and $Z$-bosons [312]. Another study considers a 750 GeV quarkonium model initially built to explain an excess at 750 GeV in diphoton events [313]. Finally, also a model with leptophobic color-singlet gauge bosons has been proposed [314], where this new $Z'$-boson decays through anomaly-cancelling "anomalons" into a $Z$-boson, missing energy and jets.

---

[16]Note that this additional constraint does not appear in the case of squark pair production (of light flavour squarks only), where no top quarks are expected to be produced. However, since this search covers only a small corner of the parameter space not already covered by the CMS search (see below), this does not change the conclusion significantly.

[17]At the time when this work was performed, no searches for soft leptons were available, while such searches might also impose constraints on the parameter space.

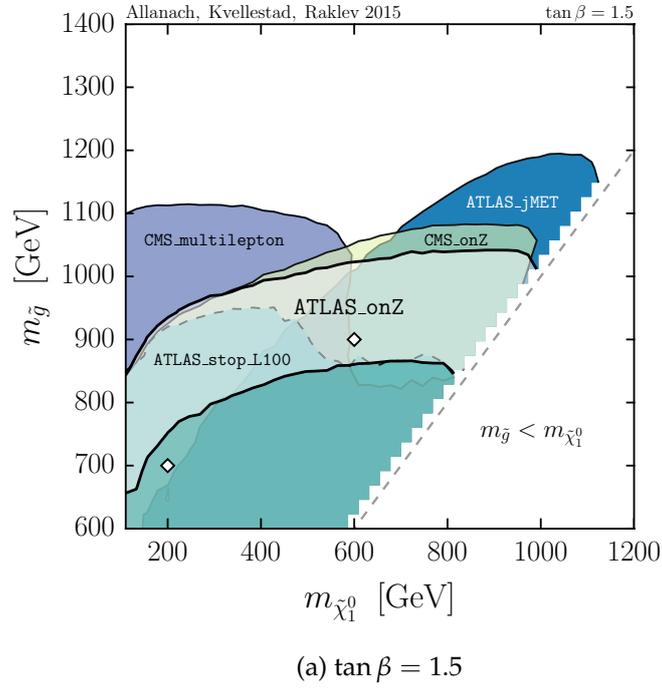

(a) $\tan\beta = 1.5$

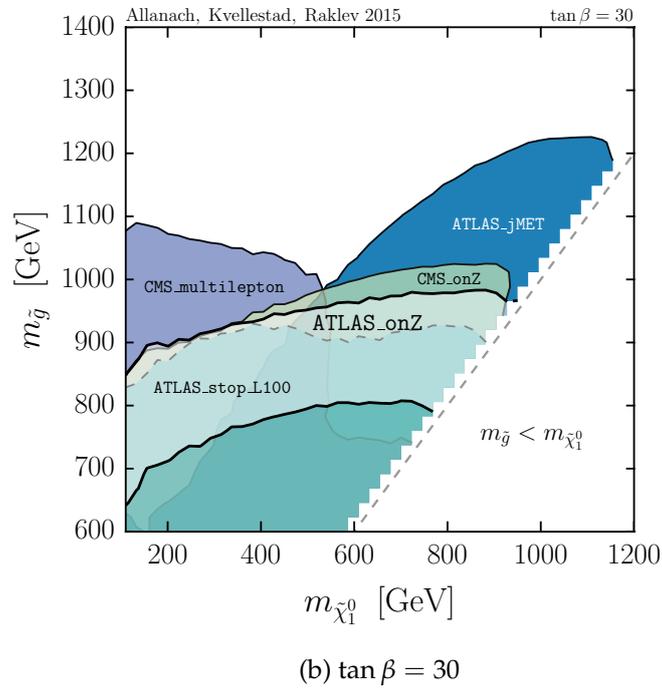

(b) $\tan\beta = 30$

Figure 2.8: Constraints from different searches on a simplified GGM model that can explain the ATLAS excess in the gluino-neutralino mass plane. The `ATLAS_onZ` region shows the 95% CL allowed region for the ATLAS excess, while the other regions show the 95% CL exclusion contours. Figure from [287].

## 2.2 The goldstini model

We have seen in the previous section that the model considered in the interpretation by ATLAS is not consistent. Moreover, if made consistent, there are significant constraints from other searches, essentially ruling out this interpretation. In the rest of this chapter, we consider an alternative model that can explain the $Z$-peaked excess seen by ATLAS, whilst remaining compatible with other searches. The model we consider is a gauge-mediation scenario where, instead of a single hidden sector which is responsible for SUSY breaking, there are now multiple secluded or hidden SUSY-breaking sectors, in our case two, the so-called goldstini model. Such a scenario is not more exotic than the existence of a single hidden sector. Due to the second hidden sector, there appears in our model a massive pseudo-goldstino. This massive pseudo-goldstino will allow our model to evade constraints from the ATLAS jets+$\not{E}_T$ search, while still explaining the ATLAS excess in a gauge-mediation model[18].

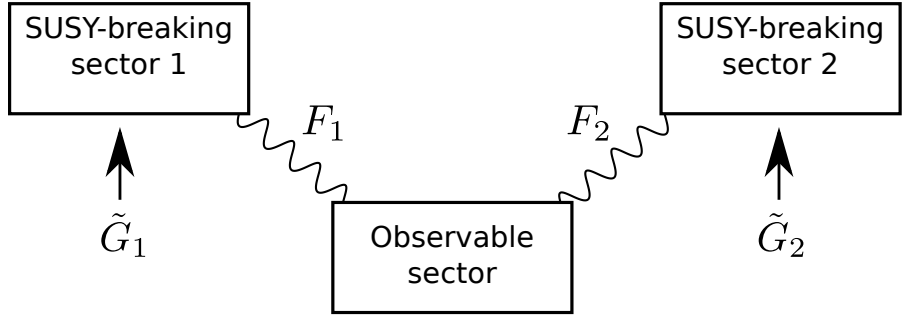

Figure 2.9: The goldstini model. Supersymmetry is broken in multiple (here two) hidden sectors, with sizes $F_1$ and $F_2$, which communicate with the observable sector through gauge interactions. Each sector has an associated goldstino when gauge interactions are turned off.

In this section, we review the theory behind the goldstini model and derive the true goldstino and the pseudo-goldstino couplings with the MSSM states; see also Refs. [315–319] and the excellent review in [320]. The application to the ATLAS excess is deferred to Section 2.3.

### 2.2.1 The setup: goldstini from multiple SUSY-breaking sectors

In our analysis, we consider a goldstini model in the context of gauge mediation, which is shown in Figure 2.9. Goldstini models, studied in [315–319, 321–328], have as a defining feature the presence of multiple SUSY-breaking sectors. When turning of gauge interactions, each SUSY-breaking sector contains a massless goldstino, i.e. the Goldstone fermion of spontaneous SUSY breaking. In a two-sector scenario, as we consider here, there are then two, initially massless, goldstini. When taking into account the mediation mechanism and the interaction of the SUSY-breaking sector with the MSSM, a linear combination of the goldstini becomes the true goldstino $\tilde{G}$ (i.e. the longitudinal component of the gravitino), while the orthogonal combination is a pseudo-goldstino $\tilde{G}'$ which can obtain an appreciable mass due to quantum corrections [325].

As Goldstone modes of supersymmetry breaking, the couplings of the individual SUSY-breaking sectors' goldstini are fixed by supersymmetry. When both sectors are connected to the MSSM, only a single true goldstino can remain, whose couplings are again determined by supersymmetry. The noteworthy feature is that while the pseudo-goldstino coupling structure is still dictated by supersymmetry (since this was true for each of the original goldstini), the sizes of the couplings are generically different with respect to the goldstino couplings. As we will see, they depend on the relative contribution to the soft masses from each of the SUSY-breaking sectors, and can be enhanced without changing the overall soft masses. This implies that the decay of SUSY particles can be dominantly into a massive pseudo-goldstino plus a Standard Model particle, drastically changing the final state topology with respect to the usual decay into the true goldstino [317–319, 326–328].

---

[18]In this way, we can thus retain its attractive features in explaining the soft masses while naturally satisfying the flavour constraints (compared to e.g. gravity-mediated SUSY-breaking models).

Moreover, while the goldstino mass after taking into account gravity is always low, the same is not true for the pseudo-goldstino. The true goldstino (or gravitino) mass is related to the SUSY-breaking scale and the Planck scale as $m_{\tilde{G}} \propto F/M_{\text{Pl}}$, and hence low-energy SUSY-breaking scenarios as gauge mediation lead to a very light goldstino. On the other hand, the mass of the pseudo-goldstino is not protected by any symmetry and generically receives relevant quantum corrections, due to the indirect interactions between the hidden sector induced by their common interactions with the MSSM particles. These corrections are proportional to the SUSY-breaking terms [325]. In this work, we consider the pseudo-goldstino mass to be of the order of a few hundred GeV.

In the next section, we will consider a model where the lowest supersymmetric particle states are the mostly-higgsino neutralinos, the pseudo-goldstino and the true goldstino. As we will see, the goldstino is, in our case, irrelevant for collider phenomenology (see Section 2.3.3). Therefore, we are interested in the decay of the neutralino into a pseudo-goldstino and an electroweak gauge boson or a Higgs boson, shown in Figure 2.10. The interaction lagrangian which is relevant for these decays is given by [315–319]

$$
\begin{aligned}
\mathcal{L}_{\tilde{G}'}^{\text{int}} = {}& i \frac{\tilde{y}_\gamma^i}{2\sqrt{2}F} \tilde{G}' \sigma^\mu \bar{\sigma}^\nu \tilde{\chi}_i^0 A_{\mu\nu} + i \frac{\tilde{y}_{Z_T}^i}{2\sqrt{2}F} \tilde{G}' \sigma^\mu \bar{\sigma}^\nu \tilde{\chi}_i^0 Z_{\mu\nu} \\
& + \frac{\tilde{y}_{Z_L}^i m_Z}{\sqrt{2}F} \tilde{\chi}_i^{\dagger 0} \bar{\sigma}^\mu \tilde{G}' Z_\mu + \frac{\tilde{y}_h^i}{\sqrt{2}F} \tilde{\chi}_i^0 \tilde{G}' h + h.c.,
\end{aligned}
\tag{2.8}
$$

where $A_{\mu\nu}$ and $Z_{\mu\nu}$ are the field strengths of the photon and the $Z$ boson, respectively, and $h$ is the lightest Higgs boson in the decoupling limit. The goldstino Lagrangian is the same, but with different coefficients, $\tilde{y} \to y$. As mentioned above, the pseudo-goldstino couplings $\tilde{y}$ can be larger than the goldstino couplings $y$, and in a simplified model approach they can be considered as free parameters [317, 318, 327]. Given a set of soft terms originating from the two SUSY-breaking sectors, one can compute these couplings. In this section, we derive expressions for the couplings of the goldstino and pseudo-goldstino which are valid as long as the pseudo-goldstino mass is not too high. While this requirement is not fulfilled in the case we will consider in the next section, so that the exact couplings need to be derived numerically, the expressions show explicitly the origin of the pseudo-goldstino phenomenology we wish to investigate and serve as a check on the numerical results. As in the previous section, we follow the conventions of [67] (see also [88]).

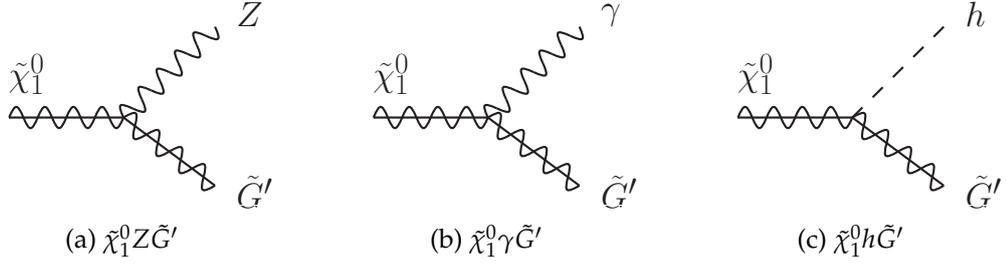

Figure 2.10: Decay modes of the lightest neutralino into a pseudo-goldstino and an electroweak gauge boson or a Higgs boson.

### 2.2.2 Goldstino coupling for one-sector SUSY breaking

First, we discuss the goldstino coupling in case of only a single hidden sector.

**General coupling structure**

In the case of a single supersymmetry-breaking sector, the interactions of the goldstino with the particles in the visible sector are determined by the supercurrent [147].

We can obtain an expression for the goldstino couplings at linear level using the supercurrent for a generic supersymmetric Lagrangian with renormalisable interactions, following the reviews in [67, 320]. As we saw in Section 1.2.3, in this case the conserved supercurrent is given by the expression

$$
\begin{aligned}
J^\mu_\alpha = {} & (\sigma^\nu \overline{\sigma}^\mu \psi_i)_\alpha \, D_\nu \phi^{*i} + i(\sigma^\mu \psi^{\dagger i})_\alpha \, W^*_i \\
& + \frac{1}{2\sqrt{2}} (\sigma^\nu \overline{\sigma}^\rho \sigma^\mu \lambda^{\dagger a})_\alpha \, F^a_{\nu\rho} + \frac{i}{\sqrt{2}} g_a \phi^* T^a \phi (\sigma^\mu \lambda^{\dagger a})_\alpha,
\end{aligned}
\tag{2.9}
$$

along with $F$- and $D$-term equations of motion

$$
F_i = -W^*_i, \tag{2.10}
$$

$$
F^{*i} = -W^i, \tag{2.11}
$$

$$
D^a = -g(\phi^* T^a \phi). \tag{2.12}
$$

Assume now that, as in most models, SUSY breaking is entirely due to $F$-term breaking. One part of the sum in the supercurrent term $i(\sigma^\mu \psi^{\dagger i})_\alpha W^*_i$ is then associated with the goldstino supermultiplet. Using the fact that the $F$-term of the goldstino supermultiplet obtains a VEV $\langle F \rangle$ (as we saw in Section 1.4.1), the conservation of the supercurrent can be split up as

$$
\partial_\mu J^\mu_\alpha = -i\langle F \rangle (\sigma^\mu \partial_\mu \tilde{G}^\dagger)_\alpha + \cdots + \partial_\mu j^\mu_\alpha = 0, \tag{2.13}
$$

where $j^\mu_\alpha$ is the supercurrent associated to all other supermultiplets (and thus not conserved on its own) and the ellipsis contains other goldstino supermultiplet terms. This equation can be interpreted as an equation of motion for the goldstino. The Lagrangian associated with this equation of motion is

$$\mathcal{L}_{\text{goldstino}} = i\tilde{G}^{\dagger}\overline{\sigma}^{\mu}\partial_{\mu}\tilde{G} - \frac{1}{F}(\tilde{G}\partial_{\mu}j^{\mu} + \text{c.c.}), \tag{2.14}$$

since we obtain the former equation by applying the Euler-Lagrange equation on this Lagrangian (and using Eq (A.22)). This coupling of the Goldstone mode to the current is an analogue of the Goldberger-Treiman relationship [329] for the bosonic case.

This coupling structure is a general result: take an effective Lagrangian $\mathcal{L}^{\text{eff}}$ which explicitly breaks a certain symmetry. It then possesses a current, still defined using the Noether procedure, which is now not conserved $\partial_{\mu}j^{\mu} \neq 0$. One can turn this theory into a spontaneously broken one by adding a Goldstone mode $\pi$, which by definition of a spontaneously broken symmetry must transform under the symmetry by a shift $\delta_{\epsilon}\pi = \epsilon v$. Adding a term in the Lagrangian of the form $-\frac{1}{v}\pi\partial_{\mu}j^{\mu}$ (along with a kinetic term for the Goldstone mode) then ensures that the Goldstone mode couples in such a way to ensure that the symmetry is restored (the variation of $\pi$ in the extra term cancels the non-zero variation of the original Lagrangian, where the latter is proportional to the non-conserved current) and the total current is conserved (as can be checked by using the Noether procedure on the new Lagrangian). Note that in both of these derivations above, higher-order corrections must be introduced using an iterative procedure to make the full Lagrangian invariant (e.g. using the above procedures, we do not obtain self-interactions or interaction terms with more than one goldstone mode).

Using the result in Eq. (2.14), we can easily identify the interactions of the goldstino with the MSSM particles, as dictated by supersymmetry, without having to specify the SUSY-breaking sector from which the goldstino originates. Note that the interaction term seems to diverge as supersymmetry is restored $\langle F \rangle \to 0$. However, using the equations of motion for the goldstino and MSSM particles (i.e. going on-shell), the interaction part of the Lagrangian is of the form

$$\mathcal{L}_{\tilde{g}} = -\frac{1}{F}(\tilde{G}^{\alpha}\Delta_{\alpha}), \tag{2.15}$$

where $\Delta_{\alpha} = \partial_{\mu}j^{\mu}\big|_{\text{e.o.m.}}$. This is then called the non-derivative form of the goldstino coupling. When using the equations of motion[19], the couplings become proportional to the squared-mass splittings $m_{\phi}^2 - m_{\psi}^2$ and $m_{\lambda}^2 - m_A^2$ (see e.g. [317, 320]), i.e. the soft masses. This immediately ensures that the interaction terms do not diverge as $\langle F \rangle \to 0$, since in that limit the soft masses vanish as well [67, 320]. The result here also agrees with our expectation, namely that the goldstino, which arises from SUSY breaking, couples to the non-conserved part of the supercurrent due to the soft masses. The couplings with the MSSM states are thus proportional to the soft supersymmetry-breaking masses (see e.g. [317] for a toy model explicitly illustrating this or the full derivation for the MSSM in general [320] and in detail [330]).

---

[19]Note that the interaction term with $\partial_{\mu}j^{\mu}$ contains two derivatives after substituting the non-goldstino part of Eq. (2.9).

By performing integration by parts, we can also express the interaction part of the Lagrangian in its derivative form[20]

$$\mathcal{L}_{\tilde{\mathcal{J}}} = \frac{1}{F}(\partial_\mu \tilde{G} j^\mu + \text{c.c.}), \qquad (2.16)$$

which is the original form in which these couplings were derived [147]. Again, there is a general argument why the coupling must be of this form: if one can always find a field redefinition in which the goldstone mode is coupled only derivatively, it is ensured that this mode is massless both classically and after adding quantum corrections. Another argument is that the theory should be invariant under a constant shift of the goldstone (by definition of a spontaneously broken symmetry), which is guaranteed if the coupling is with a derivative[21].

The advantage of using the supercurrent method to derive the goldstino couplings is that all mixing effects between the neutralino and the goldstino are automatically taken into account. In the method from Section 2.2.3, this is no longer the case and one needs to be more careful [320].

**Obtaining the $\tilde{\chi}_i^0 \tilde{G} Z$-coupling**

To explain the ATLAS signal, we are interested in the coupling between the neutralino, the (pseudo-)goldstino and the $Z$-boson[22]. In order to provide a check on our calculation in the case of multiple SUSY-breaking sectors in the next section, we calculate here already the couplings of the true goldstino, since this calculation must be valid for any number of SUSY-breaking sectors.

The gauge field appears in the first term of the supercurrent in Eq. (2.9) (through the covariant derivative) and the third term (through the gauge field strength). The former gives rise to a coupling to the higgsinos, while the latter gives a coupling to the gauginos. After rotating the (neutral) gauge bosons, $B_\mu$ and $W_\mu^3$, associated to $U(1)_Y$ and $SU(2)_L$ into their mass eigenstates, $Z_\mu$ and $A_\mu$, with the Weinberg angle[23] and rotating

---

[20]Note that the derivative-coupling Lagrangian contains operators of dimension 6, while the non-derivative coupling Lagrangian contains operators of dimension 5. Considering e.g. $2 \to 2$ scattering processes, the squared amplitudes must scale like $s^2$ and $s$ respectively (this can be obtained by considering that any $2 \to 2$ scattering amplitude must be of the same dimension as e.g. that of the dimensionless quartic scalar interaction in order to obtain the correct cross section for the cross section), with $\sqrt{s}$ the centre-of-mass energy. Therefore, since both descriptions are equivalent, SUSY must ensure that the leading order contributions $\propto s^2$ in the derivative-form cancel [317].

[21]There is yet another argument (see e.g. [331]): the vacuum is not invariant under a symmetry transformation, i.e. $\bar{Q}|0\rangle \neq 0$, with $Q = \int J^0(x) \, \mathrm{d}x$. Therefore, one has $\langle 0|J^\mu(x)|\pi\rangle \neq 0$. Using Lorentz invariance, this matrix element must be of the form $\langle 0|J^\mu(x)|\pi\rangle \propto p_\pi^\mu e^{ip_\pi \cdot x}$. Such a matrix element can be directly obtained if the Lagrangian contains a term $\mathcal{L} \sim J^\mu \partial_\mu \pi$.

[22]In the next section, we will derive also the coupling to the Higgs and the photon when treating the case of multiple SUSY-breaking sectors explicitly.

[23]For completeness, these are given by

$$\begin{cases} W_\mu^3 = \cos\theta_W Z_\mu + \sin\theta_W A_\mu \\ B_\mu = -\sin\theta_W Z_\mu + \cos\theta_W A_\mu \end{cases}. \qquad (2.17)$$

the gauginos and higgsinos into their mass eigenstates, the neutralinos, using Eq. (1.119), we find for the derivative form of the goldstino coupling

$$\mathcal{L}_{\partial}^{\tilde{\chi}_i^0 \tilde{G} Z} = i \frac{m_Z a_{Z_L}}{\sqrt{2} F} \partial_\mu \tilde{G} \sigma^\nu \overline{\sigma}^\mu \tilde{\chi}_i^0 Z_\nu + \frac{a_{Z_T}}{2\sqrt{2} F} \partial_\mu \tilde{G} \sigma^\nu \overline{\sigma}^\rho \sigma^\mu \tilde{\chi}_i^{0\dagger} F_{\nu\rho} + \text{c.c.},  \tag{2.18}$$

with[24,25]

$$a_{Z_L} = N_{13}^* \cos\beta - N_{14}^* \sin\beta,  \tag{2.19}$$

$$a_{Z_T} = -N_{11} \sin\theta_W + N_{21} \cos\theta_W.  \tag{2.20}$$

We can also obtain the non-derivative form of these couplings, by integrating by parts again (or using the general expression from the start) and using the equations of motion for the neutralinos (Eqs. (C.33) and (C.34)) and the *Z*-boson ($\partial_\mu Z^{\mu\nu} + m_Z^2 Z^\nu + \cdots = 0$), dropping the higher order interaction terms in both cases. In this way, we find[26]

$$\mathcal{L}_{\not\partial}^{\tilde{\chi}_i^0 \tilde{G} Z} = \frac{m_Z (m_{\tilde{\chi}_i^0} a_{Z_L} + m_Z a_{Z_T})}{\sqrt{2} F} \tilde{\chi}_i^{0\dagger} \overline{\sigma}^\nu \tilde{G} Z_\nu + i \frac{-m_Z a_{Z_L} + m_{\tilde{\chi}_i^0} a_{Z_T}}{2\sqrt{2} F} \tilde{\chi}_i^0 \sigma^\mu \overline{\sigma}^\nu \tilde{G} Z_{\mu\nu} + \text{c.c.}  \tag{2.21}$$

These two forms are equivalent, but the second is more useful when generalising to multiple SUSY-breaking sectors in the following.

### 2.2.3 (Pseudo-)goldstino couplings in case of multiple SUSY-breaking sectors

In the case of multiple SUSY-breaking sectors, matters are more complicated. The derivation using the supercurrent is always valid for the true goldstino once all the hidden sectors are coupled to the MSSM, regardless of the number of SUSY-breaking sectors. However, due to the mixing between the goldstini, the same is not true for the pseudo-goldstini[27]. Nevertheless, one thing we do know is the structure of the pseudo-goldstino couplings. Since they originate from the same mechanism as above for the individual goldstini, the pseudo-goldstino couplings must have the same structure as the ones of the true goldstino, except with different coefficients[28].

---

[24]Different references give slightly different expressions for these coefficients, differing in complex conjugation and order of the indices (compare for example the expressions in [317–319]). The expressions given here are all consistent with each other.

[25]The notation for the coefficients follows from the fact that these terms couple to either the transversal or longitudinal component of the *Z*-boson [317].

[26]In order to agree with expressions in the literature, we reversed the order of the fields with Eq. (A.23).

[27]Using the derivation below, one can wonder whether a similar procedure is possible using the supercurrent formalism. The guiding principle below is that both sectors individually break SUSY and do not talk directly to one another, such that both of them provide a goldstino. Therefore, one can try to start with $\mathcal{L} \sim \sum_i \frac{1}{F_i} \tilde{G}_i \partial_\mu J$, which seems promising and faster than the method below. However, this does not lead to the correct result. One way to see this, is to apply the equations of motion on this equation: one then finds that both goldstini should couple to the total soft masses, which is incorrect (they should couple proportional to their contribution to the soft masses). Redefining the current to only take into account one sector only is not possible (see e.g. Eq. (2.9)). Using the supercurrent, it is impossible to disentangle the contributions from the two sectors. Therefore, the supercurrent formalism is inherently only useful for the true goldstino. Another argument for this is the following: if the goldstini couplings could be written in this form, one could use partial integration to give both of them a derivative coupling. In that case, both goldstini would be massless, which can not be the case.

The results of the previous section suggest another way to obtain the (pseudo-)goldstino couplings, which is easily generalisable to multiple SUSY-breaking sectors. Since the goldstino appears due to the soft SUSY-breaking masses, its couplings were proportional to these masses. In the case of multiple SUSY-breaking sectors, the various goldstini couplings must then also be proportional to their respective sector's contribution to the soft masses. While the true goldstino couplings must still be proportional to the total soft masses, the pseudo-goldstino (or pseudo-goldstini in case of more than two hidden sectors) will be proportional to the relative contribution of the different sectors to the soft masses. In the previous chapter, Section 1.4.4, we already saw how the appearance of the soft masses can be treated using spurion fields. Expanding this formalism, we can also obtain the (pseudo-)goldstino couplings.

**Non-linear goldstino multiplet $X_{NL}$**

We can obtain the couplings of the goldstino by constructing the goldstino multiplet as a non-linear representation of SUSY, following the discussions in [315, 320, 332].

In order to construct the goldstino multiplet, we use the property that the Goldstone mode arises by transforming the vacuum. Consider first the analogy for the Higgs mechanism. In this case, the Higgs VEV is given by

$$\langle h \rangle = \begin{pmatrix} 0 \\ v \end{pmatrix}. \tag{2.22}$$

The Goldstone bosons can be identified by applying the gauge transformation on the vacuum

$$h_{NL} = U \begin{pmatrix} 0 \\ v \end{pmatrix}, \qquad \text{with} \quad U \equiv e^{i\pi^a T^a/v}. \tag{2.23}$$

Indeed, if we expand the operator $U$, we recover the well-known expression for the Higgs multiplet containing the would-be Goldstone bosons

$$h = \begin{pmatrix} i\pi^1/2 + \pi^2/2 \\ v - i\pi^3 \end{pmatrix}. \tag{2.24}$$

Note that this parameterisation does not include the massive excitation $h$. The expression in Eq. (2.23) forms a non-linear realisation of the $SU(2)$ symmetry. From it, we can obtain the interaction Lagrangian by constructing all possible terms with $h_{NL}$. In a similar way, the effective Lagrangian of pions can also be constructed.

So, in general, we can identify the Goldstone mode by applying the relevant symmetry transformation on the VEV. The same procedure can be applied to spontaneous

---

[28]Sometimes parameterised by multiplying the coefficients $a_{Z_L}$ and $a_{Z_T}$ in Eq. (2.19) with prefactors $K_{a_{Z_L}}$ and $K_{a_{Z_T}}$, as in [317].

SUSY breaking [332–336]. Supersymmetry breaking is triggered by a chiral superfield which obtains a VEV for its *F*-component

$$\langle X \rangle = \theta^2 F. \tag{2.25}$$

The broken SUSY transformations are given by

$$\theta^\alpha \to \theta^\alpha + \tilde{G}^\alpha, \tag{2.26}$$

$$y^\mu \to y^\mu + 2i\tilde{G}^\dagger \bar{\sigma}^\mu \theta, \tag{2.27}$$

with $\tilde{G}^\alpha$ the goldstino. If we treat $F$ as non-dynamical constant, the second line is irrelevant. A non-linear representation of SUSY is then formed by[29]

$$X_{NL} = \left( \theta + \frac{\tilde{G}}{\sqrt{2}F} \right)^2 F \tag{2.28}$$

$$= \frac{\tilde{G}^2}{2F} + \sqrt{2}\theta\, \tilde{G} + \theta^2 F, \tag{2.29}$$

where we normalised the goldstino using $\tilde{G} \to \tilde{G}/\sqrt{2}F$. This explicitly shows the content of the full goldstino multiplet.

In the case of multiple SUSY-breaking sectors, each goldstino multiplet is similarly given by

$$X_a = \frac{\tilde{G}_a^2}{2F_a} + \sqrt{2}\theta\, \tilde{G}_a + \theta^2 F_a. \tag{2.30}$$

**Goldstino couplings: general case**

As already discussed in Sections 1.2.3 and 1.4.4, we can obtain the soft masses from a spurion analysis[30]. Using the non-linear goldstino multiplet above, we can give a more physical interpretation to this mechanism and identify simultaneously the goldstino couplings. Note that we restrict ourselves to a discussion in the context of gauge mediation, where the coupling of the gravitino to matter is dominated by the goldstino, i.e. we can neglect effects of gravity.

The soft masses are generated when a chiral superfield $X_{NL}$ obtains a non-zero VEV $\langle X_{NL} \rangle = \theta^2 F$. We can then obtain the interactions between this hidden sector multiplet and the observable sector by listing all the allowed couplings between them, keeping in mind that $X_{NL}^2 = 0$. For example, for the coupling to the gauge kinetic term (Eq. (1.69)), we find

$$\mathcal{L} \supset -\int d^2\theta \, \frac{X_{NL}}{2\Lambda} \mathcal{W}^\alpha \mathcal{W}_\alpha + \text{c.c.}, \qquad (2.31)$$

which is similar to the expression we already obtained in the case of gravity mediation in Section 1.4.4, where now $X/M_P$ is replaced by $X_{NL}/\Lambda$ and we expect $\Lambda$ to be related to the messenger scale $M_{\text{messenger}}$. Performing the integration (using Eq. (1.64)), we find the Lagrangian

$$\begin{aligned}
\mathcal{L} \supset & -\frac{1}{2} M_\lambda \lambda\lambda + \frac{i M_\lambda}{2\sqrt{2} F} \lambda \sigma^\mu \overline{\sigma}^\nu \tilde{G} F_{\mu\nu} + \text{c.c.} \\
& + \frac{D}{\sqrt{2} M} \tilde{G}\lambda + \frac{D^2}{8 M F} \tilde{G}^2 + \text{c.c.},
\end{aligned} \qquad (2.32)$$

where we identify the gaugino mass as

$$M_\lambda \equiv \frac{F}{\Lambda}. \qquad (2.33)$$

---

[29]Note that this field satisfies $X_{NL}^2 = 0$. We can also instead *define* $X_{NL}$ as a constrained superfield satisfying $X_{NL}^2 = 0$ [332], which is the analogue of the constraint $UU^\dagger = 1$ in case of the Higgs or pion. Interpreted like this, we can build the most general Lagrangian and recover the Akulov-Volkov Lagrangian [101], which was the first non-linear realisation of supersymmetry. There is a subtle difference between treating $X_{NL}$ as a constrained superfield or as an expansion around a constant $\theta^2 F$ concerning the relation between the goldstino decay constant and the vacuum energy (which is realised for the constrained superfield and not for the other). However, the formalism with fixed $F$ is easiest for our purposes here [320].

[30]A useful analogy for spurion analyses for the soft SUSY-breaking masses is the following. Writing down a mass term for the quarks, while necessary, breaks the $SU(2)_L$ symmetry. However, we can restore the symmetry by thinking of these masses as transforming under $SU(2)_L$. In this way, we have effectively identified the Higgs field.

This expression agrees with the structure of Eq. (1.141). The first line[31] of Eq. (2.32) gives the soft masses for the gauginos and the coupling of the goldstino to the gauge multiplet. This latter coupling is proportional to the soft masses. This shows explicitly the result we already obtained for the non-derivative Lagrangian before.

In this way, we can account for all the soft SUSY-breaking masses in Section 1.2.5. In typical analyses, soft masses are included through the spurion $Y = \theta^2 m_{\text{soft}}$ [114]. We can then obtain the coupling to the goldstino multiplet from these simply by performing the substitution $Y \to \frac{m_{\text{soft}}}{F} X_{NL}$. We find the following Lagrangian

$$\begin{aligned}
\mathcal{L} \supset & -\int d^4\theta \, \frac{m_i^2}{F^2} X_{NL}^\dagger X_{NL} \Phi_i^\dagger e^{2V} \Phi_i - \left( \int d^2\theta \, \frac{M_a}{2F} X_{NL} \mathcal{W}^{\alpha a} \mathcal{W}_\alpha^a \right. \\
& \left. + \frac{t^i}{F} X_{NL} \Phi_i + \frac{b^{ij}}{2F} X_{NL} \Phi_i \Phi_j + \frac{a^{ijk}}{6F} X_{NL} \Phi_i \Phi_j \Phi_k + \text{c.c.} \right),
\end{aligned} \qquad (2.34)$$

which gives rise to the soft terms

$$\mathcal{L}_{\text{soft}} = -m^{i2}|\phi_i|^2 - \left( \frac{1}{2} M_a \lambda^a \lambda^a + t^i \phi_i + \frac{1}{2} b^{ij} \phi_i \phi_j + \frac{1}{6} a^{ijk} \phi_i \phi_j \phi_k + \text{c.c.} \right), \qquad (2.35)$$

---

[31]The second line, if the $D$-term acquires a VEV, also gives mass terms. The first term gives the mixing between the gaugino and the goldstino, while the second is a pure goldstino mass. The presence of either of these two terms gives mass to goldstino, but together they conspire to ensure that the goldstino remains massless [319, 320]. Its appearance is automatic using the non-linear representation $X_{NL}$, because of its scalar component. Without it, obtaining such a term requires a more detailed analysis.

and couplings to the goldstino

$$\mathcal{L}_{\tilde{G}} = \frac{1}{F}\tilde{G}\left( mi^2\psi_i\phi_i^\dagger + t^i\psi_i + b^{ij}\psi_i\phi_j + \frac{1}{2}a^{ijk}\psi_i\phi_j\phi_k \right.$$
$$\left. + \frac{iM_a}{2\sqrt{2}}\sigma^\mu\bar{\sigma}^\nu\lambda^a F^a_{\mu\nu} + \frac{M_a}{\sqrt{2}}\lambda^a D^a \right) + \text{c.c..} \tag{2.36}$$

This gives, again, the expected result that the goldstino couples proportionally to the soft masses and is suppressed by $F$.

This formalism can easily be extended to the case of multiple hidden sectors.

**Goldstino couplings for interpretation of the ATLAS excess**

We can now construct the couplings of the multiple goldstini to the observable sector fields. From the discussion above, these are determined by the same operators which generate the soft SUSY-breaking masses. The operators relevant to our discussion in the context of the ATLAS excess are those that generate the coupling of the goldstino to the $Z$-boson, the Higgs and the photon. These operators, and their expansion in component fields, are given by the bino soft mass

$$-\int d^2\theta \frac{M_{B(a)}}{2F_a}X_a\mathcal{W}\mathcal{W}, \tag{2.37}$$

and its expansion

$$-\frac{M_{B(a)}}{2}\left( \tilde{B}\tilde{B} - \frac{\sqrt{2}}{F_a}\tilde{G}_a\tilde{B}D_Y - \frac{i}{\sqrt{2}F_a}\tilde{B}\sigma^\mu\bar{\sigma}^\nu\tilde{G}_aB_{\mu\nu} \right), \tag{2.38}$$

(and analogous for the wino), the Higgs soft masses

$$-\int d^4\theta \frac{m^2_{H_{d/u(a)}}}{F_a^2}X_a^\dagger X_a H_{d/u}^\dagger H_{d/u}, \tag{2.39}$$

and their expansion

$$-m^2_{H_{d/u(a)}}\left( h_{d/u}^{0\dagger}h_{d/u}^0 - \left( \frac{1}{F_a}\tilde{G}_a\tilde{H}_{d/u}^0 h_{d/u}^{0*} + \text{c.c.} \right) \right), \tag{2.40}$$

and the $b$-term

$$-\int d^2\theta \frac{b_{(a)}}{F_a}X_a H_d H_u, \tag{2.41}$$

and its expansion

$$-b_{(a)}\left( h_d^0 h_u^0 - \frac{1}{F_a}\tilde{G}_a\left( \tilde{H}_d^0 h_u^0 + \tilde{H}_u^0 h_d^0 \right) \right). \tag{2.42}$$

where we denote with $M_{B/W(a)}$, $m^2_{H_{d/u(a)}}$ and $b_{(a)}$ the contribution to the soft term from sector $a$.

For the expansions above, we used the component expansion for chiral and anti-chiral superfields Eqs. (1.27)–(1.28), the expansion of the field-strength superfield Eq. (1.64), performed the integration after using Eq. (1.13) and we dropped the terms higher order in $1/F_a$ (which also have multiple goldstini) and the chargino terms.

These operators give rise to interactions between the goldstini and the MSSM states, as well as mixing terms between the goldstini and the neutralinos when a term contains only two fields or when one of the fields obtains a vacuum expectation value. The induced mixing between the goldstini and the neutralinos needs to be taken into account explicitly in order to obtain all the goldstini couplings, and therefore those of the goldstino and pseudo-goldstino.

### Gauge eigenstate Lagrangian

In this section, we review the gauge eigenstate Lagrangian relevant for the goldstini interactions in case of $n$ hidden sectors. The Lagrangian containing the interactions of the physical states, obtained after rotating into the mass eigenstate basis, will be discussed in the next section. The breaking of SUSY in $n$ separate hidden sectors results in the appearance of $n$ goldstini. These states are electrically neutral and have couplings with the gauginos and higgsinos of the MSSM. In particular, there appear mixing terms, such that we need to take into account the entire, now expanded, neutralino sector in order to obtain the goldstino and pseudo-goldstino interactions.

In the gauge eigenbasis $(\tilde{B}, \tilde{W}^{(3)}, \tilde{H}_d^0, \tilde{H}_u^0, \tilde{G}_1, \cdots, \tilde{G}_n)$, the usual $4 \times 4$ MSSM neutralino mass matrix is extended to a $(4+n) \times (4+n)$ symmetric mass matrix $\mathcal{M}$, given by

$$\mathcal{M} = \begin{pmatrix} M^{\tilde{\psi}} & \rho \\ \rho^T & M^{\tilde{G}} \end{pmatrix},$$ (2.43)

where $M^{\tilde{\psi}}$ is the usual $4 \times 4$ neutralino mixing matrix given by Eq. (1.118). The mixing terms between the gauginos, higgsinos and goldstini are contained in the $4 \times n$-matrix $\rho$, while the $n \times n$-matrix $M^{\tilde{G}}$ contains the interactions between the goldstini induced through the coupling with the MSSM (diagonal terms), as well as radiative corrections coupling the different hidden sectors to one another, which are important for the pseudo-goldstino mass. The goldstini part of the mass matrix will be discussed in more detail in Section 2.2.4.

Since the mixing between the usual neutralinos and the goldstini is suppressed by $1/F_a$, the neutralinos will be mostly the same as in the MSSM, while the goldstini mix mainly among each other. As a result, we can obtain the goldstini interactions by performing the rotation into the MSSM neutralinos as usual, while treating the mixing with the goldstini as insertions. Therefore, we split up the Lagrangian into a mass term for the MSSM neutralinos, a mass term for the goldstini and interactions of both of these. The Lagrangian relevant to the neutralino decay, in the gauge eigenstate basis

$(\tilde{\psi}_i = \{\tilde{B}, \tilde{W}, \tilde{H}_d, \tilde{H}_u\})$, is then given by

$$\mathcal{L} \supset -\frac{1}{2} M_{ij}^{\tilde{\psi}} \tilde{\psi}_i \tilde{\psi}_j - \rho_{ai} \tilde{G}_a \chi_i - \frac{1}{2} M_{ab}^{\tilde{G}} \tilde{G}_a \tilde{G}_b - \frac{1}{2} Y_{ij} \tilde{\psi}_i \tilde{\psi}_j h_0$$
$$+ \tau_{ai} \tilde{G}_a \tilde{\psi}_i h_0 + G_{ij} \tilde{\psi}_i^\dagger \bar{\sigma}^\mu \tilde{\psi}_j Z_\mu + i K_{ia} \tilde{\psi}_i \sigma^\mu \bar{\sigma}^\nu \tilde{G}_a A_{\mu\nu}, \tag{2.44}$$

where the indices $i$ and $j$ run over the gauginos and higgsinos, while $a$ and $b$ run over the hidden sectors.

The first term in this Lagrangian is the usual neutralino mass term, with $M_{ij}^{\tilde{\psi}}$ the standard $4 \times 4$ neutralino mass matrix given by Eq. (1.118). The second term is the linear mixing between the 4 neutralinos and the goldstini, with the mixing matrix $\rho_a$ given by

$$\rho_a = -\frac{1}{\sqrt{2}F_a} \begin{pmatrix} M_{B(a)} \langle D_Y \rangle \\ M_{W(a)} \langle D_{T^3} \rangle \\ \sqrt{2}v \left( m_{H_d(a)}^2 c_\beta - b_{(a)} s_\beta \right) \\ \sqrt{2}v \left( m_{H_u(a)}^2 s_\beta - b_{(a)} c_\beta \right) \end{pmatrix}. \tag{2.45}$$

The first two elements in $\rho_a$ come from the second term in the expansion of the bino and wino soft masses (Eq. (2.38)), where we take the VEV of the $D$-term (its dynamical part with Higgs bosons gives turns up in an interaction term). For the MSSM, with the equation of motion for $D$ in Eq. (1.89), these VEVs are given by $\langle D_Y \rangle = g'v^2 \cos 2\beta/2$ and $\langle D_{T^3} \rangle = -gv^2 \cos 2\beta/2$. The third and fourth element in $\rho_a$ are due to the second term in the expansion of the Higgs soft mass terms and $b$-term when taking the Higgs VEV.

The third term in the Lagrangian is the goldstino mass matrix, the discussion of which is deferred to Section 2.2.4.

The fourth term in the goldstini Lagrangian represents the coupling between the gauginos and higgsinos with the Higgs boson, given by

$$Y = \frac{1}{2} \begin{pmatrix} 0 & 0 & g's_\alpha & g'c_\alpha \\ 0 & 0 & -gs_\alpha & -gc_\alpha \\ g's_\alpha & -gs_\alpha & 0 & 0 \\ g'c_\alpha & -gc_\alpha & 0 & 0 \end{pmatrix}. \tag{2.46}$$

These terms are the gaugino-higgsino-Higgs couplings from the second line in Eq. (1.75), the supersymmetrised version of *Bhh* and *Whh* couplings.

The fifth term in the Lagrangian contains the coupling between a goldstini, a neutralino and the Higgs boson, with

$$\tau_a = \frac{1}{\sqrt{2}F_a} \begin{pmatrix} -m_Z M_{B(a)} s_W s_{\alpha+\beta} \\ m_Z M_{W(a)} c_W s_{\alpha+\beta} \\ -m_{H_d(a)}^2 s_\alpha - b_{(a)} c_\alpha \\ m_{H_u(a)}^2 c_\alpha + b_{(a)} s_\alpha \end{pmatrix}. \tag{2.47}$$

The first two elements arise again from the second term in the bino and wino soft mass expansion, but now using the dynamical part, i.e. using the expansion in Eq. (1.112), ignoring the higher mass Higgs states. The higgsino elements are due to the second Higgs mass term and *b*-term.

The sixth term in the Lagrangian gives the coupling between two neutralinos and a *Z*-boson, given by

$$G = \frac{g}{2c_W} \begin{pmatrix} 0 & 0 & 0 & 0 \\ 0 & 0 & 0 & 0 \\ 0 & 0 & 1 & 0 \\ 0 & 0 & 0 & -1 \end{pmatrix}. \tag{2.48}$$

This coupling is due to the higgsino kinetic term, which is given by

$$\left[ \Phi^{*i} (e^V)_i^j \Phi_j \right]_D \supset i \psi^{\dagger i} \bar{\sigma}^\mu D_\mu \psi_i \tag{2.49}$$

with the covariant derivative given by Eq. (1.78). After rotating to the mass eigenstates of the gauge bosons, this indeed gives the couplings in *G*. Similar terms for the bino and wino do not appear in the Lagrangian, since they are associated to an Abelian group or the neutral component of a non-Abelian group which have no self interactions, i.e. there are no *ZZZ* or $W^3 W^3 W^3$ couplings in the Standard model, so the corresponding supersymmetrised version is also absent.

The seventh term in the Lagrangian represents the coupling between a neutralino, a goldstino and a *Z*-boson, given by

$$L_a = \frac{1}{2\sqrt{2}F_a} \begin{pmatrix} -M_{B(a)} s_W \\ M_{W(a)} c_W \\ 0 \\ 0 \end{pmatrix}. \tag{2.50}$$

This coupling originates in the third term of the bino and wino soft mass term expansion, after rotating the gauge bosons to their mass eigenstates.

The eighth and final term in the Lagrangian gives a similar coupling to the photon

$$K_a = \frac{1}{2\sqrt{2}F_a} \begin{pmatrix} M_{B(a)} c_W \\ M_{W(a)} s_W \\ 0 \\ 0 \end{pmatrix}. \tag{2.51}$$

**Rotating to the mass eigenstates**

Given a set of soft terms and the relative contributions from the different hidden sectors, one should diagonalise the Lagrangian and write the resulting couplings in the mass eigenbasis. In this way, we obtain the Lagrangian quoted in Eq. (2.8), whose couplings can then be computed exactly for a given set of soft terms.

However, in order to gain insight into the features of the pseudo-goldstino couplings, we calculate here analytic expressions for these couplings at leading order. In order to do so, although we study a rather heavy pseudo-goldstino case in our scenario, we consider the $m_{\tilde{G}'} = 0$ limit here, such that the pseudo-goldstino behaves similar to the true goldstino.

In order to obtain the couplings of the physical mass eigenstates between the pseudo-goldstino and the neutralinos, we need to rotate both the goldstini and the gauginos and higgsinos into their mass eigenstates. As argued in the previous section, we can perform these rotations independently (which is justified at leading order and in the $m_{\tilde{G}'} = 0$-limit), while treating the mixing between goldstini and the neutralinos as interaction terms. Indeed, since the hidden sector VEVs $F_a$ are much greater than the $D$- and $F$-terms of the gauge and Higgs superfields, the true goldstino will be approximately a linear combination of only goldstini and have a small mixing with the neutralinos (see Eq. (2.45)).

First, we perform the rotation of the neutralinos and take into account the mixing terms between the goldstini and the gauginos and higgsinos. This can then be immediately compared with our previously obtained result for the non-derivative form of the goldstino couplings from the supercurrent Eq. (2.21). Afterwards, we perform the rotation into the true goldstino and pseudo-goldstino.

The $\tilde{\chi}_i^0 \tilde{G}_a Z_{\mu\nu}$- and $\tilde{\chi}_i^0 \tilde{G}_a A_{\mu\nu}$-coupling can be immediately obtained from the corresponding couplings ($\propto L_{ia}$) in the gauge eigenstate Lagrangian Eq. (2.44). Rotating the gauginos, we obtain (with $N_{ij}$ elements of the neutralino mixing matrix defined in Eq. (1.119))

$$\mathcal{L} \supset i\frac{M_{B,a}}{2\sqrt{2}F_a}\tilde{\chi}_i^0 N_{i1}^* \sigma^\mu \overline{\sigma}^\nu \tilde{G}_a\big(-\sin\theta_W Z_{\mu\nu} + \cos\theta_W A_{\mu\nu}\big)$$
$$+ i\frac{M_{W,a}}{2\sqrt{2}F_a}\tilde{\chi}_i^0 N_{i2}^* \sigma^\nu \overline{\sigma}^\nu \tilde{G}_a\big(\cos\theta_W Z_{\mu\nu} + \sin\theta_W A_{\mu\nu}\big). \tag{2.52}$$

We can check this result for the single SUSY-breaking sector case (i.e. drop the $a$ index). Using Eq. (1.120), we can reexpress the combination $M_{B,a}N_{i1}^*$ in terms of elements of the neutralino mixing matrix $N_{kl}$, the neutralino mass $m_{\tilde{\chi}_i^0}$, and combinations of $m_Z$ with Weinberg angles (and analogous for the wino soft mass). The result then agrees immediately with the $Z_{\mu\nu}$-part of the goldstino Lagrangian obtained from the supercurrent. For the $A_{\mu\nu}$ coupling, one can compare with the supercurrent calculation for a single sector in [317, 318].

The $h\tilde{\chi}_i^0 \tilde{G}_a$-coupling can be similarly obtained from the corresponding term ($\propto \tau_{ai}$) in the gauge eigenstate Lagrangian. In addition, we consider the decoupling limit $\alpha \approx \beta - \pi/2$, where the lightest Higgs boson corresponds to the SM Higgs boson and the higher mass states are decoupled. These results can again be compared with the supercurrent calculation for a single sector in [318].

Finally, there appears also an effective coupling to $Z_\mu$, i.e. $\tilde{\chi}_i^0 \tilde{G}_a Z_\mu$. In the supercurrent derivation, this coupling turned up automatically. Similarly, when numerically diagonalising the neutralino mass matrix, these also automatically appear. In our ap-proximation, however, we need to explicitly take into account the mixing between the neutralinos and the goldstinos. While these mixing terms are suppressed by $1/F_a$ compared to the neutralino mixing matrix, this induces couplings at the same order in $1/F_a$ as the ones above and must thus be included to be consistent. This is the reason why we have also written the terms containing only the neutralino gauge eigenstates in Eq. (2.44).

The goldstino appears from the higgsino coupling to $Z_\mu$ in three ways: through the higgsino-goldstino mixing, through higgsino-bino mixing followed by bino-goldstino mixing and through higgsino-wino mixing followed by wino-goldstino mixing, where we need to take into account a propagator for the internal higgsino[32], bino and wino lines. This is shown in Figure 2.11. Therefore, for the down-type higgsino, the interaction term is

$$\mathcal{L}_{\tilde{H}_d^0 Z \tilde{G}_a} = \frac{g}{2\cos\theta} \tilde{H}_d^{0\dagger} \bar{\sigma}^\mu Z_\mu \left[ \frac{-1}{\mu} \rho_{a4} + \frac{-1}{\mu} N_{14} \frac{1}{M_B} \rho_{a1} + \frac{-1}{\mu} N_{24} \frac{1}{M_W} \rho_{a2} \right] \tilde{G}_a, \qquad (2.53)$$

with a similar term appearing for $\tilde{H}_u^0$.

Afterwards, we again perform the rotation into the neutralino mass eigenstates using the neutralino mixing matrix and use Eq. (1.120). In the case of single-sector SUSY breaking, we obtain correctly the $Z_\mu$ couplings from the supercurrent derivation in Eq. (2.21), after using the minimisation conditions (Eqs. (1.110) and (1.111)).

Finally, using these expressions which give the correct result in the case of single-sector SUSY breaking, we can obtain the goldstino and pseudo-goldstino couplings in the case of multiple-sector SUSY breaking. The true goldstino is a linear combination of the different goldstini (as argued before, neglecting the mixing with the neutralinos). More precisely, it is proportional to $\tilde{G} \propto F_1 \tilde{G}_1 + F_2 \tilde{G}_2 + \cdots + F_n \tilde{G}_n$. This follows from the identification of the massless mode through the fermion mass matrix (Eq. (1.124)). Alternatively, this can be found by considering that each of the $F_a$ is an order parameter for SUSY breaking. When one of them goes to zero $F_a \to 0$, the corresponding goldstino should also drop out, while if one of the $F_a$ is dominant, the true goldstino should be mostly aligned with the goldstino corresponding to that sector. The pseudo-goldstini are then orthogonal to the true goldstino.

Consider now the case of two SUSY-breaking sectors. The goldstino and pseudo-goldstino are then given by

$$\tilde{G} = \frac{1}{F} \left( F_1 \tilde{G}_1 + F_2 \tilde{G}_2 \right), \qquad (2.54)$$

$$\tilde{G}' = \frac{1}{F} \left( -F_2 \tilde{G}_1 + F_1 \tilde{G}_2 \right), \qquad (2.55)$$

where $F = \sqrt{F_1^2 + F_2^2}$ is the total SUSY-breaking scale and the prefactor $1/F$ ensures that the true goldstino is properly normalised. One can check that this indeed gives the correct result for the true goldstino coupling. Indeed, inverting this set of equations and substituting $\tilde{G}_1$ and $\tilde{G}_2$ in the expressions for the goldstino couplings (i.e. Eq. (2.52)) and writing the bino, wino, the down/up type Higgs soft masses, and the soft $b$-parameter as

$$M_B = \sum_{a=1}^{2} M_{B(a)}, \tag{2.56}$$

$$M_W = \sum_{a=1}^{2} M_{W(a)}, \tag{2.57}$$

$$m_{H_{d/u}}^2 = \sum_{a=1}^{2} m_{H_{d/u}(a)}^2, \tag{2.58}$$

$$b = \sum_{a=1}^{2} b_{(a)}, \tag{2.59}$$

the goldstino couplings reduce to the correct form, i.e. proportional to the total soft masses.

---

[32]The off-diagonal mass term of the higgsinos means that the higgsino propagator also flips $\tilde{H}_d^0$ into $\tilde{H}_u^0$.

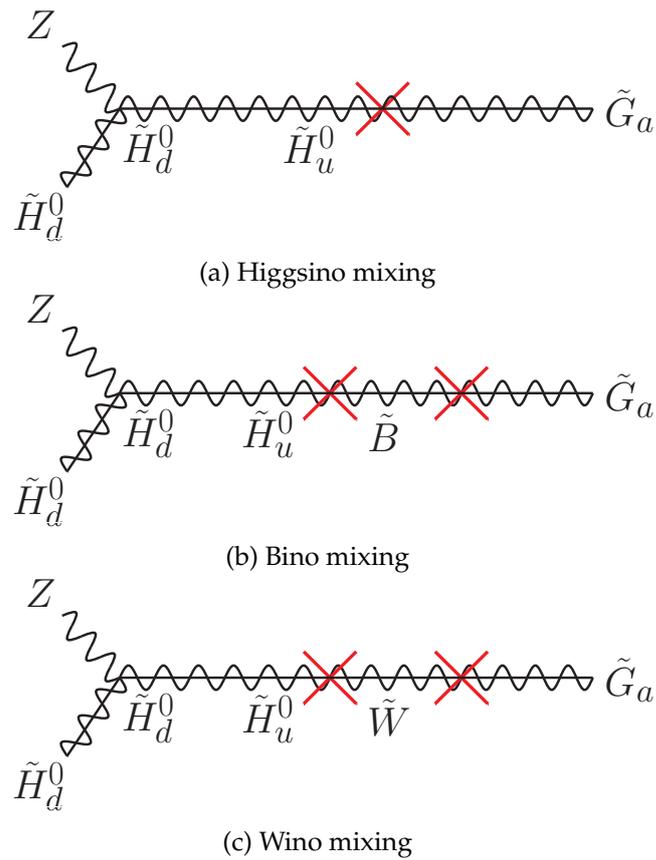

(a) Higgsino mixing

(b) Bino mixing

(c) Wino mixing

Figure 2.11: Coupling of the goldstini to the higgsino-component of the neutralino through the higgsino kinetic term and neutralino-goldstino mixing.

From this, it is easy to obtain now also the pseudo-goldstino couplings. Define, for convenience, the tilded soft masses, e.g.

$$\tilde{M}_B = -\frac{F_2}{F_1} M_{B(1)} + \frac{F_1}{F_2} M_{B(2)}. \tag{2.60}$$

These combinations appear when performing the rotation to the pseudo-goldstino. Pulling out the total soft masses from the pseudo-goldstino couplings and defining $K$-factors as $K_B = \tilde{M}_B/M_B$ and $K_W = \tilde{M}_W/M_W$ for Eq. (2.52) and analogous for the others, we can proceed as in the one-sector case, but carrying along the $K_B$ and $K_W$. In this way, we finally obtain the couplings pseudo-goldstino couplings, given by (repeated from the beginning of this section)

$$\begin{aligned}
\mathcal{L}_{\tilde{G}'}^{\text{int}} &= i\frac{\tilde{y}_\gamma^i}{2\sqrt{2}F} \tilde{G}'\sigma^\mu\bar{\sigma}^\nu\tilde{\chi}_i^0 A_{\mu\nu} + i\frac{\tilde{y}_{Z_T}^i}{2\sqrt{2}F} \tilde{G}'\sigma^\mu\bar{\sigma}^\nu\tilde{\chi}_i^0 Z_{\mu\nu} \\
&\quad + \frac{\tilde{y}_{Z_L}^i m_Z}{\sqrt{2}F} \tilde{\chi}_i^{\dagger 0}\bar{\sigma}^\mu\tilde{G}'Z_\mu + \frac{\tilde{y}_h^i}{\sqrt{2}F} \tilde{\chi}_i^0\tilde{G}'h + h.c.,
\end{aligned} \tag{2.61}$$

where we now have an expression for the coefficients appearing here. These are

$$
\begin{aligned}
\tilde{y}^i_\gamma =\ & m_{\tilde{\chi}^0_i}(K_B N_{i1} c_W + K_W N_{i2} s_W) \\
& + m_Z(N^*_{i3} c_\beta - N^*_{i4} s_\beta) s_W c_W (K_B - K_W),
\end{aligned} \tag{2.62}
$$

$$
\begin{aligned}
\tilde{y}^i_{Z_T} =\ & m_{\tilde{\chi}^0_i}(-K_B N_{i1} s_W + K_W N_{i2} c_W) \\
& + m_Z(N^*_{i4} s_\beta - N^*_{i3} c_\beta)(s^2_W K_B + c^2_W K_W),
\end{aligned} \tag{2.63}
$$

$$
\begin{aligned}
\tilde{y}^i_{Z_L} =\ & -m_Z K_\mu(-N_{i1} s_W + N_{2i} c_W) \\
& - m_{\tilde{\chi}^0_i}(K_u N^*_{4i} s_\beta - K_d N^*_{3i} c_\beta),
\end{aligned} \tag{2.64}
$$

$$
\begin{aligned}
\tilde{y}^i_h =\ & -m_Z m_{\tilde{\chi}^0_i} \cos 2\beta (-K_B N_{i1} s_W + K_W N_{i2} c_W) \\
& - |\mu|^2 (K_d N^*_{i3} c_\beta + K_u N^*_{i4} s_\beta).
\end{aligned} \tag{2.65}
$$

The *K*-factors read

$$
\begin{aligned}
K_B =\ & \frac{\tilde{M}_B}{M_B},\ K_W = \frac{\tilde{M}_W}{M_W},\ K_\mu = c^2_\beta K_d + s^2_\beta K_u, \\
K_d =\ & -\frac{1}{|\mu|^2}(\tilde{m}^2_{H_d} - \tilde{b} \tan\beta \\
& + \frac{m^2_Z}{2}(s^2_W K_B + c^2_W K_W) \cos 2\beta), \\
K_u =\ & -\frac{1}{|\mu|^2}(\tilde{m}^2_{H_u} - \tilde{b} \cot\beta \\
& - \frac{m^2_Z}{2}(s^2_W K_B + c^2_W K_W) \cos 2\beta),
\end{aligned} \tag{2.66}
$$

The goldstino Lagrangian [67, 304, 330, 332, 337] is recovered by converting all the tilded soft terms to the un-tilded ones, and using the electroweak symmetry-breaking minimisation conditions in Eqs. (1.110) and (1.111). As a result, all the *K* factors in Eq. (2.66) become unity.

From the above analytic couplings, one can already observe that for a mostly higgsino-like neutralino with $K_B = K_W$ the pseudo-goldstino coupling to the photon is suppressed, which is important for the interpretation of the ATLAS excess.

We checked that the above expressions agree with the numerics if the pseudo-goldstino mass is sufficiently smaller than the neutralino masses. For the large pseudo-goldstino mass, on the other hand, there can be deviations from these formulas once we rotate in the mass eigenbasis. Hence, in order to compute correctly the effective couplings in Eq. (2.8) for generic pseudo-goldstino masses, one is instructed to diagonalise numerically the Lagrangian Eq. (2.44), for a given set of soft terms, and write it in the mass eigenbasis.

### 2.2.4 Pseudo-goldstino mass

While there are no direct interactions between the hidden sectors, by construction, they communicate through their common interactions with the supersymmetric Standard

Model sector. As a result, only one linear combination of the goldstini can be the true goldstino, which remains massless since it is protected by supersymmetry, while the remaining linear combinations of goldstini form pseudo-goldstini and receive a mass through radiative corrections. The form and size of these masses, in the context of gauge mediation, was studied in detail in [325]. There, they find that the pseudo-goldstino mass cannot be determined from the low-energy theory only, but requires some details of the microscopic theory from the hidden sectors. Using general properties of these hidden sectors, one can estimate the pseudo-goldstino mass. Here, we review how one can estimate this mass, following the discussion in [319, 325].

As before, the goldstini mix mainly among each other, such that we can find the pseudo-goldstino mass(es) by studying only the goldstini sector of the mass matrix[33]. The $n \times n$ goldstini mass matrix receives two contributions. First, the SUSY operators giving soft masses to the bino, wino and higgsinos (Eqs. (2.38)–(2.42)) generate tree-level diagonal mass values. Indeed, taking the scalar component $\tilde{G}_a^2/2F_a$ of the goldstino multiplet $X_{NL}$ in the bino soft mass operator, along with the VEV of the auxiliary field, a diagonal mass term $M_{B(i)}\langle D_Y\rangle^2/(2F_a^2)$ is generated[34]. A similar term appears from the Higgs superfield $B$-term. However, in both cases, these contributions are suppressed by $1/F_a^2$ and proportional to auxiliary field VEVs which are negligible compared to the hidden sector VEVs, so these tree-level contributions are in general negligible compared to those from radiative corrections.

The second contribution to the goldstini mass matrix is due to radiative corrections, where the goldstini of the individual hidden sectors are connected with each other through the Standard Model gauge interactions. In [325], it was shown explicitly that the leading contributions are due to the operators in Eqs. (2.38), (2.40) and (2.42), by integrating out the gauge and Higgs superfields at one loop. The exact result depends on the details of the dynamics of the hidden sectors (e.g. the minimal gauge mediation model in Section 1.4.4). Once these are specified, one can explicitly compute the two-point functions $\langle \tilde{G}_a \tilde{G}_b \rangle$ to determine the goldstino mass matrix. However, even without specifying the dynamics of the hidden sector, it is possible to derive general results on the structure of this goldstino mass matrix $M^{\tilde{G}}$.

Since the supersymmetric Standard Model $F$- and $D$-terms are typically negligible compared to those in the hidden sector, the true goldstino will be a linear combination of the goldstini $\tilde{G} \propto F_1\tilde{G}_1 + F_2\tilde{G}_2 + \cdots + F_n\tilde{G}_n$ and is associated to a zero eigenvalue of the goldstino mass matrix. Therefore, we have $M^{\tilde{G}}\tilde{G} = 0$, which imposes $n$ conditions

---

[33]This seems counter-intuitive, since the radiative corrections giving mass to the pseudo-goldstino necessarily occur through the same interactions with the MSSM fields as those generating the mixing. However, the mixing occurs through the $D$-term VEVs, while the mass terms appear by integrating out dynamical fields, as we show below. On the other hand, for large pseudo-goldstini masses as we will consider here, this is no longer a good approximation. However, in this case, we already stressed the necessity to perform the rotation to the mass eigenbasis correctly using numerical methods. While some of the approximations obtained here to estimate the pseudo-goldstino mass might then no longer be strictly correct, the pseudo-goldstino mass still follows from the full neutralino-goldstino mass matrix.

[34]As already mentioned before, the presence of an off-diagonal mixing term between the bino and goldstino ensures that a zero eigenvalue for the goldstino sector appears [319, 320].

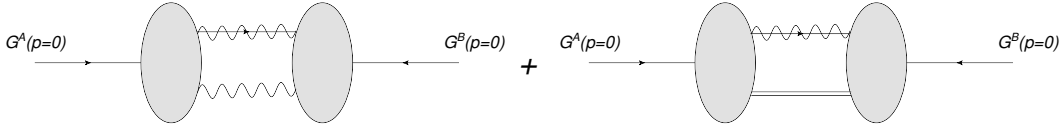

Figure 2.12: Radiative corrections to the goldstini mass matrix. At leading order, the goldstini propagator mixing the goldstini of the two sectors receives corrections involving unspecified hidden sector interactions (i.e. loops), mediated by supersymmetric Standard Model propagators, i.e. two fields from either the gauge field, gaugino or *D* auxiliary field, the latter being indicated by double lines (from *R*-parity, there needs to be a SM particle and one of its SUSY partners). Figure from [325].

on the mass matrix. As a result, we can express the diagonal entries in terms of the off-diagonal ones

$$M^{\tilde{G}} = \begin{pmatrix} -\frac{F_2\mathcal{M}_{12}+F_3\mathcal{M}_{13}+\cdots+F_n\mathcal{M}_{1n}}{F_1} & \cdots & \mathcal{M}_{1n} \\ \vdots & \ddots & \vdots \\ \mathcal{M}_{1n} & \cdots & -\frac{F_1\mathcal{M}_{1n}+F_2\mathcal{M}_{2n}+\cdots+F_{n-1}\mathcal{M}_{n-1n}}{F_n} \end{pmatrix}. \tag{2.67}$$

In the case of two SUSY-breaking sectors, the mass matrix $M^{\tilde{G}}$ is

$$M^{\tilde{G}} = \begin{pmatrix} -\frac{F_2}{F_1}\mathcal{M}_{12} & \mathcal{M}_{12} \\ \mathcal{M}_{12} & -\frac{F_1}{F_2}\mathcal{M}_{12} \end{pmatrix}. \tag{2.68}$$

All the model-dependency is now absorbed in the off-diagonal entry $\mathcal{M}_{12}$. This matrix has two eigenvalues: zero for the true goldstino mass and $m_{\tilde{G}'}$ for the pseudo-goldstino mass. This eigenvalue can be solved for explicitly, so that the pseudo-goldstino mass is

$$m_{\tilde{G}'} = \left(\frac{F_2}{F_1} + \frac{F_1}{F_2}\right)\mathcal{M}_{12}. \tag{2.69}$$

In order to estimate the size of this mass, one needs to evaluate a single matrix element: the two-point function $\langle \tilde{G}_a \tilde{G}_b \rangle$. In general, leaving the hidden sector dynamics unspecified, at leading order in $g$ this can be calculated from the diagrams in Figure 2.12. Indeed, setting the gauge coupling to zero decouples the hidden sectors from the supersymmetric Standard Model (the defining feature of gauge mediation). While connection through a single superfield (instead of two in Figure 2.12) is possible in principle, it turns out that this typically leads to unappealing models for gauge mediation (due to messenger parity, see [325]). The blobs in this figure represent unspecified hidden sector interactions. In practice, at lowest order this will include messenger loops (since these are what couples the hidden sector to the Standard Model in the first place).

Assume now that each hidden sector has a supersymmetric mass scale *M* and SUSY-breaking scales $F_1$ and $F_2$. Then, the pseudo-goldstino mass at leading order must be

$$m_{\tilde{G}'} \sim \frac{g^4}{(16\pi^2)^3} \left(\frac{F_1}{F_2} + \frac{F_2}{F_1}\right) \left(\frac{F_1}{M} + \frac{F_2}{M}\right), \tag{2.70}$$

which includes $g^4$ from the four gauge vertices in Figure 2.12 and a loop factor which accounts for the three loops (two messenger and one with gauge fields). The final factor follows from dimensional analysis: the pseudo-goldstino mass needs to disappear for $F_a \to 0$, where the only other mass scale in the theory fixes the dimensions. In case the SUSY-breaking scales are comparable, one finds

$$m_{\tilde{G}'} \sim \frac{g^4}{(16\pi^2)^3} \frac{F}{M} \sim \frac{g^2}{(16\pi^2)^2} m_{\text{soft}} \sim 1 \text{ GeV}, \qquad (2.71)$$

using the definition of the soft masses in Section 1.4.4 (this estimate includes a factor $\mathcal{O}(10)$ from the number of gauge fields).

However, if the SUSY-breaking scales are very different, e.g. $F_1 \gg F_2$, then from Eq. (2.69), the pseudo-goldstino mass is enhanced by the factor $F_1/F_2$

$$m_{\tilde{G}'} \sim \frac{F_1}{F_2} \text{ GeV}. \qquad (2.72)$$

On the other hand, this enhancement can not be arbitrarily large, since otherwise the backreaction of the visible sector on the second hidden sector becomes too large [325]. In [325], it is found that for $F_1/F_2 \ll 10^3$, these effects are under control. Therefore, the pseudo-goldstino mass can easily be at the electroweak scale.

In the case of large pseudo-goldstino masses, the mixing with the neutralinos can not be ignored. However, once the off-diagonal matrix element $\mathcal{M}_{12}$ is specified, the pseudo-goldstino mass can still be obtained by performing the full neutralino-goldstino rotation properly.

Finally, note that this discussion was performed in the context of gauge mediation, setting to zero all gravity mediation effects. This approximation is correct under the same conditions as those required to neglect the gravity mediation effects for ordinary gauge mediation models.

## 2.3 ATLAS excess in a goldstini simplified model

In this section, we describe a simplified model, based on the goldstini model, in which we will interpret the 8 TeV ATLAS excess. It includes as low-energy degrees of freedom only a gluino, higgsino-like neutralinos, a pseudo-goldstino and a goldstino. In the following, we discuss in more detail our simplified model and discuss the decay modes of the gluino and neutralinos. Afterwards, we show that this simplified model can be correctly incorporated within a full goldstini model, i.e. we can find a set of soft masses which reproduces the required masses and couplings.

### 2.3.1 Simplified model

The mass spectrum of the model is depicted in Figure 2.13a. Note that the higgsino fields include two almost degenerate neutral mass eigenstates and a charged one. Also note that the true goldstino is in the bottom of the spectrum, but is not shown in Figure 2.13a since it is irrelevant in our scenario. Since the gluinos interact strongly, our model can appear in LHC searches through gluino pair production.

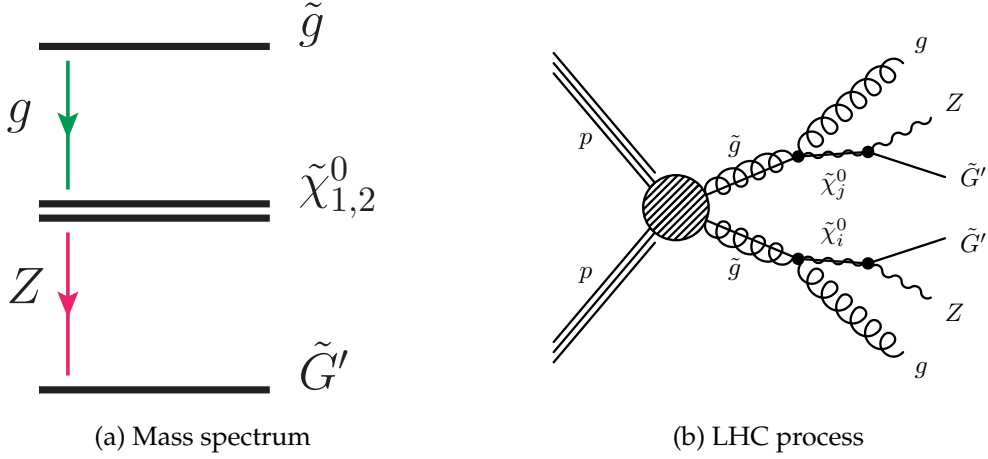

(a) Mass spectrum            (b) LHC process

Figure 2.13: The simplified model considered here to explain the 8 TeV ATLAS excess. (a) Mass spectrum for our simplified goldstini model, with the relevant decay modes. (b) The process at the LHC in our simplified goldstini model.

Given the mass spectrum we consider and the decay modes of the gluino and neutralino (discussed in Sections 2.3.2 and 2.3.3), in a certain parameter space, the gluino two-body decay chain

$$\tilde{g} \to g + \tilde{\chi}^0_{1,2} \to g + Z + \tilde{G}' \tag{2.73}$$

becomes dominant (which we verify below), and gluino pair production can contribute to the ATLAS signal. The full process relevant for LHC searches is illustrated in Figure 2.13b, where we assume a 100% branching ratio in each decay step as a good approximation. Depending on the decay of the $Z$ boson, the final state can be either SFOS lepton pair + jets + $\not{E}_T$ or jets + $\not{E}_T$.

In the following we study each decay step in detail to verify the gluino decay chain in Eq. (2.73).

### 2.3.2   Gluino decay

As already discussed in Section 2.1.3, in the spectrum presented in Figure 2.13a, the gluino can potentially have several decay channels, depending on the mass splitting between the gluino and the neutralinos. If the mass splitting is sufficiently above the top mass, i.e.

$$(m_{\tilde{g}} - m_{\tilde{\chi}^0_i}) \gg m_t, \tag{2.74}$$

the tree-level three-body decays into a pair of third-generation quarks and a chargino or neutralino are dominant [287].

On the other hand, in the regime

$$\Delta m_{\tilde{g}-\tilde{\chi}^0_{1,2}} \equiv m_{\tilde{g}} - m_{\tilde{\chi}^0_{1,2}} \leq 200 \text{ GeV},$$ (2.75)

the gluino decays predominantly into a gluon and a neutralino via a (s)top loop. The analytic expression for the $\tilde{g} \to g\tilde{\chi}^0_i$ decay can be found in [338, 339]. We find that this result is robust in the full MSSM as soon as the Bino ($M_B$) and Wino ($M_W$) masses are moderately larger than the higgsino mass ($\mu$). We checked this decay pattern with SUSY-HIT [257], for instance, by fixing $m_{\tilde{g}} = 1000$ GeV, $\mu = 800$ GeV, and $\tan\beta = 10$, we find $B(\tilde{g} \to g\tilde{\chi}^0_{1,2}) > 85\%$ as soon as $M_B$ and $M_W$ are larger than about 1.4 TeV, with squark masses of $\mathcal{O}(5)$ TeV.

The gluino decays into a pseudo-goldstino or a goldstino with a gluon are also possible. However, unless the gluino coupling to the pseudo-goldstino is largely enhanced, these decay modes are always suppressed compared with the decays into the MSSM states, since they are proportional to $\propto 1/F$.

In the following, to fit the ATLAS excess, we consider the small mass splitting in Eq. (2.75), where only the $\tilde{g} \to g\tilde{\chi}^0_{1,2}$ decay is relevant.

### 2.3.3 Neutralino decay

Once the effective couplings in Eq. (2.8) are evaluated, the neutralino decay widths into $\gamma/Z/h + \tilde{G}'$ can be computed as [316, 317]

$$\Gamma_{\tilde{\chi}^0_i \to \gamma\tilde{G}'} = \frac{(\tilde{y}^i_\gamma)^2 m^3_{\tilde{\chi}^0_i}}{16\pi F^2} \left(1 - \frac{m^2_{\tilde{G}'}}{m^2_{\tilde{\chi}^0_i}}\right)^3,$$ (2.76)

$$\begin{aligned}
\Gamma_{\tilde{\chi}^0_i \to Z\tilde{G}'} = \frac{\beta_Z m_{\tilde{\chi}^0_i}}{32\pi F^2} &\left[\left(1 - \frac{m_{\tilde{G}'}}{m_{\tilde{\chi}^0_i}}\right)^2 - \frac{m^2_Z}{m^2_{\tilde{\chi}^0_i}}\right] \\
&\times \left[(\tilde{y}^i_{Z_T})^2(2(m_{\tilde{\chi}^0_i} + m_{\tilde{G}'})^2 + m^2_Z) \right. \\
&\quad + (\tilde{y}^i_{Z_L})^2((m_{\tilde{\chi}^0_i} + m_{\tilde{G}'})^2 + 2m^2_Z) \\
&\quad \left. + 6\tilde{y}^i_{Z_T}\tilde{y}^i_{Z_L} m_Z(m_{\tilde{\chi}^0_i} + m_{\tilde{G}'})\right],
\end{aligned}$$ (2.77)

$$\Gamma_{\tilde{\chi}^0_i \to h\tilde{G}'} = \frac{\beta_h (\tilde{y}^i_h)^2 m_{\tilde{\chi}^0_i}}{32\pi F^2} \left(1 + 3\frac{m^2_{\tilde{G}'}}{m^2_{\tilde{\chi}^0_i}} - \frac{m^2_h}{m^2_{\tilde{\chi}^0_i}}\right),$$ (2.78)

where $\beta_{Z/h} = \overline{\beta}(m^2_{\tilde{G}'}/m^2_{\tilde{\chi}^0_i}, m^2_{Z/h}/m^2_{\tilde{\chi}^0_i})$ with $\overline{\beta}(a,b) = (1 + a^2 + b^2 - 2a - 2b - 2ab)^{1/2}$. The standard decays into the true (massless) goldstino are obtained from this expression by sending the tilded quantities to the non-tilded quantities and putting $m_{\tilde{G}'} \to 0$.

Although the neutralino decay can present a rich pattern, with six competing decay modes as $(Z, h, \gamma)$ plus $(\tilde{G}', \tilde{G})$, we are interested in a scenario where the neutralino predominantly decays into a pseudo-goldstino and a $Z$ boson. This scenario can be realised by enhancing the coupling parameters $\tilde{y}$, especially $\tilde{y}_{Z_T}$ and $\tilde{y}_{Z_L}$, and by assuming the higgsino-like neutralino (this latter point was already discussed in Section 2.1.3). In Section 2.3.5, we will provide the details of our illustrative benchmark point in the SUSY-breaking parameters determining the couplings $\bar{y}$ and $y$; we typically take $\mu \sim 800$ GeV and $M_B = M_W \sim 1.5$ TeV.

For the parameters we consider here, the mass splitting between the two neutralinos is of the order of a few GeV. In this scenario, the second lightest neutralino $\tilde{\chi}_2^0$ decays to the lightest neutralino $\tilde{\chi}_1^0$ with soft SM particle emissions. Indeed we checked with SUSY-HIT [257] and the formulas for the decay width above that its decay modes to the goldstino and to the pseudo-goldstino for our benchmark point are negligible compared with the $\tilde{\chi}_2^0 \rightarrow \tilde{\chi}_1^0$ decays. Hence in the following we assume $B(\tilde{\chi}_2^0 \rightarrow \tilde{\chi}_1^0 + \text{undetectable SM particles}) = 100\%$. The final state topology is then determined by the possible $\tilde{\chi}_1^0$ decays, which we now investigate.

In Figure 2.14, we show the branching ratios of the lightest neutralino as a function of the pseudo-goldstino mass, evaluated on our illustrative benchmark point described in detail in the Section 2.3.5 using the decay formulas above. The decay pattern depends on the mass splitting $\Delta m_{\tilde{\chi}_1^0 - \tilde{G}'} \equiv m_{\tilde{\chi}_1^0} - m_{\tilde{G}'}$ and on the possible kinematically allowed modes. For $\Delta m_{\tilde{\chi}_1^0 - \tilde{G}'} > m_Z$, the branching ratio into a Z-boson and a pseudo-goldstino depends on whether the mass splitting is larger or smaller than the Higgs mass. For $\Delta m_{\tilde{\chi}_1^0 - \tilde{G}'} > m_h$, $\tilde{\chi}_1^0 \rightarrow Z\tilde{G}'$ is always greater than 80%, while it saturates at 100% for $m_Z < \Delta m_{\tilde{\chi}_1^0 - \tilde{G}'} < m_h$. We note that the $\tilde{\chi}_1^0 \rightarrow \gamma\tilde{G}'$ decay is negligible in our parameter choice, due to the higgsino nature of the neutralino. In the regime of $\Delta m_{\tilde{\chi}_1^0 - \tilde{G}'} < m_Z$, the decays into a true goldstino, instead of a pseudo-goldstino, plus a $\gamma$, $Z$ or $h$ becomes dominant due to the available phase space. In the following, we consider the region

$$\Delta m_{\tilde{\chi}_1^0 - \tilde{G}'} > m_Z, \tag{2.79}$$

where the $\tilde{\chi}_1^0 \rightarrow Z\tilde{G}'$ decay is dominant, as assumed in the simplified model in Figure 2.13a and shown in Figure 2.14.

We have also verified that the total decay width in the region of interest is always larger than $2 \times 10^{-12}$ GeV, implying that the decays happen promptly in the detector.

### 2.3.4 Pseudo-goldstino decay

Finally, the pseudo-goldstino will eventually decay into a goldstino plus $\gamma$, $Z$ or $h$ [35]. However, one can verify that the pseudo-goldstino is sufficiently long-lived for this decay to happen outside the detector. The decay of the pseudo-goldstino can be computed using the same decay formulas quoted above, where the coupling between the pseudo-goldstino and the goldstino are extracted numerically[36] from the original Lagrangian, once we switch to the mass eigenbasis. With these formulas, we find that for the benchmark point considered here, the decay width is around $10^{-22} - 10^{-24}$ GeV, i.e. $\tau_{\tilde{G}'} \lesssim 1$ sec, depending on the pseudo-goldstino mass. Even though we are not addressing cosmological issues in this work, we observe that this decay is fast enough to not spoil big bang nucleosynthesis (BBN) [340].

---

[35] Additional decay modes are those into difermions or diphotons, but these modes are strongly suppressed, see e.g. [315, 316, 319].

[36] This needs to be done numerically, since the pseudo-goldstino mass originates in radiative corrections which are not included in our tree-level treatment of the (pseudo–)goldstino couplings.

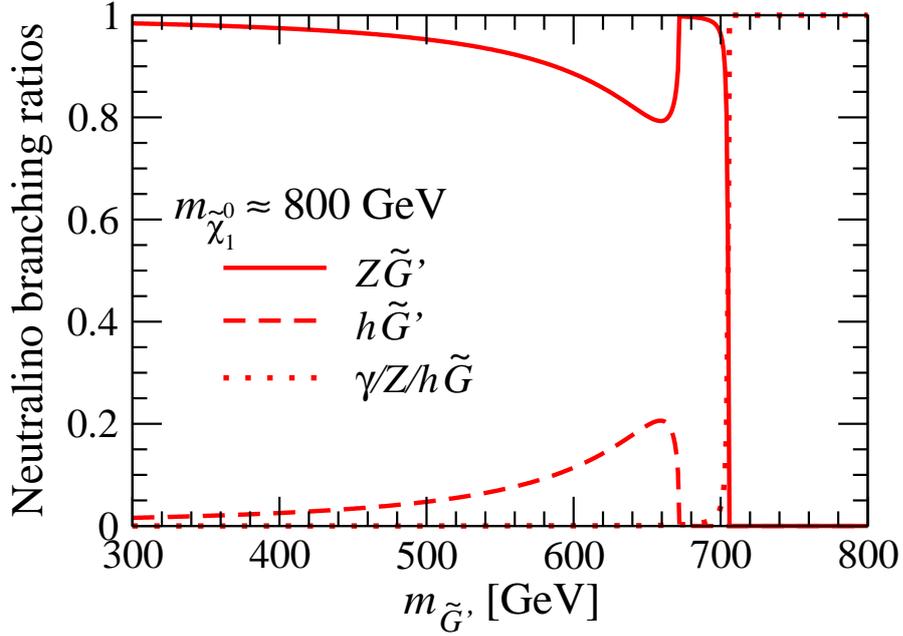

Figure 2.14: Branching ratios of the lightest neutralino as a function of the pseudo-goldstino mass for a representative benchmark point described in Section 2.3.5. The $\tilde{\chi}_1^0 \to \gamma \tilde{G}'$ decay is too small to see in the plot.

### 2.3.5 Interpretation in the full goldstini model

Here, we frame the simplified model considered in this section within the complete model described in Section 2.2, i.e. we search for a set of fundamental parameters which can give rise to the desired phenomenology. We are interested in configurations where the two lightest neutralinos are mostly higgsinos and the decay of $\tilde{\chi}_1^0$ is predominantly into a massive pseudo-goldstino and a $Z$ boson. To find whether this scenario is compatible with the Lagrangian constructed in Section 2.2, we numerically explored the parameters of the model (i.e. the soft terms) looking for a representative benchmark point satisfying these requirements. In Figure 2.14 we report the branching ratios for a configuration with $\mu = 804$ GeV, $\sqrt{b} = 800$ GeV, $M_B = M_W = 1.5$ TeV and $\tan\beta = 10$. The total soft masses for the Higgses are extracted by solving the EWSB conditions. The two SUSY-breaking scales are chosen as $\sqrt{F_1} = 1.5 \times 10^6$ GeV and $\sqrt{F_2} = 5 \times 10^4$ GeV. The gaugino masses and the Higgs soft masses are distributed in the two sectors as $M_{B/W(1)}/M_{B/W(2)} = \tan^2\theta$, $m^2_{H_{d/u}(1)}/m^2_{H_{d/u}(2)} = \cot^2\theta$. The angle $\theta$ is taken to be $\tan\theta = 0.2$. Finally, the $b$-terms are chosen as $b_{(a)} = F_a/(F_1 + F_2)b$. With these values, we obtain a scenario where the two lightest neutralinos are mostly higgsinos, with masses at $(797, 805)$ GeV. The lightest neutralino decay is predominantly into a pseudo-goldstino plus a $Z$ boson for different ranges of pseudo-goldstino mass, as we show in Figure 2.14. The pseudo-goldstino is varied by changing the off-diagonal goldstini mass term[37] $\mathcal{M}_{12}$ in $M^{\hat{G}}$, defined in Eq. (2.43), consistently with the perturbative definition in [325].

## 2.4 Analysis

In this section, we investigate whether or not the event topology of our model can successfully fit the ATLAS on-$Z$ excess without conflicting with other searches. The most stringent constraints come from the ATLAS jets+$\not{E}_T$ search [283] and the CMS dilepton search [280] in our parameter region of interest [286, 287, 310]. The latter has signal regions which look at the same final state as in the ATLAS on-$Z$ signal region and are potentially quite constraining. We have implemented these analyses as well as the ATLAS on-$Z$ signal region in the analysis package[38] `Atom` [341]. Some description and validation results of `Atom` are given in Appendix E. Note that in this section we only consider data at $\sqrt{s} = 8$ TeV, which was available for the original analysis. An update for these results at $\sqrt{s} = 13$ TeV is given in Section 2.5

The main result is that jets from the gluino decay and the hadronic $Z$ decay are softened when the mass spectrum is compressed due to the massive nature of the pseudo-goldstino. We find that there are viable points in the parameter space even after the ATLAS jets+$\not{E}_T$ constraint [283], as well as the CMS constraint on the identical final state [280], are taken into account. We also show that the two-body gluino decay in Eq. (2.73) provides a better fit to the data for the distributions with respect to the three-body gluino decay $\tilde{g} \to q\bar{q}\tilde{\chi}^0_{1,2}$.

In order to fit the excess we scan over the gluino mass, while fixing the neutralino masses at

$$m_{\tilde{\chi}^0_{1,2}} = m_{\tilde{g}} - 200 \text{ GeV}. \tag{2.80}$$

We consider three cases, featuring the pseudo-goldstino masses:

$$m_{\tilde{G}'} = \begin{cases} 0 & \text{(A)} \\ m_{\tilde{\chi}^0_i} - 200 \text{ GeV} & \text{(B)} \\ m_{\tilde{\chi}^0_i} - 100 \text{ GeV} & \text{(C)} \end{cases} \tag{2.81}$$

Case A is equivalent to the gauge mediation scenario with only one SUSY-breaking sector and a very light gravitino, while cases B and C have compressed spectra.

---

[37]In practice, one can scan over the pseudo-goldstino mass and determine the only free neutralino-goldstino mass matrix element $\mathcal{M}_{12}$ from that. This then fully determines the mass matrix, which also determines the exact content of each of the neutralinos and their couplings.

[38]This package is private and it is not clear whether at the time of writing this thesis it is sill maintained. However, other analysis packages and strategies exist which can perform the same task, see Section 1.5.3.

In order to assess the consistency between the model and data, the ideal approach would be to carry out a global fit, treating the excess and constraints in the same manner. This requires the details of the systematic uncertainties and good understanding of the correlation among different signal regions. Rather than taking this rigorous approach, in this exploratory work we instead fit the model to the excess independently from the constraints and check the exclusion individually for the signal regions using the following prescription.

To test the goodness of fit, we define the measure $R$ as

$$R \equiv N_{\text{SUSY}} / (N_{\text{obs}} - N_{\text{SM}}), \tag{2.82}$$

for the ATLAS on-$Z$ signal region, where $N_{\text{SUSY}}$ is the expected number of SUSY events, $N_{\text{obs}}$ is the number of observed events and $N_{\text{SM}}$ is the expected number of SM events in the signal region. With this definition, the best fit is given by $R = 1$.

For the other signal regions, labelled $i$, used as constraint, we instead define

$$R^i \equiv N^i_{\text{SUSY}} / N^{\text{UL},i}_{\text{BSM}}, \tag{2.83}$$

where $N^i_{\text{SUSY}}$ is the expected number of SUSY events and $N^{\text{UL},i}_{\text{BSM}}$ is the 95% CL$_s$ limit[39] obtained by the experiments for signal region $i$. Having any $R_i$ greater than one indicates that the model is disfavoured.

$N^{(i)}_{\text{SUSY}}$ can be expressed as $\sigma_{\tilde{g}\tilde{g}} \cdot \mathcal{L} \cdot \epsilon_{(i)}$, where $\mathcal{L}$ is the integrated luminosity used in the experimental analysis and $\sigma_{\tilde{g}\tilde{g}}$ is the production cross section of the gluino pair, for which we use the values reported in [225, 288]. To estimate the efficiency $\epsilon_{(i)}$, we use the following simulation chain: first the signal events are generated using `MadGraph5_-aMC@NLO` [258] and showered and hadronised by `Pythia6` [343]. The hadron-level events are then passed to `Atom` [341] to estimate the efficiency for each signal region taking the event reconstruction and detector effects into account. For our signal simulation, we extended the goldstini model [317, 344] (building on [345]) to include the effective two-body gluino decay (which is otherwise not present in the tree-level calculation of `MadGraph5_aMC@NLO`) using `FeynRules2` [250].

The results for the three cases of the goldstini scenario in Eq. (2.81) are presented in Figs. 2.15 and 2.16, where the different scenarios are confronted with the ATLAS jets+$\not{E}_T$ [283] and the CMS dilepton [280] searches, respectively. The fitting measure $R$ for the ATLAS on-$Z$ signal region [278] is shown with the solid black curve, whereas the constraints $R^i$ are shown with the other curves, corresponding to the signal regions (2jl, 2jm, 2jt, 3j, 4jl-, 4jl, 4jm, 4jt, 5j, 6j, 6jm, 6jt, 6j+) in the ATLAS jets+$\not{E}_T$ search in Figure 2.15 and (cms2jl, cms2jm, cms2jh, cms3jl, cms3jm, cms3jh, cms(C), cms(F)) in the CMS dilepton search in Figure 2.16.[40] The green (yellow) band around $R = 1$ represents the 1 (2) $\sigma$ region of the fit for the ATLAS on-$Z$ excess.

---

[39]The CL$_s$ technique is a limit-setting technique, more appropriate for exclusion intervals, while Feldman and Cousins's [342] method is more appropriate for treating established signals [274].

[40] The $nj$ in the signal region name indicates it requires more than $n$ high $p_T$ jets. The letter "l", "m", "t" for the ATLAS jets+$\not{E}_T$ search means "loose", "medium", "tight", while "l", "m", "h" for the CMS dilepton search denotes "low", "medium", "high". The cms(C) and cms(F) represent the central and forward signal regions in which the number of jets and $\not{E}_T$ are treated inclusively. See [283] and [280] for the exact definition.

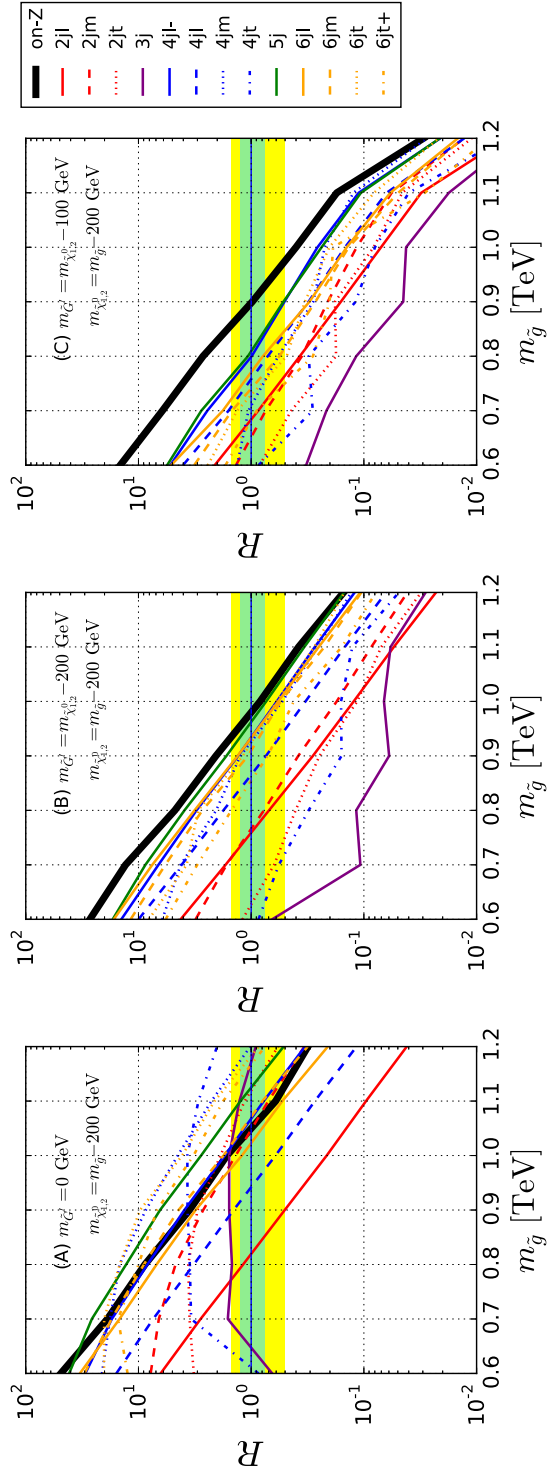

Figure 2.15: *R* values for the fit (black solid) and constraints (others) for cases A (left), B (middle) and C (right) in Eq. (2.81). The green and yellow bands correspond to the 1 and 2 *σ* regions of the fit. If the black curve is within the bands, the model provides a good fit for the ATLAS on-*Z* excess [278]. On the other hand, if there is any other curve above one, the model point is strongly disfavoured by the signal region corresponding to the curve in the ATLAS jets + *E̸_T* analyses [283].

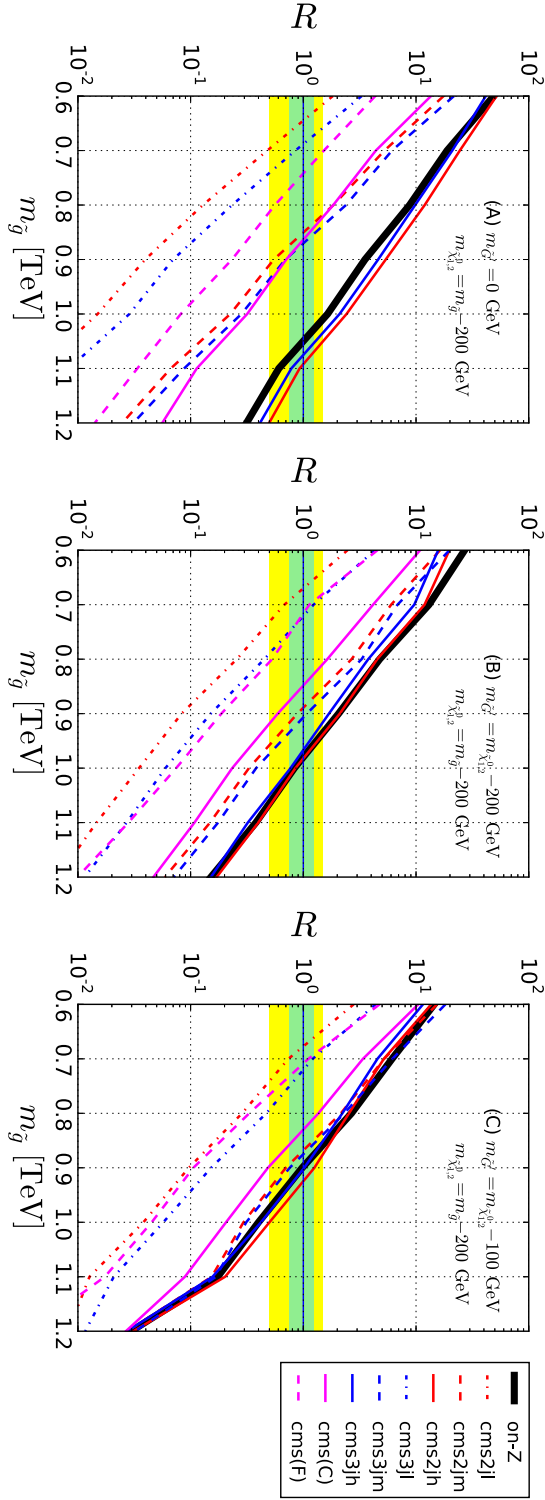

Figure 2.16: The same as in Figure 2.15 but for the constraints from the CMS on-Z analyses [280].

One can see that the entire region of case A, i.e. the massless goldstino case, is excluded by the ATLAS jets+$\not{E}_T$ search. The CMS dilepton search also excludes the region with $m_{\tilde{g}} < 1.1$ TeV. These strong limits can be attributed to the large mass gap between the gluino and the pseudo-goldstino, because of which the jets and leptons from the gluino decays tend to be hard, increasing the efficiency of the ATLAS jets+$\not{E}_T$ search as well as the CMS dilepton search.

Moving to cases B and C, one can see that the constraints are generally more relaxed than in case A, since these cases have milder mass hierarchy with the massive pseudo-goldstino. The ATLAS jets+$\not{E}_T$ search now excludes the gluino mass up to 970 (810) GeV for case B (C). Here, the best fit of the ATLAS on-$Z$ excess is given at $m_{\tilde{g}} = 980$ (900) GeV for case B (C), which is just outside of the ATLAS jets+$\not{E}_T$ exclusion limit.

The CMS dilepton search, on the other hand, still provides tight constraints, especially the cms2jh signal region, where $\not{E}_T > 300$ GeV is required, very similar to the $\not{E}_T > 225$ GeV cut in the ATLAS on-$Z$ analysis. The difference between the two analyses is that the ATLAS search additionally imposes $H_T > 600$ GeV, where $H_T$ is the scalar sum of the $p_T$ of jets and leptons. For case B, the best fit point ($m_{\tilde{g}} = 980$ GeV) is just on the exclusion boundary, while for case C the best fit point ($m_{\tilde{g}} = 900$ GeV) is just excluded by the cms2jh signal region. For case B, at $m_{\tilde{g}} = 1$ TeV, the fit for the ATLAS on-$Z$ excess is still within the $1\,\sigma$ band, and the point is still outside of the 95% CL$_s$ exclusion limits from the various constraints. We therefore choose our best fit benchmark point $P1$ as

$$P1 : (m_{\tilde{g}}, m_{\tilde{\chi}^0_{1,2}}, m_{\tilde{G}'}) = (1000, 800, 600)\ \text{GeV} \qquad (2.84)$$

for the following analysis. Even for case C, the tension between the data and the prediction observed in the ATLAS on-$Z$ signal region can be ameliorated to the $2\,\sigma$ level with benchmark point $P2$, given by

$$P2 : (m_{\tilde{g}}, m_{\tilde{\chi}^0_{1,2}}, m_{\tilde{G}'}) = (950, 750, 650)\ \text{GeV} \qquad (2.85)$$

which is outside the 95% CL$_s$ exclusion region.

In Figure 2.17 we compare the data with the signal + background for the $\not{E}_T$ (left) and jet multiplicity (right) distributions in the ATLAS on-$Z$ signal region at our best fit point $P1$ in Eq. (2.84). Here, we took the data and the SM background from Figs. 6 and 7 in the ATLAS paper [278] and combined the $ee$ and $\mu\mu$ channels. The data in the $\not{E}_T$ distribution has a preference for low $\not{E}_T$, peaking around 240 GeV and roughly falling down up to around 500 GeV. The distribution is well fitted with the signal + background at our best fit point, because the massive nature of the pseudo-goldstino reduces the $\not{E}_T$ in the event. In the jet multiplicity distribution, the data prefers $2-5$ jets and disfavors the region with $\geq 6$ jets. The distribution of the signal + background at our best fit point peaks around $3-4$ and gives a good fit to the data. This is an advantage of the radiative decay $\tilde{g} \to g\tilde{\chi}^0_{1,2}$ compared to the three-body $\tilde{g} \to q\bar{q}\tilde{\chi}^0_{1,2}$ decay considered in the experimental analysis, because the number of jets is reduced typically by two.

As a reference, in Figure 2.18 we also compare the signal with the CMS data in the on-$Z$ signal region with $N_{\text{jets}} \geq 2$ and $\geq 3$ at our best fit point $P1$ in Eq. (2.84). Here we took the data and the SM background from Figure 7 in the CMS paper [280]. As already seen in Figure 2.16, the most stringent constraint comes from the high $\not{E}_T$ region, i.e. cms2jh and cms3jh, where the $\not{E}_T > 300$ GeV bin is considered.

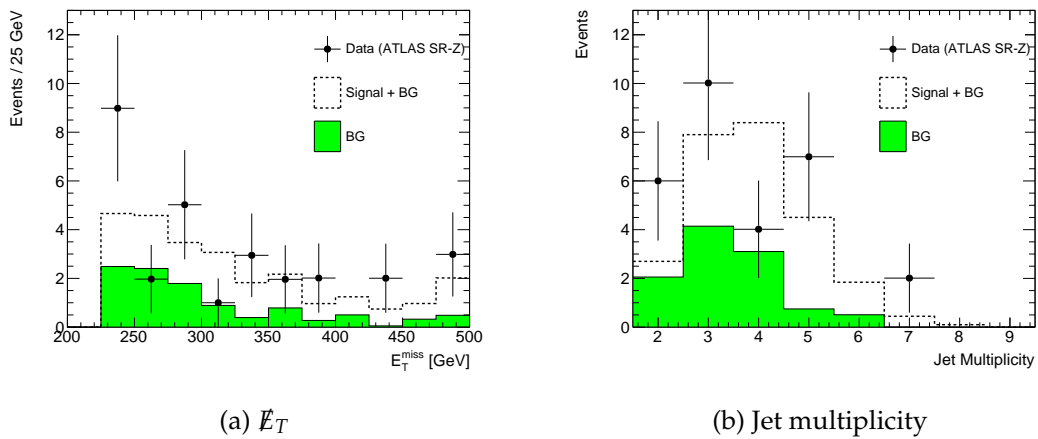

(a) $\not{E}_T$              (b) Jet multiplicity

Figure 2.17: Comparison between data and signal + background in the $\not{E}_T$ and jet multiplicity distributions in the ATLAS on-$Z$ signal region [278] at our best fit point: $(m_{\tilde{g}}, m_{\tilde{\chi}^0_{1,2}}, m_{\tilde{G}'}) = (1000, 800, 600)$ GeV.

## 2.5 Update after Run 2

Since the publication of this work, new analyses have been performed using data gathered in Run 2 of the LHC. The ATLAS analysis for events containing a same-flavour opposite-sign (SFOS) dilepton pair, jets and large missing transverse momentum has been updated twice for center of mass energy $\sqrt{s} = 13$ TeV, with 14.7 fb$^{-1}$ of data [346] and then again with 36.1 fb$^{-1}$ of data [347], both using data gathered in 2015–2016. Similarly, CMS has updated their analysis for events containing two opposite-charge, same-flavour leptons, jets and missing transverse momentum twice at $\sqrt{s} = 13$ TeV, once with 2.3 fb$^{-1}$ of data gathered during 2015 [348] and with 35.9 fb$^{-1}$ of data gathered during 2016 [349]. Finally, also the ATLAS search for squarks and gluinos in final states with jets and missing transverse momentum has been updated several times at $\sqrt{s} = 13$ TeV, the latest analysis using 36.1 fb$^{-1}$ of data [350]. All of these analyses found results consistent with the Standard Model expectation.

Therefore, the need to explain an excess has disappeared. Moreover, if the original excess was due to new physics, and not due to a statistical fluctuation, one would expect the excess to grow with more data, not disappear. As a result, it is unlikely that the scenario described in this chapter is realised in nature. However, we want to know more concretely whether it has also been ruled out by current searches.

First, we consider the updated ATLAS search for events with at least two SFOS

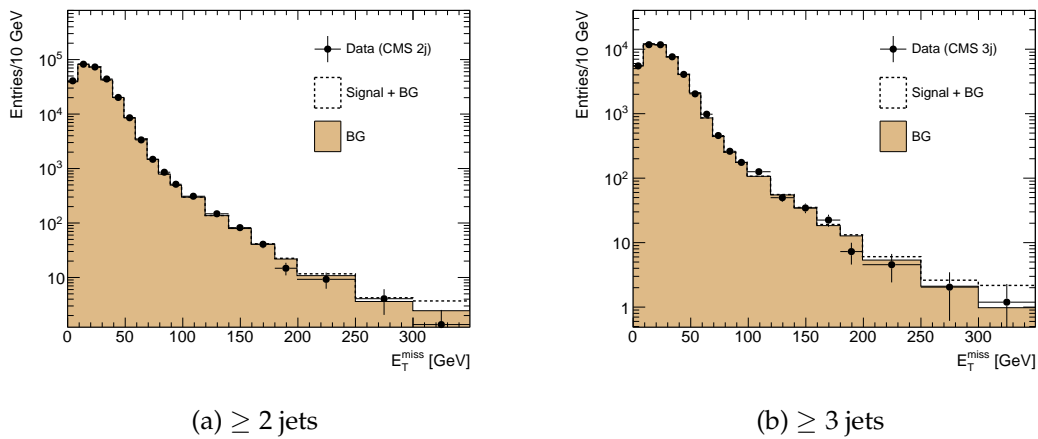

<div align="center">(a) ≥ 2 jets             (b) ≥ 3 jets</div>

Figure 2.18: Comparison between data and signal + background in the $\not{E}_T$ distribution in the CMS on-$Z$ signal region with $\geq 2$ jets and $\geq 3$ jets [280] at our best fit point: $(m_{\tilde{g}}, m_{\tilde{\chi}^0_{1,2}}, m_{\tilde{G}'}) = (1000, 800, 600)$ GeV.

leptons, jets and missing energy [347]. It is targeted at scenarios similar to the 8 TeV analysis discussed in Section 2.1.1, shown in Figure 2.19. However, the two lightest neutral states are now denoted more generally by $\tilde{\chi}^0_1$ and $\tilde{\chi}^0_2$, dropping the assumption that $\tilde{\chi}^0_1$ is the goldstino with a mass necessarily much lower than the electroweak scale. Therefore, in addition to a scenario where the lightest neutralino mass is fixed at $m_{\tilde{\chi}^0_1} = 1$ GeV while scanning over $m_{\tilde{\chi}^0_2}$, the analysis also considers a scenario where instead the mass difference between the two lightest neutralinos is fixed to $m_{\tilde{\chi}^0_2} - m_{\tilde{\chi}^0_1} = 100$ GeV while scanning over $m_{\tilde{\chi}^0_1}$.

Compared to the previous ATLAS search at $\sqrt{s} = 13$ TeV [346], this analysis extends the gluino/squark mass reach by several hundred GeV and into the compressed region due to optimisations for 13 TeV[41] and nearly doubles the reach compared to the 8 TeV search. As before, the analysis requires the presence of at least two signal leptons, the highest $p_T$ ones of which must be a SFOS lepton pair and two jets. The search targets region is separated into SR-low, SR-medium and SR-high, with different values of the hadronic activity $H_T = \sum p_T^{\text{jet}}$: $H_T > 200$ GeV, $H_T > 400$ GeV and $H_T > 1200$ GeV with corresponding $\not{E}_T$-requirements $\not{E}_T > 250$ GeV, $\not{E}_T > 400$ GeV and $\not{E}_T > 200$ GeV, targeting different values of the mass splitting $m_{\tilde{g}} - m_{\tilde{\chi}^0_1}$. The analysis considers the full $m_{ll}$-spectrum above 12 GeV by performing a profile likelihood fit a well as dividing the full distribution into a number of overlapping windows over the $m_{ll}$-distribution. For each of the three signal regions, the on-$Z$ bins with boundaries 81 GeV $< m_{ll} < 101$ GeV are the ones of interest for our specific scenario. The analysis finds that all windows are consistent with the Standard Model expectation, as shown in Figure 2.20, with the largest excess in the mass range 101 GeV $< m_{ll} <$ 201 GeV where 18 events are seen compared to the expected $7.5 \pm 3.2$, corresponding to a local significance of $2\sigma$.

---

[41]In addition, a low $p_T$ search was performed, but it is only relevant for scenarios where a kinematic endpoint is expected in the lower range of the $m_{ll}$-distribution, since for leptons from $Z$-boson decay the peak is at a fixed position and the $p_T$ of the leptons is generally high.

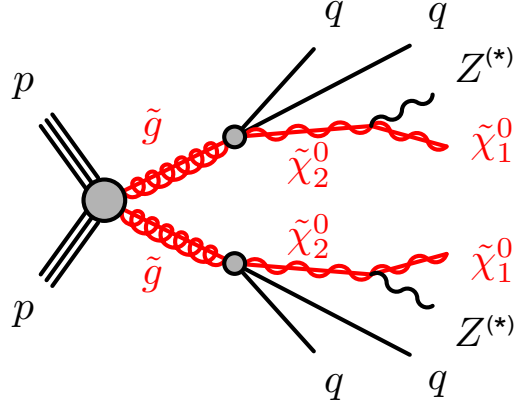

Figure 2.19: Feynman diagram for the process considered in the ATLAS search for 2 SFOS leptons, jets and $\not{E}_T$ at $\sqrt{s} = 13$ TeV relevant to our model. Note that the neutral particle in the final state is now denoted by $\tilde{\chi}_1^0$, to indicate that its mass is no longer restricted to 1 GeV as in the 8 TeV analysis. Figure from [347].

The 95% CL exclusion limits are shown in Figure 2.21. The kink in the limit is due to the signal regions with different $H_T$ requirements. Figure 2.21a shows the gauge-mediation scenario with the LSP at $m_{\tilde{\chi}_1^0} = 1$ GeV and gluino pair production leading to an on-shell $Z$-boson, similar to the original search that found an excess. In this case, gluino masses can be excluded up to 1.65 TeV (with expected limit 1.60 TeV) for $m_{\tilde{\chi}_2^0} = 1.2$ TeV. Clearly, the exclusion limit has moved beyond the region that can accomodate the excess. For illustration, the best-fit points $P1$ and $P2$ found in our previous analysis are indicated at the relevant $(m_{\tilde{g}}, m_{\tilde{\chi}_2^0})$-point. Note however, that the lightest neutralino mass is much lower than in our best-fit points, such that a direct comparison is impossible. Since our model has a very heavy pseudo-goldstino in the final state, it is not possible to draw a definitive conclusion from this exclusion contour. Instead, the second scenario considered by ATLAS, with $m_{\tilde{\chi}_2^0} - m_{\tilde{\chi}_1^0} = 100$ GeV scanning over $m_{\tilde{\chi}_1^0}$, shown in Figure 2.21b, is more representative. As can be seen from the exclusion contour, even here the region of interest is excluded by the ATLAS search. Moreover, this particular mass splitting between the neutralinos is exactly applicable to our second best-fit point $P2$ (indicated by a black star) in Eq. (2.85), with $(m_{\tilde{g}}, m_{\tilde{\chi}_{1,2}^0}, m_{\tilde{G}'}) = (950, 750, 650)$ GeV and is well within the excluded region. The best-fit point $P1$ considers a slightly different splitting between the neutral state masses (indicated by the white star), such that while the exclusion limit suggests it is ruled, this is not yet definitive from this analysis alone.

The results of the CMS update with 35.9 fb$^{-1}$ gathered in 2016 are similar [349].

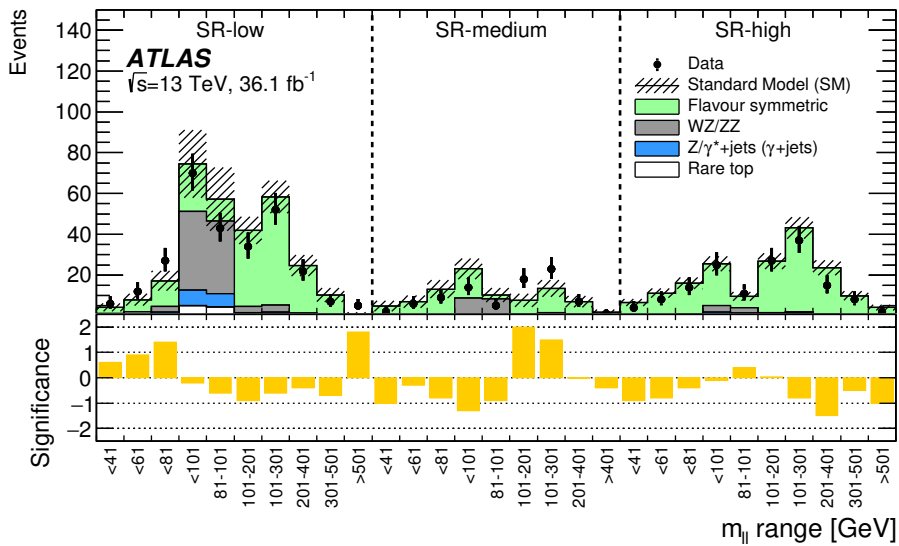

Figure 2.20: Observed number of events for different windows of the $m_{ll}$-distribution in the ATLAS search for 2 SFOS leptons, jets and $\not{E}_T$ at $\sqrt{s} = 13$ TeV for the channel relevant to our scenario. Figure from [347].

This search now targets both strongly and electroweakly produced new physics, where neutralinos appear either from gluino decay or are directly produced[42]. Again, the data are consistent with the Standard odel expectations, excluding gluino masses in the gauge-mediation scenario with a 1 GeV goldstino up to 1500–1770 GeV at 95% CL, but providing no interpretation for scenarios with higher mass final state neutralinos.

In order to draw a more definitive conclusion on our best-fit point in Eq. (2.84), with a higher mass splitting between the lightest neutralino and the pseudo-goldstino than in the scenario considered by ATLAS, we also consider the result from the search for squarks and gluinos in final states with jets and missing transverse momentum at $\sqrt{s} = 13$ TeV by ATLAS, using 36.1 fb$^{-1}$ of data [350]. This search targets several scenarios, two of which are of interest to our model, shown in Figure 2.22. In these, gluino pair production is followed by either direct decay to neutralinos ($\tilde{g} \to qq\tilde{\chi}_1^0$) shown in Figure 2.22a or by decay to the lightest neutralino $\tilde{\chi}_1^0$ through an intermediate $\tilde{\chi}_2^0$ ($\tilde{g} \to qqZ\tilde{\chi}_1^0$) shown in Figure 2.22b.

The analysis finds agreement between data and the background prediction. The most significant excess has a *p*-value of 0.02, corresponding to 2.0 standard deviations, in a method based on $m_{\text{eff}}$ (as in the 8 TeV search covered in Section 2.1.2) and 0.01, 2.5 standard deviations, in a new approach based on recursive jigsaw reconstruction[43]

---

[42]Since bino and wino cross sections are very small, this part of the analysis targets mass-degenerate higgsinos.

[43]This technique uses approximate (due to missing momenta) rest frames of intermediate particle states [351–353].

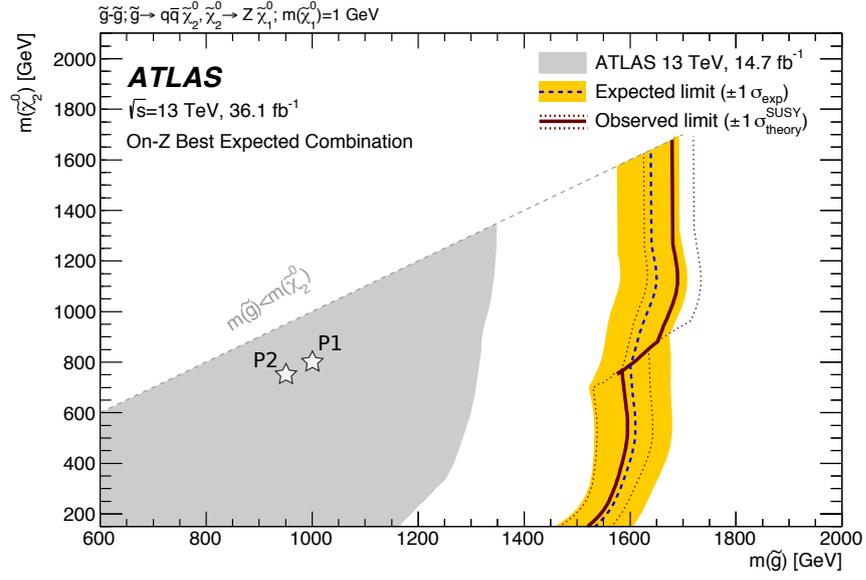

(a) $m_{\tilde{\chi}_1^0} = 1$ GeV

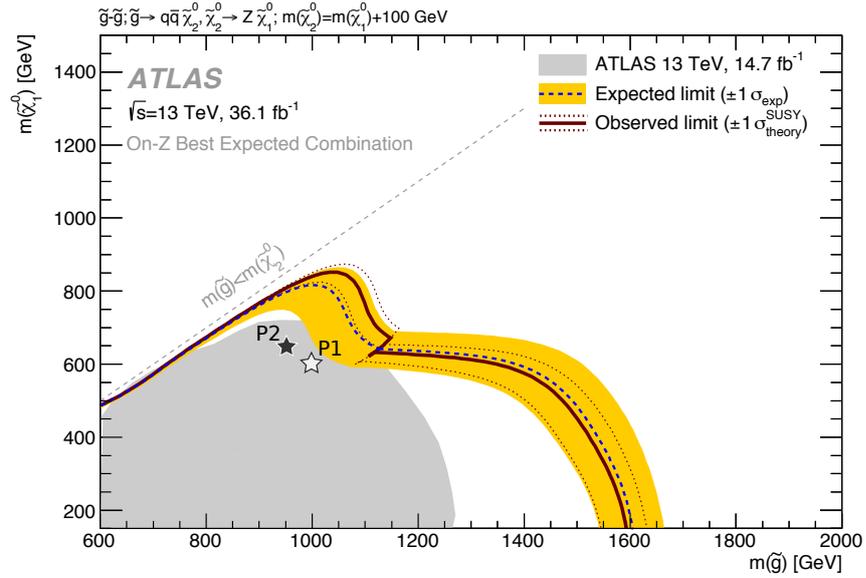

(b) $m_{\tilde{\chi}_2^0} = m_{\tilde{\chi}_1^0} + 100$ GeV

Figure 2.21: Constraints in the gluino-neutralino mass plane from the ATLAS search for 2 SFOS leptons, jets and $\not{E}_T$ at $\sqrt{s} = 13$ TeV, for two different scenarios of the neutralino masses. Indicated in white stars are our best-fit points *P1* and *P2*, at a point where the mass hierarchy is similar, but not exactly identical, to the one considered in the experimental analysis. Indicated in a black star is the parameter point *P2* where the mass hierarchy is exactly the same as in the experimental analysis. Figure from [347].

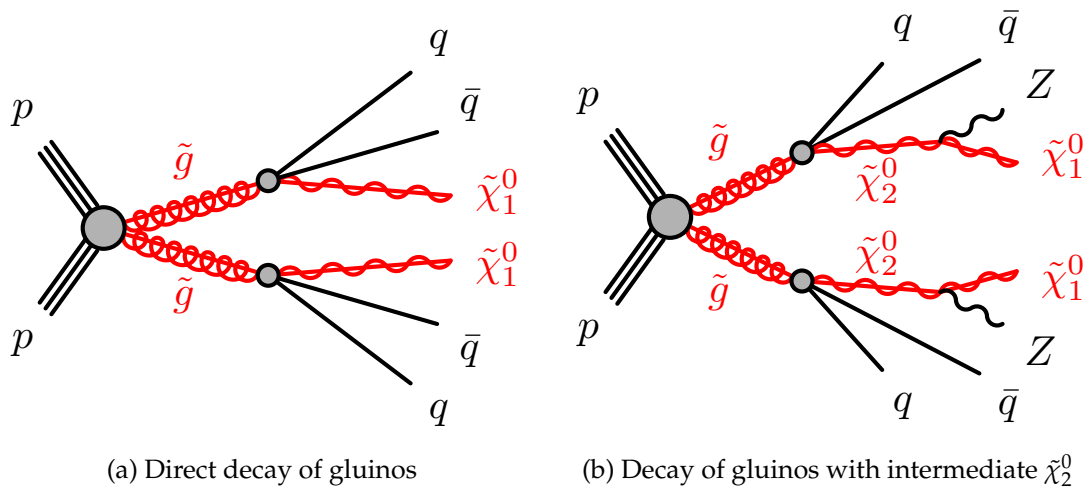

(a) Direct decay of gluinos            (b) Decay of gluinos with intermediate $\tilde{\chi}_2^0$

Figure 2.22: Feynman diagrams of the processes considered in the ATLAS search for gluinos and squark in events with jets+$\not{E}_T$ at $\sqrt{s} = 13$ TeV relevant to our scenario. Figure from [350]

which improves sensitivity to supersymmetric models with small mass splitting.

The inferred exclusion limits at 95% CL on the scenarios under consideration by the analyses are shown in Figure 2.23. In the case of a massless neutralino, gluino masses are excluded up to 2.03 TeV. Considering Figure 2.22a, our best-fit point $P1$ in Eq. (2.84), where we compare with the point $(m_{\tilde{g}}, m_{\tilde{\chi}_1^0}) = (1000, 600)$ GeV indicated in the figure, is well within the excluded region of the 13 TeV search, assuming that the presence of an additional intermediate state in our scenario has negligible effect on the analysis. We can verify this assumption by comparing with the exclusion limit in case of a two-step decay shown in Figure 2.23b, where the final state neutral particle has a mass fixed to 1 GeV. Here, we see that, except for small masses of the intermediate state, the exclusion limit is nearly independent of the intermediate particle mass, bounded only by the requirement that its mass is less than that of the gluino. Therefore, we infer that we can use the exclusion contour of the direct decay scenario to conclude that our best-fit point $P1$ is also excluded.

## 2.6   Summary

Even though the LHC at 8 TeV has not discovered a new physics signal, the Run 1 data contain a few small excesses that deserve a thorough investigation. In 2015, the ATLAS Collaboration reported a 3.0 $\sigma$ excess in the dilepton + jets + $\not{E}_T$ channel at $\sqrt{s} = 8$ TeV, with the dilepton invariant mass reconstructed at the $Z$-boson mass.

In this chapter, we proposed an explanation of such excess in a SUSY model of gauge mediation with two SUSY-breaking sectors, presenting in the SUSY spectrum an extra neutral fermion besides the MSSM neutralinos, that is the pseudo-goldstino. Our

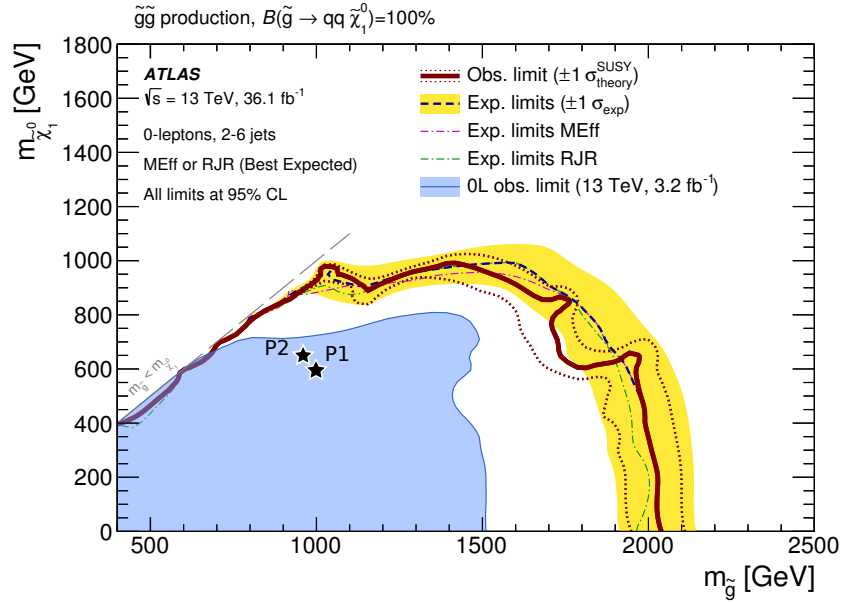

(a) Direct decay of gluinos

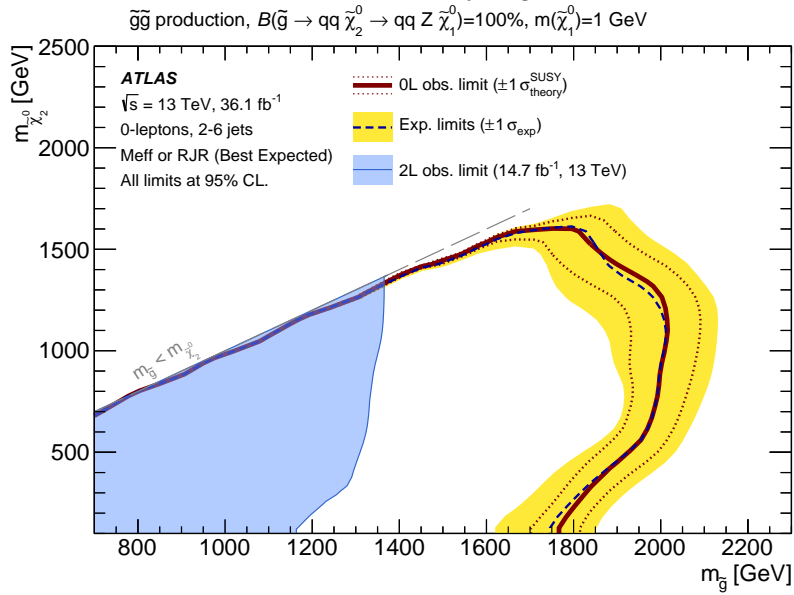

(b) Decay of gluinos with intermediate $\tilde{\chi}_2^0$

Figure 2.23: Constraints in the gluino-neutralino mass plane from the ATLAS jets+$\not{E}_T$ search. Indicated by black stars are our best-fit points $P1$ and $P2$, where the mass hierarchy is identical to the one in the experimental paper (although the decay is through an intermediate particle instead in our case). Figure from [350].

simplified model consists of a gluino, a pair of higgsino-like neutralinos, and a pseudo-goldstino. We showed that our goldstini model can explain the 8 TeV ATLAS $Z$-peaked excess without conflicting with the constraints from jets + $\not{E}_T$ as well as from the CMS analysis for the same final state. The compressed spectrum such as $m_{\tilde{g}} \sim 1000$ GeV, $m_{\tilde{\chi}^0_{1,2}} \sim 800$ GeV and $m_{\tilde{G}'} \sim 600$ GeV gives a very good fit to the data, not only for the rate, but also for the kinematic distributions.

Given the new data gathered at $\sqrt{s} = 13$ TeV since the publication of this work, we revisited the constraints on our model. The updated search in the dilepton + jets + $\not{E}_T$ channel no longer sees an excess of events and excludes the exact mass hierarchy of our second best-fit point $P2$. Moreover, we find that the updated jets+$\not{E}_T$ search also rules out both of our best-fit points. Therefore, we conclude that, for the specific scenario considered here, our model is now ruled out. However, the general result that multiple SUSY-breaking sectors can provide an interesting phenomenology distinct from standard SUSY scenarios, which can alleviate some of the typical constraints from e.g. jets+$\not{E}_T$ searches, still stands and might be relevant should another excess appear.

**Part II**

# Astroparticle physics and neutrino phenomenology

## Astroparticle physics and neutrino astronomy

The second part of the thesis covers astroparticle physics phenomenology, in particular the study of high-energy astrophysical neutrinos. In recent years, the IceCube Collaboration has established the existence of a high-energy astrophysical neutrino flux, most of which seemingly being of extragalactic origin. While a single source of these neutrinos has now been identified, flares of the blazar TXS 0506+056, it is still unknown which source class is responsible for the bulk of the neutrino flux. By applying the concepts from particle physics and using complementary information from cosmic rays, gamma rays and gravitational waves (the 'multimessenger paradigm'), we can learn more about the sources responsible for this flux. In this chapter, we review the basic concepts necessary for these studies as well as the current status of the field.

### 3.1 A short historical introduction

The history of astroparticle physics (an extended review of which can be found in [354]) starts with the discovery of cosmic rays in 1912 by Victor Hess. At the beginning of the 20$^{\text{th}}$ century, it had been found that the air is always slightly ionised, even without the presence of a source of ionising radiation. While there was the possibility that this ionisation is due to naturally occurring radioactive elements inside Earth's crust, the origin of this slight ionisation remained unclear until Victor Hess performed a series of balloon flights to measure the ionisation at different heights. His results [355], shown in Figure 3.1, indicated that, after reaching a certain height, the amount of ionisation started to increase with increasing height. This indicated that the source of the ionisation must be of extraterrestrial origin. This was later confirmed with another experiment by Werner Kolhörster (quoted in [356]), also shown in Figure 3.1. During subsequent years, it was established that the Earth is continuously bombarded by charged particles, cosmic rays, which collide with the atmosphere and create showers of secondary particles which we can detect on Earth's surface.

In subsequent years, the existence of cosmic rays and their resulting air showers became well established. By studying the ionising particles supplied by the cosmic rays, new knowledge was gained on fundamental physics as well. By observing ionising tracks in cloud chambers and bubble chambers, new particles could be detected which pass through the detector or are created in interactions within it. This led, for example, to the discovery of the positron by Anderson in 1932 [18] (shown in Figure 3.2) and the discovery of the muon by Neddermeyer and Anderson in 1936 [19]. While particle accelerators were at that point already available, they were unable to compete with the higher energies supplied by cosmic rays.

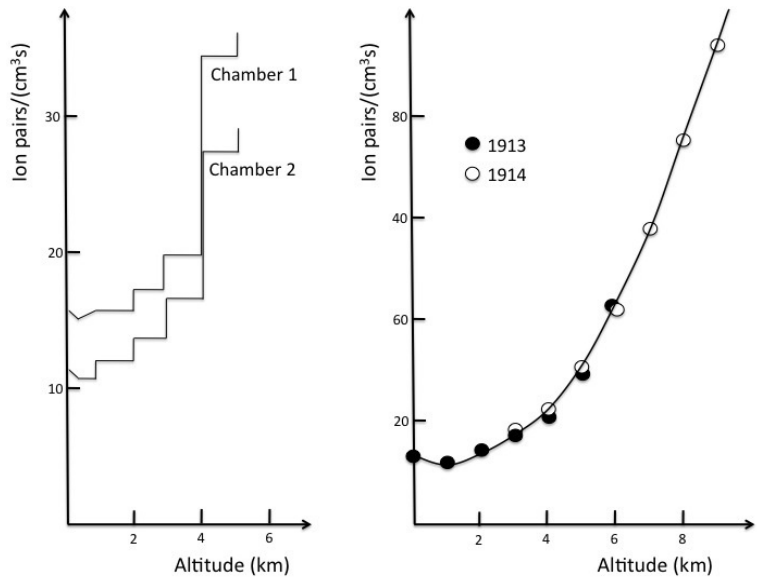

Figure 3.1: First measurements of the ionisation in the atmosphere as a function of height. By Hess [355] (left) and by Kolhörster (1913, 1914) quoted in [356] (right).

After World War II, particle accelerators came to dominate over cosmic-ray experiments for the study of particle physics. Technical developments led to the availability of accelerators capable of colliding particles at higher energy and their controlled environment presented a significant advantage over cosmic-ray experiments. As a consequence, particle physicists shifted their attention to accelerator experiments, while studies of cosmic radiation came to belong mainly in the field of astrophysics, leading to the "downfall" of cosmic-ray studies [354].

The field of astroparticle physics as we know it today can be considered to have (re)started in the 1980s [354]. This revival was due to the significant progress that had been made in particle physics (building the Standard Model [1–7]) and cosmology (discovery of the Cosmic Microwave Background [357] and establishment of the ΛCDM-model [46]) in the preceding two decades. In particular, when describing early universe physics (for example big bang nucleosynthesis [358]) or providing an explanation for dark matter [359–362], elements from both cosmology and particle physics are needed. At present, astroparticle physics is a thriving interdisciplinary field at the interface of

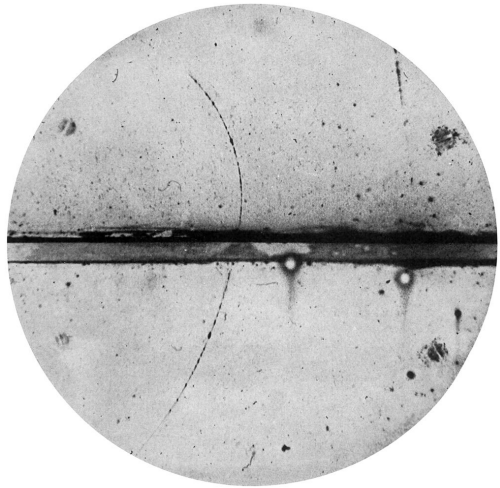

Figure 3.2: The beginning years of (astro)particle physics, where the study of cosmic rays led to the discovery of new particles. This picture shows the discovery of the positron. By observing its bending in a magnetic field and energy loss after passing through a lead plate, it was identified as a 'positive electron' [18].

particle physics, cosmology and astrophysics. Its studies concern cosmic rays across a vast energy range and includes particles like protons and antiprotons, nuclei of different masses, electrons and positrons, gamma rays and neutrinos as well as gravitational waves. These particles can come from the sun, supernova explosions in our galaxy or from more powerful objects in other galaxies (even the earliest ones), providing us with an opportunity to probe the cosmological history of the universe and the astrophysical environments from which these particles originate. The emission, propagation and detection of these (high-energy) particles are governed by particle physics. As such, there is a plethora of phenomena that can be studied with astroparticle physics. In turn, it is possible to use the particle fluxes supplied by nature to search for physics beyond the Standard Model. The highest energy cosmic rays provide particle physicists (again) with an opportunity to study fundamental physics at energies higher than those available at particle accelerators[1,2]. Other examples of beyond the Standard Model

---

[1]Comparing the energies studied in their respective fields can be a source of amusement for physicists working on accelerator experiments, cosmic-ray experiments or theories such as inflation or string theory. Concerning the experimental energies, cosmic-rays energies are indeed much higher than those achieved in LHC, as can be seen in Figure 3.3. In reality however, the difference is slightly less absurd than it seems. The observation of cosmic rays is essentially a fixed target experiment, for which the centre of mass energy is $\sqrt{s} = \sqrt{2E_{cr}m_p}$. For the highest energy cosmic rays between $E_{cr} = 10^{18-20}$ eV colliding with particles in our atmosphere, this gives $\sqrt{s} \approx 30 - 600$ TeV to be compared with $\sqrt{s} = 13$ TeV for the LHC and $\sqrt{s} = 100$ TeV for a proposed future accelerator [363]. A few decades away for colliders, but not as bad as the 7 orders of magnitude in energy suggested by the figure.

[2]Another difference between accelerator and air shower experiments is the collision angle of interest. Accelerator experiments mainly probe collisions in the central region (low rapidity), with final state particles having high momentum in the plane orthogonal to the beam direction. Air shower experiments probe the

scenarios that are being or have been investigated using astroparticle physics include, but are not limited to, neutrino oscillations (in both atmospheric neutrinos [50] and solar neutrinos [49, 364]), dark matter, monopoles and the neutrino-nucleon cross section at high energy [365, 366]. In the rest of this thesis, however, we will instead study astrophysical objects, through the use of the high-energy particles they emit. In particular, we will focus on high-energy astrophysical neutrinos and their connection to cosmic rays, gamma rays and gravitational waves (GW).

## 3.2 Cosmic rays

We start with a discussion on the observed flux of cosmic rays and its origin. By cosmic rays, we mean all charged baryonic particles like protons and nuclei, excluding electrons and positrons, neutrinos and gamma rays, the latter two of which will be discussed later. There are several reasons for splitting up the discussion and treating cosmic rays separately. First of all, this follows the historical developments: since cosmic rays were discovered first, they are also the origin of the field. More importantly, however, understanding the origin of cosmic rays serves as a theoretical introduction to astroparticle physics using neutrinos and gamma rays. Cosmic rays are also distinct from neutrinos and gamma rays, since they are charged such that their trajectories are bent by the galactic and intergalactic magnetic fields. As a consequence, cosmic rays do not point back to their sources[3]. In addition, interpreting the measurements of the highest energy cosmic rays involves hadronic physics, introducing uncertainties different from those in neutrino and gamma ray detection. This section will only discuss the cosmic-ray physics necessary to understand and motivate the rest of the thesis. Excellent reviews of cosmic ray physics can be found in e.g. [367, 368]

### 3.2.1 Energy spectrum and composition

Since the discovery of cosmic rays, many experiments have measured the cosmic-ray spectrum and composition, the results of which are summarised in Figure 3.3. The cosmic-ray spectrum spans many orders of magnitude in energy and flux, from 1 particle per m$^2$ per second at $10^9$ eV to 1 particle per km$^2$ per century at $10^{20}$ eV. While particles at the end of the spectrum are therefore rare, they contain an enormous amount of energy. A single elementary particle can have an energy of several tens of Joules, a macroscopic quantity! The spectrum follows a near-perfect power law $\frac{dN}{dE} \propto E^{-\gamma}$ across the entire energy range, with small breaks where the spectral index changes slightly as the only visible features . From the onset of the power law in the region 1–10 GeV, the

---

forward region.

[3]At the highest energies however, the deflection is less strong. Coupled with a limited horizon due to the GZK effect (see Section 3.2.5), this means that local structures are imprinted in the angular spectrum of cosmic rays at the highest energies.

spectrum follows the form [29, 369]

$$\frac{\mathrm{d}N}{\mathrm{d}E}(E) \propto \begin{cases} E^{-2.7} & (E \lesssim 10^{15-16} \text{ eV}) \\ E^{-3} & (10^{15-16} \text{ eV} \lesssim E \lesssim 10^{18.5} \text{ eV}) \\ E^{-2.7} & (10^{18.5} \text{ eV} \lesssim E \lesssim 10^{20.5} \text{ eV}). \end{cases} \tag{3.1}$$

The first break between $10^{15}$ and $10^{16}$ eV, where the spectrum steepens, is called the knee and the second break at $10^{18.5}$ eV, where the spectrum flattens again, is called the ankle. Finally, the spectrum has been measured up to energies of almost $10^{20.5}$ eV, with evidence that this is approximately where the spectrum ends and that this is due to the sources not emitting cosmic rays much above this energy (see Section 3.2.5). While this form describes the spectrum well at first sight, it is a simplification. More recent measurements of the spectrum around the knee have determined a more rich structure, as shown in Figure 3.4. First, there is the appearance of the 'second knee': after the knee, the cosmic-ray spectrum has a power law-index of $\gamma = 3$ down to the 'second knee' at $4 \times 10^{17}$ eV, where it steepens again to an index of $\gamma = 3.2$ up to the ankle. More importantly, there is a change in composition of the flux as a function of energy, with kinks around the energy of the knee, second knee and ankle. As we will see later, power laws are very generic and, apart from their normalisation and spectral index, hide all information from the system in which they originate. Therefore, in order to study cosmic rays, one really needs to study the features where there are *deviations* from a perfect power law. As an example, the influence of the solar wind is imprinted on the spectrum up to energies of 1 GeV. For this reason, when showing results, the flux are often multiplied by the overall power of the spectrum.

**Cosmic rays up to the knee**

The first part of the cosmic-ray spectrum, up to the knee, is repeated in more detail in Figure 3.4a. At these energies, the flux is high enough for direct detection and identification of cosmic-ray particles by relatively small detectors on board of balloons and satellites, above the point where the first interaction with the atmosphere occurs. This energy range also mostly overlaps with the energy range where we expect the cosmic rays to be of galactic origin. There are two reasons for why we expect a galactic origin at these energies. First of all, the galactic magnetic field is such that particles up to about $10^{15}$ eV can easily be confined[4], with higher energy cosmic rays escaping from the galaxy. Secondly, there are no known galactic objects or environments which can be the source of cosmic rays with an energy much higher than the knee (though likely possible up to the second knee), see also Section 3.2.6. The composition can be measured with high accuracy, also shown in Figure 3.4a. The abundance of elements in cosmic rays closely follows that of the elements found in the solar system, with some significant deviations for specific elements like B and Be. B is not formed through nucleosynthesis in stars, but can be formed by spallation reactions of C with the interstellar gas. Therefore, the B/C-ratio probes the average path length traversed by cosmic rays before their arrival on Earth. Similarly, $^{10}$Be is created in the spallation of C and O. This element, however, is unstable and decays to $^{10}$B, with a half-life similar to the confinement time of cosmic rays in the galaxy. Therefore, the ratios $^{10}$Be/$^9$Be and Be/B are sensitive to this confinement time of cosmic rays in the galaxy[5]. Near the end of this part of the spectrum, the composition changes towards heavier elements, since these elements have a higher charge and are therefore more efficiently confined in the source (see Section 3.2.3) and in the galaxy, i.e. they escape at higher energies. By comparing all these observables, and others, with theoretical models and simulation codes of cosmic-ray propagation in the galaxy (GALPROP [370] and DRAGON2 [371]), cosmic rays in this energy range are currently very well understood.

---

[4]Starting at $10^{15}$ eV, the proton gyroradius is roughly equal to the scale of magnetic irregularities in the galaxy, so that their propagation through the galaxy can no longer be described by the diffusion approximation and they escape the galaxy more readily. At $10^{18}$ eV, the proton gyroradius is roughly equal to the thickness of the galactic disk and protons can no longer be confined [368].

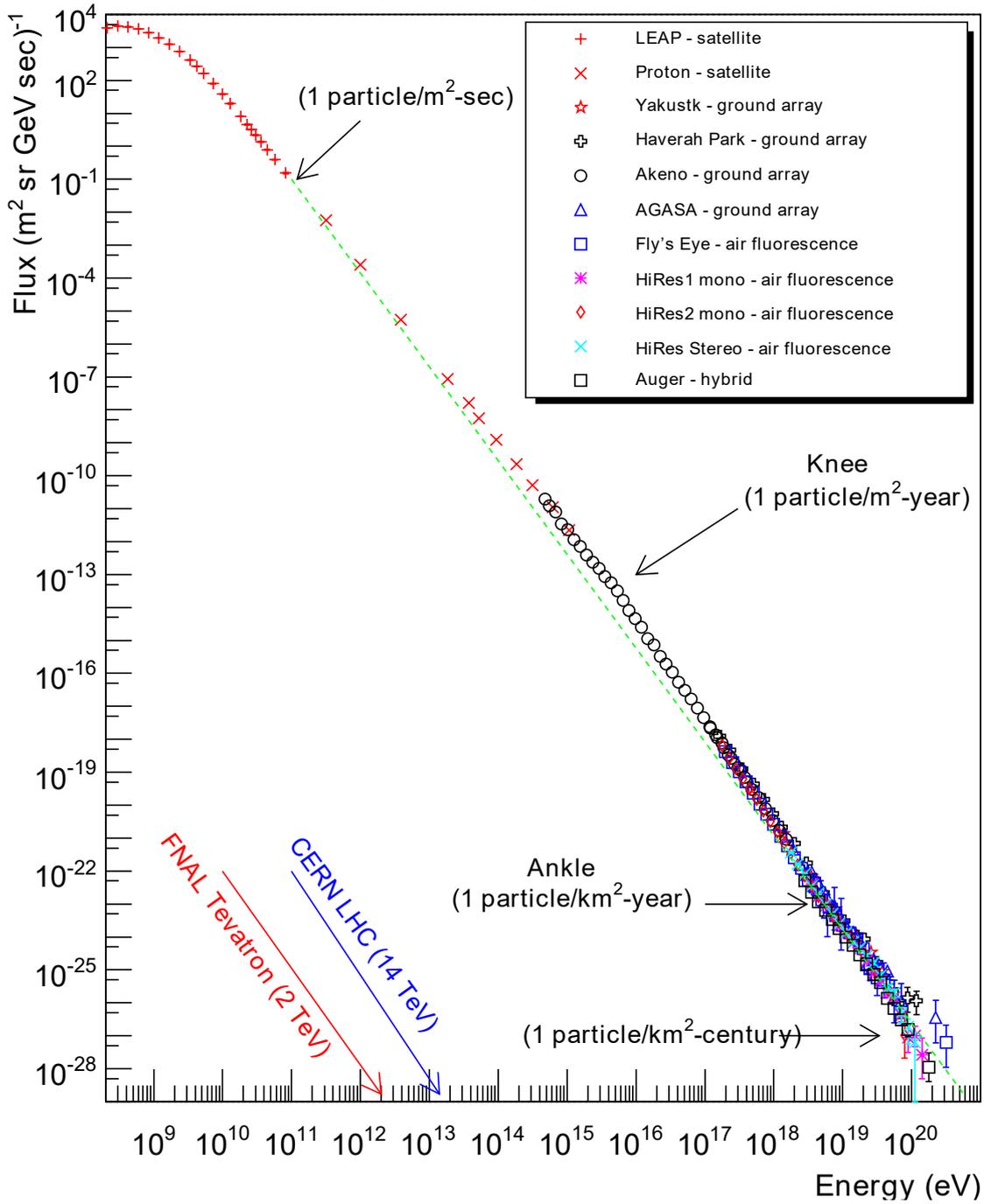

Figure 3.3: The total cosmic-ray spectrum, spanning more than 11 orders of magnitude in energy and 32 in flux, described by a near perfect power law over its entire range. Credit: W. Hanlon.

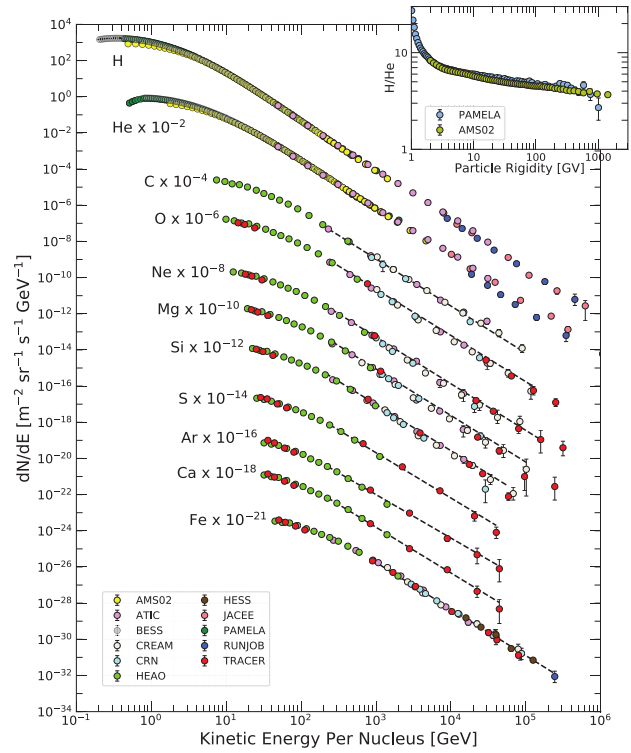

(a) Direct measurements

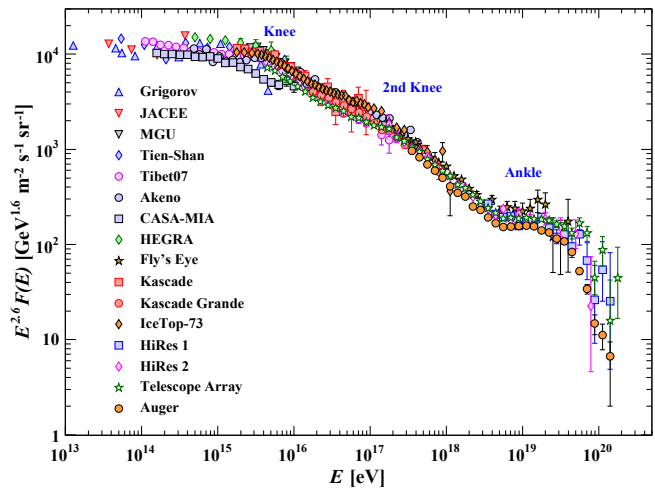

(b) Air shower measurements

Figure 3.4: The cosmic-ray spectrum focussed on two regimes. Figures from Particle Data Group [29]. (a) Spectrum of different nuclei below the knee, from direct detection of the cosmic rays where the primary nuclei can be individually identified. Inset shows the H/He-ratio. (b) Spectrum starting at the knee, observed using air showers. The flux is multiplied with $E^{2.6}$ to enhance the features of the spectrum.

**Cosmic rays beyond the knee**

Beyond the knee, the cosmic-ray flux has dropped down to such low levels that direct detection is no longer feasible. Instead, the atmosphere is now used as a calorimeter, where a cosmic ray interacts with a nucleus in the atmosphere and creates many secondary particles. Subsequently, these products can in turn interact again with particles in the atmosphere or decay, initiating a shower of particles. Experiments are designed to detect these air showers: by sparsely covering a large surface area with particle detectors, experiments are able to obtain large effective areas and perform measurements of the cosmic-ray flux above the knee, the results of which are shown in Figure 3.4b. Such measurements are more tricky to interpret than direct measurements however, since one needs to model the interactions in the atmosphere[6] in order to infer information on the original cosmic-ray particle. One difficulty is determining the energy of this original cosmic ray, leading to energy calibrations which initially differed widely between experiments. In recent years, a lot of effort went into recalibrating the energy scale of different experiments in order to find agreement between measured fluxes, with great success. More important for the discussion here, is the difficulty in determining the composition of the cosmic-ray flux. Since the original particle is broken up, its nature needs to be determined using hadronic interaction models.

For ultra-high energy cosmic rays (UHECRs), with energies above $10^{18}$ eV, the amount of particles deposited in the atmosphere by air showers is enormous and at the Earth's surface they are spread out over hundreds of metres up to kilometres. Such showers are called extensive air showers (EAS). Current experiments measuring the UHECR flux are the Pierre Auger Observatory [373] and Telescope Array (TA) [374][7].

---

[5]More details can be found in [367, 368].

[6]These models combine hadronic interaction models, discussed in Section 3.3.2, with shower development theory. The most important implementation of this is CORSIKA [372].

[7]These are hybrid experiments. In addition to particle detectors, they also use fluorescence detectors to measure the light created by the air shower. This provides an independent calibration of the energy, significantly improving the systematic uncertainties.

The most important features of the UHECR flux are the behaviour of the flux with energy (rising or falling) and the composition (proton, heavier-than-proton or iron-like dominated) at the maximum cosmic ray energy. As will be seen later, these observables are important because they contain information about the origin of the cosmic rays. From Figure 3.4b, it can be seen that the cosmic-ray flux falls down at the end of the spectrum. Figure 3.5 shows the results from composition measurements by Auger (and at lower energies by different experiments). Due to the shower-to-shower fluctuations, a composition measurement can only be made on a statistical basis. The variables of interest are the average shower depth $\langle X_{max} \rangle$ (the depth (in column density) of the shower maximum, i.e. where the number of particles in the shower is at its maximum) and fluctuations around this value $\sigma \langle X_{max} \rangle$. A comparison of the measured values with those obtained in simulations by hadronic interaction models for pure proton and pure iron spectra suggests that the cosmic rays become dominated by heavier elements at high energy. The exact results from the hadronic interaction models are quite uncertain, however, since the models are extrapolated from measurements at LHC energies and below, probing the central region of the collision. Moreover, the true composition is expected to be a mix of elements, which influences the relation between $\langle X_{max} \rangle$ and $\sigma \langle X_{max} \rangle$. More intricate, combined fits of spectrum and composition taking into account a mixed composition are available (see e.g. [375]). While the issue is not settled, they show the same trend where the composition gets heavier towards the end of the spectrum (see also [376, 377]). Telescope Array has performed similar measurements and, up to a few small deviations, the results are largely compatible between the two experiments, as determined by a joint working group [378–380][8].

For cosmic rays between the knee and the ankle, the (galactic) origin is not yet settled (see Section 3.2.6). For UHECRs, the situation is different: while the exact sources are not yet known, we do know that they need to be extragalactic. The reason for this is threefold [382]. First, there are no known objects in our galaxy which could be the source of such energetic particles (see also Section 3.2.3). Secondly, cosmic rays at these energies are no longer contained in the galaxy and would therefore escape. Finally, since cosmic rays at these energies suffer little deflection from the galactic magnetic field, they would not arrive isotropically if they came from within the galaxy. Since we do observe an isotropic spectrum[9], the UHECRs have to be extragalactic.

## 3.2.2 The origin of cosmic rays

Studying the cosmic-ray spectrum and its behaviour across a vast energy range raises an important question: from where do these particles originate? While from the energy and direction it is possible to determine a galactic or extragalactic origin, this does not yet give us information about a possible source.

---

[8]Some people like to emphasise that the TA measurements are still compatible with a pure proton spectrum. While this statement leads to animated discussions, it seems to revolve mostly about a choice of wording. On the other hand, since these two experiments are in a different hemisphere, there are scenarios where a different composition can be naturally explained, see e.g. [381]

[9]At least at first order. More recent measurements with higher statistics have shown evidence for the existence of a dipole above $8 \times 10^{18}$ eV, see [383].

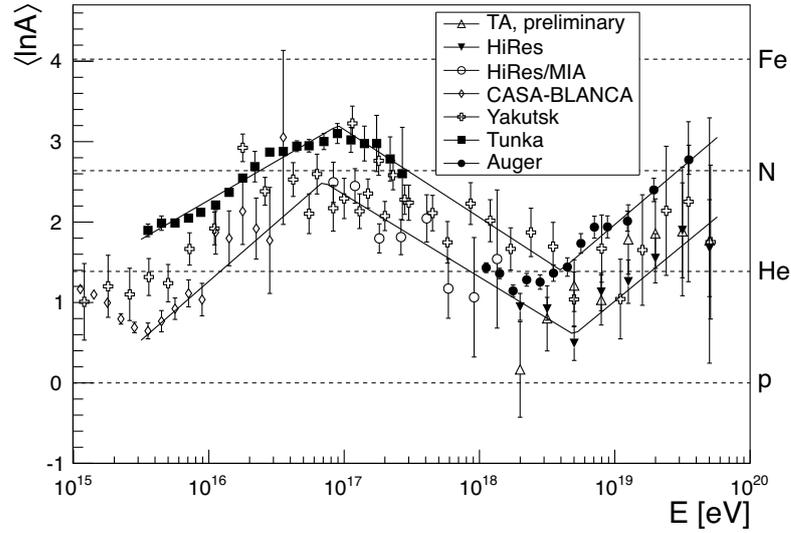

(a) Above the knee

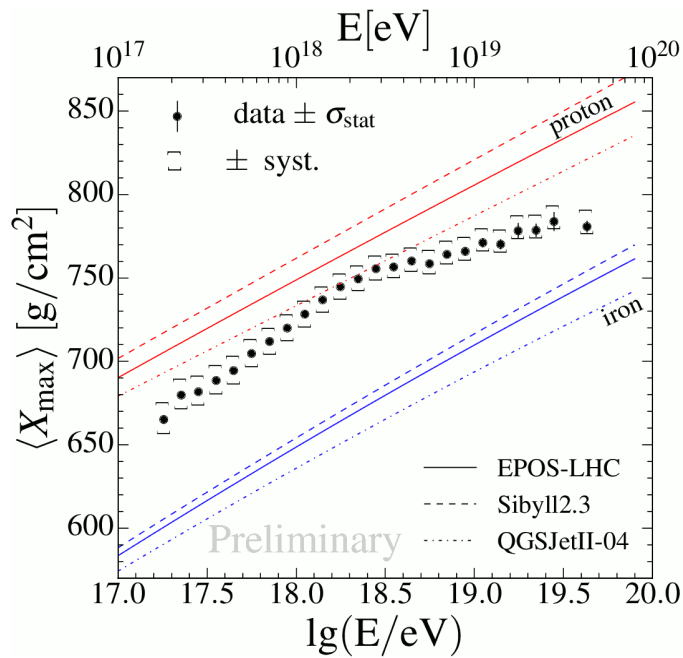

(b) UHECR

Figure 3.5: Measurement of the composition of cosmic rays above the knee. (a) Average logarithmic mass as a function of energy derived from $X_{max}$ measurements for the EPOS 1.99 interaction model for different experiments, by [384]. (b) Recent measurement of $\langle X_{max} \rangle$ of UHECR air showers by Auger [377].

There are two ways in which particles can get high, non-thermal energies[10]. The first option is the top-down mechanism. Super-heavy particles or topological defects can decay and create the (UHE)CR. Such objects appear in UV completions of the Standard Model and could be associated to the scale of Grand Unified theories with $M_{GUT} \sim 10^{24}$ GeV. For an extensive review, see [369]. Since these models also predict other high-energy particles, this possibility is already highly constrained, by searches for ultra-high energy photons [385–387] and neutrinos [388–392].

The second option, which we will consider for the rest of this thesis, is bottom-up, where low energy particles (likely from a thermal distribution) are accelerated in astrophysical sources to high energy. The mechanism through which this happens needs to be very efficient, since a macroscopic amount of energy needs to be channelled into a few particles. Considering the known astrophysical objects and forces in the universe, there are not a lot of candidate mechanisms. Let us start by considering a simple energy gain of particles due to a fundamental force. The strong force is confining, so does not seem capable of creating free high-energy particles, while the weak force is too weak to reach the required efficiencies. The universe and astrophysical objects within it are observed to be charge neutral, eliminating simple electric fields. Therefore, the only remaining option is to make use of magnetic fields. Since static magnetic fields are unable to perform work on particles to increase their energy directly, only time dependent magnetic fields (and their induced electric fields) are left as a candidate. Such a situation can occur, for example, near sunspots (magnetic fields evolving in time or moving) or in pulsars (rotating magnetic field), see e.g. [393]. A more complicated mechanism involving magnetic fields is magnetic reconnection, which occurs when the magnetic topology in a plasma is suddenly rearranged (see e.g. the review in [394]). During this rearrangement, magnetic energy can be converted into kinetic energy and particle acceleration can occur. Such a process can take place in solar flares and could be responsible for the high-energy particles [11] that have been observed from them (see e.g. [396]). All of these systems, however, seem to be able to explain only part of the flux in part of the energy range.

Another way to accelerate particles efficiently, which also immediately gives rise to a power law distribution with (almost) the correct spectral index, is the (first order) Fermi mechanism. Also known as shock acceleration, it occurs in astrophysical shocks and has been used to successfully explain non-thermal particle spectra observed in Earth's bow shock and in supernova remnants in our galaxy. Whilst not the only possible mechanism for particle acceleration, it is the most popular for the sources in which we are interested. Therefore, this mechanism will serve as the motivation for some of the assumptions in the rest of this thesis[12].

---

[10]Non-thermal means that there is no natural energy scale associated with these particles, in contrast to those following a Maxwell-Boltzmann distribution or black body spectrum with a characteristic temperature. See also the intermezzo about power laws.

[11]High energy in the context of solar flare physics means at most 10 GeV or so [395].

[12]In principle, though, we can be apathetic about the exact origin of the power law spectrum of cosmic rays for the work in this thesis, since we will use it as an input. Still, the knowledge that a good candidate

In the above discussion, we silently ignored the gravitational force. Since it is weak and couples to everything, it is not a good candidate for accelerating the particles directly. On the other hand, as we will see, it is the force that is responsible for powering the central engines, the macroscopic systems, that serve as the acceleration sites for the cosmic rays.

**Intermezzo — Power laws**

Power laws, while peculiar, occur frequently in nature in many different systems (see the standard work on power laws [397]). They are associated with scale-free systems: only power laws can satisfy the demand that a distribution is invariant under arbitrary rescaling (up to normalisation). The argument goes as follows [397]. Suppose that a distribution satisfies the property that it is scale-free, or self similar,

$$p(bx) = g(b)p(x) \quad \forall b. \tag{3.2}$$

Using this equation with $x = 1$ to find $g(b)$, we have

$$p(bx) = \frac{p(b)}{p(1)}p(x). \tag{3.3}$$

Differentiating with respect to $b$ and setting $b = 1$ gives

$$x\frac{\mathrm{d}p}{\mathrm{d}x} = \frac{p'(1)}{p(1)}p(x). \tag{3.4}$$

This differential equation can be solved to give

$$\ln p(x) = \frac{p(1)}{p'(1)}\ln x + \text{const.,} \tag{3.5}$$

where the constant term can be determined by setting $x = 1$, yielding const. $= \ln p(1)$. This gives as a final result the power law distribution

$$p(x) = p(1)x^{-\alpha}, \tag{3.6}$$

with $\alpha = -p(1)/p'(1)$.

An example of where such a power law occurs is in systems exhibiting critical behaviour, where the only relevant scale of the problem diverges. This happens in magnets, where the correlation length diverges during the (continuous) phase transition. As we will see, the Fermi mechanism is also scale-free and will therefore give rise to a power law for the cosmic-ray flux[13]. An important remark however, is that some

---

mechanism exists which moreover gives a prediction for the exact spectral index eases the mind.

[13]Obviously this is not true for arbitrarily large or small energies. As we saw, the solar wind has a natural scale and thus induces a deviation from a power law in the observed cosmic-ray flux. In the other extreme, we know that astrophysical objects are not infinite in size and energy, which unavoidably puts a limit on the energy up to which the power law can be valid.

distributions can mimic a power law, whilst not strictly being power laws[14]. This can happen for example with log-normal distributions if they are observed in a parameter range small enough[15] that the quadratic behaviour in log-space is not visible. Therefore, an observed power law for the cosmic rays does not necessarily mean that the system from which they originate needs to be scale-free.

**Fermi mechanism and shock acceleration**

The original Fermi mechanism was proposed by Fermi in 1949 [398]. He suggested that particles could be accelerated by having repeated encounters with magnetised clouds in the galaxy. In each encounter, the particle gains or loses energy depending on whether the encounter is head-on or overtaking[16]. Because the cloud is magnetised, the particles undergo collisionless scattering with the irregularities in the magnetic field, such that they do not lose energy and will not thermalise. Since there are more head-on than overtaking collisions, on average the particle will gain energy. Considering the particle and cloud velocities and averaging over all possible angles of approach, one finds that (see Appendix F) the average energy gain in each encounter is

$$\frac{\Delta E}{E} \propto \frac{v^2}{c^2} = \beta^2,\tag{3.7}$$

with $v$ the velocity of the cloud [367]. Since the velocity enters squared in the energy gain, this is called the second order Fermi mechanism.

When a set of particles has repeated encounters with magnetised clouds with a constant fractional energy gain and a fixed chance for each particle to escape the system after each encounter, the energy distribution of the particles will follow a power law. This can be understood intuitively from the discussion in the previous section, since such a system has no inherent energy scale for the particles to converge on. A more rigorous derivation is given in Appendix F.

Unfortunately, because of the $\beta^2$-dependence, the energy gain is too inefficient for this version of the mechanism to be responsible for the acceleration of galactic or extragalactic cosmic rays. Furthermore, when doing the full calculation, the resulting spectral index is dependent on several parameters which are variable, such as the cloud speed, its density and the acceleration time. Such a variable spectral index seems difficult to reconcile with the fact that the observed cosmic-ray spectral index is constant over almost the entire energy range.

Interest in the Fermi mechanism was reinvigorated when it was realised that there exists a system where particles undergo only head-on collisions: non-relativistic shock fronts moving through a collisionless plasma [399–403] as happens, for example, in supernova blast waves. In this case, a shock propagates through the interstellar medium at a speed higher than the sound or Alfvén speed. The medium in front of the shock, called the upstream region, is at rest in the frame of the interstellar medium. The shocked medium, called the downstream region, is moving in the same direction as the shock, but at a lower velocity. Due to this setup, from the rest frame of either side of the shock, the region on the other side of the shock is moving towards it. Particles that are already sufficiently relativistic[17] continuously cross the shock, each time colliding head-on with the turbulent plasma on the other side. This situation is sketched in Figure 3.6a. Each encounter now results in an energy gain, making the acceleration process very efficient. In fact, the energy gain is proportional to the shock speed $v$

$$\frac{\Delta E}{E} \propto \frac{v}{c} = \beta, \tag{3.8}$$

---

[14]Another option is that a certain distribution only shows power law behaviour in its tails.

[15]"Small enough" can still mean several orders of magnitude in the parameter of interest.

[16]Actually, it depends both on the orientation of the ingoing particle *and* of the outgoing one, see e.g. [367].

Therefore, this version is called the first order Fermi mechanism, also known as shock acceleration. After particles cross the shock, they can undergo several collisions, which will cause them to move isotropically in the rest frame of that region before recrossing the shock and gaining energy again. Because of the movement of the downstream region away from the shock in the shock rest frame, particles have a fixed chance of escaping the acceleration region, independent of their energy[18]. As already seen above, this will cause the energy distribution of the accelerated particles to follow a power law. In this case, however, the only input is the shock velocity, with the ratio of the velocities up- and downstream of the shock fixed by kinetic gas theory. As a result, shock acceleration predicts a universal power law with a spectral index of 2,

$$\frac{\mathrm{d}N}{\mathrm{d}E} \propto \left(\frac{E}{E_0}\right)^{-2}, \tag{3.9}$$

where $E_0$ indicates the initial particle energy. For a derivation of this result, see Appendix F. It turns out that this is exactly the spectral index that is needed to explain the spectrum of galactic cosmic rays, since diffusion through the galaxy will contribute an extra power of 0.7 (see e.g. [367]), such that the resulting energy distribution of galactic cosmic rays is exactly $\propto E^{-2.7}$. In reality, there can be deviations from this index, most importantly when properly taking into account the magnetic fields present in the plasma.

Shocks and the accompanying accelerated particles have been experimentally observed [406], for example in supernova remnants (see Figure 3.6b) and in Earth's bow shock, confirming that shock acceleration occurs in nature[19]. Moreover, when taking into account the lifetime of supernovae, the amount of supernovae in the galaxy at each particular time, their luminosity and their maximum energy from this mechanism, it is found that supernovae have the required energy budget to supply all of the galactic cosmic rays (see e.g. [367]).

---

[17]This is actually a non-trivial requirement and the question of how particles can enter the acceleration is called the injection problem. For a short discussion, see Appendix F.

[18]At least until their gyroradius is such that they are no longer confined, see Section 3.2.3.

[19]In fact, it seems to be an important dissipation mechanism for collisionless shocks [406], which otherwise have no other efficient way to dissipate energy. Accelerated particles can drive the magnetic instabilities, increasing the acceleration efficiency until there is an equal amount of energy in the plasma and the cosmic rays. Therefore, the assumption of efficient shock acceleration might be less fine-tuned than it first appears.

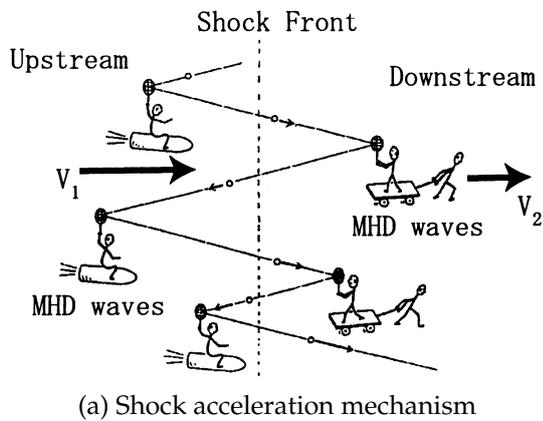

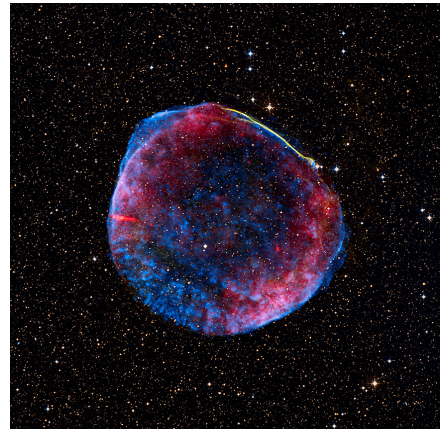

(a) Shock acceleration mechanism          (b) Supernova remnant SN 1006

Figure 3.6: Shock acceleration (a) Cartoon illustration of the mechanism behind shock acceleration. Figure taken from [404], after an original sketch by M. Scholer. (b) An image of the supernova remnant SN 1006, showing X-ray measurements (blue) and radio (red). The X-ray emission at the edge of the expanding shell provides evidence for electrons accelerated to 100 TeV within the shock front [405]. Image credit: NASA Chandra X-ray Observatory.

The objects discussed above and the derivation of the spectrum from shock acceleration in Appendix F all considered cosmic-ray acceleration at non-relativistic shocks, applicable to galactic cosmic rays. When trying to explain extragalactic cosmic rays at ultra-high energies, we need more powerful accelerators. The obvious choice is to consider relativistic shocks, where the energy gain (proportional to the shock speed $v$) gets higher. In this case however, many of the assumptions in the derivation are no longer valid. One of the more important changes, however, is that for relativistic shocks, the assumption of isotropy is no longer valid, due to relativistic beaming [407, 408]. Discussing the specifics is beyond the scope of this work. For a review of shock acceleration in relativistic outflows, see e.g. [409]. The most important result, is that acceleration is still possible (see e.g. [410]), although different works find different spectral indices: some still find spectral indices around 2 or steeper, while others find also much flatter spectra (see e.g. [411, 412]).

There are of course still alternatives for the acceleration of UHECRs, such as plasma wakefield acceleration [413] and reconnection [414]. These typically predict spectral indices between 1 and 2 or steeper. More exotic scenarios are also possible, even predicting inverted spectra [415–418]. In conclusion, there is a wide variety of acceleration mechanisms, either the Fermi mechanism and its variations or completely different ones, all of which can lead to the "standard" spectral index of 2 as well as much flatter or steeper spectra. As we will see in Section 3.2.6, many of these can explain the observed cosmic ray spectrum when superimposing the spectra of different sources.

### 3.2.3 The Hillas criterion

In reality, the particle acceleration process can not go on indefinitely. For the case of supernovae in the galaxy, for example, the maximum energy can be found by combining the expected acceleration rate with the lifetime of the supernova blast wave. The maximum energy depends on the value of the magnetic field strength, so that estimates of this maximum energy have evolved over time. Currently, it is believed that supernovae are responsible for the cosmic rays up to the knee (see e.g. [382]).

Finding the sources of the UHECRs is more tricky. The maximum energy depends on many parameters, such as the accelerator lifetime, the source environment and possible energy losses during the acceleration, which are uncertain and can vary significantly between models. However, an estimate for the absolute maximum energy is given by the Hillas criterium [419]. For the case of relativistic shocks, it expresses that in order to be accelerated, the gyroradius of the accelerated particle needs to be smaller than the size of the acceleration region, i.e. the cosmic rays need to be confined in the acceleration region in order to be accelerated. This leads to a maximum energy

$$E_{\max} = qBR,\tag{3.10}$$

with $R$ the size of the acceleration region[20], $B$ the average magnetic field strength and $q$ the charge of the accelerated particle. Therefore, heavier nuclei, which have a higher charge, can be accelerated to higher energy by the same source. Again, it should be emphasized that the true maximum energy can be due to either the criterion above, by the lifetime of the source (if the acceleration is slow) or by energy losses (due to magnetic fields or interactions with matter and radiation fields), which depend on the exact source environment. When applying Eq. (3.10) to various extragalactic objects, using their estimated sizes and magnetic field strength, one can find candidate sources for the acceleration of UHECRs. This criterion can be visualised in the, now famous, Hillas plot, shown in Figure 3.7. It shows the extragalactic objects which are potentially capable of accelerating cosmic rays up to $10^{21}$ eV.

### 3.2.4 UHECR sources

In this section, we will give an overview of some of the most important extragalactic objects which can potentially supply the observed UHECR flux. The goal is not to provide a detailed description of how these objects can accelerate the cosmic rays. Rather, we will sketch the basic properties of these sources, such as their components and luminosity, which are needed to gain an intuitive understanding of why they are candidates for UHECR acceleration. For a more in-depth overview of these sources, see e.g. [367, 368, 382]. A common property of all the systems considered here is that they are powered by gravity. They possess an outflow of matter in which shocks occur, allowing for the acceleration of cosmic rays. The kinetic energy of this outflow is sourced by the potential energy which is released by matter falling into a central gravity well. Therefore, while magnetic fields are responsible for the acceleration of cosmic rays, it is ultimately gravity which is the source of their energy.

---

[20]Which can be directly related to the maximum gyroradius of the accelerated particles

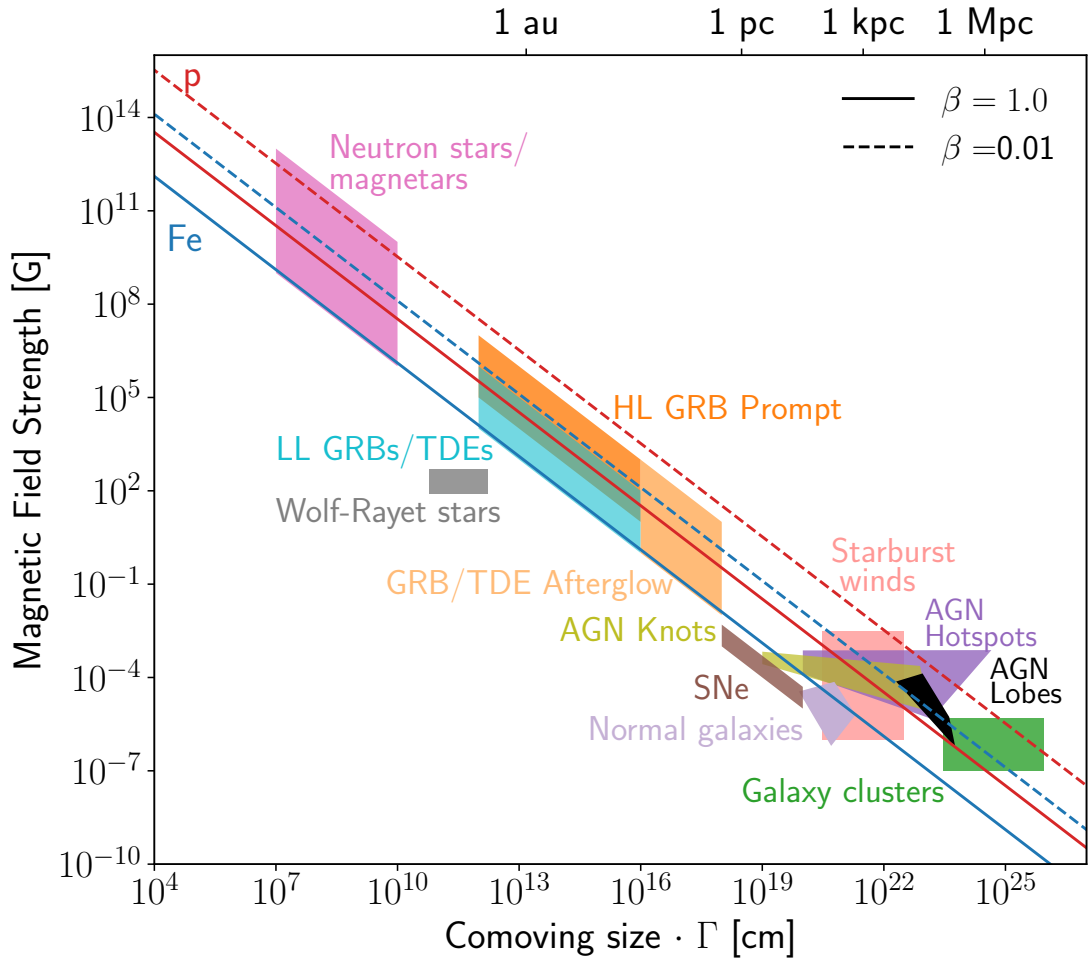

Figure 3.7: Updated version of the original Hillas plot, from [420], indicating different source classes as a function of source radius $R$ and magnetic field (in the comoving frame) $B$. Solid lines indicate the values of the product $B \cdot R$ where proton (red) and iron (blue) nuclei can be contained up to energies of $10^{21}$ eV for two values of the outflow velocity $\beta$. References for the inferred values of $B$ and $R$ for the different source classes can be found in Figure 10 of [420]. Credit: F. Oikonomou and K. Murase.

**Active galactic nuclei**

Active galactic nuclei (AGN) are compact cores of galaxies with a total luminosity exceeding that of their host galaxy. These objects can emit radiation over the entire electromagnetic spectrum, from radio waves up to gamma rays. Their emission is incompatible with a stellar origin and, due to their high luminosity and compactness, they are believed to be powered by supermassive black holes accreting matter. AGNs have a total luminosity of $L_{AGN} \sim 10^{44-47}$ erg s$^{-1}$ over a lifetime of about $10^7$ years, which gives an estimated time-integrated luminosity of $L_{AGN} \sim 10^{62}$ erg [382], which is sufficient to power the observed UHECR flux[21].

There is a great variety in the observational properties of AGNs in different galaxies. Throughout the years, using many observations in all parts of the electromagnetic spectrum, the following picture has emerged (sketched in Figure 3.8a) [421]. At the centre of an AGN, there is a supermassive black hole. This black hole accretes matter, forming an accretion disk (with a radius up to 1000 AU), which radiates the released potential energy by emitting in the optical, the UV and X-rays (see e.g. the review in [422]). Surrounding the compact core is a region of gas, divided in a broad line region (up to 1 pc from the central black hole) and, along the polar direction, a narrow line region (up to $10^4$ pc from the centre), which is excited by the accretion disk and is characterised by (forbidden) emission lines. Farther out from the accretion disk, along the equatorial plane and beyond the sublimation radius, is a dusty torus (at the parsec scale), which can obscure the core of the AGN and re-emits the energy it absorbs back in the IR (see also the review [423]). In about 10% of the AGNs, there is also a prominent radio jet (reaching up to several 100 kpc), which can emit in the X-rays and gamma-rays. In radio-weak AGNs, without a jet, the emission is typically dominated by the accretion flow, whereas for radio-loud AGNs, with a jet, it is dominated by the strong non-thermal emission from this jet.

The bulk of the observed variety in AGNs can then be explained by their radio-loudness, their total power and the viewing angle, which determines which components are observable, as indicated in Figure 3.8a. The most important AGN subclasses for the work here are the radio-loud AGNs, all connected in the unified model [424, 425]. Radio galaxies are radio-loud AGNs viewed off-axis and feature 2 lobes powered by narrow jets. They are divided in the FR-I galaxies[22], which are overall weaker but bright at the core, and the FR-II galaxies, which are more powerful and feature luminous lobes. More recently, another class has been identified, the FR-0 galaxies [426]. They possess radio-loud emission and features similar to FR-I, but have no extended radio emission. The radio emission in radio galaxies is believed to be generated by synchrotron radiation off relativistic electrons, implying that particle acceleration takes place. Particle acceleration can happen at shocks alongside the jet, especially at so-called hotspots. These same shocks are capable of accelerating cosmic rays, up to energies of $10^{21}$ eV [427]. Note that while relativistic electrons are required to explain the AGN emission, hadronic particles are not. Even though it seems logical to assume that e.g. protons are also accelerated, jet physics is not well enough understood to make this claim. For example, it might be that protons and nuclei are never efficiently injected into the acceleration process.

---

[21]To be more precise, we need to compare the injected luminosity with the density of AGNs in order to obtain the energy density of UHECRs, which can be converted to a flux.

[22]Named after Fanaroff and Riley, who performed the classification.

When viewed on-axis, radio-loud AGNs are seen as blazars [425]. The relativistic jet emission, along with accelerated particles, are beamed towards the Earth, with a typical Lorentz gamma factor of $\Gamma \sim 10$. The spectrum[23] is completely dominated by the jet emission and features two bumps (Figure 3.8b). The first bump is attributed to synchrotron emission from an accelerated population of electrons. The electrons can also create high-energy gamma rays by up-scattering photons in inverse-Compton scattering, giving rise to the second bump. The up-scattered photons can be either sourced by the synchrotron emission itself in synchrotron self-Compton (SSC) models or by external fields in external Compton (EC) models. An alternative explanation for the origin of the second bump is through hadronic processes (synchrotron emission off protons or gamma rays from pion decay, see Section 3.3.2).

Based on the presence of emission lines, blazars are classified as either flat-spectrum radio quasars (FSRQ, with emission lines) or Bl Lac objects (without emission lines). In addition, there is also a classification in luminosity, called the blazar sequence [428, 429], shown in Figure 3.8b. More luminous blazars (typically FSRQ) are peaked at lower energy ('low synchrotron peaked'), whereas less luminous blazars are peaked at higher energy ('high synchrotron peaked'), i.e. at X-rays and gamma-rays. This also ties into the unification model, where FR-I galaxies are the parent population of Bl Lacs, while the more luminous FR-II galaxies are the parent population of FSRQ. Due to the beamed emission, there sources are especially important for the emission of gamma rays and neutrinos, which are not bent by magnetic fields (Section 3.7). For a more in-depth review of AGNs, see e.g. [421] for an astronomer's point of view and [382] in the context of neutrino astronomy.

**Gamma-ray bursts**

Gamma-ray bursts (GRBs) are short, but extremely luminous, events that outshine any object in the gamma-ray sky while they are active. They typically possess a luminosity of $L_{GRB} \sim 10^{51}$ erg s$^{-1}$ [382] and last from milliseconds up to hundreds of seconds. As with AGNs, the GRB population possesses sufficient power to supply the UHECR flux. GRB can be subdivided into two populations depending on their duration: long GRBs last more than 2 s, while short GRBs last less. These two classes are believed to have different progenitors [430]. Long GRBs are thought to be powered by the collapse of a massive star to a black hole [431], while short GRBs are believed to be powered by the merger of compact massive objects [432, 433]. With the recent observation of a neutron star merger in gravitational waves along with a short GRB 1.7 s after the merger [434–436], this is confirmed to be true for at least part of the short GRBs. This can also explain the different redshift distribution of these classes, where short GRBs occur mainly at small redshift since binaries need to be formed while massive stars appear and collapse mainly in periods of high star-formation at larger redshift.

---

[23]More specifically, we mean the spectral energy density, or SED, which is denoted as $\nu F_\nu$ which has the units erg cm$^{-2}$ s$^{-1}$ Hz$^{-1}$, although sometimes $F_\nu$ is also called spectral energy density.

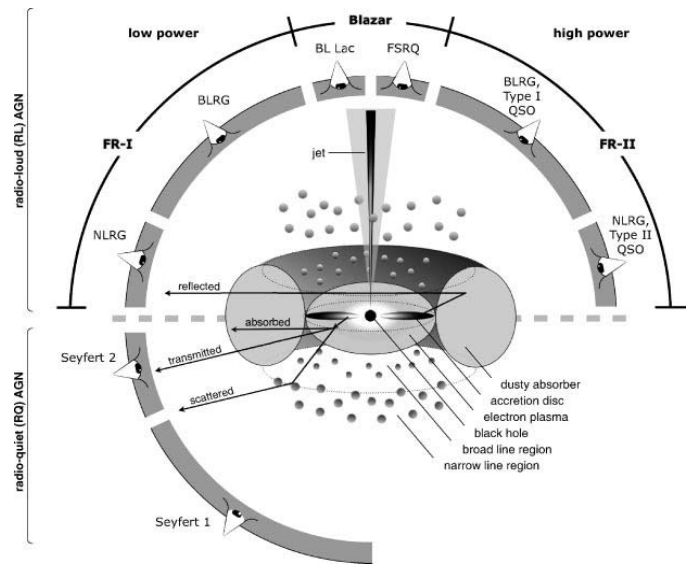

(a) AGN unification model

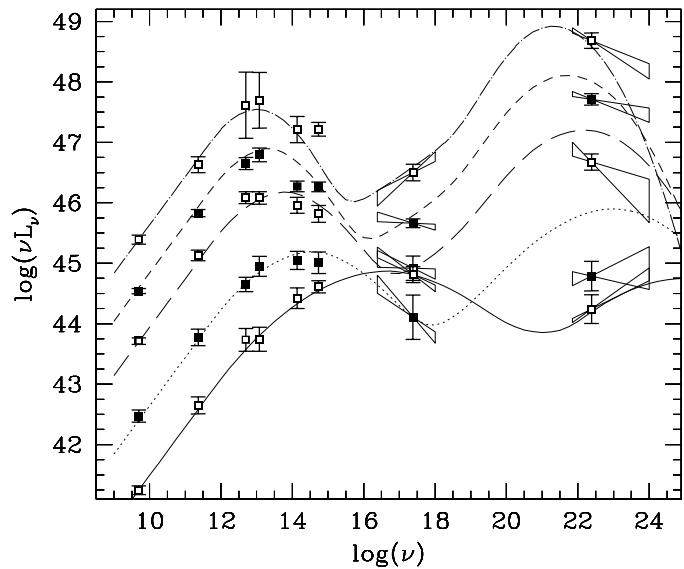

(b) Blazar sequence

Figure 3.8: Active galactic nuclei. (a) Unified model of AGNs, showing the components making up an AGN and how the radio emission, total power and viewing angle can explain the variety between observed AGNs. Figure from [421], graphic by Marie-Luise Menzel. (b) Blazar spectra are described by the blazar sequence [428], characterised by two bumps at an energy which is directly related to the total luminosity. Most high luminosity blazars are FSRQ, while lowering the luminosity results in blazars peaked at higher energy.

The most prominent model to explain the GRB emission is the fireball model [437]. In this model, a jet is formed (similar to AGNs, now with a Lorentz boost $\Gamma \sim 100 - 1000$) in which the central engine injects matter shells at different, relativistic, speeds. When these shells collide, they form shocks which can funnel the macroscopic energy into relativistic particles, as shown in Figure 3.9. This mechanism can explain the acceleration and subsequent emission of high-energy gamma-rays. The same process might also accelerate UHECRs [438, 439], which has been confirmed in detailed simulation of UHECR acceleration (see e.g. [440]).

Finally, besides the prompt emission from GRBs discussed above, they can also feature precursor and afterglow emission at lower energies. Typically the prompt emission phase is considered the most important for the acceleration of particles, although shocks (and thus particle acceleration) also occur during the precursor or afterglow [441] phase[24].

**Other candidates**

There are still other candidate UHECR sources, the more important ones of which we discuss here. While pulsars have been considered as particle acceleration sites since their discovery [444], they have historically mainly been discussed in the context of galactic very-high energy cosmic ray sources (if they accelerate hadrons at all). However, more recently, young pulsars and magnetars (highly magnetised neutron stars) have also been considered as UHECR sources, as first suggested in [416, 445]; see also the discussions in [446–449]. Another possible source class is structure formation shocks, which can accelerate particles to high energies due to their enormous size, over long times ($\sim$ age of the universe) [450]. Finally, also tidal disruption events, where a star is torn apart when accreted onto a supermassive black hole, have recently gained more attention [451–453].

### 3.2.5 The end of the UHECR spectrum

One important ingredient to understand the origin of UHECRs, is to understand the end of the cosmic-ray spectrum. From the measured spectrum, it can be inferred that the end of the spectrum is not an observational effect (i.e. experiments running out of statistics), but really a maximum energy of the cosmic-ray flux. There are various possible reasons for why the cosmic ray spectrum could end.

---

[24]Which can be relevant for neutrino production, see e.g. [442, 443]

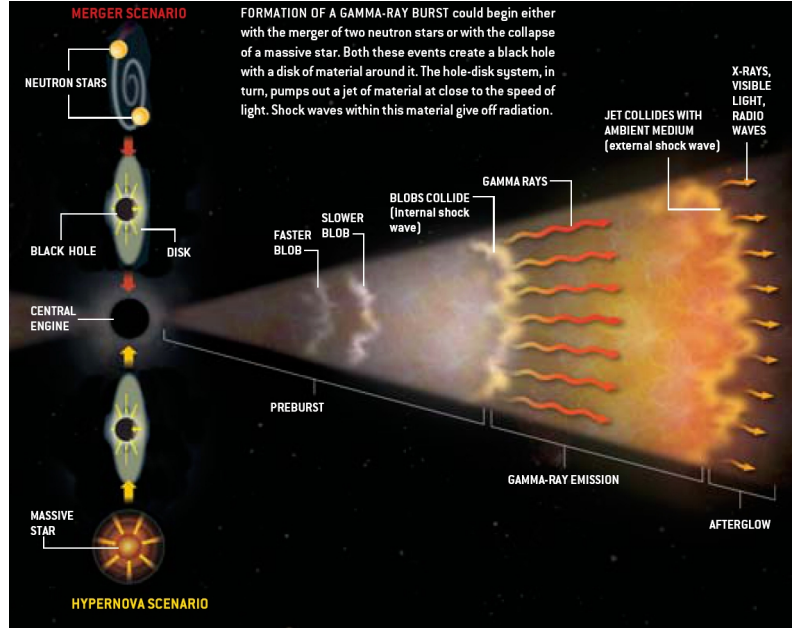

Figure 3.9: Current picture of GRB emission. The merger of two neutron stars or the collapse of a massive stars triggers a relativistic outflow. Colliding shells within this outflow cause particle acceleration and gamma-ray emission. Credit: NASA.

The maximum energy of an accelerator (ignoring losses) is proportional to the charge of the accelerated particle, according to Eq. (3.10). Heavier elements can thus be accelerated up to higher energies than lighter ones[25]. Therefore, one expects the composition to get heavier near the end of the cosmic-ray spectrum, although light elements can still be important if the nuclei get broken up by interactions with a strong radiation field (see e.g. [454]). As seen in Section 3.2.1, the composition indeed gets heavier, lending support to the idea that we are observing the end of an accelerated population[26].

Historically, there was an alternative explanation for the appearance of the ankle and the end of the UHECR spectrum, called the proton dip model. If the UHECRs were to be pure proton, the ankle is naturally explained by a dip due to Bethe-Heitler pair production [457] of the protons interacting with the cosmic microwave background (CMB) [458, 459],

$$p + \gamma_{\mathrm{CMB}} \rightarrow p + e^+ + e^-, \tag{3.11}$$

where the largest contribution is through the $\Delta^+$-resonance. The threshold for this reaction is at $E_p^{\mathrm{th}} \approx 6 \times 10^{17}$ eV. It has a high interaction cross section ($\sigma_{\mathrm{BH}} \approx 10$ mb), but in each interaction the protons only lose a small fraction of their energy. At even higher energies, protons can interact photohadronically with the photons of the CMB, with $\sigma_{p\gamma} \approx 120\ \mu$b at high energy, producing pions

$$p + \gamma_{\mathrm{CMB}} \rightarrow \binom{p}{n} + \pi^{0/+}. \tag{3.12}$$

---

[25]On the other hand, nuclei suffer additional energy loss processes due to their composite nature, so the interplay is non-trivial.

[26]Similarly, the second knee could be explained by different maximum energies for different elements [455], although this is challenged. Instead, the second knee could be due to a re-accelerated, second population of galactic cosmic rays [456].

The threshold for this process is $E_p^{\mathrm{th}} \approx 7 \times 10^{19}$ eV. In this interaction, the protons lose a large amount of energy. Therefore, pion production on the CMB makes the universe opaque to protons (and nuclei) of very high energy, limiting their horizon to $d \lesssim 100$ Mpc or $z \lesssim 0.03$. This is called the GZK-effect, after the discoverers [460, 461]. A more detailed overview of these processes and more can be found in [462].

In recent years, the pure proton model has become disfavoured, due to cosmic-ray composition measurements [463–466] and high-energy neutrino constraints [467, 468]. On the other hand, while the heavier composition suggests that we are observing the end of an accelerated population[27], the GZK-effect could still contribute. As we will see in Section 3.3.2, neutrinos from the decay of pions produced in this interaction can confirm or rule out the relevance of the GZK-effect.

### 3.2.6 Explaining the UHECR spectrum

Since we focus on extragalactic sources in this thesis, we are mostly concerned with the origin of the UHERCs. Explaining their energy spectrum and composition requires combining all the elements previously discussed. One immediate problem is the spectral index. The Fermi mechanism typically predicts a spectral index of 2 or lower for relativistic shocks. On the other hand, the observed spectral index for UHECRs is 2.7. A similar problem exists for the galactic cosmic rays, but as already mentioned, the problem there is solved by the diffusion, which brings in an extra power of 0.7 [367]. For the UHECRs, this solution no longer works, since their propagation can not be described by a diffusion process due to their energy. Instead, the spectral index of the total spectrum can differ from 2 by superimposing the energy spectra of different elements.

There exist many fits to the observed spectrum from the ankle onwards [381, 469–471], varying the spectral index, the cut-off and the relative importance of the different elements[28]. In these models, the best-fit spectral index at the source, per element, can be significantly different from the canonical value of 2, e.g. 1 in [375, 469, 470], 1.6 in [471] or 1.8 in [381]. As explained in the previous section, all of these could be explained by variations of the Fermi mechanism at relativistic shocks or other acceleration mechanisms. As an example of how such flat spectra can lead to a steep total spectrum, consider the model in [381]. There, they investigate radio galaxies as the source of UHECRs.

---

[27]In practice, photohadronic interactions of accelerated cosmic rays with radiation fields at the source can cause the heaviest nuclei to break up, also limiting the maximum energy of emitted nuclei.

[28]The propagation of these source spectra to Earth is simulated using dedicated codes, most importantly `CRPropa` [472] and `SimProp` [473].

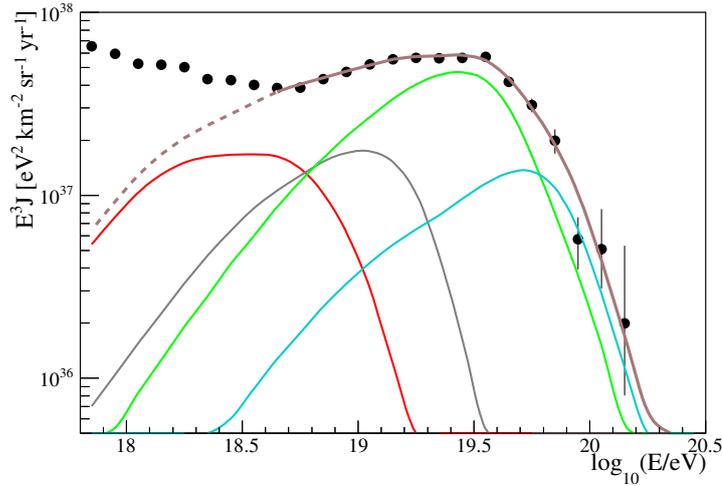

Figure 3.10: Simulated energy spectrum of UHECRs for the best fit to data, as performed by Auger [375]. The spectral index of the best-fit is around 1 and the composition is mixed, dominated by He/N/Si and no H or Fe at the sources. Nuclei are grouped according to their mass: $A = 1$ (red), $2 \leq A \leq 4$ (gray), $5 \leq A \leq 22$ (green), $23 \leq A \leq 38$ (cyan), total (brown).

Using the average radio luminosities, they derive an analytical expression for the total cosmic-ray spectrum, where the changing spectral index is explicitly visible.[29] A possibility to reconcile the *standard* Fermi mechanism with the observed cosmic-ray spectrum, is by using sources with varying maximum energy, as in [474, 475]. By integrating over the contributions of different sources with varying cut-off energies, the resulting spectrum can be steeper than the original spectra. Recently, also the Auger Observatory performed a combined fit in energy and composition [375], shown in Figure 3.10. Their model is agnostic about the exact source details and they assume a universal power law from sources which are isotropically distributed. Their best fit point gives a quite flat spectrum, with a power law index $\gamma \approx 1$. For the composition, they find that the composition at the source is dominated by intermediate mass nuclei like He, N and Si, with no H or Fe at the source (although H is produced during propagation).

Another observable which can be used to identify the UHECR sources is the anisotropy. Since the UHECRs have a limited horizon and are not significantly bent at the highest energy, local structures in the UHECR sources are expected to be imprinted in the arrival directions of the cosmic rays. A recent Auger analysis compared the observed anisotropy with skymaps of starburst galaxies, AGNs or combinations of the two [476]. Starburst galaxies are a subset of star-forming galaxies with an extremely high star formation rate. While starburst galaxies themselves are unlikely to be capable of accelerating cosmic rays to the highest energies, their high star-formation rate does imply that extreme events like GRBs, hypernovae and magnetars which can accelerate UHECRs occur more frequently. In the analysis, they found that the skymap with starburst galaxies fits better than isotropy with a significance of $4\sigma$, whereas the other models fit better than isotropy with a significance of 'only' $2.7 - 3.2\sigma$. However, their analysis did not take into account galactic and extragalactic magnetic fields and winds. Telescope array performed a similar analysis, but is not able to discriminate between the isotropic and starburst galaxies hypotheses [477].

---

[29]In fact, while each element at the source has a spectral index of only 1.8, the resulting spectrum is too steep for that model. By assuming the flux is dominated by Cygnus A and Centaurus A, however, they are able to correctly fit the UHECR spectrum.

### 3.2.7 UHECR luminosity density

In order to explain the observed UHECR flux, a source class needs to satisfy two criteria: it needs to be able to accelerate particles to the required energy and it needs to be able to supply the observed luminosity density in UHECRs $\sim 5 \times 10^{44}$ erg Mpc$^{-3}$ yr$^{-1}$ [478]. In order to check the second criterion, one needs an estimate of the potential UHECR output and the density of these sources. In fact, all of the sources on the Hillas plot feature strong electromagnetic output, such as in radio waves or gamma rays. It is then assumed that this power is also representative for their potential UHECR output. This is the simplest application of the multimessenger paradigm, where information from different messengers is combined to learn more about their source (see Section 3.3.1), although it does not yet fully exploit this connection. As it turns out, it seems that sources satisfying the first criterion also typically satisfy the second [479].

A more recent estimate of the total energy budget of several sources was made in [410], with their result shown in Figure 3.11. This figure shows the cumulative, volume-averaged luminosity density of various galaxies observed in GeV gamma rays by Fermi (see Section 3.6.2) as a function of redshift. Also indicated is the GZK radius up to which sources can be responsible for UHECRs that reach Earth. In order to supply the UHECR flux, a luminosity density of $\sim 10^{44}$ erg Mpc$^{-3}$ yr$^{-1}$ is needed [478, 480] (see also Section 3.3.5). From these results, it follows that Bl Lacs are still allowed, while FSRQs are too far away and too rare[30] to be responsible for the observed flux. A similar conclusion is true for their parent population, with FR-II galaxies disfavoured and FR-I galaxies still allowed[31]. Also starburst galaxies are allowed, although the steady emission used here is unlikely to be associated to an accelerator which can reach UHECR energies [410].

## 3.3 Neutrinos and the multimessenger connection

We will now leave behind the topic of cosmic-ray physics and instead turn our attention to the field of high-energy neutrino astrophysics and its connection to gamma rays and gravitational waves. We are interested in neutrinos from the most extreme events in the universe, which occur outside our galaxy, and are associated to the UHECRs, but have a lower energy[32].

---

[30]From the figure it might seem they are just too far away. However, the number density of FSRQ is lower than BL Lacs (see e.g. Section 3.8.3) and the number of detected FSRQ by Fermi is also lower [481, 482]

[31]Note that since these galaxies have misaligned jets, their total gamma-ray luminosity is expected to be much higher than what we can observe.

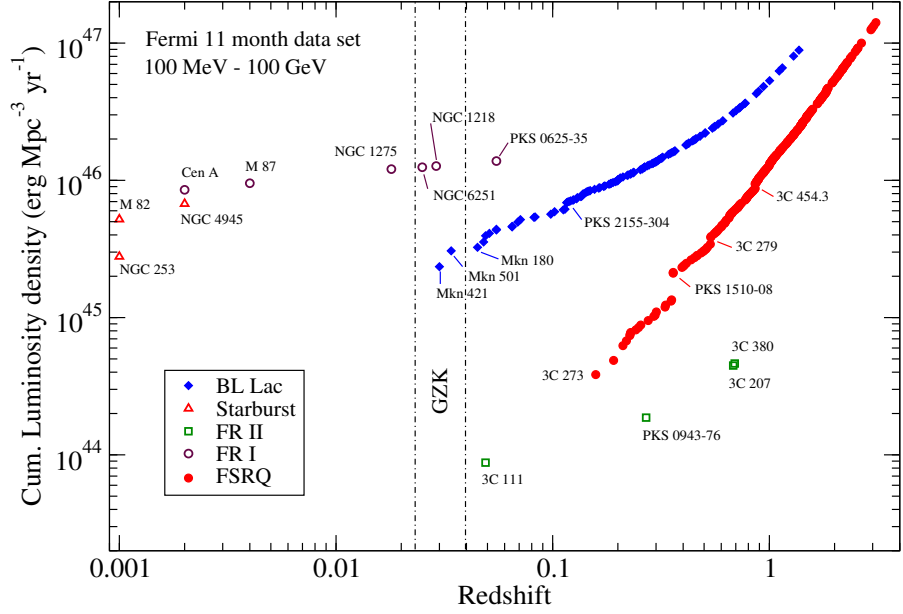

Figure 3.11: Cumulative non-thermal luminosity density of galaxies detected in GeV gamma rays from Fermi measurements, showing FSRQ, Bl Lac, FR-I and FR-II radio galaxies and star-forming galaxies. The GZK horizon for UHECRs with $E \gtrsim 10^{21}$ eV is indicated by the vertical band. Figure from [410].

### 3.3.1   Going beyond cosmic rays

UHECRs provide us with an opportunity to study the most extreme events in the universe, as well as interactions at the highest energy. One of the most important questions remains unresolved however: what is the origin of the UHECRs? Since they are charged, cosmic rays are bent by the intergalactic magnetic fields and, when observed, do not point back to their source. Therefore, it is difficult to resolve this question using cosmic rays alone[33]. However, protons and nuclei can undergo interactions with matter and radiation fields either at the source or on their way to Earth (for a review, see [483]). These interactions create many secondary, unstable particles such as muons, pions and kaons which eventually decay. We are interested in the neutrinos and gamma rays that are created in these decays.

Neutrinos and gamma rays share the property that they are electrically neutral.

---

[32]This is an important point which will be repeated in Section 3.6.4. While the neutrinos to which we are currently sensitive are associated to the UHECRs, due to their extragalactic origin, they have a lower energy. This makes a direct connection between e.g. the UHECR composition observed by Auger and the cosmic rays directly responsible for the neutrino flux more difficult.

[33]As already mentioned, however, anisotropy studies can be used to try and reveal their origin [383, 476].

Therefore, they both point back to their point of origin. When enough of these particles are detected (above background), it is possible to resolve and identify the source responsible for their production. However, gamma rays can interact electromagnetically with matter and radiation fields at the source itself or on their way to Earth, attenuating the gamma-ray flux that can be detected. On the other hand, neutrinos only interact weakly and are therefore not hindered on their way to Earth. The universe is transparent to neutrinos. Therefore, neutrinos can reach us from any source in the observable universe, allowing us to probe the cosmological history of these sources and the physical processes within them.

There is a major complication when connecting the information received in photons and neutrinos. Whereas neutrinos can only be created through weak decays, which requires the creation of unstable particles through hadronic interactions[34], high-energy photons can also be created through purely electromagnetic processes. As already mentioned in Section 3.2.4, in the presence of accelerated electrons, low-energy photons can be up-scattered to high energy by inverse-Compton scattering [367]. Using only leptonic processes, it is possible to explain the gamma-ray spectra of supernovae remnants and some blazars and blazar flares. On the other hand, it is also often possible to fit the spectra of these objects with a lepto-hadronic model or to include a subdominant hadronic component. Therefore, neutrinos are important in order to discriminate between the two possibilities and identify the sources of UHECRs[35].

More recently, with the first detection of gravitational waves [484], a new detection channel has opened. They probe a completely different part of the sources in which they are generated. While photons and neutrinos are produced in the interactions of accelerated cosmic rays with the environment, gravitational waves probe the inner engines of these sources. Gravitational waves associated to astrophysical sources[36] are created when massive, compact objects merge or deform, allowing the very stiff, but dynamical, spacetime field to fluctuate. Therefore, gravitational waves, if present, carry information that is complementary to that obtained from cosmic rays, neutrinos and gamma rays. Currently available experiments can observe the gravitational waves emitted in the merger of two black holes, a black hole and a neutron star or two neutron stars, the latter two of which are thought to power short gamma-ray bursts (true for at least part of the short GRB population, as evidenced by the recent observation of a short GRB in coincidence with a neutron star merger [435, 436]). All of these are extreme, transient events[37]. Therefore, if the bulk of the UHECRs are associated to steady sources, they might not be associated to gravitational waves at all. For a more details about gravitational waves, see Chapter 5.

---

[34]This is actually not true, since neutrinos can also be created by the weak interactions during a fusion process, such as in the sun or in supernovae or in neutron decay. However, these processes do not typically involve the high energies we are interested here.

[35]This does not necessarily imply a one-to-one relation between neutrinos and UHECRs. For example, when modelling the interaction of cosmic-ray nuclei with radiation fields in blazars, it is found that Bl Lacs can be more efficient cosmic-ray emitters, whereas FSRQs can be more efficient neutrino emitters [454].

[36]As opposed to gravitational waves associated with phase transitions in the early universe or inflation.

[37]Due to the requirement of a time-varying quadrupole for the generation of gravitational waves, all gravitational wave sources are necessarily transient. Of course, the time-scale of this event can be very large, such as in the merger of two supermassive black holes.

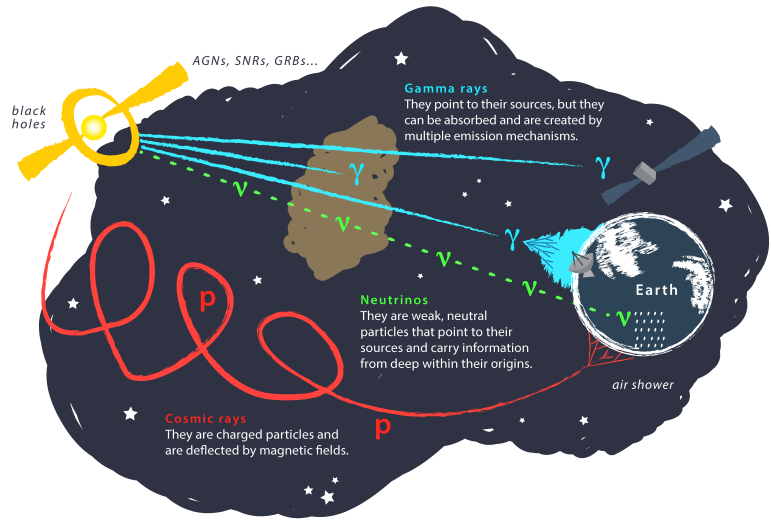

Figure 3.12: Multimessenger astrophysics. A source sends out cosmic rays, gamma rays and neutrinos, which can have a common origin. In addition, also gravitational waves can be emitted (not shown). By combining information from these different messengers, more can be learned about their sources. Credit: Juan Antonio Aguilar and Jamie Yang, IceCube/WIPAC.

By combining the information from cosmic rays, gamma rays, neutrinos and gravitational waves, one can study different aspects of the physics that govern the astrophysical objects responsible for the UHECRs, their environment and how they fit in the cosmological history. This is the core idea of multimessenger astrophysics and is illustrated in Figure 3.12.

### 3.3.2   Neutrino and gamma-ray production

Neutrinos and gamma rays are produced when cosmic rays interact with matter (assumed to be mainly $p$) or radiation fields ($\gamma$). For simplicity, when studying interaction processes of cosmic rays, we will mainly consider interactions of protons instead of nuclei. Historically, this was motivated due to the cosmic ray flux being dominated by protons for most of the energy range. As we have seen in the previous section, however, this is probably not true for the UHECR spectrum. Still, since nuclei undergo the same interactions as protons, plus additional ones such as photo-dissociation, the error will be small if the nuclei are sufficiently light.

In $p\gamma$- and $pp$-interactions, many new particles are produced, such as protons,

neutrons[38], pions, kaons and hyperons. Among the mesons, the pions are by far the most abundant. Charged pions decay almost exclusively to muons, which again decay into electrons, both accompanied by neutrinos

$$\pi^+ \to \mu^+ \nu_\mu \to e^+ \nu_e \bar{\nu}_\mu \nu_\mu, \tag{3.13}$$

$$\pi^- \to \mu^- \bar{\nu}_\mu \to e^- \bar{\nu}_e \nu_\mu \bar{\nu}_\mu. \tag{3.14}$$

In these decays, 1 pair of $\nu_\mu \bar{\nu}_\mu$, 1 $\overset{(-)}{\nu}_e$ and 1 $e^\pm$ are produced, all with roughly equal energy [480]. The electrons are assumed to further interact electromagnetically in the environment and are not of interest for the neutrino production. Neutral pions instead decay into photons,

$$\pi^0 \to \gamma\gamma, \tag{3.15}$$

through the chiral anomaly. Because of their origin in these decays, the neutrino and gamma-ray energy and flux are intimately related, through the number of neutral and charged pions that are produced in a typical interaction, which differs between $p\gamma$- and $pp$-interactions. The interaction of cosmic-ray protons with an energy $E_p$ leads to typical neutrino and photon energies of $E_\nu \simeq E_\gamma/2 \simeq E_p/20$ [487].

Interactions of cosmic rays can occur at various points between their production at the source and their observation on Earth. At the source, cosmic rays can interact with the intense radiation fields or gas associated to the astrophysical object that is responsible for the acceleration of the cosmic rays, giving rise to an *astrophysical neutrino flux*. On their way, they can interact with interstellar gas[39] as in the spallation reactions discussed in Section 3.2, or with the CMB, giving rise to the GZK-effect already mentioned in Section 3.2.5. In the latter case, the associated *cosmogenic neutrino flux* can discriminate scenarios where the end of the UHECR spectrum is due to the GZK-effect from scenarios where it is due to the end of an accelerated population. Finally, cosmic rays will interact with the nuclei in Earth's atmosphere, producing the air showers that allow for their detection and giving rise to an *atmospheric neutrino flux*. We are mainly interested in astrophysical neutrinos, although most of the discussion in this section is general.

First, we will discuss some details of neutrino and gamma-ray production in $p\gamma$- and $pp$-interactions. Their cross sections are shown in Fig. 3.13.

### $p\gamma$-interactions

High-energy protons can interact with radiation fields. The total $p\gamma$-cross section is shown in Figure 3.13a. The cross section can be approximated by a two-step function, with a cross section $\sigma_{p\gamma} \approx 340 \ \mu b$ near the threshold and a constant $\sigma_{p\gamma} \approx 120 \ \mu b$ at higher energies [490]. At threshold, this interaction is dominated by the $\Delta^+$-resonance, leading to pion production through the following channels

$$p\gamma \to \Delta^+ \to \begin{cases} p\pi^0 & \text{(fraction 2/3)} \\ n\pi^+ & \text{(fraction 1/3),} \end{cases} \tag{3.16}$$

where the relative contribution of the decay products is fixed by isospin conservation. The threshold for this process can be easily calculated using special relativity and is

$$E_p \cdot \epsilon_\gamma = \frac{m_\Delta^2 - m_p^2}{4} \frac{\Gamma}{1+z}, \tag{3.17}$$

where $E_p$ is the proton energy for an observer at Earth, $\epsilon_\gamma$ is a representative photon energy of the radiation field, $\Gamma$ is the bulk Lorentz factor of the production region and $z$ is the redshift of the source. The existence of this threshold has two important implications. First, since there is a lower limit on the energy of the interacting protons, there is also a lower limit on the energy of the neutrinos resulting from this interaction. When including a distribution in photon energy and higher order corrections to the cross section, this feature is only slightly washed out. Secondly, since the interaction is dominated by the resonance channel, higher energy protons will preferably interact with lower energy photons. This can lead to a neutrino spectrum with an inverted energy dependence compared to the photons, as is predicted for gamma-ray bursts [382].

---

[38]Neutrons can also give rise to a neutrino flux in their decay, see e.g. [485], but their energy is typically too low [486].

[39]Although this is only important for cosmic rays below the knee, which are confined and can therefore pass large integrated column densities of matter during their time in the galaxy. For UHECRs, this effect is not important.

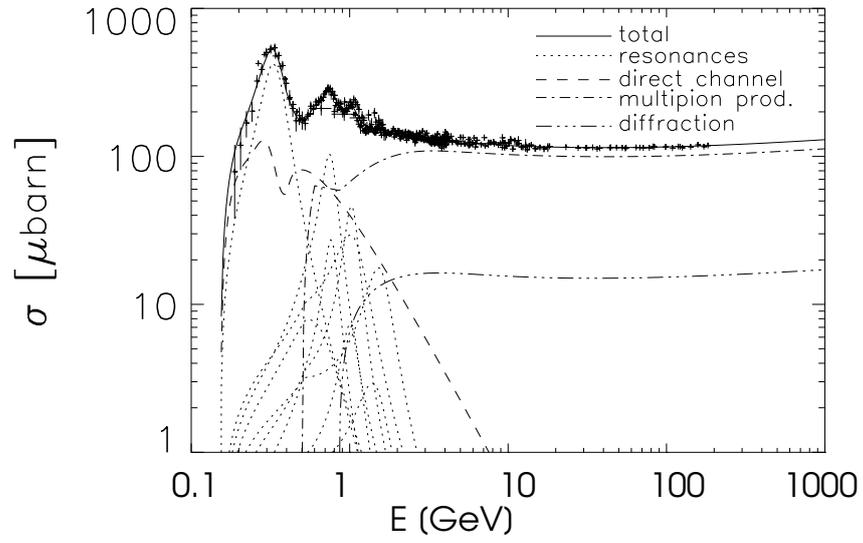

(a) $p\gamma$-cross section

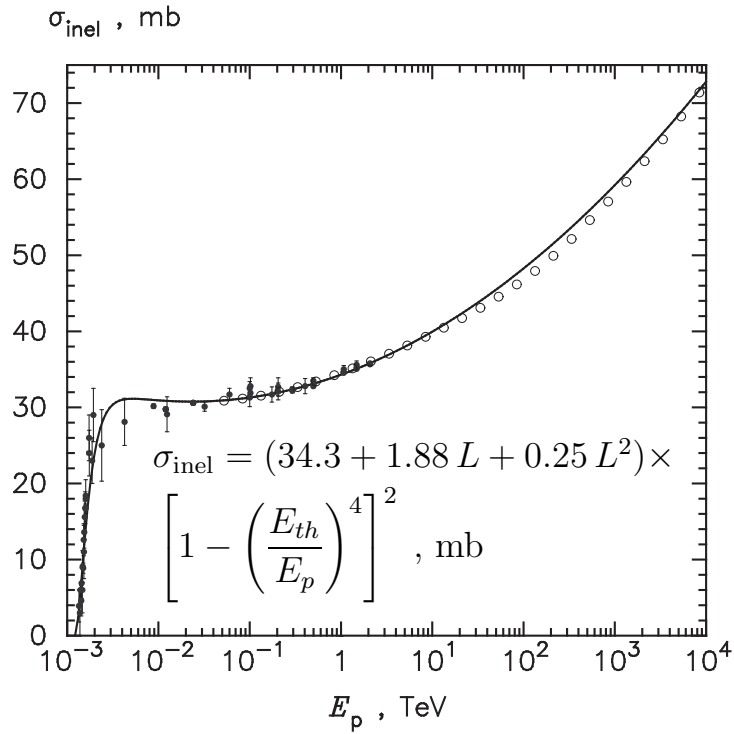

(b) $pp$-cross section

Figure 3.13: Interaction cross sections of protons, showing both experimental measurements and theoretical models. (a) Total $p\gamma$-cross section as a function of the photon energy in the rest frame of the nucleon [488]. (b) Inelastic $pp$-cross section as a function of the proton energy in a fixed target frame [489].

When modelling $p\gamma$-interactions, one often considers only the resonance channel, dubbed the $\Delta^+$-resonance approximation. While considering only this channel is often a good approximation, it does not describe the full interaction. In addition to the $\Delta^+$ resonance, the total photohadronic interaction process also includes higher order resonances, multi-pion and direct pion production (t-channel with pion exchange instead of s-channel) [491]. It is not possible to obtain the resulting particle spectra from first principles, since this would require calculating non-perturbative QCD. Instead, one uses phenomenological models, based on QCD, that are tuned to reproduce the observed physics. The Monte Carlo code `SOPHIA` [488] represents the current state-of-the art implementation of photohadronic interaction processes and its results are shown in Figure 3.13a. There exist also alternatives to this code, such as simplified models of photohadronic interactions [491] or analytical fits to the particle spectra produced by `SOPHIA` [492].

The result of taking into account the different interaction channels is that the relative number of $\pi^+$, $\pi^-$ and $\pi^0$ are different from those expected in the $\Delta^+$-resonance approximation. For example, whereas the $N_{\pi^+}/N_{\pi^0} = 1/2$ in the $\Delta^+$-resonance approximation, this ratio lies between $1/1$ and $3/2$ in the model of [491], where they find a minimum for the charged-to-neutral pion ratio

$$K_\pi \equiv \frac{N_{\pi^\pm}}{N_{\pi^0}} \sim 1.2 \tag{3.18}$$

as a function of $E_p \cdot \epsilon_\gamma$ for arbitrary input spectra. On the other hand, while some $\pi^-$ can be produced in higher order processes, the charged pions are still dominated by $\pi^+$ [491].

The prediction of the (relative) number of $\pi^+$, $\pi^-$ and $\pi^0$ is important, because it determines the relation between the neutrino and gamma-ray flux (through $N_{\pi^\pm}/N_{\pi^0}$), as well the relative flux of $\nu_e$ versus $\bar{\nu}_e$ (through $N_{\pi^+}/N_{\pi^-}$), see Eqs. (3.14) and (3.15). After charged pion decay, which is dominated by $\pi^+$, the neutrino flux has a flavour composition ($\phi_{\nu_e} : \phi_{\nu_\mu} : \phi_{\nu_\tau}$) equal to $(1 : 2 : 0)$ at the source, which is divided as $(1 : 1 : 0)$ and $(0 : 1 : 0)$ between neutrinos and anti-neutrinos respectively. This can be altered in extreme environments however. For example, when muons lose energy before they decay (the so-called damped muon scenario), high-energy neutrinos have a composition $(0 : 1 : 0)$ with no anti-neutrinos [491, 493]. For an overview on flavour ratios in UHECR sources with $p\gamma$-interactions, see [486].

### $pp$-interactions

High-energy protons can also interact with matter. For astrophysical applications, this target matter will consist mostly of protons, whereas for air showers the target consists mainly of nuclei. Since we are most interested in astrophysical applications, we will consider only the former. The cross section for $pp$-interactions, shown in Figure 3.13b for a fixed target frame, starts at $\sigma_{pp} \approx 30$ mb and rises to $\sigma_{pp} \approx 80$ mb at 100 PeV. As for $p\gamma$-interactions, $pp$-interactions will produce nucleons and hyperons, pions, kaons and charmed mesons. Amongst the mesons, the pions are dominant and they can be produced as soon as the centre of mass energy is sufficient for the production of a pion, leading to the reaction

$$pp \rightarrow \begin{cases} pp\pi^0 & \text{(fraction 2/3)} \\ pn\pi^+ & \text{(fraction 1/3)}, \end{cases} \tag{3.19}$$

from isospin conservation. For a fixed target frame collision, this requires a proton beam with a threshold energy of

$$E_p^{\text{th}} = 1.22 \text{ GeV.} \tag{3.20}$$

Since for accelerated protons in an astrophysical source this condition is trivially satisfied, protons of all energies will interact. Consequently, there is also no lower limit on the energy of neutrinos produced in subsequent pion decay, in contrast to those from $p\gamma$-interactions.

While the process above is correct for single pion production, the energy available will allow for many more pions, both charged and neutral, to be produced. In fact, from isospin symmetry, one expects each of these to be produced in equal numbers, such that the charged-to-neutral pion ratio is [483]

$$K_\pi \equiv \frac{N_{\pi^\pm}}{N_{\pi^0}} = 2. \tag{3.21}$$

This is also expected from the production of a pion cloud in thermal equilibrium, as in the original pion production theory by Fermi [494]. Up to an energy of $10^4$ GeV the pion multiplicity can be approximated as [483, 495]

$$N_{\pi^\pm}(E_p) \simeq 2 \left[ (E_p - E_{\text{th}}/\text{GeV}) \right]^{1/4}. \tag{3.22}$$

As is the case in $p\gamma$-interactions, it is not possible to calculate these interactions in full detail from first principles and one has to resort to phenomenological descriptions instead. Such models have been implemented in several numerical codes. For the high energies required for cosmic-ray studies, these codes include `SIBYLL` [496, 497], `QGSJET` [498], `DPMJET` [499] and `EPOS LHC` [500]. As an example of the uncertainty in these models, the charmed production in cosmic-ray air showers has only recently been implemented in e.g. `SIBYLL` [501, 502] and the contribution of the so-called prompt neutrino flux from this component is still uncertain (since it has not been detected yet, see e.g. [503]). As another example, the muon content measured in UHECR air showers has been found to exceed the value predicted in these Monte Carlo generators with several ten percent [504]. The origin of this anomaly is not yet understood, as it could be due to a problem in the hadronic interaction models or due to new physics (see e.g. [505]). Again, there also exist analytical fits to the output spectra predicted by these generators [489].

The total inelastic $pp$-cross section is shown in Figure 3.13b. While its magnitude is larger than the $p\gamma$-cross section, the latter often dominates in astrophysical environments since the photon density in a radiation field can be much larger than the matter density.

The neutrino flavour composition predicted from $pp$-interactions can be found immediately from the pion multiplicities above and is $(\nu_e : \nu_\mu : \nu_\tau) = (1 : 2 : 0)$ for both neutrinos and anti-neutrinos at the source. Again, this can be modified due to e.g. cooling of the muons.

**Interactions of cosmic-ray nuclei**

The end of the UHECR spectrum seems to be dominated by heavier elements. Therefore, when building a complete model, it is important to model their interactions as well. The interaction of nuclei can be approximated with the superposition model, where the nucleons are interacting independently with the target [367]. This leads to a neutrino energy related to the cosmic ray energy as $E_\nu \approx (0.03 - 0.05) E_{\rm cr} / A$, with $A$ the atomic mass number [506, 507]. However, such a description is not perfect (see the comments in [497] for hadronic and [454] for photohadronic interactions). In addition, in $N\gamma$-interactions, the nucleus can be split up in photo-disintegration and this can be a primary energy-loss mechanism. As a consequence, a full nuclear interaction/decay chain can be initiated [454]. Because of the additional energy loss mechanisms (and lower neutrino energy for same cosmic ray energy), neutrino emission models which are dominated by nuclei predict less high-energy neutrinos.

### 3.3.3   Neutrino propagation

The predicted flavour ratio of neutrinos for both $p\gamma$- and $pp$-interactions is $(\nu_e : \nu_\mu : \nu_\tau) \approx (1 : 2 : 0)$ at the source, assuming the intermediate muons do not lose too much energy before decaying. For $pp$-interactions the predicted $\nu$- and $\bar\nu$-ratios are equal, while for $p\gamma$-interactions they are $(1 : 1 : 0)$ and $(0 : 1 : 0)$ respectively. However, the mass eigenstates and flavour eigenstates of neutrinos do not coincide: each flavour state is a superposition of different mass eigenstates. While neutrinos are created and observed in their flavour eigenstate, the propagation occurs according to the mass eigenstate. Since the different mass eigenstates propagate differently, the relative importance of the three mass eigenstates, and therefore also of the three flavour eigenstates, changes along the travelled distance. As a consequence, the neutrinos oscillate between different flavours and the observed flavour composition depends on the distance from the source.

The mixing of different eigenstates is parameterised by the PMNS-matrix [32, 53]. With three neutrino species, the mixing matrix $U$ is defined by three rotations ($\theta_{12}$, $\theta_{13}$ and $\theta_{23}$) and a complex phase $\delta$ which gives $CP$-violation,

$$U = \begin{pmatrix} c_{12}c_{13} & s_{12}c_{13} & s_{13}e^{-i\delta} \\ -s_{12}c_{23} - c_{12}s_{23}s_{13}e^{i\delta} & c_{12}c_{23} - s_{12}s_{23}s_{13}e^{i\delta} & s_{23}c_{13} \\ s_{12}s_{23} - c_{12}c_{23}s_{13}e^{i\delta} & -c_{12}s_{23} - s_{12}c_{23}s_{13}e^{i\delta} & c_{23}c_{13} \end{pmatrix}, \qquad (3.23)$$

with $c_{ij} = \cos(\theta_{ij})$ and $s_{ij} = \sin(\theta_{ij})$. The flavour eigenstates, denoted by $|\nu_\alpha\rangle$ ($\alpha = e, \mu, \tau$), are then related to the mass eigenstates, denoted by $|\nu_j\rangle$ ($j = 1, 2, 3$), as

$$\nu_\alpha = \sum_i U_{\alpha i} \nu_i. \qquad (3.24)$$

The oscillation probability in vacuum[40] can be derived as follows (for a review, see e.g. [508]). At a time $t$ after being produced in an eigenstate $|\nu_\alpha\rangle$ at $t = 0$, the wave function of each flavour can be found by applying the time evolution operator on the mass eigenstates,

$$|\nu(t)\rangle = \sum_{j=1,2,3} U_{\alpha j}^* \exp(-iE_j t) |\nu_j\rangle, \qquad (3.25)$$

where $E_j$ is the eigenvalue of the Hamiltonian operator. The probability for a neutrino $\nu_\alpha$ at the source to oscillate to $\nu_\beta$ on detection is found to be

$$\begin{aligned} P_{\nu_\alpha \to \nu_\beta} = \left| \langle \nu_\beta(t) | \nu_\alpha(t=0) \rangle \right|^2 = {}& \delta_{\alpha\beta} - 4 \sum_{i<j} \mathrm{Re} \left[ U_{\alpha i} U_{\beta i}^* U_{\alpha j}^* U_{\beta j} \right] \sin^2 \left( \frac{\Delta m_{ji}^2 L}{4 E_\nu} \right) \\ & + 2 \sum_{i<j} \mathrm{Im} \left[ U_{\alpha i} U_{\beta i}^* U_{\alpha j}^* U_{\beta j} \right] \sin \left( \frac{\Delta m_{ji}^2 L}{4 E_\nu} \right). \end{aligned} \qquad (3.26)$$

The oscillation probability depends on the mixing matrix, the squared mass differences $\Delta m_{ji}^2 \equiv m_j^2 - m_i^2$ between the neutrino mass eigenstates and a combination of the propagated distance $L$ and the neutrino energy $E_\nu$. As it turns out, the neutrino mass differences are such that each of the mixing angles can be measured from completely different experimental setups (characterised by $\frac{L}{E}$). The angle $\theta_{12}$ can be measured from solar neutrinos ($\nu_e \leftrightarrow \nu_\mu$), $\theta_{23}$ from atmospheric neutrinos ($\nu_\mu \leftrightarrow \nu_\tau$) and $\theta_{13}$ from reactor experiments.

---

[40]When propagating in matter, the coupling of neutrinos to matter alters the mixing and thus the oscillation probability.

Several experiments have determined the mixing angles to be $\theta_{12} \approx \pi/6$, $\theta_{23} \approx \pi/4$, $\theta_{13} \approx 0$. While precision experiments are very sensitive to the exact values of these mixing angles, for the present purposes we can approximate the PMNS-matrix with the tri-bimaximal model [509, 510] following [511, 512],

$$U = \begin{pmatrix} \frac{\sqrt{3}}{2} & \frac{1}{2} & 0 \\ \frac{-1}{2\sqrt{2}} & \frac{\sqrt{3}}{2\sqrt{2}} & \frac{1}{\sqrt{2}} \\ \frac{1}{2\sqrt{2}} & \frac{-\sqrt{3}}{2\sqrt{2}} & \frac{1}{\sqrt{2}} \end{pmatrix},$$ (3.27)

which follows closely the experimental values [513, 514]. We have put $\delta = 0$.

For astrophysical distances, which are much longer than the oscillation length, the oscillation factor in Eq. (3.26), $\sin^2\left(\Delta m_{ij}^2 L / 4 E_\nu\right)$, averages out to $1/2$ [515]. Combining this with the tri-bimaximal mixing matrix, we can find the neutrino flavour ratio at the location of the detector $\phi^f = \mathcal{P}\phi$, with

$$\mathcal{P} = \frac{1}{18} \begin{pmatrix} 10 & 4 & 4 \\ 4 & 7 & 7 \\ 4 & 7 & 7 \end{pmatrix}.$$ (3.28)

For the total neutrino flavour composition $(1:2:0)$ at the source, as predicted from both $p\gamma$- and $pp$-interactions, applying Eq. (3.28) leads to the flavour ratio $(1:1:1)$ at the site of detection. For the case of neutrinos from $p\gamma$-interactions, this is divided as $(14:11:11)$ and $(4:7:7)$ for $\nu$'s and $\bar\nu$'s respectively, whereas for neutrinos from $pp$-interactions the two flavour ratios are equal.

### 3.3.4 A song of flux and energy density

In the following sections, we consider power law fluxes of protons, neutrinos and gamma rays and how these are generated. This requires frequent conversions between particle flux, energy flux and energy density, with similar notation and dimensions. Therefore, we collect here the required quantities in order to elucidate the different notations used.

A power law particle flux follows a distribution with the functional form

$$\Phi(E) \equiv \frac{\mathrm{d}N}{\mathrm{d}E} = A E^{-\alpha},$$ (3.29)

with dimensions[41],

$$[\Phi(E)] = E^{-1} L^{-2} T^{-1},$$ (3.30)

$$[A] = E^{\alpha-1} L^{-2} T^{-1},$$ (3.31)

typically in units of $\mathrm{GeV}^{-1}\,\mathrm{cm}^{-2}\,\mathrm{s}^{-1}$. We will use both the $\Phi(E)$- and $\frac{\mathrm{d}N}{\mathrm{d}E}$-notation, the former for brevity and the latter when necessary for clarity (as in this section). In the

---

[41]I will break the SI-convention and keep energy as a separate dimension for clarity.

special case of an $E^{-2}$-spectrum, we have

$$[A_{\alpha=2}] = EL^{-2}T^{-1}. \tag{3.32}$$

When discussing a diffuse flux, one typically adds an explicit "unit" of $\mathrm{sr}^{-1}$, unless the flux is integrated over solid angle. In order to stay general, we will not include it in this discussion.

The energy flux per logarithmic energy interval can be found by multiplying Eq. (3.29) with $E^2$

$$E^2\Phi(E) = E\frac{\mathrm{d}N}{\mathrm{d}\ln E} = AE^{-\alpha+2}, \tag{3.33}$$

with $\left[E^2\Phi(E)\right] = EL^{-2}T^{-1}$. In the case of an $E^{-2}$-spectrum, as predicted by the Fermi mechanism, the energy flux per logarithmic energy interval becomes constant,

$$E^2\Phi_{\alpha=2}(E) = A_{\alpha=2}. \tag{3.34}$$

Therefore, when an accelerated particle population follows an $E^{-2}$-distribution, the total energy is divided equally over each decade in energy. This is an important distinction from power laws with an index $\alpha \neq 2$, where this is not true, which can lead to big differences e.g. when extrapolating the observed UHECR flux. Since we will often deal with $E^{-2}$-fluxes, we will often show $E^2\frac{\mathrm{d}N}{\mathrm{d}E}$ instead of $\frac{\mathrm{d}N}{\mathrm{d}E}$. In addition to enhancing deviations from a power law, this also results in a measure for the energy budget associated to the source of the flux, as shown by the equations above and the following discussion.

When dividing a diffuse flux by $c/4\pi$, the energy flux in Eq. (3.33) is converted into an energy density

$$\left[\frac{4\pi}{c}E^2\frac{\mathrm{d}N}{\mathrm{d}E}\right] = EL^{-3}. \tag{3.35}$$

Finally, from the energy density, one can find the power requirement of the sources supplying the cosmic rays or neutrinos, also called the energy dependent or differential energy generation rate. For this, we divide the energy density by the characteristic time scale. For extragalactic sources, this is the Hubble time $t_H$, giving[42]

$$E\mathcal{Q}_E \equiv \frac{1}{t_H}\frac{4\pi}{c}E^2\frac{\mathrm{d}N}{\mathrm{d}E}, \tag{3.36}$$

with dimensions

$$[E\mathcal{Q}_E] = EL^{-3}T^{-1}, \tag{3.37}$$

typically in units of $\mathrm{erg\,Mpc^{-3}\,yr^{-1}}$. The quantity $E\mathcal{Q}_E$ denotes the energy dependent generation rate of particles per logarithmic energy interval and can be immediately connected to the estimated non-thermal output of different astrophysical source classes.

---

[42]Note that $E\mathcal{Q}_E$ is equivalent to $E\frac{\mathrm{d}\dot{N}}{\mathrm{d}\ln E}$ in the notation of [480], where there can be confusion between the meaning of $\frac{\mathrm{d}N}{\mathrm{d}E}$ (a flux, i.e. per area) versus $\frac{\mathrm{d}\dot{N}}{\mathrm{d}E}$ (an energy-dependent generation rate, i.e. per volume).

The total energy flux received from a source can be found from Eq. (3.29) by multiplying by and integrating over energy[43]

$$S = \int_{E_{min}}^{E_{max}} E \frac{dN}{dE} \, dE = \begin{cases} \frac{A}{-\alpha+2} \left( E_{max}^{-\alpha+2} - E_{min}^{-\alpha+2} \right) & (\alpha \neq 2) \\ A \ln \left( \frac{E_{max}}{E_{min}} \right) & (\alpha = 2). \end{cases}$$
(3.38)

This quantity is related to the intrinsic luminosity of the source through the luminosity distance (which also takes into account redshift effects),

$$S = \frac{L}{4\pi d_L^2}.$$
(3.39)

Similarly, integrating the differential energy generation rate $\mathcal{Q}_E$ defined in Eq. (3.36) gives the integrated energy generation rate (again typically in $erg \, Mpc^{-3} \, yr^{-1}$)

$$\mathcal{Q}_X = \int_{E_{min}}^{E_{max}} \mathcal{Q}_{E_X} \, dE_X,$$
(3.40)

where $X$ is the particle under consideration. From the two equations above, we can see that for $E^{-2}$-power laws, the total energy generation rate is related to the differential one $E \mathcal{Q}_E$ through the small factor $\ln \left( E_{max}/E_{min} \right)$. For example, for $E_{min} = 1$ GeV and $E_{max} = 10^{11}$ GeV, corresponding to the full range of the observed cosmic-ray spectrum, we have $\ln \left( E_{max}/E_{min} \right) = 27$. Moreover, this total budget can be estimated from the observation of the spectrum at an arbitrary energy (see also below Eq. (3.34)). Conversely, for $\alpha \neq 2$, the analogous factor can be much larger and grows quickly when increasing the index. For example, in the case of $\alpha = 2.2$, the total energy generation rate is related to the differential one as $E_X \mathcal{Q}_{E_X}|_{E=100 \, PeV} \approx \mathcal{Q}_X/200$ [516].

### 3.3.5 Astrophysical neutrino flux and the Waxman-Bahcall upper bound

After cosmic rays are accelerated, they can interact with matter or radiation fields close to the source according to the interactions in Section 3.3.2, creating neutrinos and gamma rays. The cumulative output of all such neutrino sources gives rise to an astrophysical neutrino flux. If the energy-loss fraction of cosmic rays due to these interactions is independent of energy, then the neutrino spectrum will follow the proton spectrum. As we have seen in Section 3.3.2, this is indeed the case: the typical neutrino, gamma-ray and proton energies are related as $E_\nu \simeq E_\gamma/2 \simeq E_p/20$ [487]. A priori it is not clear how high one should expect the astrophysical neutrino flux to be, since the sources of cosmic rays, the acceleration site within a source and the exact environment of potential sources are not or poorly known. However, a simple argument due to Waxman and

---

[43]Conventions for the symbol for flux differ per reference. Here and in the rest of this work, I will use $S$ for flux and $F$ for fluence (which is flux integrated over time). However, the reverse convention is also popular, as well as using $F$ for both.

Bahcall [480] provides us with an upper bound on the diffuse neutrino flux[44] due to extragalactic sources, associated with the UHECRs.

From the observed UHECR flux, between $10^{19}$ eV and $10^{21}$ eV, one can find the energy generation rate of protons in this interval by integrating over energy, dividing out $\frac{c}{4\pi}$ to convert from flux to energy density and dividing by the Hubble time [478], giving

$$\mathcal{Q}_{\mathrm{cr}}^{[10^{19}, 10^{21}]} \sim 5 \times 10^{44} \, \mathrm{erg\, Mpc^{-3}\, yr^{-1}}. \tag{3.41}$$

From this integrated energy generation rate of the observed extragalactic cosmic-ray flux, we can find the differential energy generation rate, assuming that the cosmic-ray spectrum follows an $E^{-2}$-behaviour (which is compatible with observation as discussed in Section 3.2.6). Combining Eq. (3.36) and the analogue of Eq. (3.38), we find

$$E_{\mathrm{CR}} \mathcal{Q}_{E_{\mathrm{CR}}} = \frac{\mathcal{Q}_{\mathrm{cr}}^{[10^{19}, 10^{21}]}}{\ln(10^{21}/10^{19})} \approx 10^{44} \, \mathrm{erg\, Mpc^{-3}\, yr^{-1}}. \tag{3.42}$$

Since for an $E^{-2}$-spectrum this quantity is independent of energy, it is valid for any energy where the extragalactic sources produce a cosmic-ray flux (in particular up to $10^{21}$ eV and down to energies far lower than the ankle). In principle, the extrapolation should be valid down to $E_{\mathrm{cr}} \sim \Gamma A m_p c^2$ with $\Gamma$ the Lorentz factor of the emitting region (see e.g. [438, 518]. For a discussion on the connection with observable quantities, see also Section 4.3.3. As mentioned previously, the energy dependent generation rate of cosmic rays $E_{\mathrm{CR}} \mathcal{Q}_{E_{\mathrm{CR}}}$ can be compared to the estimated non-thermal output of different astrophysical source classes in the universe per year. As already seen in Section 3.2, there are several source classes that can supply this level of non-thermal output and are therefore good candidate sources of UHECRs.

From this energy generation rate, one can obtain an upper limit on the neutrino flux expected from extragalactic objects[45], assuming they accelerate only protons, when the proton path length is not much larger than the mean free path under $p\gamma$- or $pp$-interactions. Assuming the protons lose a fraction $\epsilon < 1$ of their energy due to pion production, the resulting neutrino flux is

$$E_\nu^2 \Phi_\nu(E) \propto \epsilon \xi_z t_{\mathrm{H}} \frac{c}{4\pi} E_{\mathrm{CR}} \mathcal{Q}_{E_{\mathrm{CR}}} \tag{3.43}$$

where the proportionality factor is of order one and contains the charged-to-neutral pion ratio $K_\pi$ (i.e. the fraction of end products that can produce neutrinos) and the amount of energy of the pion going into neutrinos. The factor $\xi_z$ encodes the source evolution with redshift[46] and is $\xi_z = 2.4$ for an evolution following the star formation rate. A more detailed derivation of this equation will be given in Section 4.4.2.

---

[44]The prediction of the diffuse neutrino flux is sometimes called Olbers's paradox for neutrinos (see e.g. [382]) after Olber's paradox [517] which states that for a homogeneous, infinite and static universe the night sky should be as bright as the Sun. The paradox is resolved by realising that the universe is neither of these. Similarly, for neutrinos there is an interplay between the neutrino output of each source and the total diffuse neutrino flux.

[45]At first sight, it is possible that the bound can be exceeded if the high-energy protons can be confined at the source, increasing the total energy generation rate above the observed value. However, neutrons are also produced in these interactions, with a similar budget. These then decay after escaping the source and can supply the UHECR flux.

The upper bound on the single-flavour neutrino flux can then be found by setting $\epsilon = 1$ and is

$$E_{\nu_\alpha}^2 \Phi_{\nu_\alpha}(E) \sim 2 \times 10^{-8} \, \mathrm{GeV \, cm^{-2} \, s^{-1} \, sr^{-1}}. \tag{3.44}$$

This bound is valid for all energies where extragalactic cosmic rays are produced (with an $E^{-2}$-spectrum) and where the neutrino production is efficient, which depends on the source environment and relevant interaction (see Sections 3.3.2, 3.7 and 3.8.4). Following [519], the upper bound in Eq. (3.44) should rather be interpreted as a prediction, since none of the uncertainties are large enough to significantly reduce the resulting flux. For sources optically thick under photohadronic interactions, where the escaping neutrons (which decay back into protons after their energy) are instead responsible for the observed UHECR flux, the interactions need to be taken into account more carefully. On the other hand, in this case the prediction is more accurate, since the amount of observed cosmic rays is directly related to the amount of protons used to produce neutrinos. This has been discussed in [520].

As a final remark, note that for this argument, the cosmic-ray spectrum is assumed to be dominated by protons at the relevant energies. If the composition at ultra-high energies is different, the neutrino flux can change.

### 3.3.6 Neutrino/gamma-ray connection

The neutrino and gamma-ray fluxes due to cosmic-ray interactions are related, since they share their origin in pion decay. Suppose the cosmic rays need to pass through an area with total optical depth $\tau = nl\sigma$ for hadronic interactions, with $n$ the density of the target (gas or photons), $l$ the thickness and $\sigma$ the inelastic $p\gamma$- or $pp$-cross section. The combination of the inelasticity $\kappa$ and the inelastic cross section gives the attenuation cross section, which is the cross section for energy loss. The fraction of the proton energy converted into (mainly) pion energy is then $f_\pi \simeq 1 - \exp(-\kappa\tau)$. Therefore, the total energy dependent generation rates of neutrinos and gamma rays are related to the cosmic-ray generation rate[47] as [487, 518]

$$\frac{4}{3} \sum_\alpha E_{\nu_\alpha} Q_{\nu_\alpha} \simeq K_\pi E_\gamma Q_\gamma|_{E_\gamma = 2E_\nu} \simeq f_\pi \frac{K_\pi}{1 + K_\pi} E_{\mathrm{cr}} Q_{\mathrm{cr}}, \tag{3.45}$$

where $K_\pi = \frac{N_{\pi^\pm}}{N_{\pi^0}}$ is the average ratio of charged and neutral pions. The first equality is given by the fact that $E_{\pi^0}$ goes completely into gamma rays through $\pi^0 \to 2\gamma$, whereas for the neutrinos charged pion decay, which divides the pion energy approximately equally between the four final-state leptons, three of which are neutrinos, so $\sim (1/4)E_{\pi^\pm}$ goes to electrons. The second equality assumes that the total energy into $\pi^0$ is dependent only on the number of $\pi^0$ versus $\pi^\pm$. The average particle energies appearing in this equation are related as $E_\nu \simeq E_\gamma/2 \simeq E_p/20$. From the discussion in Section 3.3.2 (Eqs. (3.18) and (3.21)), we have $K_\pi \simeq 1$ ($K_\pi \simeq 2$) for $p\gamma$ ($pp$). For $pp$-interactions, we have $\kappa \approx 0.5$ [367]. The production spectrum of gamma rays associated to $pp$-interactions spans down to sub-TeV energy, while the spectrum associated to $p\gamma$-interactions has a lower bound due to the threshold for pion production.

---

[46]This is an important distinction with the analogous prediction for the UHECR flux. While UHECRs are attenuated already over distances of 100 Mpc (Section 3.2.5), neutrinos can reach us from anywhere in the observable universe, as long as there exists a source.

[47]In this equation, $E_{\mathrm{cr}}Q_{\mathrm{cr}}$ is the theoretical value, derived from models of the source or e.g. radio measurements, before interaction, as opposed to the value required to explain the total cosmic-ray flux. The difference between these is given exactly by energy losses of the cosmic rays at the source and this difference can be significant if the losses are high for a particular source class.

While the above simple relation between neutrino and gamma-ray energy generation rates (and thus also fluxes) is correct at production, this relationship can be spoiled by three effects: subsequent interactions inside the source, interactions of propagating gamma rays in the intergalactic medium and additional, non-hadronic, sources of gamma rays adding to the total observed flux.

During gamma-ray propagation, high-energy gamma rays can interact with photons from the cosmic microwave background (CMB) and the extragalactic background light (EBL) and initiate electromagnetic cascades [521]. PeV gamma rays have an absorption length of about 10 kpc due to pair production on the CMB, while 100 TeV gamma-ray interactions with the CMB limit the distance travelled to Mpc scales [487]. In turn, the produced electrons can up-scatter the background photon fields in inverse Compton scattering to produce secondary gamma rays. This process can repeat many times, forming a cascade. As a result, the original high-energy photons disappear and are regenerated at sub-TeV energies. The cascade development affects the gamma-ray spectrum from both $p\gamma$- and $pp$-interactions. In the former case, this results is the appearance of a sub-TeV gamma-ray component which was initially absent. This cascading will be treated in more detail in the next chapter, Section 4.4.3.

Due to the interactions with the CMB and EBL, the universe becomes opaque to high-energy photons (illustrated in Fig. 3.14). Since the universe remains transparent for neutrinos, these are the only messengers which carry direct information of extragalactic cosmic-ray interactions at the source, at the highest energies. On the other hand, the cascaded gamma-ray signal can also be used to study the sources of cosmic rays, as in [522].

Similarly, inside the astrophysical sources exist strong electromagnetic fields with which the gamma rays can also interact [520], such that the high-energy gamma-ray flux is depleted [518]. Finally, while there is a direct relation between neutrinos and gamma rays created from hadronic interactions, leptonic processes can also produce high-energy gamma rays, as mentioned in Sections 3.2.4 and 3.3.1.

## 3.4   Atmospheric, astrophysical and GZK-neutrinos

As mentioned in Section 3.3.2, neutrinos can be produced by cosmic rays associated to UHECRs interacting at the source (astrophysical neutrinos), during propagation (cosmogenic neutrinos) or in our atmosphere (atmospheric neutrinos). The predicted energy range and flux of these different populations, as well as others not connected to cosmic rays, are shown in Figure 3.15 as summarised in [524]. The astrophysical neutrino flux is shown for a specific model of neutrinos from AGNs (source models will be discussed in Section 3.7), at a level compatible with the Waxman-Bahcall bound (Eq. (3.44)).

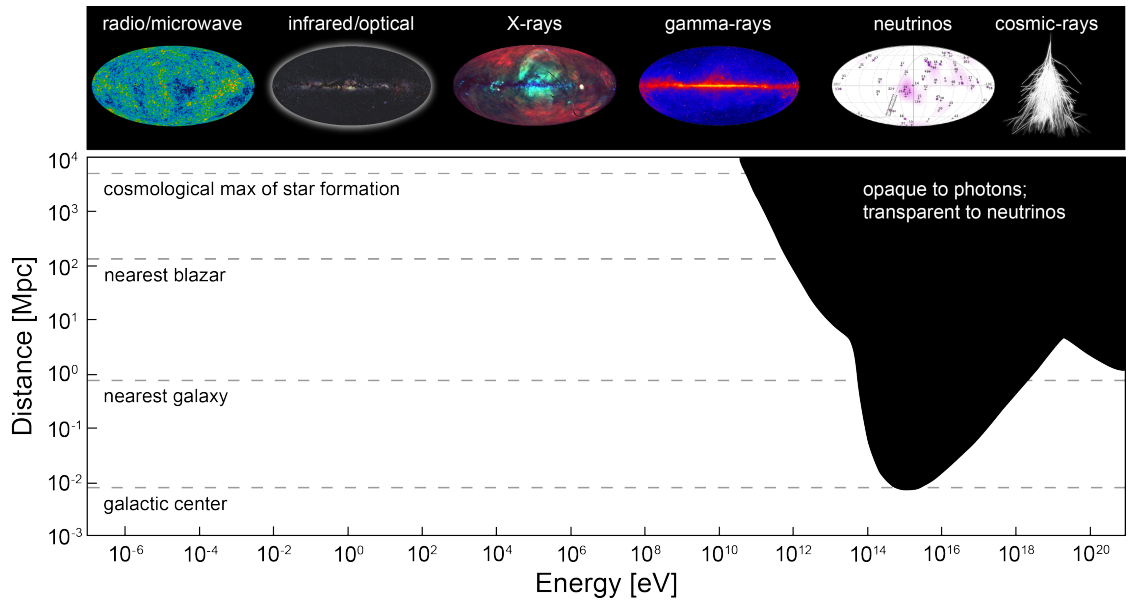

Figure 3.14: Transparency of the universe to photons. At the highest energies, the universe is opaque to photons, while neutrinos can propagate freely. Also gravitational waves propagate without attenuation, carrying different information than the neutrinos. Figure from [523].

Cosmogenic or GZK-neutrinos were discussed in Sections 3.2.5 and 3.3.2. They are directly produced by the UHECRs and are predicted to possess higher energies (due to the threshold for $p + \gamma_{CMB} \rightarrow \Delta^+$) than the astrophysical neutrinos and a lower flux (due to the lower flux of cosmic rays at high energy). The figure shows a prediction of the cosmogenic neutrino flux, which is as of yet still undetected, although the exact flux can vary significantly depending on the energy and composition at the end of the UHECR spectrum.

Finally, atmospheric neutrinos are created in the decay of mesons produced in cosmic-ray interactions with the atmosphere. The atmospheric neutrino flux is one power steeper than the cosmic rays, $\Phi_{atm}(E) \propto E^{-3.7}$ [367], since the production of neutrinos at higher energy is suppressed due to higher energy mesons interacting again with the atmospheric nuclei before they can decay. There is also an additional component of neutrinos from charmed mesons, which decay faster and are thus expected to have a harder spectrum. However, as already mentioned in Section 3.3.2, there is a large uncertainty on theoretical models predicting this flux and experiments have only been able to determine an upper bound on this component [503]. For a recent, detailed, calculation of the atmospheric neutrino flux, see [525, 526].

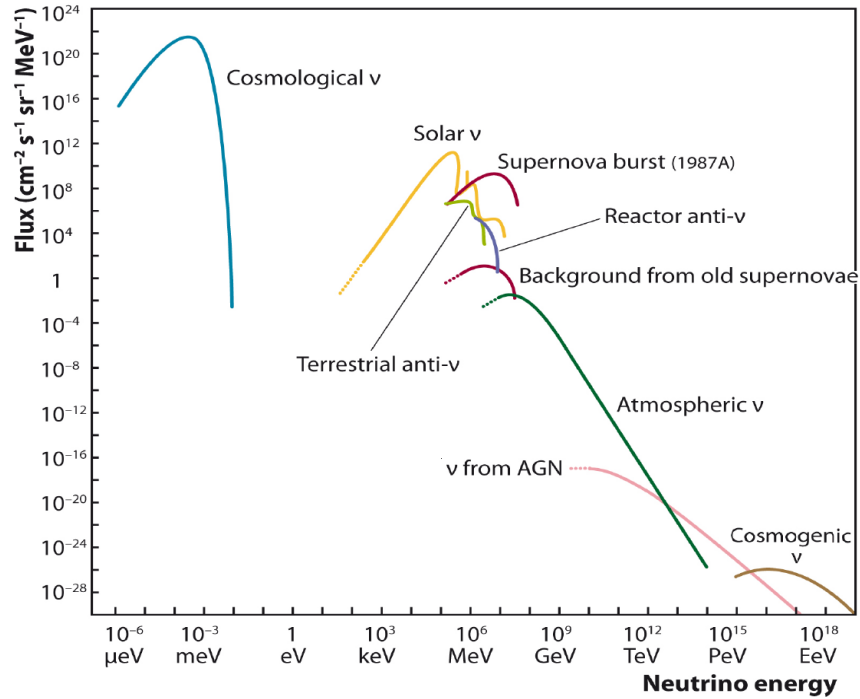

Figure 3.15: A summary of the neutrino flux from different sources and their energy range. The atmospheric, astrophysical and cosmogenic neutrinos are connected to the observed UHECRs. The other shown neutrino fluxes are either man-made (reactor), associated to fusion (solar) or decay (terrestrial), produced in supernovae or the frozen-out remainder of neutrinos from the thermal bath of the early universe (relic or cosmological neutrinos). Figure from [524].

For the remaining of this thesis, we are only interested in the astrophysical neutrinos, while the atmospheric neutrinos will serve as a background in experimental searches.

## 3.5 Astrophysical neutrino detection

Neutrinos only interact through the weak interaction[48] and are therefore hard to detect, requiring a sizeable detector volume, sensitive detectors and good background rejection. Several experiments have been built that are capable of detecting both naturally occurring neutrino fluxes at low and high energy as well as man-made neutrino beams.

In the case of high-energy neutrinos from astrophysical sources associated to the UHECRs, the flux is so low that it is not feasible to construct a typical particle detector[49].

---

[48]For an overview of neutrino cross sections at various energies, see [527].
[49]The same is true for the GZK-neutrinos, which exhibit even lower fluxes. These can be probed with the

Instead, one can instrument a huge volume of naturally occurring transparent material, as pioneered (though never realized) by the DUMAND experiment [528]. Nowadays, several experiments are running which can successfully detect high-energy neutrinos: the IceCube Neutrino Observatory [529], ANTARES [530] and Baikal NT200 [531]. IceCube instruments a volume of $1 \text{ km}^3$ of ice at the South Pole, while the other two experiments instrument a volume of water about two orders of magnitude smaller than IceCube and are located in the northern hemisphere. Because of its bigger volume, the discussion will be focussed on IceCube unless otherwise stated.

When a flux of high-energy neutrinos passes through ice, some of the neutrinos can interact with the nuclei inside the ice. In these interactions, relativistic secondaries are created which emit Cherenkov radiation as they propagate through the ice. By instrumenting a huge volume of this ice with photomultipliers (called DOMs —digital optical modules— in the case of IceCube) on vertical strings, one can collect this light and measure the energy deposited by the neutrino interaction, as well as infer the direction of the original neutrino. On the other hand, it is not possible to discriminate between $\nu_\alpha$ and $\bar{\nu}_\alpha$. With this setup, IceCube is sensitive to neutrinos between about 100 GeV up to several PeV, limited by detector spacing and statistics (detector size) respectively.

There are two types of interactions the neutrino[50] can undergo with the nuclei: charged current and neutral current interactions. In charged current (CC) interactions, the neutrino is converted into a charged lepton through the exchange of a $W^\pm$-boson

$$N \, \nu_l \xrightarrow{CC} X \, l. \tag{3.46}$$

In this interaction, part of the neutrino energy is given to the resulting lepton, the rest is deposited in hadronic debris. In neutral current (NC) interactions the neutrino scatters off a nucleus without converting, through the exchange of a $Z$-boson,

$$N \, \nu_l \xrightarrow{NC} X \, \nu_l. \tag{3.47}$$

Part of the neutrino energy is deposited in the hadronic debris, the rest being carried away by the surviving neutrino.

The different types of neutrinos and interactions give rise to distinct event topologies in the detector: showers, tracks and double bangs, as shown in Figure 3.16. Showers and tracks are the most prominent. Showers are created when a $\nu_e$ undergoes a CC interaction or any $\nu_\alpha$ undergoes a NC interaction with an ice nucleus. In the case of a $\nu_e$ in CC interactions, the created electron and hadronic debris undergo frequent interactions with the ice and the total energy of the original neutrino is quickly deposited in the ice. In the case of NC interactions, only the part of the neutrino energy that is transferred to the hadronic debris is deposited in the detector. In both cases, the created particles lose energy fast and their path length is short compared to the detector size, such that a roughly spherical region of the detector is lit up and the entire event is typically

---

same detectors as the astrophysical neutrinos or with dedicated experiments. However, we will not discuss these here.

[50]Unless otherwise stated, neutrino can mean both $\nu_\alpha$ and $\bar{\nu}_\alpha$.

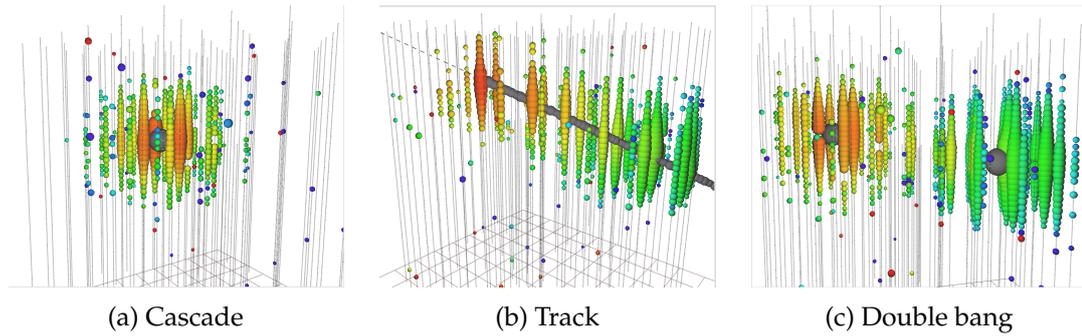

|              (a) Cascade              |              (b) Track              |              (c) Double bang              |

Figure 3.16: Different event topologies for neutrino detection. (a) Cascade (or shower) created by $\nu_e$ CC interaction or any $\nu$ NC interaction. (b) Track of a muon created in a $\nu_\mu$ CC interaction. (c) Double bang of a $\tau$ creation and subsequent decay from a $\nu_\tau$ CC interaction. Figures from [534].

contained (unless it occurs at the edge of the detector). As a result, shower events have good energy resolution but poor directional resolution of only $10°$ [532].

Tracks are created when a $\nu_\mu$ undergoes CC interaction. Because of its larger mass, the muon created in this interaction loses energy much more slowly than electrons and can travel several kilometres through the ice [532, 533]. This means that muons escape the detector before being stopped completely. As a result, it is only possible to determine a lower limit on the muon and original neutrino energy, although modelling of the muon energy losses allows one to estimate this energy. On the other hand, because of the long muon path length, the muon direction, which is approximately equal to the neutrino direction, can be determined accurately. Therefore, track events have poor energy resolution but a good directional resolution better than $1°$ [532].

Finally, $\nu_\tau$ in CC interactions give rise to a peculiar event topology, called double bang. The first bang occurs when the original neutrino interacts with the ice and deposits energy in hadronic debris, after which the so-created $\tau$ travels further in the detector. Subsequently, this $\tau$ decays and deposits the rest of its energy with a second bang in the detector (except in the case of decay to a muon, which happens in 17% of the decays [29], creating a track). The decay length of this $\tau$ is 50 m/PeV [532, 534], to be compared with a horizontal separation of detector modules of 125 m and a vertical separation of 17 m. Therefore, the double bang signature is difficult to resolve. More details about the different event topologies and how to model them can be found in e.g. the appendix of [535].

When searching for astrophysical neutrinos, there are two important backgrounds that must be taken into account: atmospheric muons and atmospheric neutrinos, both created in air showers initiated by cosmic rays. Atmospheric muons occur at a rate of about 3 kHz in the IceCube detector. In order to get rid of this background, aggressive analysis cuts are required. However, while muons can traverse a large distance before being stopped completely, they will eventually be stopped by the Earth. Therefore, in the case of IceCube, the analysis cuts are only necessary in the southern hemisphere, while the background is absent in the northern hemisphere. As a result, IceCube is most sensitive for sources in the northern sky. On the other hand, ANTARES and Baikal are most sensitive in the southern hemisphere and can even compete with IceCube in this region, despite their smaller volume. Atmospheric neutrinos represent a second, irreducible background, of about 4 mHz in the IceCube detector, since their signature is identical to astrophysical neutrinos.

As a result, this background can only be subtracted on a statistical basis. This is possible, since the calculated atmospheric neutrino spectrum $\propto E^{-3.7}$ and the predicted astrophysical neutrino spectrum $\propto E^{-2}$ have a different spectral index.

## 3.6 The high-energy sky

In Section 3.3.2, we have seen that the sky should be lit up by a diffuse flux of high-energy neutrinos and gamma rays from the extragalactic objects responsible for the UHECR flux, as well as gravitational waves from extreme events. In the last decade, several experiments have been built to observe these cosmic messengers. Here, we will summarise the main results concerning the diffuse flux, while individual sources will be left to the next section.

### 3.6.1 Diffuse astrophysical neutrino flux

Astrophysical neutrinos were first detected by IceCube in 2013 [536] as a high-energy diffuse flux. There are several analyses focussing on different energy ranges and event topologies in order to characterise this flux. Two important analyses which will be discussed here are the high-energy starting events (HESE) analysis and the through-going muon analysis.

The HESE analysis was responsible for the initial discovery of astrophysical neutrinos. It is focussed on identifying high-energy neutrino events which have a high probability to be of astrophysical origin, utilising very strict analysis cuts. The outer layers of the detector are used as a veto, which rejects the atmospheric background of muons passing through this layer while accepting events where the neutrino interaction occurs inside the detector. This technique results in a smaller effective detector volume, but in return one obtains a very pure event sample. An analysis using six years of data [534] detected 82 events (22 tracks, 58 showers and 2 events produced from a coincident pair of background muons), with an expected background of about 15 events. The best-fit per-flavour neutrino flux from this analysis, fitted in a range between 60 TeV and 10 PeV of deposited energy, is

$$E_\nu^2 \Phi_\nu(E) = 2.46 \pm 0.8 \times 10^{-8} \left( \frac{E}{100\,\text{TeV}} \right)^{-0.92} \text{GeV cm}^{-2}\,\text{s}^{-1}\,\text{sr}^{-1}, \tag{3.48}$$

and is shown in Figure 3.17a. This flux is exactly at the level that was expected from the Waxman-Bahcall prediction in Eq. (3.44). The most recent analysis at the time of writing this thesis, uses 7.5 years of data [537] and includes a separate event category for double cascade events. It fits instead the all-neutrino flux and finds (with the same form for the spectrum as above) a normalisation $\Phi_{\text{astro}}^{6\nu} = 6.45^{+1.46}_{-0.46}$ and spectral index $\gamma = 2.89^{+0.2}_{-0.19}$. Both analyses also test for a broken spectrum with two components, but find that the soft component is compatible with zero within $2\,\sigma$ (see also [538]).

Complementary to the HESE analysis is the through-going muon analysis. It searches for muon tracks, using the full detector volume without defining a veto region. Instead, the analysis uses the Earth as shielding from atmospheric muons and, as a result, can only be used to observe the northern hemisphere. The advantage of using muons without a veto layer is that $\nu_\mu$ can interact far outside the detector and still be detected as muons in IceCube, boosting the effective detector volume. Because of a higher background of atmospheric muon neutrinos and poorer energy resolution (because muons are not fully contained), this analysis is only sensitive to astrophysical neutrinos above 200 TeV, which is a higher threshold than the HESE analysis. An analysis using eight years of data [534] found 36 events above 200 TeV. The best fit muon neutrino flux is

$$E^2_{\nu_\mu + \bar\nu_\mu} \Phi_{\nu_\mu + \bar\nu_\mu}(E) = 1.01^{+0.26}_{-0.23} \times 10^{-8} \left( \frac{E}{100\,\text{TeV}} \right)^{-0.19 \pm 0.10} \quad \text{GeV cm}^{-2}\,\text{s}^{-1}\,\text{sr}^{-1}. \quad (3.49)$$

This flux is also shown in Fig. 3.17a. The measurement has a significance of $6.7\,\sigma$ with respect to the atmospheric-only hypothesis. The most recent analysis at the time of writing this thesis uses 10 years of data [537] and finds a normalisation $\Phi = 1.44^{+0.25}_{-0.24}$ and spectral index $\gamma = -2.28^{+0.08}_{-0.09}$, showing a slight softening of the spectrum compared to the previous analysis above.

The through-going muon analysis finds a harder spectrum than the HESE analysis. As a result, the single power law hypothesis has been questioned [539–543], which motivated the test of a broken power law spectrum in the HESE analysis. However, the difference between the two spectra is at this moment not significant and the issue remains as of yet unsettled. In the most recent update at the time of writing this thesis, the results remain compatible [537].

The events detected in the HESE analysis cover the whole sky[51]. Their directions are shown in Fig. 3.17b. While the accumulation of events near the galactic centre is suggestive, there is no significant clustering [534]. The observed astrophysical neutrino flux is compatible with isotropy and is not associated to the galactic plane[52]. Therefore, these events must be extragalactic, as expected. Still, several galactic interpretations of (part of) this flux exist [539, 544–548].

---

[51] Although above 100 TeV, the Earth starts to becomes opaque also to neutrinos [536], initially only for the longest pathlength through Earth (i.e. from the North Pole at a declination of $90°$, as can be seen from the Figure 1 of the HESE analysis in [534]). This does not contradict with the energy threshold for through-going muons. Indeed, checking the arrival directions in Figure 6 of the through-going muon analysis in [534], none of them come from the pole and most of them are concentrated towards the equator.

[52] Although it has been noted that above 100 TeV there is a slight "excess" of $1.8\,\sigma$ for events associated to the galactic plane, as noted in [535, 539, 544, 545]

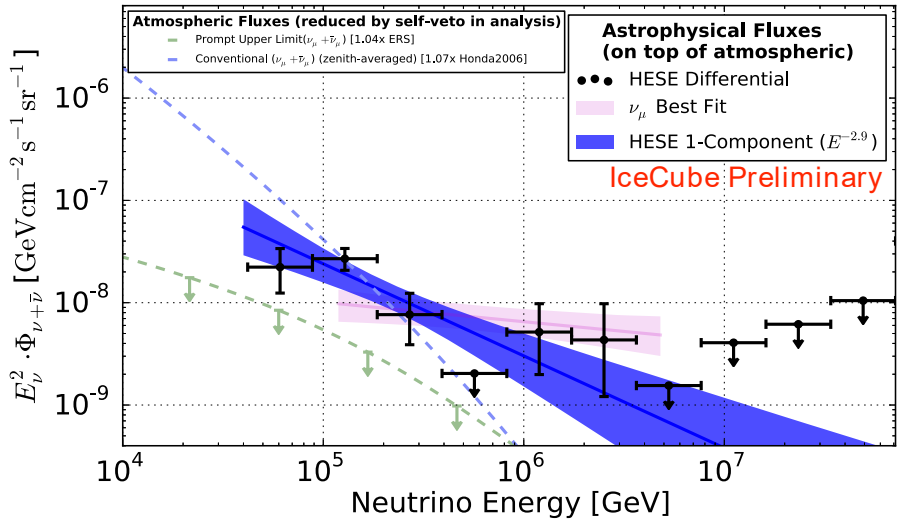

(a) Energy spectrum (HESE & $\nu_\mu$)

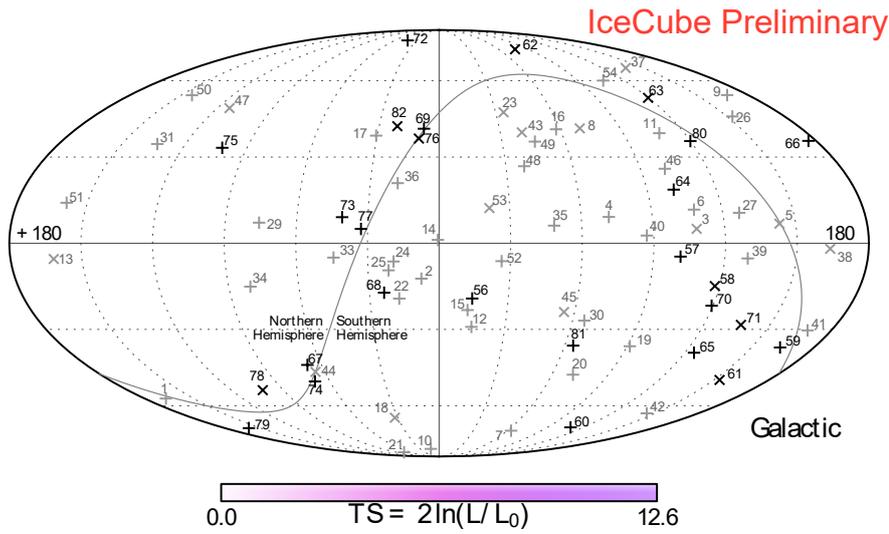

(b) HESE neutrino directions

Figure 3.17: High-energy starting events (HESE) detected by the IceCube Collaboration [534]. (a) Detected HESE events and best-fit per-flavour flux from the HESE and through-going muon analysis. The atmospheric background is subtracted. Best-fit conventional and prompt atmospheric background are shown with dotted lines. (b) Directions of detected neutrinos and test statistics for point source clustering at each location.

### 3.6.2 Extragalactic gamma-ray background

Gamma rays with an energy between 20 MeV and about 2 TeV are observed by Fermi-LAT, the main instrument aboard the Fermi Gamma-Ray Space Telescope spacecraft. Fermi-LAT is a pair conversion telescope, which detects incoming gamma rays when they are converted to $e^+e^-$-pairs inside a tracker and those deposit their energy in a calorimeter [549, 550].

Gamma rays are emitted by many sources, from both galactic and extragalactic origin, which are collected in the various catalogues constructed by the Fermi-LAT collaboration. Gamma rays from inside our galaxy are emitted from point sources such as pulsars and extended sources such as supernova remnants and pulsar wind nebulae. In addition, there is also a truly diffuse emission from cosmic rays interacting with interstellar gas and radiation fields. However, we are mainly interested in gamma rays of extragalactic origin, connected to the UHECRs and astrophysical neutrinos. Extragalactic sources of GeV gamma rays include AGNs (blazars in particular), star-forming galaxies and GRBs [551]. The total sum of this extragalactic gamma-ray flux is called the extragalactic gamma-ray background (EGB). The EGB can be decomposed in two parts: individually resolved sources and a diffuse flux. Most of the resolved EGB is made out of blazars. which are extremely bright. The remaining diffuse flux of gamma rays is called the isotropic diffuse gamma-ray background (IGRB). There are two contributions to this diffuse flux. The first of these is the collective background of unresolved point sources. This includes the extrapolation of the observed blazar population to lower fluxes, as well as misaligned AGNs and star-forming galaxies whose contribution needs to be estimated based on theoretical models, since most of these objects are individually too faint to be detected. The unresolved-source component is detector dependent: more sensitive instruments are able to resolve more point sources[53]. The second component of the IGRB is a true diffuse contribution, coming from very high-energy gamma rays cascading in the CMB and EBL [552] (as explained in Section 3.3.6) and cosmic-ray interactions with the EBL [521]. In addition, it is also possible that there is a contribution from new physics, in the form of dark matter annihilations [553–555].

The Fermi sky map using 7 years of data in the 10 GeV to 2 TeV band [556] is shown in Fig. 3.18a. In this sky map the diffuse emission from the galaxy shows up very bright, while the isotropic component is weaker. The most recent measurement of the IGRB by Fermi uses 50 months of data and includes gamma rays with an energy between 100 MeV and 820 GeV [557]. The analysis considers emission over the full sky, excluding the region of the galactic plane, and performs a multicomponent fit of the total energy spectrum, separating contributions from resolved point sources, the IGRB as well as a foreground of galactic emission.

---

[53]Since neutrino telescopes have not yet resolved any point sources, this subtlety did not arise in the previous discussion.

The resulting IGRB spectrum is [557]

$$E_\gamma^2 \Phi_\gamma(E) = (0.95 \pm 0.08) \times 10^{-6} \left( \frac{E}{100 \, \text{MeV}} \right)^{-(0.32 \pm 0.02)}$$
$$\times \exp\left( -\frac{E}{(279 \pm 52) \, \text{GeV}} \right) \, \text{GeV} \, \text{cm}^{-2} \, \text{s}^{-1} \, \text{sr}^{-1},$$
(3.50)

determined from their foreground model A[54]. The total diffuse intensity above 100 MeV from this is $(7.2 \pm 0.6) \times 10^{-6} \, \text{cm}^{-2} \, \text{s}^{-1} \, \text{sr}^{-1}$. Adding this diffuse spectrum to the flux from resolved extragalactic sources (both shown separately in Figure 3.18b), one obtains the full EGB spectrum, which is given as

$$E_\gamma^2 \Phi_\gamma(E) = (1.48 \pm 0.09) \times 10^{-6} \left( \frac{E}{100 \, \text{MeV}} \right)^{-(0.31 \pm 0.02)}$$
$$\times \exp\left( -\frac{E}{(362 \pm 64) \, \text{GeV}} \right) \, \text{GeV} \, \text{cm}^{-2} \, \text{s}^{-1} \, \text{sr}^{-1}.$$
(3.51)

The integrated extragalactic intensity above 100 MeV is $(1.13 \pm 0.07) \times 10^{-5} \, \text{cm}^{-2} \, \text{s}^{-1} \, \text{sr}^{-1}$ (for foreground model A in [557]).

Using the known sources, their luminosity distribution and their expected evolution with redshift, one can decompose the remaining diffuse emission into its components. One finds that most of the diffuse emission is most likely associated with blazars: only about 30% of the IGRB can be of non-blazar origin [558–564]. As we will see, this puts strong constraints on source populations responsible for both gamma ray and neutrino emission.

Finally, it is also possible to observe TeV gamma rays. However, due to their low flux and high energy, it is not possible to detect these with relatively small satellite detectors. Instead, like with cosmic rays above the knee, one uses the atmosphere as a detector and observes the gamma-ray induced cascade. This is done by Imaging Air Cherenkov Telescopes (IACTs) such as MAGIC [565], H.E.S.S. [566] and VERITAS [567]. Known sources of TeV gamma rays are the galactic centre, the Crab Nebula, pulsar wind nebulae, blazars (e.g. Mkn 421), nearby radio galaxies (M87, Centaurus A and NGC 1275) and starburst galaxies (M 82 and NGC 253). For a review, see [568]. However, for the purposes of this thesis, we will mostly be interested in the total (diffuse) gamma-ray flux in connection with neutrinos. The horizon for TeV gamma rays is rather small (see Section 4.4.3 in the next chapter), so that TeV gamma-ray emission from interesting sources is recycled into the 100 GeV region. Therefore, we will not further discuss these TeV gamma-ray experiments.

---

[54]This model considers cosmic-ray injection in the Milky Way following the pulsar distribution and with a constant diffusion coefficient. Alternative models feature a deviating cosmic-ray injection, additional electron sources or varying diffusion coefficient. These slightly alter the derived IGRB and EGB spectra, with a change in the total intensity of the order of 10%. For the present purposes, this can be neglected.

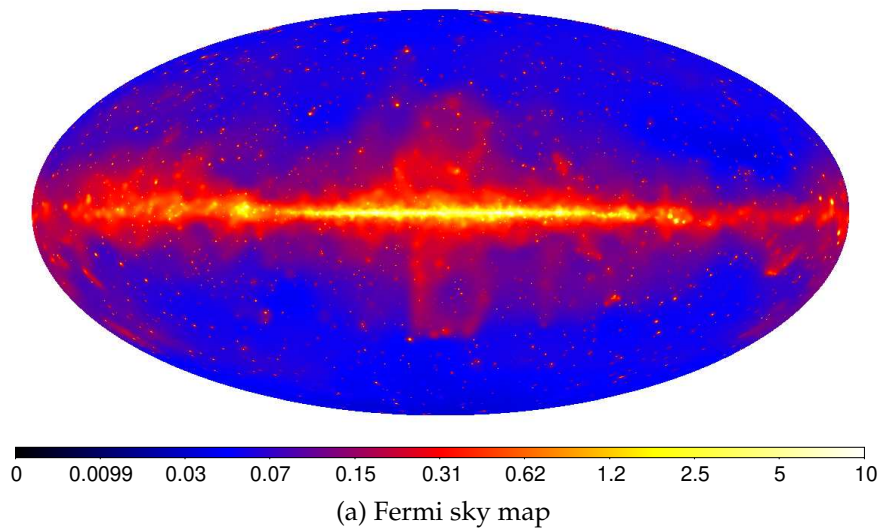

(a) Fermi sky map

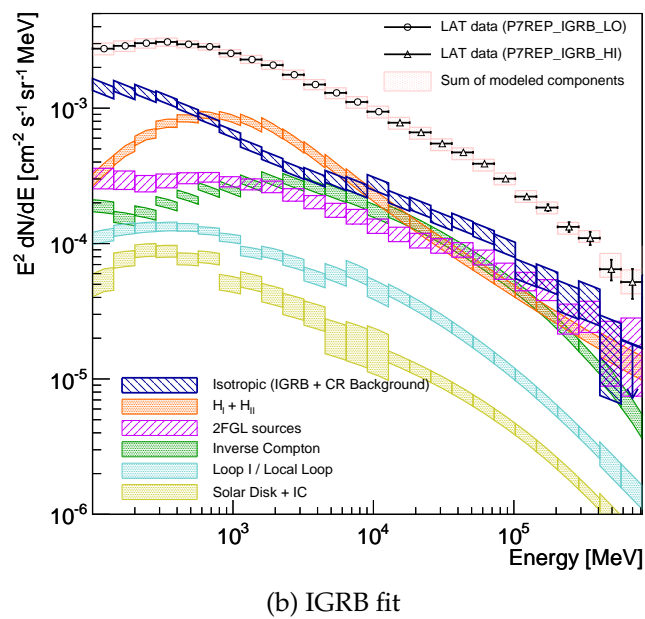

(b) IGRB fit

Figure 3.18: GeV gamma-rays observed by Fermi. (a) Sky map of gamma-ray counts in the 10 GeV-2 TeV band using 7 years of data [556]. Shows counts per $(0.1 \text{ deg})^2$, smoothed with Gaussian kernel. Logarithmic colour scale. (b) Measurement of the isotropic diffuse gamma-ray background (IGRB) —including a background from cosmic rays interacting in the detector— identified extragalactic sources and a galactic foreground with different components (model A) [557].

### 3.6.3   Gravitational waves

Finally, it is also expected that certain sources of cosmic rays, gamma rays and neutrinos are also sources of gravitational waves. Here, we are mostly interested in gravitational waves from compact binary coalescences (i.e. binary black hole mergers, neutron star mergers or neutron star-black hole mergers; we ignore gravitational waves from phase transitions in the early universe or supermassive black hole mergers).

Since gravitational waves probe the inner engine of cosmic accelerators, they can supply independent information from gamma rays and neutrinos in a multimessenger analysis [569, 570].

In keeping with the spirit of this section, we discuss here only the "gravitational wave sky", while a detailed discussion of gravitational wave astronomy and the individual sources of gravitational waves is deferred to Chapter 5. While neutrinos have only been detected as a diffuse background and gamma rays have been detected in both diffuse and point-source searches, gravitational waves have only been detected as point sources. Therefore, limits have been placed on the energy density of the gravitational wave background. The latest limit[55] is from Advanced LIGO's second observing run [573]

$$\Omega_{GW} = \frac{\rho_{GW}}{\rho_c} < 4.8 \times 10^{-8} \qquad (3.52)$$

at a frequency of $f = 25$ Hz for compact binary coalescences. Here, $\rho_c = 3H_0^2 c^2/(8\pi G) \approx 5000$ eV cm$^{-3}$ is the critical density required for a flat universe. This limit is also shown in Figure 3.19. We can compare this limit on the energy density with the energy density of cosmic rays, which is[56] $\rho_{CR} \approx 1$ eV cm$^{-3}$ [367]. We find $f_{GW/CR} \lesssim 2.3 \times 10^{-4}$. We will briefly come back to this in Chapter 5.

### 3.6.4   Summary

In Fig. 3.20, we collect the data on ultra-high energy cosmic rays, the extragalactic gamma-ray background and the astrophysical neutrino flux. The observation that these messengers have a similar energy budget (due to a similar $E^2\Phi$) and that they are all isotropic suggests that these messengers indeed have a common origin[57,58]. This indicates that the ideas which led to the original derivation of the Waxman-Bahcall bound are correct and motivates us to continue modelling the different messengers together.

---

[55]Another analysis extends the earlier analysis from run O1 [571] for different gravitational-wave polarisations [572].

[56]This is mostly driven by galactic cosmic rays, not UHECRs, so the comparison is only indicative.

[57]One subtlety here, is that UHECRs have a limited horizon due to the GZK-effect, whereas neutrinos of all energies and gamma rays sufficiently below 100 GeV can propagate (approximately) unhindered throughout the universe. This means that the amount of sources visible in cosmic rays and neutrinos/gamma rays is different.

[58]This connection need not be direct. A single model could predict that certain sources supply the UHECR flux, while others (almost) completely convert the cosmic rays in neutrinos, see e.g. [454].

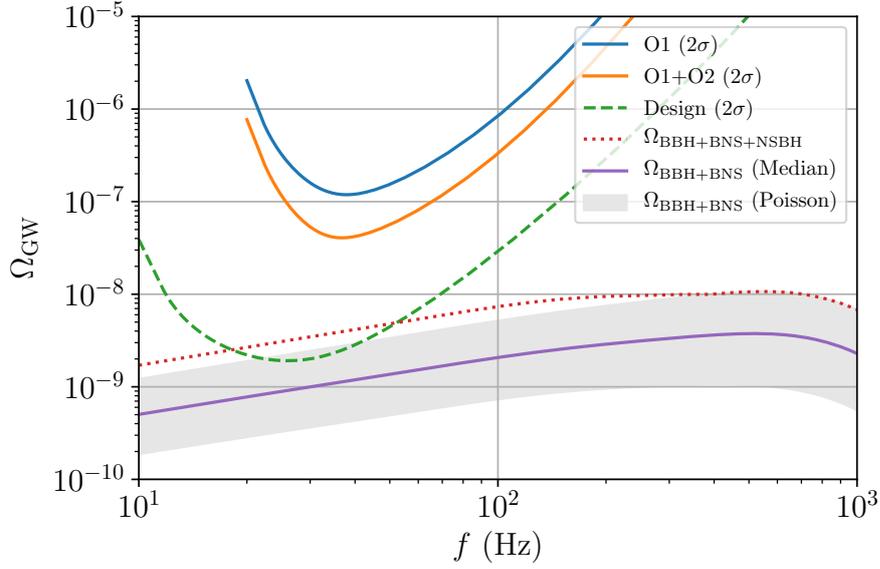

Figure 3.19: Sensitivity of LIGO to the isotropic stochastic gravitational wave background, compared against the design sensitivity and model predictions from binary black hole mergers (BBH), binary neutron star mergers (BNS) and neutron star-blck hole mergers (NSBH). Figure from [573]

On the other hand, the three messengers are relevant in different energy ranges. In the case of cosmic rays, we can only observe the ultra-high energy part (above $10^{18.5}$ eV) of the extragalactic flux, due to the appearance of the galactic cosmic rays at lower energy. Astrophysical neutrinos have been observed between about 100 TeV and a few PeV, where the lack of higher energy events could be due to the sensitivity of IceCube or due to a cut-off in the astrophysical neutrino spectrum (see discussion in Section 3.3.5 and specific models in Section 3.7). Extragalactic gamma rays connected to neutrinos are produced at the same energies, but after propagation are mainly important below 100 GeV, where the universe becomes transparent to them. Therefore, the different energy ranges have a natural explanation whilst still being compatible with a common origin (although it is possible for them to have a different origin, see Section 3.7.3). An important consequence of the different energy ranges, however, is that the observed neutrinos are produced in interactions of cosmic rays below the ankle[59]. The highest energy neutrino observed is a muon neutrino with a median inferred energy of 7.8 PeV [534], which corresponds to a parent cosmic ray with an energy of around 200 PeV. The composition of extragalactic cosmic rays at these energies is uncertain. Measurements indicate that the mass of the cosmic rays is between proton and iron, and not pure proton or iron. More importantly, however, at these energies the extragalactic cosmic ray flux needs to be extrapolated down from the observed UHECRs, such that the inferred cosmic-ray composition is model-dependent.

---

[59]As mentioned in Section 3.4, there is also a second population of neutrinos (and gamma rays which will cascade down) associated to the GZK-effect. These are produced directly by the UHECRs and are therefore more directly connected to the observed extragalactic cosmic rays.

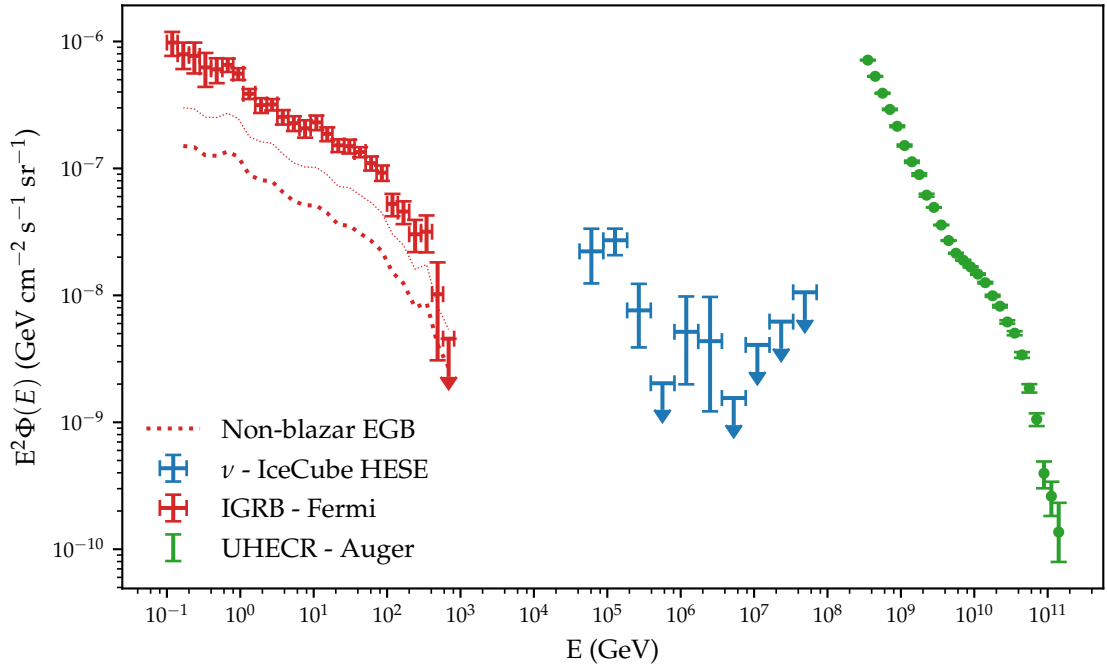

Figure 3.20: The high-energy sky using different messengers, showing the UHECR spectrum measured by Auger [377], the IGRB measured by Fermi [557], the upper limit on the non-blazar contribution to the EGB (both best fit (14%) and weakest upper limit (28%)) and the diffuse astrophysical neutrino flux measured by IceCube [534]. It can be seen that the differential energy budget of all three messengers is similar, suggesting that they are connected.

In order to elucidate the connection between the UHECRs, gamma rays and neutrinos, a diffuse measurement is not sufficient. Only by identifying the objects where neutrinos are produced, can we unambiguously establish the presence of UHECR acceleration and the interactions which they undergo. The search for astrophysical neutrino sources will be discussed in the following sections.

## 3.7 Neutrino source models

There is a great variety of astrophysical objects that could be responsible for the observed astrophysical neutrino flux. These objects need to be capable of accelerating cosmic rays to sufficient energies and allow for efficient interaction with a target radiation or matter field. Typically, these objects are also the sources of UHECRs, but this is not necessary. One additional subtlety compared to cosmic-ray sources is that sources with strong beaming might only be detectable in neutrinos if the outflow is directed towards

Earth. For cosmic rays this distinction did not arise since directionality is lost in the intergalactic magnetic fields.

There are several ways to classify the various neutrino production models. The first important distinction is between $p\gamma$- and $pp$-interaction models, as was already discussed in great detail in Section 3.3.2. This property is, usually, correlated with a second possible classification of neutrino sources, namely accelerator models versus reservoir models. Accelerator models assume that the cosmic-ray interactions take place at the same place as where they are accelerated, which typically means in a compact region around an astrophysical object. Usually, the radiation fields at these sources are so strong[60] that $p\gamma$-interactions are important and dominate over $pp$-interactions[61] (although models with dominant $pp$-interactions exist, e.g. [574]). In reservoir models, one assumes that the cosmic rays interact while they are confined within a region, integrating over a sufficient amount of target material due to their long residence time in e.g. a galaxy or galaxy cluster of sufficient size. These models typically feature $pp$-interactions, since they can have large integrated matter densities but negligible radiation fields[62]. As already discussed in Section 3.3.2, the difference between these models is reflected in the neutrino spectrum. The spectra predicted in $p\gamma$-interaction models feature a low-energy cut-off and are peaked at PeV energies [575], although the exact shape of the resulting spectrum depends also on the target photon field (see e.g. [495]). On the other hand, $pp$-interaction models predict neutrinos down to GeV energies, following the original cosmic-ray spectrum. Also the connection between the gamma-ray and neutrino spectrum is different for these two models, as outlined in Section 3.3.6. Finally, there is a distinction between steady sources, which continuously emit neutrinos, versus transient sources which emit neutrinos only at one (or several) moment(s) in time. This distinction has modelling consequences (e.g. since transient sources need to temporarily support huge luminosities) but, more importantly, implies different detection strategies.

Model parameters can be fixed with theoretical arguments and experimental observations. For this, information from different messengers is used, such as gamma rays[63], cosmic rays, X-rays, optical, infrared or radio observations and recently gravitational waves, considering either the diffuse or individual source emission. This information can then be used to either determine the input parameters of the model (which can be considered more 'physically motivated', but might predict an unrealistically high or irrelevantly low flux) or fit the resulting spectrum to agree with observation (but might require unrealistic source conditions).

In the following, we will give a short overview of some common neutrino source models and their predictions.

---

[60]Strong meaning that the energy is sufficiently high for interactions (Eq. (3.17)) and that the number density of photons is large.

[61]While $\sigma_{pp} > \sigma_{p\gamma}$ (Fig. 3.13), the number density of photons is typically much larger.

[62]Since radiation fields drop off as $\sim r^{-2}$ away from a source, while matter densities inside a galaxy vary much more slowly.

[63]Although again this connection depends on their leptonic or hadronic origin, see Section 3.3.6.

### 3.7.1 Active galactic nuclei

Active galactic nuclei were already discussed as candidate cosmic-ray sources in Section 3.2.4. As such, they are also candidate neutrino sources. There exist many different models for neutrino production from both radio-loud [576] and radio-quiet [577] AGNs. These models explore the various possibilities for the exact location and mechanism for the neutrino production, such as AGN cores [578], accretion disks [574], jets [454, 579], knots inside jets [495], lobes of FR II galaxies [495], AGN winds [580], the interaction of the jet with the interstellar medium, and others.

Most models consider $p\gamma$-interactions, due to the strong radiation fields present, although there are also several models that consider $pp$-interactions (e.g. [495, 574]) or even cosmic-ray reservoir models [581, 582]. The predicted flux can be normalised in various ways, based on electromagnetic output in for example radio waves (e.g. [495]), X-rays (e.g. [574]) or gamma rays (e.g. [583]), based on the individual source luminosity or the diffuse background in the respective band. In the case of a normalisation based on gamma rays, an important uncertainty is the relative contribution of leptonic versus hadronic gamma rays (as mentioned in Sections 3.2.4 and 3.3.6), although there exist models for e.g. blazar emission where the high-energy component is fully hadronic [583, 584].

Blazars, as a subset of (radio-loud) AGNs, are of special interest as neutrino sources. They exhibit highly beamed electromagnetic emission and, assuming that cosmic rays are accelerated in the jet such that they are similarly beamed, would give rise to enhanced neutrino emission. Neutrino emission from blazar jets can arise both from $pp$-interactions with gas inside the jet [585] and from $p\gamma$-interactions with internal radiation [576] or external radiation fields [586]. There are many models available; the ones listed here have been investigated by IceCube analyses [533, 587, 588]. Early models of neutrinos from blazars include generic ones by Mannheim [576], Halzen and Zas [589] and Protheroe [590]. But more recently, more specialised models have been developed for both BL Lac objects [583, 591–593] as well as FSRQs [579, 594].

While many models of neutrino emission from blazars assume steady emission, the electromagnetic emission from blazars is highly variable (in fact, this is one of their main characteristics) and exhibits "flaring" [595]. These sudden increases of non-thermal output indicate an increase in the amount of accelerated particles (in particular electrons). Therefore, it is reasonable to assume that neutrino emission from blazars might be mainly associated with these flares[64]. Of special interest are so-called orphan flares, where a sudden increase of gamma rays is not accompanied with increased emission in other wavelengths. Such an orphan flare, as has been observed in the TeV blazar ES 1959+650 [596], is difficult to explain in purely leptonic models where the emission in different wavelengths is all linked to the same population of accelerated electrons. If, instead, the gamma-ray emission is due to a hadronic population, this emission is not directly correlated to electromagnetic radiation at lower energies. This

---

[64]On the other hand, increased particle acceleration could lead to a neutrino flare without an accompanying gamma-ray flare, if the gamma rays are attenuated by matter or radiation. See also Section 3.8.2.

hadronic component could then also give rise to neutrino emission [597, 598]

Finally, the discovery of neutrinos from the blazar TXS 0506+056 confirmed that blazar (flares) must accelerate cosmic rays (although not necessarily up to ultra-high energies) and are responsible for at least part of the astrophysical neutrino flux. This observation lead to many new modelling efforts, which will be discussed in Section 3.8.2.

### 3.7.2 Gamma-ray bursts

Gamma-ray bursts form another important cosmic-ray source class candidate which was already discussed in Section 3.2.4. These short events are extremely luminous, implying the presence of strong radiation fields. Provided that cosmic-ray acceleration takes place, efficient $p\gamma$-interactions are then expected, giving rise to a neutrino flux [599–601]. Within the fireball model for gamma-ray bursts, these interactions can possibly convert more than 10 % of the fireball energy into high-energy neutrinos [599].

GRB emission occurs in three phases and neutrino production can take place in any of these. Most models assume that the neutrino flux is created at the same time as the gamma rays, in the prompt phase [480, 599]. However, there exist also models for neutrinos from the precursor [442] or afterglow [443] phases. These different possibilities have implications concerning the correlation between the electromagnetic and neutrino flux (see e.g. [382]), as well as the search strategies for neutrino emission, due to different timing of neutrinos compared to the prompt phase gamma rays. The two reference models for neutrino emission from GRB obtain their normalisation in slightly different ways: either the UHECR flux arises from proton escape [599], or neutron escape [602]. In both cases, the neutrino spectrum can be parameterised by a doubly-broken power law [603]. Recent models feature more detailed calculations, including e.g. the full nuclear cascade when considering a mixed composition of the cosmic rays [604] or detailed simulation of the time-dependent emission, including multiple shocks [440, 605, 606].

More recently, there has been interest in another, distinct, class of gamma-ray bursts for neutrino emission: low-luminosity gamma-ray bursts [607–611]. Compared with the typical high-luminosity gamma-ray burst, these bursts have a lower luminosity (less than $10^{49}$ erg s$^{-1}$ compared to $10^{51}$ erg s$^{-1}$ to $10^{53}$ erg s$^{-1}$) but a higher local rate ($325^{+352}_{-177}$ Gpc$^{-3}$ yr$^{-1}$ compared to $1.12^{+0.43}_{-0.20}$ Gpc$^{-3}$ yr$^{-1}$) [610, 612–614]. These events can occur if a jet launched by the progenitor star is inhibited by a dense envelope of material surrounding the star, a so-called choked jet. Due to this, the gamma-ray emission is suppressed, while neutrinos are still produced and can escape freely. These objects are an attractive UHECR and neutrino[65] source candidate, since their lower luminosity increases the survival chance of heavier nuclei (which can be an issue in high-luminosity GRBs (HL-GRBs)) and the absence of a strong gamma-ray flux can weaken constraints from multi-messenger searches. In addition, they are an interesting target for joint GW-neutrino searches, since they would still emit strong gravitational waves.

---

[65]Neutrino production might in this case occur through $pp$-interactions.

### 3.7.3   Starburst galaxies

The diffuse gamma-ray background has a guaranteed contribution from star-forming galaxies due to cosmic rays in these galaxies interacting with interstellar gas. This is similar to what happens in our own galaxy, where the gamma-ray sky is dominated by a diffuse galactic emission due to this process [615–618] (see also Figure 3.18a). Starburst galaxies (SBG) are a subset of star-forming galaxies which feature higher luminosities. They possess an increased star formation rate ($10 - 100 \, \mathrm{M_\odot \, yr^{-1}}$) compared to normal galaxies (e.g. $1 - 5 \, \mathrm{M_\odot \, yr^{-1}}$ for the Milky Way). Due to the rate with which new stars are born in these galaxies, many supernovae occur which can accelerate cosmic rays. If the cosmic rays are confined, they will eventually traverse a sufficient amount of interstellar gas to interact, producing a gamma-ray and associated neutrino flux. In fact, these sources can be completely calorimetric, converting all the energy of cosmic rays into photons and neutrinos. Starburst galaxies were first considered as interesting cosmic-ray reservoir neutrino sources in [619]. There, the observed synchrotron emission from starburst galaxies was used to calibrate the total power in relativistic electrons in these galaxies, which is directly related to the energy budget in cosmic rays. They find that the cumulative neutrino flux from such galaxies can reach the WB bound.

Due to the fact that these sources can be calorimetric, they can be described by very few parameters (as opposed to AGNs and GRBs). In general, the diffuse neutrino flux due to starburst galaxies can be written as [516, 518]

$$E_\nu^2 \Phi_{\nu_i} \approx \frac{c}{4\pi} t_H \xi_z \frac{1}{2} \min[1, f_{pp}] (E_p Q_{E_p}), \tag{3.53}$$

by combining Eq. (3.43) and Eq. (3.45). In this case, the efficiency with which proton energy is converted into (mainly) pion energy is given by $f_{pp} \approx n \kappa_p \sigma_{pp}^{\mathrm{inel}} c t_{\mathrm{int}}$, but is bounded from above by 1 due to conservation of energy. The interaction time $t_{\mathrm{int}} \approx \min[t_{\mathrm{inj}}, t_{\mathrm{esc}}]$, where $t_{\mathrm{inj}}$ is the injection time and $t_{\mathrm{esc}}$ is the escape time of cosmic rays, is the duration that the cosmic rays interact with the interstellar gas with density $n$. From this formula, we see that an energy budget of $Q_{cr} \approx 10^{46} \, \mathrm{erg \, Mpc^{-3} \, yr^{-1}}$ and efficiency $f_{pp} \approx 0.1$ is sufficient to supply the IceCube flux [516].

In order to fully explain the neutrino flux detected by IceCube, with neutrinos reaching PeV energies, cosmic rays need to be confined up to $\sim 100$ PeV. In our own galaxy, the energy scale up until which cosmic rays are (sufficiently) confined is given by the knee, well below this requirement. However, owing to their stronger magnetic field, starburst galaxies are possibly capable of containing cosmic rays up to the required $\sim 100$ PeV [619, 620]. Together, the increased energy budget of cosmic rays and the higher maximum energy scale for confinement lead to the possibility that starburst galaxies are the source of the diffuse neutrino flux [516, 619–621]. Moreover, the predicted neutrino spectrum features a clear break at PeV energies, since the parent cosmic rays are no longer confined above the corresponding energy. Therefore, the spectral shape of astrophysical neutrino around and beyond PeV energies is a key observable to verify whether starburst galaxies are good candidates sources for the neutrino flux.

It is important to note that the neutrino flux from starburst galaxies could in principle exceed the WB bound, since the energy in cosmic rays can be completely dissipated due to their confinement. The fact that the obtained energy budget in these galaxies is similar to the one inferred for UHECRs, seems coincidental, although [619] argues it might not be. Since normal galaxies and starburst galaxies make up a comparable amount of the total stellar mass reservoir in the local universe, this coincidence is automatic if starburst galaxies contain UHECR sources[66]. This also naturally explains the observed neutrino and UHECR spectra and energy budgets [622]. On the other hand, the connection to UHECRs is not required to explain neutrinos in starburst galaxy models, since cosmic rays only need to be accelerated up to 100 PeV to explain the observed neutrino flux. These energies can be reached without requiring the presence of UHECR sources.

### 3.7.4 Others

While the astrophysical objects mentioned above are the most popular neutrino source candidates, they are not the only ones. Examples of other candidates (all featuring $pp$-interactions) are halo and galaxy mergers [623], FR-0 radio galaxies [624] and large scale structures such as galaxy groups and galaxy clusters [516, 625, 626].

Recently, there has been an increased interest in another candidate class, namely tidal disruption events (TDE). These occur when a star is torn apart by a supermassive black hole, or possibly a white dwarf by an intermediate mass black hole. Half of the material is ultimately accreted and a relativistic jet can be launched. Such events provide an interesting explanation for the observed UHECR composition, due to a natural occurrence of higher mass nuclei. Moreover, due to the intense radiation fields efficient $p\gamma$-interactions can occur, giving rise to neutrino emission [453, 627–632].

## 3.8 Identifying the neutrino sources

In this section, we review the current status of experimental searches and phenomenological analyses attempting to identify the sources of the observed astrophysical neutrinos. While one particular object, a blazar, has been observed to emit neutrinos, it is not clear for how much of the total astrophysical neutrino flux this class can account. Therefore, the origin of the bulk of the neutrino flux remains unclear. However, strong constraints already exist on many source classes. In addition, it is possible to derive some general requirements on the unknown neutrino source population.

### 3.8.1 Experimental searches

Several searches have been performed by IceCube trying to identify the sources responsible for the diffuse astrophysical neutrino flux. The conceptually simplest one is to look for steady point sources in a time-integrated all-sky search for clustering of neutrino events, with no preference for a certain source class or location in the sky. While the HESE sample had already shown a lack of such clustering (Section 3.6.1), that analysis was performed with a very strict selection in order to get a high purity sample. One can increase statistics by loosening the selection, which in turn also significantly increases the background. The latest all-sky point source search[67] uses through-going muons (augmented by starting tracks in the southern sky) in seven years of data [532] over the full sky, recording 712830 events. The analysis achieves a discovery potential[68] for steady sources in the northern sky below $E_\nu^2 \Phi_\nu(E) = 10^{-9}$ GeV cm$^{-2}$ s$^{-1}$. The discovery potential over the entire sky is shown in Figure 3.21a. The results of the analysis are shown in Figure 3.21b, with no significant clustering above background being detected. In addition, the same analysis also investigates a list of promising sources in the northern and southern sky, selected on the basis of e.g. gamma-ray data, by only testing for a signal flux from those directions. The northern sky search includes mostly extragalactic objects (Bl Lacs, FSRQs and radio galaxies), while the southern sky search focuses mostly on galactic objects (pulsar wind nebulae, supernova remnants). Again, no significant signal above background was detected.

---

[66]Which is then compatible with the observed correlation between UHECR directions and starburst galaxies, as mentioned in Section 3.2.6.

The results from the all-sky clustering search indicate that the total diffuse flux is not dominated by a few individually powerful sources. Still, it might be that a single source class is responsible for the diffuse flux. In order to improve on the sensitivity to such a population, one can do a stacked search. By adding the flux from the directions of the sources expected to be the strongest in a particular class, one can increase the signal flux with respect to the background. There exists an optimal number of sources to be included in such a search, which depends on the assumed distribution of signal strength amongst the sources, since including the weakest sources would increase the background more than the signal. In the absence of a detected signal, the result of a stacked search is a limit on the total flux from the investigated set of sources. In case there is a good estimate of how the flux from the selected sources is related to the flux from the total population (i.e. how the sources are distributed in luminosity and redshift), this limit can also be converted in a limit on the total diffuse flux from this population [634].

One source class which has been extensively investigated in various analyses is blazars. One of these performs a search for neutrinos from blazars in the Fermi-LAT 2LAC catalogue [635] in three years of data [533]. The search considers both the full population (862 sources) as well as several sub-populations (e.g. FSRQs, low synchrotron-peaked Bl Lacs,...) from this catalogue, but finds no significant signal. As a result, the neutrino emission from all 2LAC blazars is limited to $< 30\%$, the exact limit depending on the spectral index and weighing scheme across the population and going as low as below 10%. This result is shown in Figure 3.22a. This same search also strongly constrains some of the models mentioned in Section 3.7. A more recent search using seven years of data [636] investigated several other blazar selections: the sources in the 2WHSP catalogue [637], a multi-frequency selected catalogue of high synchrotron peaked blazars (HSP) not relying on gamma-ray data, HSP Bl Lacs in 2FHL [638] which is a list of ∼ 360 sources detected above 50 GeV by Fermi-LAT in ∼ 6 years of data and FSRQs in the 3LAC catalogue [481]. The results are similar to the previous search, limiting the contribution of these blazars to the diffuse neutrino flux to only few percent[69].

---

[67]There exists an updated analysis using 8 years of data and improved event selection and reconstruction techniques [633]. However, since that analysis only scans over the Northern hemisphere, we discuss the 7 year search instead. The updated analysis improves the sensitivity and discovery potential with roughly a factor of 2 in the Northern hemisphere, compared to the 7 year analysis.

[68]The discovery potential is defined as a false-positive rate of $5\sigma$ or $2.87 \times 10^{-7}$ with false-negative of 50%. The sensitivity is defined as a false-positive rate of 50% with false-negative of 10% [532].

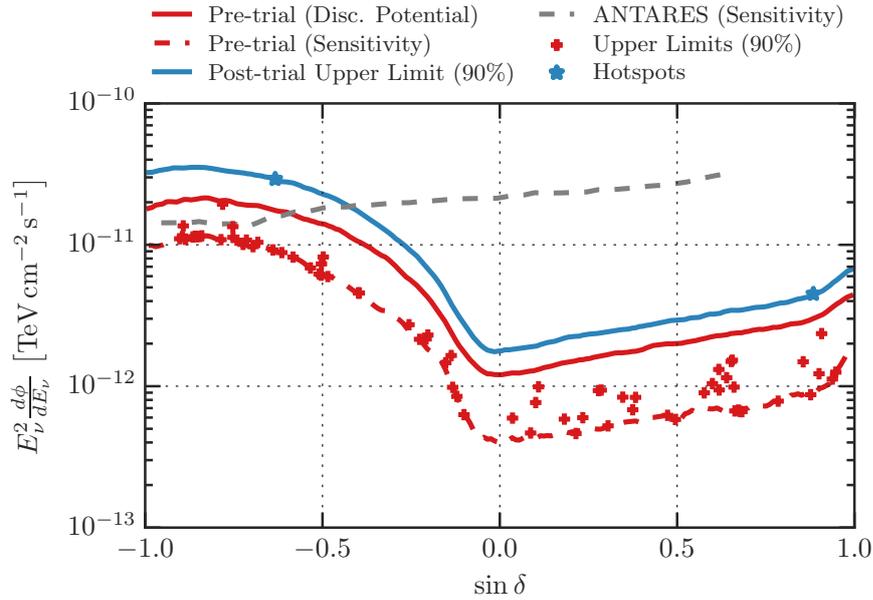

(a) Point source sensitivity

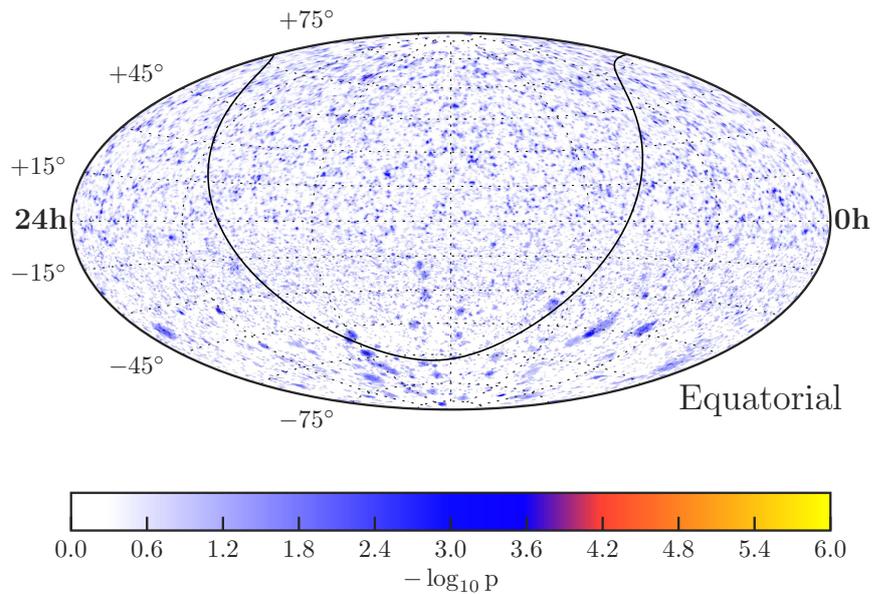

(b) All sky clustering

Figure 3.21: IceCube results from the all-sky search for integrated neutrino emission using 7 years of data [532]. (a) Point source sensitivity, discovery potential and obtained upper limit for an unbroken $\nu_\mu + \bar{\nu}_\mu$ $E^2$-spectrum as a function of declination. The most significant spots in the Northern and Southern hemisphere are indicated with stars. The sensitivity is given by the requirement that at any declination $\delta$, a flux at this level would be detected 90% of the time with a significance greater than that of the most significant spot. Red pluses are individual upper limits on a list of pre-selected sources. The sensitivity is worse in the southern hemisphere due to the atmospheric muon background. (b) Results showing the pre-trial p-value for clustering in each direction. No significant clustering has been detected.

In the case of transient sources, one can not only restrict the search towards a specific direction, but can also use the timing information in order to search only in a limited time window surrounding the event, eliminating atmospheric background. On the other hand, if the neutrino flux arrives at a time significantly different from the arrival time of the electromagnetic emission and this is not taken into account in the analysis, a possible signal might be missed completely unless an individual signal is powerful enough to trigger the detector multiple times within a short time span. This was recently applied in the archival analysis of the blazar TXS 0506+056 [639].

A search for neutrinos from gamma-ray bursts during the prompt phase[70] has been performed by IceCube using seven years of data in the northern sky and five years of data in the southern sky [603], investigating a sample of 1172 GRBs. The results are consistent with background and the total contribution from GRBs to the diffuse flux is constrained to ≤ 1%. Although it should be noted that this analysis tests for emission during the prompt phase and neutrino emission outside of this phase could be missed. An analysis with an expanding time window was performed in [640], where it is seen that at time scales larger than ∼ 30 s, the sensitivity worsens significantly. Another method looking for precursor and afterglow neutrinos has been developed in [641] and is currently still being improved. The constraints on the neutrino flux from GRBs are shown in Figure 3.22b. This also shows that typical models where GRBs are responsible for the UHECR flux (through proton escape [599] or neutron escape [602]) are already excluded. Note, however, that it is still possible that GRBs supply (part of) the UHECR flux if the neutrino flux is significantly lower than these models (or if it arrives significantly earlier or later than the gamma rays). For example, multi-zone models briefly mentioned in Section 3.7 are still below the sensitivity of this search.

Neutrinos from blazar flares, i.e. transient emission, have also been investigated. The expected sudden increase of neutrino emission from a specific location and during a restricted time window makes for a very promising target, since the background can be suppressed. A currently ongoing analysis adds to this strength by also stacking the various flares (each with their own time window) [636]. Recently, the coincident detection of a high-energy neutrino with the blazar TXS 0506+056 in a flaring state confirmed that blazar flares are not only interesting candidates, but actual emitters of neutrinos. This discovery will be discussed in more detail below.

More generally, a realtime alert system has been set up [642] which can quickly identify possibly interesting neutrino events to be followed up by other observatories (of neutrinos or electromagnetic radiation).

---

[69]In the case of hard spectra, with a spectral index of −1, these sources can still explain ∼30% of the diffuse flux at PeV energies

[70]The analysis searches for neutrinos within a time window given by the earliest reported start time and latest report stop time for each GRB individually, with the signal dropping of with Gaussian tails outside of this window.

One important population which has been absent from this discussion is the starburst galaxies. Due to their high density, but individually low luminosity, these sources are currently beyond the sensitivity of IceCube. A future upgrade of the detector, however, should be able to constrain this source class as well [622].

Finally, the search for neutrinos from sources of gravitational waves (mainly compact binaries: binary neutron stars, binary black holes and neutron star-black hole mergers) will be discussed in Chapter 5.

### 3.8.2 Neutrinos from TXS 0506+056

On September 22, 2017, IceCube detected a high-energy neutrino, designated as IceCube-170922A, from a direction coincident with the blazar TXS 0506+056, which was in a flaring state [643]. Searching for neutrinos from the same direction in archival data, another, independent, neutrino flare (not associated with a gamma-ray flare) was found [639], this time with multiple neutrinos emitted over a time window of $\sim 150$ days. Combined, these two observations represent the first definitive detection of a high-energy neutrino source, namely a blazar flare. Consequently, this is also the first identification of a very-high energy cosmic ray source[71].

The detection of the neutrino in the initial alert, IceCube-170922A, as well as multiwavelength observations from the same direction, are described in [643] and summarised here. The neutrino event IceCube-170922A has an estimated energy 290 TeV ($23.7 \pm 2.8$ TeV energy deposited in the detector) and a signalness, i.e. the likelihood of it being of astrophysical origin [642], of 50% [643]. However, by itself, at several hundred TeV an atmospheric origin of this neutrino can not be ruled out. Due to the high signalness of this event, GCN and AMON alerts were sent out to observatories of neutrinos and electromagnetic radiation.

The neutrino was coincident with the direction of the blazar TXS 0506+056, detected by Fermi. This blazar was in a flaring state, i.e. enhanced gamma-ray activity in the GeV range, at the time of the neutrino detection and for the six months prior[72]. The recently measured redshift of this blazar is $z = 0.3365$. The integrated gamma-ray flux, for $E > 0.1$ GeV, averaged over 9.5 years of Fermi observations is $(7.6 \pm 0.2) \times 10^{-8}$ cm$^{-2}$ s$^{-1}$. A statistical analysis was performed, considering all public neutrino alerts and archival events which would have passed the alert filter, and comparing this with Fermi-LAT sources. It was found that, for various correlations between the gamma-ray an neutrino emission, the chance coincidence of a neutrino with a flaring blazar is disfavoured at the $3\sigma$ level.

TXS 0506+056 was also investigated at higher energies by imaging air Cherenkov telescopes (IACTs). H.E.S.S. and VERITAS did not detect anything significant and could only put an upper limit on any possible very-high energy gamma-ray flux. MAGIC,

---

[71]This does not necessarily imply that this source can also accelerate cosmic rays to ultra-high energies.
[72]Which by itself is not unusual for a blazar and would not have received detailed follow-up.

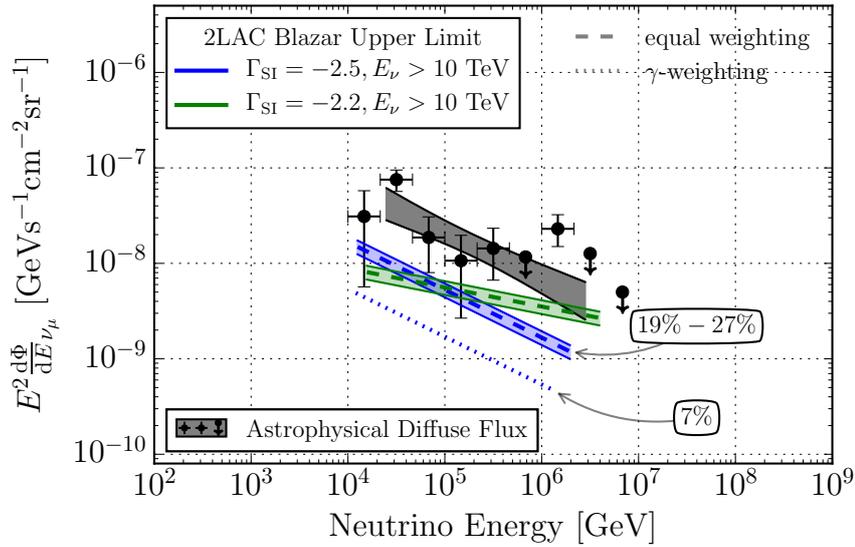

(a) Blazar limits (all 2LAC blazars)

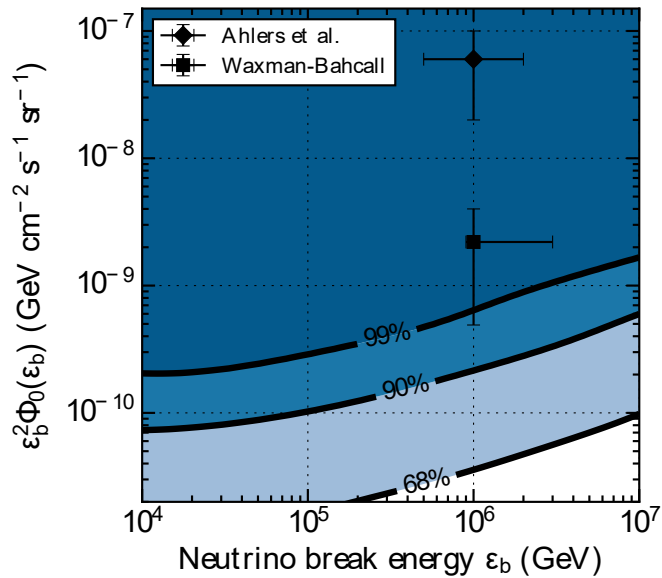

(b) GRB limits

Figure 3.22: Limits on blazar and GRB contributions to the astrophysical neutrino flux. (a) Limits on the astrophysical neutrino flux from all blazars in the 2LAC catalogue, for different assumptions on the spectral index [533]. Their contribution to the astrophysical neutrino flux is limited to below 30%. (b) Limits (for different confidence levels) on the prompt neutrino flux from GRBs as a function of the break energy [603]. The two reference models explaining the UHECR flux with GRBs (Section 3.7.2) are now excluded.

on the other hand, did detect photons a few days after[73] the IceCube alert and found $374 \pm 62$ excess photons, which represents a $6.2\sigma$ excess above expected background, between 80 GeV and 400 GeV. The inferred flux is $\mathrm{d}N/\mathrm{d}E \propto E^{-\gamma}$ with $\gamma = -3.9 \pm 0.4$ and normalisation $(2.0 \pm 0.4) \times 10^{-10}$ TeV$^{-1}$ cm$^{-2}$ s$^{-1}$ at $E = 130$ GeV. At even higher energies, above 1 TeV, HAWC did not observe any gamma rays. At lower energies, the radio emission from this object had been gradually increasing 18 months prior to the alert, while the optical emission in the V band was at its highest in recent years. Similarly, there was an increase in the X-ray emission.

Building the spectral energy distribution (SED), shown in Figure 3.23, one finds that all these components connect smoothly with each other and are consistent. Comparing with archival observations, the emission from TXS 0506+056 shows a clear increase over all wavelengths.

The isotropic gamma-ray luminosity from TXS 0506+056 between 0.1 GeV and 100 GeV is $1.3 \times 10^{47}$ erg s$^{-1}$ averaged over the $\pm 2$ weeks around the alert and $2.8 \times 10^{46}$ erg s$^{-1}$ over all Fermi-LAT observations. On the other hand, the neutrino fluence for which one expect a single event over all IceCube observations is $2.8 \times 10^{-3}$ erg cm$^{-2}$, assuming a spectrum with power-law index $-2$ between 200 TeV and 7.5 PeV. This results in a luminosity of $7.2 \times 10^{46}$ erg s$^{-1}$ if the neutrino occurs over a 6 month period (when the blazar was in a flaring state) or $4.8 \times 10^{45}$ erg s$^{-1}$ if the emission occurs over 7.5 years. Therefore, the neutrino and gamma-ray luminosities are comparable, suggesting that they are related. The associated neutrino flux for these two time windows is also shown in Figure 3.23. However, the estimate of the flux (and thus also luminosity) from this source is dependent has an important caveat. Even if the flux was much lower than this, such that the expected number of neutrino events from this source is much lower than 1, one would still expect to see a neutrino if there are many such sources. Finally, the observation of a single neutrino does not allow one to probe the production mechanism or the exact relation between the gamma-ray and neutrino luminosities.

The investigation of neutrino emission from the direction of the blazar TXS 0506+056 prior to the IceCube-170922A alert in 9.5 years of data is described in [639] and summarised here. The analysis discovered a neutrino flare in 2014-2015, with $13 \pm 5$ neutrino events above a background of 5.8 events in a one degree search radius during the 158-day time window identified during the analysis. At the time, the blazar TXS 0506+056 was not in a flaring state.

The analysis consisted of two independent analyses. The first one is a time-dependent unbinned maximum-likelihood ratio search, where a high-significance point source detection can require 2-3 neutrinos up to 30, depending on the energy spectrum and time clustering. The analysis considered both a Gaussian time window and a box-shaped time window. For the latter, a time window of 158 days was identified[74], with a fluence of $E^2 J_{100} = 2.2^{+1.0}_{-0.8} \times 10^{-4}$ TeV cm$^{-2}$ (normalised at 100 TeV) and a spectral index $\gamma = 2.2 \pm 0.2$. This leads to an average flux over this time window of $\Phi_{100} = 1.6^{+0.7}_{-0.6} \times 10^{-15} \, \mathrm{TeV}^{-1} \, \mathrm{cm}^{-2} \, \mathrm{s}^{-1}$. This excess has a significance of $3.5\sigma$. The neutrinos from the direction in TXS 0506+056 are shown in Figure 3.24.

---

[73]It already performed observations on September 24 and did not observe significant emission at that point.

[74]The analysis also finds the IceCube-170922A flare, which is driven purely by that single event.

The second analysis is a time-integrated unbinned maximum-likelihood ratio search. It identified a flux of $\Phi_{100} = 0.8^{+0.5}_{-0.4} \times 10^{-16} \, \mathrm{TeV}^{-1} \, \mathrm{cm}^{-2} \, \mathrm{s}^{-1}$ with spectral index $\gamma = 2.0 \pm 0.3$ over 9.5 years from the position of TXS 0506+056, with a significance of $4.1\sigma$. However, this is an a posteriori significance, since it includes the event which motivated the analysis in the first place. Taking into account the look-elsewhere effect, one expects two or three directions with equal or greater significance from the Northern hemisphere[75]. However, since this analysis tries to answer whether there is any evidence for additional neutrinos from the specific direction of TXS 0506+056, this is not an issue. Removing the final data taking period which contains the initial alert and performing the same analysis on only 7 years of data, the inferred flux is almost unchanged at $\Phi_{100} = 0.9^{+0.6}_{-0.5} \times 10^{-16} \, \mathrm{TeV}^{-1} \, \mathrm{cm}^{-2} \, \mathrm{s}^{-1}$ and $\gamma = 2.1 \pm 0.3$, with a significance of $2.1\sigma$. Therefore, the excess is dominated by the neutrino flare and finds a similar emission as the time-dependent analysis[76].

Assuming that the muon neutrino flux represents $1/3$ of the total neutrino fluence, the implied isotropic neutrino luminosity is $1.2^{+0.6}_{-0.4} \times 10^{47} \, \mathrm{erg} \, \mathrm{s}^{-1}$ averaged over 158 days over the energy range between 32 TeV and 3.6 PeV. This is higher than the gamma-ray luminosity, which is given by $0.28 \times 10^{47} \, \mathrm{erg} \, \mathrm{s}^{-1}$ between 0.1 GeV and 100 GeV averaged over all Fermi-LAT observations of TXS 0506+056. Therefore, since gamma rays are coproduced with neutrinos, a significant part of them must be either absorbed or arrive outside the Fermi band.

One can question why TXS 0506+056 is the first identified source of neutrino emission. While it was not previously identified as a promising target for neutrino emission, it is among the 50 brightest objects in the third catalogue of active galactic nuclei by Fermi [481] (out of 1591 AGN, most of which are blazars). Its measured redshift immediately places it among the most luminous objects out to that distance, orders of magnitude more than Mkn 421 or Mkn 501. Moreover, these latter objects are at more northern declinations, such that $\sim 300$ TeV neutrinos from their direction are 3 to 5 times more likely to be absorbed by Earth than that they reach the detector. Therefore, it seems that TXS 0506+056 was optimal both because of its high intrinsic luminosity and its favourable declination.

The identification of a neutrino from blazars is compatible with the searches for neutrinos from catalogued blazars, which put an upper limit on the contribution of Fermi blazars to the astrophysical neutrino flux[77] of 27% for a spectral index $-2.5$ fit between 10 TeV and 100 TeV and 40%–80% for a spectral index $-2$ compatible with the diffuse flux fit above $\sim 200$ TeV, depending on the spectral index and energy range under consideration. TXS 0506+056 itself is responsible for 1% of astrophysical flux.

Combined, the 2017 neutrino/gamma-ray flare with the 2014-2015 neutrino flare here

---

[75]This was expected, since this blazar was not previously identified in a time-integrated search which did not consider a preferred direction in the sky.

[76]With lower significance due to the integrated background.

[77]The contribution from all blazars is higher, since also unresolved sources contribute.

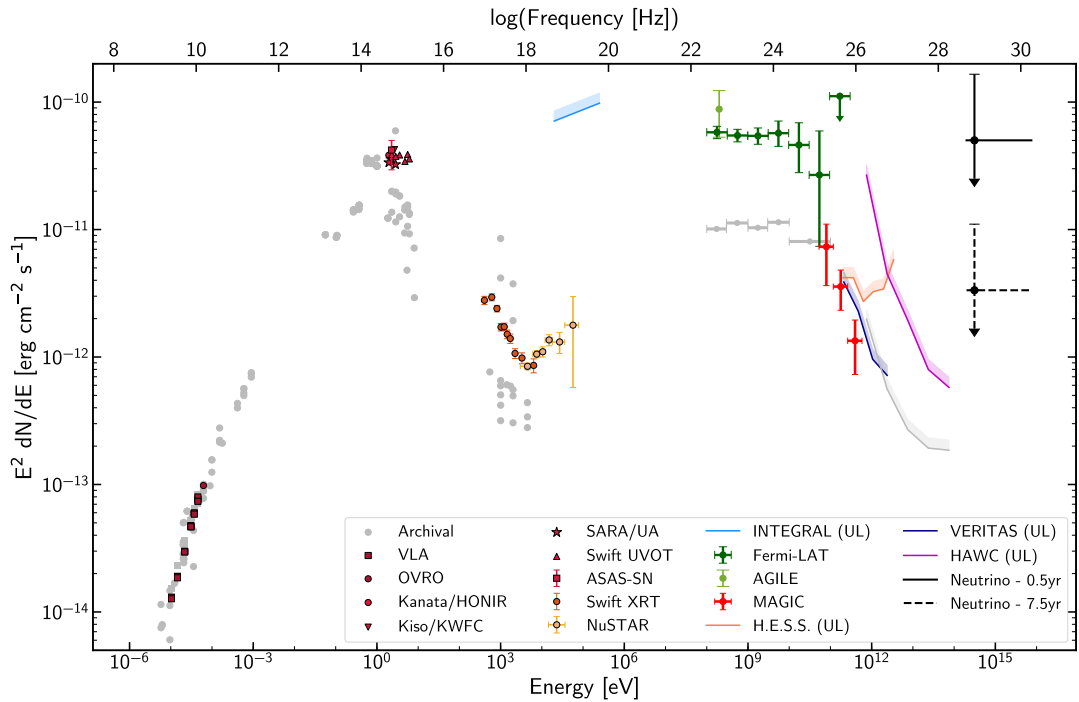

**Figure 3.23:** The spectral energy distribution (SED) of TXS 0506+056 in a $\pm 2$ week period around the detection of IceCube-170922A. All observations lead to a consistent picture. Comparing with archival observations in grey, the blazar is clearly in a state of enhanced emission. The downwards arrow on the neutrino data point indicates that for a single neutrino detection, one can not exclude that the actual flux is much lower (see text). Figure from [643].

suggest that blazar flares are indeed a source of high-energy cosmic rays and neutrinos

Together with the IceCube analysis, several other follow-up papers were published. ANTARES did not find any significant neutrino emission, either around the alert, during the neutrino flare or in a time-integrated search [644]. The MAGIC collaboration interpreted the measured SED in a one-zone lepto-hadronic model and finds that most of the gamma-ray emission must come from leptonic processes [645]. They infer a maximum proton energy of $10^{14} - 10^{18}$ eV. VERITAS [646] finds no significant gamma-ray emission ($E_\gamma > 100$ GeV) in a two week period immediately following the alert. Several interpretation papers were also published simultaneously. Dissecting the region around IceCube-170922A [647], one finds that the nearby blazar PKS 0502+049 contaminates the gamma-ray signal during the neutrino flare, but only at low energies. Above a few GeV, TXS 0506+056 dominates and is in a low but hard state. The analysis in [648] finds that the emission during the gamma-ray flare can only be explained by a hybrid scenario with both leptonic and hadronic emission, the latter of which is subdominant. In order to arrive at this conclusion, it is important to take into account the electromagnetic cascade and the resulting emission in the X-ray band. Similarly, the analysis [649] explains the 2017 flares with leptohadronic single-zone[78] models, where the hadronic component is subdominant. Another paper modelling the SED [650] finds that a moderate enhancement of the cosmic rays during the gamma-ray flare can explain the strong increase of neutrino flux. Again, it is found that the co-produced hard X-rays (and TeV gamma rays) give important constraints.

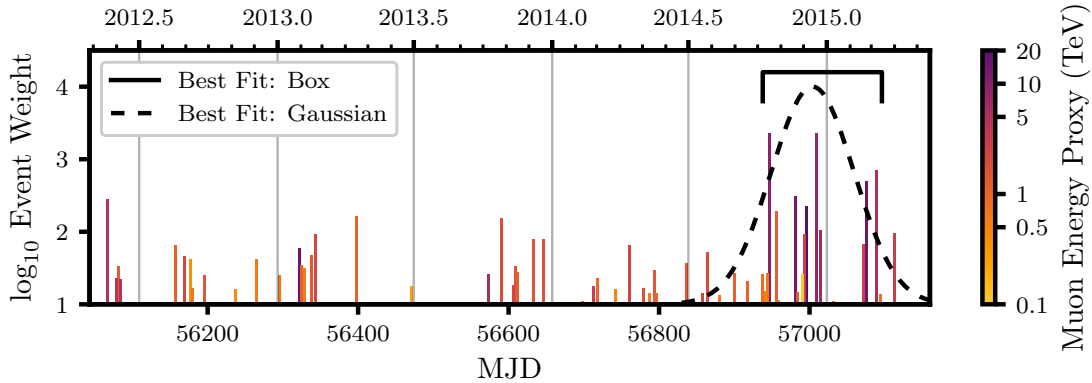

Figure 3.24: Distribution of neutrino events in time and energy from the direction of TXS 0506+056. A clear accumulation of events can be seen during the identified 2014-2015 flare. Superimposed is the inferred Gaussian time window. Figure from [639].

The detection sparked the publication of many other attempts to explain the emission from TXS 0506+056. The emissions during the gamma-ray flare [651, 652] and during the neutrino outburst [653] have been explained with $pp$-interaction models. The authors of [654] find that the constraints from electromagnetic cascades imply that neutrinos and gamma rays in the 2014-2015 flare can not originate in the same process. In [655], it was found that a leptohadronic model for the 2014-2015 neutrino flare has trouble explaining the high neutrino event number without violating the limit on gamma-rays and X-rays. However, in [656], it was found that two-zone models can naturally satisfy the X-ray constraint. In [654], it is found that TXS 0506+056 has been wrongly classified as a BL Lac object (partly due to its synchrotron peak being two orders of magnitude larger than expected from the blazar sequence). Rather, it is an FSRQ with emission lines diluted by strong Doppler-boosted jet. This has implications for detailed modelling of the SED, since the strength of the radiation fields and their place of origin are different than for BL Lac. Finally, when considering photons from radiatively inefficient accretion flows as the target fields [657], it is possible to explain the emission from TXS 0506+056, while predicting negligible emission from other high-synchrotron-peak BL Lacs (such as Mkn 421 and Mkn 501).

Finally, the consequences on the origin of the diffuse neutrino flux have also been investigated. If TXS 0506+056, with its emission dominated by a single neutrino flare, is representative for this type of sources, then it will be difficult to chase these sources without larger neutrino telescopes [658]. Considering blazars as the origin of high-energy astrophysical neutrinos [656], it was found that blazars are likely subdominant at sub-PeV energies, with flares like TXS 0506+056 making out <1–10% of the total flux. Comparing the neutrino flux with blazar catalogues [659], it was found that at most 5–15% of the diffuse flux can originate in blazars, with the significant neutrino emission from TXS 0506+056 as an extreme outlier. Similarly, the neutrino flux from blazar flares was investigated in [660]. Considering twelve blazars, they find that these only produce few neutrinos, with the two strongest sources producing three muon neutrinos over ten years, suggesting the need for larger instruments. On the other hand, the authors in [661] interpreted the diffuse neutrino flux in terms of the blazar sequence. They find that they can explain the full diffuse neutrino flux, the emission from TXS 0506+056 and satisfy the stacking limits on blazars. In their scenario, the dominant contribution to the neutrino flux must come from unresolved BL Lacs with large baryonic loading (ratio of proton to gamma-ray luminosities). They predict $\sim 0.3$ gamma-ray/neutrino associations per year from the whole population, mostly from BL Lacs.

---

[78]I.e. the enhanced particle acceleration and emission of all components occurs in a single place.

### 3.8.3 Population density and luminosity

As seen in the previous sections, the population of astrophysical sources responsible for the diffuse neutrino flux has not been identified yet. However, by combining the non-detection of such a population with the requirement that it needs to supply the detected astrophysical flux, it is possible to infer some general properties of the neutrino source population.

   In the following argument, we will consider only steady sources. The main idea is that each individual neutrino source needs to have a low neutrino luminosity in order to not have shown up in any point source search. On the other hand, this means that the population needs to be sufficiently numerous in order to supply the detected astrophysical neutrino flux[79]. As we will see, this constraint is already strong enough to rule out certain powerful, but rare, sources such as Bl Lacs. This argument has already been applied by several authors in order to constrain the source population [487, 522, 622, 663–665]. Here, we will show the analysis of the most recent one [622]. The constraints are only derived from the muon neutrino flux above 100 TeV.

   Consider a population of standard candle sources, i.e. with fixed luminosity $L_{\nu_\mu}$, with local ($z = 0$) number density of $n_0^{\mathrm{tot}}$. In the case of sources with a luminosity distribution (as is realistic), we can define instead an effective local number density $n_0^{\text{eff}}$ of sources with a constant effective luminosity $L_{\nu_\mu}^{\text{eff}}$. These quantities can be constructed such that they characterise the sources dominating the neutrino flux from this population. On the other hand, $n_0^{\text{tot}}$ is then typically determined by the weakest sources in this population.

---

[79]A similar argument can be used for transient sources. However, a transient object only leads to neutrino emission from a fixed direction in the sky for a limited time (related to this is the fact that one can only put a limit on the fluence over the observed window, and not the flux, without additional information on the time window, see also [662]). Unless the emission during this time interval is sufficiently above background, it will not show up in a clustering search integrated over time. Still, at least in principle, a similar bound on the number density of sources could be obtained. However, as time increases, the number of expected neutrino events per source/position does not increase as with steady sources. Instead, the number of sources (e.g. number of GRBs which occurred over the analysis time) increases.

The statement that no point source has been identified yet is equivalent to the statement that no neutrino multiplets have been detected above 100 TeV, where background is negligible (see also Sections 3.6.1 and 3.8.1). The expected number of sources which give rise to at least $k$ neutrino events is given by

$$N_{m \geq k} = \int \mathrm{d}\mathcal{V} \, n_s^{\text{eff}}[z] P_{m \geq k}[z], \tag{3.54}$$

where we integrate over the entire observable universe. The comoving source density at $z$ is given by $n_s^{\text{eff}}[z] = n_0^{\text{eff}} \mathcal{H}(z)$, where $\mathcal{H}(z)$ contains the redshift evolution of the source. The probability that a single source at redshift $z$ leads to at least $k$ events in the detector is given by $P_{m \geq k}[z]$. The Poissonian probability to detect at least two events is given by $P_{m \geq 2}(\lambda) = 1 - (1 + \lambda)\exp(-\lambda)$, where $\lambda[z]$ is the average number of detected events produced by a source at redshift $z$. This $\lambda$ therefore depends on the sensitivity of the detector and can be expressed using the luminosity distance $d_{\mathcal{N}=1}$ for which a source produces one event, such that $\lambda[z] = (d_{\mathcal{N}=1}/d_L[z])^2$, with $d_L[z]$ the luminosity distance to a source at redshift $z$. The luminosity distance $d_{\mathcal{N}=1}$ can be found from the point source sensitivity. As seen in Section 3.8.1, IceCube sets a 90% CL upper limit of $E_\nu^2 \Phi_{\nu_\mu} < F_{\text{lim}} \approx 10^{-9}$ GeV cm$^{-2}$ s$^{-1}$ to the muon neutrino flux produced by a possible point source. This flux limit corresponds to an expected average number of events[80] of $\lambda < 2.44$. Therefore, from the flux corresponding to 1 event, we find

$$d_{\mathcal{N}=1} \approx \left( \frac{E_\nu L_{E_{\nu_\mu}}^{\text{eff}}}{4\pi F_{\text{lim}}/2.4} \right)^{1/2} \simeq 110 \text{ Mpc} \left( \frac{E_\nu L_{E_{\nu_\mu}}^{\text{eff}}}{10^{42}\text{erg s}^{-1}} \right)^{1/2} F_{\text{lim},-9}^{-1/2},$$

where we used the source differential muon neutrino luminosity $L_{E_{\nu_\mu}}^{\text{eff}} = \mathrm{d}L_{\nu_\mu}^{\text{eff}}/\mathrm{d}E_\nu$ and $F_{\text{lim}} = 10^{-9} F_{\text{lim},-9}$ GeV cm$^{-2}$ s$^{-1}$. The quantity $d_{\mathcal{N}=1}$ then encodes all the information from the detector.

Combining all this, the average number of steady sources producing multiple events is given by

$$N_{m \geq 2} = n_0^{\text{eff}} \Delta\Omega \int \mathrm{d}z \, \frac{(c/H_0)d_L^2[z]}{(1+z)^2 \sqrt{\Omega_m(1+z)^3 + \Omega_\Lambda}} \times \left( \frac{n_s^{\text{eff}}[z]}{n_0^{\text{eff}}} \right) P_{m \geq 2}(\lambda[z]), \tag{3.55}$$

where $\Delta\Omega$ is the solid angle covered by the detector. The first factor in the integral comes from the volume element in an expanding universe, see also Section 4.4.2 in the next chapter. In the case that $d_{\mathcal{N}=1} \ll c/H_0$, we can ignore the cosmology factor and the integral approximates to

$$N_{m \geq 2} \approx \sqrt{\pi} \left( \frac{\Delta\Omega}{3} \right) n_0^{\text{eff}} d_{\mathcal{N}=1}^3. \tag{3.56}$$

---

[80]Following the Feldman-Cousins method, a 90% CL upper limit for detected 0 events with an expected background of 0 corresponds to an upper limit of 2.44 [342].

For higher multiplets, the result is similar, with a different prefactor.[81]

Setting $N_{m \geq 2} < 1$, we find a constraint on the combination of number density and luminosity

$$n_0^{\text{eff}} \left( \frac{E_\nu L_{E_{\nu\mu}}^{\text{eff}}}{10^{42}\text{erg s}^{-1}} \right)^{3/2} F_{\text{lim},-9}^{-3/2} \lesssim 1.9 \times 10^{-7} \text{Mpc}^{-3} q_L^{-1} \left( \frac{2\pi}{\Delta\Omega} \right). \tag{3.57}$$

The factor $q_L$ comes from the redshift evolution in Eq. (3.55) and is equal to $q_L = 2.0$ for $\mathcal{H}(z) \propto (1+z)^3$ and to $q_L = 0.94$ for $\mathcal{H}(z) \propto (1+z)^0$. This result is an upper limit on the number density, which depends on the luminosity. This limit is valid for any source class with steady emission, regardless of whether or not it gives a significant contribution to the total diffuse neutrino flux.

Next, we demand that the population under consideration is responsible for the observed diffuse neutrino flux. This constraint can be found by repeating the same argument as for the WB-flux in Section 3.3.5, but using now immediately the neutrino luminosity instead of the cosmic-ray luminosity. The result is

$$n_0^{\text{eff}} \left( \frac{E_\nu L_{E_{\nu\mu}}^{\text{eff}}}{10^{42}\text{erg s}^{-1}} \right) \simeq 1.6 \times 10^{-7} \text{ Mpc}^{-3} \, (3/\xi_z) \times \left( \frac{E_\nu^2 \Phi_{\nu\mu}}{10^{-8} \text{ GeV cm}^{-2} \text{ s}^{-1} \text{ sr}^{-1}} \right). \tag{3.58}$$

As seen previously, $\xi_z$ encodes the cosmological evolution of the sources. For an evolution $\propto (1+z)^m$ with $m = 0$, we have $\xi_z = 0.6$, while for $m = 3$, we have $\xi_z = 3$ [480]. For an evolution following star formation rate, we have $\xi_z = 2.4 - 2.8$ [666]. For blazars, we have FSRQ with a strong evolution $\xi_z = 8.4$ and Bl Lacs with a weak evolution $\xi_z = 0.68$ [667].

By combining the constraints from the absence of multiplets (Eq. (3.57)) with the demand of supplying the diffuse neutrino flux (Eq. (3.58)), we can independently constrain the local effective number density and luminosity of the astrophysical neutrino sources,

$$\left( \frac{E_\nu L_{E_{\nu\mu}}^{\text{eff}}}{10^{42}\text{erg s}^{-1}} \right) \lesssim 1.4 \, q_L^{-2} \left( \frac{\xi_z}{3} \right)^2 F_{\text{lim},-9}^3 \left( \frac{\Delta\Omega}{2\pi} \right)^{-2}, \tag{3.59}$$

$$n_0^{\text{eff}} \gtrsim 1.1 \times 10^{-7} \text{ Mpc}^{-3} q_L^2 \left( \frac{\xi_z}{3} \right)^{-3} F_{\text{lim},-9}^{-3} \left( \frac{\Delta\Omega}{2\pi} \right)^2. \tag{3.60}$$

---

[81]In the case of a non-negligible background, the same integral without the $P_{m \geq 2}(\lambda[z])$-factor can be used to find the total number of identifiable sources. In this case, the integral needs to be cut off at $z_{\text{lim}}$ up to which point point sources can be identified. This results in $N_{\text{lim}} = \left( \frac{\Delta\Omega}{3} \right) n_0^{\text{eff}} d_{\text{lim}}^3$ [522, 622].

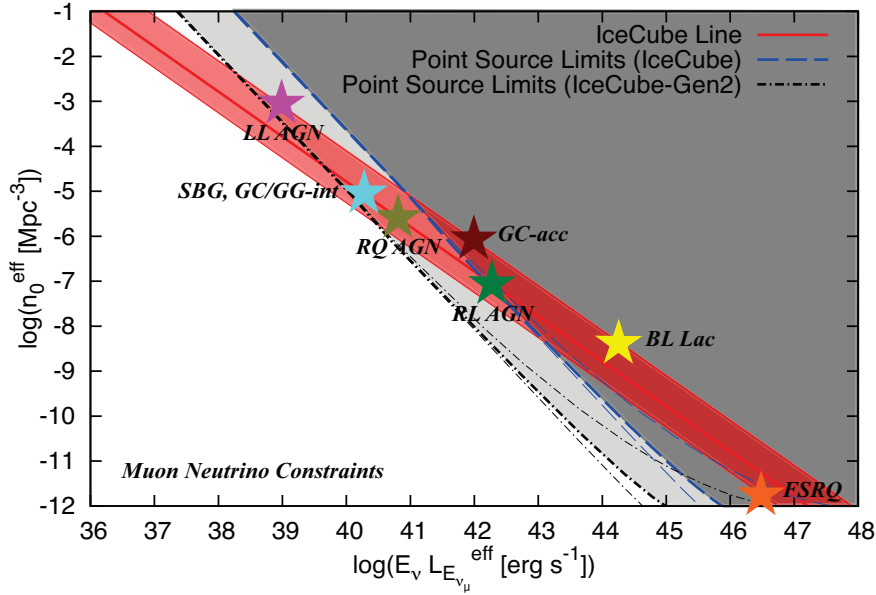

Figure 3.25: Constraints on the local source density and evolution by combining point source limits with the requirement of reproducing the astrophysical neutrino flux observed by IceCube as found by [622]. The analysis assumes an $E^{-2}$-spectrum. Constraints from the IceCube flux shown are by the red band from no source evolution (up) to strong evolution (down). The grey region shows the exclusion from six years of IceCube (dark grey) and ten years with the future IceCube-Gen2 facility (light grey), thick lines for a source evolution following star formation rate, thin lines for no evolution (up) and strong evolution (down). Several source classes are indicated by stars.

Now, in contrast to the previous luminosity-dependent upper limit, we find a lower limit on the local effective number density, together with an upper bound on the effective luminosity. This was the expected result: we need a numerous population of weak neutrino sources. Figure 3.25 shows the same constraints resulting from a numerical calculation [622] as well as the typical effective number density and luminosity associated to several often-considered neutrino sources.

These constraints already disfavour several important source classes as the (dominant) source of the observed astrophysical neutrinos. Again, we follow the calculations in [622], where typical expected neutrino luminosity distributions are considered and the effective neutrino luminosity of each source class is fixed such that the population can supply the observed diffuse flux. FSRQ have $n_0^{\mathrm{eff}} \sim 2 \times 10^{-12}\ \mathrm{Mpc}^{-3}$, significantly below the lower limit. On the other hand, their total number density $n_0^{\mathrm{tot}} \sim 10^{-9}\ \mathrm{Mpc}^{-3}$ is still allowed (by comparing with the limit on $n_0$ above when taking into account the strong evolution factor), but this requires that also FSRQ with weak electromagnetic emission contribute significantly to the neutrino flux. Bl Lac have $n_0^{\mathrm{eff}} \sim 5 \times 10^{-9}\ \mathrm{Mpc}^{-3}$ and are therefore too rare, even for weak evolution. In addition, the total number density $n_0^{tot} \sim 10^{-7}$ Mpc$^{-3}$ is also too low[82]. Therefore, the blazars seem to be ruled out as the dominant source of the diffuse neutrino flux, which is similar to the conclusion found in Section 3.8.1.

Table 3.1: Summary of the number density of important candidate neutrino source classes [622].

| | $n_0^{eff}$ (Mpc$^{-3}$) | $n_0^{tot}$ (Mpc$^{-3}$) | Constraint |
|---|---|---|---|
| FSRQ | $2 \times 10^{-12}$ | $10^{-9}$ | *disfavoured* |
| Bl Lac | $5 \times 10^{-9}$ | $10^{-7}$ | *disfavoured* |
| Starburst | $10^{-5}$ | $3 \times 10^{-5}$ | **allowed** |
| GC/GG | $10^{-5}$ | $5 \times 10^{-5}$ | **allowed** |
| RL AGN | $10^{-7}$ | $10^{-4}$ | **allowed** |
| RQ AGN | $3 \times 10^{-6}$ | $10^{-4}$ | **allowed** |
| LL AGN | $10^{-3}$ | $\geq 10^{-2}$ | **allowed** |

Starburst galaxies have $n_0^{eff} \sim 10^{-5}$ Mpc$^{-3}$, safely above the lower limit and are therefore still allowed. While such a number density can not currently be identified by IceCube, future upgrades will be sensitive to such a population [622]. Galaxy clusters and galaxy groups have $n_0^{eff} \sim 10^{-5}$ Mpc$^{-3}$ in a model where cosmic rays are accelerated by sources resided within these clusters and groups (in contrast to acceleration of cosmic rays by accretion and merger shocks). Therefore these are also still allowed.

Misaligned Radio-Loud AGNs have $n_0^{eff} \sim 10^{-7}$ Mpc$^{-3}$, which is close to the limit derived above. Therefore these candidates are still allowed and could be identified in the near future. Radio-quiet AGNs (e.g. the AGN core model in [577]) have $n_0^{eff} \sim 3 \times 10^{-6}$ Mpc$^{-3}$, which is still unconstrained but could be probed by an upgraded detector in the future. Finally, low-luminosity AGNs have $n_0^{eff} \sim 10^{-3}$ Mpc$^{-3}$, which could be constrained in the future in case of a weak source evolution. These results are summarised in Table 3.1.

### 3.8.4 Energy spectrum and flavour ratio

It is also possible to derive general properties of the neutrino production in the unknown source population by inspecting the energy spectrum and flavour composition of the observed flux. As seen in Section 3.3.2, the ratios between different neutrino flavours are fixed, since they originate from pion decay (or neutron decay), but are different between $p\gamma$- and $pp$-production. In addition, the flavour composition can be modified by, for instance, strong magnetic fields in the so-called damped muon scenario. Therefore, a precise determination of the flavour ratios can be used to probe the acceleration mechanism (see also [493]).

---

[82]Comparing with the result for FSRQ requires taking into account the cosmological evolution factor given above, which explains why BL Lac are too rare, while FSRQ aren't even though the $n_0^{tot}$ is lower for FSRQ. Note that the difference in luminosity is already folded in the definition of $L_\gamma^{eff}$ (not shown) and $n_0^{eff}$.

One important unresolved question is the determination of the flavour ratios in the presence of $\nu_\tau$. While typical neutrino production models do not predict a flux of $\nu_\tau$ at the source, propagation effects cause such a component to appear. In the case of a $1 : 2 : 0$ production ratio, as in pion decay, the observed flavour ratio is $1 : 1 : 1$ (Section 3.3.3). However, correctly identifying the $\nu_\tau$ is difficult. In the case of NC interactions, they give rise to a cascade. In the case of a CC interaction, there are in principle two depositions of energy, one for the CC interaction and one for the $\tau$-decay, giving rise to the double bang signature (Section 3.5). However, the separation between these two cascade events is only 50 m/PeV, which is comparable to the separation between detector modules in IceCube. Therefore, double bang events under 100 TeV can not be resolved. As a consequence, many of the $\nu_\tau$-events will effectively look like single cascades, such that in most flavour analyses there is a degeneracy between the $\nu_e$ and $\nu_\tau$ content.

In three recent analyses using 7.5 years of data, the first $\nu_\tau$ candidate events have been detected [537, 668]. The first of these uses an updated HESE analysis, first described in [534], where it was the first $\nu_\tau$ analysis to achieve an expected number of events more than one. In this analysis, a new, third, event topology is included ("ternary ID") for double cascade events. Using the energy asymmetry in an event, it can discriminate ordinary cascades from double cascades. As first reported in [668], two events are identified as candidate double cascade events. The first of these is "Big Bird", one of the earliest high-energy events detected, with a visible energy up to 300 TeV and an a posteriori probability of being a $\tau \sim 75\%$. The second event "Double Double", has a visible energy up to 3 PeV and a probability of being a $\tau > 97\%$. Moreover, this second event is badly described by the single-cascade hypothesis. The two other analyses [537] search for a two resolved pulses on a single sector. They find respectively 3 and 2 events, where in both cases one of them is very likely to be a $\tau$ (while the others are background). In both analyses, the identified $\tau$-candidate is again "Double Double". The new (preliminary) inferred flavour composition, using the HESE analysis with ternary ID, is $0.29 : 0.50 : 0.21$, both consistent with the previous analyses (which preferred no $\nu_\tau$) and with the expected $1 : 1 : 1$ composition. This is shown in Figure 3.26.

Additionally, there is also information in the ratio between the $\nu$- and $\bar{\nu}$-flux. As seen in Section 3.3.2, this ratio is different between $p\gamma$- and $pp$-interactions[83], since the former produces far more $\pi^+$ than $\pi^-$ and thus more $\nu_e$ than $\bar{\nu}_e$ in the ideal case. After propagation effects, a small $\bar{\nu}_e$-component is created. However, IceCube is unable to distinguish between $\nu$- and $\bar{\nu}$-events, due to identical event topologies. Still, there exists one possibility to identify the $\bar{\nu}_e$ component separately, since its interaction with $e^-$ features a resonance, absent for other neutrinos, at a neutrino energy of 6.3 PeV: the Glashow resonance [669],

$$\bar{\nu}_e + e^- \rightarrow W^-. \tag{3.61}$$

Due to the previous argument, this resonance is expected to be suppressed for $p\gamma$- compared to $pp$-interactions, as discussed in [670–672]. However, when including higher order resonances the difference between the $\nu_e$- and $\bar{\nu}_e$-flux diminishes and the discrimination power of IceCube is not expected to be sufficient [493]. In the case of environments which are optically thick to photohadronic interactions, produced neutrons can also interact with the radiation, increasing the source $\bar{\nu}_e$-flux, further decreasing the difference between $p\gamma$- and $pp$-scenarios. On the other hand, when efficient $pp$-interactions are known to be absent, the Glashow resonance can instead be used to discriminate between $A\gamma$- and $p\gamma$-scenarios. This is due to heavy nuclei containing many neutrons (typically more than protons), such that the number of expected Glashow events can be increased to even above that predicted by $pp$-scenarios [493]. Very recently, at the time of writing this thesis, a (preliminary) candidate Glashow resonance event was detected, with a deposited energy of $5.9 \pm 0.18$ PeV [673], in an analysis of PeV Energy Partially-contained Events (PEPE), which was specifically designed to search for the Glashow resonance. On the other hand, no Glashow event has been detected in the HESE analysis yet, but this is consistent with the measured spectrum [534].

---

[83]Another possible application is to investigate anti-neutrino sources from neutron decay, e.g. in [485].

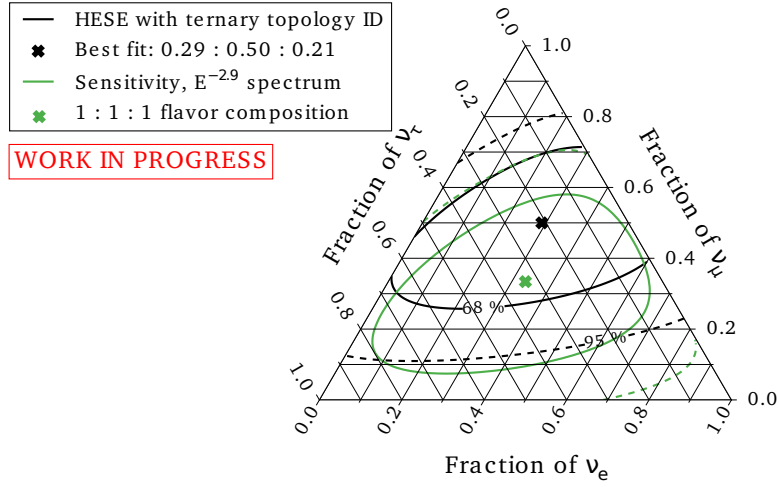

Figure 3.26: Preliminary constraints (at different confidence level contours) on the flavour ratio, from a combined analysis and a dedicated $\nu_\tau$ analysis using HESE events [537]. The first identified candidate $\nu_\tau$ events move the best fit point away from the zero $\nu_\tau$-normalisation of previous analyses.

The flavour information can also be used in more exotic ways. As an example, there is a tension when comparing the best-fit spectral index in track-based analyses, compared to cascades, HESE or a combined fit. As mentioned in Section 3.6.1, these analyses are sensitive to a slightly different energy range, but fitting the HESE spectrum with a broken power-law does not improve the fit. Instead, in [674], the data is interpreted in a model with neutrino decay, changing the expected flavour ratio and alleviating the tension.

Even taking into account these possible methods to determine the flavour composition, there is still no clear conclusion. Even stronger, different analyses can find opposing results. For example, two very recent analyses contradict each other. In the first [675], a Bayesian fit is performed, using the measured flavour ratios at Earth. They find that neutrino production via pion decay is favoured, while neutron decay is ruled out. On the other hand, a second study [676] uses the energy spectrum of HESE events and through-going muons (which are sensitive to different flavours). Assuming a broken power law, they find that the pion decay channel is disfavoured at $2\sigma$, while the neutron decay channel is favoured. Clearly then, the flavour composition of neutrinos at the source is an unresolved issue.

Another way to get information about the sources is to look at only the energy spectrum. As found in Section 3.3.2, the production mechanism has an impact on the expected energy range of the produced neutrino spectrum. Concretely, there is a lower bound associated with $p\gamma$-interactions, due to the necessity to produce at least a pion in the centre of mass frame. Conversely, in the $pp$-channel, this condition is almost trivially met and neutrinos can reach down to any energy.

The observed neutrino spectrum might then be explained by a combination of different contributions; see for example the multi-component fit in [535]. In this work, the spectrum is divided into different contributions based on their production mechanism, while being agnostic about the exact sources. The bulk of the spectrum is generated by a $pp$-component, which has a natural high-energy cut-off due to the typical requirement of confinement (see Section 3.7). The high-energy events are then explained with a separate $p\gamma$-component. Also a galactic component is included, but removing it does not worsen the quality of the fit.

### 3.8.5   Constraints from the isotropic diffuse gamma-ray background

Finally, the neutrino source population can also be constrained by comparing the astrophysical neutrino flux with the isotropic diffuse gamma-ray background (IGRB) or the total extragalactic gamma-ray background (EGB)[84]. As shown in Section 3.3.6, the neutrino and gamma-ray flux are directly related, since they are both the decay products of pions produced in $p\gamma/pp$-interactions. Therefore, by comparing the diffuse neutrino and gamma-ray fluxes, one can constrain the source environment: either the source is transparent to gamma rays or not. The weakest bound comes from the EGB, since this incorporates all sources. However, most of the EGB flux is made up from blazars, whose spectra can typically be explained with leptonic emission only and whose total neutrino emission is constrained. Therefore, when considering non-blazar sources, the neutrino emission can instead be compared by the unresolved gamma-ray flux, the IGRB. Finally, the strongest constraints can be obtained by using the same argument, but comparing the neutrino flux with the estimated (i.e. model dependent) non-blazar contribution to the EGB.

The comparison between the diffuse neutrino and gamma-ray fluxes gives very important constraints in the case of CR reservoir models, which feature $pp$-interactions and are typically transparent to gamma-rays. In these models, the cosmic rays are confined in e.g. a starburst galaxy or galaxy cluster, such that the cosmic rays effectively see a large integrated column densities of interstellar gas and undergo $pp$-interactions efficiently. On the other hand, gamma rays (and neutrinos) are not confined, such that they effectively see a much smaller column of gas. Therefore, gamma rays escape the production environment unhindered. In typical CR reservoir models, this is only true up to about $10 - 100$ TeV, above which the photons can interact with the infrared radiation field and produce $e^{\pm}$-pairs which are confined in the galaxy [677, 678].

---

[84]In principle this connection was also used in previous sections, both by using gamma-rays in order to select potential neutrino sources and by normalising the expected neutrino flux to the observed (diffuse) gamma-ray flux. Here, this connection is investigated more systematically.

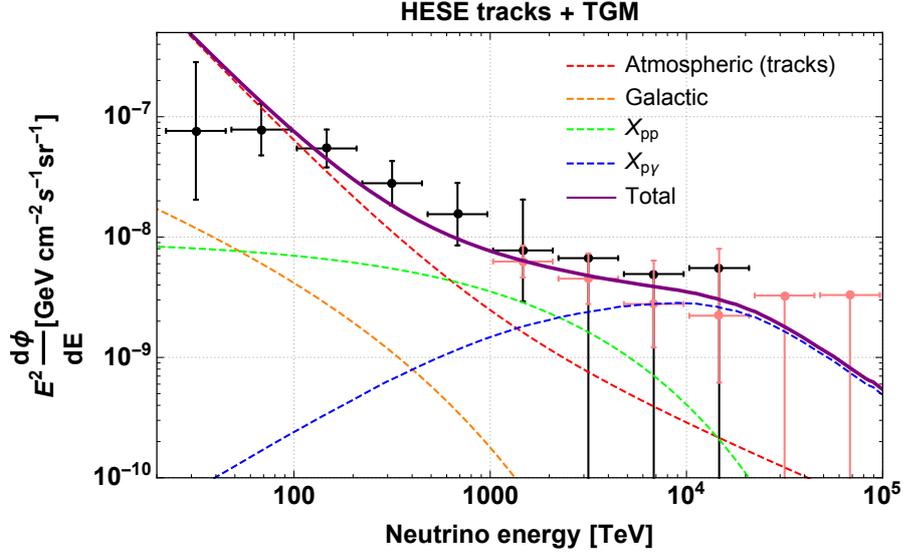

Figure 3.27: Multi-component fit to the high-energy starting events (black) and through-going muons (pink) observed by IceCube [535]. The total observed neutrino flux is decomposed into an atmospheric background contribution, a small galactic contribution, an extra-galactic component from $pp$-sources and a high-energy extra-galactic component coming from $p\gamma$-sources.

In [516], the current constraints from the neutrino flux and the IGRB were investigated for $pp$-sources. If such sources are responsible for the neutrino flux observed by IceCube, they also produce a diffuse gamma-ray flux with the same spectral index, down to low energies. Given the gamma-ray flux observed by Fermi in the $1 - 100$ GeV band, this implies a bound on the source spectral index $\alpha \lesssim 2.1 - 2.2$, which is identical for the neutrinos, gamma-rays and cosmic rays (see Section 3.3.2). This result takes into account the electromagnetic cascades initiated by high-energy gamma-rays on the CMB and EBL.

The case of star-forming galaxies was investigated in more detail in [620], where they exploit the IR luminosity function[85] measured by Herschel [679] and the IR/gamma-ray correlation measured by Fermi [680] to predict the background flux of gamma-rays and neutrinos produced by starburst galaxies (assuming the cosmic rays are accelerated and confined up to 110 PeV) as a function of the spectral index. This flux is lower than the extragalactic diffuse gamma-ray background and consistent with IceCube only when the spectral index $\alpha \leq 2.2$ and can reproduce the IceCube results when $\alpha = 2.15$. On the other hand, they find $\alpha \gtrsim 2.1$, since smaller values are ruled out by both IceCube and Fermi. Therefore, on the basis of this, a spectral index between 2.1 and 2.2 can explain the full IceCube flux, consistent with the previous argument.

---

[85]The luminosity function gives the number density of sources as a function of luminosity and redshift.

Later studies, however, disfavour star-forming galaxies as the dominant source of neutrinos [681]. These stronger constraints come from the bounds on the non-blazar contribution to the EGB, which is about 28% (see Section 3.6.2). For a spectral index of $\alpha \gtrsim 2.15$, they find results which are consistent with the IR/gamma-ray correlation of star-forming galaxies, the non-blazar EGB and individual spectra of star-forming galaxies, but underpredict the neutrino flux. On the other hand, when considering a general broken power law optimised to be the least constrained by the measured IGRB, explaining the measured IceCube flux leads to a gamma-ray flux compatible with the IGRB, but above the non-blazar contribution to the EGB (see Figure 3.28a).

Most recently, this conclusion was again reversed in a study of hadronically powered gamma-ray galaxies [682]. In such galaxies, such as starbursts and ultra-luminous infrared galaxies, gamma rays are produced by cosmic-ray interactions with gas. They find that a spectral index $\alpha < 2.12$ is compatible with all observations, including the most recent estimates of the non-blazar contribution to the EGB. They conclude that various classes of hadronically powered gamma-ray galaxies can provide the dominant contribution to the astrophysical neutrino flux.

In $p\gamma$-source scenarios, there is a lower limit on the energy of the produced gamma-rays from $\pi^0$-decay, such that their flux at lower energies, i.e. in the Fermi band, is only generated through the electromagnetic cascade. Therefore, the constraint from the above argument is slightly weaker (although dependent on specific models for the target radiation field). Still, given that the non-blazar contribution to the EGB is already constrained below 28% and bright gamma-ray blazars detected by Fermi are disfavoured as the main source of the neutrinos, a tension remains when considering the gamma-ray flux produced by $p\gamma$-neutrino sources. This tension can be resolved when considering photon-photon annihilation inside these sources, caused by the same radiation field as the one responsible for $p\gamma$-interactions. As shown in [518], the efficiency for neutrino production $f_{p\gamma}$ and the optical depth $\tau_{\gamma\gamma}$ for two-photon annihilation into $e^\pm$-pairs are then related as

$$\tau_{\gamma\gamma}(\epsilon_\gamma^c) \approx \frac{\eta_{\gamma\gamma}\sigma_{\gamma\gamma}}{\eta_{p\gamma}\hat{\sigma}_{p\gamma}} f_{p\gamma}(\epsilon_p) \sim 10 \left( \frac{f_{p\gamma}(\epsilon_p)}{0.01} \right), \tag{3.62}$$

where $\hat{\sigma}_{p\gamma} \sim 0.7 \times 10^{-28}$ cm$^2$ is the attenuation cross section, $\sigma_{\gamma\gamma} = \sigma_T \approx 6.65 \times 10^{-25}$ cm$^2$ and both $\eta_{p\gamma}(\alpha) \approx 2/(1+\alpha)$ and $\eta_{\gamma\gamma} \approx 7/[6\alpha^{5/3}(1+\alpha)]$ take into account the spectral index $\alpha$. The gamma-ray energy associated to the resonant proton energy is

$$\epsilon_\gamma^c \approx \frac{2m_e^2 c^2}{m_p \bar{\epsilon}_\Delta} \epsilon_p \sim \text{GeV} \ \left( \frac{\epsilon_\nu}{25\,\text{TeV}} \right). \tag{3.63}$$

This relates the IceCube energy range directly to that of Fermi. These relations are valid since both processes interact with the same radiation field. More precisely, the target fields relevant for these interaction are in the keV-MeV energy range, implying that efficient $p\gamma$-neutrino sources are bright in X-rays or MeV gamma-rays (assuming these can escape the system unhindered). On the other hand, given that the observed neutrino flux is close to the WB flux (Eq. (3.44)), $f_{p\gamma}$ needs to be high ($f_{p\gamma} \gtrsim 0.01$), leading to high $\tau_{\gamma\gamma}$. Therefore, such sources are expected to be dark in GeV gamma-rays. This is shown in more detail in Figure 3.28b, which shows the $f_{p\gamma}$ required by the IceCube neutrino flux and UHECR energy budget, together with the $\tau_{\gamma\gamma}$ implied by this, as given by analytical estimates and numerical simulations, for different target photon field spectra [518].

## 3.9 Final thoughts

In this chapter, we have reviewed the current status of multimessenger astronomy and in particular the search for sources responsible for the astrophysical neutrino flux. The most important candidates, blazars and GRBs, considered for their potential to accelerate UHECRs, are severely constrained. No neutrino emission has been detected from GRBs in their prompt phase, making them unlikely to be the dominant source of astrophysical neutrinos. For blazars, the situation is less clear. While one blazar has been identified as a neutrino source, TXS 0506+056, due to stacking limits and constraints from the gamma-ray background, blazars as a class are unlikely to make up the bulk of the diffuse neutrino flux. Moreover, also starburst galaxy models are in tension with observation, due to their gamma-ray emission. The emerging picture is that of a numerous population of neutrino sources with suppressed gamma-ray emission.

Several improvements are expected in the coming years. As more data is gathered, the composition of the neutrino flux can be more accurately determined and the existence of a spectral break can be confirmed or refuted. The development of a strong real-time search program will allow for immediate investigation of "interesting" events detected in electromagnetic emission, neutrinos or gravitational waves, exploiting the multimessenger connection to its fullest [636, 683]. Finally, future upgrades to the existing neutrino experiments, IceCube-Gen2 [684], KM3NeT [685] and Baikal-GVD [686] will make it possible to constrain all of the current source candidates.

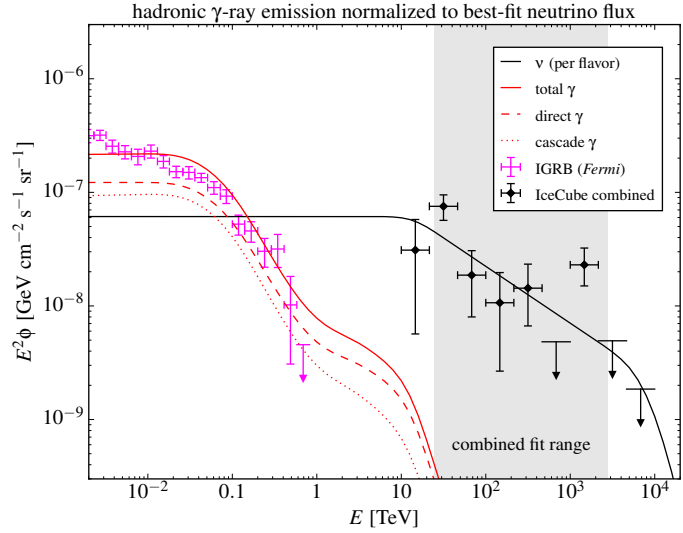

(a) cosmic-ray reservoirs

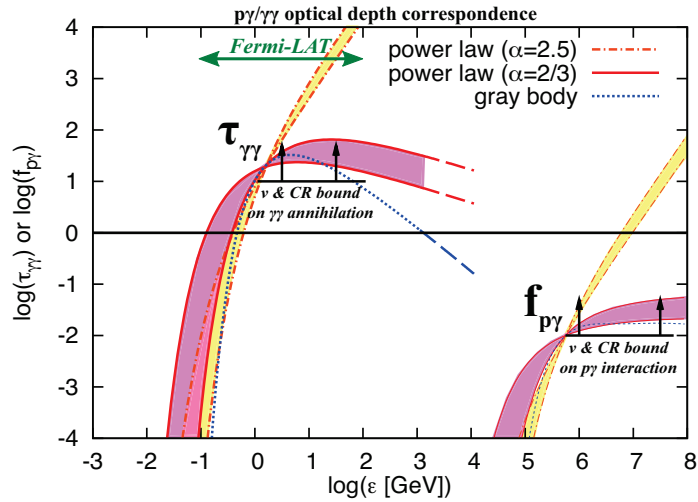

(b) $p\gamma$-sources

Figure 3.28: Constraints on neutrino sources from the diffuse GeV gamma-ray flux. (a) Neutrino and gamma-ray flux for cosmic-ray reservoirs transparent to gamma rays, assuming the optimal (least constrained) broken power law. When normalised to the neutrino flux, the gamma-ray flux is comparable to the IGRB and violates the non-blazar bound on the EGB. [681]. (b) The neutrino production efficiency $f_{p\gamma}$ from comparing the observed cosmic-ray and neutrino fluxes and the resulting two-photon annihilation optical depth $\tau_{\gamma\gamma}$, for different target photon field spectra [518].

# Neutrino sources obscured by matter

In the previous chapter, we saw that the neutrino sources responsible for the bulk of the astrophysical neutrino flux measured by IceCube are still unknown. Moreover, there exists a tension with the observed extragalactic gamma-ray background. In particular, the current limit on the non-blazar contribution to the EGB is at the same level or even lower than the expected gamma-ray flux associated with the observed neutrino flux in case the sources are transparent to gamma rays. In this chapter, we explore the possibility that the astrophysical neutrinos are produced in $pp$-interactions with a gas cloud near the source which is dense enough to significantly attenuate the gamma-ray flux through pair-production on this gas. After defining our model, we implement a Monte Carlo simulation and apply it to different cases. First, we investigate a set of objects selected for their obscuring properties. We find that currently, the expected neutrino flux in our model is below the exclusion limits, but can already constrain the amount of protons accelerated in such sources. Second, we investigate the diffuse flux generated by a population of obscured sources. We find that such a population can indeed alleviate the tension with the extragalactic background light. This work is inspired by a previous project where we performed an object selection of possible $pp$-neutrino sources obscured in X-rays, published in [687].

## 4.1 Obscured $pp$-channel neutrino sources

### 4.1.1 Model definition

In our model, we consider the setup shown schematically in Figure 4.1. A source of accelerated cosmic rays is obscured from our line of sight by a sufficiently large and dense cloud of gas near to the source. While traversing the gas cloud, cosmic rays undergo $pp$-interactions, producing gamma rays and neutrinos. The gas cloud is sufficiently thick that the gamma rays can interact again with the remaining part of the gas column after their generation, undergoing Bethe-Heitler pair production [457]. Therefore, the source can be hidden (or at the very least obscured) in gamma rays.

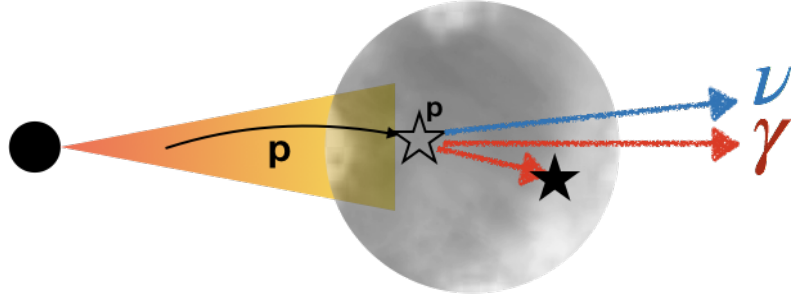

Figure 4.1: Schematic representation of the model we consider for neutrino and gamma-ray production from $pp$-interactions of cosmic rays with a dense gas cloud near the cosmic-ray source. Due to the high integrated column density of the gas cloud, the produced gamma rays are also attenuated by the same cloud through Bethe-Heitler pair production. The relative size of the cloud and the source/outflow can vary. Note that while the figure features a jet, this is not a requirement for this mechanism to work and we do not initially assume its existence in our calculations.

The amount of matter present in the gas cloud is expressed by the equivalent hydrogen column density $N_H$, defined as the line-of-sight integral of the hydrogen density $n_H$

$$N_H = \int dl\, n_H(l),\tag{4.1}$$

and denotes the amount of hydrogen atoms per cm$^2$ (this is discussed in more detail in Section 4.1.2). The column density required for the gamma-ray attenuation at GeV energies and above to be significant can be estimated from the Bethe-Heitler pair production cross section with matter. Using the approximate value $\sigma_{BH} \approx 20\,\text{mb} = 2 \times 10^{-26}\,\text{cm}^2$ (see Section 4.1.2), we see that we require a column density of approximately

$$N_H = 5 \times 10^{25}\,\text{cm}^{-2}.\tag{4.2}$$

This column density is very high and its use will be motivated in Section 4.1.4. In this chapter, we explore the possibility that neutrinos are produced in highly obscured sources, investigating two benchmark values of the column density: $N_H^{(1)} = 5 \times 10^{25}\,\text{cm}^{-2}$, which is high but easily motivated, or the more extreme value $N_H^{(2)} = 10^{26}\,\text{cm}^{-2}$.

As a consequence of this high column density, (nearly) all cosmic rays will interact with the gas before traversing the cloud, since the proton-proton cross section is a few times higher than the Bethe-Heitler pair production cross section (depending on the energy, see Figure 3.13b). Therefore, such a neutrino source would be a poor cosmic ray source, although similar but unobscured sources would still make excellent cosmic ray sources. Moreover, the remaining gas column after interaction is sufficiently thick for secondary protons to interact with the cloud again, giving a slight boost to the total neutrino and gamma-ray flux.

While the sketch in Figure 4.1 shows an AGN-like scenario featuring a jet, and we will apply the model to objects of this scale, the mechanism is not restricted to this case and we do not a priori assume the existence of a jet in our calculations. The only requirement is that there exists a compact source of accelerated cosmic rays obscured by a dense gas cloud near to the source. In principle, such a model can feature both transient (see also Section 4.1.3) and continuous neutrino production, although we will restrict ourselves to continuous emission and assume that the configuration is stable for a sufficiently long time[1].

The model considered here should be contrasted with cosmic ray reservoir models of neutrino production. There, cosmic rays have a significant interaction probability with the target gas only by integrating the gas density over the cosmic ray trajectory inside a galaxy or cluster, since cosmic rays are confined in these structures. Gamma rays, which are not confined, escape unattenuated. Instead, in our model discussed here, the interaction happens close to the source with a thin, but dense target (i.e. the cosmic rays are not confined to this region), such that the cosmic rays and gamma rays traverse the same gas column. On the other hand, cosmic ray accelerator models with a similar configuration (e.g. [495, 574], see also Section 3.7.1) feature lower column densities, such that the gamma ray flux is either unattenuated or attenuated only by invoking a strong radiation field.

Another model with more similarities to our scenario but of a different scale is one of the microquasar SS433, where a supergiant star feeds a $10 M_\odot$ black hole [688–690]. Inside the jet, accelerated particles undergo $pp$-interactions with cold matter and produce both neutrinos and gamma rays. In this model, attenuation of gamma rays by both Bethe-Heitler pair production and photomeson production are taken into account, on top of the more standard $\gamma\gamma$-annihilation.

Typical neutrino models consider sources from which strong non-thermal emission has been observed. Instead, our model features objects with obscured gamma- and X-ray emission. Therefore, strong sources can only be selected at radio or infrared wavelengths, where the presence of an accelerated particle population can also be inferred. Our model also differs from the hidden sources of [518], which require strong X-ray emission in order to attenuate the gamma rays.

### 4.1.2 Photon attenuation

Here, we briefly summarise the attenuation channels of photons, in particular X-rays and gamma rays, since photons at these energies are sensitive to the presence of surrounding matter. In addition, gamma rays can also be attenuated by interaction with a radiation field, if a sufficiently strong radiation field is available at the production site of the gamma rays. However, such an attenuation is more model dependent and we do not include it. Therefore, the gamma-ray flux in our model will be an upper bound.

At lower energies (UV, optical, infrared, radio) the attenuation of electromagnetic waves is caused by different properties (line absorption, heating of dust, synchrotron

---

[1] "Sufficiently long" depends on the specific case under consideration.

self-absorption respectively) and requires a different and more complicated modelling.

**Attenuation by matter**

The interaction of X-rays and gamma rays with matter, which is the main attenuation channel in our model, happens through different processes. The photo-electric effect is important for soft X-rays traversing an unionised[2] medium. For hard X-rays propagating through any medium or soft gamma rays through an ionised medium, Compton scattering is important. Finally, for gamma rays, Compton scattering dominates at MeV energies, while at GeV energies pair production on either the electrons or the nuclei is dominant. The strength of these interactions depends on the energy of the photons and the target composition. Figure 4.2 shows the total cross section of photons with a hydrogen target, as obtained from the XCOM database [692]. For GeV gamma rays, in which we will be interested[3], the only relevant process is pair production, either on the nuclei or on electrons.

The target gas is expected to have a composition similar to the interstellar medium, which is dominated by hydrogen [694]. When determining the hydrogen equivalent column density through spectral modelling, this composition is taken into account properly. The total amount of matter in the column is then given by summing over all elements, with a density equal to $N_H$ times the abundance $a_Z$ of that element relative to hydrogen

$$N = \sum_Z a_Z N_H, \tag{4.3}$$

see e.g. [695] and the absorption models included in `XSPEC` [693], an X-ray spectral modelling tool. However, given the dominance of hydrogen, we assume that the target is a pure proton gas in our calculations[4]. For modelling the X-ray attenuation, this introduces a rather large error. For our model, however, where we will mainly look at the gamma-ray attenuation, this is a reasonable assumption. The results for a cloud following the composition of the interstellar medium are shown in Appendix G. As expected, for gamma rays the differences are negligible.

The expression for Bethe-Heitler pair production [457] is complicated and the complete expression for pair production on protons is given as a series expansion in [696] (see also the review [462] and an earlier treatment in [697]). The threshold for pair production is $E_{\mathrm{BH}}^{\mathrm{thres.}} = 2m_e c^2$. Define $\epsilon = E_\gamma / m_e c^2$, with $E_\gamma$ the photon energy in the

---

[2]**unionised** 1. *Not ionised.* 2. *Organised into a trades union or trades unions.* [691]

[3]In order to properly model X-rays propagating through a medium with Compton scattering, multiple scattering needs to be taken into account, in particular when the densities are high. This requires the use of specialised codes, such as `XSPEC` [693].

[4]For neutrino and gamma-ray production, the additional mass (increased by a factor of 1.4) due to the heavier elements can be included by rescaling $N_H$, see for example [495] (where instead a factor 1.6 is used [483]). However, for the purpose of our work, we do not explicitly take into account this small numerical factor.

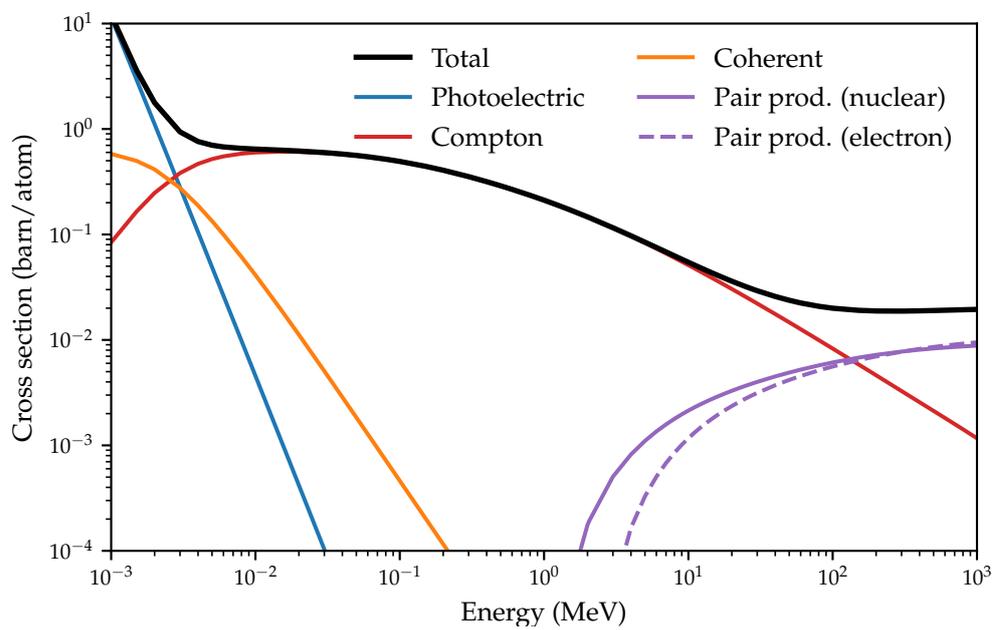

Figure 4.2: Cross section of photons interacting with hydrogen as a function of energy, showing both the total cross section and its decomposition in the different processes. Data taken from [692].

proton rest frame. In the regime $2 \leq \epsilon \leq 4$, the cross section is given by [696]

$$\sigma_{\mathrm{BH}}^{\mathrm{thr}}(\epsilon) \simeq \frac{2\pi}{3} \alpha \left( \frac{\alpha \hbar}{m_e c} \right)^2 \left( \frac{\epsilon - 2}{\epsilon} \right)^3 \left( 1 + \frac{1}{2}\eta + \frac{23}{40}\eta^2 + \frac{37}{120}\eta^3 + \frac{61}{192}\eta^4 \right), \quad (4.4)$$

with $\eta = (\epsilon - 2)/(\epsilon + 2)$. For $\epsilon > 4$, we have

$$\begin{aligned}
\sigma_{\mathrm{BH}}^{\mathrm{he}} = \alpha \left( \frac{\alpha \hbar}{m_e c} \right)^2 & \left\{ \frac{28}{9}\delta - \frac{218}{27} + \left( \frac{2}{\epsilon} \right)^2 \left[ 6\delta - \frac{7}{2} + \frac{2}{3}\delta^3 - \delta^2 - \frac{\pi^2}{3}\delta + 2\zeta(3) + \frac{\pi^2}{6} \right] \right. \\
& \left. - \left( \frac{2}{\epsilon} \right)^4 \left[ \frac{3}{16}\delta + \frac{1}{8} \right] - \left( \frac{2}{\epsilon} \right)^6 \left[ \frac{29}{9 \times 256}\delta - \frac{77}{27 \times 512} \right] \right\}.
\end{aligned}$$

$$(4.5)$$

with $\delta = \log(2\epsilon)$. For the full pair production cross section (on proton *and* electron), we multiply these expressions by 2. Since the cross section rises only logarithmically with the photon energy, the cross section is approximately constant with a value of $\sigma_{\mathrm{BH}} \approx 20$ mb. This validates our estimation of the required column density $N_H$ in Eq. (4.2).

For high-energy gamma rays in matter, the interaction cross section of high-energy gamma rays with matter starts to decrease due to the "Landau-Pomeranchuk-Migdal" (LPM) effect [698–702] (see also the PDG review [29]). This effect is due to destructive interference between amplitudes from different, nearby, scattering centres. However, for an astrophysical gas cloud, the density is too low for this effect to be relevant, even if the integrated column density is high.

Another possible source of gamma-ray attenuation would be through photohadronic $p\gamma$-interactions. However, the cross section for this process is much lower than for pair production[5] (see Section 3.3.2) and we will therefore neglect it (see also the related discussion in Section 4.1.5).

**Attenuation by a radiation field**

Gamma rays can interact with a radiation field and undergo pair production if the centre of mass energy exceeds $2m_e$, attenuating the gamma-ray flux. The optical depth to pair production depends on the number density of photons (i.e. the energy density of the radiation field). It is possible that the radiation fields in the cloud are weak enough for $\gamma\gamma$-pair production to be negligible. In case of a strongly radiating source (as will usually be the case), this means that the cloud can not be too close to the source. In the other case, attenuation of gamma rays through pair production on radiation will also happen and decreases the gamma-ray flux beyond the level of attenuation due to matter alone.

Even if there is not a sufficiently strong radiation field present at first, it can be generated by the particle cascade initiated in the $pp$-interactions. Depending on the density and location of the gas cloud, the gamma rays and $e^{\pm}$-pairs produced in pion decay can initiate an electromagnetic cascade through repeated creation of photons through synchrotron radiation, bremsstrahlung and inverse-Compton scattering and of $e^{\pm}$-pairs through pair production. As a result, the cloud can become fully ionised and the photon field can become Comptonised[6]. The cloud is then optically thick to X-rays (due to the free electrons) and the generated photon field can then act as a target for pair production for gamma rays (see e.g. [653]).

---

[5]This is the reason that $p\gamma$-interaction models require strong radiation fields (leading to gamma-ray attenuation) and much larger proton luminosities than $pp$-interaction models.

Since $\gamma\gamma$-attenuation introduces a stronger model dependency, we do not take this additional attenuation channel into account. In that sense, the amount of obscuration we obtain in our model is conservative.

### 4.1.3  Cloud dynamics

We envision a scenario where a dense gas cloud gets bombarded by accelerated cosmic rays close to the source. Often, this will be accompanied by a jet. While we assume that the configuration is stable for a sufficiently long time, either on the observing timescale or a relevant period in the cosmological history of the source, eventually such a jet will break up the configuration if it is sufficiently strong at the location of the cloud. The exact physics depends on the location of the cloud, its total mass and the strength of the jet. A full study of this is outside the scope of this work. However, similar studies exist in the literature, which we briefly discuss in the following.

The interaction of a cloud of cold gas with a jet[7] was studied in [704, 705]. Due to the pressure of the jet, some or all of the cloud material can be blown away and swept up by the jet. Initially, the interaction will cause shocks in both the jet and cloud material, which can serve as a potential site for particle acceleration. As the material is swept up by the jet, eventually it will spread out and move along with the jet, at which point significant interactions between the cloud and jet cease. Depending on the kinetic energy of the jet and the mass of the cloud, the jet might be slowed down.

Several scenarios have been studied in the literature, often with a cloud or a star interacting with an AGN jet, sketched in Figure 4.3. Due to the scale of the cloud/star compared to the jet (i.e. typically much smaller than the jet at entry), these scenarios feature transient emission and have been invoked to explain gamma-ray flares in blazars [705, 706]. These models differ from the scenario we envision mostly due to the scale of the cloud, since we focus on a case where a possible jet is obscured for an extended amount of time[8].

Jet-cloud/star models have been used to explain the neutrinos from the blazar TXS 0506+056. The bright gamma-ray flare associated with a high-energy neutrino has been modelled from clouds with $N_H > 10^{24}$ cm$^{-2}$ present in the broad-line region of the blazar [652]. The electromagnetic cascade initiated by the $pp$-interactions (modelled using the analytical fits described in Section 4.2.1) ionises the cloud, which becomes optically thick for optical to X-ray photons while the gamma rays and the neutrinos escape. The SED at lower frequencies can then be explained with a second leptonic emission zone. This scenario can explain the observed neutrino and electromagnetic emission with moderate proton luminosity, in contrast with $p\gamma$-models which have a low interaction rate in order to avoid efficient $\gamma\gamma$-annihilation and therefore require high proton luminosity. The neutrino flare while the blazar was in a quiescent state has been explained with unbound layers from a tidally-disrupted red giant[9] [653]. In this case, the target has a column density of $N_H = 5 \times 10^{25}$ cm$^{-2}$, the same as our first benchmark point. Again, the cloud becomes ionised and is optically thick to X-rays, but now the Comptonised radiation field is also sufficiently thick to efficiently attenuate gamma rays through $\gamma\gamma$-pair production.

---

[6]An equilibrium is reached where the energy is distributed between free electrons and gamma rays due to Compton scattering [703].

[7]Note that the jet and cloud have different properties: a jet is a Poynting flux (i.e. dominated by radiation) or extremely relativistic matter, while a gas cloud is dense and cold matter.

[8]In this sense, our model has some similarities with low-luminosity gamma-ray burst models, where the jet is stopped by a cocoon of matter [610].

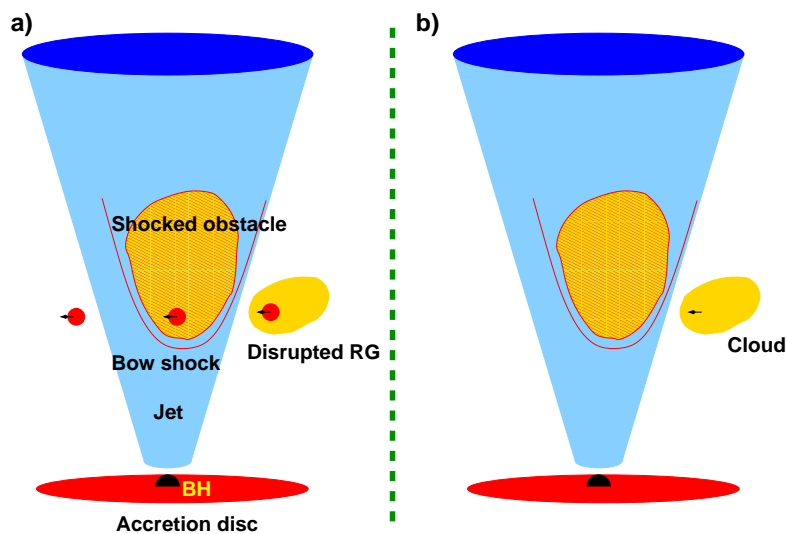

Figure 4.3: Sketch of the jet-cloud/star scenario, leading to transient emission, previously considered in the literature (RG indicates a red giant). In this case the cloud is much smaller and lighter than the jet, contrary to our scenario where a significant part of a possible jet is obscured for an exteded amount of time. Figure from [707].

More generally, other models of jet-cloud/star interactions include models of jets interacting with a BLR cloud of lower density than above to [709], particle acceleration induced by a strong star wind and loss of mass of a star in an FSRQ jet [710], M87 TeV flares from jet-cloud interaction which accelerates particles and leads to $pp$-interactions [708] and alternative scenarios where instead orphan flares are caused by interaction of accelerated particles with star radiation inside jet blobs [711]. All these models are built only to explain gamma-ray emission.

---

[9]A red giant is necessary since only there the outer layers are sufficiently weakly bound to be blown away and interact with a significant portion of the jet [708].

### 4.1.4 Motivating high column densities

The column densities required for the gamma-ray attenuation to be significant are high compared to typical astrophysical environments. In order to get a sense of scale, consider the gas in the interstellar medium of our own galaxy. With a scale height of 0.3 kpc and a gas density of 0.1 $cm^{-3}$, we find a column density $N_H^{gal} \sim 10^{20}$ $cm^{-2}$, much lower than our benchmark model. On the other hand, the Earth atmosphere has a column density of $10^3$ $g\,cm^{-2}$ or $N_H \sim 10^{26}$ $cm^{-2}$. Still, the latter is a compact object and thus not a good comparison for the viability of our scenario[10], although it is more appropriate as a comparison for the jet-cloud/star scenario. In order to have longer lasting neutrino production and obscuration, a more extended gas cloud is necessary. In that sense, high density star-forming regions with column densities $N_H \sim 10^{23}$–$10^{24}$ $cm^{-2}$ are a more relevant comparison[11].

However, though rare, astrophysical environments with a gas column of the required magnitude do exist. As already mentioned in the previous section, models of AGN jets interacting with dense clouds possessing column densities up to $5 \times 10^{25}$ $cm^{-2}$ have been invoked to explain the TXS 0506+056 neutrino and gamma-ray flares. More generally, the supermassive black hole in AGNs is surrounded by gas and dust in the broad-line region and torus (for a review, see [423, 712]). In case an AGN is observed edge-on, the central engine is frequently hidden from our view by the dusty torus. The column density of this torus varies and an AGN is considered obscured when $N_H \geq 10^{22}$ $cm^{-2}$. From figure 4.2, we can read off that below $N_H \sim 10^{24}$ $cm^{-2}$, the photon attenuation cross section in the 2–10 keV X-ray regime[12], has the correct magnitude to probe the column density[13]. If the column density is higher than the inverse of the Thompson cross section, $N_H \geq 1.5 \times 10^{24}$ $cm^{-2}$, then the AGN is called Compton-thick (see the review [713]). At these values, the density can be probed by hard X-rays ($E_{X-ray} \geq 10$ keV), where Compton scattering dominates. However, for densities above $N_H = 10^{25}$ $cm^{-2}$ (heavily Compton-thick), the X-ray emission is suppressed even above 10 keV, since the photons are down-scattered by Compton interactions and subsequently absorbed. Other (indirect) methods need to be used to probe these obscuring columns (e.g. through reflected X-rays). In this way, Compton-thick AGNs have been found with column densities exceeding $N_H = 10^{25}$ $cm^{-2}$ [713, 714]. AGNs obscured by column densities significantly higher than $10^{25}$ $cm^{-2}$ are suspiciously missing from surveys. From the above discussion, it is clear that there is an observational bias against such highly obscured objects.

---

[10]The existence of one or a few extremely dense compact objects does not imply anything about the average density, i.e. the presence of the sun does not mean our galaxy has an average column density higher than $N_H^{gal} \sim 10^{20}$ $cm^{-2}$.

[11]In this sense, one could speculate that if extremely obscured objects are indeed as rare as they appear, it is because at higher densities such structures have long since collapsed, accreted onto supermassive black holes and/or been expelled by jets. However, this is a personal perception.

[12]For reference, the full X-ray regime is between 100 eV and 100 keV.

[13]More specifically, the main indicator is the photo-electric cut-off, induced by the sharply rising cross section of photo-electric absorption towards low energies.

The Chandra X-ray Observatory [715] detects X-rays between 0.1-10 keV and has been used to detect obscured AGNs. An analysis fitting AGN spectra in Chandra Deep Fields with physical models, finds many highly obscured AGNs [716]. While their analysis corrects for observational bias, they are unable to constrain the number of AGNs with $N_H \geq 10^{25}$ cm$^{-2}$, since these sources are missed in their sample and therefore the missing number of sources cannot be determined. As such, this kind of sources might contribute to the heavily obscured AGN population. This agrees with an older result [717]. Other analyses do find multiple objects with $N_H \geq 10^{25}$ cm$^{-2}$, even up to $N_H \sim 10^{26}$ cm$^{-2}$ in Chandra surveys [718]. Another analysis studies torus model properties [719] with an ultra-hard X-ray sample (14–195 keV) of Seyfert galaxies from Swift/BAT [720], which can identify more strongly obscured objects. They find that even from such a selection, a population of the most obscured objects is still missing, agreeing with [714, 721, 722].

An example of such an obscured source is NGC 4418, a luminous infrared galaxy (LIRG). It has a core bright in IR along with the deepest known silicate absorption, but it has not been detected in X-rays [719, 723]. An analysis infers a column density $N_H > 10^{25}$ cm$^{-2}$, a spectrum consistent with AGNs as the main power source and showing similarities with ARP 220 [724] (which will end up as a source in the analysis of Section 4.3). Another analysis finds that an AGN is only allowed in case the column density exceeds this same value [725].

While the objects above do have strong obscuration, this does not yet mean that the obscuring material is bombarded by cosmic rays or blocks a jet. However, models of tilted tori where this is possible do exist [712, 726]. Such models are interesting candidates for neutrino production through the model proposed here and are part of the motivation for the work in Section 4.3.

From the examples above, it can be seen that, while extreme, column densities $N_H > 10^{25}$ cm$^{-2}$ in astrophysical environments must occur (although this does not mean that they indeed produce neutrinos or, if they do, that they are common enough to explain the bulk of the astrophysical neutrino flux). Therefore, our lower benchmark value $N_H^{(1)} = 5 \times 10^{25}$ cm$^{-2}$ for obscured neutrino sources is compatible with the conventional view of various astrophysical objects. On the other hand, values much higher than this have not been observed, such that we consider our second benchmark $N_H^{(2)} = 10^{26}$ cm$^{-2}$ as an extreme case.

### 4.1.5 Additional modelling assumptions

In addition to intrinsic features of the model above, we also make some further assumptions in order to reduce the model parameter space. The cosmic-ray composition is considered to be pure proton, which is a reasonable approximation in the energy range relevant for IceCube (see also Section 3.3.2). As already mentioned in Section 4.1.2, the target gas is also approximated as pure hydrogen. Therefore, we can model the interactions as pure *pp*-collisions, without complications from nuclear effects.

We do not model possible additional gamma rays from leptonic processes (i.e. pro-duced in inverse-Compton scattering) or neutrinos and gamma rays from $p\gamma$-interactions, since both of these channels are strongly model-dependent. Both of these processes would increase the total gamma-ray flux. On the other hand, if intense radiation fields are present inside the cloud, the gamma rays would be attenuated by those as well through $\gamma\gamma$-pair production (which we do not take into account), reducing the total gamma-ray flux. In addition, if the latter process is not dominant, it is safe to ignore photomeson production, since its interaction rate is roughly three orders of magnitude smaller than $\gamma\gamma$-annihilation for the same photon field [518, 652, 727]. Finally, $p\gamma$-interaction models typically require higher proton luminosities than $pp$-models to obtain sizeable fluxes[14].Therefore, if $pp$-interactions are present, they can be expected to be dominant compared to $p\gamma$-interactions.

We do not take into account synchrotron losses of the muons, pions and kaons. However, for the scenario we envision, these losses should be negligible: while the obscuring gas should be close to the source, it can not be too close, otherwise we need to take into account the effect of strong radiation fields on the gas. Therefore, if the gas is sufficiently far removed from the source, one can expect the synchrotron losses due to the magnetic fields to be negligible. Indeed, following the approach of [486] we can estimate above which critical energy these losses become dominant (i.e. the timescale associated to synchrotron losses is shorter than the decay time), for an AGN jet scenario with the gas cloud at parsec scales. From this, we find that this energy can easily exceed $10^6$ GeV, even for the magnetic fields associated to jets, using the magnetic field found in [728] or higher; if no (strong) jet is present at the location of the cloud, magnetic fields can be expected to be even weaker at the parsec scale.

Unless otherwise stated, the protons are assumed to follow an $E^{-2}$-spectrum, consistent with Fermi acceleration. However, this immediately implies that the predicted neutrino spectrum also follows an $E^{-2}$-spectrum, while IceCube observes a softer spectrum. Therefore, the computation here can only serve to explain the observed neutrino flux above $\sim 100$ TeV, while there must be a second, softer, component below this energy that we do not model.

With all these assumptions and the choice of two benchmark values of $N_H$, the only free parameters left are the normalisation of the injected proton flux (used in Section 4.3) or of the resulting neutrino flux (used in Section 4.4) and the energy range in which the protons are injected. The maximum energy of the protons will be fixed at $10^8$ GeV, which is sufficient to explain the neutrinos observed by IceCube without violating the limits at the highest energy. The minimum proton energy is determined by the acceleration mechanism. In the case of shock acceleration, one typically has $E_p^{\min} \sim \Gamma m_p c^2$ [599], with $\Gamma$ the Lorentz factor of the shock, which is the minimum energy with which the particles can efficiently participate in the acceleration process. Sometimes, also the value $E_p^{\min} \sim \Gamma^2 m_p c^2$ is used [518]. Since we will consider AGNs as the central engine in the following, with $\Gamma \sim 10 - 30$ (Section 3.2.4), we take $E_p^{\min} = 10^2$ GeV. For an $E^{-2}$-spectrum, the final result is not very sensitive to the exact value of the minimum

---

[14]This is due to their much lower cross section, which might not be compensated completely by the high target density of a radiation field

energy (as long as it is around the same order of magnitude), since the luminosity is per decade of energy. More concretely, we have

$$L \propto \int_{E_{\min}}^{E_{\max}} \mathrm{d}E \, E E^{-2} = \ln\left(\frac{E_{\max}}{E_{\min}}\right), \tag{4.6}$$

resulting in a logarithmic dependence only. On the other hand, varying the minimum energy with more than an order of magnitude has a significant effect on the total normalisation of the flux. However, more importantly, if the spectral index deviates from 2, the luminosity quickly becomes very sensitive to the minimum energy, see also the discussion in [729] and Section 4.3.3. Since cosmic ray experiments are only sensitive to the maximum energy of extragalactic cosmic rays (below the knee galactic cosmic rays dominate), this is an important source of uncertainty. However, the choice above is theoretically well motivated.

## 4.2 Calculating the ν and γ-ray flux

In this section, we describe the method for calculating the neutrino and gamma-ray flux for our model of neutrino production in obscured sources, before applying it to specific scenarios in the following sections. We perform two separate calculations. The first one is based on analytical fits of the neutrino and gamma-ray spectrum from $pp$-collisions. The second one is a complete Monte Carlo simulation. For our final results, we use the Monte Carlo simulation, since it includes more details, while the analytical method serves only as a consistency check on our results.

### 4.2.1 Analytical

We calculate in this section the neutrino and gamma-ray spectra from cosmic ray interactions with a cloud of integrated column density $N_H$ following the method of [489]. These authors provided analytical fits, which we describe below, to the neutrino and gamma-ray spectra using the meson spectra simulated with the Monte Carlo generators `SIBYLL` [496] and `QGSJET` [730], with their decay to photons, neutrinos and electrons being treated analytically.

Define the dimensionless quantity $L = \ln\left(\frac{E_p}{1\,\mathrm{TeV}}\right)$. The spectrum of the leptons can be described as the sum of several components. First, the spectrum of electrons from $\pi \to \mu \nu_\mu$ decay can be described by

$$F_e(x, E_p) = B_e \frac{(1 + k_e(\ln x)^2)^3}{x(1 + 0.3/x^{\beta_e})}(-\ln x)^5, \tag{4.7}$$

with $x = E_e / E_\pi$ and

$$B_e = \frac{1}{69.5 + 2.65L + 0.3L^2}, \tag{4.8}$$

$$\beta_e = \frac{1}{(0.201 + 0.062L + 0.00042L^2)^{1/4}}, \tag{4.9}$$

$$k_e = \frac{0.279 + 0.141L + 0.0172L^2}{0.3 + (2.3 + L)^2}. \tag{4.10}$$

The same function also describes the spectrum of $\nu_\mu$ from muon decay $F_{\nu_\mu^{(2)}} = F_e$, with now $x = E_{\nu_\mu} / E_p$.

Muon neutrinos directly from $\pi \to \mu \nu_\mu$ decay are described by

$$
\begin{aligned}
F_{\nu_\mu^{(1)}}(x, E_p) = B' \frac{\ln y}{y} & \left( \frac{1 - y^{\beta'}}{1 + k' y^{\beta'}(1 - y^{\beta'})} \right)^4 \\
& \times \left[ \frac{1}{\ln y} - \frac{4\beta' y^{\beta'}}{1 - y^{\beta'}} - \frac{4k' \beta' y^{\beta'}(1 - 2y^{\beta'})}{1 + k' y^{\beta'}(1 - y^{\beta'})} \right],
\end{aligned}
\tag{4.11}
$$

with $x = E_{\nu_\mu}/E_p$ and $y = x/0.427$ and

$$B' = 1.75 + 0.204L + 0.010L^2, \tag{4.12}$$

$$\beta' = \frac{1}{1.67 + 0.111L + 0.0038L^2}, \tag{4.13}$$

$$k' = 1.07 - 0.086L + 0.002L^2. \tag{4.14}$$

This spectrum has a sharp upper cut-off at $x = 0.427$, which comes from $E_{\nu,\text{max}} = \left( 1 - \frac{m_\mu^2}{m_\pi^2} \right) E_\pi$ from simple relativistic kinematics. The total muon neutrino spectrum is then given by $F_{\nu_\mu} = F_{\nu_\mu^{(1)}} + F_{\nu_\mu^{(2)}}$.

Finally, for the electron neutrinos we have $F_{\nu_e} \approx F_e$, to an accuracy better than 5%. Note that while $\pi$-decays dominate the neutrino production, the decay of other mesons, in particular $K$-mesons, is also non-negligible, boosting the total neutrino flux for power law proton spectra up to 10%.

The spectra of photons from $\pi$- and $\eta$-meson decay can be described by

$$
\begin{aligned}
F_\gamma(x, E_p) = B_\gamma \frac{\ln x}{x} & \left( \frac{1 - x^{\beta_\gamma}}{1 + k_\gamma x^{\beta_\gamma}(1 - x^{\beta_\gamma})} \right)^4 \\
& \times \left[ \frac{1}{\ln x} - \frac{4\beta_\gamma x^{\beta_\gamma}}{1 - x^{\beta_\gamma}} - \frac{4k_\gamma \beta_\gamma x^{\beta_\gamma}(1 - 2x^{\beta_\gamma})}{1 + k_\gamma x^{\beta_\gamma}(1 - x^{\beta_\gamma})} \right],
\end{aligned}
\tag{4.15}
$$

with $x = E_\gamma / E_p$ and

$$B_\gamma = 1.30 + 0.14L + 0.011L^2, \tag{4.16}$$

$$\beta_\gamma = \frac{1}{1.79 + 0.11L + 0.008L^2}, \tag{4.17}$$

$$k_\gamma = \frac{1}{0.801 + 0.049L + 0.014L^2}. \tag{4.18}$$

This fit has an accuracy better than a few percent in the region $x \gtrsim 10^{-3}$, which is the most important one. Compared to the total production rate of gamma rays, including also other meson decays, this fit can describe more than 95% of the total flux. Note that this fit is only accurate for $E_\gamma \gtrsim 1$ GeV. Therefore, it can not be used to estimate the total number of produced gamma rays.

With a proton flux given as $dN_p = \Phi_p(E_p)\, dE_p$, the total produced neutrino flux can then be found from

$$\Phi_\nu(E_\nu) = \int_{E_\gamma}^\infty \left[1 - \exp(-N_H \sigma_{\text{inel}}(E_p))\right] \Phi_p(E_p) F_{\nu_\mu}(\frac{E_{\nu_\mu}}{E_p}, E_p) \frac{dE_p}{E_p}, \tag{4.19}$$

which can be rewritten as

$$\Phi_\nu(E_\nu) = \int_{E_\gamma}^\infty \left[1 - \exp(-N_H \sigma_{\text{inel}}(E_{\nu_\mu}/x))\right] \Phi_p(E_{\nu_\mu}/x) F_{\nu_\mu}(x, \frac{E_{\nu_\mu}}{x}) \frac{dx}{x}. \tag{4.20}$$

The cross section (shown previously in Figure 3.13b) is well described by

$$\sigma_{\text{inel}}(E_p) = (34.3 + 1.88L + 0.25L^2) \left[1 - \left(\frac{E_{th}}{E_p}\right)^4\right] \text{mb}, \tag{4.21}$$

with the threshold for $\pi^0$-production $E_{th} = m_p + 2m_\pi + m_\pi^2/2m_p = 1.22$ GeV . The correction from the last factor is only relevant below 1 GeV. However, for protons below 100 GeV, the analytical fits above are no longer accurate and instead the $\delta$-functional approximation can be used. In this approximation, the $\pi$-meson production rate is given by

$$F_\pi = \tilde{n}\delta\left(E_\pi - K_\pi E_{\text{kin}}\right), \tag{4.22}$$

where $E_{\text{kin}} = E_p - m_p$. Here we used $K_\pi = \frac{\kappa}{\tilde{n}}$, with $\tilde{n} = \int d\, E_\pi F_\pi$ and $\kappa = \int dE_\pi\, E_\pi F_\pi$. However, since we are not interested in the lowest energy gamma rays[15] and because, as we will see next, we prefer the full Monte Carlo simulation anyway, we did not implement the $\delta$-functional approximation[16].

After production, the gamma-ray flux still needs to be attenuated. Using the expressions above, we need to approximate the attenuation by assuming that the gamma rays are attenuated by the full column, instead of the column remaining after production

$$\Phi_\gamma = \Phi_\gamma^0 \exp(-N_H \sigma_{\text{BH}}), \tag{4.23}$$

---

[15]Since we either do not focus on gamma rays (Section 4.3) or our results are dominated by higher energy gamma rays and their cascades (Section 4.4).

[16]This correction is important when one is interest in a detailed SED modelling of AGNs.

valid above the threshold for Bethe-Heitler pair production. This limitation is more of a design choice and can be solved by calculating the production rate instead of the totally produced flux[17] as given in [489] and integrating over the length of the dust cloud. However, since we only use the analytical approximation to verify the Monte Carlo simulation, this was not implemented.

Finally, note that using these fits, we have no access to the secondary proton spectrum and therefore can not implement the interactions of secondary protons.

### 4.2.2 Monte Carlo simulation

The analytical fits above are accurate and faster than performing Monte Carlo simulations. Nevertheless, a full Monte Carlo simulation was performed in order to model the neutrino production in more detail. The two main reasons are that this allows us to use an updated Monte Carlo generator for performing the $pp$-interactions and that in this way we have access to secondary protons, which can interact again with the gas column. We implemented our model with the Monte Carlo generator `SIBYLL` 2.3 [497], which is an updated version of the code used for the analytical fits above. In particular, it includes the contribution from charmed meson decay [501, 502], although, as we will see, this will not noticeably impact our results[18].

In the simulation, protons are propagated through a matter column of specified integrated density $N_H$ and allowed to interact using standard Monte Carlo techniques[19]. Initially, in order to build up sufficient statistics at high energies, we inject a proton with an energy which is drawn from a power law distribution with index 1. This energy is determined using

$$E_p = E_{\min} \left( \frac{E_{\max}}{E_{\min}} \right)^{\eta} \tag{4.24}$$

with $\eta$ a random number drawn from a uniform distribution between 0 and 1. Afterwards, events are reweighted to the distribution under study (usually $\propto E_p^{-2}$). The mean free path of a proton of energy $E_p$ under $pp$-interactions propagating through a medium with density $n$ is given by

$$\lambda(E) = \frac{1}{n\sigma(E_p)}, \tag{4.25}$$

which is determined using the cross section tables calculated by `SIBYLL` (which agrees with Eq. (4.21)). The number of mean free paths travelled by a proton from the start of

---

[17] The rate can be found by replacing in Eq. (4.19) the factor $(1 - \exp(-N_H \sigma_{\text{inel}}))$ with $c n_H \sigma_{\text{inel}}$.

[18] This is in contrast with the result for atmospheric neutrinos, where charmed meson decay produces a distinct, harder spectrum. However, the reason for this difference is that in atmospheric neutrino production there is a competition between the decay of the meson and its interaction with an air nucleus. The latter produces softer spectra, since it initiates a new cascade. Because charmed mesons have a shorter lifetime, their contribution produces a harder spectrum. In astrophysical scenarios, this competition is not present, as also mentioned in [574].

[19] See e.g. the `GEANT4` [264] physics reference manual.

the column to a position $x$ is given by

$$n_\lambda = \int_0^x \frac{\mathrm{d}x}{\lambda(x)} \tag{4.26}$$

and is, by definition, distributed as

$$P(n_r < n_\lambda) = 1 - e^{-n_\lambda}. \tag{4.27}$$

The interaction point of a proton can be determined by sampling from this distribution, using

$$n_\lambda = -\log(\eta) \tag{4.28}$$

with $\eta \in (0,1)$ drawn from a uniform distribution. If this point is beyond the total depth of the gas column at the respective proton energy $n_{\lambda,\mathrm{tot}}(E_p) = N_H \sigma(E_p)$, the proton escapes and is saved in the final output. Otherwise, a collision is performed using `SIBYLL` and the final state particles $\nu_\alpha$, $\bar{\nu}_\alpha$, $e^\pm$ and $\gamma$ are saved, while secondary $p$, $n$ and their antiparticles are allowed to interact again with the remaining column[20]. The decay of pions and other mesons to neutrinos is performed by the `SIBYLL` decay routines. In order to obtain a better accuracy, these decay routines are often replaced by analytical calculations or by interfacing the output to other codes like `Pythia` [260]. However, these inaccuracies are mainly important for air shower simulations, where the full particle spectrum of individual events needs to be well modelled. For our purposes, where we only care about the total neutrino spectrum, the `SIBYLL` routines should suffice. Neutrons are considered stable in this simulation (the additional neutrinos from their decay outside of the source are at low energy, since most of the energy goes towards to resulting proton).

The attenuation of photons is taken into account by reweighting each photon by[21]

$$w_\gamma(E_\gamma, E_p) = \exp\left[-\left(1 - \frac{n_\lambda}{n_{\lambda,\mathrm{tot}}(E_p)}\right) N_H \sigma_{\mathrm{BH}}(E_\gamma)\right], \tag{4.29}$$

using the expressions for the Bethe-Heitler cross section in Eqs. (4.4) and (4.5).

Finally, the overall normalisation of the produced spectra is determined in different ways, depending on the scenario under consideration.

### 4.2.3 Simulation results

In this section, we explore the basic results from our simulation and, in particular, the effect of including secondary interactions and the correct column density for the gamma-ray attenuation. In addition, the Monte Carlo simulation is compared with the analytical calculation.

---

[20]For simplicity, the cross section of interactions with protons is put equal to the proton-proton cross section for all these particles (i.e. also for $n$ and anti-$p/n$), which is a good approximation at high energies.

[21]Note that by using this formula, we implicitly reduce the simulation to a one-dimensional one, ignoring the photon momentum in the direction perpendicular to the initial proton direction. While taking this into account would increase the gas column seen by the photon in the case of an infinite "plane" of gas, this correction is minor due to the beaming. Moreover, taking into account a more realistic geometry than a flat infinite plane would decrease the column slightly.

Table 4.1: Default parameters for the simulation against which to check the effect of parameter variations.

| # events | $E_{\min}$ (GeV) | $E_{\max}$ (GeV) | $\gamma$ | $N_H$ (cm$^{-2}$) | Second. int. | Attenuation |
|---|---|---|---|---|---|---|
| $5 \times 10^4$ | 100 | $10^8$ | $-2$ | $10^{26}$ | Yes | Full expr. |

Table 4.2: Relative luminosities of the neutrinos (all flavours), gamma rays and secondary protons for the different parameter choices, compared to the standard choices in Table 4.1.

| Scenario | $L_\nu / L_{\mathrm{Inj.}}$ | $L_\gamma / L_{\mathrm{Inj.}}$ | $L_p / L_{\mathrm{Inj.}}$ |
|---|---|---|---|
| Standard | 0.15 | 0.04 | 0.04 |
| $N_H^{(1)} = 5 \times 10^{25}$ cm$^{-2}$ | 0.13 | 0.06 | 0.16 |
| No secondaries | 0.12 | 0.03 | 0.18 |
| No attenuation | 0.15 | 0.11 | 0.04 |

**Full simulation**

First, we discuss the results of the full simulation for a set of standard parameters shown in Table 4.1. Most importantly, the protons are injected with an energy $E_p \in [10^2, 10^8]$ GeV and we consider the cases $N_H^{(1)} = 5 \times 10^{25}$ cm$^{-2}$ and $N_H^{(2)} = 10^{26}$ cm$^{-2}$. The spectrum is normalised to $\Phi(E_p) = 1$ GeV $\cdot E_p^{-2}$, i.e. the normalisation $A_p = 1$ GeV.

The result of the standard simulation is shown in Figure 4.4, indicating the final total neutrino, gamma-ray and proton flux. The proton flux includes both the escaping primary protons and the produced secondary protons that do not undergo a new interaction. As expected, most of the proton flux is depleted in the thick gas column, even for $N_H^{(1)} = 5 \times 10^{25}$ cm$^{-2}$. This immediately implies that the neutrino and gamma-ray flux are not significantly enhanced by increasing the density, as is confirmed by the figure. Moreover, for the neutrinos, the difference between the two target densities is largest at lower energy, where the $pp$-cross section is lowest. On the other hand, the higher density does increase the attenuation of the gamma rays significantly, up to a factor of a few at the highest energies. The integrated luminosities, relative to the injected luminosity, are also shown in Table 4.2 for this case and several of the parameter variations of the following sections.

**Comparison with analytical results**

The flux predicted by the Monte Carlo code is compared to the one calculated analytically in Figure 4.5a, for the parameters in Table 4.1, but turning off secondary interactions and gamma-ray attenuation. In general, there is a good agreement between the two versions. Inspecting the spectra in more detail, the neutrino fluxes have a similar normalisation, but the Monte Carlo flux is slightly lower and extends to slightly higher energy, although the latter is not significant. The neutrino luminosities in the range $[10^2, 10^7]$ GeV compare as $L_\nu^{MC}/L_\nu^{Anal.} = 1.23$. On the other hand, the gamma-ray flux in the analytical calculation is consistently above the simulation, with a gamma-ray luminosity ratio $L_\gamma^{MC}/L_\gamma^{Anal.} = 0.65$. This is an important difference when calculating the diffuse flux in Section 4.4. Since the Monte Carlo code uses a more recent model, gives good agreement with the neutrino flux and includes an updated interaction model, we prefer it over the analytical calculation.

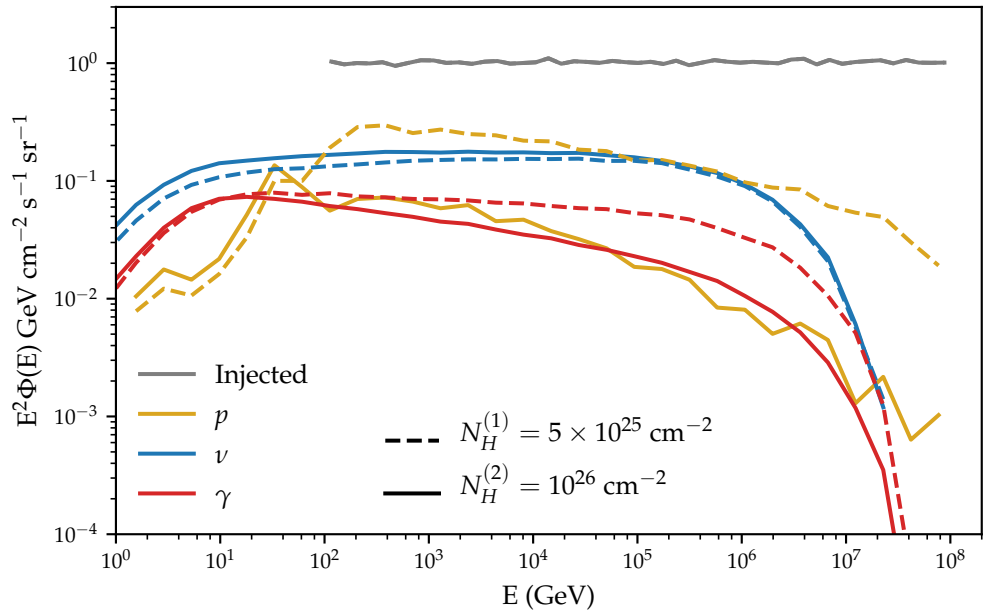

Figure 4.4: Predicted neutrino (all flavours), gamma-ray and proton fluxes (neutron flux is comparable to the proton flux, but not shown) from the obscured $pp$-neutrino production scenario for standard parameters given in Table 4.1, for column densities $N_H^{(1)} = 5 \times 10^{25}$ cm$^{-2}$ and $N_H^{(2)} = 10^{26}$ cm$^{-2}$. The injected proton spectrum is normalised to $\Phi_p(E_p) = 1$ GeV $\cdot E_p^{-2}$.

### ν/γ-ratio

In Figure 4.5b, we show the predicted ν/γ-ratio from both the Monte Carlo and analytical calculation, using the same parameters as the previous calculation. The ratio is compared with the theoretical value found from Eq. (3.45). The Monte Carlo simulation finds $L_\nu^{MC}/L_\gamma^{MC} = 1.30$, while the analytical fits give $L_\nu^{Anal}/L_\gamma^{Anal.} = 0.69$. The Monte Carlo code agrees better with the expected value of $2\frac{3}{4}$ (see Section 3.3.6).

### Contribution from secondary interactions

Figure 4.6 shows the influence of including secondary interactions, for the parameters in Table 4.1. By showing the effect for the highest column density benchmark point to be investigated, we maximise the effect of parameter variations.

Including secondary interactions depletes the proton flux, which consisted mainly of secondary protons created in $pp$-interactions. While the secondary protons traverse a reduced column density compared to the primary protons, a large fraction manages to interact again, since the total gas column is several mean free paths thick. However, the secondary proton flux in the case of no secondary interactions is irrelevant compared to the primary proton flux[22]. Therefore, turning on secondary interactions does not significantly increase the neutrino and gamma-ray flux. The luminosity ratios for the case of no secondary interactions are also shown in Table 4.2.

The equivalent figure and table for $N_H^{(1)} = 5 \times 10^{25}$ cm$^{-2}$ can be found in Appendix H.1.1.

### Effect of the gamma-ray attenuation

The effect of including the gamma-ray attenuation is shown in Figure 4.7a, for the parameters in Table 4.1, varying only the attenuation. The no-attenuation case is compared with different methods of calculating the attenuation: using the full expression in Eqs. (4.4) and (4.5), using the approximation $\sigma_{BH} = 20$ mb and including only the attenu-ation by the protons $\sigma_{\mathrm{BH}} = 10$ mb[23]. Including the attenuation significantly reduces the total gamma-ray flux. Moreover, using the full expression has a large additional effect at the highest energies. This is particularly important when considering the cascading of high-energy gamma rays to lower energy in order to calculate gamma-ray bounds from the EGB (see Section 4.4). The luminosity ratios in the case of no attenuation are also shown in Table 4.2.

---

[22]This is a simple consequence of the power law proton spectrum: the $N$ secondary protons produced by a proton of energy $E_p$ carry on average a fraction $x$ of the parent proton energy and are dominated by the primary protons at the lower energy $xE_p$, which are more numerous by a factor $\frac{(xE_p)^{-2}}{NE_p^{-2}} = \frac{1}{x^2 N}$. Since the sum of all secondary energies (including leptons and gamma rays) needs to total $E_p$, we have $xN < 1$ and $x < 1$, so the primary protons are indeed dominant.

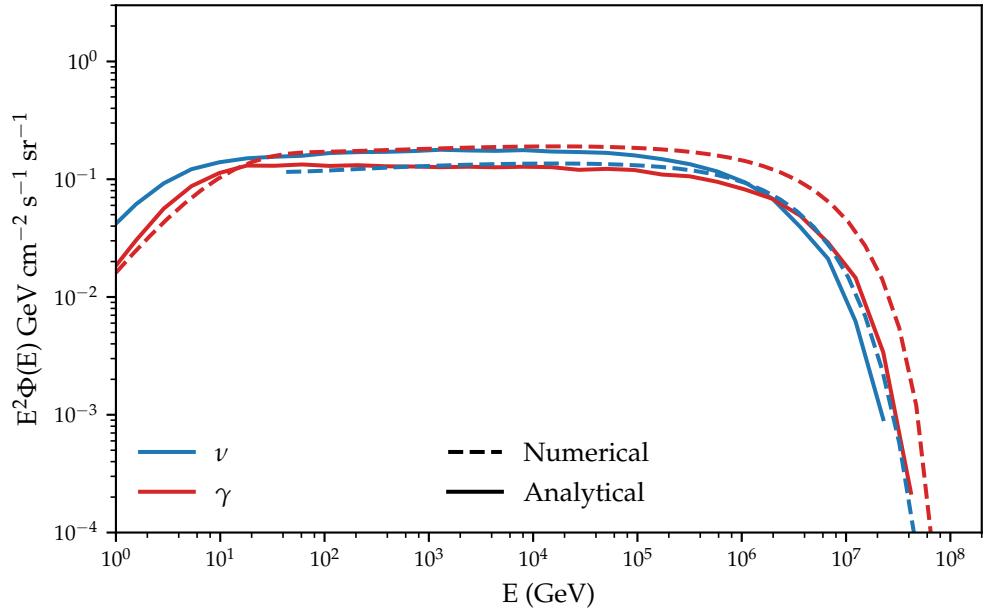

(a) ν and γ-ray flux

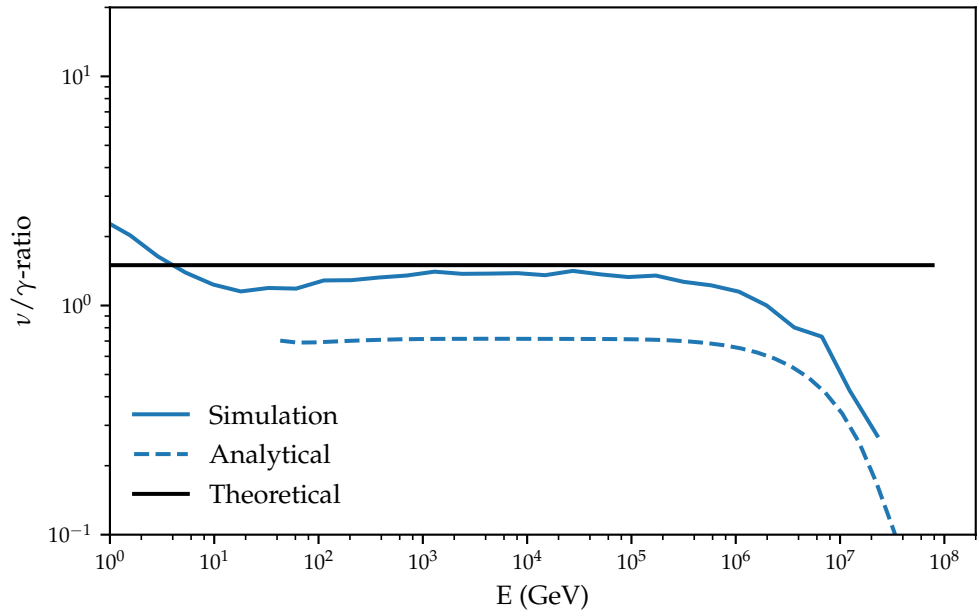

(b) Predicted ν/γ-ratio

Figure 4.5: Comparison of the analytical result with the Monte Carlo code on the neutrino (all flavour) and gamma-ray flux, for the parameters in Table 4.1, but turning off secondary interactions and gamma-ray attenuation.

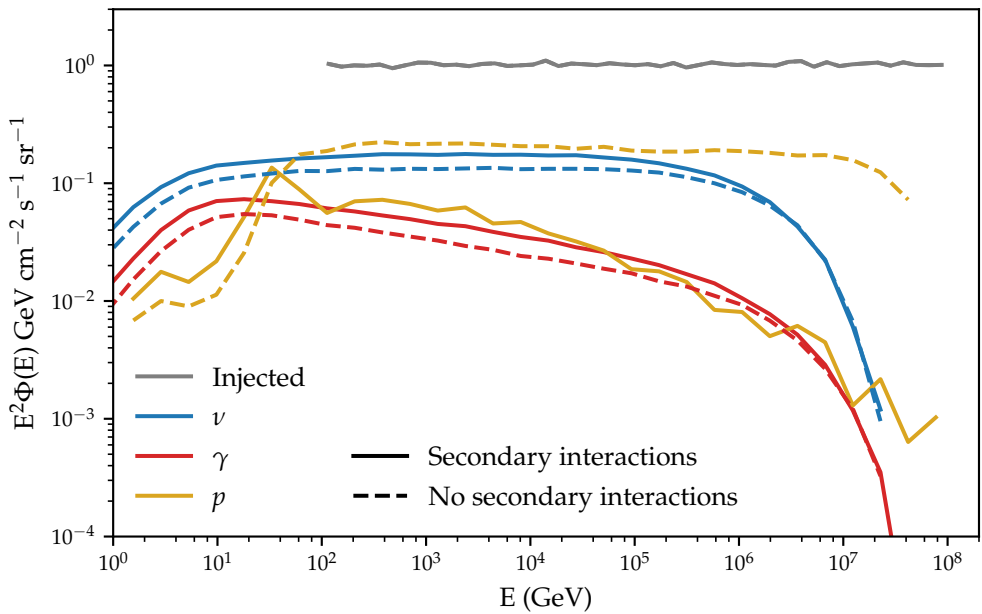

Figure 4.6: The effect of including secondary interactions on the neutrino (all flavour) and gamma-ray flux, for the parameters in Table 4.1.

In addition to the above, in Figure 4.7b we also show the effect of taking into account the correct column density remaining after creation of the photons compared to using the full density, as well as ignoring the attenuation. Since the cloud is several proton interaction depths thick, the remaining part of the column is sufficient for the photons to be significantly attenuated. In fact, the attenuation is less than a factor of 2 lower than in the approximation using the full column.

The equivalent figures and table for $N_H^{(1)} = 5 \times 10^{25}$ cm$^{-2}$ are found in Appendix H.1.2.

---

[23]This would be relevant if there were a significant charge separation between the protons and electrons, but this is unlikely.

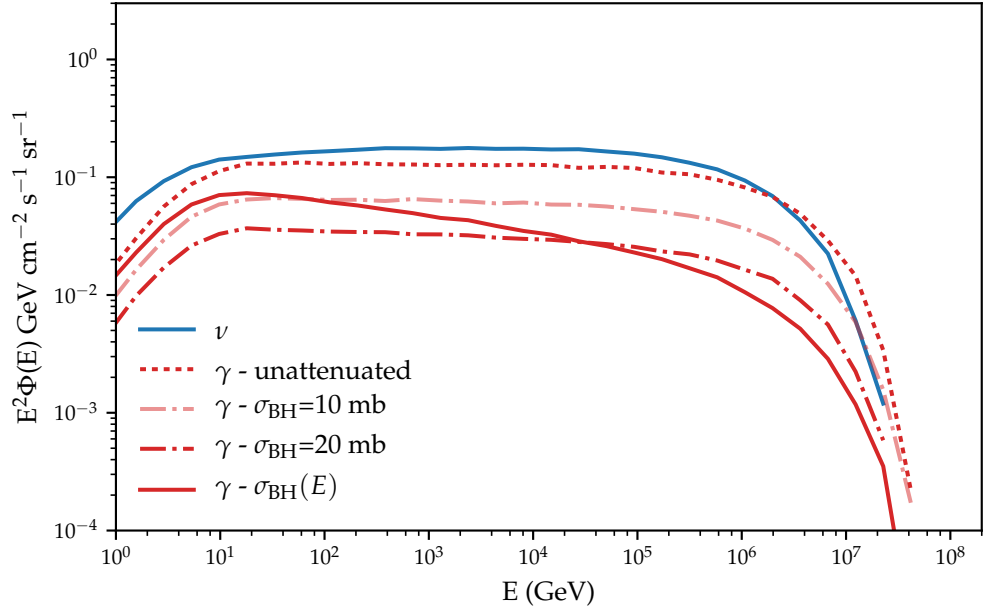

(a) Including pair production

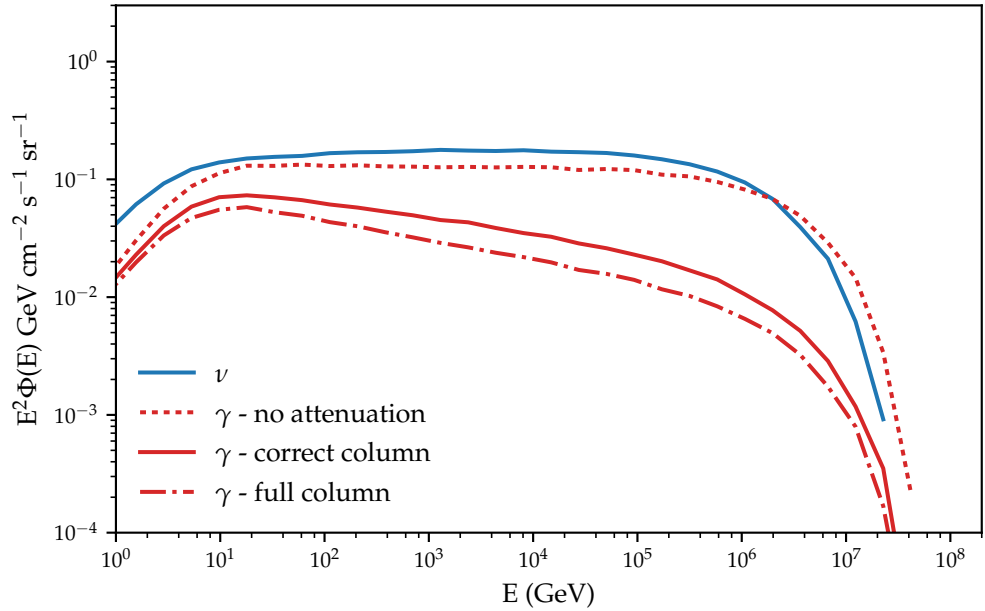

(b) Column density determination

Figure 4.7: Effect of including attenuation of the gamma rays, compared to several other cases. Parameters for the simulation are given in Table 4.1. (a) The effect of including gamma-ray attenuation by pair production, for different approximations of the Bethe-Heitler cross section. The full expression is given by Eqs. (4.4) and (4.5). (b) The effect of including the correct column density after creation of the photons ins $pp$-interactions, compared to several approximations.

## 4.3   Obscured flat-spectrum radio AGNs

In this section, our model is applied to a set of AGNs selected for their possible obscuration by matter. We calculate the neutrino flux and, using existing limits on their neutrino emission from IceCube [636], derive limits on the cosmic ray content of these sources. Our selection, performed in [687], targets nearby sources which are candidate cosmic ray accelerators, feature beamed emission (i.e. a jet, so that all emission is boosted towards Earth) and which exhibit signs of obscuration by matter. As already explained in Section 4.1.2, this is best done using X-rays. On the other hand, radio waves can propagate through gas unimpeded. Moreover, radio emission is usually explained by synchrotron emission from a non-thermal population of electrons. Therefore, the radio emission from these sources characterises the strength of the inner engine. As such, we select objects with a lower-than-expected X-ray flux, relative to the radio flux. Under the assumption that this reduced X-ray flux is (mainly) due to attenuation by matter, these objects are potential strong neutrino sources through $pp$-interactions (in addition to the $p\gamma$-interactions, which we ignore). This selection will be briefly described in Section 4.3.2.

   The original motivation for this object selection is the possibility of having AGNs with tilted tori [712, 726]. In this case, the jet can penetrate the dust torus, which has a considerable integrated column density, and produce neutrinos efficiently while being obscured in X-rays. However, the exact origin of the gas cloud is not important for the details of the calculation. We assume the cloud is stable on observation timescales (i.e. at least years). Due to the strong electromagnetic radiation from the jet, the cloud can (and will, see next section) become ionised. At the same time, these radiation fields also attenuate the gamma rays through $\gamma\gamma$ pair production, which we do not model. The gamma ray flux predicted is therefore an upper limit on the hadronic gamma rays. On the other hand, leptonic processes might create additional gamma rays. Since we are only interested in the neutrino emission from these objects and do not model their complete SED, this is not an issue.

### 4.3.1   Jet-matter interaction

In order to estimate the fraction of interacting protons from the X-ray obscuration, we need to know the column density of the obscuring matter. This can be determined from the observed X-ray flux if we know which process is responsible for X-ray attenuation. At higher X-ray energies, this is always Compton scattering (which we already discussed in Section 4.1.2), but at 1.24 keV photo-electric absorption dominates if the electrons are still bound in atoms. Due to the strong radiation of these sources, however, the obscuring gas cloud is completely ionised for natural geometries. This can be seen as follows. The degree of ionisation is determined from the balance between ionising reactions and recombinations. Define the ionisation parameter [731]

$$U = \int_{E_1}^{E_2} \mathrm{d}E \; \frac{L_E/E}{4\pi r^2 c n_N},$$  (4.30)

with $L_E$ the differential luminosity, $n_N = N_H/d$ the nucleon number density, $d$ the thickness of the cloud and $r$ the distance of the cloud to the central engine. This quantity represents the ratio of ionising photons to the gas density. The factor of $c$ is included to make $U$ dimensionless. The integral is performed over the energy range of the photon flux relevant for the ionisation, in this case from $\sim 0.1$ keV to $\sim 10$ keV. For AGNs, typical values range from $10^{-3}$ to 1. For $U_X \approx 10^{-3}$, the cloud is partially ionised, while for $U_X > 0.1$ the cloud is highly ionised. The ionisation parameter $U$ is related to the overall ionisation as

$$\frac{N_{H^+}}{N_{H^0}} \simeq 10^{5.3} U \tag{4.31}$$

in the case of hydrogen for BLR clouds (this relation depends on the element and cloud under consideration).

We now estimate the cloud parameters $r$ and $d$, assuming the cloud leads to an attenuation of the X-ray flux by 90%. In case of an ionised cloud, where Compton scattering dominates, the column density leading to this level of obscuration is $N_H \sim 10^{26}$ cm$^{-2}$. For typical values of $L_E$ and $U_X = 0.1$, this corresponds to a cloud of thickness $d$ and distance $r$ from the central engine between 1 pc and 10 pc. This coincides with the typical location and size of the dust torus, at a different orientation, although we do not restrict ourselves to this case. Therefore, applying our model to the case of AGN jets, we naturally find the required column density for significant neutrino production with a reasonable physical configuration for ionised clouds.

### 4.3.2 Object selection

The object selection targets nearby potential cosmic ray sources and identifies objects whose X-ray emission is low compared to their radio emission. As a basis, we use two catalogues. The first catalogue is the Van Velzen catalogue [732], which is a volume-limited catalogue of radio sources that could be responsible for the acceleration of the observed UHECRs[24]. From this catalog, we retain only the sources with a morphology compatible with a jet pointing towards Earth (i.e. star-forming galaxies and unresolved point sources, as opposed to sources with jets and lobes or an unknown morphology). The second catalogue is a subsample of the second Fermi-Large Area Telescope AGN catalogue (2LAC) [635], which contains gamma-ray sources[25] like blazars, radio galaxies and others. After combining the two catalogues, we obtain 735 unique galaxies with redshift information in the NASA/IPAC extragalactic database (NED) [733].

Far-away objects could be detected with a lower-than-expected X-ray flux due to an observational bias or due to attenuation by the intergalactic medium. Therefore, we perform a redshift selection in order to get an unbiased set of objects. Since the Van Velzen catalogue is already limited to nearby objects with $z < 0.1$, this selection is only performed for the 2LAC sample. In order to determine the redshift value on which to cut, we inspect the cumulative flux distribution. For a uniform source density and generic luminosity, this distribution behaves as $N(F > F_0) \propto F_0^{-3/2}$ (the number of sources increases with $r^3$, while their flux diminishes as $r^{-2}$). Inspecting this distribution (for the 1.24 keV flux) for the objects in our starting sample, we observe this behaviour down to a flux $\log_{10}(F/Jy) = -5.5$, indicating a deficit of sources below this value. From the redshift distribution for the objects above this threshold, we choose to limit our sample to objects with $z < 0.17$ (without a requirement on the luminosity), such that 209 objects remain.

---

[24]Although, since this catalog is based on an optical catalog, some strong radio galaxies are missing.

[25]While we do expect sources obscured by matter to have reduced gamma-ray emission, this does not imply they could not be detected by Fermi-LAT.

Next, we require that the presence of a jet pointing towards Earth. This can be inferred from the spectral index of the radio flux, which is flat in the case of a jet[26]. We fit the radio flux with a power law in the range $\nu = 0.843 - 5$ GHz and require $\alpha_R + \sigma_{\alpha_R} > -0.5$. With this requirement, we retain 98 nearby flat spectrum radio AGNs.

Finally, we select for X-ray obscuration relative to the radio. More specifically, we compare the 1.24 keV X-ray luminosity with the 1.4 GHz radio luminosity[27]. Only 62 objects in our sample have sufficient X-ray data in NED. Of these, 49 objects are in the North, 13 in the South; 14 are FSRQ, 3 Ultra-luminous infrared galaxies (ULIRG) and 45 BL Lac. One of the objects, NGC 3628, has an X-ray flux compatible with background. Since our selection requires X-ray emission in order to estimate the amount of obscuration, it is not considered in our selection. However, its completely suppressed X-ray emission makes it a potentially interesting target for neutrino production.

For FSRQ, there is a relation between X-ray luminosity $L_X$ (from the hotter parts of the accretion disk) and the radio luminosity $L_R$,

$$L_R = L_X^\beta, \tag{4.32}$$

with $\beta = 0.6 - 0.7$ [735, 736]. In our sample, we indeed find this relationship for the FSRQ, with $\beta \approx 0.73$. Therefore, we correct for this correlation and will perform the selection on the corrected relative luminosity $I = L_X^{0.73}/L_R$. Ultra-luminous infrared galaxies (ULIRG) are galaxies exhibiting an enormous infrared bump in their spectral energy distribution. This indicates the presence of a lot of dust and a strong engine. Moreover, while all ULIRGs are powered by a starburst, (some) ULIRGs might also harbour an AGN (for more details, see Section 4.4.1). In any case, their radio emission, in particular for the objects in this selection, implies that particle acceleration takes place, making them an extremely interesting target. For these objects, the same correlation between the X-ray and radio luminosity as for FSRQ is observed. BL Lac show no clear correlation between the X-ray and radio luminosity, which is verified in our sample. Therefore, we do not include a correction and base our final selection purely on the X-ray luminosity.

Finally, we search for obscured objects in our sample. In order to determine the amount of obscuration, knowledge of the X-ray luminosity before attenuation is needed.

---

[26]The flatness of the radio spectrum is typically explained by a superposition of synchrotron emission spectra with different peak frequencies. In a conical jet, the peak frequency diminishes with radius of the emitting region. When pointing towards the observer, the emitted flux originates from different parts of the jet, with a different radius and therefore naturally leads to this superposition (see e.g. [734]).

[27]Using data where available, which is the case for most objects, or determined from the radio-flux fit.

Since this is not available from observations, we assume that our sources are generic (either in $I$ defined above for FSRQs and ULIRGs or in $L_X$ for BL Lac) and that the observed spread in X-ray intensity is due to presence of gas and dust. In that case, we can estimate the generic X-ray luminosity pre-attenuation from unobscured sources from the 25% strongest sources (i.e. we take the first quartile of the distribution as $I_X^0$). We select the 25% weakest objects as interesting targets for neutrino production from obscured sources. This leaves 15 objects in the final selection (3 FSRQ or FSRQ-like, 1 ULIRG, 11 BL Lac), which was published in [687].

From the attenuation of X-rays as $I_X^{\mathrm{obs}} = I_X^0 \exp(-X_{\mathrm{tot}}/\lambda_X)$, the amount of interacting protons can then be calculated as

$$\frac{I_p^{\mathrm{int}}}{I_p^0} = 1 - \left( \frac{I_X^{\mathrm{obs}}}{I_X^0} \right)^{\lambda_X/\lambda_{pN}}. \qquad (4.33)$$

Here, the X-ray attenuation depth due to Compton scattering in units $\mathrm{g\,cm}^{-2}$ is given by $\lambda_X = \frac{N_A m_A}{\sigma}$, which can be easily generalised to compounds (which were also studied in [687]) and is obtained from the XCOM database [692]. The fraction of interacting protons thus obtained varies between 0.80 and 0.99. With the proton cross section $\sigma_{pp}$ and assuming again a pure proton target, this corresponds to column densities $N_H \sim 5 \times 10^{25}$–$10^{26}\ \mathrm{cm}^{-2}$, exactly those we consider in our model.

An IceCube analysis was done for 14 objects. Two objects from the final selection above were not analysed, since they are located in the southern sky where IceCube has a lower sensitivity. On the other hand, NGC 3628 was added back in the analysis, since it was previously omitted for its lack of X-ray emission above background, making it, however, an interesting target. The analysis finds no significant signal and gives an upper limit on the $E^2\Phi$-flux for each of these objects, assuming an $E^{-2}$-flux between 1 TeV and 1 PeV. The final list of objects, their classification and the limits on their neutrino emission are shown in Table 4.3.

Finally, it is important to remark that since almost all objects in this selection are a subset of blazars, it is not possible that sources from this class (i.e. "obscured blazars") are responsible for the bulk of the neutrino flux (from the source density arguments in Section 3.8.3). However, it is still possible that the same model applied to a different source class *does* produce the required flux (e.g. ULIRGs, see Section 4.4.1). Moreover, this selection in particular is still interesting for two reasons. First, given the luminosity of these objects, they are good targets to test whether the model occurs in nature. Second, if the scenario is indeed applicable to these objects, the neutrino flux thus produced allows us to directly probe the amount of accelerated hadrons in these blazars.

### 4.3.3   Normalising the neutrino flux

In the following we calculate the neutrino flux expected in our model for the objects in the final IceCube analysis cited above. However, we do not use the values of the column densities for each object individually inferred from our selection. Instead, we assume the benchmark values $N_H^{(1,2)}$, which are compatible with the inferred values, and investigate the resulting neutrino production.

Table 4.3: Final objects in the selection for which an IceCube analysis exists and the upper limit on their neutrino emission in units of $[10^{-9}\,\mathrm{GeV\,cm^{-2}\,s^{-1}}]$, from [636].

| Name | RA (°) | Dec (°) | $E^2\Phi_\nu^{90\%}$ | Classification |
|------|--------|---------|----------------------|----------------|
| PKS1717+177 | 259.80 | 17.75 | 0.754 | BL Lac |
| CGCG186-048 | 176.84 | 35.02 | 0.856 | BL Lac |
| RGBJ1534+372 | 233.70 | 37.27 | 0.899 | BL Lac |
| NGC3628 | 170.07 | 13.59 | 0.719 | Radio gal. |
| SBS1200+608 | 180.76 | 60.52 | 1.090 | BL Lac |
| GB6J1542+6129 | 235.74 | 61.50 | 1.070 | BL Lac |
| 4C+04.77 | 331.07 | 4.67 | 0.650 | BL Lac |
| MRK0668 | 211.75 | 28.45 | 0.879 | FSRQ |
| 3C371 | 271.71 | 69.82 | 1.180 | BL Lac |
| B21811+31 | 273.40 | 31.74 | 0.850 | BL Lac |
| SBS0812+578 | 124.09 | 57.65 | 1.090 | BL Lac |
| 2MASXJ05581173+5328180 | 89.55 | 53.47 | 1.080 | FSRQ |
| 1H1720+117 | 261.27 | 11.87 | 0.695 | BL Lac |
| ARP220 | 233.74 | 23.50 | 0.746 | ULIRG |

Given the choice of benchmark values of $N_H$ in our model, the only parameter left to determine in order to predict the neutrino flux from the set of objects selected above, is the normalisation of the flux. The first selection criterion for the object selection was strong radio emission. Moreover, the radio flux is unattenuated by matter in between the source and the observer, giving a direct view of the inner engine, a feature which was also already exploited in the analysis above. Therefore, it is natural to normalise the expected neutrino flux based on the radio flux.

The radio emission from astrophysical objects is typically attributed to synchrotron emission from accelerated electrons in the magnetic field of the source. Therefore, it is expected that the radio and electron luminosity are comparable in size. The exact relation between the radio and electron luminosities was derived in [495]. It is obtained by integrating the synchrotron emission from electrons of energy $E_e$, or equivalently $\gamma_e$, over the electron spectrum. The result for $\chi = L_e/L_R$ is shown in Figure 4.8, assuming an electron spectrum $\frac{\mathrm{d}N_e}{\mathrm{d}\gamma_e} \propto \gamma_e^{-2}$. The minimum energy $\gamma_e^{\min}$ is estimated from efficient cooling (i.e. strong radio emission) of the electrons and is shown for $\gamma_e^{\min} = \{1, 10\}$. The maximum energy $\gamma_e^{\max} = 10^9$–$10^{11}$ corresponds to the assumption that protons are co-accelerated up to energies $10^{18}$–$10^{21}$ eV. For these limits of the electron energy, we find that

$$\chi = \frac{L_e}{L_R} \approx 100, \tag{4.34}$$

over a large range of magnetic field strengths. Deviations from this value occur fastest for $\gamma_e^{min} = 10$, with significant changes starting at 10 G. Modelling the properties of bright Fermi blazars, one finds that typical magnetic field strengths are between $0.1 - 2$ G for BL Lacs and 1–10 G for FSRQs [421, 737]. Since the objects in our selection are assumed to be typical, apart from their obscuration, we use $\chi = 100$.

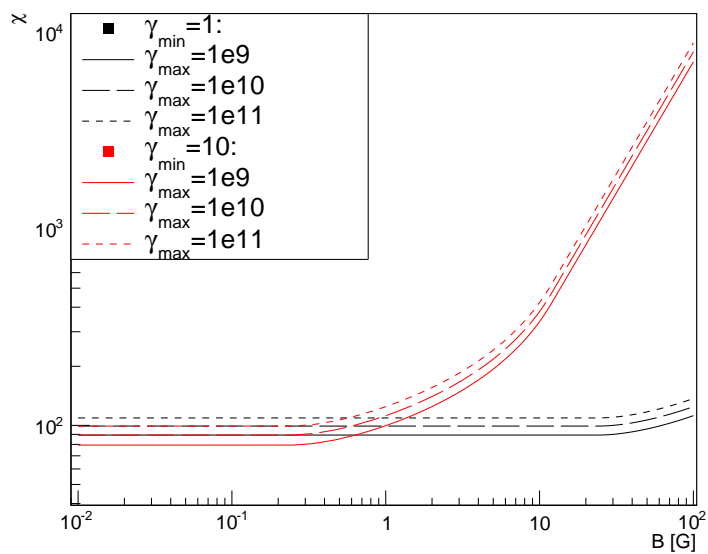

Figure 4.8: Radio-electron luminosity correlation $\chi = L_e/L_R$ as a function of magnetic field strength, with an accelerated electron population $\frac{dN_e}{d\gamma_e} \propto \gamma_e^{-2}$ for different energy ranges of the electron population. Figure from [495].

From the electron luminosity, we can obtain the proton luminosity, since these two species are co-accelerated. Typically, one assumes that the number of accelerated protons and electrons are equal[28], $N_e = N_p$ and that their spectral indices are the same. It is then straightforward to calculate the electron-proton luminosity ratio and this results in $f_e = \frac{L_e}{L_p} \approx 1/100$ [729, 738], which is true also for the differential luminosity (i.e. independent of energy). This result is supported by observations from the galaxy: when comparing the observed electrons with cosmic rays up to the knee, one obtains a luminosity ratio 1/100. On the other hand, for extragalactic sources, this value is estimated to be closer to 1/10, obtained by comparing the observed radio luminosity with that of cosmic rays above the ankle. Since obtaining a more precise value requires extrapolating the extragalactic cosmic ray flux to energies below the ankle, with an unknown energy spectrum and minimum energy, this value is quite uncertain. Moreover, this extrapolation is dependent on the exact spectral index of extragalactic cosmic rays,

---

[28]From charge balance, this is true for the total number of electrons and protons. If the only requirement for initiating the acceleration is sufficient energy, this is then also true for the number of particles above the energy threshold, since in a plasma the particle energies follow a Maxwell-Boltzmann distribution which is independent of the particle mass [729]

which is still uncertain (since there is a degeneracy with composition and maximum energy, see Section 3.2.6). In the case of an index deviating from 2, the luminosity integral becomes very sensitive on the minimum energy, making a strong constraint on $f_e$ difficult (for a recent discussion, see [729]). However, for sure $f_e \ll 1$. Note that from particle-in-cell simulations, it is found that the assumption of equal spectral indices might not be true. In this case the luminosity ratio becomes energy dependent, further complicating the conversion from electron to proton luminosity (again, see [729]). In the following, we will assume a fixed ratio

$$f_e = \frac{L_e}{L_p} = \frac{1}{10},$$ (4.35)

which is conservative (i.e. relatively little protons).

Summarising, in order to determine the expected neutrino flux from the objects in the selection above, we integrate the observed radio flux from each object individually and convert this to a proton luminosity with

$$L_p = \frac{\chi \cdot L_R}{f_e}.$$ (4.36)

A proton population with this total luminosity is then allowed to interact with a gas cloud of column density $N_H = N_H^{(1,2)}$. Note that it is not needed to convert the observed flux to luminosity at the source, since the same factor $d_L^2$ appears when propagating the obtained neutrino flux at the source back to Earth.

### 4.3.4 Results

Now, we show the neutrino flux predicted by our model for the objects in our obscured flat-spectrum AGN selection. The parameters for the simulation are the same as in Table 4.1, varying only the column density for the two cases under consideration. As already stated, we include only gamma-ray attenuation at the source by matter, not radiation fields. While propagating to Earth, the gamma-ray flux is also attenuated by interaction with the EBL and CMB (this is discussed in more detail in Section 4.4.3). This attenuation, but not the full EM cascade, is included in the final gamma-ray flux[29].

The hybrid spectral energy distribution (SED) for the object with the highest neutrino emission, 3C371, is shown in Figure 4.9, with $N_H^{(2)} = 10^{26} \, \text{cm}^{-2}$. The figure includes the measured photon SED across all wavelengths, the predicted gamma rays and muon neutrinos[30] and the IceCube limit on the muon neutrino flux from this object. The SEDs of the other objects are shown in Appendix H.2.1. The predicted neutrino flux for 3C371 is much below the limit from IceCube, so that our model is not ruled out and could only be observable with IceCube Gen2. The gamma-ray flux from $pp$-interactions in this model is far below the observed flux from this object, leaving the model also unconstrained here.

---

[29]This is the usual approach. For a point source, including the full cascade would require more detailed modelling than the simple approach of Section 4.4.3. Since the predicted gamma-ray flux turns out to be very low compared to the observed flux, the additional modelling is not important.

[30]The muon neutrino flux is 1/3 of the total neutrino flux, assuming full mixing as in Section 3.3.3. This is the reason why the neutrino flux is now below the gamma ray flux, instead of above as in the figures in Section 4.2.3.

Table 4.4: Summary of the predicted muon neutrino flux compared to the upper limits on their muon neutrino flux (from [636]), for the objects in the obscured flat-spectrum radio AGN selection in units of $10^{-9}$ GeV cm$^{-2}$ s$^{-1}$, as well as the corresponding limits on $f_e$. Protons are injected within the energy range $E_p \in [10^2,\ 10^8]$ GeV and the considered column densities are $N_H^{(1)} = 5 \times 10^{25}$ cm$^{-2}$ and $N_H^{(2)} = 10^{26}$ cm$^{-2}$.

| Name | $E^2\Phi_\nu^{90\%}$ | $E^2\Phi_\nu^{(1)}$ | $E^2\Phi_\nu^{(2)}$ | $f_e^{(1)}$ | $f_e^{(2)}$ |
|---|---|---|---|---|---|
| PKS1717+177 | 0.754 | 0.062 | 0.070 | $8.1 \times 10^{-3}$ | $9.2 \times 10^{-3}$ |
| CGCG186-048 | 0.856 | 0.032 | 0.037 | $3.7 \times 10^{-3}$ | $4.2 \times 10^{-3}$ |
| RGBJ1534+372 | 0.899 | 0.002 | 0.002 | $1.9 \times 10^{-4}$ | $2.2 \times 10^{-4}$ |
| NGC3628 | 0.719 | 0.039 | 0.045 | $5.4 \times 10^{-3}$ | $6.1 \times 10^{-3}$ |
| SBS1200+608 | 1.090 | 0.012 | 0.014 | $1.1 \times 10^{-3}$ | $1.2 \times 10^{-3}$ |
| GB6J1542+6129 | 1.070 | 0.009 | 0.011 | $8.7 \times 10^{-4}$ | $9.9 \times 10^{-4}$ |
| 4C+04.77 | 0.650 | 0.069 | 0.079 | $1.0 \times 10^{-2}$ | $1.2 \times 10^{-2}$ |
| MRK0668 | 0.879 | 0.155 | 0.176 | $1.7 \times 10^{-2}$ | $2.0 \times 10^{-2}$ |
| 3C371 | 1.180 | 0.240 | 0.274 | $2.0 \times 10^{-2}$ | $2.3 \times 10^{-2}$ |
| B21811+31 | 0.850 | 0.015 | 0.017 | $1.7 \times 10^{-3}$ | $1.9 \times 10^{-3}$ |
| SBS0812+578 | 1.090 | 0.008 | 0.009 | $7.0 \times 10^{-4}$ | $8.0 \times 10^{-4}$ |
| 2MASXJ05581173+5328180 | 1.080 | 0.028 | 0.032 | $2.6 \times 10^{-3}$ | $2.9 \times 10^{-3}$ |
| 1H1720+117 | 0.695 | 0.007 | 0.008 | $1.0 \times 10^{-3}$ | $1.1 \times 10^{-3}$ |
| ARP220 | 0.746 | 0.031 | 0.035 | $4.0 \times 10^{-3}$ | $4.6 \times 10^{-3}$ |

The same conclusions are also true for the other objects in the selection. Their calculated neutrino fluxes can be found in Table 4.4, for both $N_H^{(1)} = 5 \times 10^{25}$ cm$^{-2}$ and $N_H^{(2)} = 10^{26}$ cm$^{-2}$ and are also shown in Figure 4.10. The figure also includes the flux in case the minimum proton energy is lowered to 1 GeV. Similar figures showing the effect of secondary interactions can be found in Appendix H.2.2. For all the objects, the predicted neutrino flux is below the limit placed by IceCube, with natural choices for the values of the parameters $\chi$ and $f_e$. Given the expected sensitivity of IceCube Gen2, only some of these objects could be observable in the near future in this model. The gamma-ray flux is in each case much below the observed value, putting no constraint on the model. This also immediately implies that, for this class of objects, there is no constraint from the EGB, since their contribution is irrelevant compared to the blazar contribution already present. On the other hand, this also immediately implies that, even if this scenario were applicable to all blazars (which is certainly not true), the neutrino flux would not be high enough to explain the diffuse flux observed by IceCube.

The above results already allow us to put meaningful constraints on the only param-eter which is not fixed by our model[31]: the electron-proton luminosity ratio, for which we took $f_e = 1/10$. Translating the IceCube limit to a lower bound on $f_e$ (i.e. maximum amount of accelerated protons) in our model, we find a bound between about 0.001 and 0.02. This is also shown in Table 4.4 and Figure 4.11. Again, additional figures can be found in Appendix H.2.2. Since the galactic value of the electron-proton luminosity ratio is $f_e \approx 1/100$, these constraints on $f_e$ within the currently considered model are already quite strong. The reason that these bounds are so strong, is that for the column densities considered here, the full proton population is depleted to produce neutrinos, as opposed to typical scenarios where only part of the proton flux interacts. The advantage of this was already discussed: even if such blazars are not the dominant source of astrophysical neutrinos, identifying a few obscured blazars gives independent constraints on the amount of accelerated protons in blazars.

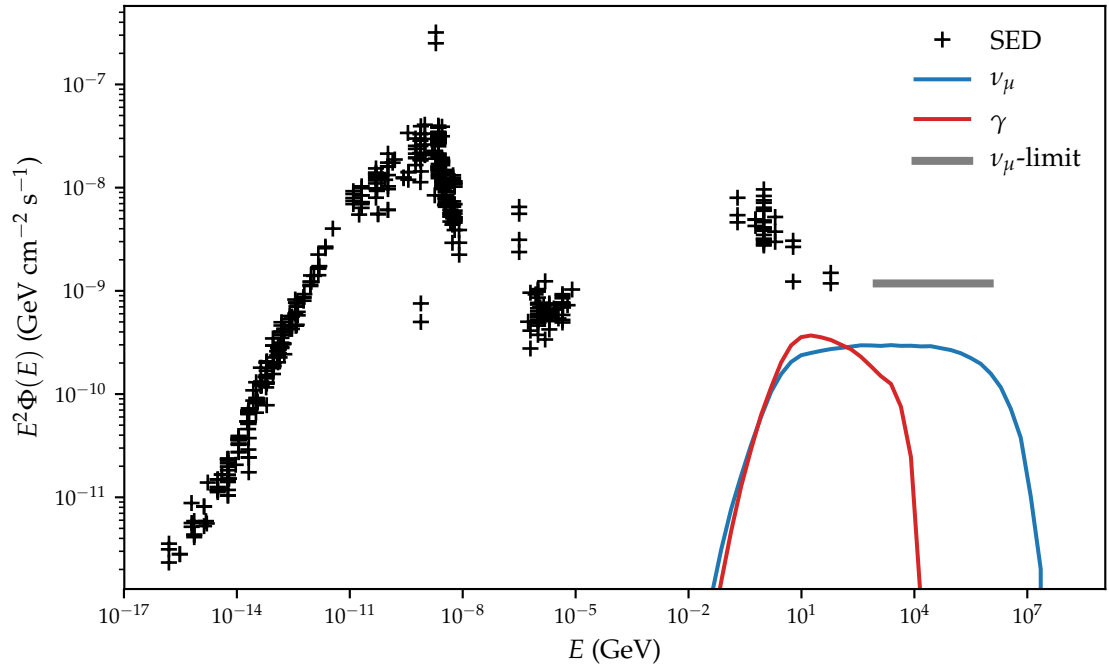

Figure 4.9: Hybrid SED for 3C371, one of the objects with a predicted muon neutrino flux closest to the current upper limit [636], showing the measured electromagnetic data together with the predicted muon neutrino flux and gamma-ray flux in our obscured neutrino source model. Here we assumed $N_H^{(2)} = 10^{26}$ cm$^{-2}$ and $E_p \in [10^2, 10^8]$ GeV. Electromagnetic spectrum data from [739–765] retrieved using the SSDC SED Builder [766].

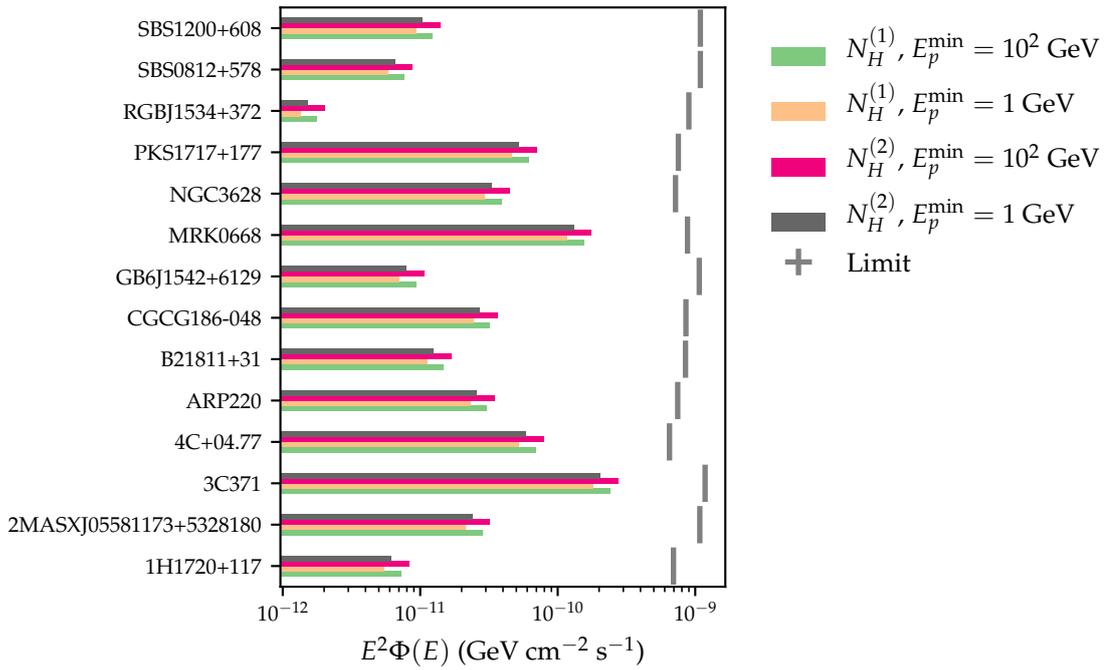

Figure 4.10: Summary of the predicted muon neutrino flux in our obscured neutrino source model and the IceCube upper limits (from [636]) for the objects in the obscured flat-spectrum radio AGN selection.

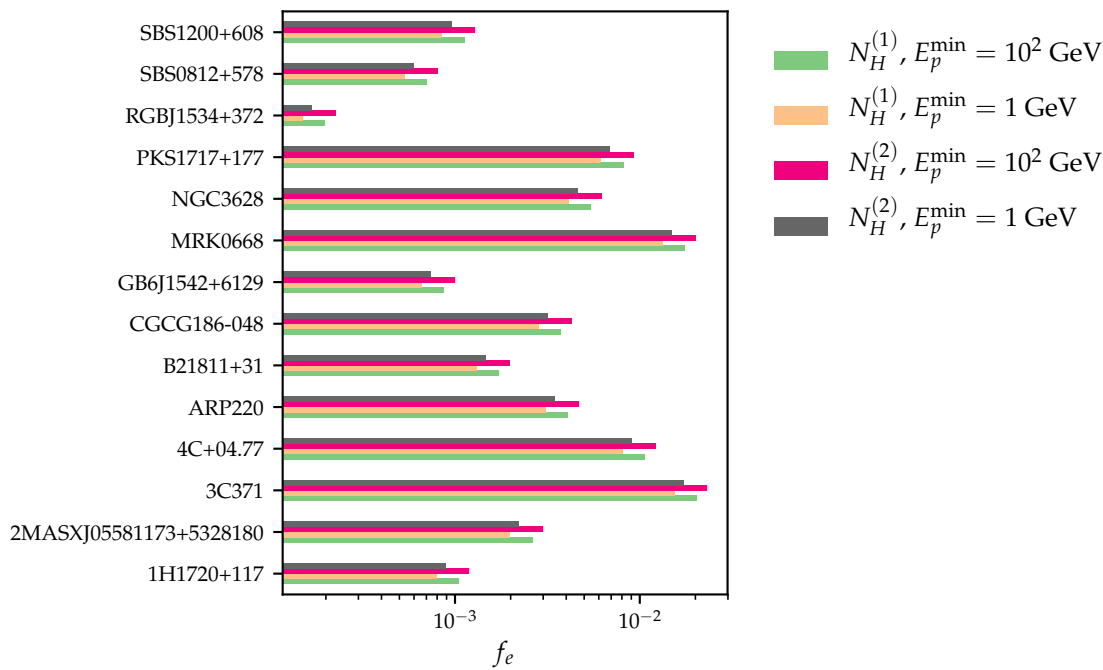

Figure 4.11: Summary of the lower limits on $f_e$ for the objects in the obscured flat-spectrum radio AGN selection.

Finally, we can also investigate a variant of the model above, normalising the predicted fluxes to the gamma-ray flux instead of to the radio flux. In the case of 3C371, this would lead to a neutrino emission comparable to the IceCube limit. However, in this case more detailed SED modelling is needed in order to explain the complete emission from this object. Moreover, not all objects in our selection have sufficient data to perform this analysis.

---

[31]Since we consider the value of $\chi$ to be quite robust.

## 4.4   Diffuse flux from a generic obscured population

In this section, we calculate the diffuse neutrino and gamma-ray flux from a generic population of obscured sources. In particular, we investigate whether sources obscured by a gas with column density $N_H^{(1,2)}$ can be responsible for the astrophysical neutrino flux measured by IceCube, without violating the bound on the non-blazar contribution to the EGB.

For this calculation, we will not be limiting ourselves to sources similar to those in the selection of Section 4.3. Instead, we consider a generic population of obscured sources characterised only by their evolution with redshift, while their total energy budget will be fitted to reproduce the observed neutrino flux. From the arguments in Chapter 3, in particular Sections 3.2.3 and 3.3.5, a source class with a power similar to the UHECR sources[32] can reproduce the flux observed by IceCube. The only requirement on the source configuration is that at each time throughout the history of the universe, there is a population of obscured sources and each individual source is obscured for a time relevant compared to the cosmic history. In our calculations, we do not assume the presence of a jet.

In addition to the generic analysis, we also consider in more detail a particular class of objects called ultra-luminous infrared galaxies (ULIRG), which already appeared in the previous analysis. From their expected cosmological evolution, individual power and number density, we estimate whether they represent a possible neutrino source class under the obscured $pp$-interaction model.

### 4.4.1   Populations

In this section, we show the different redshift evolution scenarios that are investigated in the context of obscured neutrino production through $pp$-interactions. In particular, we consider the possibilities of redshift evolution following star formation, no evolution and the case of ULIRGs. There are still many alternative evolutions possible, as already seen in Section 3.8.3. However, in order to reduce the number of scenarios, we limit ourselves to these three.

**Star formation rate**

First, we consider a redshift evolution following the star formation rate. This is a natural choice, since it follows the evolution of galaxies throughout the cosmic history. This is particularly relevant for GRBs and supernovae, since these events occur more frequently during star formation, when many large stars with short lifetimes are formed. It is also relevant for AGN- and blazar-like scenarios, since during intense star formation there could also be efficient accretion around supermassive black holes in the centres of galaxies. Even if the true rate deviates from the star formation rate, it is still a good first approximation for the amount of activity in galaxies as a function of redshift.

---

[32] Although, as we will see, we will end up requiring a maximum proton energy of $E_p \sim 10^8$ GeV in order to fit the IceCube flux and its upper limits at the higher energies. Therefore, these sources would not supply the UHECRs, but possess a similar energy budget.

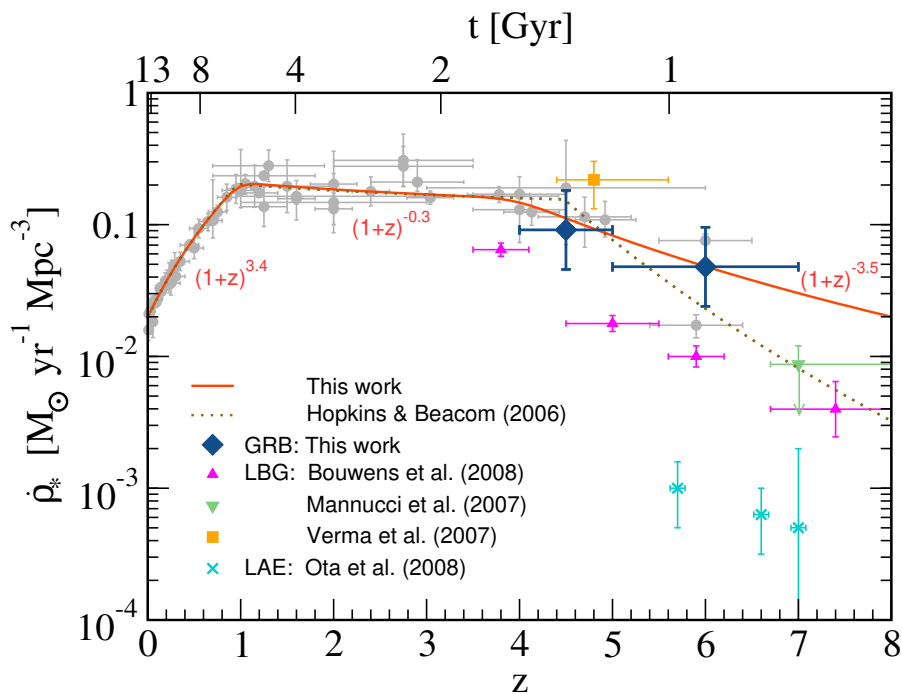

Figure 4.12: Compilation of measurements of the cosmic star formation history. Figure from [767].

The star formation rate is measured using different observables, such as the supernova rate, luminosity densities, limits on the diffuse neutrino background from supernovae and GRBs. The star formation rate is well-known, up to redshifts of $z \approx 1$. At higher redshifts, measurements deviate, although the knowledge is improving. In the following, we will use the star formation history derived in [666, 767], which is also shown in Figure 4.12. Its form is

$$\mathcal{H}_{\text{SFR}} \propto (1+z)^{n_i}, \tag{4.37}$$

with

$$n_i = \begin{cases} 3.4 & z < 1 \\ -0.3 & 1 < z < 4 \\ -3.5 & z > 4. \end{cases} \tag{4.38}$$

The normalisation is such that $\mathcal{H}_{\text{SFR}}(z = 0) = 1$ and the function is continuous.

**Flat evolution**

As an alternative to cosmic star formation, we consider also the simplest case of no evolution

$$\mathcal{H}_{\text{flat}} = 1. \tag{4.39}$$

**Ultra-luminous infrared galaxies**

In addition to the above generic scenarios, we will also consider in more detail the possibility that ultra-luminous infrared galaxies (ULIRG) could be responsible for the astrophysical neutrino flux detected by IceCube through our model. ULIRGs showed up already in the object selection in Section 4.3.2, where ARP 220 was a candidate object in the final selection. ARP 220 is the most well-known and best studied ULIRG and is an object with a very high IR luminosity formed by the merger of two galaxies, with a very high column density of at least $10^{25}$ cm$^{-2}$. Its nucleus is possibly powered by an AGN, which must however be obscured and Compton-thick. [768].

More generally, ULIRGs are defined as galaxies with extreme infrared luminosities $L_{\text{IR}} > 10^{12} \, L_{\odot}$ (see e.g. the review [769]). They are the mergers of gas rich galaxies, with the central regions harbouring huge amounts of gas and dust. The emission is caused by starburst activity (i.e. high star formation rate), and possibly also AGN activity, triggered by the merger [770]. It is believed that submillimetre galaxies are the high-redshift counterparts of local ULIRGs.

The abundance and importance of AGN activity compared to starburst activity in ULIRGs is still not completely clear. Surveys indicate that ULIRGs contain radio cores which are due to AGN activity [771]. In studies of local ULIRGs [772], it was found that all of them require starburst activity to explain their emission, while only half require an AGN. Moreover, in 90% of the cases, the starburst activity provides over half of the IR luminosity, with an average fractional luminosity of 82%. The AGN contribution does not increase with luminosity. Other studies have found that over half of the ULIRGs contain an AGN, with the fraction increasing with total IR luminosity [773]. In [774], it was found that only few ULIRGs are dominated by AGNs (5% at $z \sim 1$ and 12% at $z \sim 2$) although, at a given luminosity, the fraction of AGN activity is lower in high-$z$ ULIRGs. In a $5 - 8 \, \mu$m analysis [775] of local ULIRGs, signatures of AGN activity were found in $\sim 70\%$ of the sample. While most of the luminosity is due to the starburst activity, $\sim 23\%$ is due to the AGN, increasing with luminosity. More general, part of the emission of star-forming galaxies at high redshifts can be explained by AGN activity [776]. So, while AGNs are in general definitely not the dominant component of ULIRGs emission, their contribution is not negligible. Moreover, high obscuration of the central core may lead to underestimation of the AGN power [769].

The exact interplay between the central AGN and the starburst activity is still unclear. Several scenarios are still possible: one could evolve from the other, trigger the other or their coexistance might be coincidental. Typically, the AGN and ULIRG activity is unified in an evolutionary scenario. Early on in the merger, starburst activity is high when there is still plenty of gas. This merger might also relate to the growth of a supermassive black hole and AGN activity at the core [777]. This was confirmed by observations of stellar kinematics [778, 779]. Later, when the gas is concentrated in a compact region in the centre, a starburst-ULIRG phase occurs. During the late merger state, there is a high accretion rate at the centre, giving rise to an obscured QSO[33]/AGN-powered ULIRGs, due to the huge amount of gas and dust driven to the centre. Afterwards, the galaxy enters its most luminous phase with an optically-visible QSO which drives out the remaining material [769, 780]. In this sense, neutrino emission from ULIRGs would have an interesting interplay with galaxy formation and evolution.

Estimates of the space densities of ULIRGs have been made by several groups. In [781], they found at $z = 0.15$ that $n = 3 \times 10^{-7}$ Mpc$^{-3}$ for $L_{IR} = 1.6 \times 10^{12} \, L_\odot$ and $n = 9 \times 10^{-8}$ Mpc$^{-3}$ for $L_{IR} = 2.5 \times 10^{12} \, L_\odot$. This decreases with a factor 1.5 to lower redshift ($z = 0.04$), while it increase to $\Phi(L > 10^{11} \, L_\odot) = 1 - 3 \times 10^{-2}$ Mpc$^{-3}$ at $z \sim 1$–3. A full luminosity function was derived in [782], although their analysis is only sensitive to $L_{IR} > 10^{12.3} \, L_\odot$. In the same redshift range $z \sim 1$–3, they find $n > 6 \times 10^{-6}$ Mpc$^{-3}$. This leads to a redshift evolution[34]

$$\mathcal{H}_{ULIRG} \propto \begin{cases} (1+z)^4 & z <= 1 \\ \text{const.} & 1 < z < 4. \end{cases} \tag{4.40}$$

We will consider the central AGN, obscured by gas and dust driven to the centre by the merger, as a potential target for the obscured $pp$-neutrino production mechanism. ULIRGs have been considered before as the source of astrophysical neutrinos [682, 784], but in the context of cosmic-ray reservoir models, through confined cosmic rays interacting with gas in the galaxy. Instead, we consider only neutrino production in a compact region near the core. Typically, starbursts only have a surface gas density of about $1 \, \text{g cm}^{-2}$, or $N_H \sim 10^{23} \, \text{cm}^{-2}$, much lower than the densities required by our mechanism. However, this measurement is only valid for the overall gas density[35] and does not exclude the existence of local, compact objects or regions with higher densities. Finally, in our accelerator model, the cut-off of the proton and neutrino spectra can be at higher energy than in reservoir models, since there is no confinement criterion, although we will keep the cut-off at $10^8$ GeV in order to not overshoot the IceCube flux at high energies.

In order to investigate the possibility that ULIRGs could provide the neutrino flux without violating the non-blazar EGB bound, we therefore adopt their redshift evolution $\propto (1+z)^4$ up to $z = 1$, followed by a flattening (up to $z = 4$). To estimate whether ULIRGs can provide the required luminosity, we use the number densities given above along with their minimum luminosity $L_{IR} = 10^{12} \, L_\odot$ to estimate their local energy generation rate in the IR

$$\mathcal{Q}_{IR} = n(z=0) \cdot L_{IR}, \tag{4.41}$$

---

[33]Quasi-stellar object or quasar, see the unified model in Figure 3.8a.

[34]Although other studies found more extreme evolution, up to $\propto (1+z)^7$ for $z = 0 - 1.5$ [783].

[35]It is derived using the Kennicutt-Schmidt law [785–787] which relates the star-formation rate with the gas density.

where we adopt the conservative value $n \approx 5 \times 10^{-7}$ Mpc$^{-3}$ (integrating over luminosity). Afterwards, this IR luminosity is converted to radio luminosity using the radio-IR relation [788] for ULIRGs. Using the thermal infrared luminosity $L_{\text{TIR}} \equiv L(8 - 1000\,\mu\text{m})$, the TIR/radio flux ratio is defined as

$$q_{\text{TIR}} = \log\left(\frac{L_{\text{TIR}}}{3.75 \times 10^{12}\,\text{W}}\right) - \log\left(\frac{L_{1.4\,\text{GHz}}}{\text{W}\,\text{Hz}^{-1}}\right). \tag{4.42}$$

On average, this ratio has the value $\langle q_{\text{TIR}} \rangle = 2.6$, with no evolution in redshift. After obtaining $L_{1.4\,\text{GHz}}$ from this relation, we estimate the total radio luminosity as $L_R = 1.4$ GHz $\times L_{1.4\,\text{GHz}}$. Using the relations in Section 4.3.3, the radio luminosity can be converted into the proton luminosity $\mathcal{Q}_p = \frac{\chi \cdot \mathcal{Q}_R}{f_e}$. We find

$$\mathcal{Q}_p^{\text{ULIRG}} \approx 10^{43.4}\,\text{erg}\,\text{Mpc}^{-3}\,\text{yr}^{-1}. \tag{4.43}$$

This luminosity is slightly below the value estimated in the Waxman-Bahcall calculation in Section 3.3.5, such that ULIRGs do not initially seem capable of explaining the full astrophysical neutrino flux. Moreover, in this calculation it was assumed that all of the ULIRG luminosity is related to AGN activity, while in reality the AGN contribution to the luminosity is at least one order of magnitude lower (integrated over all ULIRGs). On the other hand, we assumed that all ULIRGs have exactly $L_{\text{IR}} = 10^{12}\,L_\odot$, while many have higher luminosity. In addition, we used the standard value for the parameter $f_e$, while it can easily be lower by at least an order of magnitude.

### 4.4.2 Diffuse neutrino flux

In order to obtain the total diffuse flux of neutrinos from all sources in the observable universe, we need to perform an integral over cosmic history or, equivalently, over cosmological distance. The derivations below follow those in the reviews [789, 790].

Consider first the flux from a single source from a cosmological distance. The relation between the received bolometric flux $S$ and the source bolometric luminosity $L$ is given by

$$S = \frac{L}{4\pi d_L^2}, \tag{4.44}$$

where the luminosity distance $d_L$ is defined such that this formula incorporates the distribution of the flux over a surface area (in coordinate distance) centred at the source as well as two times a redshift factor $\frac{1}{1+z}$ from the reduced rate of arrival from the expansion of the universe and the energy loss due to redshift. Therefore, $d_L = a(t_0)\chi(1 + z)$, with $a(t_0)$ the current scale factor and $\chi$ the coordinate distance (or, for $a(t_0) = 1$, also the comoving distance) to the source.

Compared to this, the differential flux $S_E$ and luminosity $L_E$ require an extra correction, because the received flux was emitted at a different energy. In astronomy, this is known as the k-correction. The exact correction depends on the spectrum and is unnecessary for an $E^{-2}$-spectrum, where $EL_E = $ const. In general, the correction is

$$S_E = (1+z)\frac{L_{(1+z)E}}{L_E}\frac{L_E}{4\pi d_L^2}, \tag{4.45}$$

where the factor $(1+z)$ accounts for the redshifting of the "bandwidth" and the luminosity ratio equalises the difference in luminosity between the observed and emitted bands. By rewriting, this equation obtains the simpler form

$$ES_E = \frac{\varepsilon L_\varepsilon}{4\pi d_L^2}\bigg|_{\varepsilon = (1+z)E}. \tag{4.46}$$

The comoving distance to an object which emitted its light at a time $t_e$ and is observed at a time $t$ can be found by integrating over the path of the photon, or equivalently its flight time through $ds^2 = 0$, and dividing out the scale factor of the expanding universe (this is correct for $a(t_0) = 1$)

$$d_C = c\int_{t_e}^{t}\frac{dt'}{a(t')}, \tag{4.47}$$

or when rewriting using the Hubble constant $H = \dot{a}/a$

$$d_C = \frac{c}{H_0}\int_0^z\frac{dz}{E(z)}. \tag{4.48}$$

with $H_0$ the current Hubble constant and the evolution of the Hubble constant given by $E(z) = \sqrt{\Omega_m(1+z)^3 + \Omega_\Lambda}$. For a flat cosmology, we then have $d_L = (1+z)d_C$

Another distance measure is the angular diameter distance $d_A$, which gives the (physical) transversal size of a patch on the sky with a certain angular size. Its definition can be read off immediately from the metric and it is related to the comoving distance as[36] $d_A = \frac{d_C}{1+z}$, giving $d_L = (1+z)^2 d_A$ from the previously derived relation between $d_L$ and $d_C$.

The comoving volume $V_C = V(1+z)^3$ can then be found by integrating the volume element

$$dV_C = \frac{c}{H_0}\frac{(1+z)^2 d_A^2}{E(z)}\,d\Omega\,dz \tag{4.49}$$

$$= \frac{c}{H_0}\frac{d_L^2}{(1+z)^2 E(z)}\,d\Omega\,dz. \tag{4.50}$$

Which is built out of the differential of the comoving distance and a factor $(1+z)^2$ from converting the proper area of a patch with solid angle $d\Omega$ into comoving area.

---

[36]Note that for this formula I am already assuming a flat universe.

The total diffuse flux is obtained by combining Eq. (4.46) and Eq. (4.49), integrating the flux from each comoving element

$$ES_E(E) = \int dV_C \frac{\varepsilon L_\varepsilon(\varepsilon, z)|_{\varepsilon = (1+z)E}}{4\pi d_L^2} \tag{4.51}$$

$$= \int d\Omega \, dz \, \frac{c}{H_0} \frac{d_L^2}{(1+z)^2 E(z)} \frac{\varepsilon L_\varepsilon(\varepsilon, z)|_{\varepsilon = (1+z)E}}{4\pi d_L^2} \tag{4.52}$$

$$= \frac{c}{4\pi} \frac{1}{H_0} \int \frac{dz}{(1+z)^2 E(z)} \varepsilon L_\varepsilon(\varepsilon, z)|_{\varepsilon = (1+z)E} \int d\Omega, \tag{4.53}$$

where $L_\varepsilon(\varepsilon, z)$ is the differential luminosity per comoving volume. Assuming that the neutrino luminosity of the population evolves with redshift as $\mathcal{H}(z)$, we can write $L_\varepsilon(\varepsilon, z) = \mathcal{H}(z) L_\varepsilon(\varepsilon, 0)$, with $L_\varepsilon(\varepsilon, 0)$ the local differential luminosity per comoving volume, which contains the spectrum integrated over sources of potentially differing luminosities[37,38].

Finally, switching notation back to our usual for the neutrino flux (as in Section 3.3.4), such that $S_E = E_\nu \Phi_\nu^{\text{diffuse}}$ and $L_E = E_\nu \mathcal{Q}_{E_\nu}$ and considering the differential flux per solid angle we find

$$E_\nu^2 \Phi_\nu^{\text{diffuse}}(E_\nu) = \frac{c}{4\pi} \frac{1}{H_0} \int \frac{dz}{(1+z)^2 E(z)} \mathcal{H}(z) \, \varepsilon_\nu \mathcal{Q}_{\varepsilon_\nu}(\varepsilon_\nu)|_{\varepsilon_\nu = (1+z)E_\nu}. \tag{4.54}$$

The factor $\mathcal{Q}_{E_\nu}(E_\nu)$, defined as $\mathcal{Q}_{E_\nu}(E_\nu) = E_\nu \Phi_\nu(E_\nu)$ such that $\int dE_\nu \, E_\nu \Phi_\nu(E_\nu) = \mathcal{Q}_\nu$ with $\mathcal{Q}_\nu$ the total injected neutrino luminosity per comoving volume, is obtained from our simulation described in Section 4.2.2. The normalisation of $\mathcal{Q}_{E_\nu}$ is free to be determined either by fitting the final diffuse flux $\Phi_\nu^{\text{diffuse}}$ or by fixing the total injected proton luminosity.

The above formula can be further simplified, by pulling out the $E$-dependence from the integral. In general, we can write

$$E_\nu^2 \Phi_\nu^{\text{diffuse}}(E) = \frac{c}{4\pi} \frac{1}{H_0} E_\nu \mathcal{Q}_{E_\nu}(E_\nu) \int \frac{dz}{(1+z)^2 E(z)} \mathcal{H}(z) \frac{\varepsilon_\nu \mathcal{Q}_{\varepsilon_\nu}(\varepsilon_\nu)|_{\varepsilon_\nu = (1+z)E_\nu}}{E_\nu \mathcal{Q}_{E_\nu}(E_\nu)}, \tag{4.55}$$

and define

$$\xi_z(E_\nu) = \int \frac{dz}{(1+z)^2 E(z)} \mathcal{H}(z) \frac{\varepsilon_\nu \mathcal{Q}_{\varepsilon_\nu}(\varepsilon_\nu)|_{\varepsilon_\nu = (1+z)E_\nu}}{E_\nu \mathcal{Q}_{E_\nu}(E_\nu)}. \tag{4.56}$$

The $z$-integral becomes energy-independent for the case of a power law spectrum[39]

---

[37]In other words, the redshift evolution can contain both a change in number density and a change in luminosity, although we will typically only consider the former

[38]This differs slightly from the definition in [487], where $\mathcal{H}(z)$ includes the full source density and not just its evolution with redshift, while the luminosity per source is considered constant as $\mathcal{Q}(E)$.

[39]I.e. this is no longer true once we include a cut-off (e.g. an exponential) in the spectrum. This agrees with the intuition that the contribution from the end of the spectrum changes with redshift. However, as long as we calculate the flux at an energy sufficiently far from the cut-off, the calculation for a power law is valid. For our own result, we use the full integral, which does not suffer from this subtlety.

$\mathcal{Q}_{E_\nu} \propto E_\nu \cdot E_\nu^{-\gamma}$ and the previous assumption that $L_\varepsilon(\varepsilon, z)$ was factorisable. We then get

$$\xi_z = \int \frac{\mathrm{d}z}{E(z)} \mathcal{H}(z)(1+z)^{-\gamma}. \tag{4.57}$$

This integral can be solved numerically. The simplified formula for the final flux then becomes

$$E_\nu^2 \Phi_\nu^{\text{diffuse}}(E_\nu) = \frac{c}{4\pi} \xi_z \frac{1}{H_0} E_\nu \mathcal{Q}_{E_\nu}(E_\nu). \tag{4.58}$$

For the case of $\gamma = 2$, we find $\xi_z = 2.4$ for a redshift evolution following star formation $\mathcal{H}(z) = \mathcal{H}_{\text{SFR}}(z)$ (Eqs. (4.37) and (4.38)). This then fully explains the formula we used before in Section 3.3.5. For no evolution we have $\xi_z = 0.53$ and for the case of ULIRGs, we find $\xi_z = 3.6$.

### 4.4.3 Diffuse gamma-ray flux

In order to obtain the diffuse gamma-ray background from the neutrino sources considered here, an extra effect needs to be taken into account. During propagation, gamma rays can interact with the extragalactic background light (EBL) and the cosmic microwave background (CMB), producing an $e^+e^-$-pair. The gamma-ray flux from redshift $z$ is thus cut off above the energy where the optical depth in the EBL becomes equal to 1. In turn, these electrons can up-scatter EBL and CMB photons back to gamma-ray energies through inverse-Compton scattering or emit synchrotron radiation, initiating an electromagnetic cascade. In this way photons of energies above $\sim 100$ GeV are reprocessed and accumulate at energies at and below 100 GeV, significantly increasing the flux at these energies.

#### Gamma rays in the EBL and CMB

The EBL is the extragalactic background light, ranging from the far-infrared to ultraviolet energies. It is the integrated emission from stars and dust throughout the history of the universe. At $z = 0$, the observed EBL features two bumps, the optical/near-infrared is associated to emission from stars, while the far-infrared bump is due to reprocessed emission from dust (see e.g. [791]). Measuring the EBL can be done in different ways, either through direct measurements (where foregrounds are a problem), galaxy counts (giving lower bounds) or indirect, e.g. through measuring its effect on the gamma-ray spectrum of distant blazars (where the energy range and behaviour of the emitted spectrum is not known). Moreover, in order to model the propagation of gamma rays, the evolution of the EBL with redshift needs to be known. This requires input from galaxy and star evolution, which is non-trivial. Due to all these difficulties, different EBL models exist, such as [791–798], which are shown in Figure 4.13a. The CMB is the background radiation of photons after decoupling from matter in the early universe when it was merely 380 000 years old ($z \sim 1100$) and had a temperature of $\sim 3000$ K or $\sim 0.26$ eV. At present, the CMB has a temperature of 2.725 K or 0.2348 meV.

Gamma rays interact with EBL or CMB photons depending on their energy. Pair production turns on when the centre of mass energy is sufficient to create two electrons, i.e. $\epsilon_\gamma \approx 0.25 \left(\frac{\text{TeV}}{E_\gamma}\right)$ eV. From the EBL, the optical depth of gamma rays due to pair production can be calculated, the results of which are shown in Figure 4.13b. For an evolution following star formation or close to it, the dominant contribution of injected gamma rays comes from $z \sim 1$ (see the integrandum of Eq. (4.54) and the evolution in Eq. (4.37)). Inspecting the optical depth in Figure 4.13b, we see that for the "average" photon, the cut-off will be at $E_\gamma \sim 100$ GeV.

Often (e.g. for blazar spectra), only the attenuation, not the full cascade, is taken into account, by multiplying the predicted flux with a factor $\exp(-\tau(E_\gamma, z))$, since this is a guaranteed effect, while the cascade is less established[40].

**Analytical description of the cascade**

We follow the procedure outlined in [522] in order to calculate the gamma ray spectrum after cascading in the EBL analytically. As described in [521, 552], after the EM cascade has sufficiently developed, it attains a universal form given by (using $G_{E_\gamma} = E_\gamma \Phi_\gamma$ in order to denote the cascade)

$$G_{E_\gamma} \propto \begin{cases} \left(\frac{E_\gamma}{E_\gamma^{\text{br}}}\right)^{-1/2} & (E_\gamma < E_\gamma^{\text{br}}) \\ \left(\frac{E_\gamma}{E_\gamma^{\text{br}}}\right)^{1-\beta} & (E_\gamma^{\text{br}} < E_\gamma < E_\gamma^{\text{cut}}), \end{cases} \tag{4.59}$$

normalised to $\int dE_\gamma \, G_{E_\gamma} = 1$. Typically, $\beta \approx 2$. The cut-off energy $E_\gamma^{\text{cut}}$ is the energy where suppression due to pair production occurs. It can be obtained from the requirement[41] $\tau(E_\gamma^{\text{cut}}, z) = 1$, for which we use the tables provided in [791]. The break energy is given by[42] $E_\gamma^{\text{br}} \approx \frac{4}{3} \left(\frac{E_\gamma^{\text{cut}}}{2m_ec^2}\right)^2 \varepsilon_{CMB} \approx 0.034$ GeV $\left(\frac{E_\gamma^{\text{cut}}}{0.1 \text{ TeV}}\right)^2 \left(\frac{1+z}{2}\right)^2$, where $\varepsilon_{CMB}$ is the typical CMB energy. Above the cut-off energy $E_\gamma^{\text{cut}}$, but below $\min\left[\frac{E_\gamma^{\max}}{2}, \frac{4}{3}\left(\frac{E_\gamma^{\prime\max}}{2m_ec^2}\right)^2 \varepsilon_{CMB}\right]$, the cascade is not sufficiently developed and its exact form depends on the details of the injection. With far away sources, for gamma rays scattering in the Thompson regime one can assume a simple exponential cut-off $e^{-\tau_{\gamma\gamma}}$ for $d > \lambda_{\gamma\gamma}$ with $\tau_{\gamma\gamma}(E_\gamma, z) = d(z)/\lambda_{\gamma\gamma}(E_\gamma)$. On the other hand, in the Klein-Nishina regime pairs are continuously supplied, giving a shape $\frac{1-e^{-\tau_{\gamma\gamma}}}{\tau_{\gamma\gamma}}$ as long as the pair injection length ($\lambda_{BH}$) is longer than $d$. For our purposes, we use the exponential form, although the exact details do not significantly influence our conclusion. For more details with a full numerical calculation of the cascade, see [799–801].

---

[40]E.g. the importance of magnetic fields and their effect during propagation is very uncertain

[41]As opposed to the formula in [552], where an extra factor (1+z) is included in the energy.

[42]This break energy is due to the lowest energy at which electrons are created that can up-scatter photons from the CMB through inverse-Compton scattering. From the energy loss rate of an electron through IC scattering and the number of photons scattered per unit time, one finds the average energy of scattered photons as $\epsilon_\gamma \approx \frac{4}{3}\gamma_e^2 \epsilon_{CMB}$, with $\gamma_e$ the Lorentz factor of the electron. This roughly corresponds to the handwaving argument that the photons gains two Lorentz factors of energy: one from transforming to the electron frame (where scattering is easy) and one from transforming back.

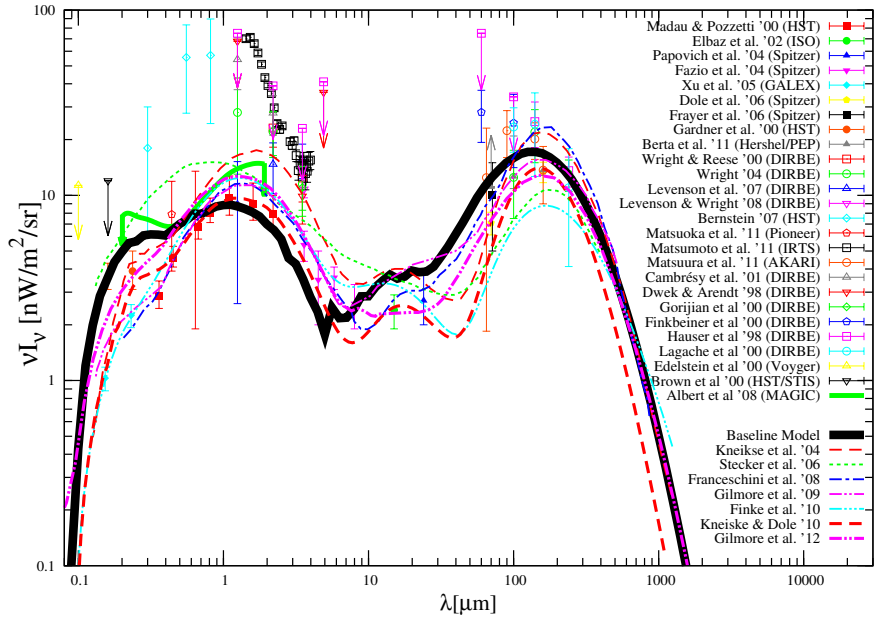

(a) EBL

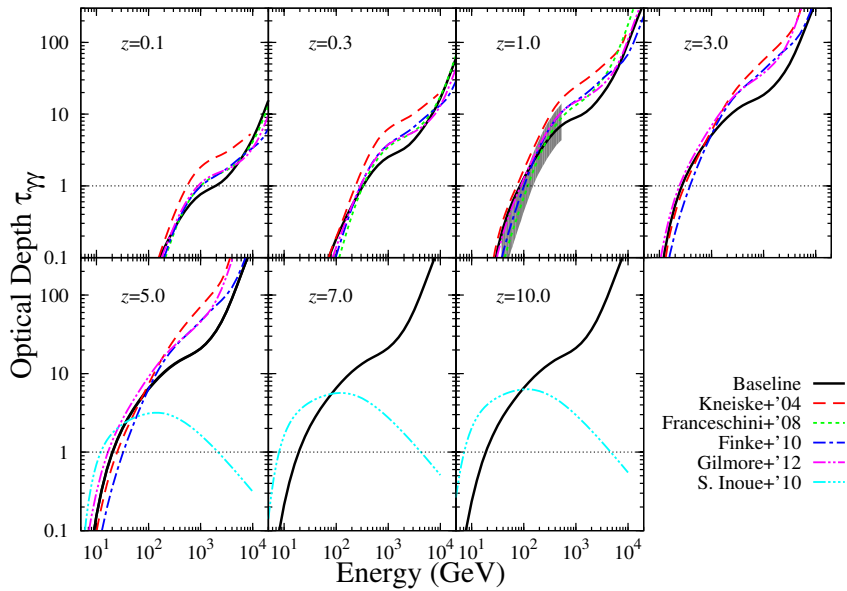

(b) $\tau_{\gamma\gamma}$

Figure 4.13: Extragalactic background light measurements. (a) Various models of the extragalactic background light [791–798] (references for the different measurements can be found in [791]). (b) The optical depth $\tau_{\gamma\gamma}(E_\gamma, z)$ of gamma rays in the EBL as a function of observed gamma ray energy, for sources at various redshifts. Figures from [791].

Finally, the full diffuse spectrum is obtained by integrating the cascaded spectrum from each redshift and weighting with the injected luminosity from each distance, giving

$$E_\gamma^2 \Phi_\gamma(E_\gamma) = \frac{c}{4\pi} \frac{1}{H_0} \int \frac{\mathrm{d}z}{(1+z)^2 E(z)} \mathcal{H}(z) E_\gamma G_{E_\gamma} \mathcal{Q}_\gamma. \tag{4.60}$$

The factor $\mathcal{Q}_\gamma$ is the total integrated luminosity in gamma rays per comoving volume injected by the sources (after attenuation by the gas column) and is obtained from our simulation.

Since the dominant injection is from sources at $z \sim 1$, as previously explained, we can estimate the diffuse gamma-ray flux as [522]

$$E_\gamma^2 \Phi_\gamma(E_\gamma) \approx \frac{c}{4\pi} \frac{1}{H_0} \xi_z \, E_\gamma G_{E_\gamma} \big|_{z=1} \mathcal{Q}_\gamma, \tag{4.61}$$

with $E_\gamma G_{E_\gamma} \big|_{z=1} \approx 0.1$ between the break and the cut-off. Since the cut-off energy at $z=1$ is about 100 GeV, there will be an accumulation[43] of gamma rays at this energy, which is then also typically the point most strongly constrained by the Fermi measurements.

In order to get the full spectrum, however, this simple approximation is not sufficient and we need to do the full integration numerically. As a consequence of this complicated integration over redshift, the gamma-flux is sensitive to the details of the redshift evolution, in contrast to the neutrino flux where there is a degeneracy between the evolution (through $\xi_z$) and the normalisation.

The results of this numerical calculation were verified with a numerical simulation using `CRPropa 3` [472] and its DINT module, which follows the description in [802]. The behaviour of the spectrum agrees with the analytical description, while the overall normalisation differs by several orders of magnitude. Since this difference is present even in the absence of interactions, the correct result could be obtained by rescaling the result without interactions to its correct value. The same rescaling factor then also brings the cascaded result to its analytical value.[44]

### 4.4.4 Results

The diffuse flux for the different redshift evolutions above and the assumed column densities $N_H^{(1)} = 5 \times 10^{25}$ cm$^{-2}$ and $N_H^{(2)} = 10^{26}$ cm$^{-2}$ are calculated following the description in Section 4.2.2. The injected proton energy budget is then normalised by fitting the neutrino spectrum (taking into account a flavour factor $1/3$ after oscillation) to the single-flavour neutrino flux observed by IceCube [534]. The results of this calculation are shown in Figure 4.14, where the gamma ray flux can be compared to the bound on the non-blazar contribution to the EGB (both the best fit (14%) and its weakest upper limit (28%)). These results can be compared with Figure 4.15, which shows the result for the same cases in case there is no attenuation of the gamma rays. This comparison seems a bit artificial (although it serves only to show the effect of obscuration in our model), since these large column densities automatically imply that attenuation is present. Physically, it might be more relevant to compare with the case $N_H = 10^{24}$ cm$^{-2}$ (shown in Figure H.10 in Appendix H.3) if one is interested in the effect of the higher density. This change lowers both the attenuation of gamma rays and the fraction of protons that interact[45], the latter of which needs to be compensated for by increasing the injected proton luminosity to fit the IceCube flux. In the end, only the normalisation of the neutrino luminosity and its correlation with the gamma ray luminosity are then important. On the other hand, in cosmic ray reservoir models, the protons effectively cross a high density target, while the gamma rays escape immediately. This then corresponds to the high density case shown here with gamma-ray attenuation turned off.

---

[43]This accumulation does indeed rise above the hypothetical flux in the case of no attenuation, since the non-attenuated spectrum can be substituted in Eq. (4.61) by replacing the factor $E_\gamma G_{E_\gamma}$ by $1/\ln\left(\frac{E_\gamma^{\max}}{E_\gamma^{\min}}\right)$. With an injected energy range between $E_p^{\max}/20$ and $\ll 1$ GeV from pion decays, this is smaller than 0.1.

[44]The issue was reported to the developers, but the cause had not been found. In the future, the DINT module will be completely replaced, which might solve the problem.

We find that in the case of a redshift evolution following star formation $\mathcal{H}_{\mathrm{SFR}}$ or the ULIRG evolution $\mathcal{H}_{\mathrm{ULIRG}}$, the bounds from the non-blazar contribution to the EGB can be satisfied, as opposed to the case in which there is no attenuation, although in the case $N_H^{(1)} = 5 \times 10^{25}$ cm$^{-2}$ the improvement is small. In the case of a flat evolution $\mathcal{H}_{\mathrm{Flat}}$, in which the contribution from low-$z$ sources is more important, this bound can not be satisfied for $N_H^{(1)} = 5 \times 10^{25}$ cm$^{-2}$ and only marginally for $N_H^{(2)} = 10^{26}$ cm$^{-2}$.

The fitted proton energy budgets for all the cases are shown in Table 4.5. For the case of ULIRGs, these can be compared with the optimistic estimate from Section 4.4.1. We find that even this estimate, which did not take into account that most of the ULIRG luminosity is due to starburst activity instead of a central engine, is about one order of magnitude too low. In order to explain the diffuse neutrino flux with objects like ULIRGs, other object classes are therefore necessary. The slightly less luminous LIRG[46] are candidates for this. On the other hand, the estimate also depended on the electron-proton luminosity ratio $f_e$. Lowering this ratio can already increase the contribution from ULIRGs significantly.

Finally, we show in Figure 4.16 results of the same calculation for a proton index $\gamma = 2.1$. In this case, only $N_H^{(2)} = 10^{26}$ cm$^{-2}$ can marginally satisfy the constraints from the non-blazar contribution to the EGB for $\mathcal{H}_{\mathrm{SFR}}$ and $\mathcal{H}_{\mathrm{ULIRG}}$.

## 4.5 Conclusion

In this chapter, we investigated the possibility that objects obscured by matter emit high energy neutrinos created in $pp$-interactions. In the case of gas clouds with sufficient integrated column density, neutrino production can be efficient, whilst the produced gamma rays can at the same time be attenuated through pair production by the same gas cloud. Here, we took as benchmark values for the column density $N_H^{(1)} = 5 \times 10^{25}\,\mathrm{cm^{-2}}$ and $N_H^{(2)} = 10^{26}\,\mathrm{cm^{-2}}$.

---

[45]It also is more sensitive to the energy dependence of the proton-proton cross section, as seen in Figure H.10.

[46]Galaxies with luminosities $L > 10^{11}\,L_\odot$, which are more numerous than e.g. starburst galaxies.

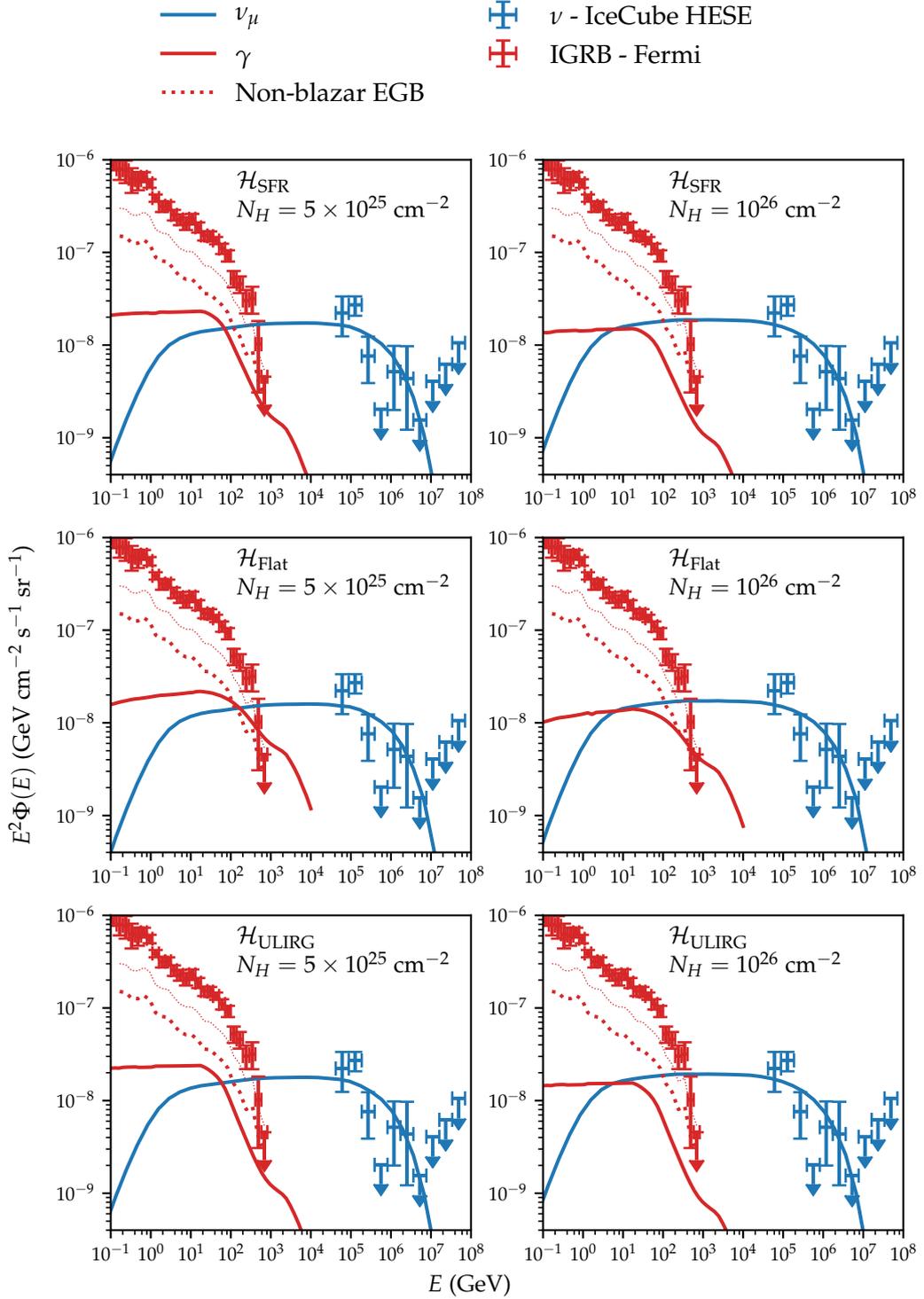

Figure 4.14: Results for the diffuse neutrino and gamma-ray flux, for different evolutions and column densities for the obscured $pp$-neutrino scenario, fitted to the IceCube single-flavour neutrino flux (HESE) [534]. The other parameters of the calculation are the same as in Table 4.1. The non-blazar contribution to the EGB shows both the best fit value (14% of the EGB measured by Fermi [557]) and the weakest upper limit (28%).

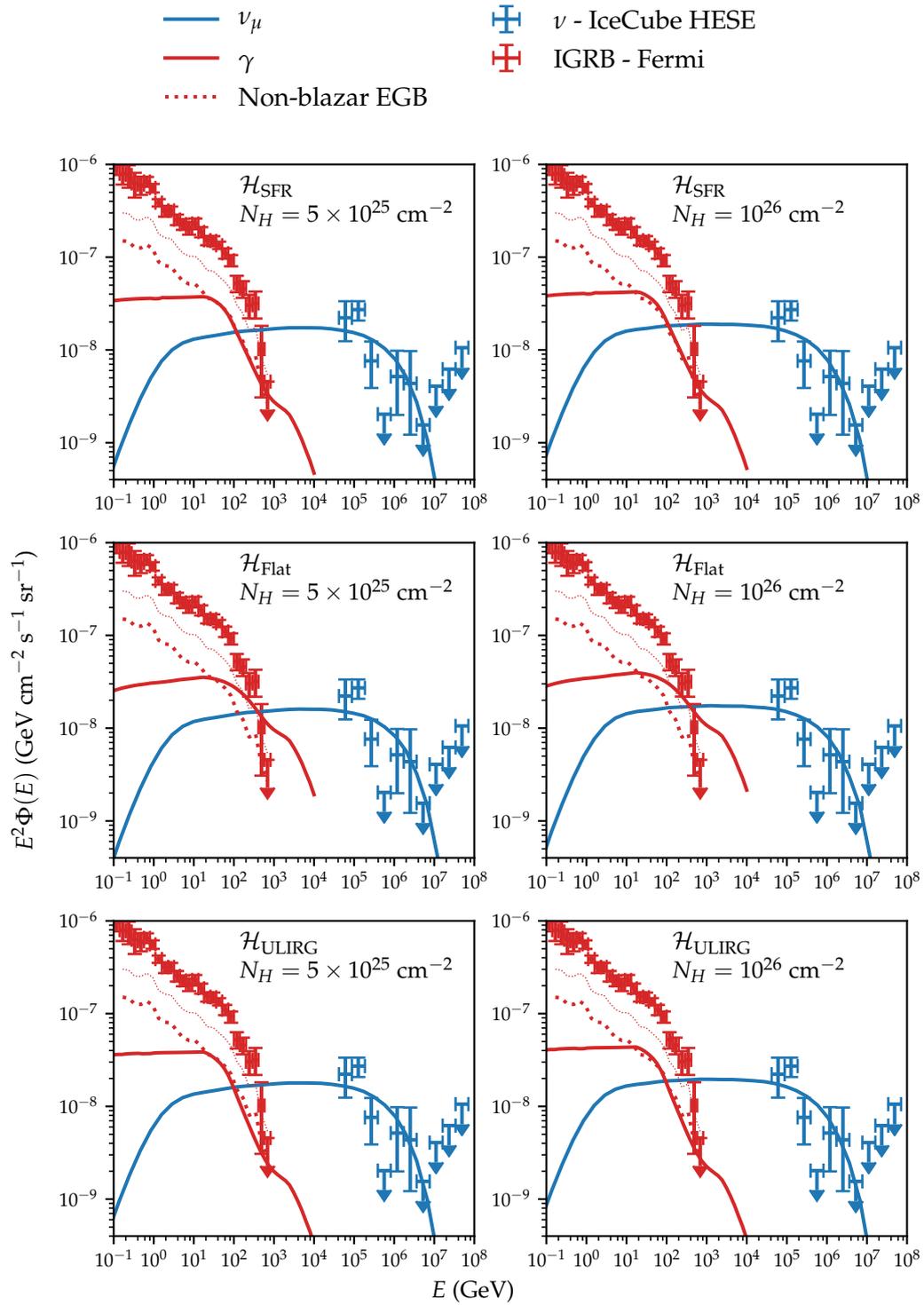

Figure 4.15: Same as Figure 4.14, now without attenuation.

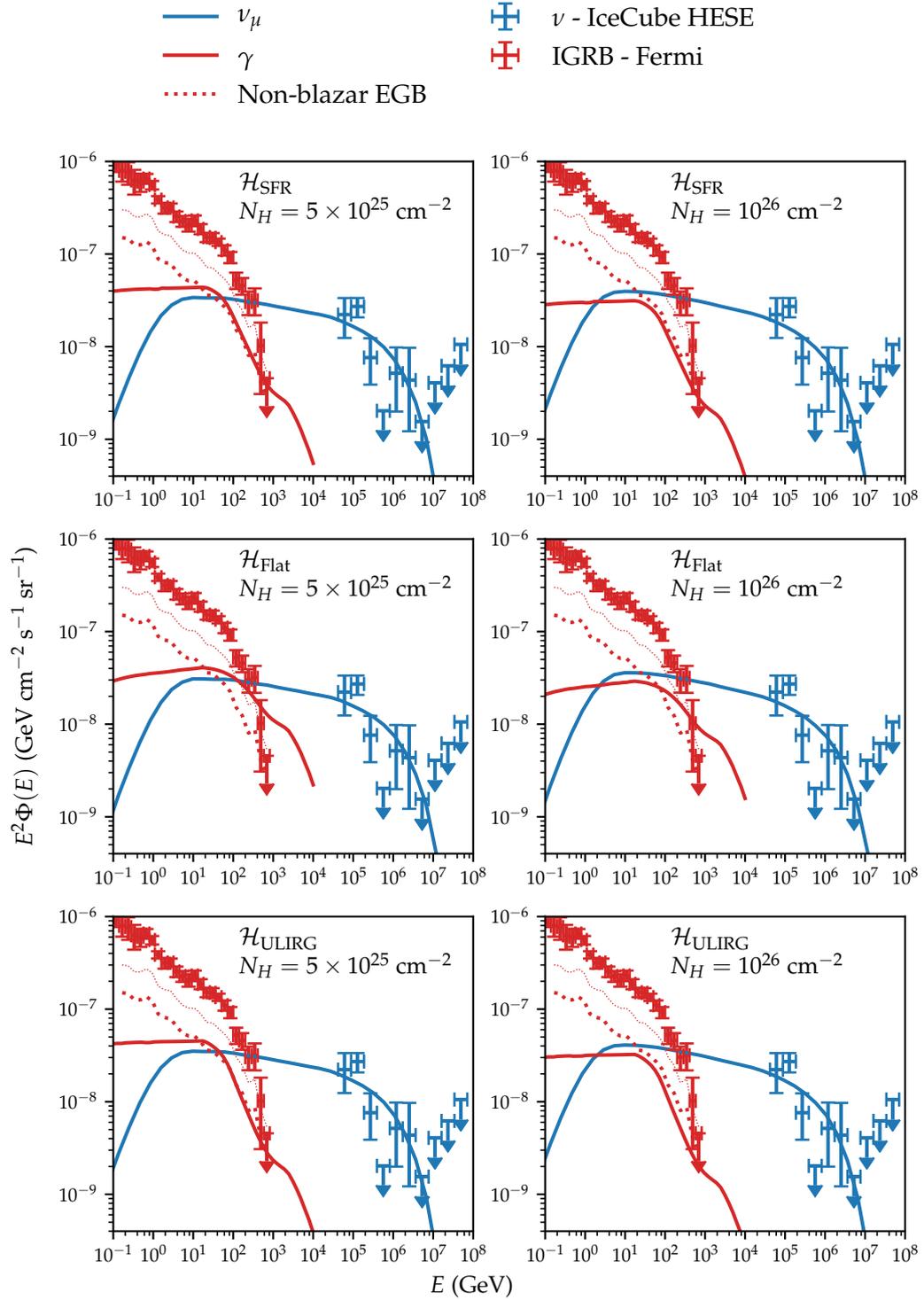

Figure 4.16: Same as Figure 4.14, now with proton spectral index $\gamma = 2.1$.

Table 4.5: Required proton luminosities $\mathcal{Q}_p$ from the fit to the observed IceCube single-flavour neutrino flux, in units of $10^{45}\,\mathrm{erg\,Mpc^{-3}\,yr^{-1}}$.

| Evolution | $\mathcal{H}_{\mathrm{SFR}}$ | $\mathcal{H}_{\mathrm{Flat}}$ | $\mathcal{H}_{\mathrm{ULIRG}}$ |
|---|---|---|---|
| **Column density** | | | |
| $N_H^{(1)} = 5 \times 10^{25}\,\mathrm{cm^{-2}}$ | 2.30 | 9.71 | 1.60 |
| $N_H^{(2)} = 10^{26}\,\mathrm{cm^{-2}}$ | 2.17 | 9.18 | 1.51 |

First, we calculated the neutrino spectra emitted from a set of objects whose electromagnetic spectrum can be explained with obscuration by matter. The selection, performed in [687], searches for strong radio-emitting galaxies with a lower-than-expected X-ray flux. The resulting objects are mainly blazars, along with one radio galaxy and one ULIRG. The candidate class of obscured blazars, being a subset of blazars, can not be responsible for the bulk of the IceCube flux, since their number density is too low. Still, they are interesting targets to test the viability of our model. Moreover, neutrino emission from such objects would provide a simple measurement of the accelerated proton content of blazars. We found that the predicted neutrino emission from these objects is below the limit set by IceCube, leaving the model unconstrained. For several of these objects, the scenario can be tested in the future upgrade of IceCube, while for many of the predicted neutrino flux is too low to be detected in the near future. On the other hand, current limits already allow to constrain the amount of accelerated protons in blazars, with values $f_e > 0.001$–$0.02$.

Second, we investigated the diffuse neutrino and gamma-ray flux from an unspecified population of neutrino sources operating under our model. In particular, we tested whether obscured $pp$-neutrino sources can be the source of the IceCube flux without violating the bounds on the non-blazar contribution to the EGB. In the case of a redshift evolution following star formation or for ULIRGs, this is indeed possible for both $N_H^{(1)} = 5 \times 10^{25}\,\mathrm{cm^{-2}}$ and $N_H^{(2)} = 10^{26}\,\mathrm{cm^{-2}}$, although in the former case the gamma-ray flux is close to the EGB bound. In the case of no evolution with redshift, the EGB bound is violated. This conclusion is valid for a neutrino spectrum $\propto E_\nu^{-2}$, which has trouble explaining the low energy events recorded by IceCube (although this can be solved with a second distinct population). In the case of steeper spectra, the constraints are more tight. Even for the thickest gas clouds considered here, the gamma-ray flux is just barely below the EGB limit.

We also discussed in more detail whether ULIRGs could be responsible for the Ice-Cube flux. Since ULIRGs occur when two galaxies merge, neutrino emission from such objects could have an interesting interplay with galaxy formation and evolution. While the neutrino and gamma-ray flux are compatible with observation[47], a simple, optimistic, estimate of their luminosity budget undershoots the required value. However, this conclusion depends on the electron-proton luminosity ratio $f_e$, for which we assumed a conservative value $1/10$. Taking lower values boosts the amount of accelerated protons, increasing the provided luminosity. Another possibility is that also the slightly less luminous LIRG contribute through the same scenario. In this sense, a separate analysis of the highly obscured ULIRG NGC 4418 would be interesting.

Finally, the same scenario could also be applicable to other objects. One intriguing, though speculative, possibility is that AGNs in the early universe can produce neutrinos through our model when they first turn on. At this time, a lot of gas and dust is still surrounding the centres of galaxies, potentially providing an ideal target. While these early AGNs might not explain the full IceCube neutrino flux (even more so because then the connection with the observed total gamma-ray energy budget is less obvious), they could make up part of it. This high-$z$ flux would an be unresolvable component of the total flux. Moreover, gamma-rays from such a high redshift would be cascaded down to lower energy than from $z = 1$-sources, such that the Fermi bounds from such a population are much less stringent. Moreover, like ULIRGs, such a scenario would tie neutrino production to galaxy formation and evolution.

---

[47]The gamma-ray flux might even be more attenuated by interactions at high energy with the IR field present in these galaxies, see e.g. [518].

# Neutrinos from binary black hole mergers

On September 14, 2015, LIGO observed gravitational waves emitted by a binary black hole (BBH) merger. This observation was the first direct detection of gravitational waves, kicking off a new era of gravitational wave astronomy and providing a new window in the context of multimessenger astronomy. After the detection of this event —referred to as GW150914— several follow-up searches for coincident emission in the electromagnetic spectrum and in neutrinos were performed, but no such signal was detected. Here, we investigate the implication of the non-detection of counterpart neutrinos, using generic arguments. We find that currently, searches for neutrinos from GW150914 can not yet rule out a significant contribution of BBH mergers to the astrophysical neutrino flux. In addition, we show how to interpret our result in specific models and provide estimates of the limit on the amount of matter present around BBH mergers. This work was published in [803]. Moreover, in the final section, we provide an update of this work, given the updated results by LIGO/Virgo. We find that, in the near future, neutrino detections coincident with BBH merger events will be able to constrain the contribution of BBH mergers to the astrophysical neutrino flux.

## 5.1 Gravitational waves

First, we give a quick introduction to the basic theory of gravitational waves, in particular for the case of gravitational waves generated by compact binaries. The goal is not to provide a comprehensive overview of the theory of gravitational waves, but rather to introduce the subject, sketch an intuitive picture of the physics involved and motivate some of the choices and approximations of the work in the subsequent sections.

The discussion below is heavily based on the classic text [804] as well as the more recent book [805]. In addition, it draws inspiration from the basic physics papers [806, 807], which discuss the case of GW150914 in particular, and [808].

### 5.1.1 Introduction to the theory of gravitational waves

In general relativity, gravity is not an ordinary force, but rather a consequence of the geometry of spacetime itself. Instead of using the flat Minkowski metric $\eta_{\mu\nu}$ as in special relativity, the line element is now given by

$$ds^2 = g_{\mu\nu}(x^\mu)\,dx^\mu\,dx^\nu, \tag{5.1}$$

where $g_{\mu\nu}$ is a $4 \times 4$ symmetric matrix and is a function of the coordinates $x^\mu$. As a consequence of this $x^\mu$-dependence, spacetime is generally curved and becomes dynamical.

The curvature of spacetime is dictated by the Einstein field equations, which relate the curvature of spacetime (given by the Einstein tensor $G_{\mu\nu}$, which itself is formed out of the Riemann curvature tensor $R_{\mu\nu}$) to the local energy and momentum densities through the stress-energy tensor $T_{\mu\nu}$,

$$G_{\mu\nu} = \frac{8\pi G}{c^4} T_{\mu\nu}. \tag{5.2}$$

The above equation, while seemingly simple, hides a lot of complexity (in the Einstein tensor) and is difficult to solve in general. Different methods have been developed to solve them, including linearised theory (for gravitational waves, see below), perturbation theory (for small perturbations of known black hole solutions, in extreme mass-ratio inspirals or ringdown of a black hole after a merger), post-Newtonian expansion (expansion in $v/c$ characterising the deviation from the Newtonian solution, used to solve the general relativistic two-body problem) and more recently computers have become sufficiently powerful for numerical solutions (for e.g. the late-stage merger between two black holes).

Due to the dynamical nature of spacetime in general relativity, disturbances caused by matter can propagate and create gravitational waves. However, since general relativity is non-linear[1], there is in general no clear distinction between a wave and the rest of the metric. The concept of a wave is only well-defined in certain regimes (coinciding with the regimes where linearised gravity, perturbation theory and the post-Newtonian expansion are valid).

For the study of gravitational waves from compact sources, one can work in linearised theory. In the weak field regime[2], the metric can be decomposed in the Minkowski one plus a small perturbation $h_{\mu\nu}$,

$$ds^2 = (\eta_{\mu\nu} + h_{\mu\nu})\,dx^\mu\,dx^\nu, \qquad |h_{\mu\nu}| \ll 1. \tag{5.3}$$

---

[1]Analogous to the difference between QED and QCD: since QCD is a non-Abelian theory, gluons are charged under the symmetry group themselves, making the theory more difficult to solve. Gravity is associated to the Poincaré group, which is also non-Abelian. In other words, deformations like gravitational waves carry energy themselves.

[2]I.e., this is not appropriate during the final phase of the merger of two black holes. Here, the non-linearity is important and the field equations need to be solved numerically.

Defining the trace-reversed metric perturbation $\bar{h}_{\mu\nu} = h_{\mu\nu} - \frac{1}{2}\eta_{\mu\nu}\eta^{\alpha\beta}h_{\alpha\beta}$ and working in Lorentz gauge[3] $\bar{h}_{,\nu}^{\mu\nu} = 0$, the Einstein field equations can be expanded in powers of $\bar{h}_{\mu\nu}$. Keeping only linear terms[4] in $\bar{h}_{\mu\nu}$, they reduce to a set of (decoupled) wave equations

$$\left(-\frac{\partial^2}{\partial t^2} + \nabla^2\right)\bar{h}_{\mu\nu} = -\frac{16\pi G}{c^4}T_{\mu\nu}. \tag{5.4}$$

The general solution to the previous equation is the retarded integral (with $\mathbf{x}$ and $\mathbf{x}'$ three-dimensional vectors)

$$\bar{h}_{\mu\nu}(\mathbf{x}, t) = 4\frac{G}{c^4}\int d^3x' \frac{T_{\mu\nu}(\mathbf{x}', t - |\mathbf{x} - \mathbf{x}'|)}{|\mathbf{x} - \mathbf{x}'|}. \tag{5.5}$$

The mathematics for solving the equations for gravitational radiation from a compact source are analogous to the case of electromagnetic radiation, so we will use our knowledge of this case to obtain an intuitive understanding of gravitational waves. In the electromagnetic case, the electric field for radiation is proportional to the second time derivative of the electric dipole moment $\ddot{\mathbf{p}}$, and drops off inversely with a single power of distance $|\mathbf{E}| \sim \frac{|\ddot{\mathbf{p}}|}{r}$. This results in an integrated luminosity $L_{EM} \sim |\ddot{\mathbf{p}}|^2$. However, in the case of gravitational radiation a similar term does not appear. More specifically, the analogue of the electric dipole moment is the mass dipole moment $\sum_A m_A \mathbf{x}_A$. Since the first time derivative of this is the total linear momentum, which is conserved for a closed system, the second time derivative of the mass dipole moment is zero. The leading contribution to the gravitational radiation field will then come from the quadrupole term[5,6]. The expression for the gravitational wave amplitude needs to contain the prefactor from the Einstein equations and, like the electric field strength, falls of as $\sim \frac{1}{r}$ due to the inverse-square law for the luminosity. Finally, it needs to be proportional to a certain order time derivative of the quadrupole $Q$, with the number of time derivatives set by dimensional analysis, leading to

$$h \sim \frac{G}{c^4}\frac{1}{r}\frac{d^2Q}{dt^2}. \tag{5.6}$$

More details of this derivation can be found in Appendix I.1.

In order to expose the physical content in the previous equation, it is necessary to fix the remaining gauge freedom by going to the transverse-traceless (TT) gauge. Using all the gauge conditions (Lorentz, transverse and tracelessness), 8 out of the 10 components of $h_{\mu\nu}$ can be eliminated, leaving 2 polarisations as physical degrees of freedom. These polarisations are designated as $h_+$ and $h_\times$, which are equal to $h_+ = h_{xx} = -h_{yy}$ due to the tracelessness condition and $h_\times = h_{xy} = h_{yx}$ due to symmetry of the metric for a system observed from the $z$-axis. Again, this is the expected result for a massless non-zero spin field [95].

---

[3] The subscript $_{,\nu}$ indicates the derivative w.r.t. $x^\nu$.

[4] The approximations in linearised theory seem evident. However, in order to obtain, for example, orbits in Newtonian systems, an interaction between the field and the gravitating body is required, which is a second order term and thus not present in linearised theory. Still, for the generation of gravitational waves, the use of linearised theory is justified.

[5] Actually, we have ignored that in electromagnetism, the magnetic dipole is also important at this order. However, the gravitational analogue of this is the angular momentum $\sum_A \mathbf{x}_A \times (m_A \mathbf{v}_A)$, which is again conserved and thus does not give radiation.

[6] More general, the radiation from a field which is associated with massless particles of spin $S$ has as lowest multipole component $l = S$ [809] and for slow moving sources the lowest component dominates.

In order to obtain the fully correct expression, define the trace-free part of the second moment of the mass distribution or the reduced quadrupole, the definition of which is chosen to simplify formulas[7]

$$Q_{ij} = \int \mathrm{d}^3x\, \rho(x_i x_j - \tfrac{1}{3}\delta_{ij}r^2).$$

(5.7)

Since the metric perturbation was put in the TT gauge, we also need to select only the TT part of the reduced quadrupole[8]. We then finally obtain the expression for the gravitational wave strain $h_{ij}^{TT}$ due to a varying quadrupole [810]

$$h_{ij}^{TT} = \frac{G}{c^4}\frac{2}{d_L}\frac{\mathrm{d}^2 Q_{ij}^{TT}}{\mathrm{d}t^2},$$

(5.8)

where $d_L$ represents the luminosity distance. The appearance of the TT part on the left and right side of this equation has a simple interpretation. From the left side, we see that the action of the wave is in the transverse plane. Therefore, a detector follows the source movement projected onto the plane of the sky. From the right side, we see that the generation of gravitational waves is only due to the transverse distribution of the masses, i.e. the motions transverse to the line of sight.

The energy loss due to gravitational waves can again be found from analogy with electromagnetism. There, the quadrupole luminosity[9] is given by $L = \frac{1}{20}\frac{\mathrm{d}^3 Q_{ij}}{\mathrm{d}t^3}\frac{\mathrm{d}^3 Q_{ij}}{\mathrm{d}t^3}$. Therefore, we anticipate here $L_{\mathrm{GW}} \sim \frac{\mathrm{d}^3 Q_{ij}}{\mathrm{d}t^3}\frac{\mathrm{d}^3 Q_{ij}}{\mathrm{d}t^3}$. The correct expression is

$$L_{\mathrm{GW}} = \frac{1}{5}\frac{G}{c^5}\langle\frac{\mathrm{d}^3 Q^{ij}}{\mathrm{d}t^3}\frac{\mathrm{d}^3 Q_{ij}}{\mathrm{d}t^3}\rangle,$$

(5.9)

where the average is performed over several wavelengths, since the energy of a gravitational wave can not be localised more precisely than this. For a more detailed discussion, see Appendix I.3.

---

[7]There is possible confusion between different definitions of the quadrupole moment. For example, the quadrupole which appears in the theory of spherical harmonics is 3/2 times larger than the reduced quadrupole defined here.

[8]Actually, since we go to TT gauge anyway, there was no reason to insist on using the trace-free part of the second moment instead of the full second moment before. One conceptual reason to prefer the former is that it is the trace-free one which appears in the expansion of the Newtonian potential $\Phi \sim h_{00}$ and is thus the observable from a pure gravity perspective.

[9]The same analogy fails with the gravitational wave amplitude and the electric field (which is proportional to the third time derivative of the quadrupole).

The energy scale of gravitational waves is set by the prefactor $c^5/G = 3 \times 10^{59} \, \mathrm{erg \, s^{-1}}$, which is larger than the total luminosity of the observable universe[10]. Very relativistic systems manage to come close to this bound and a compact binary can reach a peak exceeding the luminosity of the universe during the inspiral and merger. Conversely, plugging in typical energy releases in explosive astrophysical events at typical distances, we find that merging binaries are associated with very small metric perturbations at Earth, of the order $h \sim 10^{-21}$.

### 5.1.2 Gravitational waves from compact binaries

In this section, we focus on gravitational waves from binaries, during the inspiral phase. Two compact massive bodies (such as neutron stars or black holes) orbit each other with angular frequency $\omega_{\mathrm{orbit}}$. This system has a varying mass quadrupole moment, therefore it radiates gravitational waves, with a frequency $\omega_{\mathrm{GW}} = 2\omega_{\mathrm{orbit}}$, since the quadrupole is symmetric under a rotation by $\pi$. Due to the emission of gravitational waves and the associated loss of energy, the binary's orbit will tighten[11].

Consider a binary system orbiting in the $xy$-plane. Its quadrupole moment is then given by

$$Q_{\mathrm{binary}} = \begin{pmatrix} Q_{xx} & Q_{xy} & 0 \\ Q_{xy} & Q_{yy} & 0 \\ 0 & 0 & 0 \end{pmatrix}. \tag{5.10}$$

When observing this system face-on, this quadrupole moment is already in the TT gauge and can be immediately applied in Eq. (5.8). This results in gravitational waves with both polarisations $h_+$ and $h_\times$ present out of phase, giving circular polarisation. When observing the system edge-on, for example from the $x$-direction, we need to project away the $x$-components and make the entire object traceless again. As a result, the quadrupole will have the form

$$Q_{\mathrm{binary,edge}}^{TT} = \begin{pmatrix} 0 & 0 & 0 \\ 0 & \frac{1}{2}Q_{yy} & 0 \\ 0 & 0 & -\frac{1}{2}Q_{yy} \end{pmatrix}. \tag{5.11}$$

Therefore, only the $h_+$-polarisation will be present, with an amplitude only half the one of each polarisation in the face-on case. Using Eq. (5.9), this means that the energy flux in the plane of motion is only $\frac{1}{8}$ of that along the rotation axis (one factor $\left(\frac{1}{2}\right)^2$ from the amplitude and one factor $\frac{1}{2}$ from averaging the single polarisation over a full orbit). The complete expression for the gravitational wave amplitude, including the inclination dependence explicitly, is given in Appendix I.2 for completeness.

When combining Eq. (5.9) with an expression for the quadrupole moment, the orbital energy ($E_{\mathrm{orb}} = -\frac{GM\mu}{2r}$ with $\mu$ the reduced mass $m_1 m_2/(m_1 + m_2)$) and Kepler's law ($r^3 = \frac{GM}{\omega^2}$), one obtains a relationship between the orbital frequency and the change of the gravitational wave frequency from the binary

$$\mathcal{M} = \frac{c^3}{G} \left( \left( \frac{5}{96} \right)^3 \pi^{-8} f_{\mathrm{GW}}^{-11} \dot{f}_{\mathrm{GW}}^3 \right)^{1/5}, \tag{5.12}$$

where $\mathcal{M}$ is the chirp mass, defined as

$$\mathcal{M} = \frac{(m_1 m_2)^{3/5}}{(m_1 + m_2)^{1/5}} = \mu^{3/5} M^{2/5}. \tag{5.13}$$

---

[10]In fact, it is an upper bound to the luminosity of any physical system.

[11]In the final revolutions during the merger of two black holes, when the radius is smaller than the innermost stable circular orbit (ISCO), the orbital separation decreases due to non-existence of circular orbits making the objects plunge in towards each other, not due to gravitational wave emission.

Therefore, under the assumption of equal masses, the binary's parameters are fully determined by observing the time-frequency behaviour of the amplitude. Relaxing this assumption, a more detailed analysis is necessary. One consequence of the above is that it is immediately possible to determine the distance of the binary by inspecting its gravitational wave signal: the measured amplitude $h$ is the apparent brightness, while the absolute energy loss can be found from the binary masses, extracted from the $f$-$\dot{f}$ relationship above. Note that because of the viewing angle dependence of the gravitational wave amplitude, there is a degeneracy between the distance of the binary and its inclination (see also [811–813]).

We can estimate the total energy released in gravitational waves, assuming it is mostly due to the loss of orbital energy. In the case of inspiralling black holes, the merger will happen when the total separation is equal to the sum of the Schwarzschild radii. This gives

$$E_{\mathrm{GW}} = \frac{GM\mu}{2R}, \tag{5.14}$$

with the individual Schwarzschild radii given by [814, 815]

$$r_{\mathrm{Schwarz}} = \frac{2Gm}{c^2}. \tag{5.15}$$

### 5.1.3   Detection of gravitational waves

Gravitational waves are detected using Michelson interferometers[12], illustrated in Figure 5.1a. As a gravitational wave travels along the detectors, the effective path length the light travels in the detector changes as $\Delta L = \delta L_x - \delta L_y$, with $L_x = L_y = L$ the length of the two arms. The deformation is related to the gravitational wave strain $h$ as $\Delta L = hL$, where $h$ is the linear combination of the polarisations $h_+$ and $h_\times$ (see e.g. [817]).

Currently, the main operating gravitational wave detectors are Advanced LIGO[13], operating two specialised Michelson interferometers with 4 km long[14] arms [818] and

---

[12]For brevity, I ignore pulsar timing arrays, of which the currently active experiments are currently collaborating under the international pulsar timing array project (IPTA) [816].

[13]Laser Interferometer Gravitational-Wave Observatory

[14]Using mirrors, the light is reflected multiple times inside the arms (creating Fabry-Perot cavities), effectively making the "true" arm length 1120 km.

Advanced Virgo [819], which is equally powerful but with different design choices. During the detection of GW150914, only the two LIGO detectors were in operation. The location of the two LIGO sites are such that the travel time of gravitational waves between these two sites is at most 10 ms, depending on the direction of the wave. Figure 5.1b shows the locations of the currently operating detectors, along with the locations of GEO600 (a smaller detector used to experiment with new technologies), the under-construction KAGRA [820] (the first underground detector and using cryogenic mirrors) and the planned LIGO-India [821]. While one detector is sufficient to measure a gravitational wave signature, it can not determine the direction from which the wave originates. Multiple detectors are necessary to localise the event and disentangle the polarisations[15]. With 2 detectors, events can be localised within a broken annulus, where the annulus is set by the time delay between detections and extra information comes from the amplitude and phase of the signal. With more detectors, the localisation precision improves (see [823] and references therein).

LIGO and Virgo are sensitive to signals with a frequency from 30 Hz to several kHz [817], and most sensitive between 100 Hz and 300 Hz, set by the presence of various sources of noise[16]. In this frequency band, these experiments are expected to be mainly sensitive to compact binary inspirals and mergers [17]. Different sources of noise include seismic, thermal, quantum (from the discrete nature of light), gas, scattered light and electronic noise (from the measurement electronics) [827]. During the first observing run (O1) of the Advanced LIGO detectors (from September 12, 2015, to January 19, 2016), the strain sensitivity[18] at 100 Hz was about $10^{-23}/\sqrt{\text{Hz}}$. Around the detection of GW150914, LIGO could observe 30 $M_\odot c^2$ binaries with a signal-to-noise ratio of 10 up to $z \sim 0.4$ [817].

## 5.2 The LIGO discovery of GW150914

In this section, we describe the detection of gravitational waves from the BBH merger GW150914. We discuss the properties of this merger and its implications. Finally, we also discuss the multimessenger observations triggered by this detection.

---

[15]Since the two LIGO detectors have a similar orientation, the Virgo detector is necessary to gain information on the polarisation [822].

[16]It is not set by the requirement arm length $\sim$ wavelength, which is the case for e.g. antennae. The arm length is determined by the desire to sample the expansion/contraction over as long a distance as possible, but cannot be so long that the light is still travelling while the wave contracts/expands again.

[17]Conversely, the space-based future detector eLISA [825] with arms of 2.5 million km long is sensitive to much lower frequencies and expected to detect galactic binaries, extreme mass ratio inspirals and possibly a background from inflation or first order phase transitions. Pulsar timing arrays with even longer arms and sensitive to lower frequencies can detect the stochastic background of supermassive black hole mergers. Figures comparing the sensitivity of these instruments can be made using `GWPlotter` [826].

[18]The units $\text{Hz}^{-1/2}$ are such because the noise is determined as the amount of power delivered on average by disturbances per frequency band, called the power spectral density. Since the power goes as amplitude squared, the associated strain is the square root of this quantity. For different conventions, see [826].

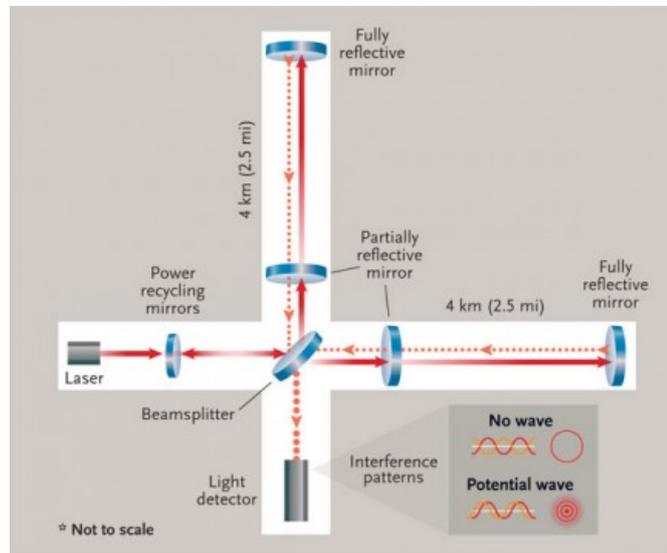

(a) Detection technology

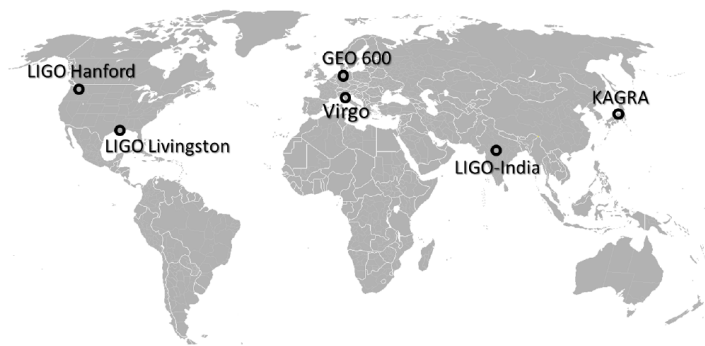

(b) Network

Figure 5.1: Gravitational wave detection (a) Sketch of the interferometer used to detect gravitational waves in LIGO. Credit: Sky & Telescope Leah Tiscione. (b) Location of the current (LIGO, Virgo, GEO600) and planned (LIGO-India, KAGRA) gravitational wave detectors. Figure from [824].

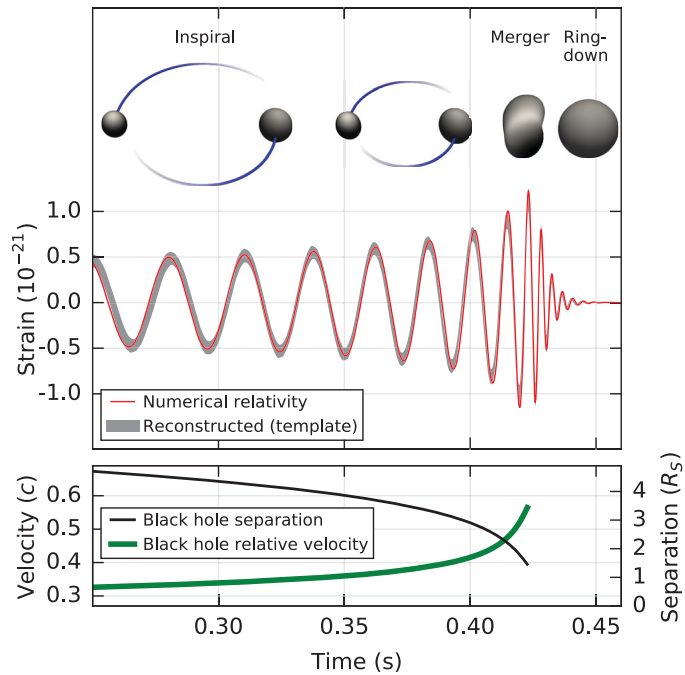

Figure 5.2: The binary black hole merger GW150914, showing numerical relativity models of the merger, the estimated gravitational-wave strain projected onto one of the detectors and the black hole separation and velocity. Figure from [484].

### 5.2.1 The binary black hole merger GW150914

**Properties of GW150914**

The two LIGO detectors observed GW150914 as a transient gravitational wave signal on September 14th 2015[19] with a significance of more than 5.1 $\sigma$ [484]. The measured signal, shown in Figure 5.2, lasted for $\sim 0.2$ s and is consistent with the gravitational wave signal expected from the merger of two black holes with a mass[20] of $36^{+5}_{-4}\ M_\odot$ and $29^{+4}_{-4}\ M_\odot$. The total energy release from such a system is $3^{+0.5}_{-0.5}\ M_\odot$, or $5.4^{+0.9}_{-0.9} \times 10^{54}$ erg. From the measured amplitude, the distance of the binary has been determined to be $410^{+160}_{-180}$ Mpc, corresponding to a redshift of $z = 0.09^{+0.03}_{-0.04}$. Initially, the event was localised within an area in the sky of 600 deg$^2$, although this was later improved to 230 deg$^2$ [823], which is set by the time delay of $6.9^{+0.5}_{-0.4}$ ms between the two LIGO detectors [828]. (The probability density function for the sky localisation is shown later, in Figure 5.3.)

In order to identify the binary and estimate the physical parameters of the binary from the measured waveform, the analysis fits the measured signal to analytical waveform predictions. In order to predict the full waveform, different techniques are necessary in different stages of the merger. While the early inspiral can be described in a post-

---

[19]Two days after calibration was complete and four days before the scheduled start of the observations.
[20]I am now switching back to natural units where $c = 1$.

Newtonian expansion, later stages are based on a phenomenological extension of this. The merger itself can only be described numerically. Finally, the ringdown is described by perturbation theory [828]. However, one can already deduce the main properties of the binary from the time-frequency behaviour of its measured gravitational wave signal and some basic physics arguments [806], some of which will be repeated here.

The system producing the gravitational waves has to be oscillating, because of the oscillating signal. Since the amplitude is not decreasing[21] and the frequency is increasing, the source can not be equilibrating (which would show as a constant frequency with decaying amplitude). The only conceivable system with the observed behaviour is an orbiting binary, where the energy losses cause the binary to tighten and the frequency and amplitude to increase. The sudden peak and subsequent dying out of the signal must then occur when the objects collide. Using Eq. (5.12), the frequency and its time derivative during the inspiral can be used to extract the chirp mass. Under the assumption of equal masses, this gives the individual objects[22] masses of about $\sim 35\,\text{M}_\odot$. Assuming that Newtonian physics is accurate enough for most of the inspiral, the orbital radius can be found from Kepler's law using the frequency of the signal. An estimate of the separation at the merger itself can then be found from the frequency at maximum amplitude, leading to $R = \left(\frac{GM}{\omega_{\text{orbit,max}}^2}\right)^{1/3} \sim 350$ km.

This quantity can be compared with the Schwarzschild radius for objects of these masses. In this case, each of the objects has a Schwarzschild radius of about $R_S \sim 100$ km. Since this is close to the total separation between the two objects, the objects must be very compact, of the order of their Schwarzschild radius. According to the hoop conjecture any non-spinning mass compressed to within the Schwarzschild radius in all directions forms a black hole [829]. Moreover, no other object is known that can support such a high mass while being smaller than the separation obtained above (neutron stars are the most compact stars and have not been observed with a mass higher than $\sim 3\,\text{M}_\odot$; even when allowing for unequal masses, the lightest object needs to be much heavier than this [806]). Therefore, these objects are likely to be black holes.

The distance of the binary can be found by comparing the observed amplitude with the calculated one for a binary with the above parameters. A naive estimate can be found as follows: the measured gravitational wave amplitude is $h \sim 10^{-21}$, which decreases as $h \propto 1/d_L$ with distance. However, going back to the source, this scaling must break down at the combined Schwarzschild radius $R \sim 200$ km, giving

$$d_L < 10^{21} \times 200 \text{ km} \sim 6 \text{ Gpc}. \tag{5.16}$$

A better simple estimate can be found by the observation that the peak luminosity is independent of the mass for an equal mass binary [806].

---

[21]Note that in some figures showing the gravitational wave amplitude, the amplitude rises rapidly from zero and undergoes only a few oscillations before the merger. However, this is an artefact of the sensitivity band of the instrument and the filtering applied in the analysis. The full signal undergoes many oscillations with an amplitude which is only slowly increasing towards the merger.

[22]Information about the total mass can be obtained from the late stage coalescence which, combined with the chirp mass, gives the individual masses [828]. Even with basic physics arguments, it is possible to constrain the mass ratio [806].

Estimating the energy released from the inspiral using Eq. (5.14), we find $E_{\mathrm{GW}} = \frac{GM\mu}{2R} \sim 3\,\mathrm{M_\odot}$. In reality, a comparable amount of energy is released during the merger phase as during the inspiral, although this can only be calculated using numerical methods [828]. This also gives a direct estimate of the final black hole mass ($M_f = m_1 + m_2 - E_{\mathrm{GW}}$), which can be checked against the value obtained separately from the ringdown.

More details of the full and proper parameter extraction from GW150914 can be found in [823]. Most importantly, while most parameters relevant for the work here are estimated with sufficient accuracy, the uncertainty on the distance is significant, due to the already-mentioned degeneracy between distance and inclination of the binary [811–813]. Due to the inclination-dependent amplitude, there is a preference for the detected binary to be either face-on or face-off (angular momentum vector parallel or anti-parallel to the line of sight), which then also corresponds to the highest distance, compared to edge-on. In the case of GW150914, there is, in addition, a slight preference for the binary to be face-off compared to face-on [823].

**Subsequent detections in O1**

After the detection of GW150914, more signals have been detected in LIGO's first observing run O1, the properties of which are summarised in [823]. These binary systems have been found in the compact binary coalescence search, which searches for objects with masses between $1\mathrm{M_\odot}$ and $99\mathrm{M_\odot}$, a total mass below $100\mathrm{M_\odot}$, and a dimensionless spin (spin over mass) up to 0.99. In total, two events have been detected with a significance above $5\sigma$: GW150914 and GW151226. A third, candidate event, was detected with a significance $\lesssim 2\sigma$: LVT151012. All of these are binary black hole mergers. The properties of these binaries are shown in Table 5.1. Of these events, GW150914 is both the most powerful and the closest and (as a result) its location has been most accurately determined, which is important for multimessenger searches.

**Inferred merger rate**

Using the number of detected events in run O1 of Advanced LIGO and the sensitivity of the instruments, it is possible to determine the rate of BBH mergers detectable by LIGO in the local universe. The 90% credible interval[23] is given by [823]

$$R = 9 - 240\,\mathrm{Gpc^{-3}\,yr^{-1}}. \tag{5.17}$$

This range is determined by the union of results for different black hole mass distributions, with the requirement $5\,\mathrm{M_\odot} \leq m_2 \leq m_1$ and $m_1 + m_2 \leq 100\,\mathrm{M_\odot}$. In case of a mass distribution flat in log mass, given by $p(m_1, m_2) \propto \frac{1}{m_1 m_2}$, the inferred rate is

$$R_{\mathrm{flatlog}} = 31^{+42}_{-21}\,\mathrm{Gpc^{-3}\,yr^{-1}}. \tag{5.18}$$

---

[23]This is the Bayesian equivalent of a confidence interval.

Table 5.1: Summary of the properties of binary black hole mergers detected in run O1 of LIGO, after the similar table in [823]. The table shows median values with 90% credible intervals including both statistical and systematic errors.

|  | GW150914 | GW151226 | LVT151012 |
|---|---|---|---|
| Primary mass $m_1/M_\odot$ | $36.2^{+5.2}_{-3.8}$ | $14.2^{+8.3}_{-3.7}$ | $23^{+18}_{-6}$ |
| Secondary mass $m_2/M_\odot$ | $29.1^{+3.7}_{-4.4}$ | $7.5^{+2.3}_{-2.3}$ | $13^{+4}_{-5}$ |
| Final mass $M_f/M_\odot$ | $62.3^{+3.7}_{-3.1}$ | $20.8^{+6.1}_{-1.7}$ | $35^{+14}_{-4}$ |
| Radiated energy $E_{rad}/M_\odot c^2$ | $3.0^{+0.5}_{-0.4}$ | $1.0^{+0.1}_{-0.2}$ | $1.5^{+0.3}_{-0.4}$ |
| Source redshift $z$ | $0.09^{+0.03}_{-0.04}$ | $0.09^{+0.03}_{-0.04}$ | $0.20^{+0.09}_{-0.09}$ |
| Luminosity distance $D_L/Mpc$ | $420^{+150}_{-180}$ | $440^{+180}_{-190}$ | $1000^{+500}_{-500}$ |
| Sky localisation $\Delta\Omega/deg^2$ | 230 | 850 | 1600 |

In case of a mass distribution where the heaviest mass follows a power law $p(m_1) \propto m_1^{-2.35}$ and with $m_2$ uniform, the inferred rate becomes

$$R_{powerlaw} = 97^{+135}_{-67} \, \text{Gpc}^{-3} \, \text{yr}^{-1}. \tag{5.19}$$

The rate for events with black hole masses similar to GW150914 is given by

$$R_{GW150914} = 3.4^{+8.8}_{-2.8} \text{Gpc}^{-3} \text{yr}^{-1}. \tag{5.20}$$

### 5.2.2 Astrophysical and fundamental physics implications

The detection of a binary black hole merger in gravitational waves allows to study previously unexplored areas in both astrophysics and fundamental physics.

Before the detection of GW150914, binary neutron star (BNS) mergers were expected to be the most promising source class for a first gravitational wave detection[24]. Therefore, the fact that a BBH merger was the first event to be detected, was a surprise. Moreover, the masses of the black holes involved were larger than expected for stellar mass black holes. The observation of GW150914 has several astrophysical implications [831]. It demonstrates that heavy black holes ($> 25 \, M_\odot$) can form in nature and form binaries. In order to create such massive black holes, one requires weak massive-star winds, which implies low metallicity[25,26]. Finally, the inferred merger rate is at the upper end of what is predicted in typical models.

There are also implications for fundamental physics. First, the gravitational wave signal from a binary black hole merger makes it possible to test general relativity in a previously unexplored regime, namely the relativistic, strong field regime[27] [832], recently joined by the first picture of a (supermassive) black hole event horizon [833]. From the recorded waveform, one can constrain the coefficients of the post-Newtonian expansion up to the 3.5 PN term in the expansion [834], whereas the observation of binary pulsar orbits only constrains up to 1 PN. Other tests involve the nature of the final black hole or constraints on the graviton mass, both of which are consistent with general relativity [484].

---

[24]See for example the living review in relativity [830] (version of 2013), were the focus is on BNS mergers, although the rate of BBH was expected to be similar (more rare but detectable up to larger distances).

[25]Since winds are mainly driven by atomic transition lines.

[26]However, it is also possible that such massive black holes are the result of mergers of less massive black holes. In this case, the LIGO mergers primarily give us information about lighter black hole mergers and not about weak massive star winds.

There has also been interest in this event in the context of dark matter physics: the high mass of the black holes involved (and also their misaligned spin) might be easier to explain with a primordial origin instead of a stellar one. In this case, a sufficiently large population of primordial black holes could act as dark matter, since they only interact gravitationally. While this possibility had been investigated before, interest in the scenario was reinvigorated after the announcement of GW150914 (see e.g. [835, 836]). Such a scenario is already constrained by many different observations from astrophysics and cosmology[28], each relevant in a different mass range, seemingly ruling out stellar mass primordial black holes as a candidate for dark matter. However, some of these constraints rely on astrophysical assumptions or require complicated modelling, such that some of them were called into question and this mass window was reopened. Recently, the constraints within the window $10^{-18}$–$10^4 M_\odot$ have been reassessed, taking into account a mass distribution of the black holes[29] [837]. At the moment, the mass window around stellar masses for primordial black holes to constitute the full dark matter abundance is still open if only well-established bounds are taken into account. On the other hand, using all available constraints, they can only constitute around 10% of all dark matter.

### 5.2.3 Multimessenger searches

The LIGO detection triggered a large follow-up campaign searching for coincident emission in the electromagnetic spectrum and in neutrinos. In general, no such emission is expected from BBH mergers, since the BBH environment is expected to be devoid of matter to give rise to a relativistic outflow: contrary to neutron stars, black holes can not supply matter themselves and (in the case of stellar black holes) most of the matter would be blown away by the supernova explosion of the progenitor stars and any remnant matter is expected to be swept up long before the merger. Therefore, any identified counterpart would reveal the existence of matter in the binary environment. Consequently, in order to test the no-counterpart hypothesis, it is useful to search for counterparts.

Moreover, the number of galaxies contained within the 90% credible area (630 deg$^2$

---

[27]Compared to other tests of solar system dynamics and binary pulsar orbits, which are in the low velocity, weak field regime.

[28]Constraints on e.g. the evaporation of black holes, microlensing, CMB anisotropy due to accretion, wide binaries and the survival of stars in dwarf galaxies.

[29]Since most constraints were previously available only for monochromatic mass functions, although in the end the difference ends up to be minor.

during the initial analysis) and the 90% confidence interval distance, corresponding to a volume of about $10^{-2}$ Gpc$^3$, and above a certain reference luminosity is $\sim 10^5$, so a concrete identification of the GW source (host galaxy and environment) using only gravitational waves is impossible [838]. Note that the addition of Virgo to the gravitational wave detector network would change the localisation to few tens of deg$^2$, a huge improvement (in particular for follow-up searches) but still not sufficient to identify a source with gravitational waves only.

**Broadband electromagnetic follow-up of GW150914**

A description of the broadband follow-up campaign can be found in [838] (some of which will be summarised here), while physics results are reported by the individual observatories. For run O1 of Advanced LIGO onwards, the LIGO and Virgo collaborations set up an extensive EM follow-up program, which was joined by 74 groups, including both ground- and space-based observatories, of which 63 were operational during O1[30].

Two days after the initial detection of the gravitational wave signal, an alert was sent out (to restricted parties only) on the gamma-ray coordinates network (GCN), reporting the observation of a gravitational wave event and providing an initial sky map of the localisation of the source. In total, 25 observatories responded[31] and performed observations of this region over a three month period, spanning 19 orders of magnitude in the electromagnetic spectrum. The identification of the source as a BBH merger was communicated only three weeks later, possibly influencing the area surveyed by the various observatories; some focussing their search on our galaxy or the Large Magellanic Clouds. The final skymap, sent out in January, covers a 630 deg$^2$ 90% credible area, spanning an arc in the Southern Hemisphere, along with a small arc at the equator (see also Figure 5.3).

While gamma-ray observatories can perform all-sky searches[32], experiments at other wavelengths can only probe part of it or only up to a certain flux (as is the case for X-rays).

Most of the optical candidates in the reported credible region were identified as supernovae, as well as a few AGNs and others, none of which significant enough to be related to the gravitational wave signal. Note that the sensitivity of the various experiments was sufficient to identify the afterglow of a binary neutron star merger, which would have been observable by LIGO up to a distance of 70 Mpc.

On the other hand, Fermi-GBM has reported a sub-threshold transient event above 50 keV with a significance of $2.9\sigma$, consistent with a weak short GRB [839], occuring 0.4 s after[33] the gravitational wave signal arrived and lasting 1 s. While the location of the event is ill-constrained, it is consistent with that of GW150914. The derived luminosity of this detection between 1 keV and 10 MeV is given as $1.8^{+1.5}_{-1.0} \times 10^{49}$ erg s$^{-1}$ (this includes the uncertainty on the distance and assumes isotropic emission), an order of magnitude dimmer than the peak luminosity of the dimmest bursts [839]. However, no signal was reported by other experiments: between 75 keV and 2 MeV by INTEGRAL [841] (putting a limit on the isotropic equivalent luminosity of $E_\gamma < 2 \times 10^{48}$ erg), in the 50 MeV to 10 GeV range by AGILE [842] and GeV gamma rays by Fermi-LAT [840]. In addition, no GRB-like afterglow was found in X-rays by Swift XRT [843], or in the UV/optical by Swift UVOT [843]. Of these, the INTEGRAL upper limits in particular cast doubt on the possibility of gamma-ray emission, since they have an overlapping energy range [841]. In [844], it was argued that the discrepancy can be solved by properly taking into account low count statistics, which causes the Fermi-GBM "signal" to be compatible with a background fluctuation. However, the authors of the Fermi-GBM search have shown [845] that this analysis is faulty and the original result stands.

---

[30]More information on this program can be found at `https://www.ligo.org/scientists/GWEMalerts.php`.

[31]The (now public) GCN circulars, where GW150914 was first reported, along with the responses of the various observatories, can be found at `https://gcn.gsfc.nasa.gov/other/GW150914.gcn3`. Note that the event was initially named G184098.

[32]E.g. Fermi-GBM covers at each moment 70% of the sky and observed 75% of the localisation area during the detection of GW150914 and had observed the full area after 25 minutes [839]. Similarly, Fermi-LAT views 1/5 of the sky at any moment and the full sky every $\sim 3$ hours, observing the full GW150914 probability area (even though it was initially out of view) within 4200 seconds [840].

### Neutrino follow-up

Coincident, transient, emission in neutrinos was searched for by various experiments in different energy ranges: $\bar{\nu}_e$ of a few tens of MeV by KamLAND [846], neutrinos between 3.5 MeV and 100 PeV by Super-Kamiokande [847] and ultrahigh-energy neutrinos (above 100 PeV) by the Pierre Auger Observatory [848]. None of these found evidence for neutrino emission from GW150914.

Here, we focus on the follow-up search by IceCube and ANTARES [849], sensitive to neutrinos above $\sim 100$ GeV. The analysis searches for neutrinos in a time window[34] of $\pm 500$ s around GW150914, which is a conservative upper limit on the plausible coincident emission of neutrinos and gravitational waves from GRBs. No other timing information within this time window is used.

IceCube searches for both up- and down-going neutrinos (i.e. from the Northern and Southern sky respectively), where the energy threshold for down-going neutrinos is increased to reduce the atmospheric muon background. The main search utilises the online event stream, which uses an event selection similar to the one in point source searches[850]. The background is mainly induced by reaction products of cosmic rays interactions with the atmosphere (muons in the South, neutrinos in the North). The selection cuts are designed to have a constant number of events per solid angle [851]. As a result, the expected number of events is 2.2 per 1000 s in both the Northern and Southern sky. The analysis found two events in the Northern sky and one event in the Southern sky (the latter one is, at 175 TeV, the most energetic of the three, but 12.5% of the background has an even higher energy). This number is consistent with background. In addition, their energy does not make them significant with respect to background and, more important, they are not spatially coincident with the localisation region[35] of GW150914, shown in Figure 5.3. In addition, IceCube also searched for neutrinos using the high-energy starting events selection and for MeV neutrinos using the photomultiplier noise rate (designed for supernovae neutrinos [852]). Both of these searches found no evidence for neutrino emission.

---

[33]Gamma-ray satellites take data continuously and can analyse archival data of the full sky, in contrast to e.g. observatories in the optical, which need to be pointed in the right direction, delaying the observation.

[34]Note that the time difference between the arrival of gravitational waves and neutrinos due to propagation is expected to be $\ll 1$ s, although alternative gravity models could change this [849].

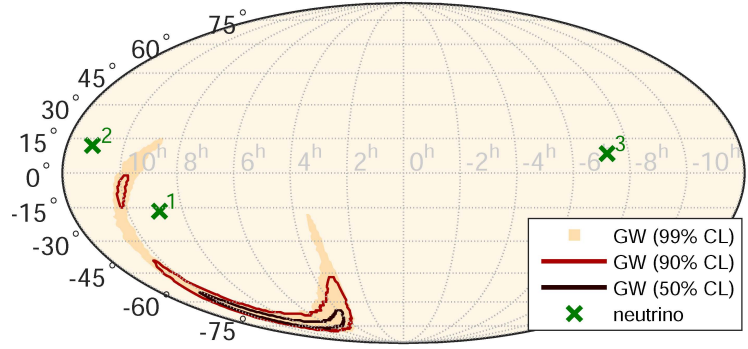

Figure 5.3: Localisation of GW150914 (set by the time delay of the signal between the two LIGO detectors) at various credible levels and the neutrino events detected within a time window of $\pm 500$ s, in equatorial coordinates. Figure from [849].

ANTARES selects only up-going events in their analysis. Their online pipeline, along with an additional minimum requirement on the energy of selected events, produces an event rate of 1.2 per day. Therefore, the number of expected events is 0.014 within the chosen time window. The analysis found none, consistent with the background expectation.

These null results were converted in a frequentist limit on the neutrino fluence[36] from GW150914, as a function of direction and for IceCube and Antares separately (since these experiments perform different in North and South and have optimal sensitivity at different energies[37]). This limit varies between $10^{-1} - 10$ GeV cm$^{-2}$. The average upper limits for the two hemispheres are $E_\nu^2 \Phi_\nu(E) = 1.2^{+0.25}_{-0.36}$ GeV cm$^{-2}$ (Southern) and $E_\nu^2 \Phi_\nu(E) = 0.10^{+0.12}_{-0.06}$ GeV cm$^{-2}$ (Northern) for an $E^{-2}$-spectrum. Note that GW150914 has a credible region mostly in the South.

Converting this flux into an energy limit, integrating the flux between 100 GeV and 100 PeV, results in a limit varying between $E_{\nu,\text{tot}} = 5.4 \times 10^{51} - 1.3 \times 10^{54}$ erg (also taking into account the uncertainty of the distance). This can be compared to the energy release in gravitational waves of $3 \text{M}_\odot$, which equals $5 \times 10^{54}$ erg or the isotropic-equivalent electromagnetic energy of typical GRBs, which is $\sim 10^{51}$ erg for long and $\sim 10^{49}$ erg for short bursts (with a neutrino energy release which is expected to be similar or higher).

---

[35]The region quoted has a solid angle of 590 deg$^2$ (140 deg$^2$) at 90% (50%) credible level.

[36]The fluence that would lead to, on average, 2.3 detected coincident events.

[37]For ANTARES, 90% of the signal neutrinos are between 3 TeV and 1 PeV, while for IceCube this is between 200 TeV and 100 PeV. Both for an $E^{-2}$-spectrum.

## 5.3 Multimessenger interpretations of the BBH merger

The possible existence of a gamma-ray signal coincident with GW150914 has been interpreted in several different models, all of which are ultimately based on a variation of the traditional GRB model[38] (reviewed in e.g. [857, 858] and already briefly discussed in Section 3.2.4). The essence of the GRB fireball model is simple and robust: a large injection of energy/radiation inside a small volume with few baryons will lead to a relativistic outflow, which gives rise to a gamma-ray burst (through internal or external shocks which convert the kinetic energy into accelerated particles and radiation). This model itself is blind to the inner engine which produces the injection of energy (which is immediately the reason why gravitational waves from a GRB are the only way to identify this inner engine unambiguously). There exist various GRB progenitor models, but they always lead to BH+debris torus systems [857]. In such a system, there are two reservoirs available to power the GRB: either through accretion or rotational energy. In the accretion scenario, the gravitational binding energy of orbiting debris is tapped, releasing up to 42% of the disk mass $M_{disk}$ for maximally rotating black holes[39]. On the other hand, it is also possible to tap the spin energy of the black hole, releasing up to 29% of the black hole mass $M_{BH}$ for a maximally rotating black hole. This can happen if magnetic fields in the torus thread the black hole, through the Blandford & Znajek (B-Z) mechanism [859]. In both cases, there is an additional magnetohydrodynamical efficiency factor to convert the energy into a jet. Typically, one assumes an efficiency factor of $\sim 10\%$ for the conversion of accretion power, to outflow, to radiation.

Possible models then mostly differ in the way they explain the presence of matter around the BBH. In [860], a scenario was considered where a previously dead disk of matter is reactivated upon the merger of the black holes, restarting accretion and creating a GRB. They show that such a disk can be created from two high-mass, low-metallicity stars ($Z < 0.1Z_\odot$), where the outer layer of the envelope of the last exploding star remains bound at large radii, which cools and becomes neutral, i.e. dead. After it has cooled (and initial accretion has thus stopped), the disk has a mass of $5.5 \times 10^{-4} M_\odot$, which is sufficient for the observed GRB candidate assuming the typical 10% efficiency factor. This model will be used as a benchmark in our study, in Section 5.5. In another study of the same model by different authors [861], they find lower masses for the dead disk.

Another model [862] considers the possibility that the merger originates from a tight binary of a massive star and a black hole, which leads to the collapse to a black hole and the merger. The final black hole gets kicked into and captures part of the envelope of surrounding material, the resulting accretion creating a GRB. In the model of [863], the core collapse of a rapidly rotating massive star forms two clumps in a dumbbell configuration, which later collapse into black holes and merge. A jet is created from the outflow of material due to the merger or due to accretion of surrounding material. In [864], BBH mergers inside the accretion disk of AGNs are investigated. Black holes are expected to be numerous inside galactic nuclei, with a large fraction of them inside binaries. It is shown that these can migrate to and merge in the disk. There, they can accrete gas, leading to a possible X-ray or gamma-ray signal. On the other hand, in [865], it is shown that the electromagnetic counterpart can be explained by magnetic reconnection creating a relativistic outflow, in case the collapsing objects have a non-vanishing electric charge. Another model [866] envisions a mini-disk around the BBH; in particular, they study the EM consequences of the disk wind due to accretion (in the optical and radio). This model does not depend on the GRB candidate, although it might explain it if a jet component is developed.

---

[38]Naively, another possibility is that the gravitational waves themselves accelerate particles to produce high energy neutrinos. While this would certainly not be possible en route to Earth, where $h \to 10^{-21}$, one could entertain this possibility in the neighbourhood of the source, where $h \to 1$. However, this possibility has been investigated for electromagnetic emission in [853], and was found to be negligible. Since neutrino emission would require the production and decay of $Z$- and $W^\pm$-bosons, their production would be even more suppressed [854]. On the other hand, in [855], it is argued that a significant effect *would* be present in a highly magnetised environment such as magnetars. Finally, in [856], it is argued that the direct conversion of gravitons into photons could take place in an external magnetic field. Such a counterpart to a binary neutron star merger would then be visible out to a distance of a few Mpc with future radio detectors.

[39]In the case of a binary neutron star, the disk mass is predicted to be $M_{disk} \sim 0.1 M_\odot$.

If a GRB would be confirmed, the obtained signal can be used to constrain the GRB parameter space, which was investigated in [867]. They find that the signal is mildly inconsistent with an internal shock model, while explaining the candidate burst with a dissipative photosphere is no problem and an external shock model provides a natural explanation.

Finally, the consequences for other emission besides electromagnetic have also been investigated. In [868] the possibility of BBH mergers as sources of UHECRs is considered[40], through the B-Z mechanism. More important for the discussion in this chapter, in [869] neutrino emission associated with the candidate GRB signal is studied. Using the observed luminosity from GW150914-GBM, the associated neutrino emission is calculated in a short GRB model, varying the jet parameters (Lorentz factor and the variability time scale) and the ratio of proton to electromagnetic energy $L_p/L_\gamma$. Using the non-detection of HESE events by IceCube, they obtain constraints on these parameters.

## 5.4 Constraints on the neutrino emission from BBH mergers

Given the first detection of gravitational waves from a binary black hole merger and existing searches for coincident neutrino emission from GW150914[41], and motivated by a possible —though unlikely— gamma-ray signal, we investigate the current and future potential to probe neutrino emission from gravitational-wave sources. Without restricting ourselves to specific models, we constrain, for a given type of merger, the amount of energy released in neutrinos relative to that in gravitational waves, using searches for neutrino emission coincident with gravitational wave signals, as well as the already existing bounds from the diffuse astrophysical neutrino flux. This work was published in [803].

---

[40]Note that due to their charge and resulting deflection, UHECRs would not be detected in coincidence with the gravitational waves.

[41]While this work was in progress, three BBH mergers had been observed, but a neutrino search was only available for GW150914.

We focus on neutrino emission in an energy range accessible to the IceCube and ANTARES neutrino observatories, above 100 GeV up to 100 PeV[42]. At energies below 100 GeV, IceCube loses sensitivity, while its DeepCore extension extends the sensitivity down to GeV energies [870]. With KamLAND, it is even possible to probe neutrino emission down at the MeV-scale. However, it is likely that neutrinos at these energies would be produced by a completely different mechanism and matching the two results requires a detailed modelling of the neutrino spectrum.

### 5.4.1 Neutrino emission fraction

We define[43] the neutrino emission fraction $f_{source}^\nu$ as the amount of energy emitted in neutrinos[44] ($E_\nu^{tot} = \int_{E_{min}}^{E_{max}} dE_\nu\, E_\nu$) within a certain energy range relative to the amount of energy emitted in gravitational waves ($E_{GW}$) for a certain source class, given by

$$f_{source}^\nu = \frac{E_\nu^{tot}}{E_{GW}}. \tag{5.21}$$

In this work, we focus on the energy range $[E_{min},\, E_{max}] = [100\,\text{GeV},\, 100\,\text{PeV}]$ and on the case of BBH mergers ($f_{BBH}^\nu$).

In general, $f_{BBH}^\nu$ could have any value. However, in the following, we will assume that $f_{BBH}^\nu < 1$. Given that a large $f_{BBH}^\nu$ would imply a significant change in the physical conditions (i.e. a large amount of matter compared to the released binding energy of the system close to the merger), the agreement between the measured gravitational wave amplitude and the theoretical waveform would likely be spoiled. Therefore, this assumption is justified (in [863], a similar argument is made). As will be shown in this work, current constraints already imply $f_{BBH}^\nu < 1$.

While we do not assume a specific source for the neutrino emission (unless stated otherwise), in any conceivable scenario the energy emitted in neutrinos is from a separate energy reservoir and not extracted from the black holes or gravitational waves[45].

### 5.4.2 Modelling assumptions

In addition to the definition of the model above, we also make certain assumptions in order to reduce the parameter space to be investigated. These assumptions are described here.

---

[42]In practice, it is extremely unlikely to detect a 100 PeV neutrino, since the neutrino flux drop down significantly at these energies (even though the effective area remains high except for the most Northern declinations; see Section 5.4.4 and [850]). Still, we include energies up to 100 PeV, because the effective area is reported up to this value.

[43]Some others (e.g. [841]) have also reported their results in terms of a ratio $E_i/E_{GW}$.

[44]In Chapter 3, we used $Q_\nu$ instead of $E_\nu^{tot}$. We use the latter notation here to be consistent with the common notation for the total gravitational wave energy $E_{GW}$.

[45]As already mentioned in Section 5.3, energy can not be easily be transferred to matter from the gravitational waves themselves. Even if this were the case, this would likely be compensated by a shift in the estimated distance and the subsequent calculations are unaffected.

**Neutrino spectrum**

We consider two benchmark scenarios for the neutrino emission: a monochromatic and an $E^{-2}$-spectrum. The first scenario can be used when the neutrino spectrum is dominated by a single energy. It also allows for a direct convolution with any user defined spectrum. In practice, we will mainly use this case to illustrate the sensitivity of IceCube and ANTARES. The second scenario is the standard power law distribution that follows from Fermi acceleration. While the spectral index could deviate from 2, we will not consider such deviations. In both cases, the spectrum is normalised to $E_\nu^{\text{tot}}$.

In the monochromatic case, we perform a scan over the energy range between 100 GeV and 100 PeV, which is the energy range of interest for IceCube and ANTARES. Since the number of neutrinos produced for monochromatic production scales like $1/E_\nu$ in order to provide the same $E_\nu^{\text{tot}}$, while the interaction cross section with nuclei for detection in this energy range increases with $E$ [527], one expects that, up to detector effects, the amount of neutrinos detected at Earth is roughly constant[46]. The same argument shows that, for general input spectra with equal total luminosity, the total number of detected neutrinos should be independent of the exact shape of the spectrum.

In both cases, we assume full mixing between the neutrino flavours, so that all three flavours arrive at Earth in equal amounts.

**Orientation**

Gravitational waves from two merging black holes which are spiralling into each other, are emitted in all directions, with a slightly stronger flux along the angular momentum vector of the binary system (as shown in Section 5.1.2). Therefore, the most likely orientation of a detected event is either face-on or face-off, with GW150914 having a slight preference for face-off (as discussed in Section 5.2.1). In the case of jet formation, one expects the electromagnetic and neutrino emission to be beamed along this same direction. Therefore, while not guaranteed, there is an increased chance for emission other than gravitational waves, if it is present, to also be directed towards the observer. However, to stay general, we initially perform all calculations assuming isotropic emission. In the case of beaming, the flux will be enhanced with the beaming factor, under the assumption that it is directed towards Earth, and the corresponding result can be directly obtained by rescaling from the isotropic case. When looking at multiple merger events (as we will do in Section 5.5), it is also possible to include an additional correction factor to take into account the possible different orientations of the source system. On average, for a set of mergers, the enhanced flux and the decreased detection probability cancel each other and the result agrees with the isotropic case.

---

[46]More concretely: below $\sim$ TeV energies, the cross section goes linearly with neutrino energy $\sigma_{\nu N} \propto E_\nu$, while at higher energies the cross section can be approximated by a power law $\sigma_{\nu N} \propto E_\nu^\alpha$, with $\alpha \simeq 0.363$ [527].

**Gamma rays**

As discussed in previous chapters, high-energy neutrino emission is typically associated with gamma-ray emission through their common origin in pion decay. However, whereas neutrinos can propagate unhindered towards Earth, gamma rays can be attenuated in a multitude of ways on their journey. To take this into account, one needs to consider a specific model for the source environment (presence of matter and radiation). Therefore, in order to stay model-independent, only the neutrino emission is treated, ignoring any constraints from gamma-ray emission.

### 5.4.3 Limits from the diffuse astrophysical neutrino flux

First, we discuss the bounds on the neutrino emission fraction $f_{\rm BBH}^{\nu}$ from the diffuse astrophysical neutrino flux for the case of an $E^{-2}$-spectrum, following the approach in [487, 665]. Under the assumption that BBH mergers emit neutrinos throughout the history of the universe, the maximally allowed $f_{\rm BBH}^{\nu}$ is the one which saturates the astrophysical neutrino flux.

We consider the diffuse neutrino flux produced by a set of BBH mergers with properties similar to GW150914. They produce gravitational waves with an energy of 3 $\rm M_{\odot}$ along with an associated neutrino flux that follows an $E^{-2}$-spectrum between 100 GeV and 100 PeV. The corresponding rate of this class is given by $\dot{R}$ (in units Gpc$^{-3}$ yr$^{-1}$), for now unspecified. The consequent single-flavour diffuse neutrino flux is directly given by [480],

$$E_\nu^2 \frac{dN_\nu}{dE_\nu}\bigg|_{\rm obs} = \frac{1}{3}\left(f_{\rm BBH}^{\nu} t_H \frac{c}{4\pi}\xi_z\right) E_\nu \; \mathcal{Q}_{E_\nu}\big|_{{\rm inj},f_{\rm BBH}^{\nu}=1}\,, \qquad (5.22)$$

where the factor $\frac{1}{3}$ is after oscillation[47] and the energy generation rate is determined by (for $f_{\rm BBH}^{\nu}=1$, $E_\nu^{\rm tot}=E_{\rm GW}$)

$$\int dE_\nu \; \mathcal{Q}_{E_\nu}\big|_{{\rm inj},f_{\rm BBH}^{\nu}=1} = \dot{R} E_{\rm GW}, \qquad (5.23)$$

with $\mathcal{Q}_{E_\nu} \propto E_\nu \cdot E_\nu^{-2}$. As in previous chapters, the cosmic evolution of the sources is contained in $\xi_z$, given by [487]

$$\xi_z(E) = \int_0^\infty dz \, \frac{H_0}{H(z)} \frac{\mathcal{L}_\nu(z,(1+z)E)}{\mathcal{L}_\nu(0,E)}, \qquad (5.24)$$

where $H(z)$ is the redshift dependent Hubble parameter and $\mathcal{L}_\nu(z,E) = \mathcal{H}(z)\mathcal{Q}_\nu(E)$ is the spectral emission rate density. $\mathcal{H}(z)$ is the source evolution, with $\mathcal{H}(0) = 1$, while $\mathcal{Q}_\nu(E)$ is the emission per source multiplied with the merger rate rate. For a power law ($\mathcal{L} \propto E^{-\gamma}$), $\xi_z$ is energy-independent. In the following, we will assume that the BBH merger rate follows the star formation rate evolution, with $\xi_z = 2.4$ (see

---

[47]This is a correction on our published paper [803], where the factor 1/3 was not properly taken into account for the astrophysical constraint.

Sections 4.4.1 and 4.4.2). This is a reasonable assumption for models of isolated binaries where the merger occurs shortly after formation (see [871] and references therein).

The single-flavour diffuse astrophysical neutrino flux measurement by IceCube which will be used to calculate the constraint of $f_{\mathrm{BBH}}^{\nu}$ is given by [872]

$$E^2 \Phi_{\nu_\alpha}(E) = 0.84 \pm 0.3 \times 10^{-8} \, \mathrm{GeV \, cm^{-2} \, s^{-1} \, sr^{-1}}, \tag{5.25}$$

fitted with a fixed spectral index of 2 in the range between 60 TeV and 3 PeV of deposited energy in the detector. While there is a more up-to-date estimate of the flux [873], which was fitted with a free spectral index, we choose to perform the analysis using the standard spectral index of 2. With this fixed spectral index, the upper bound on $f_{\mathrm{BBH}}^{\nu}$ is obtained simply by equating the normalisation of the calculated neutrino spectrum (Eq. (5.22)) to the one measured by IceCube (Eq. (5.25)).

The astrophysical bound on $f_{\mathrm{BBH}}^{\nu}$ is calculated for two source classes. First, we consider only the neutrino flux emitted by BBH mergers with properties identical to GW150914, the rate of which is given by Eq. (5.20). This results in a bound

$$f_{\mathrm{GW150914}}^{\nu} \lesssim 1.09^{+5.09}_{-0.79} \times 10^{-2}. \tag{5.26}$$

This constraint on $f_{\mathrm{BBH}}^{\nu}$ will be used when the bounds from GW150914 itself are investigated, in Section 5.4.4. It is also possible to consider the full population of BBH mergers. Since the mass distribution of this population is not known precisely, LIGO reports a range of possible merger rates (Eq. (5.17)). As a simplification, since the real mass distribution (and thus emission strength) of binary black holes is not known, we still assume all of these mergers emit 3 $\mathrm{M_\odot}$ of energy in gravitational waves. The resulting bound on $f_{\mathrm{BBH}}^{\nu}$ is

$$f_{\mathrm{BBH}}^{\nu} \lesssim 1.54 \times 10^{-4} - 4.12 \times 10^{-3}. \tag{5.27}$$

This approach does not take into account properly the variation between different mergers. In Section 5.6, we discuss this issue in more detail. It should be noted that as further GW events are detected by LIGO and Virgo, the BBH mass distribution and typical $E_{\mathrm{GW}}$ will be known with more precision. This will allow the present bound to be calculated more accurately. This second constraint will be used when investigating the prospective bound from a population of detected BBH mergers in Section 5.5.

If BBH mergers emit neutrinos with a monochromatic spectrum, the results change. The diffuse neutrino spectrum from these BBH mergers will follow the redshift evolution of the source, instead of a simple power-law spectrum (this is shown in Appendix J.1). Therefore, the simplest constraint is to require that the predicted flux overshoots the IceCube measurement nowhere. Since anyway such a scenario does not seem very motivated at the moment, we restrict ourselves to an $E^{-2}$ emission scenario for the astrophysical bound.

### 5.4.4 Limits from GW150914

Now, we consider the limit on $f_{\mathrm{BBH}}^{\nu}$ from the non-detection of counterpart neutrinos to GW150914. In order to convert the emitted neutrino flux to the neutrino flux at Earth, both the angular distribution of the emission and the distance to the source need to be taken into account. Initially, we perform the calculations assuming isotropic emission. Afterwards, the results can be rescaled to the beamed case, assuming neutrino emission towards Earth. The distance to the merger is given by LIGO, but has an associated uncertainty of about a factor of two. In the following, for simplicity, only the result of the central value of $d_L = 410$ Mpc will be shown, although the error on this value implies an uncertainty of a factor $\sim 2$. Redshift effects on the flux of individual events will be ignored in the following, which is reasonable in view of the current distance probed by LIGO.

We then calculate, from the neutrino flux at Earth, the number of detected neutrinos by IceCube [529] and ANTARES [874], which can detect high energy neutrinos between 100 GeV and 100 PeV. We use the IceCube declination dependent effective area[48] reported in [850], which is similar to the one used in the follow-up search of GW150914 [849]. This search selects only muon neutrinos, because of their excellent pointing. Assuming full mixing between the neutrino species, this means that the flux of interest is reduced by a factor of 3. The IceCube effective area is given for three declination bands in the Southern Sky ($-90° < \delta < -60°$, $-60° < \delta < -30°$ and $-30° < \delta < 0°$) and the analysis cuts are chosen such that the background rate is uniform over the entire sky. In a time window of 1000 s around GW150914, which can be assumed to contain the full neutrino signal, the expected background is 2.2 events over the full Southern Sky [849] (as previously discussed in Section 5.2.3). Similarly, the ANTARES effective area presented in [876] is used, given for two declination bands in the Southern Sky ($-90° < \delta < -45°$ and $-45° < \delta < 0°$). From this, ANTARES expected to see 0.014 neutrino events in the Southern Sky in a time window of 1000 s around GW150914 [849]. The localisation of GW150914 is given by a credible region of $\Delta\Omega = 600$ deg$^2$, spread out over the Southern Sky, up to the equator. For the signal neutrino flux, we only assume that the total neutrino flux is emitted within the analysis time window, such that we do not need to take timing into account explicitly[49]. The detected energy spectrum is then given by

$$N_\nu(E_{\mathrm{mono}} = E_\nu) = A_{\mathrm{eff}}\Delta\Omega \frac{f_{\mathrm{BBH}}^\nu 3\,\mathrm{M}_\odot}{E_\nu} \frac{1}{4\pi d_L^2},\tag{5.28}$$

for the case of monochromatic emission, and by

$$\Phi_\nu^{\mathrm{detected}}(E_\nu) = A_{\mathrm{eff}}\Delta\Omega \frac{f_{\mathrm{BBH}}^\nu 3\,\mathrm{M}_\odot}{\ln\left(100\,\mathrm{PeV}/100\,\mathrm{GeV}\right)} \frac{1}{4\pi d_L^2} E_\nu^{-2},\tag{5.29}$$

for emission following an $E^{-2}$-spectrum, with $A_{\mathrm{eff}}$ the effective area appropriate for the declination band.

---

[48]The neutrino effective area of a detector is the equivalent area for which all neutrinos of a given neutrino flux impinging on the Earth would be observed [875].

[49]However, to convert the number of detected neutrinos back into a flux, the duration of the neutrino burst *does* need to be taken into account.

**Monochromatic emission**

First, in Figure 5.4, we show the number of detectable neutrinos from GW150914 in the case of a monochromatic spectrum with $f^\nu_{BBH} = 10^{-2}$, for different declination bands, scanning over the energy range between 100 GeV and 100 PeV. The irreducible background is not shown here, since it is irrelevant for a signal peaked at a single energy. As expected, the IceCube sensitivity drops towards the more southern declination bands, since the atmospheric muon background becomes increasingly large for this part of the sky. On the other hand, since ANTARES is located in the Northern Hemisphere, it is shielded from atmospheric muons for this part of the sky and only has to cope with the irreducible atmospheric neutrino background. As such, for the most southern part of the sky, at energies below $\sim 100$ TeV, ANTARES is the most sensitive.

The figure shows that, for monochromatic neutrino emission and for constant $f^\nu_{BBH}$, the number of detectable neutrinos varies little between $10^4$ GeV and about $10^7$ GeV, as previously argued. Outside this range, the sensitivity is affected by detection efficiency in the lower end and limited statistics in the upper end of the energy range. In order to estimate the limit resulting from the non-detection of counterpart neutrinos, we highlight a single detected neutrino on the figure. It follows that

$$f^\nu_{BBH} \lesssim 10^{-2}, \tag{5.30}$$

in an energy range between $10^4$ GeV and $10^7$ GeV for monochromatic emission, when considering the effective area near the horizon. For the more southern effective area, the limit is weakened.

**$E^{-2}$-emission**

Next, the constraints for the more standard case of an $E^{-2}$-spectrum are investigated. LIGO localises GW150914 in an area spread out over the Southern Sky, where the effective area varies over the the different declination bands. Since performing an analysis taking this into account properly induces an unnecessary complication (splitting the credible region by LIGO in the declination bands) and would still be an approximation (since the "real" effective area is not just a step function), we instead perform the analysis for two extreme cases. First, we consider the effective area corresponding to declination band $-30° < \delta < 0°$, where IceCube has the largest effective area due to the still relatively small atmospheric muon background in this region. The resulting limit for this region is optimistic and is used to test whether a neutrino signal could have been seen even in the best case scenario for viable models of neutrino emission from BBH mergers. Second, we consider in a second analysis also the effective area in the declination band $-90° < \delta < -60°$, in order to obtain a conservative limit on $f^\nu_{BBH}$. Here, ANTARES has a larger effective area in the low energy range, while the one of IceCube is larger in the high energy range. Since a combined analysis is beyond the scope of this work, the energy range is instead split in two regimes and in each regime we use the experiment with the largest effective area.

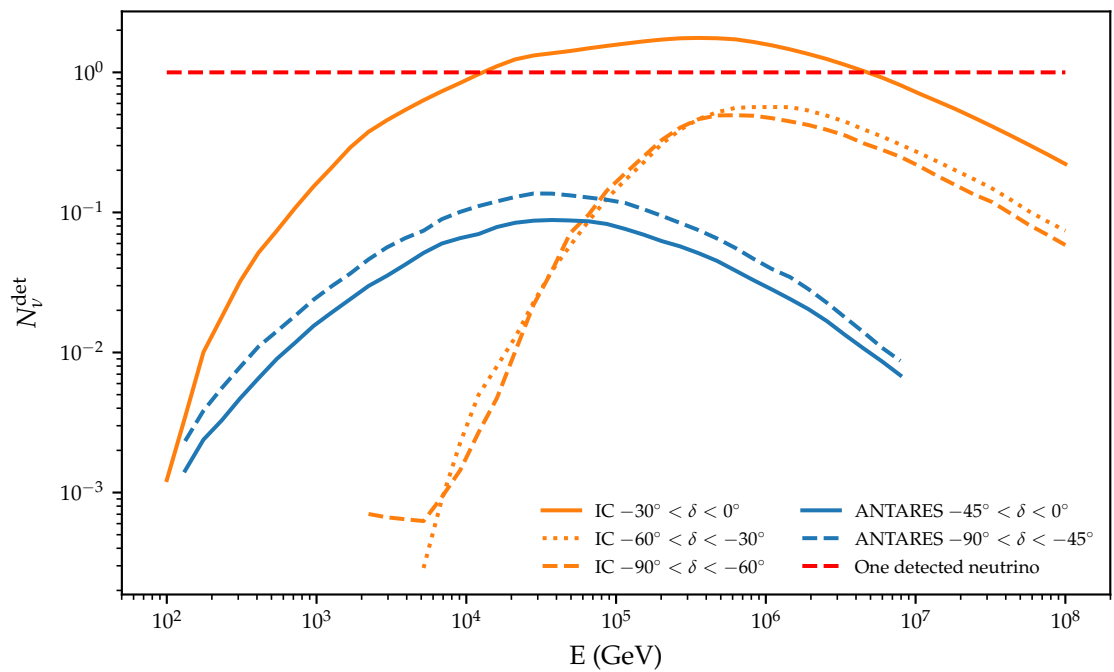

Figure 5.4: The amount of neutrinos detectable from GW150914 in the case of isotropic monochromatic emission for $f_{BBH}^{\nu} = 10^{-2}$, for a neutrino energy between 100 GeV and 100 PeV. The results are shown for both IceCube (orange) and ANTARES (blue) and for different declination bands. The red dashed line indicates one detected event.

For the limit on $f_{\text{BBH}}^{\nu}$ from the astrophysical neutrino flux, we take into account only the sub-class of binary black hole mergers with the same masses as GW150914. All of these mergers emit 3 $M_{\odot}$ of energy in gravitational waves, with a rate as in Eq. (5.20). The resulting bound on $f_{\text{BBH}}^{\nu}$ was already given in Eq. (5.26). Since the contribution from BBH mergers with different properties are not taken into account, this leads to a conservative bound on $f_{\text{BBH}}^{\nu}$.

In Figure 5.5, we show the predicted differential neutrino count which would be detected, assuming an isotropically emitted neutrino flux following an $E^{-2}$ spectrum, for $f_{\text{BBH}}^{\nu}$ varying between $10^{-7}$ and 1, indicated by the blue bands. The red dashed line indicates the threshold where a single neutrino would be detected, integrated over the entire energy range. The resulting limit from the non-detection of a neutrino counterpart to GW150914 puts an optimistic bound

$$f_{\text{BBH}}^{\nu} \lesssim 1.24 \times 10^{-2}, \tag{5.31}$$

using the effective area in the declination band $-30° < \delta < 0°$ where IceCube is the most sensitive, and a conservative bound

$$f_{\text{BBH}}^{\nu} \lesssim 5.89 \times 10^{-2}, \tag{5.32}$$

using the effective area in the declination band $-90° < \delta < -60°$ where we used a combination of IceCube and ANTARES. The limits for the different cases show little difference. These limits are comparable to the more sophisticated ones obtained by IceCube and ANTARES, reported in Section 5.2.3, which can be translated to $f_{\text{BBH}}^{\nu} \sim 10^{-3}$–0.2 (and also take into account the uncertainty on the distance). As previously stated, the astrophysical bound for mergers exactly like GW150914 has a value of $f_{\text{BBH}}^{\nu} \lesssim 1.09^{+5.09}_{-0.79} \times 10^{-2}$ (Eq. (5.26)), putting a similar limit on $f_{\text{BBH}}^{\nu}$ as the non-detection. In other words, if BBH mergers emit neutrinos and follow the star formation rate, one would not have expected a neutrino signal.

Figure 5.6 shows instead the integrated number of detectable events one expects, again from a source class with the properties of GW150914 as a function of the neutrino energy fraction $f_{\text{BBH}}^{\nu}$, this time for both isotropic (full blue line) and beamed (dashed blue line) emission. Since we want to investigate which $f_{\text{BBH}}^{\nu}$ could have lead to a visible neutrino signal in the most optimistic case, we use the effective area of IceCube near the horizon. The expected number of background events is 2.2 in a time window of 1000 s around GW150914 for the entire Southern Sky [850]. This number can then be rescaled to the expected number of background events within a solid angle of 600 deg$^2$, which corresponds to the localisation of GW150914. The resulting background, indicated by the full black line, is negligible for a single event. The bound from non-detection can be read off from the intersection of the blue lines with the one detected event threshold given by the red dashed line. In the case of beamed emission directed towards Earth, the received flux can be enhanced. For example, if a jet emits in a patch of $\Delta\Omega = 0.2 \times 0.2$ in solid angle, the flux would be enhanced with a factor $\frac{4\pi}{0.2 \times 0.2}$. As can be read off from Figure 5.6, the one detectable event threshold in the case of a beamed $E^{-2}$-spectrum then

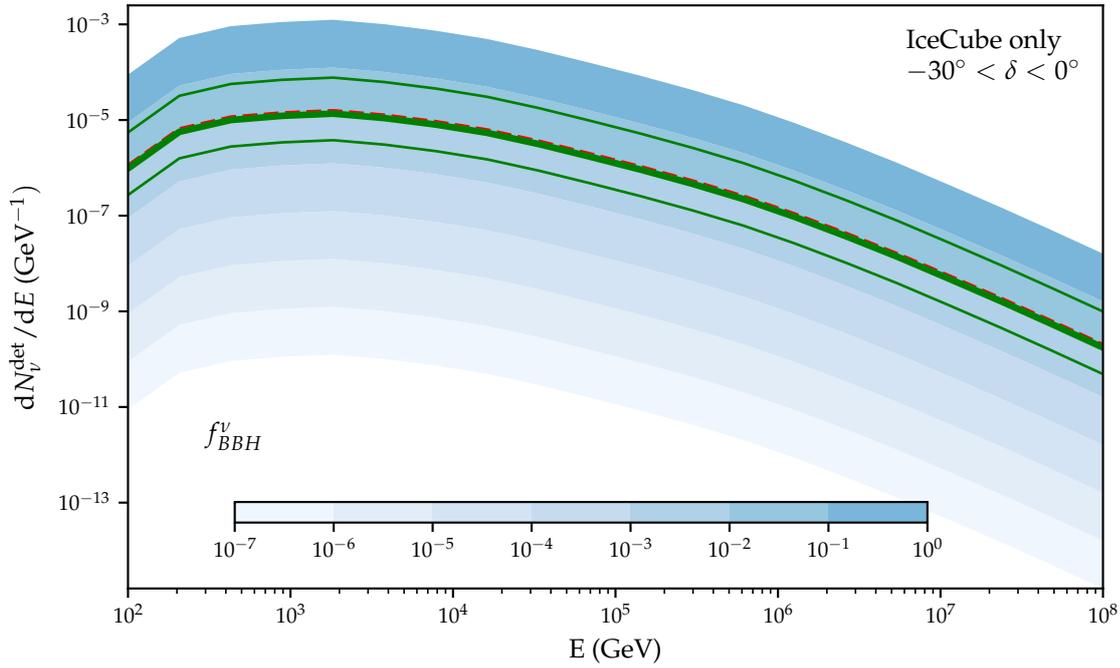

(a) Using effective area for $-30° < \delta < 0°$ (most sensitive; optimistic bound on $f^{\nu}_{\text{BBH}}$)

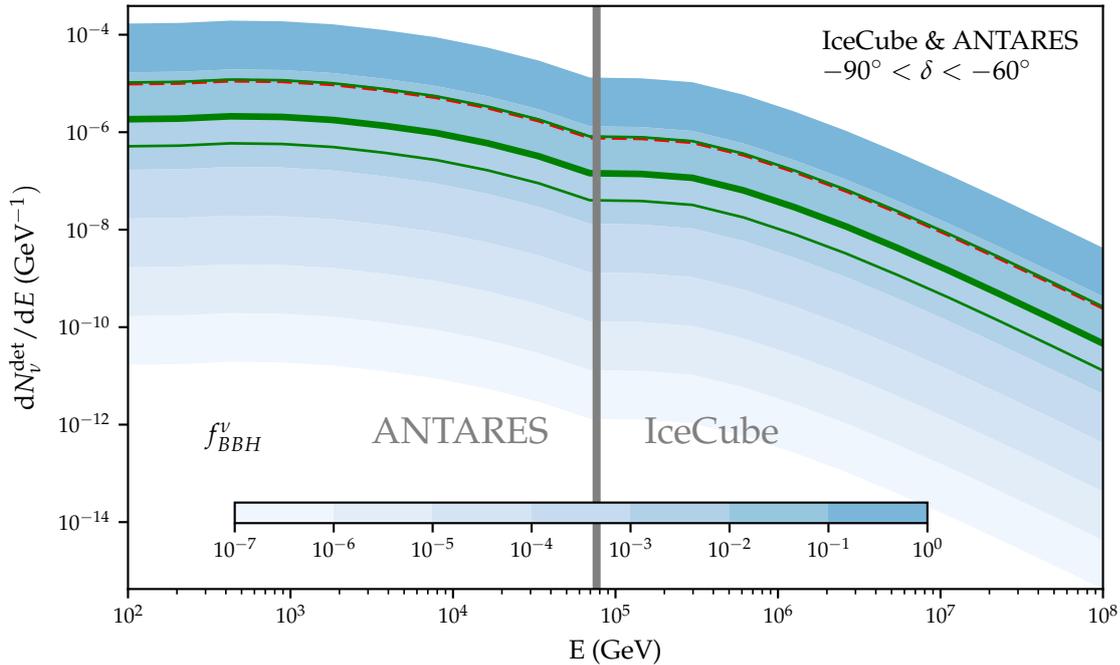

(b) Using effective area for $-90° < \delta < -60°$ (least sensitive; conservative bound on $f^{\nu}_{\text{BBH}}$)

Figure 5.5: The detectable differential neutrino count from GW150914 from a region in solid angle of 600 deg$^2$ where GW150914 is localised. The red dashed line shows the flux for which one neutrino event is detectable for this BBH merger event. The green lines show the upper bound from the astrophysical neutrino flux and its uncertainty for the class of BBH mergers with the same masses as GW150914 (Eq. (5.20)). (a) Effective area in the most sensitive region (from IceCube). (b) ANTARES and IceCube effective areas in the energy range where the respective experiment is the more sensitive.

becomes

$$f_{\mathrm{BBH}}^{\nu} = 3.96 \times 10^{-5} \times \frac{\Delta\Omega}{0.2 \times 0.2}. \tag{5.33}$$

The astrophysical flux would not change when individual sources have a beamed emission, since enhanced emission is cancelled by the reduced rate of events directed towards Earth. We find that the limit on $f_{\mathrm{BBH}}^{\nu}$ obtained from the non-detection of counterpart neutrinos from GW150914 is stronger than the one obtained from the astrophysical flux in case of a beamed emission, with the source located close to the equator. This immediately implies that in case BBH mergers with the same masses as GW150914 are responsible for the astrophysical neutrino flux, that the emission from GW150914 was not beamed towards Earth.

Since no neutrino has been detected so far and the single event detection threshold for isotropic emission is comparable to the astrophysical bound, currently all source populations are still allowed. In case of a neutrino detection in the near future, $f_{\mathrm{BBH}}^{\nu}$ would be at the level where only mergers similar to GW150914 would constitute nearly the full astrophysical neutrino flux, under the current assumptions of rate and source evolution and that $f_{\mathrm{BBH}}^{\nu}$ is universal for these mergers. Clearly, such a situation would be unnatural, constraining these assumptions.

## 5.5 Detection prospects

Since this work was performed before and during the Second Observing Run 2 of LIGO, we investigated how a stacked search for neutrino emission of expected events in run O2 would constrain $f_{\mathrm{BBH}}^{\nu}$.

We assume now that $f_{\mathrm{BBH}}^{\nu}$ is a universal constant for all mergers (for deviations on this, see the discussion in Section 5.6). For estimating the effectiveness of the stacked search, we assume that each gravitational wave event will be similar to GW150914, radiating 3 $M_{\odot}$ in gravitational waves from a distance of 410 Mpc, but with an arbitrary direction in the sky. Since GW150914 is expected to be among the more powerful BBH mergers to be detected by LIGO and it is relatively close by, this means that the bound on $f_{\mathrm{BBH}}^{\nu}$ we obtain below will be optimistic (i.e. we find the smallest $f_{\mathrm{BBH}}^{\nu}$ that could potentially be probed). Alternatively, one could use an appropriate "average event" for this analysis. However, since the mass distribution is still unknown and an appropriate average requires taking into account the LIGO detector response, we restrict ourselves to using GW150914. On the other hand, for the astrophysical bound on $f_{\mathrm{BBH}}^{\nu}$, we do not restrict ourselves to the rate associated with GW150914, but use the full estimated range of merger rates from Eq. (5.17). Thus, we compare the stacked search against the range of constraints[50] in Eq. (5.27). The assumption that all mergers are like GW150914 will be further discussed in Section 5.6 and an alternative prediction using an "average event" can be found in Appendix J.3. Again, we consider an $E^{-2}$-spectrum for the emitted neutrinos.

---

[50] Recall that while this constraint uses the full range of plausible merger rates, it still assumes all mergers emit 3$M_{\odot}$ of energy in gravitational waves.

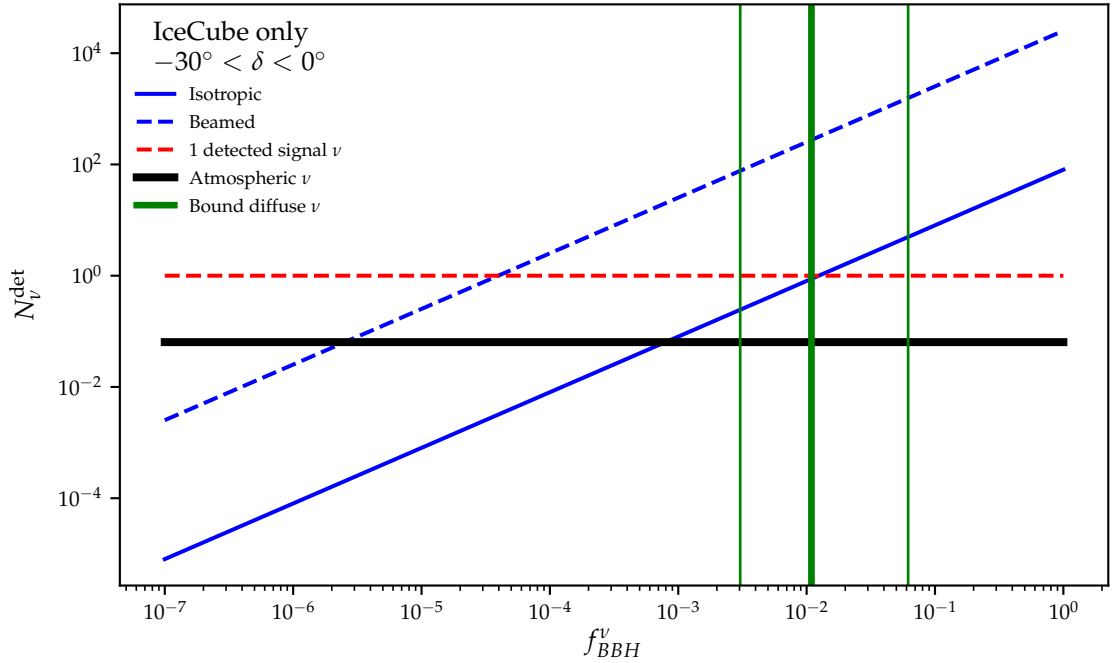

Figure 5.6: The integrated number of detectable neutrinos from GW150914, for emission following an $E^{-2}$-spectrum, as a function of $f_{\text{BBH}}^{\nu}$. Isotropic emission is indicated by the full blue line and beamed emission by the dashed blue line. The fat black line is the atmospheric neutrino flux, which is integrated over a time window of 1000 s and a solid angle of 600 deg$^2$. The red dashed line shows the one detectable event threshold. The green lines show the upper bound from the astrophysical neutrino flux and its uncertainty for the class of BBH mergers similar to GW150914 (Eq. (5.20)).

The details of the analysis are similar to the one in the previous section, with minor adjustments. We restrict ourselves to IceCube only, because of its superior sensitivity for most of the energy range in most of the sky. Since we allow the mergers to occur anywhere in the sky, the IceCube effective area is averaged over the full sky. Finally, the localisation of GW events is expected to improve with the improvements to the detector, calibration and analysis and with the enlargement of the LIGO-Virgo network [877]. In that case, neutrino observatories will be able to limit their search to a smaller solid angle in the sky, resulting in a reduced background. In order to show this improvement, the calculation is performed for a localisation ranging from 600 deg$^2$ to 20 deg$^2$. We now only take into account the irreducible background from atmospheric neutrinos [526]. This simplification corresponds to the case of an ideal analysis where all the atmospheric muon background can be removed. It is also representative for the near-future situation where KM3NeT [685] and Baikal-GVD [686] will be online, and both the Northern and Southern Sky will be optimally observed. This background is integrated over 1000 s and the localisation region of the event.

In order to estimate the sensitivity of future searches, we follow the approach from [878]. Given a number of expected background neutrinos $n_b$ (obtained from the analysis above) and the number of observed neutrinos $n_{obs}$, one can calculate the Feldman-Cousins upper limit [342] $\mu_\alpha(n_{obs}, n_b)$ at a confidence level (CL) $\alpha$. However, for a future experiment, the number of observed events is not yet know. Instead, we can calculate the "average upper limit" [878] as the sum of all upper limits $\mu_\alpha(n_{obs}, n_b)$, weighted by their Poisson probability

$$\bar{\mu}_\alpha(n_b) = \sum_{n_{obs}=0}^{\infty} \mu_\alpha(n_{obs}, n_b) \frac{(n_b)^{n_{obs}}}{(n_{obs})!} \exp(-n_b). \quad (5.34)$$

The average upper limit on $f_{BBH}^\nu$ at confidence level $\alpha$, as a function of the amount of BBH mergers being stacked $N_{GW}$, is then given by

$$f_{BBH}^\nu(N_{GW})|_\alpha = \frac{\bar{\mu}_\alpha(N_{GW} \times n_b^1)}{N_{GW} \times n_s^1(f_{BBH}^\nu = 1)}, \quad (5.35)$$

where $n_b^1$ is the number of detected background neutrinos expected in the search for neutrinos from one BBH merger and $n_s^1(f_{BBH}^\nu = 1)$ is the number of detected signal neutrinos from one merger for $f_{BBH}^\nu = 1$.

Figure 5.7 shows the average upper limits on $f_{BBH}^\nu$ at 68%, 95% and 99% confidence level (blue bands) that can be expected as a function of the number of detected BBH merger events ($N_{GW}$) by LIGO/Virgo. The bands indicate the effect of improving the BBH merger localisation from 600 deg$^2$ to 20 deg$^2$. The red dashed line indicates threshold $f_{BBH}^\nu$ at which at least one signal neutrino can be detected, integrated over all BBH merger events. At first, the limit on $f_{BBH}^\nu$ drops proportionally to the single event detection threshold, since the detection is purely signal limited. Starting at around 10 BBH mergers, however, the background starts to become relevant and the limit drops less fast. At this point, the improved localisation becomes important for neutrino observatories[51].

The obtained values for $f_{BBH}^\nu$ can be compared with the astrophysical bounds corresponding to the upper and lower limits on the BBH merger rates given in Eq. (5.17), which, following Eq. (5.27), are equal to $f_{BBH}^\nu \lesssim 1.54 \times 10^{-4}$ and $f_{BBH}^\nu \lesssim 4.12 \times 10^{-3}$, indicated by the hatched green lines. As more BBH mergers are observed, the estimate of the rate will improve and these two astrophysical bounds will get closer. We find that the average upper limits on $f_{BBH}^\nu$ reach the highest astrophysical bound at

$$N_{GW} \gtrsim 4, \, 9, \, 11, \tag{5.36}$$

at 68%, 95% and 99% CL respectively, with small differences between the different uncertainties in the localisation. The interpretation is the following: even with the lowest BBH merger rate, i.e. the highest possible neutrino emission per merger, $f_{BBH}^\nu$ can not exceed this constraint under the assumption of a universal $f_{BBH}^\nu$ with a source evolution following star formation rate. If counterpart neutrinos would be detected before reaching this number of BBH mergers, either $f_{BBH}^\nu$ is not a universal fraction for all BBH mergers, or the source population (merger rate, typical energy release or cosmic evolution of the sources) behaves differently than assumed. In either case, this would give non-trivial information on the neutrino emission from BBH mergers. The average upper limit from a search for counterpart neutrinos only reaches the lowest astrophysical bound for

$$N_{GW} \gtrsim 100, \tag{5.37}$$

at 68% CL and for a localisation of 20 deg². This constraint has a slightly different interpretation: even at the highest allowed merger rate, for $f_{BBH}^\nu$ below this value BBH mergers can not be responsible for the full astrophysical neutrino flux. Therefore, with the current available information on the merger rate, many mergers need to be observed before BBH mergers can be excluded as the main source of astrophysical neutrinos (under the assumptions on $f_{BBH}^\nu$ and $\xi_z$ above). On the other hand, the merger rate will be better constrained before reaching this number of events, such that in reality BBH mergers will be excluded faster as the source of the astrophysical neutrino flux. The vertical band indicates the expected number of BBH merger observations at the end of LIGO run O2, which is between $10 - 35$ [879]. A wider estimate puts this number between $2 - 100$, which covers the whole figure. It follows that the number of GW events needed to constrain the lowest astrophysical bound is (currently) well outside the reach of LIGO run O2. By the end of run O2, if 10 BBH mergers would be observed, it would be possible to limit $f_{BBH}^\nu$ down to about

$$f_{BBH}^\nu \approx 1 \times 10^{-3}, \, 4 \times 10^{-3}, \, 6 \times 10^{-3}, \tag{5.38}$$

at 68%, 95% and 99% CL respectively. If indeed 35 BBH mergers would be observed, it would be possible to limit $f_{BBH}^\nu$ down to about

$$f_{BBH}^\nu \approx 5 \times 10^{-4}, \, 1 \times 10^{-3}, \, 2 \times 10^{-3}, \tag{5.39}$$

---

[51] Obviously, for optical follow-up, improved localisation is important immediately, since it limits the area of the sky that needs to be surveyed.

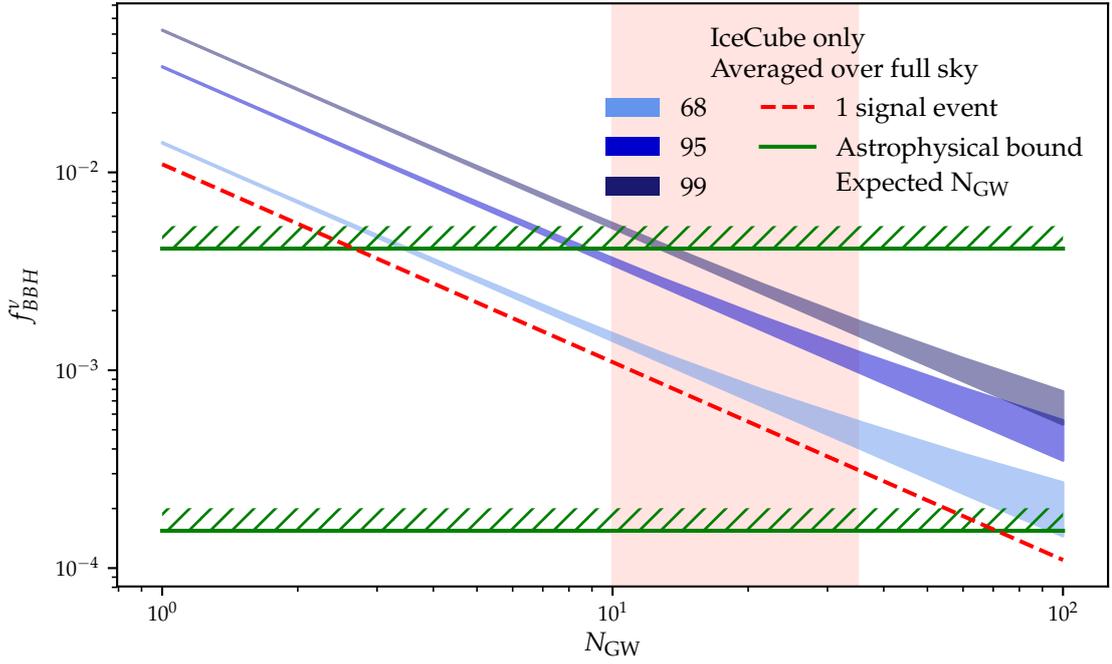

Figure 5.7: The expected average upper limits on $f_{\text{BBH}}^{\nu}$ at 68%, 95% and 99% CL (from bottom to top) as a function of number of BBH mergers observed by LIGO/Virgo emitting $3\,\text{M}_{\odot}$ in gravitational waves. The neutrino background is integrated over a solid angle between $600\,\text{deg}^2$ and $20\,\text{deg}^2$ (indicated by the bands), reflecting the expected improvement in localisation of the LIGO/Virgo network. The IceCube effective area is averaged over the full sky. The green hatched lines show the upper bounds from the astrophysical neutrino flux for the upper and lower limit of the BBH merger rate for the full population of BBH mergers (Eq. (5.17)). The vertical band shows the expected number of BBH mergers seen in LIGO run O2.

at 68%, 95% and 99% CL respectively.

Following Eq. (5.22), there is a degeneracy between the neutrino energy fraction $f_{\text{BBH}}^{\nu}$, and the source evolution parameter $\xi_z$ in determining astrophysical flux. To illustrate this degeneracy, we show in Figure 5.8 the astrophysical constraint (green lines), i.e. when neutrinos from BBH mergers saturate the astrophysical neutrino flux, on the $\xi_z$-$f_{\text{BBH}}^{\nu}$ plane. On the top axis, we indicate the amount of mergers necessary in a stacked search to constrain $f_{\text{BBH}}^{\nu}$ to the corresponding value on the lower axis. Hence, the current constraint from the non-detection of a neutrino counterpart from GW150914 is given by $N_{GW} = 1$ and the possible constraints after LIGO run O2 are indicated by the red band. From this, we find that if a single counterpart neutrino event would have been observed, or is observed within 10 GW events, the astrophysical flux can only be explained for weak source evolutions with $\xi_z < 3$. Given the current uncertainties on the BBH merger rate, to rule out BBH mergers as the main sources for the astrophysical neutrino flux, one needs to detect at least 300 BBH mergers. Nevertheless, assuming that the BBH merger rate is determined accurately to its central value after several detections, this might already be achieved after the 10–35 events predicted for LIGO run O2. To illustrate the level at which BBH mergers can be excluded as the source for the diffuse high-energy astrophysical neutrino flux, we also show the combination of $(f^\nu_{\rm BBH}, \xi_z)$ at which the neutrino flux from BBH mergers corresponds to 1% of the astrophysical flux (thin green lines).

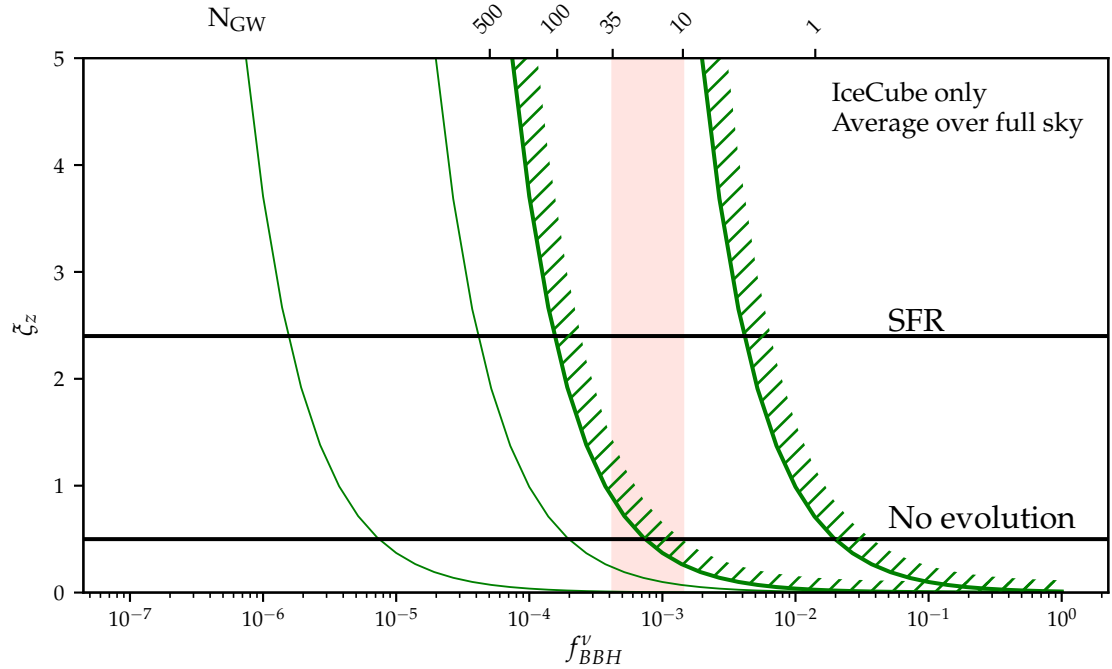

Figure 5.8: Constraints on BBH mergers as neutrino sources in the $\xi_z$-$f^\nu_{\rm BBH}$ plane. The constraints from the astrophysical neutrino flux are indicated with the hatched green lines. The top axis shows the number of BBH mergers similar to GW150914 ($N_{\rm GW}$) necessary to constrain $f^\nu_{\rm BBH}$ to the corresponding value on the lower axis at 68% CL (as in Figure 5.7) and with a 100 deg$^2$ resolution. Also shown are the bounds when BBH mergers can only be responsible for 1% of the diffuse astrophysical neutrino flux (thin green lines). The vertical band shows the expected number of BBH mergers seen in LIGO run O2. The two values of $\xi_z$ corresponding to a source evolution following the star formation rate and no evolution are indicated by black lines.

## 5.6 Extension to different scalings

The arguments presented in this work should be robust, general and lead to order of magnitude estimates for the neutrino emission fraction $f^\nu_{\rm BBH}$. In order to extend the predictions on $f^\nu_{\rm BBH}$ to BBH mergers of varying black hole masses (and thus varying $E_{GW}$), one has to make an assumption on the scaling of the neutrino emission for these different masses. The simplest assumption is that $E_\nu^{tot} \propto E_{GW}$, so that $f_{BBH}^\nu$ is a *universal* fraction for all binary black hole mergers. This assumption is valid, for example, if both $E_{GW}$ and $E_\nu^{tot}$ are proportional to the sum of the masses of the black holes. This is a reasonable approximation for $E_{GW}$, since it is true for equal-mass non-spinning black holes in the inspiral phase, as the released energy is proportional to the reduced mass of the binary[52]. The validity of this approximation was checked using fits to numerical simulations of non-spinning binary black hole mergers [880, 881] and is discussed in more detail in Appendix J.2.4. For $E_\nu^{tot}$, this scaling depends on the origin of the neutrino emission. In the case of a GRB-like scenario, the matter that seeds the neutrino production is a remnant of the original stars that formed the black holes. There, the assumption that the amount of matter available scales linearly with the star (and black hole) masses is reasonable.

It is also possible to consider more general relations between $E_{GW}$ and $E_\nu^{tot}$. This is illustrated by decomposing $f_{BBH}^\nu$,

$$f_{BBH}^\nu = f_0 \times g\left(\frac{M_{BH}}{M_{GW150914}}\right). \tag{5.40}$$

Here the normalised mass function $g\left(\frac{M_{BH}}{M_{GW150914}}\right)$ includes the amount of matter available to produce neutrinos, where in this work only a dependence on the combined mass of the black holes $M_{BH} = m_1 + m_2$ is considered.

To illustrate the effect of such a scaling, one can consider the situation described in the previous sections. To obtain the diffuse neutrino flux, the emission per source has to be convoluted with the black hole mass distribution. By considering a black hole mass distribution flat in log mass ($p(m_1, m_2) \propto \frac{1}{m_1 m_2}$), in combination with a neutrino emission proportional to $M_{BH}$ (i.e. $g = 1$ and $f_{BBH}^\nu$ universal), every mass combination has a similar contribution to the diffuse neutrino flux (this is checked in Appendix J.2). Therefore, a simple average mass over all BBH mergers will give an approximately correct result for the diffuse emission. Since we currently only have GW150914 available, we use this as the average. Therefore, the results presented in Figure 5.7, assuming all BBH mergers would be similar to GW150914, resemble the realistic situation of a flat in log mass BBH distribution, in combination with a neutrino emission that scales linearly with the black hole mass. The energy fraction is now calibrated by $f_{BBH}^\nu(M_{BH} = M_{GW150914}) = f_0$.

Consider as an extreme opposite case the situation of an inverse mass scaling such that $E_\nu^{tot} \propto 1/M_{BH}$, which leads to

$$f_{BBH}^\nu(M_{BH}) = f_0 \times \left(\frac{M_{GW150914}}{M_{BH}}\right)^2. \tag{5.41}$$

Such a proportionality could be explained in the case of a GRB-like model by more massive stars blowing away more of the surrounding matter necessary for neutrino emission (before forming the second black hole, i.e. long before the merger). In case of a flat in log mass distribution, the total contribution to the BBH neutrino emission from high-mass black hole binaries is further suppressed by their rate. Hence, the neutrino emission in this situation would be dominated by low-mass black hole mergers. Therefore, a simple average over all events will not suffice and one needs to explicitly weigh the events with their expected neutrino emission.

---

[52]Combining the inspiral energy release (Eq. (5.14)) with the definition of the Schwarzschild radius (Eq. (5.15)) gives $E_{GW} \propto \mu$, which is proportional to the total mass for equal-mass black holes.

It is possible that there are various sub-classes of BBH mergers (i.e. some have a lot of matter available, others not), with a different normalisation or scaling of the neutrino energy fraction $f_{\mathrm{BBH}}^{\nu}$. An example of a specific sub-class is the previously mentioned model of [864], where BBH mergers in AGNs might have a gas-rich environment and a one might therefore expect an enhanced neutrino energy fraction $f_{\mathrm{BBH}}^{\nu}$ for this source class. Taking into account such differences in a stacked search, the internal neutrino emission properties, as well as the source environment can be directly probed.

## 5.7   Implications

The results presented here can be used to draw more model-specific conclusions on the neutrino production. In general, no neutrino emission is expected from BBH mergers, since the black holes should have cleared the environment of all matter long before the merger occurs. This hypothesis can be tested, by assuming neutrinos are produced by accelerated matter around the BBH. One can then decompose $f_{\mathrm{BBH}}^{\nu}$ as

$$f_{\mathrm{BBH}}^{\nu} = f_{\mathrm{matter}} \times f_{\mathrm{engine}} \times \epsilon_{\mathrm{p,acc}} \times \epsilon_{\nu}. \tag{5.42}$$

Herein $f_{\mathrm{matter}}$ denotes the amount of matter present around the BBH relative to the amount of energy emitted in gravitational waves. It represents the environment and formation history of the binary and determines the possible different scaling relations discussed in the previous section. The acceleration model is contained in $f_{\mathrm{engine}}$ and determines the fraction of energy from accreting matter which is channelled into an acceleration engine. The factor $\epsilon_{\mathrm{p,acc}}$ represents the fraction of energy which is channelled into accelerated protons (or more generally, any baryons) and performs a similar role to $f_e$ of Section 4.3.3. Finally, $\epsilon_{\nu}$ denotes the fraction of accelerated proton energy which ends up in neutrinos and depends on the type of interaction producing the neutrinos ($p\gamma$ and $pp$, see Section 3.3.2) and the radiation and matter density (which determines its efficiency).

As an example, assume that during the merger, an accretion disk is activated and we end up with a system similar to the GRB-fireball model [599]. The conversion factor from accretion disk mass to fireball energy is expected to be of order $f_{\mathrm{engine}} = 1/10$ [857] (see also the discussion in Section 5.3). The amount of energy from the fireball that goes into the neutrinos is given by $\epsilon_{\mathrm{p,acc}} \times \epsilon_{\nu} = 1/10$ [599]. Hence[53]

$$f_{\mathrm{BBH}}^{\nu} \approx f_{\mathrm{matter}} \times 10^{-2}. \tag{5.43}$$

---

[53]This corrects the formula in the publication, where an additional factor of $1/20$ for $\epsilon_{\nu}$ was erroneously included.

This allows then to immediately constrain the amount of matter surrounding the two black holes. Using the non-detection of counterpart neutrinos in GW150914, which was in the most optimistic case at $f^{\nu}_{BBH} = 3.96 \times 10^{-5}$ for beamed emission in a typical solid angle $\Delta\Omega = 0.2 \times 0.2$ directed towards Earth (beamed since we are considering GRBs), this results in

$$f^{GW150914}_{matter} \lesssim 4.0 \times 10^{-3} \times \frac{\Delta\Omega}{0.2 \times 0.2}. \tag{5.44}$$

We can also calculate the expected limit after 10 and 35 BBH mergers detected by LIGO. From the analysis in Section 5.5 and rescaling the result for beamed emission, the expected limit on the amount of matter in the black hole binary environment is

$$f^{N_{GW}=10}_{matter} \lesssim 3.2 \times 10^{-4} \times \frac{\Delta\Omega}{0.2 \times 0.2}, \tag{5.45}$$

$$f^{N_{GW}=35}_{matter} \lesssim 1.6 \times 10^{-4} \times \frac{\Delta\Omega}{0.2 \times 0.2}, \tag{5.46}$$

at 68% CL. Since the astrophysical limits are weaker than the limits obtained from the non-detection of counterpart neutrinos after this many events if the emission is beamed towards Earth (in the optimistic case of all BBH equally powerful as GW150914), only the latter is considered for the limit on $f_{matter}$.

Conversely, it is also possible to estimate the value of $f^{\nu}_{BBH}$ within the GRB-fireball framework, given a model for the amount of mass. Consider the dead-disk model of [860]. There, the authors predict an amount of matter of $10^{-4}$–$10^{-3}$ $M_{\odot}$ in a non-active accretion disk around one of the black holes, coming from a massive progenitor star with low metallicity. Upon the merger, this disk is then reactivated and leads to a burst. Using these values[54], one gets $f^{\nu}_{BBH} \approx 10^{-6}$ for the fireball model. This should be compared with the reach in Figure 5.7, rescaled to smaller values of $f^{\nu}_{BBH}$ with a beaming factor. For a beaming factor of $\frac{4\pi}{0.2 \times 0.2}$, this $f^{\nu}_{BBH}$ is still below the estimated reach in Figure 5.7. As a sanity check, this estimate can also be compared with the result from Fermi-GBM, which reported a burst $L_{iso} \approx 2 \times 10^{49}$ erg s$^{-1}$ lasting for one second and the limit from INTEGRAL $E_{\gamma} < 2 \times 10^{48}$ erg, which corresponds to $f_{\gamma} < 10^{-6}$ [841]. The gamma-ray limits are stronger than the neutrino ones, since the former are easier to detect. Assuming that the neutrino and gamma-ray emission are comparable (as in Section 3.3.6), these limits are compatible[55] with the estimate of $f^{\nu}_{BBH}$ from the dead-disk model above. This was expected, since the dead-disk model for the amount of matter has been partially constructed to explain these observations.

## 5.8 Update after new detections

Since the publication of this work [803], several more gravitational wave events have been detected. The full results of the search for compact binary mergers with masses between a few and a few hundred solar masses in the first observing run (run O1, Advanced LIGO from September 12, 2015 until January 19, 2016) and the second observing run (run O2, Advanced LIGO (O2) from November 30, 2016 until August 25, 2017, Advanced Virgo joined on August 1, 2017) are found in GWTC-1: the gravitational-wave catalogue of compact binary mergers observed by LIGO and Virgo [882].

---

[54]Other models (e.g. [861]) predict even less matter around the BBH.

[55]More correctly, one should probably use the fireball energy (i.e. the first two factors in Eq. (5.42)) to compare with the X-ray energy and only include the final factor to neutrinos for the gamma-ray energy. However, since both experiments cover both X-rays and gamma rays, we do not make this distinction here.

Table 5.2: Summary of gravitational wave events in run O1 and O2 of LIGO and VIRGO. All are BBH mergers, except for GW170817. Errors show 90% credible intervals. Data taken from [882].

| Name | $m_1\ (M_\odot)$ | $m_2\ (M_\odot)$ | $E_{GW}\ (M_\odot)$ | $d_L$ (Mpc) | $z$ | $\Delta\Omega$ (deg$^2$) |
|---|---|---|---|---|---|---|
| GW150914 | $35.6^{+4.8}_{-3.0}$ | $30.6^{+3.0}_{-4.4}$ | $3.1^{+0.4}_{-0.4}$ | $430^{+150}_{-170}$ | $0.09^{+0.03}_{-0.03}$ | 180 |
| GW151012 | $23.3^{+14.0}_{-5.5}$ | $13.6^{+4.1}_{-4.8}$ | $1.5^{+0.5}_{-0.5}$ | $1060^{+540}_{-480}$ | $0.21^{+0.09}_{-0.09}$ | 1555 |
| GW151226 | $13.7^{+8.8}_{-3.2}$ | $7.7^{+2.2}_{-2.6}$ | $1.0^{+0.1}_{-0.2}$ | $440^{+180}_{-190}$ | $0.09^{+0.04}_{-0.04}$ | 1033 |
| GW170104 | $31.0^{+7.2}_{-5.6}$ | $20.1^{+4.9}_{-4.5}$ | $2.2^{+0.5}_{-0.5}$ | $960^{+430}_{-410}$ | $0.19^{+0.07}_{-0.08}$ | 924 |
| GW170608 | $10.9^{+5.3}_{-1.7}$ | $7.6^{+1.3}_{-2.1}$ | $0.9^{+0.05}_{-0.1}$ | $320^{+120}_{-110}$ | $0.07^{+0.02}_{-0.02}$ | 396 |
| GW170729 | $50.6^{+16.6}_{-10.2}$ | $34.3^{+9.1}_{-10.1}$ | $4.8^{+1.7}_{-1.7}$ | $2750^{+1350}_{-1320}$ | $0.48^{+0.19}_{-0.2}$ | 1033 |
| GW170809 | $35.2^{+8.3}_{-6.0}$ | $23.8^{+5.2}_{-5.1}$ | $2.7^{+0.6}_{-0.6}$ | $990^{+320}_{-380}$ | $0.2^{+0.05}_{-0.07}$ | 340 |
| GW170814 | $30.7^{+5.3}_{-3.0}$ | $25.6^{+2.7}_{-4.1}$ | $2.8^{+0.4}_{-0.3}$ | $560^{+140}_{-210}$ | $0.12^{+0.03}_{-0.04}$ | 87 |
| GW170817 | $1.46^{+0.12}_{-0.1}$ | $1.27^{+0.09}_{-0.09}$ | $\geq 0.04$ | $40^{+10}_{-10}$ | $0.01^{+0.0}_{-0.0}$ | 16 |
| GW170818 | $35.5^{+7.5}_{-4.7}$ | $26.8^{+4.3}_{-5.2}$ | $2.7^{+0.5}_{-0.5}$ | $1020^{+430}_{-360}$ | $0.2^{+0.07}_{-0.07}$ | 39 |
| GW170823 | $39.6^{+10.0}_{-6.6}$ | $29.4^{+6.3}_{-7.1}$ | $3.3^{+0.9}_{-0.8}$ | $1850^{+840}_{-840}$ | $0.34^{+0.13}_{-0.14}$ | 1651 |

In total, 11 sources of gravitational waves were detected, 10 of which were BBH mergers and one of which was a binary neutron star (BNS) merger. These events are GW150914, GW151012[56], GW151226 [823, 883], GW170104 [884], GW170608 [885], GW170814 (first event detected also by Virgo) [822], GW170817 (BNS) [434], GW170729 (highest mass to date), GW170809, GW170818 (second triple-coincidence) and GW170823 [882]. Their properties are shown in Table 5.2. Important for multimessenger searches is the vastly improved localisation for the triple-coincidence events. In addition to these events, also 14 GW candidates were detected, with a false alarm rate less than 1 in 30 days, but whose astrophysical origin could not be established or excluded.

With the additional detections, the BBH merger rate has been better constrained. Due to improved modelling on the analysis side [871] and the realisation of the existence of astrophysical processes which cut off the mass distribution of black holes [886], the mass distributions considered in the rate fit now cut off the primary black hole mass at $50\text{M}_\odot$, instead of constraining the total mass to $\leq 100\text{M}_\odot$ as in Section 5.2.1. Therefore, the results can not be immediately compared. They find[57] 90% confidence intervals for the merger rate [882] $R = 18.1^{+13.9}_{-8.7}$ Gpc$^{-3}$ yr$^{-1}$ for a mass distribution uniform in log for the primary mass and $R = 56^{+44}_{-27}$ Gpc$^{-3}$ yr$^{-1}$ for the case of a power law distribution $\propto m_1^{-2.3}$ for the primary mass. The union of these intervals gives an estimate

$$R = 9.7 - 101 \text{ Gpc}^{-3} \text{ yr}^{-1}. \tag{5.47}$$

---

[56]Previously LVT151012. The LVT nomenclature was dropped: everything with a false alarm rate less than 1 per 30 days and an astrophysical origin probability > 0.5 gets denoted as GW, everything else as marginal [882].

A more detailed analysis, which allows the power law index, the maximum and minimum black hole mass and the mass ratio to vary gives instead an estimate (again 90% credible intervals, model B in the paper) [871]

$$R = 53^{+59}_{-29} \text{ Gpc}^{-3} \text{ yr}^{-1} \tag{5.48}$$

with a primary mass distribution proportional to $\propto m_1^{-\alpha}$. The masses are bound as $\alpha = 1.6^{+1.5}_{-1.7}$ and $m_{min} = 7.9^{+1.2}_{-2.5}$M$_\odot$, $m_{max} = 42.0^{+15.0}_{-5.7}$M$_\odot$ and $m_2$ is distributed such that the mass ratio is peaked at equal masses (for more details: see Appendix J.2). All models favour maximum masses below 45M$_\odot$[58,59], disfavouring the older models in Section 5.2.1 and the one above with a maximum mass of 50M$_\odot$. For all these models, it is assumed that the rate is uniform in comoving volume. However, an analysis taking into account a redshift dependence has also been performed. The results are consistent with a flat or positive evolution, allowing for both no evolution and for an evolution following star formation rate.

On the other hand, as a comparison, the inferred rate of BNS mergers is $R = 110$–3840 Gpc$^{-3}$ yr$^{-1}$ (but the distance probed is lower than with BBH due to the weaker signal). Finally, neutron star-black hole mergers (NSBH) have not been detected yet, constraining their merger rate to below 610 Gpc$^{-3}$ yr$^{-1}$, which is already stronger than high rate estimates.

Fermi-GBM and Fermi-LAT have performed searches for gamma-ray emission coincident with the other run O1 events, but found no such emission [889]. Similarly, a search for high-energy neutrino emission from the run O1 events GW151226 and LVT151012 was performed with ANTARES and IceCube, detecting no neutrinos [890]. They constrain the isotropic-equivalent high-energy neutrino emission from GW151226 to $< 2 \times 10^{51} - 2 \times 10^{54}$ erg. In another analysis, a combined search was performed for gravitational waves and high-energy neutrinos in observing run 1 with Advanced LIGO, ANTARES and IceCube, focusing on events where a single messenger is not enough to detect the event. No such events were found [891]. At lower energies, KamLAND has performed a search also for the other run O1 events, finding no evidence for neutrino emission [846].

---

[57]The result is quoted for the GstLAL pipeline, one of the two matched-filter searches using relativistic models of gravitational waves from compact binary coalescences, this one based on GstLAL library [882, 887, 888]. However, the results of the two pipelines are compatible.

[58]One event, GW170729 has a mass above this value, although its error bar puts it in this region and its central value is still within the error on $m_{max}$. Since it is also the least significant event, the analysis was repeated without it, producing a fit with lower $m_{max}$.

[59]This influences the merger rate estimate dramatically even compared to the 50M$_\odot$ cut-off above.

With these new data available, we can update the results of the previous analysis. Using the new fit for the merger rate of the most complicated model (Eq. (5.48)), and the average gravitational wave emission corresponding to that black hole mass distribution $\langle E_{GW} \rangle = 1.52 M_\odot$ instead of $3 M_\odot$ (details in Appendix J.2.3), the astrophysical constraint becomes

$$f^\nu_{BBH} \leq 1.4 \times 10^{-3}. \tag{5.49}$$

The analysis in Figure 5.7 was repeated, now using the information from the events detected in run O1 and O2. We assume that for all events no neutrinos were detected coincident in both time and direction with the merger (since results have only been published for the first three mergers). The resulting Feldman-Cousins upper limits for $n_{obs} = 0$ (i.e. not the average upper limits, which are at higher $f^\nu_{BBH}$) are shown in Figure 5.9. The actual limit performs worse than the prediction. This result was expected, since GW150914 was considered to be among the stronger events to be detected. Indeed, it is still by far the most significant event. The next most significant events are GW170814 and GW170608, which drive the limit on $f^\nu_{BBH}$ down the most, as reflected in the figure.

The final limit after all BBH mergers is

$$f^\nu_{BBH} \leq 3.8 \times 10^{-3}, \, 1.1 \times 10^{-2}, \, 1.7 \times 10^{-2} \tag{5.50}$$

at 68%, 95% and 99% CL respectively. This value already reaches the highest astrophysical bound inferred at the detection of GW150914, although it is still far removed from the updated bound. We can also use this result to calculate a new constraint on $f_{matter}$, giving

$$f_{matter} \lesssim 1.2 \times 10^{-3} \times \frac{\Delta\Omega}{0.2 \times 0.2}. \tag{5.51}$$

We can also make a new prediction of the limit that will be reached in the future. Now, we define an appropriate average event to calculate the expected limit from a stacked search. This average event is not the same as the one used for the astrophysical limit, since here we also need to take into account the LIGO detector response. However, taking into account this response properly would be too complicated for the current work. Instead, we can take the value of $\frac{E_{GW}}{4\pi d_L^2}$ as a proxy[60] for the significance of an event. It was checked that this correlates extremely well with the signal-to-noise ratio of the events reported by LIGO, as it should be for constant detection response. The average detected event parameters can then be calculated from all mergers with a significance larger than the least significant event currently detected

$$\langle X \rangle = \iiint \theta\left(\frac{E_{GW}}{4\pi d_L^2} - F_{min}\right) X(m_1, m_2) p(m_1, m_2) p(R) \, dR \, dm_1 \, dm_2, \tag{5.52}$$

where $E_{GW}(m_1, m_2)$ is given in Equation (J.4) and $\theta$ is the Heaviside step function. However, as a further simplification, we instead take the average of the currently detected events, since they represent a sample of detectable events and this estimate should coincide with the estimate above. As discussed in Appendix J.3, the averages of the different observables $E_{GW}$, $d_L$, $m_1$, etc. are no longer related consistently and we need to use the average of the observable relevant for the study. We find for the parameters of the average merger $\langle m_1 \rangle = 30.61 M_\odot$, $\langle m_2 \rangle = 21.95 M_\odot$, $\langle E_{GW} \rangle = 2.5 M_\odot$ and $\langle d_L \rangle = 1038$ Mpc. However, in order to make a new prediction on the neutrino flux, we need the average energy flux at Earth[61] $\langle \frac{E_{GW}}{4\pi d_L^2} \rangle$. The energy flux received from the current events is comparable to the emission of $\sim 5 M_\odot$ from a distance of 1000 Mpc from an equal-mass BBH with $m_1 = m_2 \simeq 55 M_\odot$ (when calculated using the correct relation between $E_{GW}$, $m_1$ and $m_2$). Note that these masses are higher than the maximum mass allowed from the fit above, although this is artificial since we fixed the distance to an arbitrary value (which coincides with the actual average distance). The resulting new prediction is shown in Figure 5.10. We show again the expected average upper limit (as opposed to the limit for $n_{obs} = 0$ as for the detected events, which explains the jump from the current limit). We find that after $\sim 100$ events, the stacked search is competitive with the updated astrophysical bound, again assuming a universal $f^\nu_{BBH}$. A neutrino signal before this implies a more complicated scenario (e.g. only a small fraction of mergers emit neutrinos, but do so strongly). The performance of this estimate of the average event was checked in Appendix J.3 and it performs well. However, the improved sensitivity of LIGO and Virgo means that more weak events will be detected, which means that more events will need to be detected to reach the astrophysical constraint.

---

[60]Given the discussion about gravitational wave energy in Appendix I.3, the quantity $E_{GW}/4\pi d_L^2$ might not make sense for gravitational waves locally. However, in practice, it is a good proxy for the significance of the gravitational wave event.

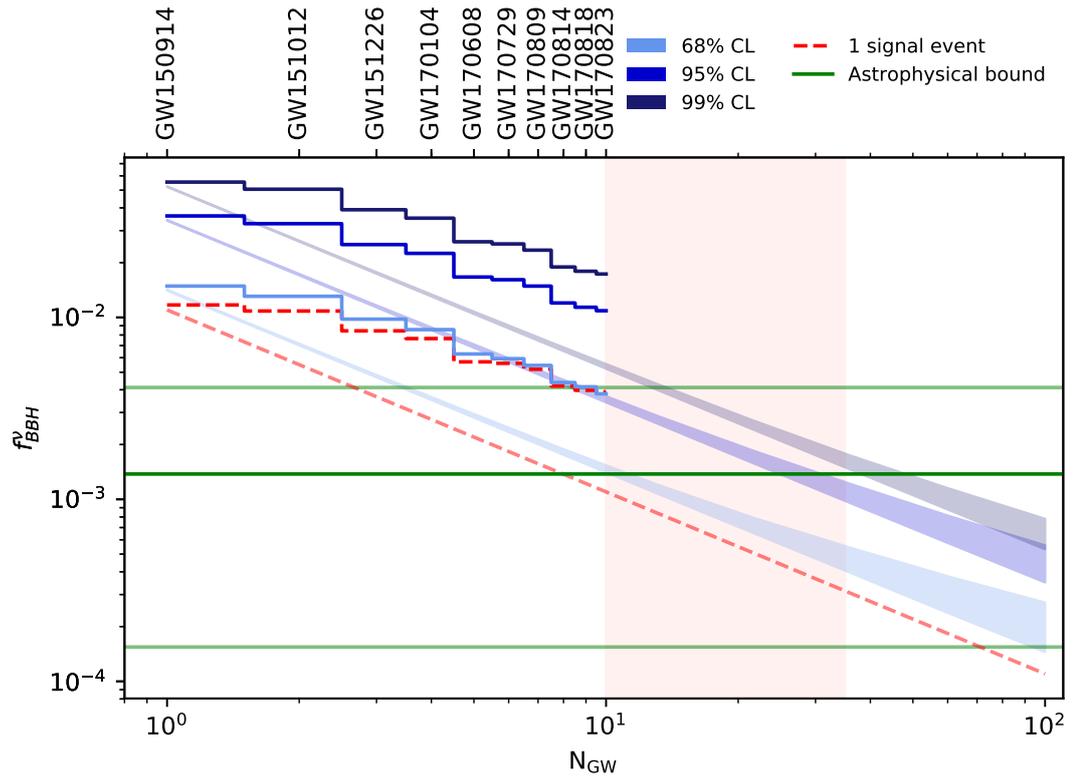

Figure 5.9: Comparison of the predicted upper limits in Figure 5.7 with the Feldman-Cousins limit from a stacked search from the actual events, assuming that no neutrino has been detected in coincidence with any merger. The astrophysical constraint is given by Eq. (5.49).

Finally, unlike BBH mergers, a multimessenger signal from BNS mergers is expected and has indeed been observed from GW170817 across the entire electromagnetic spectrum (although not yet in neutrinos) [435, 436, 892]. This detection confirms BNS mergers as (one of) the source(s) of gamma-ray bursts. The prompt gamma-ray burst was also followed by transient emission in the UV, optical and infrared, a kilonova, due to radioactive decay of heavy elements (formed by neutron capture) and a delayed X-ray and radio counterpart. Besides the information they provide on the astrophysical processes occuring during and after the merger, the electromagnetic detections also allow to probe the possible difference between the speed of light and gravity[62] or Lorentz invariance violation [434]. In addition, the absolute measurement of the distance of the merger using gravitational waves (as explained in Section 5.1.2) in combination with a redshift measurement from the optical provide an independent measurement (since it is a "standard siren" and does not require the use of the "distance ladder") of the Hubble constant[63], giving $H_0 = 70.0^{+12.0}_{-8.0}$ km s$^{-1}$ Mpc$^{-1}$, consistent with existing measurements [896].

---

[61] In this case, we are actually interested in the average neutrino flux of events detected by LIGO/Virgo, which we calculate with $\langle f^\nu_{BBH} \frac{E_{GW}}{4\pi d_L^2} \rangle$, so the issue with a local definition for the gravitational wave energy disappears.

[62] This requires an assumption on the time delay between the emission of the gamma-ray and gravitational-wave signals. The LIGO analysis [435] obtains an upper bound on $\Delta v$ by assuming they are emitted at the same time with the observed time lag of $(+1.74 \pm 0.05)$ s due to slower gravitational waves. A lower bound is obtained by assuming a reasonable time delay of 10 s, after which the photons from the GRB catch up. Of course, one can consider longer time delays. Even then, these constraints are orders of magnitudes stronger than direct measurements (from the time delay of gravitational waves between the two detectors [893]) or indirect measurements [894], which allow for time differences of more than 1000 years.

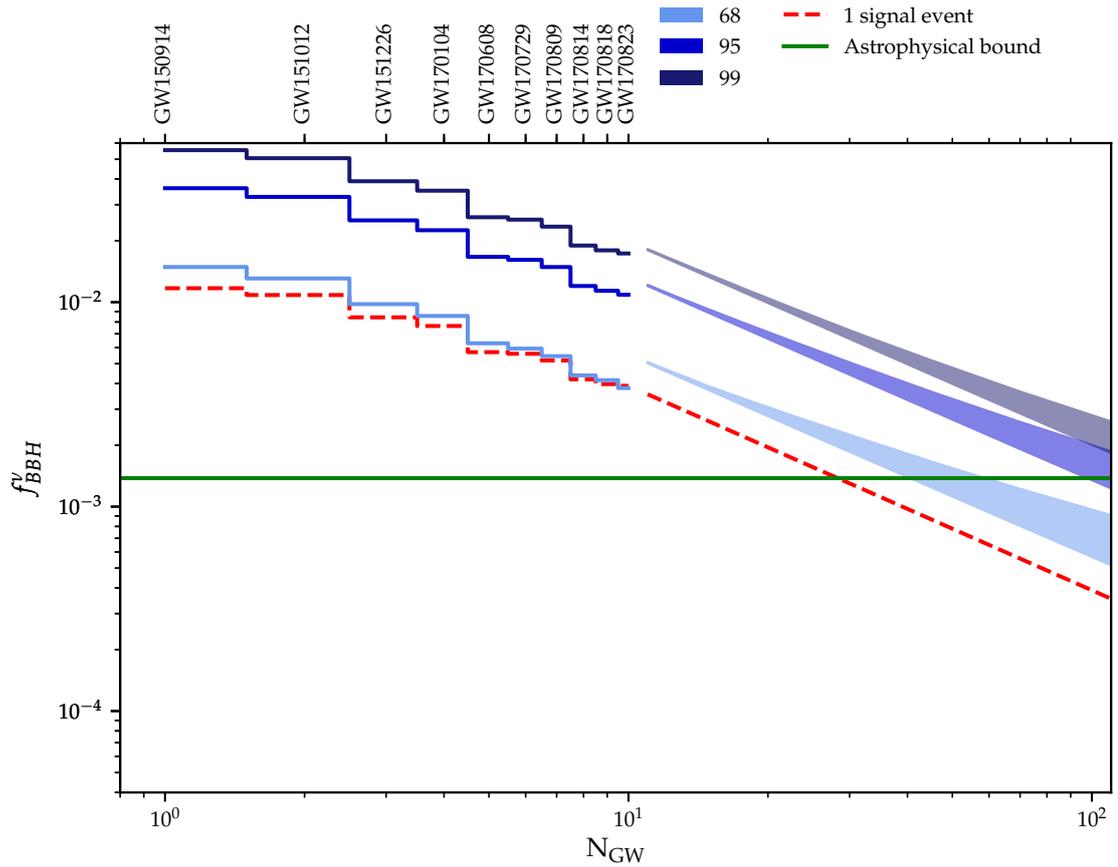

Figure 5.10: Updated prediction of the limit on $f^\nu_{\mathrm{BBH}}$ from future searches, using the average event from run O1 and O2.

## 5.9 Conclusion

With the detection of the binary black hole merger GW150914, gravitational wave astronomy has truly started and has been added as a component in multimessenger studies. No electromagnetic or neutrino emission is generally expected from binary black hole mergers, due to their environment being devoid of matter. However, given the interest in this event and a tentative detection of a faint gamma-ray burst consistent with the timing and direction of GW150914 (although this is likely to be a background fluctuation), several models have been constructed which can explain such emission. Searches for high-energy neutrino emission have been performed by IceCube and ANTARES, but no such emission has been found. In this work, we defined the neutrino emission fraction $f^\nu_{\mathrm{BBH}} = E^{\mathrm{tot}}_\nu / E_{\mathrm{GW}}$ and investigated, using only energetic model-independent arguments, the constraints on neutrino emission from BBH mergers and how these constraints are expected to evolve in the future. In particular, we considered both the constraints from the astrophysical neutrino flux (using the estimated merger rate from BBH mergers in run O1) and those from non-detection of coincident emission with GW150914. In this way, we can interpret the significance of the non-detection of neutrinos from GW150914. From the excellent fit to the expected signal from BBH mergers, $f^\nu_{\mathrm{BBH}}$ is expected to be small. Our work differs from [869], where the authors also investigate neutrino emission from BBH mergers within a GRB model and constrain the GRB parameter space, because our work does not refer to a specific model of neutrino emission (although we do provide a link to it).

The order of magnitude limits on the neutrino emission fraction $f^\nu_{\mathrm{BBH}}$ are summarised in Table 5.3. The limits on $f^\nu_{\mathrm{BBH}}$ obtained from the non-detection of counterpart neutrinos from GW150914 are similar to those obtained from the astrophysical flux assuming a isotropic emission. In case of beamed emission, the non-detection limit could be stronger than the astrophysical limit, depending on the beaming factor. This immediately implies that in case BBH mergers similar to GW150914 are responsible for the astrophysical neutrino flux, that the emission from GW150914 was not beamed towards Earth.

The same technique was also used to provide an estimate of the lowest $f^\nu_{\mathrm{BBH}}$ that can be probed in run O2 of LIGO, by assuming all events have the same properties as GW150914. It was found that after $N_{\mathrm{GW}} \gtrsim 4, 9, 11$ at 68%, 95% and 99% CL respectively, $f^\nu_{\mathrm{BBH}}$ can reach the highest astrophysical constraint. At the time, estimates for the number of BBH mergers in LIGO run O2 were between 10 and 35 events. The average upper limits that can be reached after these numbers of events are also shown in Table 5.3. Furthermore, it was shown how a possible detection in the near future provides direct information about the source evolution and BBH mass distribution, as well as the neutrino emission properties.

---

[63]A similar measurement not making use of the identification of the host galaxy, but rather considering each galaxy within the localisation region as a potential host exits. While the result is inferior, it still shows that such a method is also viable [895].

Table 5.3: Summary of the strongest bounds on $f_{BBH}^{\nu}$ (order of magnitude), assuming an $E^{-2}$-power law neutrino spectrum.

| | $f_{BBH}^{\nu}$ |
|---|---|
| GW150914 non-detection[1] | $10^{-2} \times \frac{\Delta\Omega}{4\pi}$ |
| Astrophysical flux (GW150914-like) | $10^{-3} - 10^{-2}$ |
| Astrophysical flux (All LIGO mergers) | $10^{-4} - 10^{-3}$ |
| Update: Astrophysical flux | $10^{-3}$ |
| Prospects ($N_{BBH} = 10$) at 68% CL | $10^{-3} \times \frac{\Delta\Omega}{4\pi}$ |
| Prospects ($N_{BBH} = 35$) at 68% CL | $10^{-4} - 10^{-3} \times \frac{\Delta\Omega}{4\pi}$ |
| Update: All BBH mergers in run O1 & O2 at 68% CL | $3 \times 10^{-3} \times \frac{\Delta\Omega}{4\pi}$ |
| Expectation: Fireball + dead acc. disk[2] | $10^{-6}$ |

[1] This bound is similar for the monochromatic case.

[2] This should be compared with the bounds on $f_{BBH}^{\nu}$ for the beamed case.

The analysis was updated to obtain the limit from all actually detected events in run O1 and run O2, assuming that no neutrinos are detected coincident with the mergers (also shown in Table 5.3). As expected, the actual limit after run O2 is worse than our prediction, which was optimistic. The limit on $f_{BBH}^{\nu}$ from the stacked search now reaches the original highest constraint from the astrophysical flux, but does not reach yet the updated constraint. An updated estimate for future detections was made using the average of the currently detected mergers, which reaches the updated astrophysical constraint within 100 BBH mergers. Therefore, currently it is not possible to exclude a significant contribution of BBH mergers to the astrophysical neutrino flux.

The results for a more model-dependent analysis were also presented. First, assuming the GRB-fireball model, the current and expected bounds on $f_{BBH}^{\nu}$ were used to put a bound on the amount of matter present in the BBH environment at the time of the merger. The results of this are presented in Table 5.4 (including the updated result after stacking all currently detected events, assuming all of them are beamed towards Earth). Second, the GRB-fireball model is combined with a model for a dead accretion disk around one of the black holes. The neutrino energy fraction expected in this situation, $f_{BBH}^{\nu} \approx 10^{-6}$, is below the reach of LIGO run O2. Finally, it should be noted that while for BBH mergers no neutrino emission is typically expected, realistic models of neutrino production can not be ruled out at the moment. In the future, it will be possible to use searches for neutrino emission to probe the black hole binary environment, independently from searches for gamma-ray emission. In addition, the same approach can be used for other source classes, such as neutron star-black hole and binary neutron star mergers[64], where one does expect neutrino emission.

---

[64]From typical GRB luminosities of $10^{49}$ erg (assuming it is similar for neutrinos) and the emission from

Table 5.4: Summary of the strongest bounds on $f_{\text{matter}}$.

|  | $f_{\text{matter}}$ |
|---|---|
| GW150914 non-detection | $4.0 \times 10^{-3} \times \frac{\Delta\Omega}{0.2 \times 0.2}$ |
| Prospects ($N_{\text{BBH}} = 10$) at 68% CL | $3.2 \times 10^{-4} \times \frac{\Delta\Omega}{0.2 \times 0.2}$ |
| Prospects ($N_{\text{BBH}} = 35$) at 68% CL | $1.6 \times 10^{-4} \times \frac{\Delta\Omega}{0.2 \times 0.2}$ |
| Update: All BBH mergers run O1 & O2 at 68% CL | $1.2 \times 10^{-3} \times \frac{\Delta\Omega}{0.2 \times 0.2}$ |
| Expectation: Dead acc. disk | $10^{-4}$–$10^{-3}$ |

Another avenue which was not pursued here is the potential to relate the diffuse astrophysical neutrino flux and the (thus far unobserved) gravitational wave background directly, using the same $f^{\nu}_{\text{BBH}}$. If the gravitational wave background is detected, a comparison of their respective energy density and their number of resolved sources could provide more information about this multimessenger connection. In Section 3.6.3, a naive estimate was given of the ratio between the energy density of gravitational waves and all cosmic rays. There, it was found that $f_{\text{GW/CR}} \lesssim 2.3 \times^{-4}$.

On April 1, 2018, LIGO and Virgo started their third observing run (O3), with an increased sensitivity (40% increase in sensitivity for LIGO, almost a factor of 2 for Virgo) and this run is expected to last for one year. During this period, many more gravitational wave events are expected to be detected (with alerts made public). Multimessenger studies will be able to quickly follow up on detected events (either triggered by gravitational waves, gamma rays or neutrinos). In particular, the time delay due to filtering and reconstruction of a neutrino event after the recorded trigger is 3-5s for ANTARES [897], 20-30s for IceCube [898] and about 1 minute for LIGO-Virgo. On the longer term, with the upgrades to IceCube and eventually IceCube-Gen2 [684] and the construction of KM3NeT [685] and of Baikal-GVD [686], neutrino emission from BBH mergers will be much better constrained (or detected).

---

GW170817 of $7 \times 10^{52}$ erg, one can estimate $f^{\text{BNS}}_{\nu} \sim 10^{-4}$.

# Conclusion and outlook

In this thesis, we studied two aspects of particle physics: beyond the Standard Model physics and high-energy neutrino astronomy. While these are very different, seemingly unrelated, topics, they both require the toolkit of particle physics to solve current issues.

The first part concerned the study of beyond the Standard Model physics. More specifically, we studied the phenomenology of supersymmetric model. Supersymmetry is a well-motivated model which can solve important issues from which the Standard Model suffers. Despite very strong constraints on its parameter space, it remains as one of the prime candidates for BSM physics. In particular, more exotic signatures of supersymmetry are now worth exploring.

In this, we discussed in detail a gauge-mediation model with multiple hidden sectors, the presence of which is natural in the context of UV models. While such a scenario represents only a small deviation from conventional models, it still features a distinct phenomenology, but remains very predictive. In the spectrum, there appears a massive pseudo-goldstino, with a coupling structure dictated by supersymmetry. This new degree of freedom allows such a model to evade constraints from which conventional models suffer. We applied a specific realisation of this goldstini model to interpret an ATLAS excess in leptons+jets+$\not{E}_T$, which was present in the $\sqrt{s} = 8$ TeV data. From our analysis, it followed that such a model was capable of explaining the excess, even in the face of very strong constraints from complementary jets+$\not{E}_T$ searches, which strongly constrain the presence of gluinos and squarks at low masses. However, with new data at $\sqrt{s} = 13$ Tev, the excess has disappeared and this specific implementation of the model is ruled out.

However, this does not mean that all was for naught. The specific features of this model remain. The presence of the pseudo-goldstino allows any implementation of this model to evade constraints from typical searches for supersymmetry, such as the jets+$\not{E}_T$ searches above. In practice, the focus of experimental analyses has shifted away from supersymmetry, or at least broadened their scope, towards more general exotic signatures. Still, the same model remains useful. For example, if more than one pseudo-goldstino is present, which occurs when there are more than two hidden sectors, the decay chain of particles to the real goldstino can be accompanied by the emission of soft particles, which can escape detection. In this way, such a scenario can be constrained by existing searches for compressed spectra, which are also accompanied by soft particles.

However, more important than this specific model, is the general toolset used in this study, which can be applied to any model. Given that there are no clear indications towards a specific theory, it is wise to remain agnostic about the exact form that new physics might take near the electroweak scale. Moreover, this toolset does not only concern numerical tools. Indeed, many of the alternative searches (such as dark matter searches and model developments) still use the language of supersymmetry or use supersymmetry as a benchmark. In this sense, supersymmetry is, at the very least, a useful framework to guide the exploration for new physics.

In the second part of the thesis, we discussed high-energy neutrino astronomy, in particular the search for the sources of high-energy neutrinos and their connection to other messengers such as gamma rays and gravitational waves. With the detection of neutrinos coincident with the flaring blazar TXS 0506+056, the first neutrino source has now been identified. However, the bulk of the source population remains unknown. Constraints from the gamma-ray background lead us to consider alternative models and maybe even abandon the connection with bright GeV sources altogether. Instead, it is possible that neutrinos are associated to sources bright in MeV gamma rays, X-ray, or, at the other side of the spectrum, TeV blazars. However, since neutrinos are always accompanied by gamma rays through their shared origin in hadronic collisions, we need to motivate the disappearance of GeV gamma rays from such sources. One possibility is that the gamma rays are attenuated by strong radiation fields present at the source.

In this thesis, we considered another possibility: that of attenuation by a dense column of gas, with column densities up to $N_H = 10^{26} \, \mathrm{cm^{-2}}$. In our model, the gamma rays are attenuated by pair production on the same obscuring gas that is responsible for the production of neutrinos in the first place through $pp$-interactions. First, we considered a set of blazars which was selected for indications of obscuration by matter in X-rays. We find that IceCube is not yet sensitive to the neutrino flux predicted by our model. However, the parameter which determines the proton luminosity in the source is very uncertain. Therefore, currently IceCube can already put relevant constraints on this parameter (shown in Figure 4.11).

In this sense, such an analysis is mainly interesting as an independent constraint on the proton luminosity in blazars or other possible neutrino sources: as soon as there is a sufficient amount of matter present close to the source, the system becomes a complete proton beam dump. This allows for a direct measurement of or constraint on the proton content of the source. While this is also possible in $p\gamma$-models, such a determination is more complicated, since it requires modelling the full spectral energy distribution.

We also investigated the diffuse neutrino flux from an obscured population in our model. We find that our model can alleviate constraints with the extragalactic gamma-ray background (Figure 4.14). On the other hand, the prediction is still close enough to the current limit on gamma rays that the model will be constrained in the near future. While increasing the column density would allow the tension to be reduced even further, the current value of $N_H = 10^{26}$ cm$^{-2}$ is already at limit of motivated column densities.

In the future, different avenues are worth pursuing. In our analysis, we identified Ultra-luminous infrared galaxies as interesting targets, due to their high dust content. A dedicated analysis for such objects by IceCube is currently ongoing. On the other hand, as a population these objects are on the edge of possessing the required luminosity to reproduce the astrophysical neutrino flux (although this depends again strongly on the proton luminosity normalisation). In face of this, an extension to luminous infrared galaxies can be interesting as well, since then the total population is increased. Moreover, we identified the object NGC 4418 as a highly obscured object (coincidentally a LIRG), which deserves further study. On the other hand, the same model might also be relevant for young AGN (at higher redshift) which are just turning on. In this case, a lot of matter can be present and provide for an initial boost of the neutrino flux. While we do not envision such a scenario to be responsible for the bulk of the astrophysical neutrino flux, it might provide a contribution free from gamma rays and has an interesting connection to cosmology. Finally, with the planned upgrades of IceCube and other neutrino observatories, observatories will become sensitive to fainter populations of numerous sources. With this, it will be possible to detect or exclude most reasonable models currently considered.

We also discussed neutrino astronomy in connection to gravitational waves. Gravitational waves are the final messenger to be included in the multimessenger picture. Interestingly, these probe the inner engine of their sources, providing a completely new look at the objects which can emit them. Here, we investigated neutrino emission from binary black hole mergers, in particular from the first such detected event: GW150914. While no neutrino emission is expected from BBH mergers, it is still necessary to test this hypothesis. We find that, currently, we can not rule out a large contribution of BBH mergers to the total astrophysical neutrino flux and also provide a limit on the amount of matter surrounding a BBH merger in the context of gamma-ray burst models (Table 5.4). However, in the near future (Figure 5.10), this will become possible. Moreover, with run O3 of LIGO/Virgo currently ongoing, many more BBH merger events will be detected. With this wealth of new information, the scenarios considered in our work can be investigated in more detail. In particular, a more complicated relation between the BBH merger properties and possible neutrino emission can then be explored.

# Summary

The Standard Model of particle physics describes all currently known particles and the interactions between them. Its success at describing the world around us within an elegant mathematical framework is both astonishing and mysterious. Even more perplexing is its ability to describe phenomena occuring at vastly different scales, from the smallest particles probed at collider experiments to the largest objects known in the universe. In this thesis, we use the common language of particle physics to study phenomena at both of these scales.

In the first part of the thesis, we study particle physics at particle accelerators, in particular the search for beyond the Standard Model physics. While the Standard Model is very successful, the small mass of the Higgs remains unexplained (and unnatural) and it provides no explanation for dark matter, the predominance of matter compared to antimatter or even the nature of the neutrino masses. One promising candidate for beyond the Standard Model physics is supersymmetry, which is the (only) natural extension of the spacetime symmetries we already know. For every fermion in the Standard Model, it predicts a partner boson, and vice versa. If supersymmetry exists, it must be broken at the electroweak scale. This breaking is, in a sense, also the source of most of its phenomenology.

While supersymmetry is well-motivated, current constraints from the Large Hadron Collider put pressure on conventional scenarios. Therefore, we are prompted to investigate more exotic signatures. In this thesis, we investigated a model of supersymmetry, where supersymmetry is broken in multiple hidden sectors. Such model predicts the appearance of a new massive particle at the bottom of the spectrum, which is weakly coupled to all other superpartners. We apply this model in the context of an excess seen by ATLAS in 2015 at $\sqrt{s} = 8$ TeV in events with two leptons, jets and missing transverse momentum. We find that our model provides a better fit to the signal than conventional models, while also suffering less from constraints due to complementary searches. However, an update to our analysis using new data at $\sqrt{s} = 13$ TeV rules out this model.

In the second part of the thesis, we study neutrino astronomy. While the presence of a high-energy neutrino flux has been firmly established in recent years by the IceCube Observatory, the sources of these neutrinos remain unknown. Recently, the first neutrino source has been identified: the flaring blazar TXS 0506+056. However, sources such as this one are unlikely to be responsible for the bulk of the observed diffuse neutrino flux. Moreover, possible neutrino sources are now strongly constrained by measurements of the extragalactic gamma-ray background, since both share a common origin in proton interactions. As a result, conventional sources like gamma-ray bursts and active galactic nuclei are now less favoured. Instead, neutrino sources are likely dim in GeV gamma rays. In this context, we investigate a model of neutrino sources which are obscured by dense gas. We apply our model first to a selection of objects which are possibly obscured. While IceCube is not currently sensitive to the flux predicted in our model, it is capable of putting relevant constraints on the parameter space of our model. Similarly, we also investigate whether a population of obscured neutrino sources can explain the diffuse neutrino flux without exceeding the measured extragalactic gamma-ray background. We find that our model indeed alleviates some of the tension. In this context, ultra-luminous infrared galaxies are a promising source category.

Finally, triggered by the recent first detection of a gravitational wave source, the binary black hole merger GW150914, we also investigate the neutrino-gravitational wave connection. In particular, we calculate the current constraints on neutrino emission from binary black hole mergers from both direct searches and measurements of the diffuse astrophysical neutrino flux. While currently such mergers are still allowed to supply the bulk of the astrophysical neutrino flux, this possibility will be constrained in the near future.

# Samenvatting

Het Standaardmodel van de deeltjesfysica beschrijft alle deeltjes die momenteel gekend zijn en hun interacties. De succesvolle beschrijving van de wereld rondom ons door het Standaardmodel in een elegant mathematisch kader is zowel verbazingwekkend als mysterieus. Nog meer opvallend is de vaststelling dat het Standaardmodel fenomenen kan beschrijven die op compleet verschillende schalen gebeuren; van de interacties tussen de kleinste deeltjes tot de grootste objecten in het universum. In deze thesis gebruiken we de gemeenschappelijke taal van de deeltjesfysica om fenomenen op beide schalen te bestuderen.

In het eerste deel van de thesis bestuderen we deeltjesfysica in deeltjesversneller, in het bijzonder de zoektocht naar fysica voorbij het Standaardmodel. Hoewel het Standaardmodel zeer succesvol is, kan het niet verklaren waarom de massa van het Higgs boson zo onnatuurlijk klein is en geeft het geen uitleg voor het bestaan van donkere materie, de dominantie van materie over antimaterie of de aard van neutrino massa's. Eén veelbelovende kandidaat voor fysica voorbij het Standaardmodel is supersymmetrie: de (enige) natuurlijke uitbreiding van de ruimtetijdsymmetrieën die we al kennen. Supersymmetrie introduceert een boson partner voor elk fermion in het Standaardmodel en vice versa. Als supersymmetrie bestaat, moet het gebroken zijn op de schaal van elektrozwakke interacties. Deze breking is, in zekere zin, verantwoordelijk voor het grootste deel van de supersymmetriefenomenologie.

Hoewel supersymmetrie een goed gemotiveerde theorie is, zetten huidige limieten van de Large Hadron Collider conventionele modellen sterk onder druk. Als gevolg hiervan zijn we gedwongen om meer exotische signalen te onderzoeken. In deze thesis onderzoeken we een supersymmetrisch model waar supersymmetrie gebroken is in meerdere verborgen sectoren. Dit model voorspelt het bestaan van een nieuw massief deeltje onderaan het spectrum, met zwakke interacties met de andere superpartners. We passen dit model toe in de context van een afwijking in de ATLAS data van 2015 met $\sqrt{s} = 8$ TeV in evenementen met twee leptonen, jets en missend transversaal momentum. Na onderzoek concluderen we dat ons model het signaal beter kan beschrijven dan conventionele modellen en tegelijk de beperkingen van complementaire analyses kan ontwijken. Na een update van onze analyse met meer recente data met $\sqrt{s} = 13$ TeV vinden we echter dat ons model nu uitgesloten is.

In het tweede deel van de thesis bestuderen we neutrino-astronomie. Hoewel het bestaan van een hoge-energie neutrino flux nu vaststaat door observaties van IceCube, zijn de bronnen van deze flux nog steeds onbekend. Recent werd de eerste neutrinobron geïdentificeerd: the flarende blazar TXS 0506+056. Het is echter onwaarschijnlijk dat dit type bron de meerderheid van de diffuse neutrino flux kan verklaren. Anderzijds leggen metingen van de extragalactische achtergrond van gammastralen sterke beperkingen op mogelijke neutrinobronnen op, aangezien zowel neutrinos als gammastralen het gevolg zijn van protoninteracties. Als gevolg zijn conventionele kandidaten zoals gamma-ray bursts en kernen van actieve galaxieën nu minder waarschijnlijk als neutrinobronnen. In plaats daarvan is het waarschijnlijk dat neutrinobronnen zwak zijn in GeV gammastralen. In deze context onderzoeken we een model van neutrinobronnen die geöbscureerd zijn door een dens gas. We passsen ons model eerst toe op een set objecten die geselecteerd zijn voor mogelijke obscuratie. We vinden dat IceCube momenteel niet de gevoeligheid heeft om de neutrinoflux in ons model te detecteren, maar wel reeds relevante limieten kan zetten op de parameterruimte van ons model. We onderzoeken ook of een populatie van geöbscureerde bronnen de diffuse neutrinoflux kan verklaren zonder de gemeten extragalactische achtergrond van gammastralen te overschrijden. Ons model kan inderdaad de spanning tussen deze twee fluxen verminderen. In deze context vormen ultralumineuze infraroodgalaxiën een belovende kandidaat voor de neutrinobronnen.

Tenslotte bekijken we ook de link tussen neutrino- en zwaartekrachtgolfemissie, geïnspireerd door de eerste detectie van een bron van zwaartekrachtgolven, GW150914, waarbij twee zwarte gaten zijn samengesmolten. We berekenen de huidige limieten op neutrino-emissie uit samensmeltingen van zwarte gaten, gebruik makend van zowel directe metingen als de meting van de astrofysische neutrino flux. Momenteel kunnen zo'n samensmeltingen nog steeds de meerderheid van de neutrinoflux verklaren, maar in de nabije toekomst zullen experimenten de gevoeligheid hebben om dit uit te sluiten.

# Contributions

In this part, I will highlight for each chapter, excluding the introductory chapters, my own contributions.

**Chapter 2: *Z*-peaked excess**

My own work starts in Section 2.2 and encompasses the rest of the chapter, published in [277] and updated in Section 2.5 for this thesis. My personal contribution was mainly focussed on the modelling part of sections 2.2 and 2.3 as well as the update in Section 2.5. Various aspects of the model of multiple SUSY-breaking sectors discussed in Section 2.2 can be found in earlier literature in several works (cited at the relevant places). However, we have derived here in a consistent way the specific couplings necessary for our scenario and, in particular, included a high mass for the pseudo-goldstino and investigated its phenomenological consequences.

**Chapter 4: Obscured neutrino sources**

Except for the small parts where I explicitly discuss existing literature by different authors or introduce the theoretical basis, this entire chapter is original work. However, in this text, there is included a previous publication performing an object selection of possibly obscured flat-spectrum radio AGN, published in [687] of which I am co-author but not the main author (contained to Sections 4.3.1 and 4.3.2).

My most important personal work concerns the rest of this chapter: the model development in Sections 4.1 and 4.2, the calculation of the neutrino flux from obscured flat-spectrum radio AGN in Sections 4.3.3 and 4.3.4, and the investigation of the neutrino and gamma-ray flux from an obscured population in Section 4.4.

**Chapter 5: Neutrinos from BBH mergers**

The original work starts in Section 5.3 and encompasses the rest of the chapter, published in [803] and updated in Section 5.8 for this thesis. I have made a significant contribution to all aspects of this research.

# Appendix

## Conventions and useful identities

In this appendix, we define the conventions used throughout the main text and list some of the properties and relations of spinors.

## A.1  Metric signature

The contravariant Lorentz four-vectors for position and momentum are given by

$$x^\mu = (t, \mathbf{x}), \qquad p^\mu = (E, \mathbf{p}), \tag{A.1}$$

while the four-vector derivative is given by

$$\partial_\mu = (\partial/\partial t, \nabla). \tag{A.2}$$

The signature for the metric is an arbitrary choice. Here, we choose the mostly-negative metric

$$\eta_{\mu\nu} = \text{diag}(+1, -1 - 1, -1), \tag{A.3}$$

with

$$\eta_{\mu\nu}\eta^{\nu\rho} = \delta_\mu{}^\rho. \tag{A.4}$$

Using the metric, we can raise and lower indices as

$$x_\mu = \eta_{\mu\nu}x^\nu, \qquad \partial^\mu = \eta^{\mu\nu}\partial_\nu. \tag{A.5}$$

Contracting the indices, we can define the norm of a four-vector as

$$x^2 \equiv x^\mu x_\mu = \eta_{\mu\nu}x^\mu x^\nu. \tag{A.6}$$

With this signature of the metric, the relation between the norm of the four-momentum and the mass is $p^2 = m^2$.

## A.2    Poincaré algebra

The Poincaré algebra is given by

$$\big[P_\mu, P_\nu\big] = 0, \tag{A.7}$$

$$\big[M_{\mu\nu}, M_{\rho\sigma}\big] = -i(\eta_{\mu\rho}M_{\nu\sigma} + \eta_{\nu\sigma}M_{\mu\rho} - \eta_{\mu\sigma}M_{\nu\rho} - \eta_{\nu\rho}M_{\mu\sigma}), \tag{A.8}$$

$$\big[M_{\mu\nu}, P_\rho\big] = -i(\eta_{\mu\rho}P_\nu - \eta_{\nu\rho}P_\mu), \tag{A.9}$$

where $P_\mu$ generates translations in space-time and $M_{\mu\nu}$ Lorentz rotations.

The generators $M_{\mu\nu}$, which form the $SO(1,3)$ Lorentz algebra, group together rotations in space, generated by $J_i = \frac{1}{2}\epsilon_{ikl}M^{kl}$ or $M_{ij} = \epsilon_{ijk}J_k$, and boosts, given by $K_i = M_{i0}$. These satisfy additionally

$$\big[J_i, J_j\big] = i\epsilon_{ijk}J_k, \tag{A.10}$$

$$\big[J_i, K_j\big] = i\epsilon_{ijk}K_k, \tag{A.11}$$

$$\big[K_i, K_j\big] = -i\epsilon_{ijk}J_k. \tag{A.12}$$

The first of these is the algebra of $SU(2)$, while the three together are a rewriting of the commutation relations of $SO(1,3)$.

## A.3    Pauli matrices

The Pauli matrices, which form a representation of the $SU(2)$ algebra and are unitary and hermitian, are given by

$$\sigma^0 = \overline{\sigma}^0 = \begin{pmatrix} 1 & 0 \\ 0 & 1 \end{pmatrix}, \qquad \sigma^1 = -\overline{\sigma}^1 = \begin{pmatrix} 0 & 1 \\ 1 & 0 \end{pmatrix},$$

$$\sigma^2 = -\overline{\sigma}^2 = \begin{pmatrix} 0 & -i \\ i & 0 \end{pmatrix}, \qquad \sigma^3 = -\overline{\sigma}^3 = \begin{pmatrix} 1 & 0 \\ 0 & -1 \end{pmatrix}, \tag{A.13}$$

such that $\sigma^\mu = (1, \sigma^i)$ and $\overline{\sigma}^\mu = (1, -\sigma^i)$.

In addition, we also define

$$\sigma^{\mu\nu} = \frac{i}{4}(\sigma^\mu\overline{\sigma}^\nu - \sigma^\nu\overline{\sigma}^\mu), \qquad \overline{\sigma}^{\mu\nu} = \frac{i}{4}(\overline{\sigma}^\mu\sigma^\nu - \overline{\sigma}^\nu\sigma^\mu). \tag{A.14}$$

## A.4   Gell-Mann matrices

One possible representation of the generators for $SU(3)$ transformations is given by the Gell-Mann matrices. They are traceless and hermitian and are given by

$$\lambda_1 = \begin{pmatrix} 0 & 1 & 0 \\ 1 & 0 & 0 \\ 0 & 0 & 0 \end{pmatrix}, \quad \lambda_2 = \begin{pmatrix} 0 & -i & 0 \\ i & 0 & 0 \\ 0 & 0 & 0 \end{pmatrix}, \quad \lambda_3 = \begin{pmatrix} 1 & 0 & 0 \\ 0 & -1 & 0 \\ 0 & 0 & 0 \end{pmatrix}, \tag{A.15}$$

$$\lambda_4 = \begin{pmatrix} 0 & 0 & 1 \\ 0 & 0 & 0 \\ 1 & 0 & 0 \end{pmatrix}, \quad \lambda_5 = \begin{pmatrix} 0 & 0 & -i \\ 0 & 0 & 0 \\ i & 0 & 0 \end{pmatrix}, \tag{A.16}$$

$$\lambda_6 = \begin{pmatrix} 0 & 0 & 0 \\ 0 & 0 & 1 \\ 0 & 1 & 0 \end{pmatrix}, \quad \lambda_7 = \begin{pmatrix} 0 & 0 & 0 \\ 0 & 0 & -i \\ 0 & i & 0 \end{pmatrix}, \quad \lambda_8 = \frac{1}{\sqrt{3}} \begin{pmatrix} 1 & 0 & 0 \\ 0 & 1 & 0 \\ 0 & 0 & -2 \end{pmatrix}. \tag{A.17}$$

## A.5   Spinors

In this thesis, we use two-component Weyl spinors to represent the Standard Model and supersymmetry fermions. This is a more natural convention for the Standard Model and its extensions compared to four-component Dirac or Majorana spinors, since it treats left-handed and right-handed particles, which transform differently under the Standard Model gauge groups, separately from the start. Moreover, Weyl spinors appear as components of chiral superfields.

The Lorentz group $SO(3,1)$ can be rewritten[1] as $SU(2) \times SU(2)^*$, i.e. representations of the Lorentz group can be classified by two $SU(2)$ spins. The simplest (non-scalar) representations are then the left-handed and right-handed two-component Weyl spinors $\psi_\alpha$ and $\psi_{\dot{\alpha}}^\dagger$, which are complex and anti-commuting.

The hermitian conjugate, indicated in our convention with a dagger, of a left-handed Weyl spinor is a right-handed Weyl spinor

$$\psi_{\dot{\alpha}}^\dagger \equiv (\psi_\alpha)^\dagger = (\psi^\dagger)_{\dot{\alpha}}. \tag{A.18}$$

By convention, fields are named such that left-handed spinors carry no †, while right-handed spinors do (see Section 1.3). An alternative notation for the hermitian conjugate uses barred spinors[2] and was popularised by [86]. The equivalence between these notations is given by $\overline{\psi}_{\dot{\alpha}} \equiv \psi_{\dot{\alpha}}^\dagger \equiv (\psi_\alpha)^\dagger$.

---

[1] Simplified, the generators of rotations $J_i$ and of boosts $K_i$ can be combined as $J_i^\pm = \frac{1}{2}(J_i \pm iK_i)$ which satisfy the same commutation relations as those of angular momentum/spin $\left[ J_i^\pm, J_j^\mp \right] = i\epsilon_{ijk} J_k^\pm$, while the commutator with opposite $\pm$ vanishes.

[2] This bar should not be confused with that used for Dirac spinors (see Eq. (A.28) below), which is more than just a conjugation, or the barred notation in Section 1.3 used for conjugates of the right-handed part of a Dirac spinor.

Multiplication of spinors and the raising/lowering of indices are performed with the antisymmetric symbol $\epsilon^{\alpha\beta}$, with $\epsilon^{12} = -\epsilon^{21} = \epsilon_{21} = -\epsilon_{12} = 1$. Note that by convention, we contract indices like

$$\phantom{}^{\alpha}_{\phantom{\alpha}\alpha} \quad \text{and} \quad \phantom{}_{\dot{\alpha}}^{\phantom{\dot{\alpha}}\dot{\alpha}}. \tag{A.19}$$

We thus have (since spinors are anti-commuting, re-ordering them induces a minus sign)

$$\chi\xi = \chi^{\alpha}\xi_{\alpha} = \chi^{\alpha}\epsilon_{\alpha\beta}\xi^{\beta} = -\xi^{\beta}\epsilon_{\alpha\beta}\chi^{\alpha} = \xi^{\beta}\epsilon_{\beta\alpha}\chi^{\alpha} = \xi^{\beta}\chi_{\beta} = \xi\chi. \tag{A.20}$$

In contractions of left-handed with right-handed Weyl spinors, there appear Pauli matrices $(\sigma^{\mu})_{\alpha\dot{\alpha}}$ and $(\overline{\sigma}^{\mu})^{\dot{\alpha}\alpha}$ to make vectors.

Using the above, we have some identities for spinors and Pauli matrices which are useful. Here we gather only those needed to understand the equations in this thesis. First, we have

$$\xi\chi = \chi\xi, \tag{A.21}$$

since both the spinors and $\epsilon^{\alpha\beta}$ are anti-symmetric. Note that while spinors are anti-symmetric, we still have $\xi\xi = \epsilon^{\alpha\beta}\xi_{\alpha}\xi_{\beta} = 2\xi_{2}\xi_{1} \neq 0$. Similarly, we find

$$\xi^{\dagger}\overline{\sigma}^{\mu}\chi = -\chi\sigma^{\mu}\xi^{\dagger}, \tag{A.22}$$

and

$$\xi\sigma^{\mu}\overline{\sigma}^{\nu}\chi = \chi\sigma^{\nu}\overline{\sigma}^{\mu}\xi. \tag{A.23}$$

Finally, we have the reduction identities

$$[\sigma^{\mu}\overline{\sigma}^{\nu} + \sigma^{\nu}\overline{\sigma}^{\mu}]_{\alpha}^{\beta} = 2\eta^{\mu\nu}\delta_{\alpha}^{\beta}, \tag{A.24}$$

$$\sigma^{\mu}\overline{\sigma}^{\nu}\sigma^{\rho} = \eta^{\mu\nu}\sigma^{\rho} + \eta^{\nu\rho}\sigma^{\mu} - \eta^{\mu\rho}\sigma^{\nu} + i\epsilon^{\mu\nu\rho\kappa}\sigma_{\kappa}. \tag{A.25}$$

More identities can be found in [67, 88].

Two-component notation is related to four-component notation as follows. Instead of Pauli matrices, $\gamma^{\mu}$-matrices appear. Consider the representation where

$$\gamma^{\mu} = \begin{pmatrix} 0 & \sigma^{\mu} \\ \overline{\sigma}^{\mu} & 0 \end{pmatrix}. \tag{A.26}$$

In this case, instead of treating left-handed and right-handed objects separately, projection operators $P_{L,R} = (1 \pm \gamma_{5})/2$ appear which select the left-handed and right-handed parts of a 4-component spinor to build the Lagrangian[3]. A four-component Dirac spinor can then be decomposed into Weyl spinors as

$$\Psi_{D} = \begin{pmatrix} \xi_{\alpha} \\ \chi^{\dagger\dot{\alpha}} \end{pmatrix} \tag{A.27}$$

and

$$\overline{\Psi}_{D} = \Psi_{D}^{\dagger} \begin{pmatrix} 0 & 1 \\ 1 & 0 \end{pmatrix} = (\chi^{\alpha} \, \xi_{\dot{\alpha}}^{\dagger}). \tag{A.28}$$

---

[3]Giving rise to the familiar $V - A$ interactions of the Standard Model.

Projection operators then project the top or bottom component out of these.

For completeness, we also give the free-field Lagrangian in terms of Dirac spinors

$$\mathcal{L}_{\text{Dirac}} = i\overline{\Psi}_D \gamma^\mu \partial_\mu \Psi_D - M\overline{\Psi}_D \psi_D, \tag{A.29}$$

which is equivalent to the two-component notation.

$$\mathcal{L}_{\text{Dirac}} = i\xi^\dagger \overline{\sigma}^\mu \partial_\mu \xi + \chi^\dagger \overline{\sigma}^\mu \partial_\mu \chi - M(\xi\chi + \xi^\dagger \chi^\dagger). \tag{A.30}$$

## A.6   Fermionic superspace coordinates

The superspace coordinates $\theta^\alpha$ and $\theta^\dagger_{\dot\alpha}$ have been defined in Section 1.2.2. For integration over superspace, we define

$$\int \mathrm{d}^2\theta = -\frac{1}{4}\,\mathrm{d}\theta^\alpha\,\mathrm{d}\theta^\beta \epsilon_{\alpha\beta}, \qquad \int \mathrm{d}^2\theta^\dagger = -\frac{1}{4}\,\mathrm{d}\theta^\dagger_{\dot\alpha}\,\mathrm{d}\theta^\dagger_{\dot\beta}\epsilon^{\dot\alpha\dot\beta}. \tag{A.31}$$

With these, we find

$$\int \mathrm{d}^2\theta\,\theta\theta = 1, \qquad \int \mathrm{d}^2\theta^\dagger\theta^\dagger\theta^\dagger = 1, \tag{A.32}$$

while integration with a different number of $\theta$ gives zero.

Some useful identities include

$$\theta_\alpha\theta_\beta - \frac{1}{2}\epsilon_{\alpha\beta}\theta\theta, \qquad \theta^\dagger_{\dot\alpha}\theta^\dagger_{\dot\beta} = \frac{1}{2}\epsilon_{\dot\alpha\dot\beta}\theta^\dagger\theta^\dagger, \qquad \theta_\alpha\theta^\dagger_{\dot\beta} = \frac{1}{2}\sigma^\mu_{\alpha\dot\beta}(\theta\sigma_\mu\theta^\dagger), \tag{A.33}$$

and

$$(\theta\xi)(\theta\chi) = -\frac{1}{2}(\theta\theta)(\xi\chi), \qquad\qquad (\theta^\dagger\xi^\dagger)(\theta^\dagger\chi^\dagger) = -\frac{1}{2}(\theta^\dagger\theta^\dagger)(\xi^\dagger\chi^\dagger), \tag{A.34}$$

$$(\theta\xi)(\theta^\dagger\chi^\dagger) = \frac{1}{2}(\theta\sigma^\mu\theta^\dagger)(\xi\sigma_\mu\chi^\dagger), \tag{A.35}$$

$$\theta^\dagger\overline{\sigma}^\mu\theta = -\theta\sigma^\mu\theta^\dagger = (\theta^\dagger\overline{\sigma}^\mu\theta)^*, \tag{A.36}$$

$$\theta\sigma^\mu\overline{\sigma}^\nu\theta = \eta^{\mu\nu}\theta\theta, \qquad\qquad \theta^\dagger\overline{\sigma}^\mu\sigma^\nu\theta^\dagger = \eta^{\mu\nu}\theta^\dagger\theta^\dagger. \tag{A.37}$$

## A.7   Switching between metric signatures

While switching between the mostly-negative and the mostly-positive metric in Lorentz space is simple in principle, it can lead to confusion. A good reference on how to do this systematically can be found in Appendix A of [88]. The basic idea is that typically, the following quantities are defined independently of the metric

$$x^\mu,\ p^\mu,\ \partial_\mu,\ \sigma^\mu,\ \overline{\sigma}^\mu,\ J^\mu,\ A^\mu,\ D_\mu,\ F^\mu_\nu,\ \delta^\mu_\nu,\ \epsilon^{\mu\nu\rho\sigma},\ \epsilon_{\mu\nu\rho\sigma}. \tag{A.38}$$

On the other hand, the next set of quantities is defined using the metric and thus change sign when switching convention

$$x_\mu,\ p_\mu,\ \partial^\mu,\ \sigma_\mu,\ \overline{\sigma}_\mu,\ J_\mu,\ A_\mu,\ D^\mu,\ F^{\mu\nu},\ F_{\mu\nu},\ \eta_{\mu\nu},\ \eta^{\mu\nu}. \tag{A.39}$$

Sometimes, this sign change is absorbed by switching from $\theta\sigma^\mu\theta^\dagger$ to $\theta^\dagger\overline{\sigma}^\mu\theta$ and $\theta\sigma_\mu\theta^\dagger$ to $\theta^\dagger\overline{\sigma}^\mu\theta$.

Alternatively, the LaTeX sources of [67] and [88] have a switch to choose between the two signatures, serving as useful references.

## A supersymmetry cheat sheet

In this appendix, we gather the most important elements of the introduction to supersymmetry in Section 1.2, to make clear what are the main ingredients of a supersymmetric theory or to be used as a reference.

Supersymmetry is the unique extension of the usual space-time symmetries. In a supersymmetric theory, each fermion is related to a corresponding boson, which carries the same gauge quantum numbers. These particles can be gathered in supermultiplets or, in a geometric picture, in superfields. Using superfields, which are functions of $x^\mu$, $\theta$ and $\theta^\dagger$ where the latter two are spinor coordinates, one can build Lagrangians which are manifestly invariant under supersymmetry.

A chiral superfield contains a left-handed two-component Weyl fermion $\psi$ and its scalar partner $\phi$. Therefore, the Standard Model fermions and Higgs boson will reside in such multiplets. The chiral superfield can be expressed in a simplified way using the coordinate $y^\mu = x^\mu - i\theta\sigma^\mu\theta^\dagger$ and $\theta$ (the full expression is given in Appendix C.2.2), leading to the following expression

$$\Phi = \phi(y) + \sqrt{2}\theta\psi(y) + \theta\theta F(y), \tag{B.1}$$

$$\Phi^* = \phi^*(y^*) + \sqrt{2}\theta^\dagger\psi^\dagger(y^*) + \theta^\dagger\theta^\dagger F^*(y^*), \tag{B.2}$$

where the second expression is an anti-chiral superfield which contains a right-handed Weyl fermion.

The real (or vector) superfield contains a real vector $A_\mu$ and a two-component Weyl fermion $\lambda$. Therefore, the Standard Model gauge bosons will reside in such a multiplet. As with gauge bosons, the vector superfields is determined up to supergauge transformations. In Wess-Zumino gauge, the real superfield for a non-Abelian gauge field (with evident reduction to the Abelian case) can be expressed in components as

$$(V^a)_{\text{WZ gauge}} = \theta\sigma^\mu\theta^\dagger A_\mu^a + \theta^\dagger\theta^\dagger\theta\lambda^a + \theta\theta\theta^\dagger\lambda^{\dagger a} + \frac{1}{2}\theta\theta\theta^\dagger\theta^\dagger D^a, \tag{B.3}$$

which still leaves the possibility for ordinary gauge transformations. Furthermore, one can build a field strength superfield out of this to construct Lagrangians, with $\mathcal{W}_\alpha = -\frac{1}{4}\overline{D}\overline{D}D_\alpha V$ (and similar for its hermitian conjugate). This field strength superfield is a chiral superfield and can be decomposed as

$$(\mathcal{W}_\alpha^a)_{\text{WZ gauge}} = \lambda_\alpha^a + \theta_\alpha D^a - \frac{i}{2}(\sigma^\mu\overline{\sigma}^\nu\theta)_\alpha F_{\mu\nu}^a + i\theta\theta(\sigma^\mu D_\mu\lambda^{\dagger a})_\alpha. \tag{B.4}$$

where the usual field strength and covariant derivative appears.

In addition to the expected components $\phi_i$, $\psi_i$, $\lambda^a$ and $A_\mu^a$, there also appear auxiliary fields $F_i$ and $D^a$ in the expressions for both $\phi_i$ and $V^a/\mathcal{W}_\alpha$. These latter fields, which always appear as the highest component of a superfield, are only necessary off-shell: using the equations of motion, these can always be integrated out exactly, giving rise to interaction terms.

Generally, the highest components of superfields, the $F$-component of chiral superfields and $D$-component of real superfields, can be used to build Lagrangians which are invariant (or give a total derivative) under supersymmetry transformations. Since the $\theta$ and $\theta^\dagger$ are anti-commuting variables, selecting the $F$-component of a chiral superfield is equivalent to

$$[\Phi]_F = \int d^2\theta\, \Phi = F, \tag{B.5}$$

while the $D$-component of a real superfield $V$ can be selected with

$$[V]_D = \int d^2\theta\, d^2\theta^\dagger\, V = \frac{1}{2}D + \dots, \tag{B.6}$$

where the ellipsis denotes a total derivative which is irrelevant.

It is possible to build new build chiral superfields out of other chiral superfields through multiplication, e.g. $\Phi_i\Phi_j$, while real superfields can be built out of combinations like $\Phi^{i*}\Phi_i$. Using all these ingredients, the most general Lagrangian containing only renormalisable terms is given by (this is the form of Eq. (1.87), where the gauge coupling has been absorbed into the real superfield, but we have dropped the hats)

$$\mathcal{L} = \frac{1}{4}\left[\tau_a\mathcal{W}^{a\alpha}\mathcal{W}_\alpha^a\right]_F + \text{c.c.} + \left[\Phi^{*i}(e^{2T^aV^a})_i{}^j\Phi_j\right]_D + \left([W(\Phi_i)]_F + \text{c.c.}\right). \tag{B.7}$$

where

$$\tau_a = \frac{1}{g_a^2} - i\frac{\Theta_a}{8\pi^2}. \tag{B.8}$$

The gauge kinetic term (ignoring $\Theta_a$) is given by

$$\mathcal{L}_{\text{gauge}} = \frac{1}{4g_a^2}\left[\mathcal{W}^{a\alpha}\mathcal{W}_\alpha^a\right]_F + \text{c.c.} = -\frac{1}{4g_a^2}F_{\mu\nu}^a F^{\mu\nu a} + \frac{1}{g_a^2}i\lambda^{\dagger a}\overline{\sigma}^\mu D_\mu\lambda^a + \frac{1}{2g_a^2}D^a D^a, \tag{B.9}$$

and the kinetic term for chiral superfields by

$$\begin{aligned}\left[\Phi^{*i}(e^V)_i{}^j\Phi_j\right]_D = {}&F^{*i}F_i + D_\mu\phi^{*i}D^\mu\phi_i + i\psi^{\dagger i}\overline{\sigma}^\mu D_\mu\psi_i \\ &- \sqrt{2}(\phi^*T^a\psi)\lambda^a - \sqrt{2}\lambda^\dagger(\psi^\dagger T^a\phi) + (\phi^*T^a\phi)D^a.\end{aligned} \tag{B.10}$$

In this Lagrangian, there appears the superpotential $W(\Phi_i)$, which is a chiral superfield built out of other chiral superfields and contains mass terms and Yukawa interactions.

$$W(\Phi_i) = L^i \Phi_i + \frac{1}{2} M^{ij} \Phi_i \Phi_j + \frac{1}{6} y^{ijk} \Phi_i \Phi_j \Phi_k. \tag{B.11}$$

Its $F$-component is given by

$$[W(\Phi_i)]_F = W^i F_i - \frac{1}{2} W^{ij} \psi_i \psi_j, \tag{B.12}$$

where $W^i = \frac{\delta W}{\delta \Phi_i}\big|_{\Phi_i \to \phi_i}$ and $W^{ij} = \frac{\delta^2 W}{\delta \Phi_i \delta \Phi_j}\big|_{\Phi_i \to \phi_i}$.

Including the interaction between the chiral and real superfield, the supersymmetry transformations for the components of a chiral superfield are

$$\delta \phi_i = \epsilon \psi_i, \tag{B.13}$$

$$\delta(\psi_i)_\alpha = -i(\sigma^\mu \epsilon^\dagger)_\alpha D_\mu \phi^i + \epsilon_\alpha F_i, \tag{B.14}$$

$$\delta F_i = -i \epsilon^\dagger \overline{\sigma}^\mu D_\mu \psi_i + \sqrt{2}(T^a \phi)_i \epsilon^\dagger \lambda^{\dagger a}, \tag{B.15}$$

where also components of the real superfield appear through the covariant derivatives and a part with the gaugino for the $F$-component.

From the $F$- and $D$-terms, we can find a scalar potential[1]

$$V(\phi, \phi^*) = F^{*i} F_i + \frac{1}{2g_a^2} \sum_a D^a D^a = W_i^* W^i + \frac{1}{2} \sum_a g_a^2 (\phi^* T^a \phi)^2, \tag{B.16}$$

which is important for supersymmetry breaking. The full expression (including only renormalisable terms) is given in Eq. (C.36).

Since the Lagrangian constructed above is invariant under supersymmetry, there is also an associated supercurrent

$$J_\alpha^\mu = (\sigma^\nu \overline{\sigma}^\mu \psi_i)_\alpha D_\nu \phi^{*i} + i(\sigma^\mu \psi^{\dagger i})_\alpha W_i^*$$
$$+ \frac{1}{2\sqrt{2}} (\sigma^\nu \overline{\sigma}^\rho \sigma^\mu \lambda^{\dagger a})_\alpha F_{\nu\rho}^a + \frac{i}{\sqrt{2}} g_a \phi^* T^a \phi (\sigma^\mu \lambda^{\dagger a})_\alpha. \tag{B.17}$$

Finally, including non-renormalisable terms is achieved by turning the couplings appearing in front of each term into arbitrary functions of the fields

$$\mathcal{L} = \left[ K(\Phi_i, (\Phi^* e^V)^j) \right]_D + \left( \left[ \frac{1}{4} f_{ab}(\Phi_i) \hat{\mathcal{W}}^{a\alpha} \hat{\mathcal{W}}_\alpha^b + W(\Phi_i) \right]_F + \text{c.c.} \right). \tag{B.18}$$

---

[1] Note that the $g_a^2$ appears correctly again, coming now from the kinetic term of the $D^a$-field, instead of the final term in Eq. (B.10) where it was originally (Eq. (1.75)).

# Supersymmetry

In this chapter, we provide, for completeness, some additional definitions and equations which complement the discussion in Chapter 1. While not necessary to follow the discussion, they may help elucidate some of the steps.

## C.1  Hierarchy problem

Here, we calculate more explicitly the corrections to the Higgs mass, showing explicitly how the quadratic divergence is cancelled by the appearance of scalars, following the discussion in [899].

Consider a $N_f$ fermions with Yukawa couplings $\lambda_f = \sqrt{2} m_f / v$. Neglecting the external Higgs momentum, the correction to the squared Higgs mass parameter, due to the loop in Figure C.1a, is

$$\Delta m_h^2 = N_f \frac{\lambda_f}{8\pi^2} \left[ -\Lambda_{\mathrm{UV}}^2 + 6 m_f^2 \log \frac{\Lambda_{\mathrm{UV}}}{m_f} - 2 m_f^2 \right] + \mathcal{O}\left( \frac{1}{\Lambda_{\mathrm{UV}}^2} \right) \tag{C.1}$$

Let us add now $N_s$ scalars $\phi$, with masses $m_s$ and trilinear and quadrilinear couplings to the Higgs boson $v\lambda_s$ and $\lambda_s$. There are now two diagrams[1] contributing to the Higgs squared-mass parameter, shown in Figures C.1b and C.1b. Their contributions are given by

$$\Delta m_h^2 = \frac{\lambda_s N_s}{16\pi^2} \left[ -\Lambda_{\mathrm{UV}}^2 + 2 m_s^2 \log \left( \frac{\Lambda_{\mathrm{UV}}}{m_s} \right) \right]$$
$$- \frac{\lambda_s N_s}{16\pi^2} v^2 \left[ -1 + 2 \log \left( \frac{\Lambda_{\mathrm{UV}}}{m_s} \right) \right] + \mathcal{O}\left( \frac{1}{\Lambda_{\mathrm{UV}}^2} \right). \tag{C.2}$$

---

[1] Notice that we previously only included the loop using the quartic coupling, since this is sufficient to cancel the quadratic divergence.

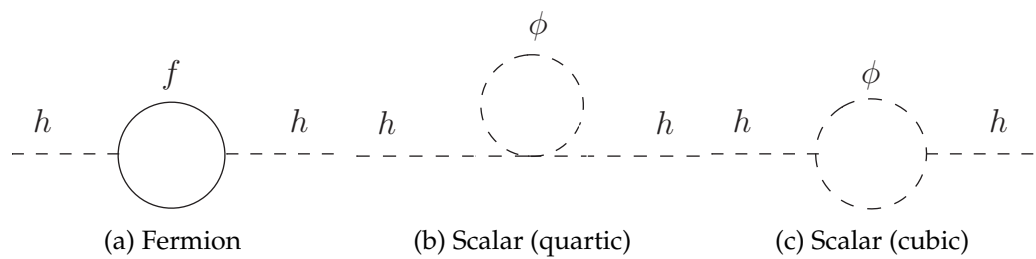

Figure C.1: Corrections to the Higgs mass due to fermions and scalars.

If we relate the coupling constants as $\lambda_f^2 = 2m_f^2/v^2 = -\lambda_s$ and take $N_s = 2N_f$, then the total correction to the Higgs squared-mass parameter is

$$\Delta m_h^2 = \frac{\lambda_f^2 N_f}{4\pi^2} \left[ (m_f^2 - m_s^2) \log\left(\frac{\Lambda_{\mathrm{UV}}}{m_s}\right) + 3m_f^2 \log\left(\frac{m_s}{m_f}\right) \right] + \mathcal{O}\left(\frac{1}{\Lambda_{\mathrm{UV}}^2}\right). \qquad (\text{C.3})$$

This shows explicitly that the quadratic divergence disappears. Only the logarithmic divergence remains. Moreover, in case of degenerate masses $m_s = m_f$ even this divergence disappears.

Supersymmetry imposes exactly the relations above, such that it is a good candidate to solve the hierarchy problem. If supersymmetry is broken, the soft masses split the degeneracy between the fermions and the scalars. The resulting correction can immediately be read of from Eq. (C.3)

## C.2 Superfields

### C.2.1 General superfield

The supersymmetry transformation of the component fields of a general superfield are given by

$$\sqrt{2}\delta_\epsilon a = \epsilon\xi + \epsilon^\dagger\chi^\dagger, \tag{C.4}$$

$$\sqrt{2}\delta_\epsilon\xi_\alpha = 2\epsilon_\alpha b + (\sigma^\mu\epsilon^\dagger)_\alpha(v_\mu - i\partial_\mu a), \tag{C.5}$$

$$\sqrt{2}\delta_\epsilon\chi^{\dagger\dot\alpha} = 2\epsilon^{\dagger\dot\alpha}c - (\overline{\sigma}^\mu\epsilon)^{\dot\alpha}(v_\mu + i\partial_\mu a), \tag{C.6}$$

$$\sqrt{2}\delta_\epsilon b = \epsilon^\dagger\zeta^\dagger - \frac{i}{2}\epsilon^\dagger\overline{\sigma}^\mu\partial_\mu\xi, \tag{C.7}$$

$$\sqrt{2}\delta_\epsilon c = \epsilon\eta - \frac{i}{2}\epsilon\sigma^\mu\partial_\mu\chi^\dagger, \tag{C.8}$$

$$\sqrt{2}\delta_\epsilon v^\mu = \epsilon\sigma^\mu\zeta^\dagger - \epsilon^\dagger\overline{\sigma}^\mu\eta - \frac{i}{2}\epsilon\sigma^\nu\overline{\sigma}^\mu\partial_\nu\xi + \frac{i}{2}\epsilon^\dagger\overline{\sigma}^\nu\sigma^\mu\partial_\nu\chi^\dagger, \tag{C.9}$$

$$\sqrt{2}\delta_\epsilon\eta_\alpha = 2\epsilon_\alpha d - i(\sigma^\mu\epsilon^\dagger)_\alpha\partial_\mu c + \frac{i}{2}(\sigma^\nu\overline{\sigma}^\mu\epsilon)_\alpha\partial_\mu v_\nu, \tag{C.10}$$

$$\sqrt{2}\delta_\epsilon\zeta^{\dagger\dot\alpha} = 2\epsilon^{\dagger\dot\alpha}d - i(\overline{\sigma}^\mu\epsilon)^{\dot\alpha}\partial_\mu b - \frac{i}{2}(\overline{\sigma}^\nu\sigma^\mu\epsilon^\dagger)^{\dot\alpha}\partial_\mu v_\nu, \tag{C.11}$$

$$\sqrt{2}\delta_\epsilon d = -\frac{i}{2}\epsilon^\dagger\overline{\sigma}^\mu\partial_\mu\eta - \frac{i}{2}\epsilon\sigma^\mu\partial_\mu\zeta^\dagger. \tag{C.12}$$

### C.2.2 Chiral superfields

The full expansion of the chiral and anti-chiral superfield in terms of the coordinates $x^\mu, \theta, \theta^\dagger$ is given by

$$\begin{aligned}
\Phi = {} & \phi(x) - i\theta\sigma^\mu\theta^\dagger\partial_\mu\phi(x) - \frac{1}{4}\theta\theta\theta^\dagger\theta^\dagger\partial_\mu\partial^\mu\phi(x) + \sqrt{2}\theta\psi(x) \\
& - \frac{i}{\sqrt{2}}\theta\theta\theta^\dagger\overline{\sigma}^\mu\partial_\mu\psi(x) + \theta\theta F(x),
\end{aligned} \tag{C.13}$$

$$\begin{aligned}
\Phi^* = {} & \phi^*(x) + i\theta\sigma^\mu\theta^\dagger\partial_\mu\phi^*(x) - \frac{1}{4}\theta\theta\theta^\dagger\theta^\dagger\partial_\mu\partial^\mu\phi^*(x) + \sqrt{2}\theta^\dagger\psi^\dagger(x) \\
& - \frac{i}{\sqrt{2}}\theta^\dagger\theta^\dagger\theta\sigma^\mu\partial_\mu\psi^\dagger(x) + \theta^\dagger\theta^\dagger F^*(x).
\end{aligned} \tag{C.14}$$

### C.2.3  Vector superfields

The supersymmetry transformations of a vector superfield are given by

$$\sqrt{2}\delta_\epsilon a = \epsilon\xi + \epsilon^\dagger\xi^\dagger, \tag{C.15}$$

$$\sqrt{2}\delta_\epsilon \xi_\alpha = 2\epsilon_\alpha b + (\sigma^\mu\epsilon^\dagger)_\alpha(A_\mu - i\partial_\mu a), \tag{C.16}$$

$$\sqrt{2}\delta_\epsilon b = \epsilon^\dagger\lambda^\dagger - i\epsilon^\dagger\overline{\sigma}^\mu\partial_\mu\xi, \tag{C.17}$$

$$\sqrt{2}\delta_\epsilon A^\mu = -i\epsilon\partial^\mu\xi + i\epsilon^\dagger\partial^\mu\xi^\dagger + \epsilon\sigma^\mu\lambda^\dagger - \epsilon^\dagger\overline{\sigma}^\mu\lambda, \tag{C.18}$$

$$\sqrt{2}\delta_\epsilon \lambda_\alpha = \epsilon_\alpha D - \frac{i}{2}(\sigma^\mu\overline{\sigma}^\nu\epsilon)_\alpha(\partial_\mu A_\nu - \partial_\nu A_\mu), \tag{C.19}$$

$$\sqrt{2}\delta_\epsilon D = -i\epsilon\sigma^\mu\partial_\mu\lambda^\dagger - i\epsilon^\dagger\overline{\sigma}^\mu\partial_\mu\lambda. \tag{C.20}$$

Once again, the variation of the auxiliary field $D$ is proportional to the equation of motion of the fermion.

The supergauge transformations of the real superfield components as defined in Eq. (1.34) is given by

$$a \rightarrow a + i(\phi^* - \phi), \tag{C.21}$$

$$\xi_\alpha \rightarrow \xi_\alpha - i\sqrt{2}\psi_\alpha, \tag{C.22}$$

$$b \rightarrow b - iF, \tag{C.23}$$

$$A_\mu \rightarrow A_\mu - \partial_\mu(\phi + \phi^*), \tag{C.24}$$

$$\lambda_\alpha \rightarrow \lambda_\alpha, \tag{C.25}$$

$$D \rightarrow D. \tag{C.26}$$

In Wess-Zumino gauge, where the additional auxiliary fields are gauged away, the variation of the fields under supersymmetry transformations are instead given by

$$\delta A^a_\mu = -\frac{1}{\sqrt{2}}(\epsilon^\dagger\overline{\sigma}^\mu\lambda^a + \lambda^{\dagger a}\overline{\sigma}_\mu\epsilon), \tag{C.27}$$

$$\delta \lambda^a_\alpha = -\frac{i}{2\sqrt{2}}(\sigma^\mu\overline{\sigma}^\nu\epsilon)_\alpha F^a_{\mu\nu} + \frac{1}{\sqrt{2}}\epsilon_\alpha D^a, \tag{C.28}$$

$$\delta D^a = \frac{i}{\sqrt{2}}(-\epsilon^\dagger\overline{\sigma}^\mu D_\mu\lambda^a + D_\mu\lambda^{\dagger a}\overline{\sigma}^\mu\epsilon), \tag{C.29}$$

which are non-linear, since covariant derivatives appear.

## C.3   Building Lagrangians

### C.3.1   Chiral superfield Lagrangians

The full expansion of the product $\Phi^{*i}\Phi_j$ is given by

$$
\begin{aligned}
\Phi^{*i}\Phi_j = {}& \phi^{*i}\phi_j + \sqrt{2}\theta\psi_j\phi^{*i} + \sqrt{2}\theta^\dagger\psi^{\dagger i}\phi_j + \theta\theta\phi^{*i}F_j + \theta^\dagger\theta^\dagger\phi_j F^{*i} \\
& + \theta\sigma^\mu\theta^\dagger \left[ -i\phi^{*i}\partial_\mu\phi_j + i\phi_j\partial_\mu\phi^{*i} - \psi^{\dagger i}\overline{\sigma}_\mu\psi_j \right] \\
& + \frac{i}{\sqrt{2}}\theta\theta\theta^\dagger\overline{\sigma}^\mu(\psi_j\partial_\mu\phi^{*i} - \partial_\mu\psi_j\phi^{*i}) + \sqrt{2}\theta\theta\theta^\dagger\psi^{\dagger i}F_j \\
& + \frac{i}{\sqrt{2}}\theta^\dagger\theta^\dagger\theta\sigma^\mu(\psi^{\dagger i}\partial_\mu\phi_j - \partial_\mu\psi^{\dagger i}\phi_j) + \sqrt{2}\theta^\dagger\theta^\dagger\theta\psi_j F^{*i} \\
& + \theta\theta\theta^\dagger\theta^\dagger \left[ F^{*i}F_j + \frac{1}{2}\partial^\mu\phi^{*i}\partial_\mu\phi_j - \frac{1}{4}\phi^{*i}\partial^\mu\partial_\mu\phi_j - \frac{1}{4}\phi_j\partial^\mu\partial_\mu\phi^{*i} \right. \\
& \left. + \frac{i}{2}\psi^{\dagger i}\overline{\sigma}^\mu\partial_\mu\psi_j + \frac{i}{2}\psi_j\sigma^\mu\partial_\mu\psi^{\dagger i} \right],
\end{aligned} \tag{C.30}
$$

where all fields are a function of $x^\mu$.

The composite superfields appearing in the superpotential are given by

$$
\Phi_i\Phi_j = \phi_i\phi_j + \sqrt{2}\theta(\psi_i\phi_j + \psi_j\phi_i) + \theta\theta(\phi_i F_j + \phi_j F_i - \psi_i\psi_j), \tag{C.31}
$$

$$
\begin{aligned}
\Phi_i\Phi_j\Phi_k = {}& \phi_i\phi_j\phi_k + \sqrt{2}\theta(\psi_i\phi_j\phi_k + \psi_j\phi_i\phi_k + \psi_k\phi_i\phi_j) \\
& + \theta\theta(\phi_i\phi_j F_k + \phi_i\phi_k F_j + \phi_j\phi_k F_i - \psi_i\psi_j\phi_k - \psi_i\psi_k\phi_j - \psi_j\psi_k\phi_i). \tag{C.32}
\end{aligned}
$$

Notice that these functions are symmetric in $i, j, k$.

The equations of motion for the components of the chiral superfield in the Wess-Zumino model (i.e. Eq. (1.52)) are

$$
\partial^\mu\partial_\mu\phi_i = -M^*_{ik}M^{kj}\phi_j + \dots, \tag{C.33}
$$

$$
i\overline{\sigma}^\mu\partial_\mu\psi_i = M^*_{ij}\psi^{\dagger j} + \dots, \qquad\qquad i\sigma^\mu\partial_\mu\psi^{\dagger i} = M^{ij}\psi_j + \dots \tag{C.34}
$$

### C.3.2   Adding gauge interactions

The real superfield in Wess-Zumino gauge, expressed in the coordinates $y^\mu, \theta, \theta^\dagger$ is given by

$$
\begin{aligned}
V(y^\mu, \theta, \theta^\dagger) = {}& \theta\sigma^\mu\theta^\dagger A_\mu(y) + \theta^\dagger\theta^\dagger\theta\lambda(y) + \theta\theta\theta^\dagger\lambda^\dagger(y) \\
& + \frac{1}{2}\theta\theta\theta^\dagger\theta^\dagger \left[ D(y) + i\partial_\mu A^\mu(y) \right]. \tag{C.35}
\end{aligned}
$$

### C.3.3 Scalar potential

The full scalar potential, substituting the most general superpotential with only renormalisable terms from Eq. (1.51), is

$$
\begin{aligned}
V(\phi, \phi^*) &= F^{*i} F_i + \frac{1}{2} \sum_a D^a D^a \\
&= M^*_{ik} M^{kj} \phi^{*i} \phi_j + \frac{1}{2} M^{in} y^*_{jkn} \phi_i \phi^{*j} \phi^{*k} \\
&\quad + \frac{1}{2} M^*_{in} y^{jkn} \phi^{*i} \phi_j \phi_k + \frac{1}{4} y^{ijn} y^*_{kln} \phi_i \phi_j \phi^{*k} \phi^{*l} \\
&\quad + \frac{1}{2} \sum_a g_a^2 (\phi^* T^a \phi)^2.
\end{aligned}
\tag{C.36}
$$

## Experimental limits on BSM physics

In this appendix, we show some additional summary plots on searches for beyond the Standard Model physics, namely exotic searches and searches for long-lived particles specifically.

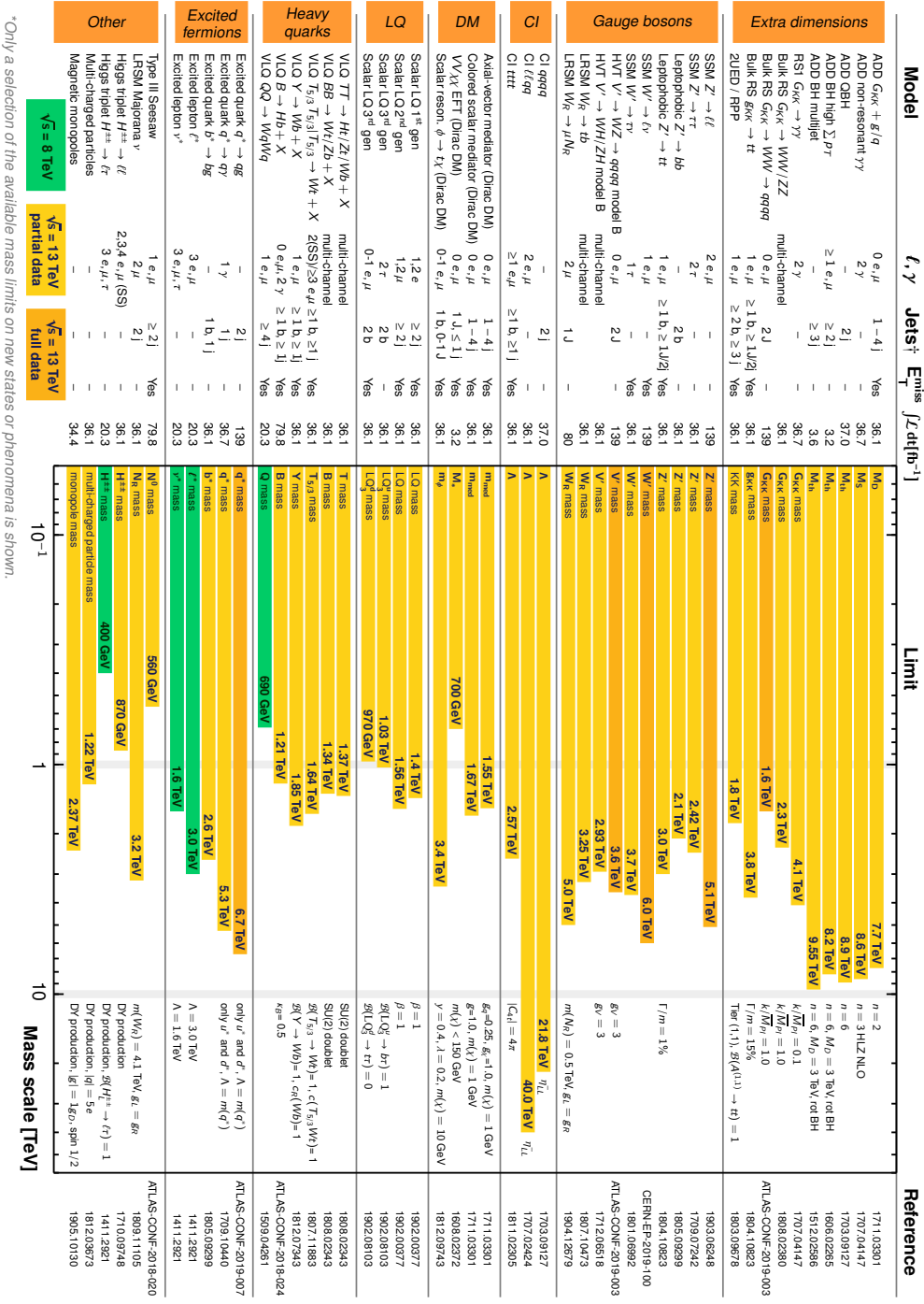

Figure D.1: ATLAS summary plot on searches for new phenomena other than SUSY. Figure from [900].

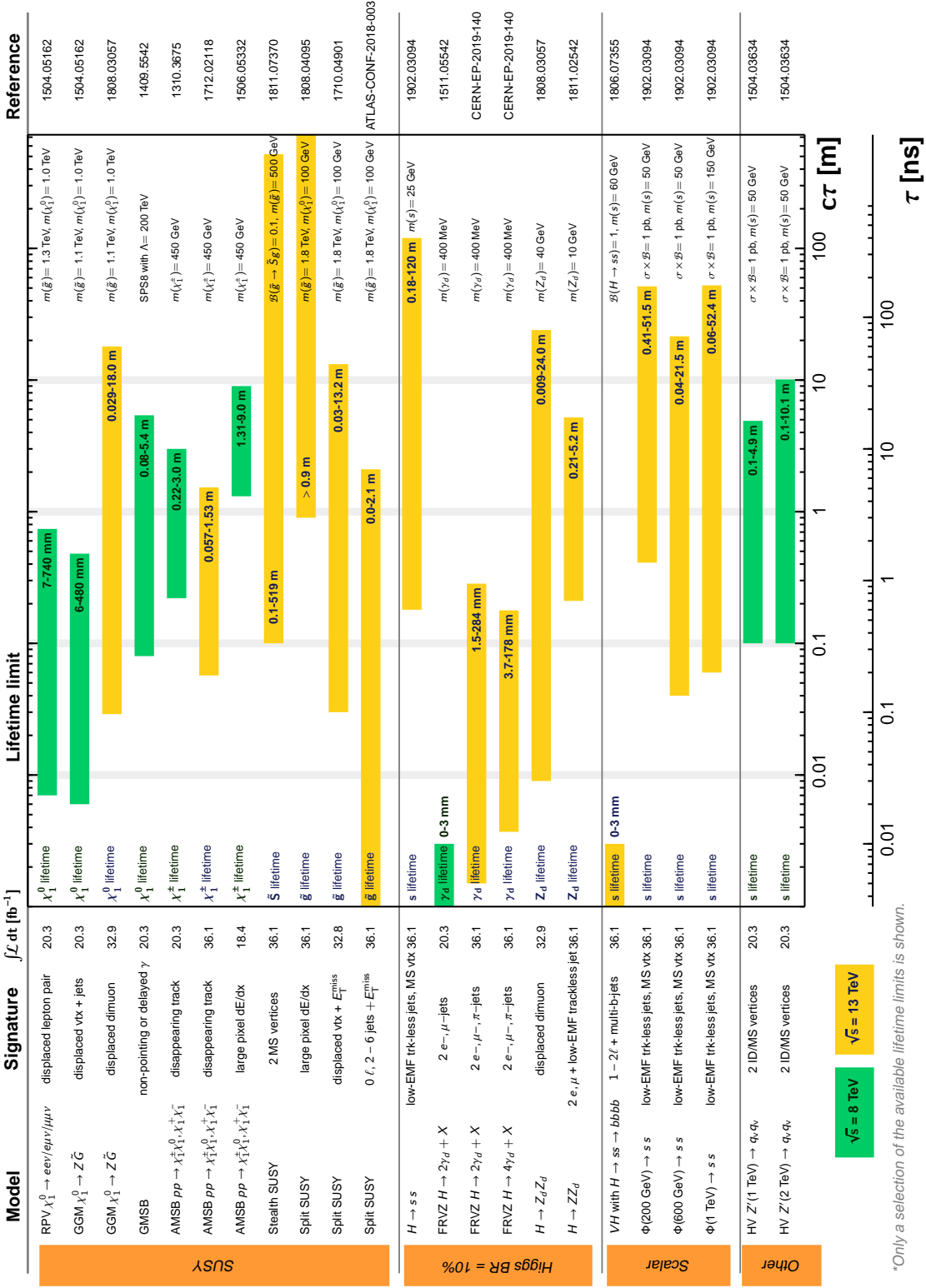

Figure D.2: ATLAS summary plot for new phenomena involving long-lived particles. Figure from [900].

## Goldstini analysis implementation

For our study, we implemented several ATLAS and CMS analyses in the `Atom` framework [341]. `Atom` is a program based on `Rivet` [269] and maps the truth level particles into the reconstructed objects such as isolated electrons and *b*-jets according to the detector performances reported by ATLAS and CMS. The validation and some application of the code can be found in [901–905].

In Table E.1 we show some of the validation results as an example. The numbers in the second column represent the expected signal events for each step of the cut used in the 5j signal region (SR) reported by the ATLAS jets+$\not{E}_T$ analysis [283], based on the $\tilde{q}_L \rightarrow q \tilde{\chi}_1^\pm \rightarrow q W^\pm \tilde{\chi}_1^0$ topology with $(m_{\tilde{q}}, m_{\tilde{\chi}_1^\pm}, m_{\tilde{\chi}_1^0}) = (665, 465, 265)$ GeV. The right column shows the ratios between the `Atom` and ATLAS results. One can see that these ratios are close to one within $\sim 20\%$ accuracy, indicating a good agreement between the `Atom` and ATLAS simulations.

Table E.1: "5j" SR validation table for the implementation of the ATLAS jets+$\not{E}_T$ analysis in `Atom`. The decay chain for validation in consideration is $\tilde{q}_L \to q\tilde{\chi}_1^{\pm} \to qW^{\pm}\tilde{\chi}_1^0$.

| 5j SR cuts | $N_{\text{SUSY}}^{\text{Exp}}$ | Atom/Exp |
|---|---|---|
| $\not{E}_T > 160$, $p_T^{j_{1(2)}} > 130(60)\,\text{GeV}$ | 317.3 | 1.17 |
| $p_T^{j_3} > 60\,\text{GeV}$ | 306.2 | 1.12 |
| $p_T^{j_4} > 60\,\text{GeV}$ | 247.6 | 1.04 |
| $p_T^{j_5} > 60\,\text{GeV}$ | 141.8 | 1.00 |
| $\Delta\phi(j_{1,2,3}, \not{E}_T) > 0.4$ | 118.6 | 1.01 |
| $\Delta\phi(j_{i>3} > 40\,\text{GeV}, \not{E}_T) > 0.2$ | 103.1 | 1.01 |
| $\not{E}_T/m_{\text{eff}}(N_j) > 0.2$ | 85.6 | 1.04 |
| $m_{\text{eff}}(\text{incl.}) > 1200\,\text{GeV}$ | 20.5 | 1.18 |

# Fermi mechanism

In order to understand how the Fermi mechanism gives rise to a power law, both in the first and second order version, we will follow the derivation found in [367, 368]. Consider a system where $N$ particles with an energy $E_0$ can move around freely, and undergo energy-gaining encounters. Suppose that with each encounter, a particle receives a fractional energy gain $\frac{\Delta E}{E} = \xi$. After $n$ encounters, a particle has an energy

$$E_n = E_0(1 + \xi)^n. \tag{F.1}$$

If after each encounter, a particle has a chance $P_{esc}$ to escape the accelerator, the number of particles reaching at least an energy $E$ is[1]

$$N(\geqslant E_n) = N(1 - P_{esc})^n. \tag{F.2}$$

When combining this with Eq. (F.1), we find

$$N(\geqslant E) = N\left(\frac{E}{E_0}\right)^{-\sigma}, \tag{F.3}$$

with

$$\sigma = \frac{\ln\left(\frac{1}{1 - P_{esc}}\right)}{\ln(1 + \xi)} \approx \frac{P_{esc}}{\xi}, \tag{F.4}$$

or as a differential energy spectrum

$$\frac{\mathrm{d}N}{\mathrm{d}E} = -\sigma N\left(\frac{E}{E_0}\right)^{-(\sigma+1)}. \tag{F.5}$$

This proves that the Fermi mechanism very generally leads to a power law spectrum $\frac{\mathrm{d}N}{\mathrm{d}E} \propto E^{-\alpha}$ with $\alpha = \sigma + 1$. From the discussion in Section 3.2.2, this was expected, since there is no inherent energy scale present in the description of this system.

---

[1] This differs from the derivation in [367, 368], where instead $N(\geqslant E) \propto \sum_{m=n}^{\infty}(1 - P_{esc})^m$ is written. While technically correct, in order to correctly track all the $P_{esc}$ that appear, an extra $P_{esc}$ is needed (since each term in this sum represents the fraction of particles that stay *exactly* $m$ encounters). This then makes the results agree.

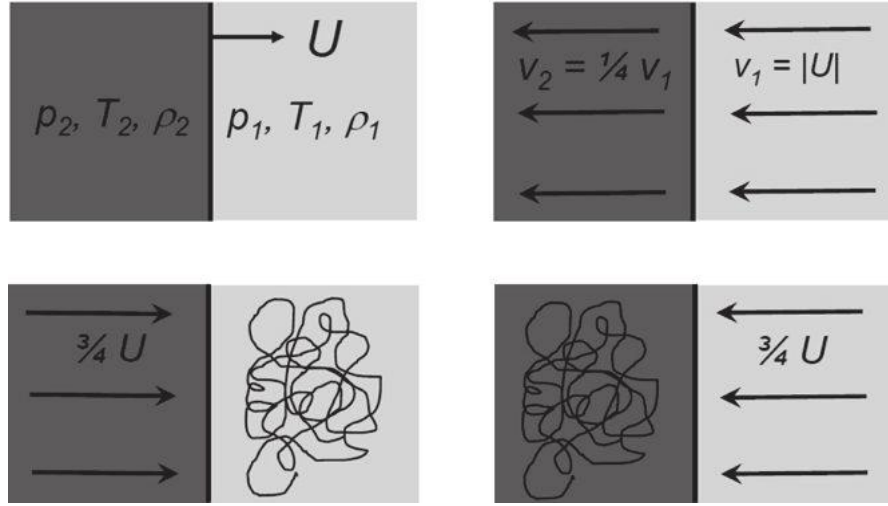

Figure F.1: Representation of a shock approximated as a plane wave. (Top left) Shock propagating through the interstellar medium at speed $U$. (Top right) In the reference frame of the shock, the interstellar medium approaches the shock and the shocked medium moves away from it. (Bottom left) In the upstream frame, accelerated particles upstream of the shock move isotropically. (Bottom right) In the downstream frame, accelerated particles downstream of the shock move isotropically.

In order to find the exact power law index, we need to find expressions for $P_{\text{esc}}$ and $\xi$. For the first order Fermi mechanism, this is straightforward and we will follow the derivations in [367, 368]. Consider the situation in Figure F.1. A shock propagates through an undisturbed medium with speed $U$, with the swept-up medium following the shock. Particles move isotropically in the up- and downstream frames through multiple scatterings. However, each time the particle crosses the shock, it will collide with the turbulence inside the plasma that is moving towards the particle at a speed $v$ (with a value derived below). In the reference frame of that plasma, this is a collision that conserves energy. The energy gain can then simply be found by performing a Lorentz transformation of the particle energy from one plasma rest frame to the other

$$E' = \Gamma(E - p_x v) = \Gamma E(1 - \beta \cos \theta),$$ (F.6)

for relativistic particles with $E = pc$ and $p_x = p \cos \theta$, where $\beta = U/c$. The acceleration is thus achieved by consecutive Lorentz transformations. One complete cycle consists of two shock crossings: from upstream to downstream and back. The energy gain for such a cycle, found by chaining two such transformations, gives

$$\frac{\Delta E}{E} = -\beta \cos \theta_1 + \beta \cos \theta_2',$$ (F.7)

for non-relativistic shocks with $\beta \ll 1$, $\theta_1$ defined in the upstream frame and $\theta_2'$ defined in the downstream frame. We need to average this over the $\theta_i$. For both of these, we have an isotropic angular distribution, so proportional to $\sin \theta_i \, \mathrm{d}\theta_i$, arriving at the shock at a rate $c \cos \theta_i$. When properly normalising, this gives $\langle \cos \theta_i \rangle = \pm \frac{2}{3}$, so that

$$\xi = \frac{4}{3}\beta. \tag{F.8}$$

The escape probability can also be derived. Since we approximated the shock as an infinite plane wave, particles can only escape the shock downstream, since there they are swept away from the shock by the plasma. The integrated rate of approach of an isotropic flux onto a plane shock (on both sides) is given by

$$\int_0^1 \mathrm{d}\cos\theta \int_0^{2\pi} \mathrm{d}\phi \, \frac{c\rho}{4\pi} \cos\theta = \frac{c\rho}{4}. \tag{F.9}$$

Of these, a fraction is swept away due to the convection downstream. The rate of convection downstream is given simply by $u_2\rho$. Therefore, the escape probability is $P_{\mathrm{esc}} = \frac{4u_2}{c}$. The resulting spectral index from Eqs. (F.4) and (F.8) is

$$\sigma = \frac{3}{u_1/u_2 - 1}, \tag{F.10}$$

since $u_1 = U$. Finally, from continuity and the kinetic theory of gasses, we have for strong shocks

$$\frac{u_1}{u_2} = \frac{\rho_2}{\rho_1} = \frac{\gamma + 1}{\gamma - 1} = 4, \tag{F.11}$$

with $\gamma = 5/3$ the ratio of specific heat capacities of a gas. Therefore, our final result is the famous $E^{-2}$ power law

$$\frac{\mathrm{d}N}{\mathrm{d}E} = N \left( \frac{E}{E_0} \right)^{-2}. \tag{F.12}$$

While shock acceleration is well understood, there is one important problem that was not mentioned here. The mechanism supposes that the accelerated particles start out sufficiently relativistic. However, at the moment there exists no good theory for how particles would achieve this. This is called the injection problem. Many important questions in cosmic ray theory, such as the composition of cosmic rays and the value of the electron-proton ratio are related to this, since injection could prefer high or low masses, preventing some particles from being accelerated at all.

For the second order mechanism, the situation is a bit different. The energy gain can be calculated in a way analogous to the derivation above, but for a cloud $\theta'_2$ is isotropically distributed and $\langle \cos \theta'_2 \rangle = 0$, while the $\theta_1$-distribution depends on the relative velocity and gives $\langle \cos \theta'_1 \rangle = -\frac{\beta}{3}$. This conspires to $\xi \propto \beta^2$. More complicated, $P_{\mathrm{esc}}$ depends on the cloud parameters, such as its density, speed and the cross section as well as on the confinement time in the galaxy. Therefore, the second order Fermi mechanism gives no unique prediction for $P_{\mathrm{esc}}$ and for the spectral index $\alpha$.

## Photon cross sections: H vs ISM-composition

Throughout the calculations in Chapter 4, we assumed that the gas cloud target is pure hydrogen, even though one would naturally expect a composition closer to that of the interstellar medium[1]. We calculate here the photon cross sections using the XCOM database [692], for a gas that is either pure hydrogen or follows the interstellar medium composition. For the latter, we follow the abundances given in [694], shown in Figure G.1. While more up-to-date measurements are available[2], the differences are negligible for the calculations here.

The cross sections are shown in Figure G.2, in units $\mathrm{cm}^2\,\mathrm{g}^{-1}$ instead of barn/atom, since these are more natural for gases with a mixed composition. As can be seen, there is little difference in the cross section for gamma rays since the interstellar medium is dominated by hydrogen[3].

---

[1]Although not completely the same, since locally the composition may be influenced by the extreme environment, causing the presence of heavier elements to be enhanced.

[2]See e.g. the datatables bundled with the XSpec program [693].

[3]This does not mean that the composition is always irrelevant, since different elements are still detected by their emission or absorption lines in the spectrum, which reveal the physics of the environment.

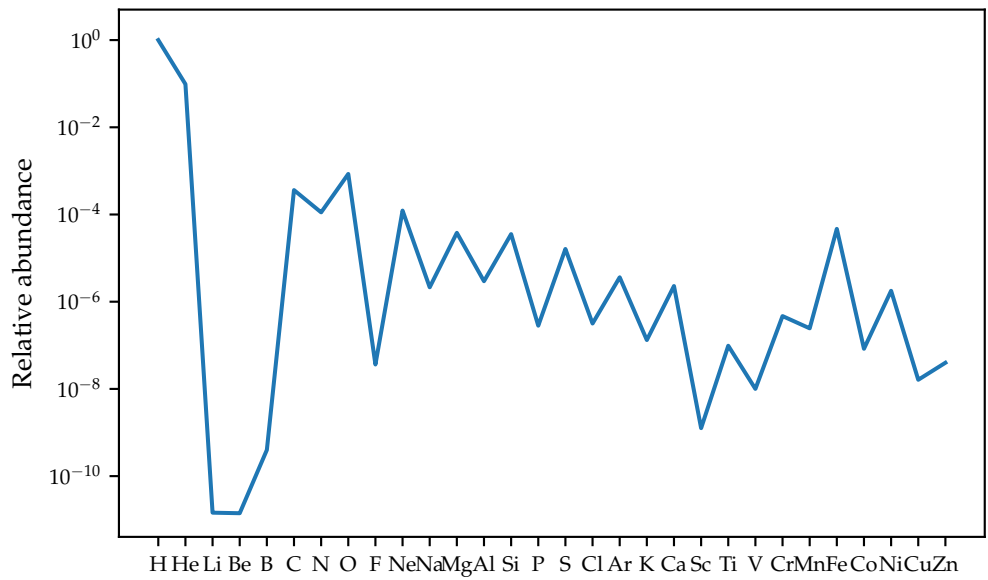

Figure G.1: Interstellar medium composition, following [694].

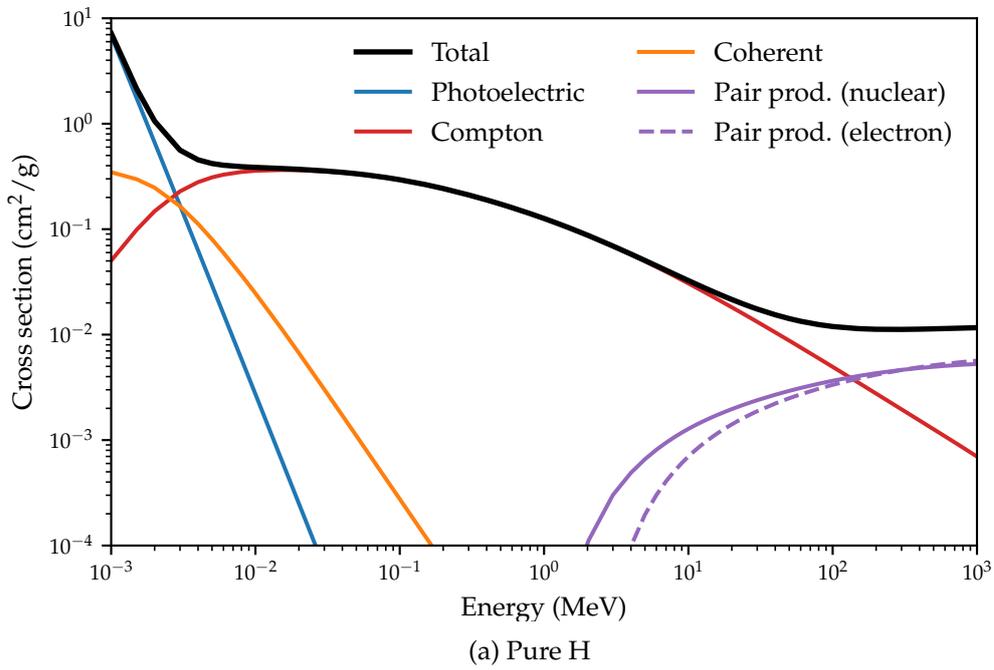

(a) Pure H

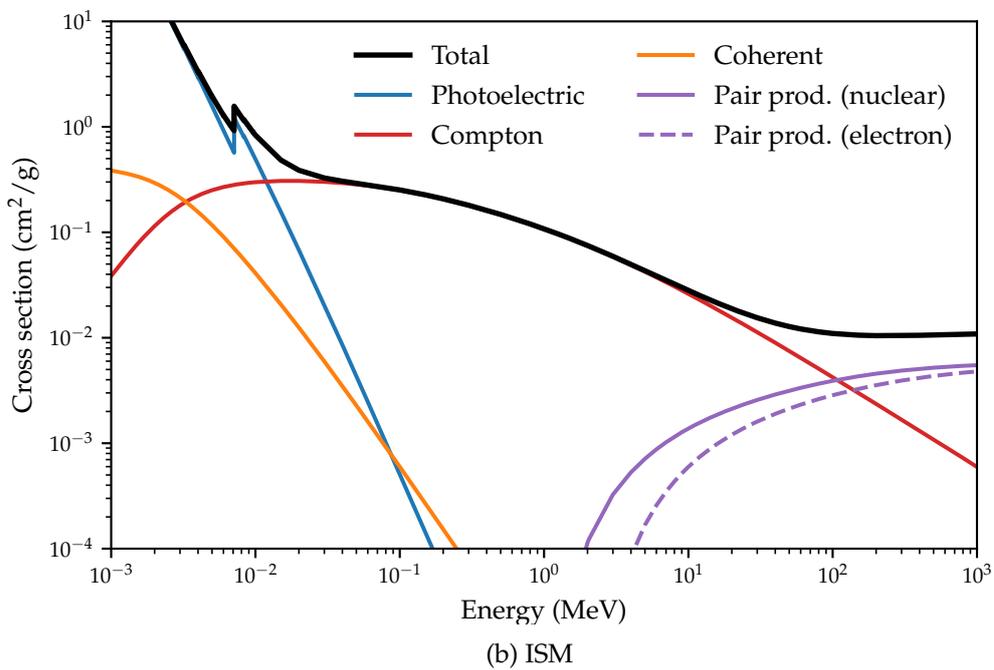

(b) ISM

Figure G.2: Cross sections of X-rays interacting with a gas, either pure hydrogen or following the interstellar medium composition, showing both the total cross section and its decomposition in the different processes.

# Additional results on neutrino production from obscured $pp$-sources

In this appendix, we gather some additional results corresponding to the calculations in Sections 4.2.3, 4.3 and 4.4.

## H.1  Simulation results

We show in Table H.1 additional validations of the Monte Carlo simulation, this time for $N_H^{(1)} = 5 \times 10^{25}$ cm$^{-2}$.

### H.1.1  Contribution from secondary interactions

Figure H.1 shows the contribution of secondary interactions, as in Section 4.2.3, but now for $N_H = 5 \times 10^{25}$ cm$^{-2}$.

Table H.1: Relative luminosities of the neutrinos (all flavours), gamma rays and secondary protons for the different parameter choices, compared to the standard choices in Table 4.1, but now with $N_H^{(1)} = 5 \times 10^{25}$ cm$^{-2}$ for the final two entries.

| Scenario | $L_\nu / L_{\text{Inj.}}$ | $L_\gamma / L_{\text{Inj.}}$ | $L_p / L_{\text{Inj.}}$ |
|---|---|---|---|
| Standard | 0.15 | 0.04 | 0.04 |
| $N_H^{(2)} = 5 \times 10^{25}$ cm$^{-2}$ | 0.13 | 0.06 | 0.16 |
| No second. | 0.11 | 0.05 | 0.25 |
| No atten. | 0.13 | 0.10 | 0.16 |

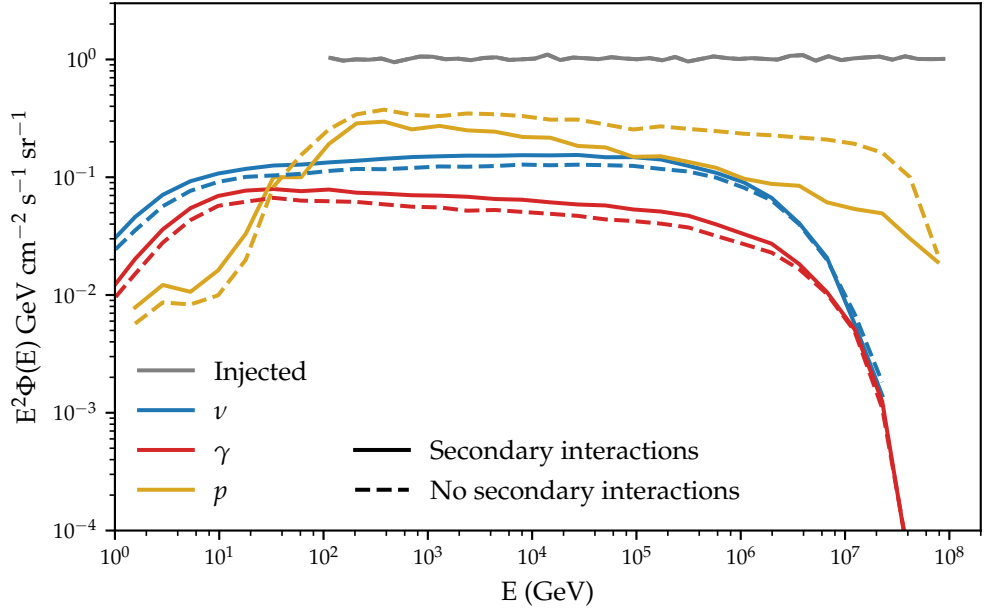

Figure H.1: The effect of including secondary interactions, for the parameters in Table 4.1 and $N_H = 5 \times 10^{25}$ cm$^{-2}$.

### H.1.2 Effect of the gamma-ray attenuation

Figures H.2a and H.2b show the effect of changing the way attenuation is taken into account, as in Section 4.2.3, but now for $N_H = 5 \times 10^{25}$ cm$^{-2}$.

## H.2 Obscured flat-spectrum radio AGN

We show additional results of our model for the obscured flat-spectrum radio AGN selection.

### H.2.1 SED of all the objects

Figures H.3, H.4 and H.5 show the hybrid SED for all other objects in our selection, as shown for 3C371 in Figure 4.9, for $N_H^{(2)} = 10^{26}$ cm$^{-2}$.

### H.2.2 Parameter variation effects on limits

Next, we show additional results when varying the parameters of our model.

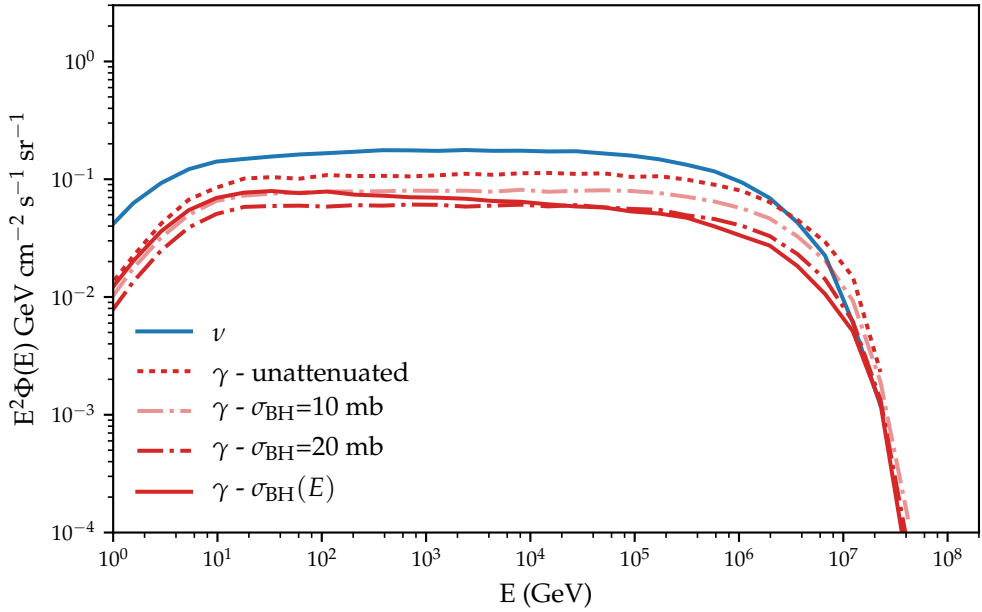

(a) Including pair production

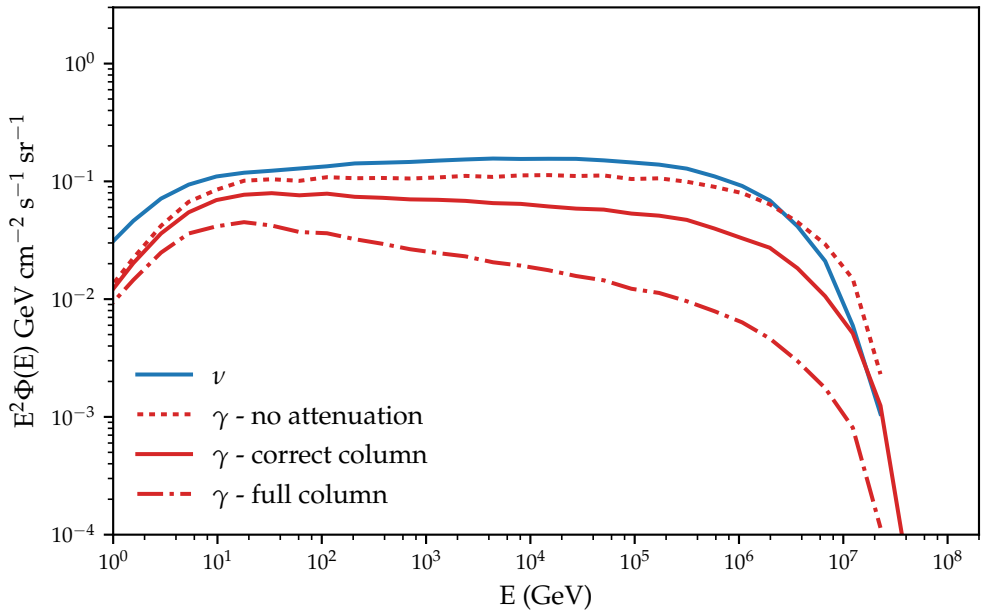

(b) Column density determination

Figure H.2: Effect of including attenuation of the gamma rays, compared to several other cases. Parameters for the simulation are given in Table 4.1, but with $N_H = 5 \times 10^{25}$ cm$^{-2}$. (a) The effect of including gamma-ray attenuation by pair production, for different approximations of the Bethe-Heitler cross section. The full expression is given by Eqs. (4.4) and (4.5). (b) The effect of including the correct column density after creation of the photons ins $pp$-interactions, compared to several approximations.

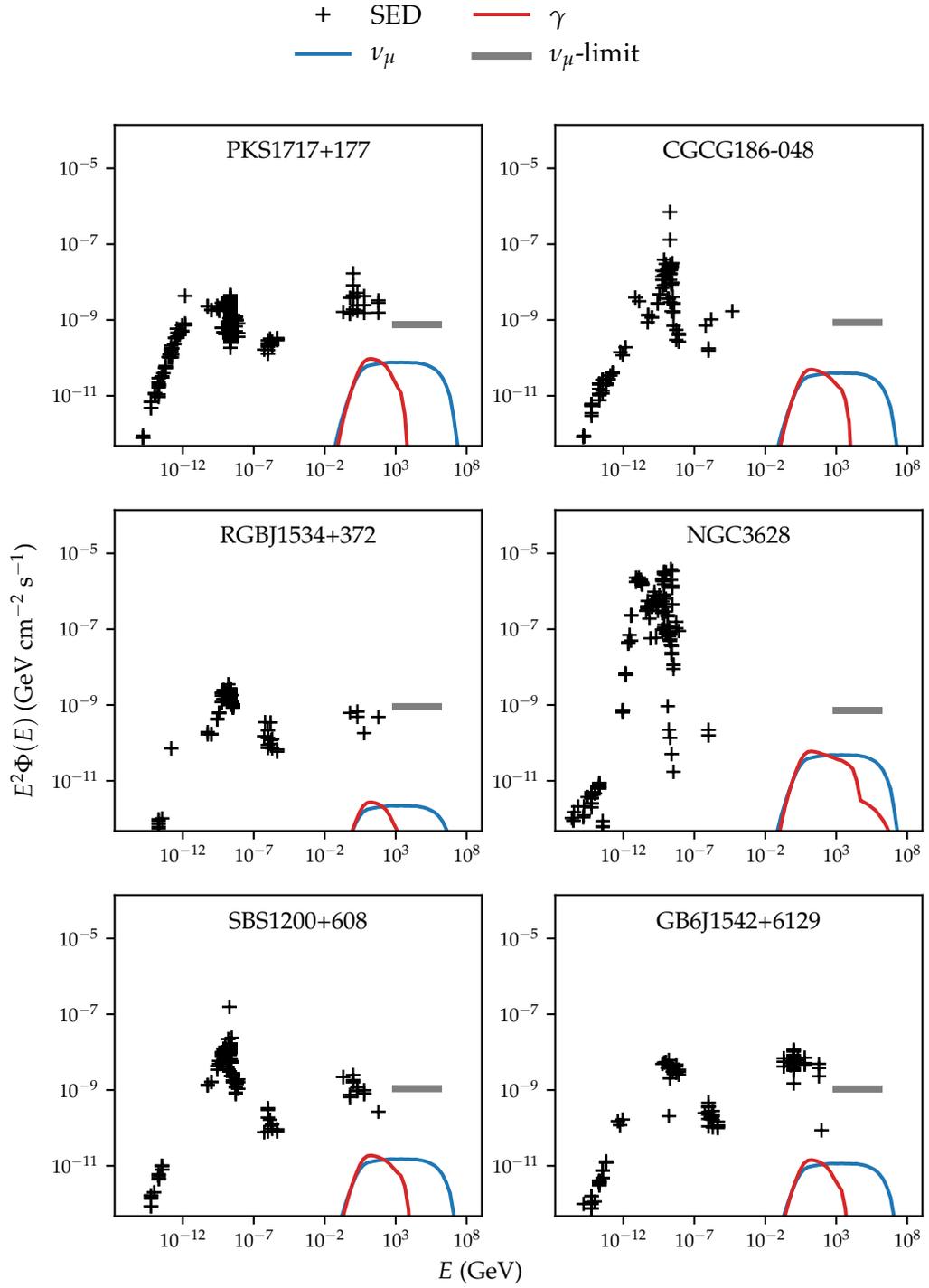

Figure H.3: Hybrid SED for all the objects in the selection besides 3C371, similar to Figure 4.9 for 3C371. Limit on the muon neutrino flux from [636]. Electromagnetic data (citations in Table H.2) retrieved using the `SSDC SED Builder` [766].

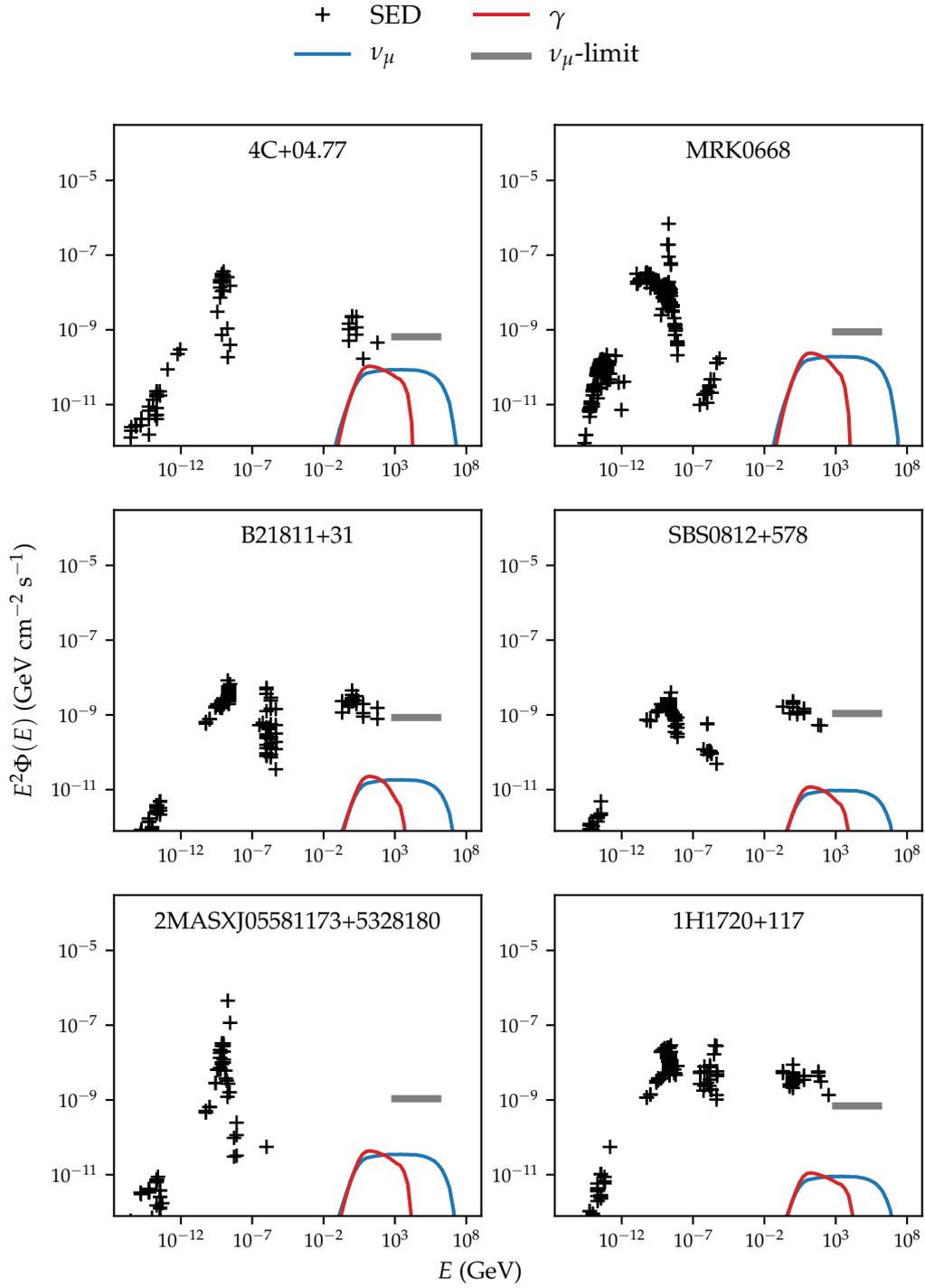

Figure H.4: Hybrid SED for all the other objects in the selection, continued.

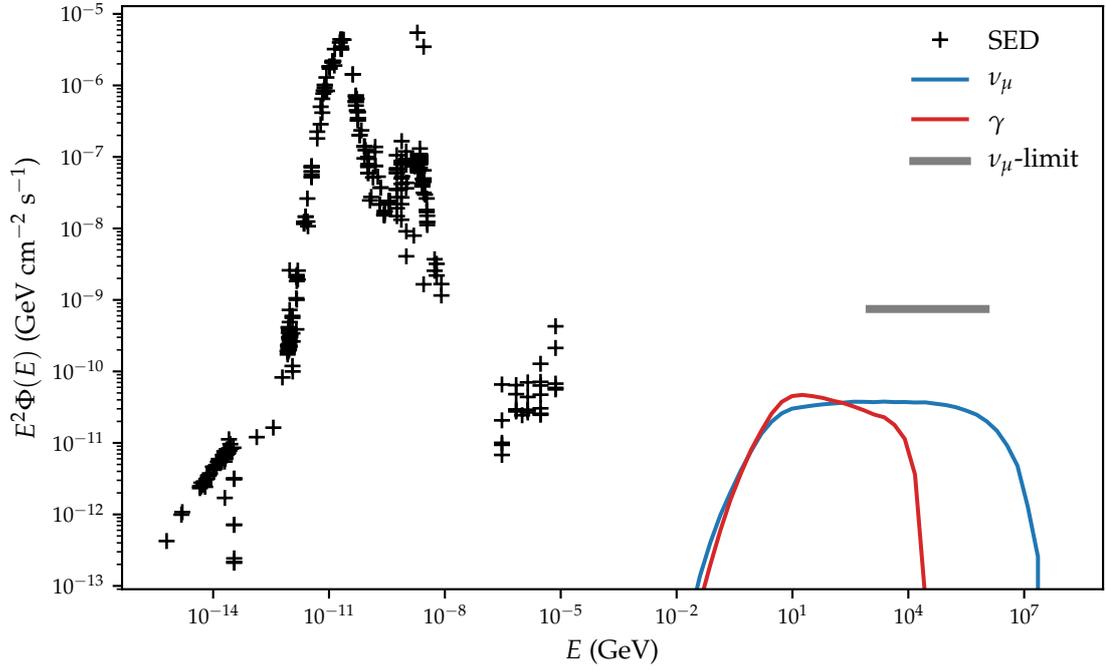

Figure H.5: Hybrid SED for all the other objects in the selection, continued. Note the different scale.

Table H.2: Citations for the electromagnetic data used in Figures H.3, H.4 and H.5.

| Object | Citations |
|---|---|
| PKS1717+177 | [740–742, 744, 745, 748–752, 754–757, 759, 760, 762–765, 906, 907] |
| CGCG186-048 | [740–742, 745, 749, 750, 754, 755, 759–761, 906, 908] |
| RGBJ1534+372 | [740, 744, 745, 749, 754, 756, 757, 759–761, 763–765, 906, 908] |
| NGC3628 | [740, 741, 745–752, 908] |
| SBS1200+608 | [740, 741, 745, 749, 750, 754–757, 759, 760, 762–765, 906, 908] |
| GB6J1542+6129 | [740, 741, 745, 749, 750, 752, 756, 757, 759, 760, 762–765, 908] |
| 4C+04.77 | [742, 751, 762–764] |
| MRK0668 | [740–743, 745–747, 749, 750, 754–757, 906, 908, 909] |
| 3C371 | [739–765] |
| B21811+31 | [740–742, 744, 745, 749, 750, 754, 756, 757, 759, 760, 762–765, 906] |
| SBS0812+578 | [740, 745, 749, 754–757, 759, 760, 762–765, 908] |
| 2MASXJ05581173+5328180 | [740, 741, 744, 745, 749, 750, 754, 755, 906] |
| 1H1720+117 | [740, 741, 744, 745, 749, 754–757, 759–765, 906, 910, 911] |
| ARP220 | [740, 741, 745–752, 754, 906, 908, 909] |

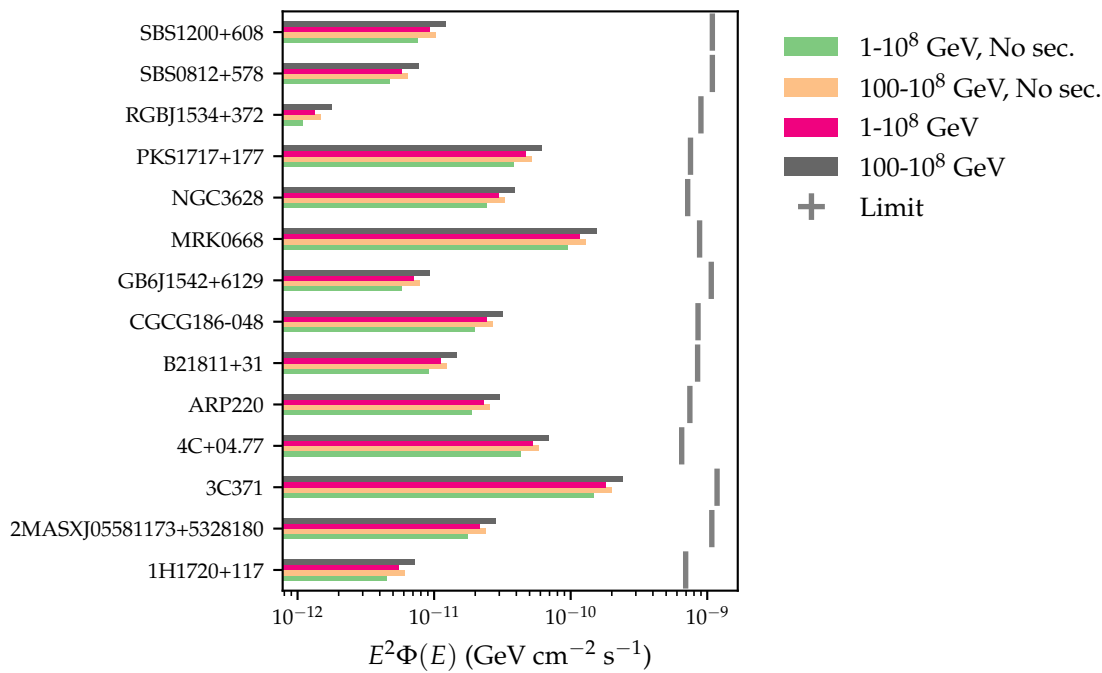

Figure H.6: Summary of the predicted flux and the upper limits for the objects in the obscured flat-spectrum radio AGN selection, for $N_H^{(1)} = 5 \times 10^{25}$ cm$^{-2}$.

**Neutrino flux**

First, we show in Figures H.6 and H.7 the predicted flux of the objects in our selection and their IceCube upper limits (equivalent of Figure 4.10), now for $N_H^{(1)} = 5 \times 10^{25}$ cm$^{-2}$ and $N_H^{(2)} = 10^{26}$ cm$^{-2}$ separately, showing additional parameter variations.

**Limits on $f_e$**

Figures H.8 and H.9 show the lower limits on $f_e$ for the objects in our selection (equivalent of Figure 4.11), now for $N_H^{(1)} = 5 \times 10^{25}$ cm$^{-2}$ and $N_H^{(2)} = 10^{26}$ cm$^{-2}$ separately, showing additional parameter variations.

## H.3 Diffuse ν and γ-ray flux for $N_H = 10^{24}$ cm$^{-2}$

In Figure H.10, we show the results for the diffuse neutrino and gamma-ray flux using our code of Section 4.4. This corresponds to a thin target in front of e.g. an AGN. Note that due to the lower density of the target, protons at lower energy do not experience a full beam dump, while those at higher energy do. This explains the inclined neutrino spectrum.

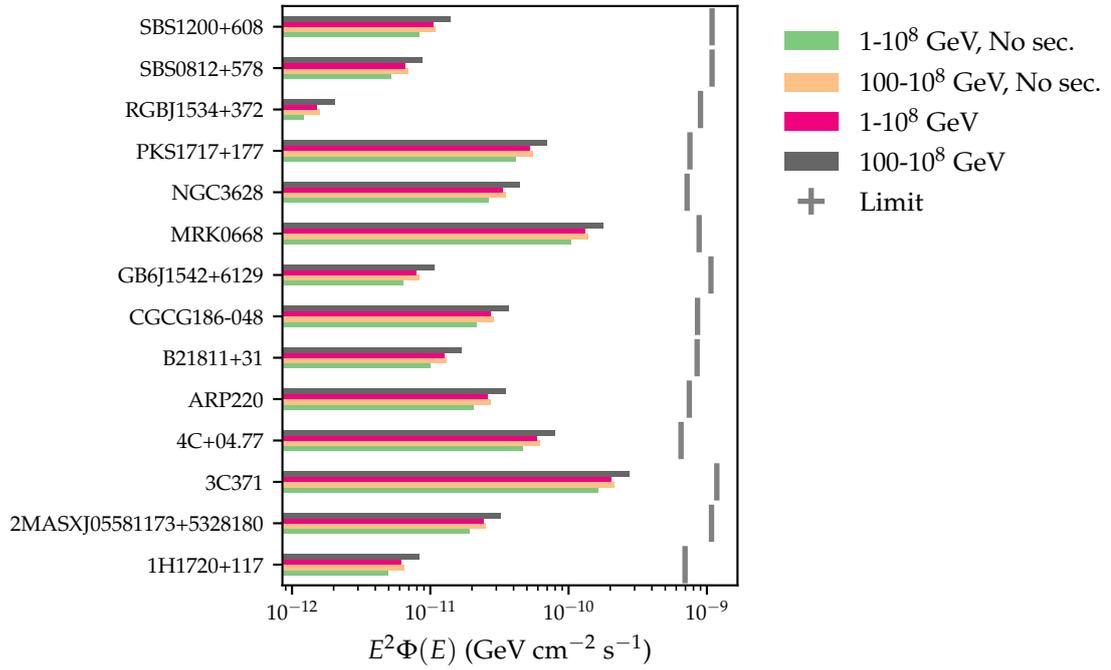

Figure H.7: Summary of the predicted flux and the upper limits for the objects in the obscured flat-spectrum radio AGN selection, for $N_H^{(2)} = 10^{26}$ cm$^{-2}$.

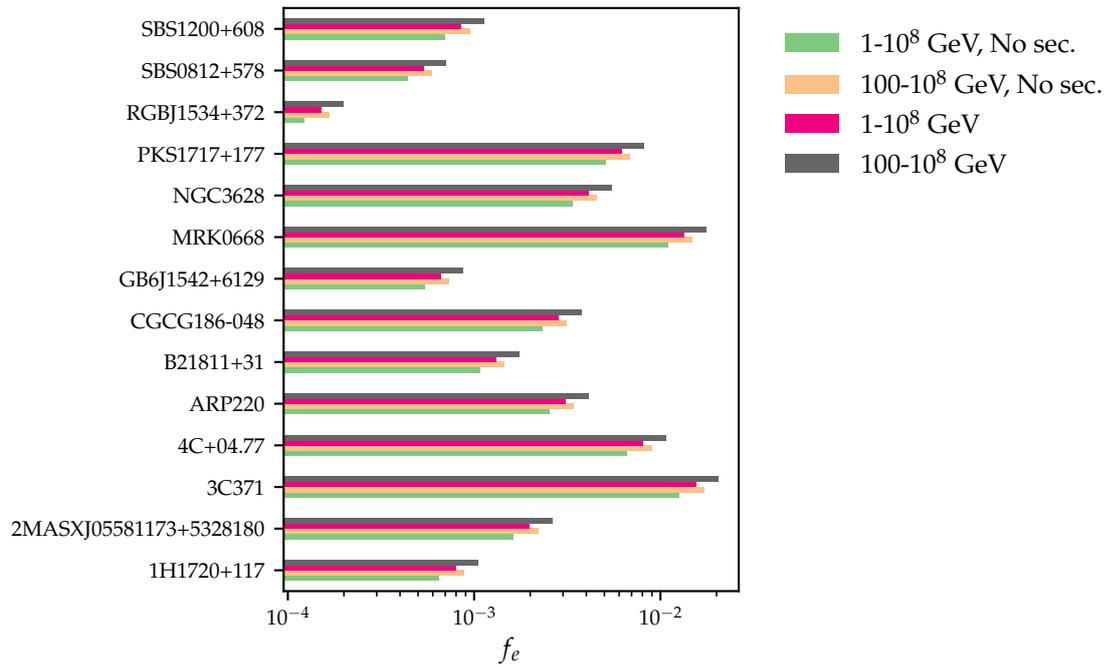

Figure H.8: Summary of the lower limits on $f_e$ for the objects in the obscured flat-spectrum radio AGN selection, for $N_H^{(1)} = 5 \times 10^{25}$ cm$^{-2}$.

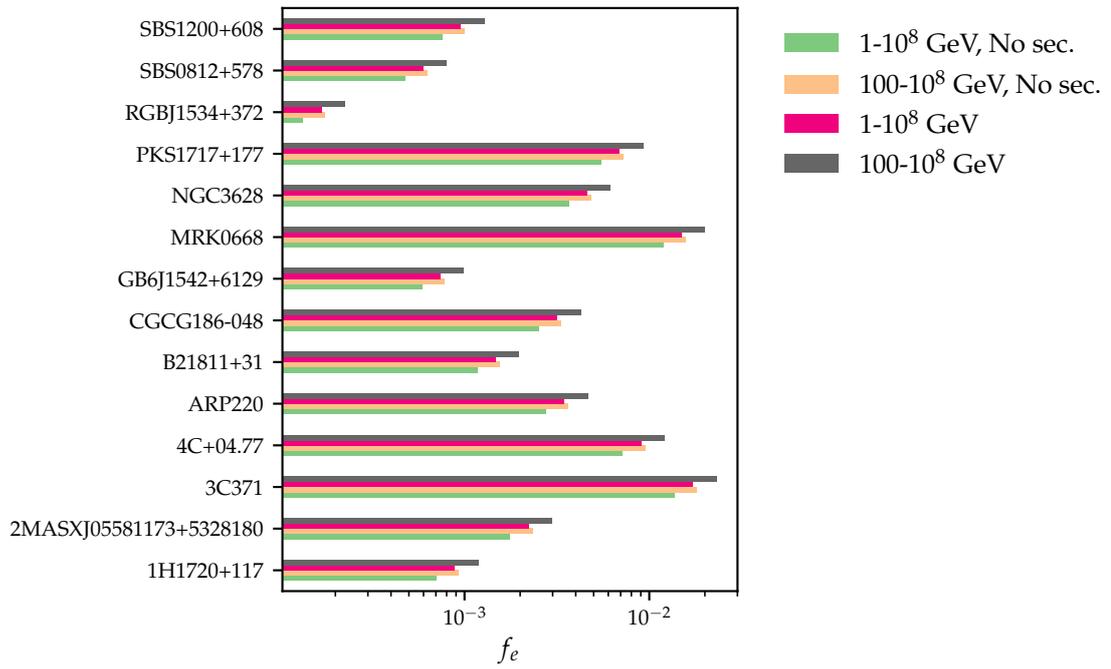

Figure H.9: Summary of the lower limits on $f_e$ for the objects in the obscured flat-spectrum radio AGN selection, for $N_H^{(2)} = 10^{26}$ cm$^{-2}$.

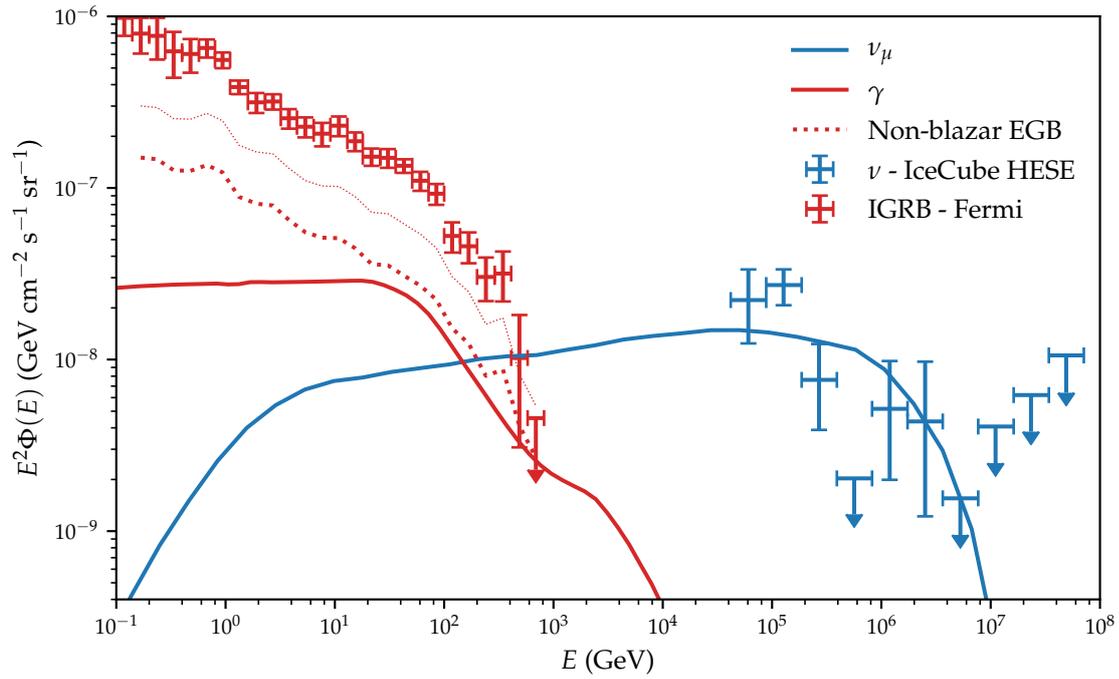

Figure H.10: Diffuse $\nu$ and $\gamma$-ray flux for $N_H = 10^{24}$ cm$^{-2}$, using the same procedure as Section 4.4, for a cosmological evolution following star-formation rate. The neutrino flux is fitted to the IceCube single-flavour neutrino flux (HESE) [534]. The non-blazar contribution to the EGB shows both the best fit value (14% of the EGB measured by Fermi [557]) and the weakest upper limit (28%).

## Deriving the gravitational wave equations

Here, we gather some additional elements of gravitational wave theory. Just as in Chapter 5, the discussion below is heavily based on the classic text [804] as well as the more recent book [805] and draws inspiration from the basic physics papers [806, 807], which discuss the case of GW150914 in particular, and [808].

## I.1   Gravitational wave amplitude

We repeat here the Einstein equations in linearised theory which need to be solved

$$\left(-\frac{\partial^2}{\partial t^2} + \nabla^2\right) \bar{h}_{\mu\nu} = -\frac{16\pi G}{c^4} T_{\mu\nu}.$$  (I.1)

The general solution to this equation is the retarded integral

$$\bar{h}_{\mu\nu}(\mathbf{x}, t) = 4\frac{G}{c^4} \int \mathrm{d}^3 x' \, \frac{T_{\mu\nu}\left(\mathbf{x}', t - |\mathbf{x} - \mathbf{x}'|\right)}{|\mathbf{x} - \mathbf{x}'|}.$$  (I.2)

For field points sufficiently far from the source ($r \equiv |\mathbf{x}| \gg |\mathbf{x}'|$), the integrand can be expanded in terms of $\mathbf{x}'/r$. As a result, on the right hand side there appear moments of mass, momentum and stress (i.e. integrals of $T_{\mu\nu}$ with powers of $\mathbf{x}'$) and their time derivatives.

In order to interpret the resulting radiation, it is easiest to fix the residual gauge freedom by going to the transverse-traceless (TT) gauge[1]. For a pure wave, imposing the TT gauge condition leaves only the pure spatial components $h_{ij}^{TT}$ as non-zero. However, for the metric perturbation $h_{\mu\nu}$ due to a matter source, there are three contributions to the metric: a pure gauge part, a physical non-radiating part (i.e. the Newtonian potential) and a physical radiating part (see also the discussion in [912]). It is the radiating part which can be put in the TT gauge and thus is purely spatial and transverse. By Eq. (I.2), these components are related to $T_{ij}$. From the conservation law $T^{\mu\nu}{}_{,\nu} = 0$ (applying it multiple times and integrating over volume), the stress, and therefore the gravitational wave amplitude, is related to the second time derivative of the second moment of mass. This gives the relation $h \sim \ddot{Q}$ as was anticipated in the main text.

---

[1] Going to TT gauge is only possible in a region of spacetime where $T^{\mu\nu} = 0$, i.e. outside a sphere containing the source.

Note that by going to the TT gauge, the difference between $h_{\mu\nu}^{TT}$ and $\bar{h}_{\mu\nu}^{TT}$ has disappeared.

Formally, the TT part of the reduced quadrupole is given by

$$Q_{ij}^{TT} = P_{ik}Q_{kl}P_{lj} - \frac{1}{2}P_{ij}P_{kl}Q_{kl},$$  (I.3)

with $P^{kl} = \delta^{kl} - n^k n^l$ the projection operator and $n^k = x^k/r$ the unit vector to field point **x**.

## I.2 Inclination-dependent gravitational wave amplitude

The full expression for the gravitational wave amplitude, showing explicitly the inclination dependence, is [828]

$$h_+(t) = A_{\text{GW}}(t)(1 + \cos^2 i)\cos\phi_{\text{GW}}(t),$$  (I.4)

$$h_\times(t) = -2A_{\text{GW}}(t)\cos i \sin\phi_{\text{GW}}(t),$$  (I.5)

with $i$ the inclination angle ($\cos i = \pm 1$ for face-on, $\cos i = 0$ for edge-on), $A_{\text{GW}}(t)$ the gravitational wave amplitude and $\phi_{\text{GW}}(t)$ the phase.

## I.3 Energy of gravitational waves

The energy associated to gravitational radiation is only well-defined in certain regimes: namely when it is possible to separate the wave from the background. In this case, however, it is possible to associate an effective stress-energy tensor to the gravitational wave[2]: the Isaacson tensor [913, 914]

$$T_{\mu\nu}^{(\text{GW})} = \frac{1}{32\pi}\frac{c^3}{G}h_{\alpha\beta,\mu}^{TT}h_{,\nu}^{TT\alpha\beta}.$$  (I.6)

In order to obtain a meaningful result for the energy associated to the gravitational wave, this tensor needs to be averaged over several wavelengths, since the energy cannot be localised more precisely than this. The luminosity of the source, or the rate at which

---

[2]Defining the concept of energy for a gravitational wave is non-trivial. Since a wave, and thus the metric, is time-dependent, there is no conservation of energy. Moreover, due to the equivalence principle, in a local frame there is not even a wave.

energy is carried away, is then given by[3]

$$L_{GW} = \frac{1}{32\pi} \frac{c^3}{G} \iint dS \, \langle \dot{h}_{\mu\nu}^{TT} \dot{h}^{TT\mu\nu} \rangle = \frac{1}{5} \frac{G}{c^5} \langle \frac{d^3 Q^{ij}}{dt^3} \frac{d^3 Q_{ij}}{dt^3} \rangle. \tag{I.7}$$

---

[3]Note that the reduced quadrupole which appears is *not* in the TT gauge. The factor $1/5$ comes from tensor calculus, since the transverse traceless quadrupole moment in Eq. (5.8) contains projection operators and thus unit vectors, which are integrated over $d\Omega$.

# Additional calculations on BBH neutrinos

In this appendix, we show additional results from our study of neutrinos from BBH mergers.

## J.1   Diffuse flux for monochromatic emission

The diffuse flux in case of monochromatic emission can be easily derived. Starting from Eq. (4.54), which we rewrite as

$$\frac{\mathrm{d}N_\nu}{\mathrm{d}E_\nu}(E_\nu) = \frac{c}{4\pi} \frac{1}{H_0} \int \frac{\mathrm{d}z}{E(z)} \mathcal{H}(z) \left. \frac{\mathrm{d}N_\nu}{\mathrm{d}\epsilon_\nu}(\epsilon_\nu) \right|_{\epsilon_\nu = (1+z)E_\nu}. \tag{J.1}$$

Now, for monochromatic emission at energy $E_0$, the spectrum injected at the source $\frac{\mathrm{d}N_\nu}{\mathrm{d}E_\nu}(E_\nu)$ is given by

$$\frac{\mathrm{d}N_\nu}{\mathrm{d}E_\nu}(E_\nu) = \frac{f^\nu_{\mathrm{BBH}} E_{\mathrm{GW}}}{E_0} \delta(E_\nu - E_0). \tag{J.2}$$

Performing the integration, we obtain

$$\frac{\mathrm{d}N^{\mathrm{mono}}_\nu}{\mathrm{d}E_\nu}(E_\nu) = \begin{cases} \frac{c}{4\pi} \frac{1}{H_0} \frac{\mathcal{H}\left(\frac{E_0}{E}-1\right)}{E\left(\frac{E_0}{E}-1\right)} \frac{f^\nu_{\mathrm{BBH}} E_{\mathrm{GW}}}{E_0} \frac{1}{E} & \frac{E_0}{1+z_{\mathrm{max}}} \leq E_\nu \leq E_0 \\ 0 & \text{otherwise} \end{cases}, \tag{J.3}$$

where we take $z_{\mathrm{max}} = 7$. In case of a single-flavour flux, an additional factor of $1/3$ needs to be taken into account. This flux is shown in Figure J.1 for $f^\nu_{\mathrm{BBH}} = 10^{-2}$. Note that since the spectrum follows the cosmic evolution of the sources, the flux corresponding to a single injection is made up of three distinct parts on a log-log plot when following the star formation rate (and this is not a plotting artefact). The figure shows three such possibilities for the injection energy.

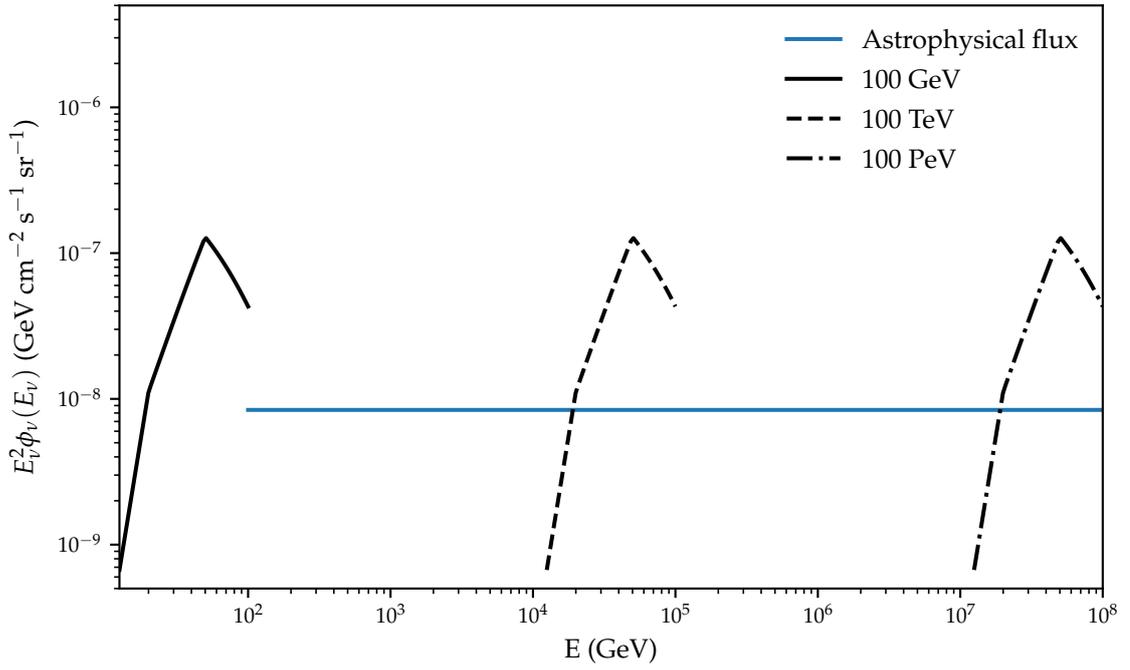

Figure J.1: Comparing single-flavour diffuse flux from monochromatic injection at different energies, each for $f^{\nu}_{\mathrm{BBH}} = 10^{-2}$, with the astrophysical flux measured by IceCube (Eq. (5.25)).

## J.2   Validating assumptions on mass distribution

Here, we show the various possible mass distributions of black hole binaries, compute the average event and check the contribution of each mass combination to the total amount of energy released into gravitational waves, all of which are important for Sections 5.6 and 5.8.

### J.2.1   $E_{\mathrm{GW}}(m_1, m_2)$

We show the energy emitted in gravitational waves as a function of the black hole masses. The energy is estimated using the fit of the final black hole mass after a merger in numerical simulations [881]. This fit gives

$$E_{\mathrm{GW}} = m_1 + m_2 - M_f(m1, m2), \tag{J.4}$$

with $M_f$ the final black hole mass

$$M_f = \left[1 + \left(\sqrt{\frac{8}{9}} - 1\right)\eta - 0.498\eta^2\right](m_1 + m_2), \tag{J.5}$$

and $\eta$ the symmetric mass ratio

$$\eta = \frac{m_1 m_2}{(m_1 + m_2)^2}. \tag{J.6}$$

This result is shown in Figure J.2, for a mass window corresponding to the distributions in the next section. We see that the amount of energy emitted increases with larger black hole masses and with a larger mass ratio $m_2/m_1$ (since this increases both the total mass and the binding energy which can be released, see the estimate in Eq. (5.14)).

### J.2.2   Original mass distributions

Consider the mass distributions considered originally at the detection of GW150914 [484]. For both of these, the black hole masses are restricted to $5\,\mathrm{M}_\odot \leq m_2 \leq m_1$ and $m_1 + m_2 \leq 100\,\mathrm{M}_\odot$. This range was inspired both by the sensitivity of the analysis and the expected masses of astrophysical black holes.

The first distribution is flat in log mass, given by

$$p(m_1, m_2) \propto \frac{1}{m_1 m_2}, \tag{J.7}$$

and shown in Figure J.3. The figure also shows the amount of gravitational waves emitted by each mass combination, weighted by their probability. This gives the importance of each mass combination to the total gravitational wave background. As can be seen, in this case all mass combinations contribute at a comparable level. If the neutrino emission is proportional with $M_{\mathrm{tot}}$, then this implies a near-equal contribution of all mass combinations to a possible diffuse neutrino flux.

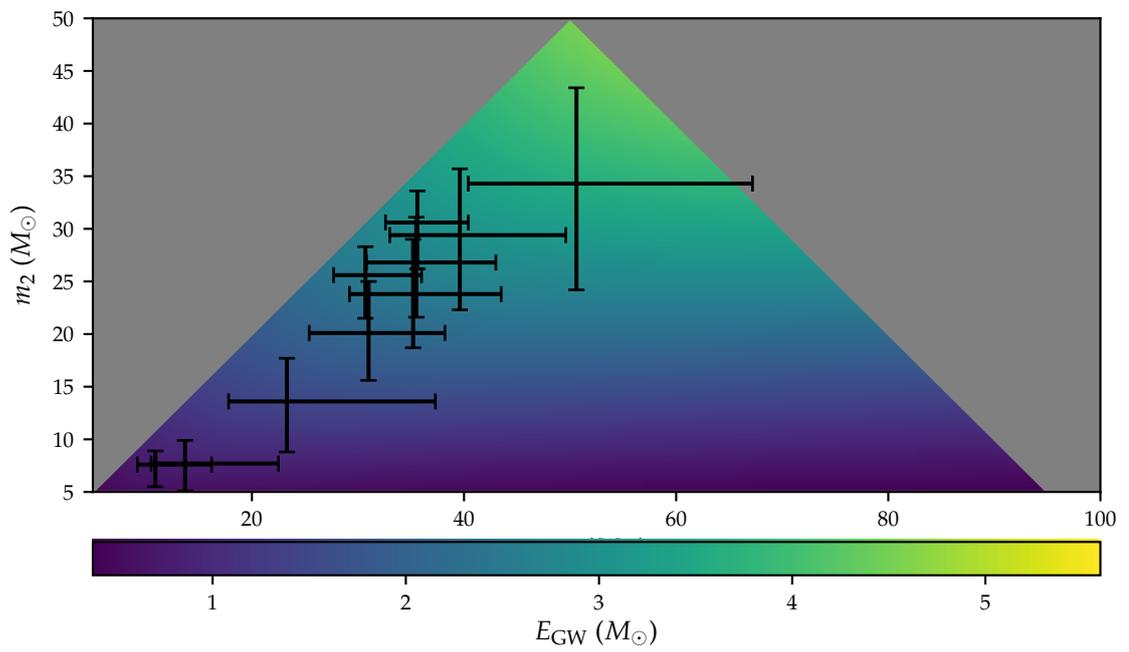

Figure J.2: $E_{\mathrm{GW}}$ as a function of $m_1$ and $m_2$, following Eq. (J.4) [881]. Also shown are the BBH events detected in run O1 and O2.

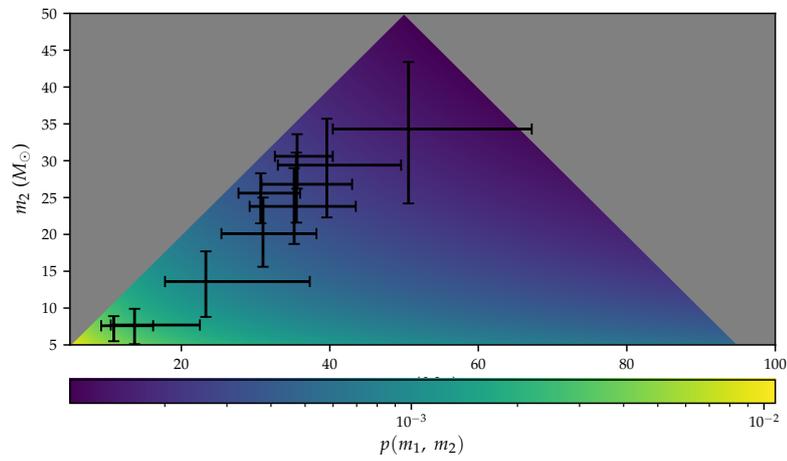

(a) $p(m_1, m_2)$

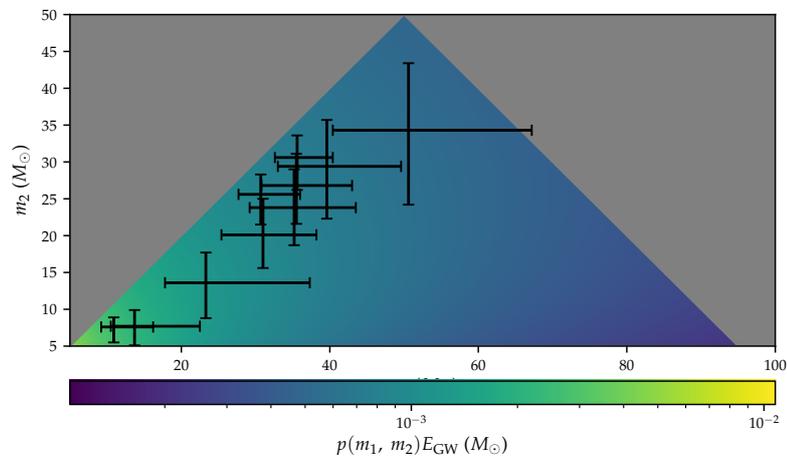

(b) $p(m_1, m_2)E_{\mathrm{GW}}$

Figure J.3: Mass distribution (left) and $E_{\mathrm{GW}}$ distribution (right) for the black hole mass distribution flat in log mass. Also shown are the BBH events detected in run O1 and run O2.

We also show in Figure J.4 the power law distribution, given by

$$p(m_1) \propto m_1^{-2.35}, \tag{J.8}$$

with $m_2$ distributed uniformly. In this case, the equal mass binaries are more important for the gravitational wave emission, although all mass ratios occur equally. This distinction is important, since it means that equal mass binaries are detected more frequently (influencing the limit from the stacked search), while for possible diffuse emission, the lower mass ratios might also be relevant (influencing the astrophysical limit).

### J.2.3 Updated mass distribution

From the previous section, it can already be suspected that the mass distributions are not accurately represented by the models above. Indeed, the models have been updated after run O1 and run O2 [871]. The mass distribution used for their fit, is given by

$$p(m_1, m_2 | m_{\min}, m_{\max}, \alpha, \beta_q) \propto \begin{cases} C(m_1) m_1^{-\alpha} \left(\frac{m_2}{m_1}\right)^{\beta_q} & \text{if } m_{\min} \leq m_2 \leq m_1 \leq m_{\max} \\ 0 & \text{Otherwise,} \end{cases} \tag{J.9}$$

where $C(m_1)$ is chosen such that the marginal distribution $p(m_1) \propto m_1^{-\alpha}$. The analysis focuses on three models, two of which we consider here[1]. In model A, they fix $m_{\min} = 5M_\odot$ and $\beta_q = 0$, such that $C(m_1) \propto 1/(m_1 - m_{\min})$. In model B, all four parameters are fitted. The resulting fit finds for model A $\alpha = 0.4^{+1.3}_{-1.9}$ and $m_{\max} = 41.6^{+9.0}_{-4.5}$. For model B, they find $\alpha = 1.6^{+1.5}_{-1.7}$, $m_{\min} = 7.9^{+1.2}_{-2.5}M_\odot$, $m_{\max} = 42.0^{+15.0}_{-5.7}M_\odot$ and $\beta_q = 6.7^{+4.8}_{-5.9}$. Note that one event, the most massive but least significant, has a central value for the mass higher than $m_{\max}$, but still within the error on both $m_{\max}$ and $m_1$ of this event. Leaving this event out of the fit only lightly changes $m_{\max}$ [871]. The resulting distributions are shown in Figures J.5 and J.6 respectively. For model A, we see that all mergers occur frequently and that they all contribute similarly to the total gravitational wave emission. For model B, which has a preference for equal mass binaries, gravitational waves are mainly emitted by equal mass binaries. This means that the sample of detected events is representative of the events potentially contributing to a diffuse neutrino flux (at least if $f_{\text{BBH}}^\nu \propto M_{\text{tot}}$ approximately).

Since model B is the most sophisticated and represents the detected events well, we used this model to calculate the "average event" (giving $\langle E_{\text{GW}} = 1.52 M_\odot \rangle$) used to calculate the astrophysical constraint in the updated prospects in Section 5.8. On the other hand, the average merger from the detected set of mergers has parameters $\langle m_1 \rangle = 30.61 M_\odot$, $\langle m_2 \rangle = 21.95 M_\odot$, $\langle E_{\text{GW}} \rangle = 2.5 M_\odot$ and $\langle d_L \rangle = 1038$ Mpc. Note however, that $\langle E_{\text{GW}} \rangle \neq E_{\text{GW}}(\langle m_1 \rangle, \langle m_2 \rangle)$, although the values are close. In order to predict the limit from the stacked search, we instead used the average energy flux at Earth. This flux corresponds to the emission of $\sim 5 M_\odot$ from two $\sim 55 M_\odot$ black holes from a distance of 1000 Mpc, calculated using the correct relation $E_{\text{GW}}(m_1, m_2)$.

---

[1]The third adds an additional high mass component.

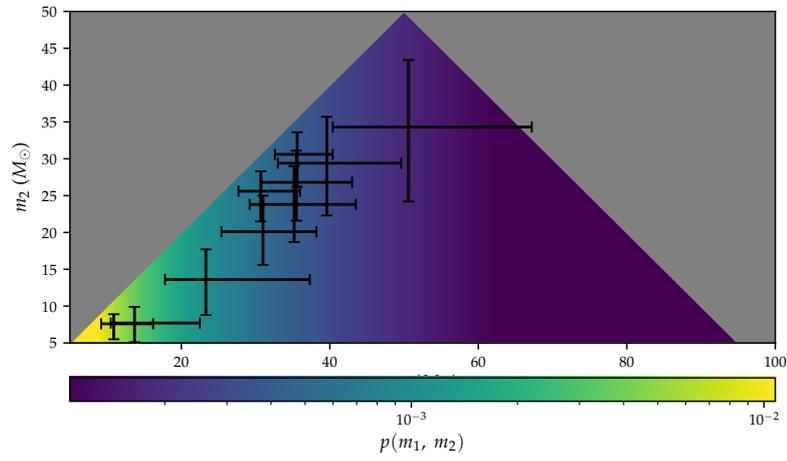

(a) $p(m_1, m_2)$

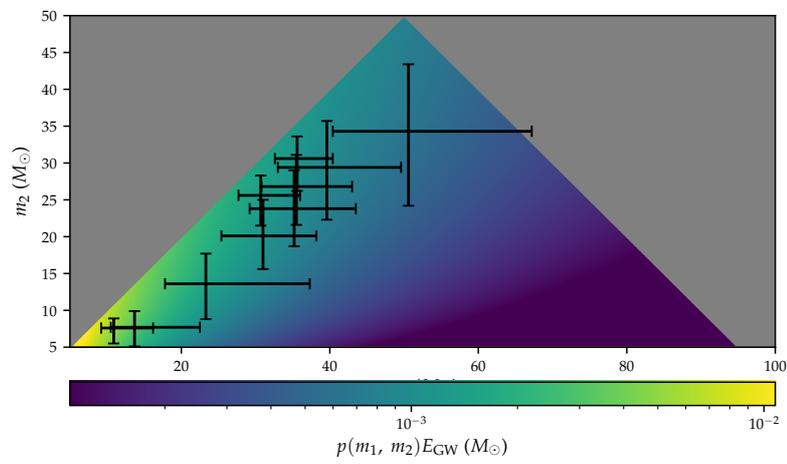

(b) $p(m_1, m_2)E_{\mathrm{GW}}$

Figure J.4: Same as Figure J.3, but for a power law mass distribution.

The distance is chosen arbitrarily and corresponds to the average BBH distance of the detected events. The average $m_1$ and $m_2$ for the astrophysical constraint, with $m_1 = 17.80\text{M}_\odot$ and $m_2 = 15.4\text{M}_\odot$, are marked on the Figure J.6. The masses corresponding to the average detected event are outside the bounds of the figure for the chosen distance.

### J.2.4   *E* **versus** $M_{\text{tot}}$

In Section 5.6, the claim was made that $E_{\text{GW}} \propto M$, since it is true for equal-mass non-spinning black hole binaries, as can be seen from Equations (5.14) and (5.15). This relation is well satisfied for the detected events, as shown in Figure J.7. This is reasonable, since for all these events the black holes are close in mass. If this is true for all detected events, which is reasonable since at the very least equal mass binaries emit stronger in gravitational waves (Section J.2.1), and if $E_\nu^{\text{tot}} \propto M_{\text{tot}}$, then $f_{\text{BBH}}^\nu$ would be universal for at least the stacked search. If all binaries are close to equal mass, then $f_{\text{BBH}}^\nu$ is universal for all mergers and applies also to the astrophysical bound.

## J.3   Check updated prospects

Here we check whether the use of an average event, defined from the currently detected BBH mergers, is appropriate for the stacked search. At the same time, this check also shows whether we could have given a better estimate in the original prospects after GW150914 by defining an appropriate average.

We use the average of events in run O1 and run O2 and repeat the analysis from Section 5.5 in order to obtain a fake prediction of the other run O1 and run O2 events. In the first check, all events in the stacked search are put equal to this average. In the second, we fix the first to GW150914 and let all other events be equal to the average. This is shown in Figure J.8. In the first version, the stacked search outperforms the "expectation" from an average event, since GW150914 is exceptionally powerful. However, after fixing the first event to GW150914, the limits show the same behaviour. In order to show that using $\langle E_{\text{GW}} \rangle$ and $\langle d_L \rangle$ separately leads to a wrong prediction, we repeat the same analysis using these values, shown in Figure J.9. In this case, the stacked search is stronger than the "expectation", even after fixing the first event to GW150914. This deviation is mainly driven by the two most significant events, since the slope of the actual limit and the expectation is similar except for these large drops.

Therefore, defining the appropriate average, it is possible to improve the prediction. However, since this average is here defined using detected events, this was not possible at the time of the writing our original study [803]. Defining an appropriate average from the full distribution of black holes, as it was constrained at the time, requires taking into account the full sensitivity of LIGO. An alternative would have been to use the averaging procedure in Eq. (5.52). However, even there, we need to define a minimum energy flux to be detected, which at the time would have been the one of GW150914, such that only powerful events would have been taken into account.

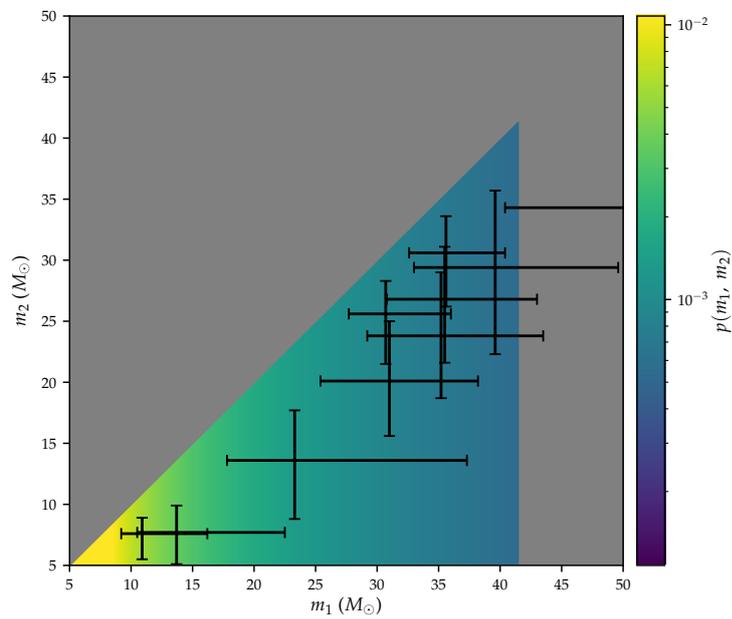

(a) $p(m_1, m_2)$

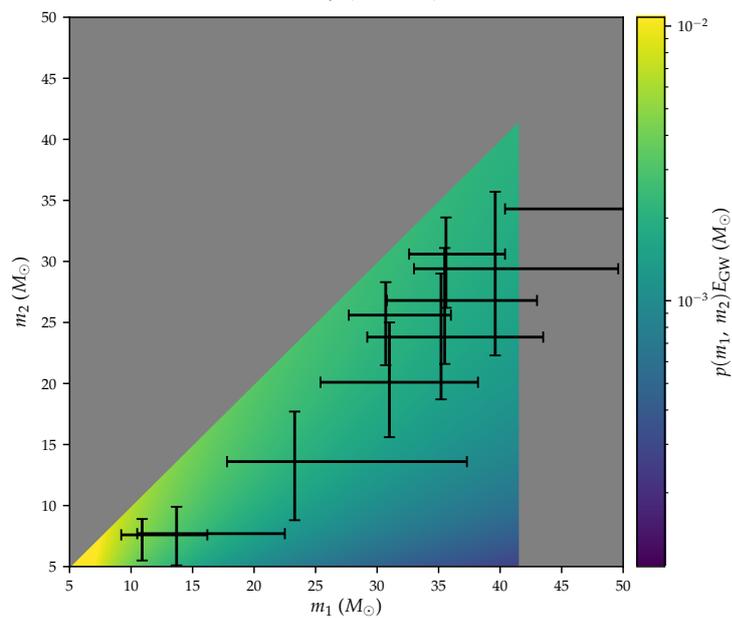

(b) $p(m_1, m_2) E_{\mathrm{GW}}$

Figure J.5: Same as Figure J.3, but for model A.

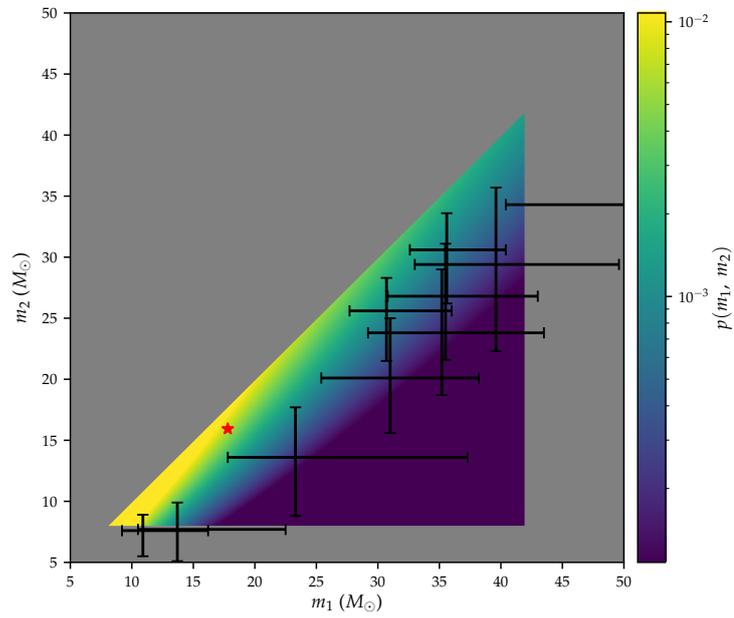

(a) $p(m_1, m_2)$

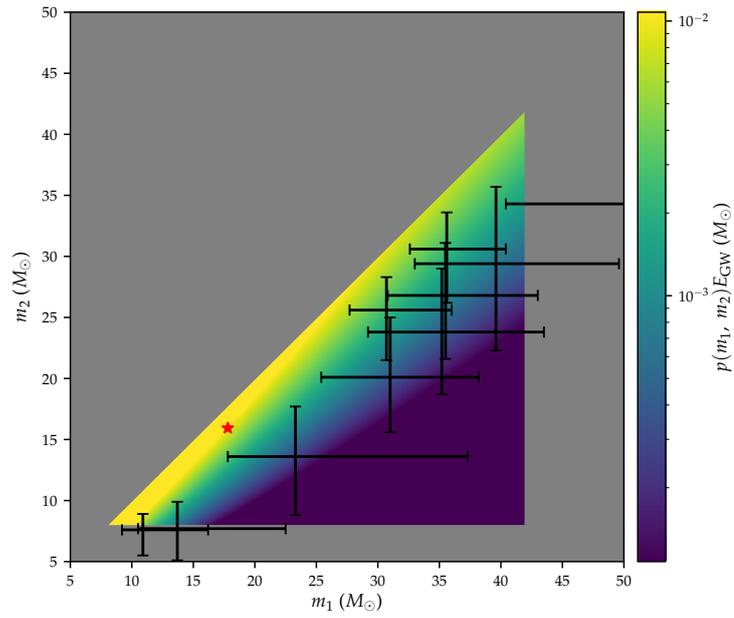

(b) $p(m_1, m_2)E_{\mathrm{GW}}$

Figure J.6: Same as Figure J.3, but for model B. Also shown are the average $m_1$ and $m_2$ used for the astrophysical constraint.

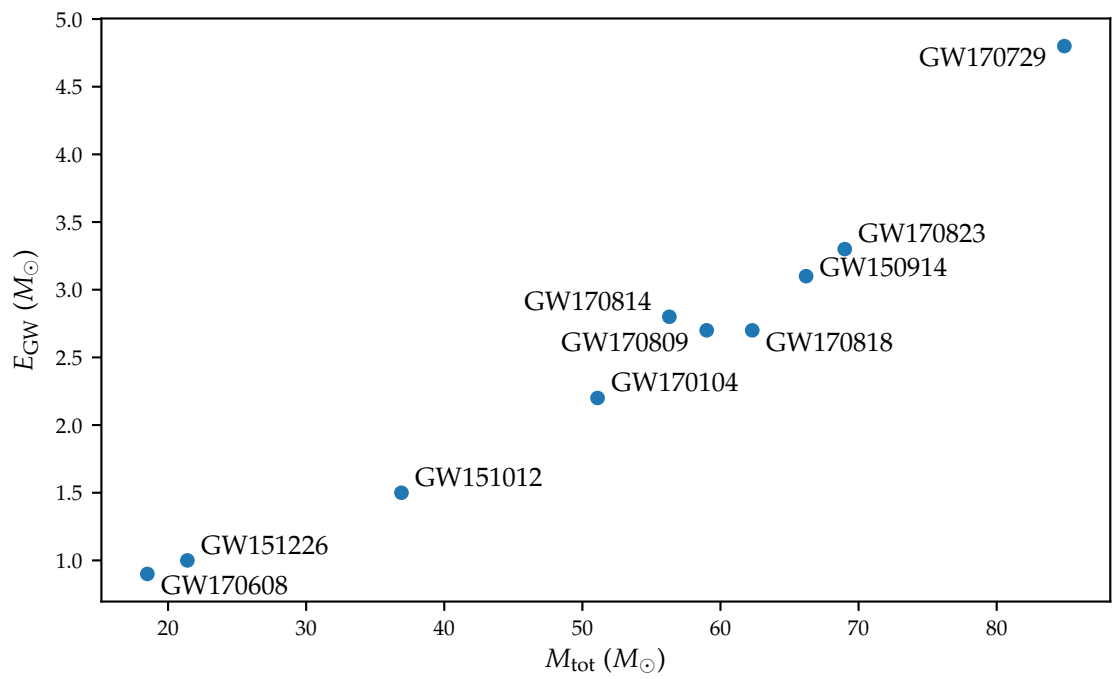

Figure J.7: Comparison of $E_{GW}$ with $M_f = m_1 + m_2$ for the observed events in run O1 and O2.

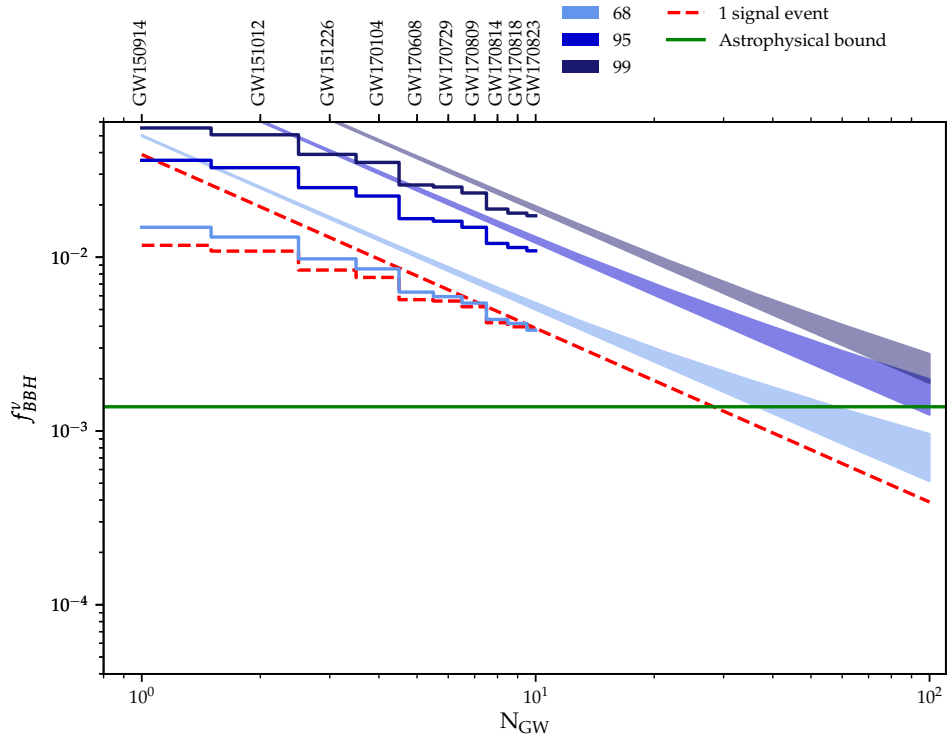

(a) All events average

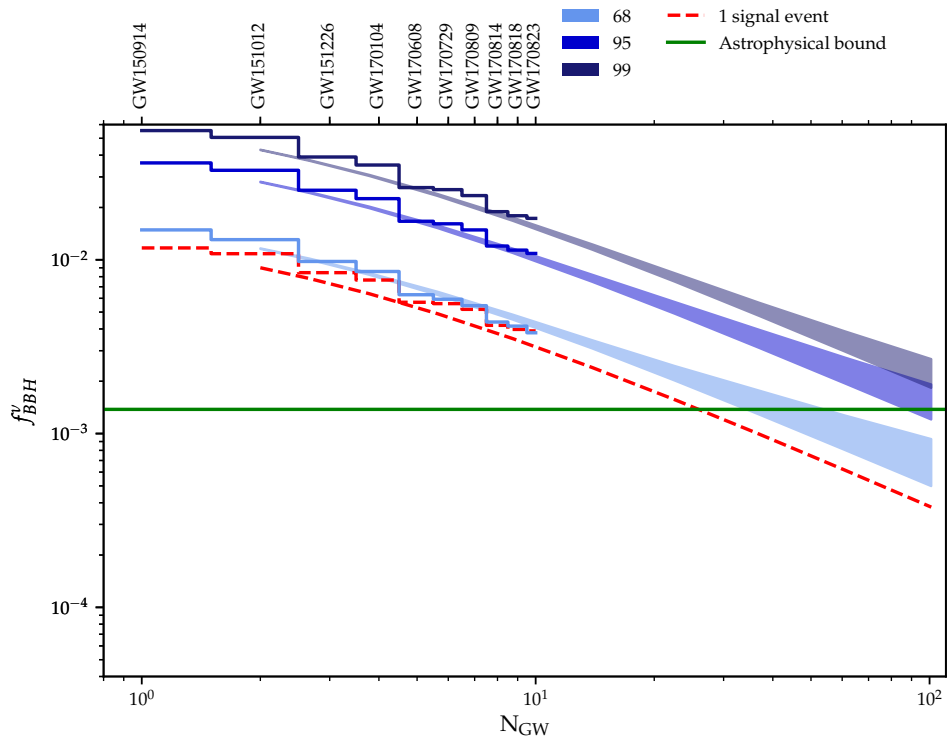

(b) First event fixed to GW150914

Figure J.8: A check on the original and the updated prospects of Sections 5.5 and 5.8, calculating the expected average upper limit for an average event $\langle E_{\mathrm{GW}}/4\pi d_L^2 \rangle$, instead of using the parameters of GW150914.

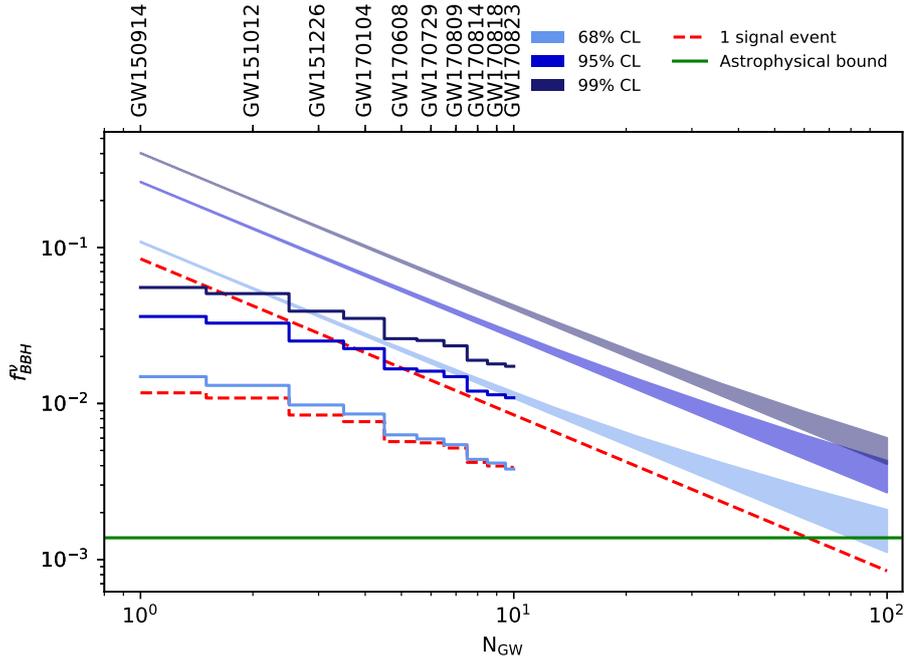

(a) All events average

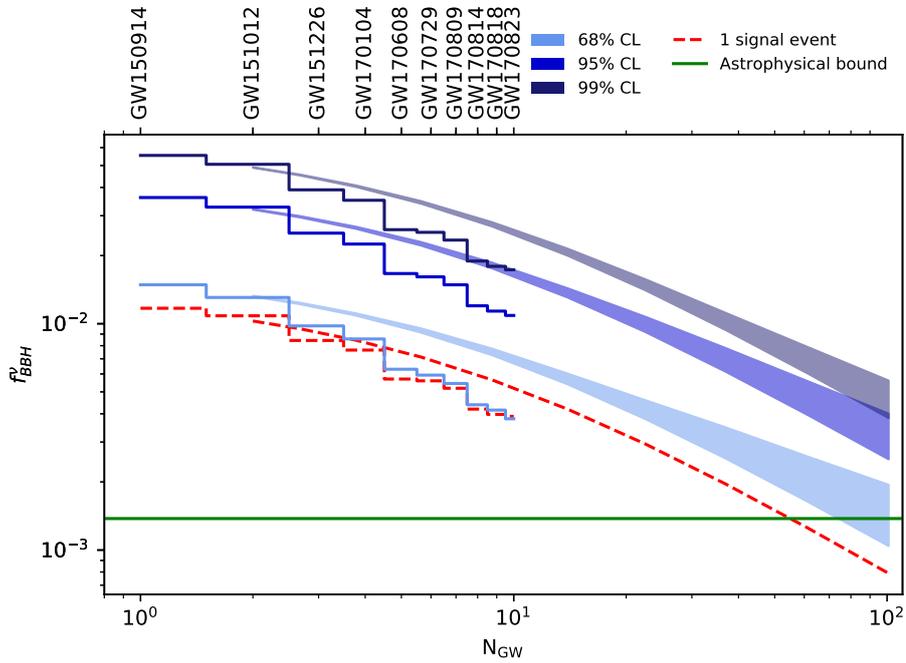

(b) First event fixed to GW150914

Figure J.9: Same as Figure J.8, but using as event parameters $\langle E_{\mathrm{GW}} \rangle$ and $\langle d_L \rangle$ separately.

# Bibliography

[1] Chen-Ning Yang and Robert L. Mills. 'Conservation of Isotopic Spin and Isotopic Gauge Invariance'. In: *Phys. Rev.* 96 (1954). [,150(1954)], pp. 191–195. DOI: `10.1103/PhysRev.96.191`.

[2] S. L. Glashow. 'Partial Symmetries of Weak Interactions'. In: *Nucl. Phys.* 22 (1961), pp. 579–588. DOI: `10.1016/0029-5582(61)90469-2`.

[3] Steven Weinberg. 'A Model of Leptons'. In: *Phys. Rev. Lett.* 19 (1967), pp. 1264–1266. DOI: `10.1103/PhysRevLett.19.1264`.

[4] Abdus Salam. 'Weak and Electromagnetic Interactions'. In: *Conf. Proc.* C680519 (1968), pp. 367–377.

[5] F. Englert and R. Brout. 'Broken Symmetry and the Mass of Gauge Vector Mesons'. In: *Phys. Rev. Lett.* 13 (1964). [,157(1964)], pp. 321–323. DOI: `10.1103/PhysRevLett.13.321`.

[6] Peter W. Higgs. 'Broken Symmetries and the Masses of Gauge Bosons'. In: *Phys. Rev. Lett.* 13 (1964). [,160(1964)], pp. 508–509. DOI: `10.1103/PhysRevLett.13.508`.

[7] G. S. Guralnik, C. R. Hagen and T. W. B. Kibble. 'Global Conservation Laws and Massless Particles'. In: *Phys. Rev. Lett.* 13 (1964). [,162(1964)], pp. 585–587. DOI: `10.1103/PhysRevLett.13.585`.

[8] Murray Gell-Mann. 'Symmetries of baryons and mesons'. In: *Phys. Rev.* 125 (1962), pp. 1067–1084. DOI: `10.1103/PhysRev.125.1067`.

[9] Murray Gell-Mann. 'A Schematic Model of Baryons and Mesons'. In: *Phys. Lett.* 8 (1964), pp. 214–215. DOI: `10.1016/S0031-9163(64)92001-3`.

[10] David J. Gross and Frank Wilczek. 'Ultraviolet Behavior of Nonabelian Gauge Theories'. In: *Phys. Rev. Lett.* 30 (1973). [,271(1973)], pp. 1343–1346. DOI: `10.1103/PhysRevLett.30.1343`.

[11] D. J. Gross and Frank Wilczek. 'Asymptotically Free Gauge Theories - I'. In: *Phys. Rev.* D8 (1973), pp. 3633–3652. DOI: `10.1103/PhysRevD.8.3633`.

[12] D. J. Gross and Frank Wilczek. 'ASYMPTOTICALLY FREE GAUGE THEORIES. 2.' In: *Phys. Rev.* D9 (1974), pp. 980–993. DOI: `10.1103/PhysRevD.9.980`.

[13] H. David Politzer. 'Reliable Perturbative Results for Strong Interactions?' In: *Phys. Rev. Lett.* 30 (1973). [,274(1973)], pp. 1346–1349. DOI: `10.1103/PhysRevLett.30.1346`.

[14] H. David Politzer. 'Asymptotic Freedom: An Approach to Strong Interactions'. In: *Phys. Rept.* 14 (1974), pp. 129–180. DOI: 10.1016/0370-1573(74)90014-3.

[15] J. J. Thomson. 'Cathode rays'. In: *Phil. Mag. Ser.5* 44 (1897), pp. 293–316. DOI: 10.1080/14786449708621070.

[16] E. Rutherford. 'The scattering of alpha and beta particles by matter and the structure of the atom'. In: *Phil. Mag. Ser.6* 21 (1911), pp. 669–688. DOI: 10.1080/14786440508637080.

[17] E. Rutherford. 'Collision of $\alpha$ particles with light atoms. IV. An anomalous effect in nitrogen'. In: *Phil. Mag. Ser.6* 37 (1919). [Phil. Mag.90,no.sup1,31(2010)], pp. 581–587. DOI: 10.1080/14786431003659230.

[18] Carl D. Anderson. 'The Positive Electron'. In: *Phys. Rev.* 43 (6 Mar. 1933), pp. 491–494. DOI: 10.1103/PhysRev.43.491. URL: https://link.aps.org/doi/10.1103/PhysRev.43.491.

[19] S. H. Neddermeyer and C. D. Anderson. 'Note on the Nature of Cosmic Ray Particles'. In: *Phys. Rev.* 51 (1937), pp. 884–886. DOI: 10.1103/PhysRev.51.884.

[20] Frederick Reines and Clyde L. Cowan. 'The neutrino'. In: *Nature* 178 (1956), pp. 446–449. DOI: 10.1038/178446a0.

[21] Elliott D. Bloom et al. 'High-Energy Inelastic e p Scattering at 6-Degrees and 10-Degrees'. In: *Phys. Rev. Lett.* 23 (1969), pp. 930–934. DOI: 10.1103/PhysRevLett.23.930.

[22] Martin Breidenbach et al. 'Observed Behavior of Highly Inelastic electron-Proton Scattering'. In: *Phys. Rev. Lett.* 23 (1969), pp. 935–939. DOI: 10.1103/PhysRevLett.23.935.

[23] G. Arnison et al. 'Experimental Observation of Isolated Large Transverse Energy Electrons with Associated Missing Energy at s**(1/2) = 540-GeV'. In: *Phys. Lett.* B122 (1983). [,611(1983)], pp. 103–116. DOI: 10.1016/0370-2693(83)91177-2.

[24] G. Arnison et al. 'Experimental Observation of Lepton Pairs of Invariant Mass Around 95-GeV/c**2 at the CERN SPS Collider'. In: *Phys. Lett.* B126 (1983). [,7.55(1983)], pp. 398–410. DOI: 10.1016/0370-2693(83)90188-0.

[25] F. Abe et al. 'Observation of top quark production in $\bar{p}p$ collisions'. In: *Phys. Rev. Lett.* 74 (1995), pp. 2626–2631. DOI: 10.1103/PhysRevLett.74.2626. arXiv: hep-ex/9503002 [hep-ex].

[26] S. Abachi et al. 'Observation of the top quark'. In: *Phys. Rev. Lett.* 74 (1995), pp. 2632–2637. DOI: 10.1103/PhysRevLett.74.2632. arXiv: hep-ex/9503003 [hep-ex].

[27] Georges Aad et al. 'Observation of a new particle in the search for the Standard Model Higgs boson with the ATLAS detector at the LHC'. In: *Phys. Lett.* B716 (2012), pp. 1–29. DOI: 10.1016/j.physletb.2012.08.020. arXiv: 1207.7214 [hep-ex].

[28]  Serguei Chatrchyan et al. 'Observation of a New Boson at a Mass of 125 GeV with the CMS Experiment at the LHC'. In: *Phys. Lett.* B716 (2012), pp. 30–61. DOI: 10.1016/j.physletb.2012.08.021. arXiv: 1207.7235 [hep-ex].

[29]  M. Tanabashi et al. 'Review of Particle Physics'. In: *Phys. Rev.* D98.3 (2018), p. 030001. DOI: 10.1103/PhysRevD.98.030001.

[30]  Nicola Cabibbo. 'Unitary Symmetry and Leptonic Decays'. In: *Phys. Rev. Lett.* 10 (1963). [,648(1963)], pp. 531–533. DOI: 10.1103/PhysRevLett.10.531.

[31]  Makoto Kobayashi and Toshihide Maskawa. 'CP Violation in the Renormalizable Theory of Weak Interaction'. In: *Prog. Theor. Phys.* 49 (1973), pp. 652–657. DOI: 10.1143/PTP.49.652.

[32]  Ziro Maki, Masami Nakagawa and Shoichi Sakata. 'Remarks on the unified model of elementary particles'. In: *Prog. Theor. Phys.* 28 (1962). [,34(1962)], pp. 870–880. DOI: 10.1143/PTP.28.870.

[33]  B. Pontecorvo. 'Inverse beta processes and nonconservation of lepton charge'. In: *Sov. Phys. JETP* 7 (1958). [Zh. Eksp. Teor. Fiz.34,247(1957)], pp. 172–173.

[34]  Izaak Neutelings. *How to draw diagrams in LaTeX with TikZ*. https://wiki.physik.uzh.ch/cms/latex:tikz. Accessed on 08-08-2019. July 2017.

[35]  Travis E. Oliphant. 'Python for Scientific Computing'. In: *Computing in Science & Engineering* 9.3 (2007), pp. 10–20. DOI: 10.1109/MCSE.2007.58. eprint: \url{https://aip.scitation.org/doi/pdf/10.1109/MCSE.2007.58}.

[36]  K. Jarrod Millman and Michael Aivazis. 'Python for Scientists and Engineers'. In: *Computing in Science & Engineering* 13.2 (2011), pp. 9–12. DOI: 10.1109/MCSE.2011.36. eprint: \url{https://aip.scitation.org/doi/pdf/10.1109/MCSE.2011.36}.

[37]  Eric Jones, Travis Oliphant, Pearu Peterson et al. *SciPy: Open source scientific tools for Python*. 2001–. URL: http://www.scipy.org/.

[38]  Travis E Oliphant. *A guide to NumPy*. Vol. 1. Trelgol Publishing USA, 2006.

[39]  S. van der Walt, S. C. Colbert and G. Varoquaux. 'The NumPy Array: A Structure for Efficient Numerical Computation'. In: *Computing in Science Engineering* 13.2 (Mar. 2011), pp. 22–30. ISSN: 1521-9615. DOI: 10.1109/MCSE.2011.37.

[40]  John D. Hunter. 'Matplotlib: A 2D Graphics Environment'. In: *Computing in Science & Engineering* 9.3 (2007), pp. 90–95. DOI: 10.1109/MCSE.2007.55. eprint: \url{https://aip.scitation.org/doi/pdf/10.1109/MCSE.2007.55}.

[41]  Wes McKinney. 'Data Structures for Statistical Computing in Python'. In: *Proceedings of the 9th Python in Science Conference*. Ed. by Stéfan van der Walt and Jarrod Millman. 2010, pp. 51–56.

[42]  R. Brun and F. Rademakers. 'ROOT: An object oriented data analysis framework'. In: *Nucl. Instrum. Meth.* A389 (1997), pp. 81–86. DOI: 10.1016/S0168-9002(97)00048-X.

[43] D. Binosi and L. Theußl. 'JaxoDraw: A graphical user interface for drawing Feynman diagrams'. In: *Computer Physics Communications* 161.1 (2004), pp. 76–86. ISSN: 0010-4655. DOI: https://doi.org/10.1016/j.cpc.2004.05.001. URL: http://www.sciencedirect.com/science/article/pii/S0010465504002115.

[44] Adam G. Riess et al. 'Observational evidence from supernovae for an accelerating universe and a cosmological constant'. In: *Astron. J.* 116 (1998), pp. 1009–1038. DOI: 10.1086/300499. arXiv: astro-ph/9805201 [astro-ph].

[45] S. Perlmutter et al. 'Measurements of Omega and Lambda from 42 high redshift supernovae'. In: *Astrophys. J.* 517 (1999), pp. 565–586. DOI: 10.1086/307221. arXiv: astro-ph/9812133 [astro-ph].

[46] Edward W. Kolb and Michael S. Turner. 'The Early Universe'. In: *Front. Phys.* 69 (1990), pp. 1–547.

[47] Gerard Jungman, Marc Kamionkowski and Kim Griest. 'Supersymmetric dark matter'. In: *Phys. Rept.* 267 (1996), pp. 195–373. DOI: 10.1016/0370-1573(95)00058-5. arXiv: hep-ph/9506380 [hep-ph].

[48] A. D. Sakharov. 'Violation of CP Invariance, C asymmetry, and baryon asymmetry of the universe'. In: *Pisma Zh. Eksp. Teor. Fiz.* 5 (1967). [Usp. Fiz. Nauk161,no.5,61(1991)], pp. 32–35. DOI: 10.1070/PU1991v034n05ABEH002497.

[49] Raymond Davis Jr., Don S. Harmer and Kenneth C. Hoffman. 'Search for neutrinos from the sun'. In: *Phys. Rev. Lett.* 20 (1968), pp. 1205–1209. DOI: 10.1103/PhysRevLett.20.1205.

[50] Y. Fukuda et al. 'Evidence for oscillation of atmospheric neutrinos'. In: *Phys. Rev. Lett.* 81 (1998), pp. 1562–1567. DOI: 10.1103/PhysRevLett.81.1562. arXiv: hep-ex/9807003 [hep-ex].

[51] Y. Abe et al. 'Indication of Reactor $\bar{\nu}_e$ Disappearance in the Double Chooz Experiment'. In: *Phys. Rev. Lett.* 108 (2012), p. 131801. DOI: 10.1103/PhysRevLett.108.131801. arXiv: 1112.6353 [hep-ex].

[52] B. Pontecorvo. 'Mesonium and anti-mesonium'. In: *Sov. Phys. JETP* 6 (1957). [Zh. Eksp. Teor. Fiz.33,549(1957)], p. 429.

[53] B. Pontecorvo. 'Neutrino Experiments and the Problem of Conservation of Leptonic Charge'. In: *Sov. Phys. JETP* 26 (1968). [Zh. Eksp. Teor. Fiz.53,1717(1967)], pp. 984–988.

[54] G. W. Bennett et al. 'Final Report of the Muon E821 Anomalous Magnetic Moment Measurement at BNL'. In: *Phys. Rev.* D73 (2006), p. 072003. DOI: 10.1103/PhysRevD.73.072003. arXiv: hep-ex/0602035 [hep-ex].

[55] Richard H. Parker et al. 'Measurement of the fine-structure constant as a test of the Standard Model'. In: *Science* 360.6385 (2018), pp. 191–195. ISSN: 0036-8075. DOI: 10.1126/science.aap7706. eprint: https://science.sciencemag.org/content/360/6385/191.full.pdf. URL: https://science.sciencemag.org/content/360/6385/191.

[56] A. Falkowski. *Both g-2 anomalies*. `https://resonaances.blogspot.com/2018/06/alpha-and-g-minus-two.html`. Last accessed on 03-10-19.

[57] Michel Davier et al. 'Reevaluation of the hadronic vacuum polarisation contributions to the Standard Model predictions of the muon $g - 2$ and $\alpha(m_Z^2)$ using newest hadronic cross-section data'. In: *Eur. Phys. J.* C77.12 (2017), p. 827. DOI: `10.1140/epjc/s10052-017-5161-6`. arXiv: `1706.09436 [hep-ph]`.

[58] F. Jegerlehner. 'The Muon $g - 2$ in Progress'. In: *Acta Physica Polonica B* 49.6 (2018), p. 1157. ISSN: 1509-5770. DOI: `10.5506/aphyspolb.49.1157`. URL: `http://dx.doi.org/10.5506/APhysPolB.49.1157`.

[59] Ying Li and Cai-Dian Lü. 'Recent Anomalies in B Physics'. In: *Sci. Bull.* 63 (2018), pp. 267–269. DOI: `10.1016/j.scib.2018.02.003`. arXiv: `1808.02990 [hep-ph]`.

[60] H. Georgi and S. L. Glashow. 'Unity of All Elementary Particle Forces'. In: *Phys. Rev. Lett.* 32 (1974), pp. 438–441. DOI: `10.1103/PhysRevLett.32.438`.

[61] Luis Alvarez-Gaume and Edward Witten. 'Gravitational Anomalies'. In: *Nucl. Phys.* B234 (1984). [,269(1983)], p. 269. DOI: `10.1016/0550-3213(84)90066-X`.

[62] John R. Ellis. 'The Superstring: Theory of Everything, or of Nothing?' In: *Nature* 323 (1986), pp. 595–598. DOI: `10.1038/323595a0`.

[63] Steven Weinberg. 'Implications of Dynamical Symmetry Breaking'. In: *Phys. Rev.* D13 (1976). [Addendum: Phys. Rev.D19,1277(1979)], pp. 974–996. DOI: `10.1103/PhysRevD.19.1277,10.1103/PhysRevD.13.974`.

[64] Eldad Gildener. 'Gauge Symmetry Hierarchies'. In: *Phys. Rev.* D14 (1976), p. 1667. DOI: `10.1103/PhysRevD.14.1667`.

[65] Leonard Susskind. 'Dynamics of Spontaneous Symmetry Breaking in the Weinberg-Salam Theory'. In: *Phys. Rev.* D20 (1979), pp. 2619–2625. DOI: `10.1103/PhysRevD.20.2619`.

[66] Gerard 't Hooft et al. 'Recent Developments in Gauge Theories. Proceedings, Nato Advanced Study Institute, Cargese, France, August 26 - September 8, 1979'. In: *NATO Sci. Ser. B* 59 (1980), pp.1–438. DOI: `10.1007/978-1-4684-7571-5`.

[67] Stephen P. Martin. 'A Supersymmetry primer'. In: *Adv.Ser.Direct.High Energy Phys.* 21 (2010), pp. 1–153. DOI: `10.1142/9789814307505_0001`. arXiv: `hep-ph/9709356 [hep-ph]`.

[68] Savas Dimopoulos and Stuart Raby. 'Supercolor'. In: *Nucl. Phys.* B192 (1981), pp. 353–368. DOI: `10.1016/0550-3213(81)90430-2`.

[69] Edward Witten. 'Dynamical Breaking of Supersymmetry'. In: *Nucl. Phys.* B188 (1981), p. 513. DOI: `10.1016/0550-3213(81)90006-7`.

[70] Michael Dine, Willy Fischler and Mark Srednicki. 'Supersymmetric Technicolor'. In: *Nucl. Phys.* B189 (1981), pp. 575–593. DOI: `10.1016/0550-3213(81)90582-4`.

[71] Savas Dimopoulos and Howard Georgi. 'Softly Broken Supersymmetry and SU(5)'. In: *Nucl. Phys.* B193 (1981), pp. 150–162. DOI: 10.1016/0550-3213(81)90522-8.

[72] N. Sakai. 'Naturalness in Supersymmetric Guts'. In: *Z. Phys.* C11 (1981), p. 153. DOI: 10.1007/BF01573998.

[73] Romesh K. Kaul and Parthasarathi Majumdar. 'Cancellation of Quadratically Divergent Mass Corrections in Globally Supersymmetric Spontaneously Broken Gauge Theories'. In: *Nucl. Phys.* B199 (1982), p. 36. DOI: 10.1016/0550-3213(82)90565-X.

[74] Michael J. Dugan, Howard Georgi and David B. Kaplan. 'Anatomy of a Composite Higgs Model'. In: *Nucl. Phys.* B254 (1985), pp. 299–326. DOI: 10.1016/0550-3213(85)90221-4.

[75] Kaustubh Agashe, Roberto Contino and Alex Pomarol. 'The Minimal composite Higgs model'. In: *Nucl. Phys.* B719 (2005), pp. 165–187. DOI: 10.1016/j.nuclphysb.2005.04.035. arXiv: hep-ph/0412089 [hep-ph].

[76] Oliver Witzel. 'Review on Composite Higgs Models'. In: *PoS* LATTICE2018 (2019), p. 006. DOI: 10.22323/1.334.0006. arXiv: 1901.08216 [hep-lat].

[77] Z. Chacko, Hock-Seng Goh and Roni Harnik. 'The Twin Higgs: Natural electroweak breaking from mirror symmetry'. In: *Phys. Rev. Lett.* 96 (2006), p. 231802. DOI: 10.1103/PhysRevLett.96.231802. arXiv: hep-ph/0506256 [hep-ph].

[78] T. Appelquist, A. Chodos and P. G. O. Freund, eds. *MODERN KALUZA-KLEIN THEORIES*. 1987.

[79] Nima Arkani-Hamed, Savas Dimopoulos and G. R. Dvali. 'The Hierarchy problem and new dimensions at a millimeter'. In: *Phys. Lett.* B429 (1998), pp. 263–272. DOI: 10.1016/S0370-2693(98)00466-3. arXiv: hep-ph/9803315 [hep-ph].

[80] Lisa Randall and Raman Sundrum. 'A Large mass hierarchy from a small extra dimension'. In: *Phys. Rev. Lett.* 83 (1999), pp. 3370–3373. DOI: 10.1103/PhysRevLett.83.3370. arXiv: hep-ph/9905221 [hep-ph].

[81] Peter W. Graham, David E. Kaplan and Surjeet Rajendran. 'Cosmological Relaxation of the Electroweak Scale'. In: *Phys. Rev. Lett.* 115.22 (2015), p. 221801. DOI: 10.1103/PhysRevLett.115.221801. arXiv: 1504.07551 [hep-ph].

[82] Kiwoon Choi and Sang Hui Im. 'Realizing the relaxion from multiple axions and its UV completion with high scale supersymmetry'. In: *JHEP* 01 (2016), p. 149. DOI: 10.1007/JHEP01(2016)149. arXiv: 1511.00132 [hep-ph].

[83] David E. Kaplan and Riccardo Rattazzi. 'Large field excursions and approximate discrete symmetries from a clockwork axion'. In: *Phys. Rev.* D93.8 (2016), p. 085007. DOI: 10.1103/PhysRevD.93.085007. arXiv: 1511.01827 [hep-ph].

[84] Gian F. Giudice and Matthew McCullough. 'A Clockwork Theory'. In: *JHEP* 02 (2017), p. 036. DOI: 10.1007/JHEP02(2017)036. arXiv: 1610.07962 [hep-ph].

[85] M. F. Sohnius. 'Introducing Supersymmetry'. In: *Phys. Rept.* 128 (1985), pp. 39–204. DOI: 10.1016/0370-1573(85)90023-7.

[86] J. Wess and J. Bagger. *Supersymmetry and supergravity*. Princeton, NJ, USA: Princeton University Press, 1992. ISBN: 9780691025308.

[87] Steven Weinberg. *The quantum theory of fields. Vol. 3: Supersymmetry*. Cambridge University Press, 2013. ISBN: 9780521670555, 9781139632638, 9780521670555.

[88] Herbi K. Dreiner, Howard E. Haber and Stephen P. Martin. 'Two-component spinor techniques and Feynman rules for quantum field theory and supersymmetry'. In: *Phys. Rept.* 494 (2010), pp. 1–196. DOI: 10.1016/j.physrep.2010.05.002. arXiv: 0812.1594 [hep-ph].

[89] Philip C. Argyres. 'An Introduction to Global Supersymmetry'. In: (2001).

[90] Riccardo Argurio. 'Introduction to supersymmetry'. In: (2017).

[91] Daniel Z. Freedman and Antoine Van Proeyen. *Supergravity*. Cambridge, UK: Cambridge Univ. Press, 2012. ISBN: 9781139368063, 9780521194013. URL: http://www.cambridge.org/mw/academic/subjects/physics/theoretical-physics-and-mathematical-physics/supergravity?format=AR.

[92] Sidney R. Coleman and J. Mandula. 'All Possible Symmetries of the S Matrix'. In: *Phys. Rev.* 159 (1967), pp. 1251–1256. DOI: 10.1103/PhysRev.159.1251.

[93] Edward Witten. 'Introduction to Supersymmetry'. In: *THE UNITY OF THE FUNDAMENTAL INTERACTIONS. PROCEEDINGS, 19TH COURSE OF THE INTERNATIONAL SCHOOL OF SUBNUCLEAR PHYSICS, ERICE, ITALY, JULY 31 - AUGUST 11, 1981*. 1983, pp. 305–371. DOI: 10.1007/978-1-4613-3655-6_7. URL: http://link.springer.com/chapter/10.1007/978-1-4613-3655-6_7.

[94] Rudolf Haag, Jan T. Lopuszanski and Martin Sohnius. 'All Possible Generators of Supersymmetries of the s Matrix'. In: *Nucl. Phys.* B88 (1975). [,257(1974)], p. 257. DOI: 10.1016/0550-3213(75)90279-5.

[95] Eugene P. Wigner. 'On Unitary Representations of the Inhomogeneous Lorentz Group'. In: *Annals Math.* 40 (1939). [Reprint: Nucl. Phys. Proc. Suppl.6,9(1989)], pp. 149–204. DOI: 10.2307/1968551.

[96] Pierre Ramond. 'Dual Theory for Free Fermions'. In: *Phys. Rev.* D3 (1971), pp. 2415–2418. DOI: 10.1103/PhysRevD.3.2415.

[97] A. Neveu and J. H. Schwarz. 'Factorizable dual model of pions'. In: *Nucl. Phys.* B31 (1971), pp. 86–112. DOI: 10.1016/0550-3213(71)90448-2.

[98] Jean-Loup Gervais and B. Sakita. 'Field Theory Interpretation of Supergauges in Dual Models'. In: *Nucl. Phys.* B34 (1971). [,154(1971)], pp. 632–639. DOI: 10.1016/0550-3213(71)90351-8.

[99] Yu. A. Golfand and E. P. Likhtman. 'Extension of the Algebra of Poincare Group Generators and Violation of p Invariance'. In: *JETP Lett.* 13 (1971). [Pisma Zh. Eksp. Teor. Fiz.13,452(1971)], pp. 323–326.

[100] J. Wess and B. Zumino. 'Supergauge Transformations in Four-Dimensions'. In: *Nucl. Phys.* B70 (1974). [,24(1974)], pp. 39–50. DOI: 10.1016/0550-3213(74)90355-1.

[101] D. V. Volkov and V. P. Akulov. 'Is the Neutrino a Goldstone Particle?' In: *Phys. Lett.* 46B (1973), pp. 109–110. DOI: 10.1016/0370-2693(73)90490-5.

[102] John R. Ellis, S. Kelley and Dimitri V. Nanopoulos. 'Probing the desert using gauge coupling unification'. In: *Phys. Lett.* B260 (1991), pp. 131–137. DOI: 10.1016/0370-2693(91)90980-5.

[103] Ugo Amaldi, Wim de Boer and Hermann Furstenau. 'Comparison of grand unified theories with electroweak and strong coupling constants measured at LEP'. In: *Phys. Lett.* B260 (1991), pp. 447–455. DOI: 10.1016/0370-2693(91)91641-8.

[104] Paul Langacker and Ming-xing Luo. 'Implications of precision electroweak experiments for $M_t$, $\rho_0$, $\sin^2 \theta_W$ and grand unification'. In: *Phys. Rev.* D44 (1991), pp. 817–822. DOI: 10.1103/PhysRevD.44.817.

[105] C. Giunti, C. W. Kim and U. W. Lee. 'Running coupling constants and grand unification models'. In: *Mod. Phys. Lett.* A6 (1991), pp. 1745–1755. DOI: 10.1142/S0217732391001883.

[106] M. Gell-Mann. 'The interpretation of the new particles as displaced charge multiplets'. In: *Nuovo Cim.* 4.S2 (1956), pp. 848–866. DOI: 10.1007/BF02748000.

[107] Abdus Salam and J. A. Strathdee. 'Supergauge Transformations'. In: *Nucl. Phys.* B76 (1974), pp. 477–482. DOI: 10.1016/0550-3213(74)90537-9.

[108] S. Ferrara, J. Wess and B. Zumino. 'Supergauge Multiplets and Superfields'. In: *Phys. Lett.* 51B (1974), p. 239. DOI: 10.1016/0370-2693(74)90283-4.

[109] J. Wess and B. Zumino. 'Supergauge Invariant Extension of Quantum Electrodynamics'. In: *Nucl. Phys.* B78 (1974), p. 1. DOI: 10.1016/0550-3213(74)90112-6.

[110] Pierre Fayet and J. Iliopoulos. 'Spontaneously Broken Supergauge Symmetries and Goldstone Spinors'. In: *Phys. Lett.* 51B (1974), pp. 461–464. DOI: 10.1016/0370-2693(74)90310-4.

[111] Emmy Noether. 'Invariant Variation Problems'. In: *Gott. Nachr.* 1918 (1918). [Transp. Theory Statist. Phys.1,186(1971)], pp. 235–257. DOI: 10.1080/00411457108231446. arXiv: physics/0503066 [physics].

[112] J. Wess and B. Zumino. 'A Lagrangian Model Invariant Under Supergauge Transformations'. In: *Phys. Lett.* 49B (1974), p. 52. DOI: 10.1016/0370-2693(74)90578-4.

[113] J. Iliopoulos and B. Zumino. 'Broken Supergauge Symmetry and Renormalization'. In: *Nucl. Phys.* B76 (1974), p. 310. DOI: 10.1016/0550-3213(74)90388-5.

[114] L. Girardello and Marcus T. Grisaru. 'Soft Breaking of Supersymmetry'. In: *Nucl. Phys.* B194 (1982), p. 65. DOI: 10.1016/0550-3213(82)90512-0.

[115] Pierre Fayet. 'Supersymmetry and Weak, Electromagnetic and Strong Interactions'. In: *Phys. Lett.* 64B (1976), p. 159. DOI: 10.1016/0370-2693(76)90319-1.

[116] Pierre Fayet. 'Spontaneously Broken Supersymmetric Theories of Weak, Electromagnetic and Strong Interactions'. In: *Phys. Lett.* 69B (1977), p. 489. DOI: 10.1016/0370-2693(77)90852-8.

[117] Pierre Fayet. 'Relations Between the Masses of the Superpartners of Leptons and Quarks, the Goldstino Couplings and the Neutral Currents'. In: *Phys. Lett.* 84B (1979), p. 416. DOI: 10.1016/0370-2693(79)91229-2.

[118] Glennys R. Farrar and Pierre Fayet. 'Phenomenology of the Production, Decay, and Detection of New Hadronic States Associated with Supersymmetry'. In: *Phys. Lett.* 76B (1978), pp. 575–579. DOI: 10.1016/0370-2693(78)90858-4.

[119] Gerard 't Hooft. 'Symmetry Breaking Through Bell-Jackiw Anomalies'. In: *Phys. Rev. Lett.* 37 (1976). [,226(1976)], pp. 8–11. DOI: 10.1103/PhysRevLett.37.8.

[120] N. Sakai and Tsutomu Yanagida. 'Proton Decay in a Class of Supersymmetric Grand Unified Models'. In: *Nucl. Phys.* B197 (1982), p. 533. DOI: 10.1016/0550-3213(82)90457-6.

[121] Savas Dimopoulos, Stuart Raby and Frank Wilczek. 'Proton Decay in Supersymmetric Models'. In: *Phys. Lett.* 112B (1982), p. 133. DOI: 10.1016/0370-2693(82)90313-6.

[122] Lawrence J. Hall and Mahiko Suzuki. 'Explicit R-Parity Breaking in Supersymmetric Models'. In: *Nucl. Phys.* B231 (1984), pp. 419–444. DOI: 10.1016/0550-3213(84)90513-3.

[123] Graham G. Ross and J. W. F. Valle. 'Supersymmetric Models Without R-Parity'. In: *Phys. Lett.* 151B (1985), pp. 375–381. DOI: 10.1016/0370-2693(85)91658-2.

[124] Vernon D. Barger, G. F. Giudice and Tao Han. 'Some New Aspects of Supersymmetry R-Parity Violating Interactions'. In: *Phys. Rev.* D40 (1989), p. 2987. DOI: 10.1103/PhysRevD.40.2987.

[125] Herbert K. Dreiner. 'An Introduction to explicit R-parity violation'. In: (1997). [Adv. Ser. Direct. High Energy Phys.21,565(2010)], pp. 462–479. DOI: 10.1142/9789814307505_0017. arXiv: hep-ph/9707435 [hep-ph].

[126] B. C. Allanach, A. Dedes and H. K. Dreiner. 'R parity violating minimal supergravity model'. In: *Phys. Rev.* D69 (2004). [Erratum: Phys. Rev.D72,079902(2005)], p. 115002. DOI: 10.1103/PhysRevD.69.115002, 10.1103/PhysRevD.72.079902. arXiv: hep-ph/0309196 [hep-ph].

[127] R. Barbier et al. 'R-parity violating supersymmetry'. In: *Phys. Rept.* 420 (2005), pp. 1–202. DOI: 10.1016/j.physrep.2005.08.006. arXiv: hep-ph/0406039 [hep-ph].

[128] Savas Dimopoulos and David W. Sutter. 'The Supersymmetric flavor problem'. In: *Nucl. Phys.* B452 (1995), pp. 496–512. DOI: 10.1016/0550-3213(95)00421-N. arXiv: hep-ph/9504415 [hep-ph].

[129] Yoichiro Nambu. 'Quasiparticles and Gauge Invariance in the Theory of Super-conductivity'. In: *Phys. Rev.* 117 (1960). [,132(1960)], pp. 648–663. DOI: 10.1103/PhysRev.117.648.

[130] J. Goldstone. 'Field Theories with Superconductor Solutions'. In: *Nuovo Cim.* 19 (1961), pp. 154–164. DOI: 10.1007/BF02812722.

[131] Jeffrey Goldstone, Abdus Salam and Steven Weinberg. 'Broken Symmetries'. In: *Phys. Rev.* 127 (1962), pp. 965–970. DOI: 10.1103/PhysRev.127.965.

[132] John M. Cornwall, David N. Levin and George Tiktopoulos. 'Derivation of Gauge Invariance from High-Energy Unitarity Bounds on the s Matrix'. In: *Phys. Rev.* D10 (1974). [Erratum: Phys. Rev.D11,972(1975)], p. 1145. DOI: 10.1103/PhysRevD.10.1145,10.1103/PhysRevD.11.972.

[133] Benjamin W. Lee, C. Quigg and H. B. Thacker. 'Weak Interactions at Very High-Energies: The Role of the Higgs Boson Mass'. In: *Phys. Rev.* D16 (1977), p. 1519. DOI: 10.1103/PhysRevD.16.1519.

[134] Edward Witten. 'Constraints on Supersymmetry Breaking'. In: *Nucl. Phys.* B202 (1982), p. 253. DOI: 10.1016/0550-3213(82)90071-2.

[135] Markus A. Luty. '2004 TASI lectures on supersymmetry breaking'. In: *Physics in D >= 4. Proceedings, Theoretical Advanced Study Institute in elementary particle physics, TASI 2004, Boulder, USA, June 6-July 2, 2004.* 2005, pp. 495–582. arXiv: hep-th/0509029 [hep-th].

[136] G. F. Giudice and R. Rattazzi. 'Theories with gauge mediated supersymmetry breaking'. In: *Phys. Rept.* 322 (1999), pp. 419–499. DOI: 10.1016/S0370-1573(99)00042-3. arXiv: hep-ph/9801271 [hep-ph].

[137] Pran Nath and Richard L. Arnowitt. 'Generalized Supergauge Symmetry as a New Framework for Unified Gauge Theories'. In: *Phys. Lett.* 56B (1975), pp. 177–180. DOI: 10.1016/0370-2693(75)90297-X.

[138] Richard L. Arnowitt, Pran Nath and B. Zumino. 'Superfield Densities and Action Principle in Curved Superspace'. In: *Phys. Lett.* 56B (1975), pp. 81–84. DOI: 10.1016/0370-2693(75)90504-3.

[139] Daniel Z. Freedman, P. van Nieuwenhuizen and S. Ferrara. 'Progress Toward a Theory of Supergravity'. In: *Phys. Rev.* D13 (1976), pp. 3214–3218. DOI: 10.1103/PhysRevD.13.3214.

[140] Stanley Deser and B. Zumino. 'Consistent Supergravity'. In: *Phys. Lett.* B62 (1976). [,335(1976)], p. 335. DOI: 10.1016/0370-2693(76)90089-7.

[141] Daniel Z. Freedman and P. van Nieuwenhuizen. 'Properties of Supergravity Theory'. In: *Phys. Rev.* D14 (1976), p. 912. DOI: 10.1103/PhysRevD.14.912.

[142] E. Cremmer et al. 'Spontaneous Symmetry Breaking and Higgs Effect in Super-gravity Without Cosmological Constant'. In: *Nucl. Phys.* B147 (1979), p. 105. DOI: 10.1016/0550-3213(79)90417-6.

[143]  Jonathan A. Bagger. 'Coupling the Gauge Invariant Supersymmetric Nonlinear Sigma Model to Supergravity'. In: *Nucl. Phys.* B211 (1983), p. 302. DOI: 10.1016/0550-3213(83)90411-X.

[144]  E. Cremmer et al. 'Yang-Mills Theories with Local Supersymmetry: Lagrangian, Transformation Laws and SuperHiggs Effect'. In: *Nucl. Phys.* B212 (1983). [,413(1982)], p. 413. DOI: 10.1016/0550-3213(83)90679-X.

[145]  D. V. Volkov and V. A. Soroka. 'Higgs Effect for Goldstone Particles with Spin 1/2'. In: *JETP Lett.* 18 (1973). [Pisma Zh. Eksp. Teor. Fiz.18,529(1973)], pp. 312–314.

[146]  Stanley Deser and B. Zumino. 'Broken Supersymmetry and Supergravity'. In: *Phys. Rev. Lett.* 38 (1977), pp. 1433–1436. DOI: 10.1103/PhysRevLett.38.1433.

[147]  Pierre Fayet. 'Mixing Between Gravitational and Weak Interactions Through the Massive Gravitino'. In: *Phys. Lett.* 70B (1977), p. 461. DOI: 10.1016/0370-2693(77)90414-2.

[148]  R. Casalbuoni et al. 'High-Energy Equivalence Theorem in Spontaneously Broken Supergravity'. In: *Phys. Rev.* D39 (1989), p. 2281. DOI: 10.1103/PhysRevD.39.2281.

[149]  E. Cremmer et al. 'Super-higgs effect in supergravity with general scalar interactions'. In: *Phys. Lett.* 79B (1978), pp. 231–234. DOI: 10.1016/0370-2693(78)90230-7.

[150]  S. Ferrara, L. Girardello and F. Palumbo. 'A General Mass Formula in Broken Supersymmetry'. In: *Phys. Rev.* D20 (1979), p. 403. DOI: 10.1103/PhysRevD.20.403.

[151]  Jonathan Bagger, Erich Poppitz and Lisa Randall. 'The R axion from dynamical supersymmetry breaking'. In: *Nucl. Phys.* B426 (1994), pp. 3–18. DOI: 10.1016/0550-3213(94)90123-6. arXiv: hep-ph/9405345 [hep-ph].

[152]  Ann E. Nelson and Nathan Seiberg. 'R symmetry breaking versus supersymmetry breaking'. In: *Nucl. Phys.* B416 (1994), pp. 46–62. DOI: 10.1016/0550-3213(94)90577-0. arXiv: hep-ph/9309299 [hep-ph].

[153]  L. O'Raifeartaigh. 'Spontaneous Symmetry Breaking for Chiral Scalar Superfields'. In: *Nucl. Phys.* B96 (1975), pp. 331–352. DOI: 10.1016/0550-3213(75)90585-4.

[154]  Kenneth A. Intriligator and Nathan Seiberg. 'Lectures on Supersymmetry Breaking'. In: *Class. Quant. Grav.* 24 (2007). [Les Houches87,125(2008)], S741–S772. DOI: 10.1088/0264-9381/24/21/S02, 10.1016/S0924-8099(08)80020-0. arXiv: hep-ph/0702069 [hep-ph].

[155]  Michael Dine and John D. Mason. 'Supersymmetry and Its Dynamical Breaking'. In: *Rept. Prog. Phys.* 74 (2011), p. 056201. DOI: 10.1088/0034-4885/74/5/056201. arXiv: 1012.2836 [hep-th].

[156] Pierre Fayet. 'Supergauge Invariant Extension of the Higgs Mechanism and a Model for the electron and Its Neutrino'. In: *Nucl. Phys.* B90 (1975), pp. 104–124. DOI: 10.1016/0550-3213(75)90636-7.

[157] Lawrence J. Hall, Joseph D. Lykken and Steven Weinberg. 'Supergravity as the Messenger of Supersymmetry Breaking'. In: *Phys. Rev.* D27 (1983), pp. 2359–2378. DOI: 10.1103/PhysRevD.27.2359.

[158] E. Cremmer, Pierre Fayet and L. Girardello. 'Gravity Induced Supersymmetry Breaking and Low-Energy Mass Spectrum'. In: *Phys. Lett.* 122B (1983), p. 41. DOI: 10.1016/0370-2693(83)91165-6.

[159] Riccardo Barbieri, S. Ferrara and Carlos A. Savoy. 'Gauge Models with Spontaneously Broken Local Supersymmetry'. In: *Phys. Lett.* 119B (1982), p. 343. DOI: 10.1016/0370-2693(82)90685-2.

[160] Ali H. Chamseddine, Richard L. Arnowitt and Pran Nath. 'Locally Supersymmetric Grand Unification'. In: *Phys. Rev. Lett.* 49 (1982), p. 970. DOI: 10.1103/PhysRevLett.49.970.

[161] Sanjeev K. Soni and H. Arthur Weldon. 'Analysis of the Supersymmetry Breaking Induced by N=1 Supergravity Theories'. In: *Phys. Lett.* 126B (1983), pp. 215–219. DOI: 10.1016/0370-2693(83)90593-2.

[162] Hans Peter Nilles, M. Srednicki and D. Wyler. 'Weak Interaction Breakdown Induced by Supergravity'. In: *Phys. Lett.* 120B (1983), p. 346. DOI: 10.1016/0370-2693(83)90460-4.

[163] Riccardo Barbieri. 'Looking Beyond the Standard Model: The Supersymmetric Option'. In: *Riv. Nuovo Cim.* 11N4 (1988), pp. 1–45. DOI: 10.1007/BF02725953.

[164] Howard E. Haber and Gordon L. Kane. 'The Search for Supersymmetry: Probing Physics Beyond the Standard Model'. In: *Phys. Rept.* 117 (1985), pp. 75–263. DOI: 10.1016/0370-1573(85)90051-1.

[165] Hans Peter Nilles. 'Supersymmetry, Supergravity and Particle Physics'. In: *Phys. Rept.* 110 (1984), pp. 1–162. DOI: 10.1016/0370-1573(84)90008-5.

[166] Savas Dimopoulos, Scott D. Thomas and James D. Wells. 'Sparticle spectroscopy and electroweak symmetry breaking with gauge mediated supersymmetry breaking'. In: *Nucl. Phys.* B488 (1997), pp. 39–91. DOI: 10.1016/S0550-3213(97)00030-8. arXiv: hep-ph/9609434 [hep-ph].

[167] Luis E. Ibanez. 'Locally Supersymmetric SU(5) Grand Unification'. In: *Phys. Lett.* 118B (1982), pp. 73–78. DOI: 10.1016/0370-2693(82)90604-9.

[168] Nobuyoshi Ohta. 'GRAND UNIFIED THEORIES BASED ON LOCAL SUPERSYMMETRY'. In: *Prog. Theor. Phys.* 70 (1983), p. 542. DOI: 10.1143/PTP.70.542.

[169] John R. Ellis, Dimitri V. Nanopoulos and K. Tamvakis. 'Grand Unification in Simple Supergravity'. In: *Phys. Lett.* 121B (1983), pp. 123–129. DOI: 10.1016/0370-2693(83)90900-0.

[170] Luis Alvarez-Gaume, J. Polchinski and Mark B. Wise. 'Minimal Low-Energy Supergravity'. In: *Nucl. Phys.* B221 (1983), p. 495. DOI: `10.1016/0550-3213(83)90591-6`.

[171] Heinz Pagels and Joel R. Primack. 'Supersymmetry, Cosmology and New TeV Physics'. In: *Phys. Rev. Lett.* 48 (1982), p. 223. DOI: `10.1103/PhysRevLett.48.223`.

[172] T. Moroi, H. Murayama and Masahiro Yamaguchi. 'Cosmological constraints on the light stable gravitino'. In: *Phys. Lett.* B303 (1993), pp. 289–294. DOI: `10.1016/0370-2693(93)91434-O`.

[173] Gordon L. Kane et al. 'Study of constrained minimal supersymmetry'. In: *Phys. Rev.* D49 (1994), pp. 6173–6210. DOI: `10.1103/PhysRevD.49.6173`. arXiv: `hep-ph/9312272 [hep-ph]`.

[174] John R. Ellis and Dimitri V. Nanopoulos. 'Flavor Changing Neutral Interactions in Broken Supersymmetric Theories'. In: *Phys. Lett.* 110B (1982), pp. 44–48. DOI: `10.1016/0370-2693(82)90948-0`.

[175] Riccardo Barbieri and Raoul Gatto. 'Conservation Laws for Neutral Currents in Spontaneously Broken Supersymmetric Theories'. In: *Phys. Lett.* 110B (1982), p. 211. DOI: `10.1016/0370-2693(82)91238-2`.

[176] John S. Hagelin, S. Kelley and Toshiaki Tanaka. 'Supersymmetric flavor changing neutral currents: Exact amplitudes and phenomenological analysis'. In: *Nucl. Phys.* B415 (1994), pp. 293–331. DOI: `10.1016/0550-3213(94)90113-9`.

[177] F. Gabbiani et al. 'A Complete analysis of FCNC and CP constraints in general SUSY extensions of the standard model'. In: *Nucl. Phys.* B477 (1996), pp. 321–352. DOI: `10.1016/0550-3213(96)00390-2`. arXiv: `hep-ph/9604387 [hep-ph]`.

[178] Riccardo Barbieri and L. J. Hall. 'Signals for supersymmetric unification'. In: *Phys. Lett.* B338 (1994), pp. 212–218. DOI: `10.1016/0370-2693(94)91368-4`. arXiv: `hep-ph/9408406 [hep-ph]`.

[179] Riccardo Barbieri, Lawrence J. Hall and Alessandro Strumia. 'Violations of lepton flavor and CP in supersymmetric unified theories'. In: *Nucl. Phys.* B445 (1995), pp. 219–251. DOI: `10.1016/0550-3213(95)00208-A`. arXiv: `hep-ph/9501334 [hep-ph]`.

[180] Michael Dine and Willy Fischler. 'A Phenomenological Model of Particle Physics Based on Supersymmetry'. In: *Phys. Lett.* 110B (1982), pp. 227–231. DOI: `10.1016/0370-2693(82)91241-2`.

[181] Chiara R. Nappi and Burt A. Ovrut. 'Supersymmetric Extension of the SU(3) x SU(2) x U(1) Model'. In: *Phys. Lett.* 113B (1982), pp. 175–179. DOI: `10.1016/0370-2693(82)90418-X`.

[182] Luis Alvarez-Gaume, Mark Claudson and Mark B. Wise. 'Low-Energy Supersymmetry'. In: *Nucl. Phys.* B207 (1982), p. 96. DOI: `10.1016/0550-3213(82)90138-9`.

[183] Michael Dine and Ann E. Nelson. 'Dynamical supersymmetry breaking at low-energies'. In: *Phys. Rev.* D48 (1993), pp. 1277–1287. DOI: 10.1103/PhysRevD.48.1277. arXiv: hep-ph/9303230 [hep-ph].

[184] Michael Dine, Ann E. Nelson and Yuri Shirman. 'Low-energy dynamical supersymmetry breaking simplified'. In: *Phys. Rev.* D51 (1995), pp. 1362–1370. DOI: 10.1103/PhysRevD.51.1362. arXiv: hep-ph/9408384 [hep-ph].

[185] Michael Dine et al. 'New tools for low-energy dynamical supersymmetry breaking'. In: *Phys. Rev.* D53 (1996), pp. 2658–2669. DOI: 10.1103/PhysRevD.53.2658. arXiv: hep-ph/9507378 [hep-ph].

[186] G. F. Giudice and R. Rattazzi. 'Extracting supersymmetry breaking effects from wave function renormalization'. In: *Nucl. Phys.* B511 (1998), pp. 25–44. DOI: 10.1016/S0550-3213(97)00647-0. arXiv: hep-ph/9706540 [hep-ph].

[187] Patrick Meade, Nathan Seiberg and David Shih. 'General Gauge Mediation'. In: *Prog. Theor. Phys. Suppl.* 177 (2009), pp. 143–158. DOI: 10.1143/PTPS.177.143. arXiv: 0801.3278 [hep-ph].

[188] Oliver S. Bruning et al. 'LHC Design Report Vol.1: The LHC Main Ring'. In: (2004).

[189] 'LEP Design Report Vol.1'. In: (1983).

[190] 'LEP Design Report: Vol.2. The LEP Main Ring'. In: (1984).

[191] D. Decamp et al. 'ALEPH: A detector for electron-positron annnihilations at LEP'. In: *Nucl. Instrum. Meth.* A294 (1990). [Erratum: Nucl. Instrum. Meth.A303,393(1991)], pp. 121–178. DOI: 10.1016/0168-9002(90)91831-U.

[192] P. A. Aarnio et al. 'The DELPHI detector at LEP'. In: *Nucl. Instrum. Meth.* A303 (1991), pp. 233–276. DOI: 10.1016/0168-9002(91)90793-P.

[193] F. Abe et al. 'The CDF Detector: An Overview'. In: *Nucl. Instrum. Meth.* A271 (1988), pp. 387–403. DOI: 10.1016/0168-9002(88)90298-7.

[194] S. Abachi et al. 'The D0 Detector'. In: *Nucl. Instrum. Meth.* A338 (1994), pp. 185–253. DOI: 10.1016/0168-9002(94)91312-9.

[195] Fabienne Marcastel. 'CERN's Accelerator Complex. La chaîne des accélérateurs du CERN'. In: (Oct. 2013). General Photo. URL: https://cds.cern.ch/record/1621583.

[196] Corinne Pralavorio. *Final lap of the LHC track for protons in 2018.* https://home.cern/news/news/accelerators/final-lap-lhc-track-protons-2018. Accessed on 05-08-2019. Oct. 2018.

[197] The HL-LHC project. *LHC/HL-LHC Plan.* Accessed on 04-08-2019.

[198] Rende Steerenberg for the Operations group. *LHC report: full house for the LHC.* https://home.cern/news/news/accelerators/lhc-report-full-house-lhc. Accessed on 05-08-2018. July 2017.

[199] 'Search for direct pair production of supersymmetric partners to the $\tau$ lepton in proton-proton collisions at $\sqrt{s} = 13$ TeV'. In: (2019). arXiv: 1907.13179 [hep-ex].

[200] *Illustration of an LHC proton-proton collision*. https://imperialhep.blogspot.com/2011/08/. Accessed on 05-08-2019.

[201] Stirling, W.J. *Parton luminosity and cross section plots*. http://www.hep.ph.ic.ac.uk/~wstirlin/plots/plots.html. Accessed on 04-08-2019.

[202] John M. Campbell, J. W. Huston and W. J. Stirling. 'Hard Interactions of Quarks and Gluons: A Primer for LHC Physics'. In: *Rept. Prog. Phys.* 70 (2007), p. 89. DOI: 10.1088/0034-4885/70/1/R02. arXiv: hep-ph/0611148 [hep-ph].

[203] S. Chatrchyan et al. 'The CMS Experiment at the CERN LHC'. In: *JINST* 3 (2008), S08004. DOI: 10.1088/1748-0221/3/08/S08004.

[204] G. Aad et al. 'The ATLAS Experiment at the CERN Large Hadron Collider'. In: *JINST* 3 (2008), S08003. DOI: 10.1088/1748-0221/3/08/S08003.

[205] A. Augusto Alves Jr. et al. 'The LHCb Detector at the LHC'. In: *JINST* 3 (2008), S08005. DOI: 10.1088/1748-0221/3/08/S08005.

[206] K. Aamodt et al. 'The ALICE experiment at the CERN LHC'. In: *JINST* 3 (2008), S08002. DOI: 10.1088/1748-0221/3/08/S08002.

[207] G. Anelli et al. 'The TOTEM experiment at the CERN Large Hadron Collider'. In: *JINST* 3 (2008), S08007. DOI: 10.1088/1748-0221/3/08/S08007.

[208] James Pinfold et al. 'Technical Design Report of the MoEDAL Experiment'. In: (2009).

[209] O. Adriani et al. 'The LHCf detector at the CERN Large Hadron Collider'. In: *JINST* 3 (2008), S08006. DOI: 10.1088/1748-0221/3/08/S08006.

[210] *CMS detector*. https://cms.cern/detector. Accessed on 04-08-2018.

[211] *ATLAS detector and technology*. https://atlas.cern/discover/detector. Accessed on 04-08-2019.

[212] Siona Ruth Davis. 'Interactive Slice of the CMS detector'. In: (Aug. 2016). URL: https://cds.cern.ch/record/2205172.

[213] Albert M Sirunyan et al. 'Measurement of the $t\bar{t}$ production cross section, the top quark mass, and the strong coupling constant using dilepton events in pp collisions at $\sqrt{s} = 13$ TeV'. In: *Eur. Phys. J.* C79.5 (2019), p. 368. DOI: 10.1140/epjc/s10052-019-6863-8. arXiv: 1812.10505 [hep-ex].

[214] Morad Aaboud et al. 'Measurement of the inclusive cross-sections of single top-quark and top-antiquark $t$-channel production in $pp$ collisions at $\sqrt{s} = 13$ TeV with the ATLAS detector'. In: *JHEP* 04 (2017), p. 086. DOI: 10.1007/JHEP04(2017)086. arXiv: 1609.03920 [hep-ex].

[215] Morad Aaboud et al. 'Measurements of $t\bar{t}$ differential cross-sections of highly boosted top quarks decaying to all-hadronic final states in $pp$ collisions at $\sqrt{s} = 13$ TeV using the ATLAS detector'. In: *Phys. Rev.* D98.1 (2018), p. 012003. DOI: 10.1103/PhysRevD.98.012003. arXiv: 1801.02052 [hep-ex].

[216] Albert M Sirunyan et al. 'Measurement of differential cross sections and charge ratios for $t$-channel single top quark production in proton-proton collisions at $\sqrt{s} = 13$ TeV'. In: (2019). arXiv: 1907.08330 [hep-ex].

[217] Albert M Sirunyan et al. 'Search for the Production of Four Top Quarks in the Single-Lepton and Opposite-Sign Dilepton Final States in Proton-Proton Collisions at $\sqrt{s} = 13$ TeV'. In: (2019). arXiv: 1906.02805 [hep-ex].

[218] Mihailo Backović. 'A Theory of Ambulance Chasing'. In: (2016). arXiv: 1603.01204 [physics.soc-ph].

[219] The ATLAS collaboration. 'Search for resonances decaying to photon pairs in 3.2 fb$^{-1}$ of $pp$ collisions at $\sqrt{s} = 13$ TeV with the ATLAS detector'. In: (2015).

[220] CMS Collaboration. 'Search for new physics in high mass diphoton events in proton-proton collisions at 13TeV'. In: (2015).

[221] *Search for resonant production of high mass photon pairs using 12.9 fb$^{-1}$ of proton-proton collisions at $\sqrt{s} = 13$ TeV and combined interpretation of searches at 8 and 13 TeV*. Tech. rep. CMS-PAS-EXO-16-027. Geneva: CERN, 2016. URL: https://cds.cern.ch/record/2205245.

[222] *Search for scalar diphoton resonances with 15.4 fb$^{-1}$ of data collected at $\sqrt{s}$=13 TeV in 2015 and 2016 with the ATLAS detector*. Tech. rep. ATLAS-CONF-2016-059. Geneva: CERN, Aug. 2016. URL: https://cds.cern.ch/record/2206154.

[223] Vardan Khachatryan et al. 'Search for high-mass diphoton resonances in proton–proton collisions at 13 TeV and combination with 8 TeV search'. In: *Phys. Lett.* B767 (2017), pp. 147–170. DOI: 10.1016/j.physletb.2017.01.027. arXiv: 1609.02507 [hep-ex].

[224] Morad Aaboud et al. 'Search for new phenomena in high-mass diphoton final states using 37 fb$^{-1}$ of proton–proton collisions collected at $\sqrt{s} = 13$ TeV with the ATLAS detector'. In: *Phys. Lett.* B775 (2017), pp. 105–125. DOI: 10.1016/j.physletb.2017.10.039. arXiv: 1707.04147 [hep-ex].

[225] LHC SUSY Cross Section Working Group. *SUSY cross sections*. https://twiki.cern.ch/twiki/bin/view/LHCPhysics/SUSYCrossSections. Last accessed on 12-08-18.

[226] Christoph Borschensky et al. 'Squark and gluino production cross sections in pp collisions at $\sqrt{s} = 13$, 14, 33 and 100 TeV'. In: *Eur. Phys. J.* C74.12 (2014), p. 3174. DOI: 10.1140/epjc/s10052-014-3174-y. arXiv: 1407.5066 [hep-ph].

[227] Wim Beenakker et al. 'NNLL-fast: predictions for coloured supersymmetric particle production at the LHC with threshold and Coulomb resummation'. In: *JHEP* 12 (2016), p. 133. DOI: 10.1007/JHEP12(2016)133. arXiv: 1607.07741 [hep-ph].

[228] Benjamin Fuks et al. 'Precision predictions for electroweak superpartner production at hadron colliders with Resummino'. In: *Eur. Phys. J.* C73 (2013), p. 2480. DOI: 10.1140/epjc/s10052-013-2480-0. arXiv: 1304.0790 [hep-ph].

[229] Benjamin Fuks et al. 'Revisiting slepton pair production at the Large Hadron Collider'. In: *JHEP* 01 (2014), p. 168. DOI: 10.1007/JHEP01(2014)168. arXiv: 1310.2621.

[230] Juri Fiaschi and Michael Klasen. 'Slepton pair production at the LHC in NLO+NLL with resummation-improved parton densities'. In: *JHEP* 03 (2018), p. 094. DOI: 10.1007/JHEP03(2018)094. arXiv: 1801.10357 [hep-ph].

[231] Giuseppe Bozzi, Benjamin Fuks and Michael Klasen. 'Threshold Resummation for Slepton-Pair Production at Hadron Colliders'. In: *Nucl. Phys.* B777 (2007), pp. 157–181. DOI: 10.1016/j.nuclphysb.2007.03.052. arXiv: hep-ph/0701202 [hep-ph].

[232] Jonathan Debove, Benjamin Fuks and Michael Klasen. 'Threshold resummation for gaugino pair production at hadron colliders'. In: *Nucl. Phys.* B842 (2011), pp. 51–85. DOI: 10.1016/j.nuclphysb.2010.08.016. arXiv: 1005.2909 [hep-ph].

[233] Benjamin Fuks et al. 'Gaugino production in proton-proton collisions at a center-of-mass energy of 8 TeV'. In: *JHEP* 10 (2012), p. 081. DOI: 10.1007/JHEP10(2012)081. arXiv: 1207.2159 [hep-ph].

[234] Juri Fiaschi and Michael Klasen. 'Neutralino-chargino pair production at NLO+NLL with resummation-improved parton density functions for LHC Run II'. In: *Phys. Rev.* D98.5 (2018), p. 055014. DOI: 10.1103/PhysRevD.98.055014. arXiv: 1805.11322 [hep-ph].

[235] Nathaniel Craig. 'The State of Supersymmetry after Run I of the LHC'. In: *Beyond the Standard Model after the first run of the LHC Arcetri, Florence, Italy, May 20-July 12, 2013*. 2013. arXiv: 1309.0528 [hep-ph].

[236] Philip Bechtle et al. 'Killing the cMSSM softly'. In: *Eur. Phys. J.* C76.2 (2016), p. 96. DOI: 10.1140/epjc/s10052-015-3864-0. arXiv: 1508.05951 [hep-ph].

[237] Daniele Alves. 'Simplified Models for LHC New Physics Searches'. In: *J. Phys.* G39 (2012). Ed. by Nima Arkani-Hamed et al., p. 105005. DOI: 10.1088/0954-3899/39/10/105005. arXiv: 1105.2838 [hep-ph].

[238] Johan Alwall, Philip Schuster and Natalia Toro. 'Simplified Models for a First Characterization of New Physics at the LHC'. In: *Phys. Rev.* D79 (2009), p. 075020. DOI: 10.1103/PhysRevD.79.075020. arXiv: 0810.3921 [hep-ph].

[239] CMS Collaboration. *CMS SUSY Summary Plots.* http://cms-results.web.cern.ch/cms-results/public-results/publications/SUS/index.html. Accessed on 04-08-2019.

[240] CMS Collaboration. *CMS EXO summary plots.* http://cms-results.web.cern.ch/cms-results/public-results/publications/EXO/index.html. Accessed on 04-08-2019.

[241] ATLAS Collaboration. *ATLAS SUSY Summary plots.* https://atlas.web.cern.ch/Atlas/GROUPS/PHYSICS/CombinedSummaryPlots/SUSY/. Accessed on 04-08-2019.

[242] A. Djouadi et al. 'The Minimal supersymmetric standard model: Group summary report'. In: *GDR (Groupement De Recherche) - Supersymetrie Montpellier, France, April 15-17, 1998*. 1998. arXiv: hep-ph/9901246 [hep-ph].

[243] Carola F. Berger et al. 'Supersymmetry Without Prejudice'. In: *JHEP* 02 (2009), p. 023. DOI: 10.1088/1126-6708/2009/02/023. arXiv: 0812.0980 [hep-ph].

[244] Georges Aad et al. 'Summary of the ATLAS experiment's sensitivity to supersymmetry after LHC Run 1 — interpreted in the phenomenological MSSM'. In: *JHEP* 10 (2015), p. 134. DOI: 10.1007/JHEP10(2015)134. arXiv: 1508.06608 [hep-ex].

[245] Vardan Khachatryan et al. 'Phenomenological MSSM interpretation of CMS searches in pp collisions at sqrt(s) = 7 and 8 TeV'. In: *JHEP* 10 (2016), p. 129. DOI: 10.1007/JHEP10(2016)129. arXiv: 1606.03577 [hep-ex].

[246] Hsin-Chia Cheng and Ian Low. 'TeV symmetry and the little hierarchy problem'. In: *JHEP* 09 (2003), p. 051. DOI: 10.1088/1126-6708/2003/09/051. arXiv: hep-ph/0308199 [hep-ph].

[247] Georges Aad et al. 'Search for pair-produced third-generation squarks decaying via charm quarks or in compressed supersymmetric scenarios in $pp$ collisions at $\sqrt{s} = 8$ TeV with the ATLAS detector'. In: *Phys. Rev.* D90.5 (2014), p. 052008. DOI: 10.1103/PhysRevD.90.052008. arXiv: 1407.0608 [hep-ex].

[248] Georges Aad et al. 'Search for massive, long-lived particles using multitrack displaced vertices or displaced lepton pairs in pp collisions at $\sqrt{s} = 8$ TeV with the ATLAS detector'. In: *Phys. Rev.* D92.7 (2015), p. 072004. DOI: 10.1103/PhysRevD.92.072004. arXiv: 1504.05162 [hep-ex].

[249] Vardan Khachatryan et al. 'Search for dark matter, extra dimensions, and unparticles in monojet events in proton–proton collisions at $\sqrt{s} = 8$ TeV'. In: *Eur. Phys. J.* C75.5 (2015), p. 235. DOI: 10.1140/epjc/s10052-015-3451-4. arXiv: 1408.3583 [hep-ex].

[250] Adam Alloul et al. 'FeynRules 2.0 - A complete toolbox for tree-level phenomenology'. In: *Comput.Phys.Commun.* 185 (2014), pp. 2250–2300. DOI: 10.1016/j.cpc.2014.04.012. arXiv: 1310.1921 [hep-ph].

[251] Celine Degrande et al. 'UFO - The Universal FeynRules Output'. In: *Comput. Phys. Commun.* 183 (2012), pp. 1201–1214. DOI: 10.1016/j.cpc.2012.01.022. arXiv: 1108.2040 [hep-ph].

[252] F. Staub. 'SARAH'. In: (2008). arXiv: 0806.0538 [hep-ph].

[253] Werner Porod. 'SPheno, a program for calculating supersymmetric spectra, SUSY particle decays and SUSY particle production at e+ e- colliders'. In: *Comput. Phys. Commun.* 153 (2003), pp. 275–315. DOI: 10.1016/S0010-4655(03)00222-4. arXiv: hep-ph/0301101 [hep-ph].

[254] W. Porod and F. Staub. 'SPheno 3.1: Extensions including flavour, CP-phases and models beyond the MSSM'. In: *Comput. Phys. Commun.* 183 (2012), pp. 2458–2469. DOI: 10.1016/j.cpc.2012.05.021. arXiv: 1104.1573 [hep-ph].

[255] B. C. Allanach. 'SOFTSUSY: a program for calculating supersymmetric spectra'. In: *Comput. Phys. Commun.* 143 (2002), pp. 305–331. DOI: 10.1016/S0010-4655(01)00460-X. arXiv: hep-ph/0104145 [hep-ph].

[256] Abdelhak Djouadi, Jean-Loic Kneur and Gilbert Moultaka. 'SuSpect: A Fortran code for the supersymmetric and Higgs particle spectrum in the MSSM'. In: *Comput. Phys. Commun.* 176 (2007), pp. 426–455. DOI: 10.1016/j.cpc.2006.11.009. arXiv: hep-ph/0211331 [hep-ph].

[257] A. Djouadi, M.M. Muhlleitner and M. Spira. 'Decays of supersymmetric particles: The Program SUSY-HIT (SUspect-SdecaY-Hdecay-InTerface)'. In: *Acta Phys.Polon.* B38 (2007), pp. 635–644. arXiv: hep-ph/0609292 [hep-ph].

[258] J. Alwall et al. 'The automated computation of tree-level and next-to-leading order differential cross sections, and their matching to parton shower simulations'. In: *JHEP* 07 (2014), p. 079. DOI: 10.1007/JHEP07(2014)079. arXiv: 1405.0301 [hep-ph].

[259] Alexander Belyaev, Neil D. Christensen and Alexander Pukhov. 'CalcHEP 3.4 for collider physics within and beyond the Standard Model'. In: *Comput. Phys. Commun.* 184 (2013), pp. 1729–1769. DOI: 10.1016/j.cpc.2013.01.014. arXiv: 1207.6082 [hep-ph].

[260] Torbjörn Sjöstrand et al. 'An Introduction to PYTHIA 8.2'. In: *Comput. Phys. Commun.* 191 (2015), pp. 159–177. DOI: 10.1016/j.cpc.2015.01.024. arXiv: 1410.3012 [hep-ph].

[261] Paolo Nason. 'A New method for combining NLO QCD with shower Monte Carlo algorithms'. In: *JHEP* 11 (2004), p. 040. DOI: 10.1088/1126-6708/2004/11/040. arXiv: hep-ph/0409146 [hep-ph].

[262] Stefano Frixione, Paolo Nason and Carlo Oleari. 'Matching NLO QCD computations with Parton Shower simulations: the POWHEG method'. In: *JHEP* 11 (2007), p. 070. DOI: 10.1088/1126-6708/2007/11/070. arXiv: 0709.2092 [hep-ph].

[263] Simone Alioli et al. 'A general framework for implementing NLO calculations in shower Monte Carlo programs: the POWHEG BOX'. In: *JHEP* 06 (2010), p. 043. DOI: 10.1007/JHEP06(2010)043. arXiv: 1002.2581 [hep-ph].

[264] S. Agostinelli et al. 'GEANT4: A Simulation toolkit'. In: *Nucl. Instrum. Meth.* A506 (2003), pp. 250–303. DOI: 10.1016/S0168-9002(03)01368-8.

[265] J. de Favereau et al. 'DELPHES 3, A modular framework for fast simulation of a generic collider experiment'. In: *JHEP* 02 (2014), p. 057. DOI: 10.1007/JHEP02(2014)057. arXiv: 1307.6346 [hep-ex].

[266] Eric Conte, Benjamin Fuks and Guillaume Serret. 'MadAnalysis 5, A User-Friendly Framework for Collider Phenomenology'. In: *Comput. Phys. Commun.* 184 (2013), pp. 222–256. DOI: 10.1016/j.cpc.2012.09.009. arXiv: 1206.1599 [hep-ph].

[267] Eric Conte et al. 'Designing and recasting LHC analyses with MadAnalysis 5'. In: *Eur. Phys. J.* C74.10 (2014), p. 3103. DOI: 10.1140/epjc/s10052-014-3103-0. arXiv: 1405.3982 [hep-ph].

[268] B. Dumont et al. 'Toward a public analysis database for LHC new physics searches using MADANALYSIS 5'. In: *Eur. Phys. J.* C75.2 (2015), p. 56. DOI: 10.1140/epjc/s10052-014-3242-3. arXiv: 1407.3278 [hep-ph].

[269] Andy Buckley et al. 'Rivet user manual'. In: *Comput. Phys. Commun.* 184 (2013), pp. 2803–2819. DOI: 10.1016/j.cpc.2013.05.021. arXiv: 1003.0694 [hep-ph].

[270] Daniel Dercks et al. 'CheckMATE 2: From the model to the limit'. In: *Comput. Phys. Commun.* 221 (2017), pp. 383–418. DOI: 10.1016/j.cpc.2017.08.021. arXiv: 1611.09856 [hep-ph].

[271] Matteo Cacciari, Gavin P. Salam and Gregory Soyez. 'FastJet User Manual'. In: *Eur. Phys. J.* C72 (2012), p. 1896. DOI: 10.1140/epjc/s10052-012-1896-2. arXiv: 1111.6097 [hep-ph].

[272] Matteo Cacciari and Gavin P. Salam. 'Dispelling the $N^3$ myth for the $k_t$ jet-finder'. In: *Phys. Lett.* B641 (2006), pp. 57–61. DOI: 10.1016/j.physletb.2006.08.037. arXiv: hep-ph/0512210 [hep-ph].

[273] Matteo Cacciari, Gavin P. Salam and Gregory Soyez. 'The anti-$k_t$ jet clustering algorithm'. In: *JHEP* 04 (2008), p. 063. DOI: 10.1088/1126-6708/2008/04/063. arXiv: 0802.1189 [hep-ph].

[274] Alexander L. Read. 'Presentation of search results: The CL(s) technique'. In: *J. Phys.* G28 (2002). [,11(2002)], pp. 2693–2704. DOI: 10.1088/0954-3899/28/10/313.

[275] G. Belanger et al. 'MicrOMEGAs 2.0: A Program to calculate the relic density of dark matter in a generic model'. In: *Comput. Phys. Commun.* 176 (2007), pp. 367–382. DOI: 10.1016/j.cpc.2006.11.008. arXiv: hep-ph/0607059 [hep-ph].

[276] Federico Ambrogi et al. 'MadDM v.3.0: a Comprehensive Tool for Dark Matter Studies'. In: *Phys. Dark Univ.* 24 (2019), p. 100249. DOI: 10.1016/j.dark.2018.11.009. arXiv: 1804.00044 [hep-ph].

[277] Seng Pei Liew et al. 'Z-peaked excess in goldstini scenarios'. In: *Phys. Lett.* B750 (2015), pp. 539–546. DOI: 10.1016/j.physletb.2015.09.035. arXiv: 1506.08803 [hep-ph].

[278] Georges Aad et al. 'Search for supersymmetry in events containing a same-flavour opposite-sign dilepton pair, jets, and large missing transverse momentum in $\sqrt{s} = 8$ TeV pp collisions with the ATLAS detector'. In: *Eur. Phys. J.* C75.7 (2015), p. 318. DOI: 10.1140/epjc/s10052-015-3518-2. arXiv: 1503.03290 [hep-ex].

[279] I. Hinchliffe et al. 'Precision SUSY measurements at CERN LHC'. In: *Phys. Rev.* D55 (1997), pp. 5520–5540. DOI: 10.1103/PhysRevD.55.5520. arXiv: hep-ph/9610544 [hep-ph].

[280] Vardan Khachatryan et al. 'Search for physics beyond the standard model in events with two leptons, jets, and missing transverse momentum in pp collisions at sqrt(s) = 8 TeV'. In: *JHEP* 1504 (2015), p. 124. DOI: 10.1007/JHEP04(2015)124. arXiv: 1502.06031 [hep-ex].

[281] Joshua T. Ruderman and David Shih. 'General Neutralino NLSPs at the Early LHC'. In: *JHEP* 08 (2012), p. 159. DOI: 10.1007/JHEP08(2012)159. arXiv: 1103.6083 [hep-ph].

[282] Patrick Meade, Matthew Reece and David Shih. 'Prompt Decays of General Neutralino NLSPs at the Tevatron'. In: *JHEP* 05 (2010), p. 105. DOI: 10.1007/JHEP05(2010)105. arXiv: 0911.4130 [hep-ph].

[283] Georges Aad et al. 'Search for squarks and gluinos with the ATLAS detector in final states with jets and missing transverse momentum using $\sqrt{s} = 8$ TeV proton–proton collision data'. In: *JHEP* 1409 (2014), p. 176. DOI: 10.1007/JHEP09(2014)176. arXiv: 1405.7875 [hep-ex].

[284] Serguei Chatrchyan et al. 'Search for new physics in the multijet and missing transverse momentum final state in proton-proton collisions at $\sqrt{s}$= 8 TeV'. In: *JHEP* 06 (2014), p. 055. DOI: 10.1007/JHEP06(2014)055. arXiv: 1402.4770 [hep-ex].

[285] G. Barenboim et al. 'METing SUSY on the Z peak'. In: (2015). arXiv: 1503.04184 [hep-ph].

[286] Ulrich Ellwanger. 'Possible explanation of excess events in the search for jets, missing transverse momentum and a Z boson in pp collisions'. In: (2015). arXiv: 1504.02244 [hep-ph].

[287] Ben Allanach, Are Raklev and Anders Kvellestad. 'Consistency of the recent ATLAS $Z + E_T^{\mathrm{miss}}$ excess in a simplified GGM model'. In: *Phys.Rev.* D91 (2015), p. 095016. DOI: 10.1103/PhysRevD.91.095016. arXiv: 1504.02752 [hep-ph].

[288] Michael Kramer et al. 'Supersymmetry production cross sections in $pp$ collisions at $\sqrt{s} = 7$ TeV'. In: (2012). arXiv: 1206.2892 [hep-ph].

[289] W. Beenakker et al. 'Squark and gluino production at hadron colliders'. In: *Nucl. Phys.* B492 (1997), pp. 51–103. DOI: 10.1016/S0550-3213(97)80027-2. arXiv: hep-ph/9610490 [hep-ph].

[290] A. Kulesza and L. Motyka. 'Threshold resummation for squark-antisquark and gluino-pair production at the LHC'. In: *Phys. Rev. Lett.* 102 (2009), p. 111802. DOI: 10.1103/PhysRevLett.102.111802. arXiv: 0807.2405 [hep-ph].

[291] A. Kulesza and L. Motyka. 'Soft gluon resummation for the production of gluino-gluino and squark-antisquark pairs at the LHC'. In: *Phys. Rev.* D80 (2009), p. 095004. DOI: 10.1103/PhysRevD.80.095004. arXiv: 0905.4749 [hep-ph].

[292] Wim Beenakker et al. 'Soft-gluon resummation for squark and gluino hadroproduction'. In: *JHEP* 12 (2009), p. 041. DOI: 10.1088/1126-6708/2009/12/041. arXiv: 0909.4418 [hep-ph].

[293] W. Beenakker et al. 'Squark and Gluino Hadroproduction'. In: *Int. J. Mod. Phys.* A26 (2011), pp. 2637–2664. DOI: 10.1142/S0217751X11053560. arXiv: 1105.1110 [hep-ph].

[294] W. Beenakker et al. 'Stop production at hadron colliders'. In: *Nucl. Phys.* B515 (1998), pp. 3–14. DOI: 10.1016/S0550-3213(98)00014-5. arXiv: hep-ph/9710451 [hep-ph].

[295] Wim Beenakker et al. 'Supersymmetric top and bottom squark production at hadron colliders'. In: *JHEP* 08 (2010), p. 098. DOI: 10.1007/JHEP08(2010)098. arXiv: 1006.4771 [hep-ph].

[296] Wim Beenakker et al. 'NLO+NLL squark and gluino production cross-sections with threshold-improved parton distributions'. In: *Eur. Phys. J.* C76.2 (2016), p. 53. DOI: 10.1140/epjc/s10052-016-3892-4. arXiv: 1510.00375 [hep-ph].

[297] Georges Aad et al. 'Search for direct top-squark pair production in final states with two leptons in pp collisions at $\sqrt{s} = 8$TeV with the ATLAS detector'. In: *JHEP* 06 (2014), p. 124. DOI: 10.1007/JHEP06(2014)124. arXiv: 1403.4853 [hep-ex].

[298] Georges Aad et al. 'Search for top squark pair production in final states with one isolated lepton, jets, and missing transverse momentum in $\sqrt{s}$ =8 TeV $pp$ collisions with the ATLAS detector'. In: *JHEP* 11 (2014), p. 118. DOI: 10.1007/JHEP11(2014)118. arXiv: 1407.0583 [hep-ex].

[299] Vardan Khachatryan et al. 'Search for top-squark pairs decaying into Higgs or Z bosons in pp collisions at $\sqrt{s}$=8 TeV'. In: *Phys. Lett.* B736 (2014), pp. 371–397. DOI: 10.1016/j.physletb.2014.07.053. arXiv: 1405.3886 [hep-ex].

[300] Ran Ding et al. 'The ATLAS Leptonic-Z Excess from Light Squark Productions in the NMSSM Extension with a Heavy Dirac Gluino'. In: (2015). arXiv: 1508.07452 [hep-ph].

[301] Junjie Cao et al. 'Explanation of the ATLAS Z-peaked excess by squark pair production in the NMSSM'. In: *JHEP* 10 (2015), p. 178. DOI: 10.1007/JHEP10(2015)178. arXiv: 1507.08471 [hep-ph].

[302] M. Cahill-Rowley et al. 'The ATLAS Z + MET Excess in the MSSM'. In: (2015). arXiv: 1506.05799 [hep-ph].

[303] Ben Allanach, Are R. Raklev and Anders Kvellestad. 'Interpreting a CMS $lljjp_T^{\text{miss}}$ Excess With the Golden Cascade of the MSSM'. In: (2014). arXiv: 1409.3532 [hep-ph].

[304] S. Ambrosanio et al. 'Search for supersymmetry with a light gravitino at the Fermilab Tevatron and CERN LEP colliders'. In: *Phys. Rev.* D54 (1996), pp. 5395–5411. DOI: 10.1103/PhysRevD.54.5395. arXiv: hep-ph/9605398 [hep-ph].

[305] Serguei Chatrchyan et al. 'Search for anomalous production of events with three or more leptons in $pp$ collisions at $\sqrt{(s)} = 8$ TeV'. In: *Phys. Rev.* D90 (2014), p. 032006. DOI: 10.1103/PhysRevD.90.032006. arXiv: 1404.5801 [hep-ex].

[306] Junjie Cao et al. 'Explanation of the ATLAS Z-Peaked Excess in the NMSSM'. In: *JHEP* 06 (2015), p. 152. DOI: 10.1007/JHEP06(2015)152. arXiv: 1504.07869 [hep-ph].

[307] Keisuke Harigaya, Masahiro Ibe and Teppei Kitahara. 'ATLAS on-Z Excess via gluino-Higgsino-singlino decay chains in the NMSSM'. In: *JHEP* 01 (2016), p. 030. DOI: 10.1007/JHEP01(2016)030. arXiv: 1510.07691 [hep-ph].

[308] Archil Kobakhidze et al. 'ATLAS Z-peaked excess in MSSM with a light sbottom or stop'. In: (2015). arXiv: 1504.04390 [hep-ph].

[309] Jack H. Collins, Jeff Asaf Dror and Marco Farina. 'Mixed Stops and the ATLAS on-Z Excess'. In: *Phys. Rev.* D92.9 (2015), p. 095022. DOI: 10.1103/PhysRevD.92.095022. arXiv: 1508.02419 [hep-ph].

[310] Xiaochuan Lu, Satoshi Shirai and Takahiro Terada. 'ATLAS Z Excess in Minimal Supersymmetric Standard Model'. In: (2015). arXiv: 1506.07161 [hep-ph].

[311] Motoi Endo and Yoshitaro Takaesu. 'ATLAS on-Z Excess Through Vector-Like Quarks'. In: *Phys. Lett.* B758 (2016), pp. 355–358. DOI: 10.1016/j.physletb.2016.05.030. arXiv: 1602.05075 [hep-ph].

[312] Natascia Vignaroli. 'Z-peaked excess from heavy gluon decays to vector-like quarks'. In: *Phys.Rev.* D91.11 (2015), p. 115009. DOI: 10.1103/PhysRevD.91.115009. arXiv: 1504.01768 [hep-ph].

[313] Koichi Hamaguchi and Seng Pei Liew. 'Models of a 750 GeV quarkonium and the LHC excesses'. In: *Phys. Rev.* D94.3 (2016), p. 035012. DOI: 10.1103/PhysRevD.94.035012. arXiv: 1604.07828 [hep-ph].

[314] Bogdan A. Dobrescu. 'Leptophobic boson signals with leptons, jets and missing energy'. In: (2015). arXiv: 1506.04435 [hep-ph].

[315] Clifford Cheung, Yasunori Nomura and Jesse Thaler. 'Goldstini'. In: *JHEP* 1003 (2010), p. 073. DOI: 10.1007/JHEP03(2010)073. arXiv: 1002.1967 [hep-ph].

[316] Jesse Thaler and Zachary Thomas. 'Goldstini Can Give the Higgs a Boost'. In: *JHEP* 1107 (2011), p. 060. DOI: 10.1007/JHEP07(2011)060. arXiv: 1103.1631 [hep-ph].

[317] Riccardo Argurio et al. 'Collider signatures of goldstini in gauge mediation'. In: *JHEP* 1206 (2012), p. 096. DOI: 10.1007/JHEP06(2012)096. arXiv: 1112.5058 [hep-ph].

[318] Tao Liu, Lin Wang and Jin Min Yang. 'Higgs decay to goldstini and its observability at the LHC'. In: *Phys.Lett.* B726 (2013), pp. 228–233. DOI: 10.1016/j.physletb.2013.08.001. arXiv: 1301.5479 [hep-ph].

[319] Gabriele Ferretti et al. 'Multiphoton signatures of goldstini at the LHC'. In: *JHEP* 1404 (2014), p. 126. DOI: 10.1007/JHEP04(2014)126. arXiv: 1312.1698 [hep-ph].

[320] Daniele Bertolini, Jesse Thaler and Zachary Thomas. 'Super-Tricks for Superspace'. In: *Proceedings, Theoretical Advanced Study Institute in Elementary Particle Physics: Searching for New Physics at Small and Large Scales (TASI 2012): Boulder, Colorado, June 4-29, 2012.* 2013, pp. 421–496. DOI: 10.1142/9789814525220_0009. arXiv: 1302.6229 [hep-ph].

[321] Karim Benakli and Cesar Moura. 'Brane-Worlds Pseudo-Goldstinos'. In: *Nucl.Phys.* B791 (2008), pp. 125–163. DOI: 10.1016/j.nuclphysb.2007.09.010. arXiv: 0706.3127 [hep-th].

[322] Clifford Cheung et al. 'A Definitive Signal of Multiple Supersymmetry Breaking'. In: *JHEP* 1007 (2010), p. 035. DOI: 10.1007/JHEP07(2010)035. arXiv: 1004.4637 [hep-ph].

[323] Nathaniel Craig, John March-Russell and Matthew McCullough. 'The Goldstini Variations'. In: *JHEP* 1010 (2010), p. 095. DOI: 10.1007/JHEP10(2010)095. arXiv: 1007.1239 [hep-ph].

[324] Hsin-Chia Cheng et al. 'Goldstini as the decaying dark matter'. In: *JHEP* 1103 (2011), p. 019. DOI: 10.1007/JHEP03(2011)019. arXiv: 1012.5300 [hep-ph].

[325] Riccardo Argurio, Zohar Komargodski and Alberto Mariotti. 'Pseudo-Goldstini in Field Theory'. In: *Phys.Rev.Lett.* 107 (2011), p. 061601. DOI: 10.1103/PhysRevLett.107.061601. arXiv: 1102.2386 [hep-th].

[326] Clifford Cheung, Francesco D'Eramo and Jesse Thaler. 'The Spectrum of Goldstini and Modulini'. In: *JHEP* 1108 (2011), p. 115. DOI: 10.1007/JHEP08(2011)115. arXiv: 1104.2600 [hep-ph].

[327] Ken-ichi Hikasa et al. 'Pseudo-goldstino and electroweak gauginos at the LHC'. In: *JHEP* 1407 (2014), p. 065. DOI: 10.1007/JHEP07(2014)065. arXiv: 1403.5731 [hep-ph].

[328] Tao Liu, Lin Wang and Jin Min Yang. 'Pseudo-goldstino and electroweakinos via VBF processes at LHC'. In: *JHEP* 1502 (2015), p. 177. DOI: 10.1007/JHEP02(2015)177. arXiv: 1411.6105 [hep-ph].

[329] M. L. Goldberger and S. B. Treiman. 'Conserved Currents in the Theory of Fermi Interactions'. In: *Phys. Rev.* 110 (1958), pp. 1478–1479. DOI: 10.1103/PhysRev.110.1478.

[330] Feng Luo, Keith A. Olive and Marco Peloso. 'The Gravitino coupling to broken gauge theories applied to the MSSM'. In: *JHEP* 1010 (2010), p. 024. DOI: 10.1007/JHEP10(2010)024. arXiv: 1006.5570 [hep-ph].

[331] Steven Weinberg. *The quantum theory of fields. Vol. 2: Modern applications*. Cambridge University Press, 2013. ISBN: 9781139632478, 9780521670548, 9780521550024.

[332] Zohar Komargodski and Nathan Seiberg. 'From Linear SUSY to Constrained Superfields'. In: *JHEP* 0909 (2009), p. 066. DOI: 10.1088/1126-6708/2009/09/066. arXiv: 0907.2441 [hep-th].

[333] M. Rocek. 'Linearizing the Volkov-Akulov Model'. In: *Phys. Rev. Lett.* 41 (1978), pp. 451–453. DOI: 10.1103/PhysRevLett.41.451.

[334] U. Lindstrom and M. Rocek. 'CONSTRAINED LOCAL SUPERFIELDS'. In: *Phys. Rev.* D19 (1979), pp. 2300–2303. DOI: 10.1103/PhysRevD.19.2300.

[335] Stuart Samuel and Julius Wess. 'A Superfield Formulation of the Nonlinear Realization of Supersymmetry and Its Coupling to Supergravity'. In: *Nucl. Phys.* B221 (1983), pp. 153–177. DOI: 10.1016/0550-3213(83)90622-3.

[336] R. Casalbuoni et al. 'Nonlinear Realization of Supersymmetry Algebra From Supersymmetric Constraint'. In: *Phys. Lett.* B220 (1989), pp. 569–575. DOI: 10.1016/0370-2693(89)90788-0.

[337] I. Antoniadis et al. 'Non-linear MSSM'. In: *Nucl.Phys.* B841 (2010), pp. 157–177. DOI: 10.1016/j.nuclphysb.2010.08.002. arXiv: 1006.1662 [hep-ph].

[338] Howard Baer, Xerxes Tata and Jeffrey Woodside. 'Phenomenology of Gluino Decays via Loops and Top Quark Yukawa Coupling'. In: *Phys.Rev.* D42 (1990), pp. 1568–1576. DOI: 10.1103/PhysRevD.42.1568.

[339] P. Gambino, G.F. Giudice and P. Slavich. 'Gluino decays in split supersymmetry'. In: *Nucl.Phys.* B726 (2005), pp. 35–52. DOI: 10.1016/j.nuclphysb.2005.08.011. arXiv: hep-ph/0506214 [hep-ph].

[340] Masahiro Kawasaki, Kazunori Kohri and Takeo Moroi. 'Big-Bang nucleosynthesis and hadronic decay of long-lived massive particles'. In: *Phys.Rev.* D71 (2005), p. 083502. DOI: 10.1103/PhysRevD.71.083502. arXiv: astro-ph/0408426 [astro-ph].

[341] Ian-Woo Kim et al. *ATOM: Automated Tests Of Models*. in preparation.

[342] Gary J. Feldman and Robert D. Cousins. 'A Unified approach to the classical statistical analysis of small signals'. In: *Phys. Rev.* D57 (1998), pp. 3873–3889. DOI: 10.1103/PhysRevD.57.3873. arXiv: physics/9711021 [physics.data-an].

[343] Torbjorn Sjostrand, Stephen Mrenna and Peter Z. Skands. 'PYTHIA 6.4 Physics and Manual'. In: *JHEP* 05 (2006), p. 026. DOI: 10.1088/1126-6708/2006/05/026. arXiv: hep-ph/0603175 [hep-ph].

[344] Kentarou Mawatari. 'Associated production of light gravitinos at future linear colliders'. In: (2012). arXiv: 1202.0507 [hep-ph].

[345] K. Mawatari and Y. Takaesu. 'HELAS and MadGraph with goldstinos'. In: *Eur.Phys.J.* C71 (2011), p. 1640. DOI: 10.1140/epjc/s10052-011-1640-3. arXiv: 1101.1289 [hep-ph].

[346] Morad Aaboud et al. 'Search for new phenomena in events containing a same-flavour opposite-sign dilepton pair, jets, and large missing transverse momentum in $\sqrt{s} = 13$ $pp$ collisions with the ATLAS detector'. In: *Eur. Phys. J.* C77.3 (2017), p. 144. DOI: 10.1140/epjc/s10052-017-4700-5. arXiv: 1611.05791 [hep-ex].

[347] Morad Aaboud et al. 'Search for new phenomena using the invariant mass distribution of same-flavour opposite-sign dilepton pairs in events with missing transverse momentum in $\sqrt{s} = 13$ TeV pp collisions with the ATLAS detector'. In: *Eur. Phys. J.* C78.8 (2018), p. 625. DOI: 10.1140/epjc/s10052-018-6081-9. arXiv: 1805.11381 [hep-ex].

[348] Vardan Khachatryan et al. 'Search for new physics in final states with two opposite-sign, same-flavor leptons, jets, and missing transverse momentum in pp collisions at sqrt(s) = 13 TeV'. In: *JHEP* 12 (2016), p. 013. DOI: 10.1007/JHEP12(2016)013. arXiv: 1607.00915 [hep-ex].

[349] Albert M Sirunyan et al. 'Search for new phenomena in final states with two opposite-charge, same-flavor leptons, jets, and missing transverse momentum in pp collisions at $\sqrt{s} = 13$ TeV'. In: *JHEP* 03 (2018), p. 076. DOI: 10.1007/s13130-018-7845-2,10.1007/JHEP03(2018)076. arXiv: 1709.08908 [hep-ex].

[350] Morad Aaboud et al. 'Search for squarks and gluinos in final states with jets and missing transverse momentum using 36 fb$^{-1}$ of $\sqrt{s} = 13$ TeV pp collision data with the ATLAS detector'. In: *Phys. Rev.* D97.11 (2018), p. 112001. DOI: 10.1103/PhysRevD.97.112001. arXiv: 1712.02332 [hep-ex].

[351] Matthew R. Buckley et al. 'Super-Razor and Searches for Sleptons and Charginos at the LHC'. In: *Phys. Rev.* D89.5 (2014), p. 055020. DOI: 10.1103/PhysRevD.89.055020. arXiv: 1310.4827 [hep-ph].

[352] Paul Jackson, Christopher Rogan and Marco Santoni. 'Sparticles in motion: Analyzing compressed SUSY scenarios with a new method of event reconstruction'. In: *Phys. Rev.* D95.3 (2017), p. 035031. DOI: 10.1103/PhysRevD.95.035031. arXiv: 1607.08307 [hep-ph].

[353] Paul Jackson and Christopher Rogan. 'Recursive Jigsaw Reconstruction: HEP event analysis in the presence of kinematic and combinatoric ambiguities'. In: *Phys. Rev.* D96.11 (2017), p. 112007. DOI: 10.1103/PhysRevD.96.112007. arXiv: 1705.10733 [hep-ph].

[354] Vanessa Cirkel-Bartelt. 'History of Astroparticle Physics and its Components'. In: *Living Reviews in Relativity* 11.1 (May 2008), p. 2. ISSN: 1433-8351. DOI: 10.12942/lrr-2008-2. URL: https://doi.org/10.12942/lrr-2008-2.

[355] V. F. Hess. 'Über Beobachtungen der durchdringenden Strahlung bei sieben Freiballonfahrten'. In: *Physikalische Zeitschrift* 13 (Nov. 1912), pp. 1084–1091.

[356] John G. Wilson. *Cosmic rays*. Vol. 40. The Wykeham science series. Wykeham Publications, 1976.

[357] Arno A. Penzias and Robert Woodrow Wilson. 'A Measurement of excess antenna temperature at 4080-Mc/s'. In: *Astrophys. J.* 142 (1965), pp. 419–421. DOI: 10.1086/148307.

[358] R. A. Alpher, H. Bethe and G. Gamow. 'The origin of chemical elements'. In: *Phys. Rev.* 73 (1948), pp. 803–804. DOI: 10.1103/PhysRev.73.803.

[359] P. J. E. Peebles. 'Large scale background temperature and mass fluctuations due to scale invariant primeval perturbations'. In: *Astrophys. J.* 263 (1982). [,85(1982)], pp. L1–L5. DOI: 10.1086/183911.

[360] J. R. Bond, A. S. Szalay and Michael S. Turner. 'Formation of Galaxies in a Gravitino Dominated Universe'. In: *Phys. Rev. Lett.* 48 (1982), p. 1636. DOI: 10.1103/PhysRevLett.48.1636.

[361] George R. Blumenthal, Heinz Pagels and Joel R. Primack. 'GALAXY FORMATION BY DISSIPATIONLESS PARTICLES HEAVIER THAN NEUTRINOS'. In: *Nature* 299 (1982), pp. 37–38. DOI: 10.1038/299037a0.

[362] George R. Blumenthal et al. 'Formation of Galaxies and Large Scale Structure with Cold Dark Matter'. In: *Nature* 311 (1984). [,96(1984)], pp. 517–525. DOI: 10.1038/311517a0.

[363] Michelangelo Mangano. 'Physics at the FCC-hh, a 100 TeV pp collider'. In: (2017). DOI: 10.23731/CYRM-2017-003. arXiv: 1710.06353 [hep-ph].

[364] W. C. Haxton. 'The solar neutrino problem'. In: *Ann. Rev. Astron. Astrophys.* 33 (1995), pp. 459–503. DOI: 10.1146/annurev.aa.33.090195.002331. arXiv: hep-ph/9503430 [hep-ph].

[365] M. G. Aartsen et al. 'Measurement of the multi-TeV neutrino cross section with IceCube using Earth absorption'. In: *Nature* (2017). DOI: 10.1038/nature24459. arXiv: 1711.08119 [hep-ex].

[366] Mauricio Bustamante and Amy Connolly. 'Measurement of the Energy-Dependent Neutrino-Nucleon Cross Section Above 10 TeV Using IceCube Showers'. In: (2017). arXiv: 1711.11043 [astro-ph.HE].

[367] T.K. Gaisser, R. Engel and E. Resconi. *Cosmic Rays and Particle Physics*. Cambridge University Press, 2016. ISBN: 9781316598436. URL: https://books.google.be/books?id=nHFNDAAAQBAJ.

[368] M.S. Longair. *High Energy Astrophysics*. Cambridge University Press, 2011. ISBN: 9781139494540. URL: https://books.google.be/books?id=KGe3FVbDNk4C.

[369] Pijushpani Bhattacharjee and Gunter Sigl. 'Origin and propagation of extremely high-energy cosmic rays'. In: *Phys. Rept.* 327 (2000), pp. 109–247. DOI: 10.1016/S0370-1573(99)00101-5. arXiv: astro-ph/9811011 [astro-ph].

[370] A. E. Vladimirov et al. 'GALPROP WebRun: An internet-based service for calculating galactic cosmic ray propagation and associated photon emissions'. In: *Computer Physics Communications* 182 (2011), pp. 1156–1161. DOI: 10.1016/j.cpc.2011.01.017. arXiv: 1008.3642 [astro-ph.HE].

[371] Carmelo Evoli et al. 'Cosmic-ray propagation with *DRAGON*2: I. numerical solver and astrophysical ingredients'. In: *JCAP* 1702.02 (2017), p. 015. DOI: 10.1088/1475-7516/2017/02/015. arXiv: 1607.07886 [astro-ph.HE].

[372] D. Heck et al. *CORSIKA: a Monte Carlo code to simulate extensive air showers*. 1998.

[373] Alexander Aab et al. 'The Pierre Auger Cosmic Ray Observatory'. In: *Nucl. Instrum. Meth.* A798 (2015), pp. 172–213. DOI: 10.1016/j.nima.2015.06.058. arXiv: 1502.01323 [astro-ph.IM].

[374] M. Fukushima. 'Telescope array project for extremely high energy cosmic rays'. In: *Prog. Theor. Phys. Suppl.* 151 (2003), pp. 206–210. DOI: 10.1143/PTPS.151.206.

[375] Alexander Aab et al. 'Combined fit of spectrum and composition data as measured by the Pierre Auger Observatory'. In: *JCAP* 1704.04 (2017). [Erratum: JCAP1803,no.03,E02(2018)], p. 038. DOI: 10.1088/1475-7516/2018/03/E02,10.1088/1475-7516/2017/04/038. arXiv: 1612.07155 [astro-ph.HE].

[376] A. Aab et al. 'Depth of maximum of air-shower profiles at the Pierre Auger Observatory. II. Composition implications'. In: *Phys. Rev.* D90.12 (2014), p. 122006. DOI: 10.1103/PhysRevD.90.122006. arXiv: 1409.5083 [astro-ph.HE].

[377] Alexander Aab et al. 'The Pierre Auger Observatory: Contributions to the 35th International Cosmic Ray Conference (ICRC 2017)'. In: *Proceedings, 35th International Cosmic Ray Conference (ICRC 2017): Bexco, Busan, Korea, July 12-20, 2017*. 2017. arXiv: 1708.06592 [astro-ph.HE]. URL: http://inspirehep.net/record/1617990/files/arXiv:1708.06592.pdf.

[378] Michael Unger. 'Report of the Working Group on the Composition of Ultra-High Energy Cosmic Rays'. In: *PoS* ICRC2015 (2016), p. 307.

[379] R. Abbasi et al. 'Report of the Working Group on the Composition of Ultra High Energy Cosmic Rays'. In: *JPS Conf. Proc.* 9 (2016), p. 010016. DOI: 10.7566/JPSCP.9.010016. arXiv: 1503.07540 [astro-ph.HE].

[380] 'Pierre Auger Observatory and Telescope Array: Joint Contributions to the 35th International Cosmic Ray Conference (ICRC 2017)'. In: *Proceedings, 35th International Cosmic Ray Conference (ICRC 2017): Bexco, Busan, Korea, July 12-20, 2017*. 2018. arXiv: 1801.01018 [astro-ph.HE]. URL: https://inspirehep.net/record/1645900/files/arXiv:1801.01018.pdf.

[381] Björn Eichmann et al. 'Ultra-High-Energy Cosmic Rays from Radio Galaxies'. In: *JCAP* 1802.02 (2018), p. 036. DOI: 10.1088/1475-7516/2018/02/036. arXiv: 1701.06792 [astro-ph.HE].

[382] Julia K. Becker. 'High-energy neutrinos in the context of multimessenger physics'. In: *Phys. Rept.* 458 (2008), pp. 173–246. DOI: 10.1016/j.physrep.2007.10.006. arXiv: 0710.1557 [astro-ph].

[383] Alexander Aab et al. 'Observation of a Large-scale Anisotropy in the Arrival Directions of Cosmic Rays above $8 \times 10^{18}$ eV'. In: *Science* 357.6537 (2017), pp. 1266–1270. DOI: 10.1126/science.aan4338. arXiv: 1709.07321 [astro-ph.HE].

[384] Karl-Heinz Kampert. 'Ultrahigh-Energy Cosmic Rays: Results and Prospects'. In: *Braz. J. Phys.* 43 (2013), pp. 375–382. DOI: 10.1007/s13538-013-0150-1. arXiv: 1305.2363 [astro-ph.HE].

[385] J. Abraham et al. 'Upper limit on the cosmic-ray photon fraction at EeV energies from the Pierre Auger Observatory'. In: *Astropart. Phys.* 31 (2009), pp. 399–406. DOI: 10.1016/j.astropartphys.2009.04.003. arXiv: 0903.1127 [astro-ph.HE].

[386] J. Abraham et al. 'Upper limit on the cosmic-ray photon flux above $10^{19}$ eV using the surface detector of the Pierre Auger Observatory'. In: *Astropart. Phys.* 29 (2008), pp. 243–256. DOI: 10.1016/j.astropartphys.2008.01.003. arXiv: 0712.1147 [astro-ph].

[387] J. Abraham et al. 'An upper limit to the photon fraction in cosmic rays above $10^{19}$-eV from the Pierre Auger Observatory'. In: *Astropart. Phys.* 27 (2007), pp. 155–168. DOI: 10.1016/j.astropartphys.2006.10.004. arXiv: astro-ph/0606619 [astro-ph].

[388] Dmitry V. Semikoz and Gunter Sigl. 'Ultrahigh-energy neutrino fluxes: New constraints and implications'. In: *JCAP* 0404 (2004), p. 003. DOI: 10.1088/1475-7516/2004/04/003. arXiv: hep-ph/0309328 [hep-ph].

[389] J. Abraham et al. 'Upper limit on the diffuse flux of UHE tau neutrinos from the Pierre Auger Observatory'. In: *Phys. Rev. Lett.* 100 (2008), p. 211101. DOI: 10.1103/PhysRevLett.100.211101. arXiv: 0712.1909 [astro-ph].

[390] P. Abreu et al. 'A Search for Ultra-High Energy Neutrinos in Highly Inclined Events at the Pierre Auger Observatory'. In: *Phys. Rev.* D84 (2011). [Erratum: Phys. Rev.D84,029902(2011)], p. 122005. DOI: 10.1103/PhysRevD.85.029902, 10.1103/PhysRevD.84.122005. arXiv: 1202.1493 [astro-ph.HE].

[391] P. Abreu et al. 'Search for point-like sources of ultra-high energy neutrinos at the Pierre Auger Observatory and improved limit on the diffuse flux of tau neutrinos'. In: *Astrophys. J.* 755 (2012), p. L4. DOI: 10.1088/2041-8205/755/1/L4. arXiv: 1210.3143 [astro-ph.HE].

[392] Pedro Abreu et al. 'Ultrahigh Energy Neutrinos at the Pierre Auger Observatory'. In: *Adv. High Energy Phys.* 2013 (2013), p. 708680. DOI: 10.1155/2013/708680. arXiv: 1304.1630 [astro-ph.HE].

[393] C. Grupen. *Astroparticle Physics*. SpringerLink: Springer e-Books. Springer, 2005. ISBN: 9783540253129. URL: https://books.google.be/books?id=ueqGAKjt0-MC.

[394] Masaaki Yamada, Russell Kulsrud and Hantao Ji. 'Magnetic reconnection'. In: *Rev. Mod. Phys.* 82 (1 Mar. 2010), pp. 603–664. DOI: 10.1103/RevModPhys.82.603. URL: https://link.aps.org/doi/10.1103/RevModPhys.82.603.

[395] M. Ackermann et al. 'High-energy Gamma-Ray Emission from Solar Flares: Summary of Fermi Large Area Telescope Detections and Analysis of Two M-class Flares'. In: *Astrophys. J.* 787 (2014), p. 15. DOI: 10.1088/0004-637X/787/1/15. arXiv: 1304.3749 [astro-ph.HE].

[396] Edward L. Chupp and James M. Ryan. 'High energy neutron and pion-decay gamma-ray emissions from solar flares'. In: *Research in Astronomy and Astrophysics* 9.1 (2009), p. 11. URL: http://stacks.iop.org/1674-4527/9/i=1/a=003.

[397] M. E. J. Newman. 'Power laws, Pareto distributions and Zipf's law'. In: *Contemporary Physics* 46 (2005), pp. 323–351. DOI: 10.1080/00107510500052444. eprint: cond-mat/0412004.

[398] ENRICO Fermi. 'On the Origin of the Cosmic Radiation'. In: *Phys. Rev.* 75 (8 Apr. 1949), pp. 1169–1174. DOI: 10.1103/PhysRev.75.1169. URL: https://link.aps.org/doi/10.1103/PhysRev.75.1169.

[399] G. F. Krymskii. 'A regular mechanism for the acceleration of charged particles on the front of a shock wave'. In: *Akademiia Nauk SSSR Doklady* 234 (1977), pp. 1306–1308.

[400] W. I. Axford, E. Leer and G. Skadron. 'The acceleration of cosmic rays by shock waves'. In: *International Cosmic Ray Conference* 11 (1977), pp. 132–137.

[401] A. R. Bell. 'The acceleration of cosmic rays in shock fronts. I'. In: MNRAS 182 (1978), pp. 147–156. DOI: 10.1093/mnras/182.2.147.

[402] A. R. Bell. 'The acceleration of cosmic rays in shock fronts. II'. In: MNRAS 182 (1978), pp. 443–455. DOI: 10.1093/mnras/182.3.443.

[403] R. D. Blandford and J. P. Ostriker. 'Particle acceleration by astrophysical shocks'. In: ApJ 221 (1978), pp. L29–L32. DOI: 10.1086/182658.

[404] Masahiro Hoshino. 'Nonthermal Particle Acceleration in Shock Front Region: "Shock Surfing Accelerations"'. In: *Progress of Theoretical Physics Supplement* 143 (2001), pp. 149–181. DOI: 10.1143/PTPS.143.149. eprint: /oup/backfile/content_public/journal/ptps/143/10.1143/ptps.143.149/2/143-149.pdf. URL: +%20http://dx.doi.org/10.1143/PTPS.143.149.

[405] K. Koyama et al. 'Evidence for shock acceleration of high-energy electrons in the supernova remnant SN1006'. In: *Nature* 378 (1995), pp. 255–258. DOI: 10.1038/378255a0.

[406] F. C. Jones and D. C. Ellison. 'The plasma physics of shock acceleration'. In: Space Sci. Rev. 58 (1991), pp. 259–346. DOI: 10.1007/BF01206003.

[407] J. A. Peacock. 'Fermi acceleration by relativistic shock waves'. In: MNRAS 196 (1981), pp. 135–152. DOI: 10.1093/mnras/196.2.135.

[408] J. G. Kirk and P. Schneider. 'Particle acceleration at shocks - A Monte Carlo method'. In: ApJ 322 (1987), pp. 256–265. DOI: 10.1086/165720.

[409] Andrei Bykov et al. 'Particle acceleration in relativistic outflows'. In: *Space Sci. Rev.* 173 (2012), pp. 309–339. DOI: 10.1007/s11214-012-9896-y. arXiv: 1205.2208 [astro-ph.HE].

[410] Charles D. Dermer and Soebur Razzaque. 'Acceleration of Ultra-High Energy Cosmic Rays in the Colliding Shells of Blazars and GRBs: Constraints from the Fermi Gamma ray Space Telescope'. In: *Astrophys. J.* 724 (2010), pp. 1366–1372. DOI: 10.1088/0004-637X/724/2/1366. arXiv: 1004.4249 [astro-ph.HE].

[411] Athina Meli, Julia K. Becker and John J. Quenby. 'Cosmic ray acceleration in subluminal and superluminal relativistic shock environments'. In: (2007). [Astron. Astrophys.492,323(2008)]. DOI: 10.1051/0004-6361:20078681. arXiv: 0709.3031 [astro-ph].

[412] A. Meli et al. 'Particle acceleration in relativistic subluminal shock environments'. In: *Proceedings, 30th International Cosmic Ray Conference (ICRC 2007): Merida, Yucatan, Mexico, July 3-11, 2007.* Vol. 2. 2007, pp. 287–292. arXiv: 0708.1438 [astro-ph]. URL: http://indico.nucleares.unam.mx/contributionDisplay.py?contribId=1060&confId=4.

[413] Pisin Chen, Toshiki Tajima and Yoshiyuki Takahashi. 'Plasma wakefield acceleration for ultrahigh-energy cosmic rays'. In: *Phys. Rev. Lett.* 89 (2002), p. 161101. DOI: 10.1103/PhysRevLett.89.161101. arXiv: astro-ph/0205287 [astro-ph].

[414] Fan Guo et al. 'Formation of Hard Power-laws in the Energetic Particle Spectra Resulting from Relativistic Magnetic Reconnection'. In: *Phys. Rev. Lett.* 113 (2014), p. 155005. DOI: 10.1103/PhysRevLett.113.155005. arXiv: 1405.4040 [astro-ph.HE].

[415] Tobias Winchen and Stijn Buitink. 'Efficient Second Order Fermi Accelerators as Sources of Ultra-High-Energy Cosmic Rays'. In: (2016). arXiv: 1612.03675 [astro-ph.HE].

[416] Pasquale Blasi, Richard I. Epstein and Angela V. Olinto. 'Ultrahigh-energy cosmic rays from young neutron star winds'. In: *Astrophys. J.* 533 (2000), p. L123. DOI: 10.1086/312626. arXiv: astro-ph/9912240 [astro-ph].

[417] Kumiko Kotera, Elena Amato and Pasquale Blasi. 'The fate of ultrahigh energy nuclei in the immediate environment of young fast-rotating pulsars'. In: *JCAP* 1508.08 (2015), p. 026. DOI: 10.1088/1475-7516/2015/08/026. arXiv: 1503.07907 [astro-ph.HE].

[418] K. Ptitsyna and A. Neronov. 'Particle acceleration in the vacuum gaps in black hole magnetospheres'. In: *Astron. Astrophys.* 593 (2016), A8. DOI: 10.1051/0004-6361/201527549. arXiv: 1510.04023 [astro-ph.HE].

[419] A. M. Hillas. 'The Origin of Ultra-High-Energy Cosmic Rays'. In: ARA&A 22 (1984), pp. 425–444. DOI: 10.1146/annurev.aa.22.090184.002233.

[420] Rafael Alves Batista et al. 'Open Questions in Cosmic-Ray Research at Ultrahigh Energies'. In: *Front. Astron. Space Sci.* 6 (2019), p. 23. DOI: 10.3389/fspas.2019.00023. arXiv: 1903.06714 [astro-ph.HE].

[421] V. Beckmann and C. Shrader. *Active Galactic Nuclei.* Physics textbook. Wiley, 2012. ISBN: 9783527410910. URL: https://books.google.be/books?id=c44yvicEX70C.

[422] A. Koratkar and O. Blaes. 'The Ultraviolet and Optical Continuum Emission in Active Galactic Nuclei: The Status of Accretion Disks'. In: PASP 111 (1999), pp. 1–30. DOI: 10.1086/316294.

[423] Cristina Ramos Almeida and Claudio Ricci. 'Nuclear obscuration in active galactic nuclei'. In: *Nature Astronomy* 1 (Oct. 2017), pp. 679–689. DOI: 10.1038/s41550-017-0232-z. arXiv: 1709.00019 [astro-ph.GA].

[424] R. Antonucci. 'Unified models for active galactic nuclei and quasars'. In: ARA&A 31 (1993), pp. 473–521. DOI: 10.1146/annurev.aa.31.090193.002353.

[425] C. Megan Urry and Paolo Padovani. 'Unified schemes for radio-loud active galactic nuclei'. In: *Publ. Astron. Soc. Pac.* 107 (1995), p. 803. DOI: 10.1086/133630. arXiv: astro-ph/9506063 [astro-ph].

[426] Ranieri D. Baldi, Alessandro Capetti and Gabriele Giovannini. 'The new class of FR 0 radio galaxies'. In: *Astron. Nachr.* 337.1/2 (2017), pp. 114–119. DOI: 10.1002/asna.201512275. arXiv: 1510.04272 [astro-ph.GA].

[427] P. L. Biermann and P. A. Strittmatter. 'Synchrotron emission from shock waves in active galactic nuclei'. In: *Astrophys. J.* 322 (1987), pp. 643–649. DOI: 10.1086/165759.

[428] G. Fossati et al. 'A Unifying view of the spectral energy distributions of blazars'. In: *Mon. Not. Roy. Astron. Soc.* 299 (1998), pp. 433–448. DOI: 10.1046/j.1365-8711.1998.01828.x. arXiv: astro-ph/9804103 [astro-ph].

[429] Gabriele Ghisellini. 'The blazar sequence 2.0'. In: *Galaxies* 4.4 (2016), p. 36. DOI: 10.3390/galaxies4040036. arXiv: 1609.08606 [astro-ph.HE].

[430] Omer Bromberg et al. 'Short vs Long and Collapsars vs. non-Collapsar: a quantitative classification of GRBs'. In: *Astrophys. J.* 764 (2013), p. 179. DOI: 10.1088/0004-637X/764/2/179. arXiv: 1210.0068 [astro-ph.HE].

[431] S. R. Kulkarni et al. 'Radio emission from the unusual supernova 1998bw and its association with the gamma-ray burst of 25 April 1998'. In: *Nature* 395 (1998), pp. 663–669. DOI: 10.1038/27139.

[432] David Eichler et al. 'Nucleosynthesis, Neutrino Bursts and Gamma-Rays from Coalescing Neutron Stars'. In: *Nature* 340 (1989). [,682(1989)], pp. 126–128. DOI: 10.1038/340126a0.

[433] Ramesh Narayan, Bohdan Paczynski and Tsvi Piran. 'Gamma-ray bursts as the death throes of massive binary stars'. In: *Astrophys. J.* 395 (1992), pp. L83–L86. DOI: 10.1086/186493. arXiv: astro-ph/9204001 [astro-ph].

[434] B. P. Abbott et al. 'GW170817: Observation of Gravitational Waves from a Binary Neutron Star Inspiral'. In: *Phys. Rev. Lett.* 119.16 (2017), p. 161101. DOI: 10.1103/PhysRevLett.119.161101. arXiv: 1710.05832 [gr-qc].

[435] B. P. Abbott et al. 'Gravitational Waves and Gamma-rays from a Binary Neutron Star Merger: GW170817 and GRB 170817A'. In: *Astrophys. J.* 848.2 (2017), p. L13. DOI: 10.3847/2041-8213/aa920c. arXiv: 1710.05834 [astro-ph.HE].

[436] B. P. Abbott et al. 'Multi-messenger Observations of a Binary Neutron Star Merger'. In: *Astrophys. J.* 848.2 (2017), p. L12. DOI: 10.3847/2041-8213/aa91c9. arXiv: 1710.05833 [astro-ph.HE].

[437] M. J. Rees and P. Meszaros. 'Relativistic fireballs - energy conversion and time - scales'. In: *Mon. Not. Roy. Astron. Soc.* 258 (1992), pp. 41–43.

[438] Eli Waxman. 'Cosmological gamma-ray bursts and the highest energy cosmic rays'. In: *Phys. Rev. Lett.* 75 (1995), pp. 386–389. DOI: 10.1103/PhysRevLett.75.386. arXiv: astro-ph/9505082 [astro-ph].

[439] Mario Vietri. 'On the acceleration of ultrahigh-energy cosmic rays in gamma-ray bursts'. In: *Astrophys. J.* 453 (1995), pp. 883–889. DOI: 10.1086/176448. arXiv: astro-ph/9506081 [astro-ph].

[440] Noemie Globus et al. 'UHECR acceleration at GRB internal shocks'. In: *Mon. Not. Roy. Astron. Soc.* 451.1 (2015), pp. 751–790. DOI: 10.1093/mnras/stv893. arXiv: 1409.1271 [astro-ph.HE].

[441] P. Meszaros and M. J. Rees. 'Optical and long wavelength afterglow from gamma-ray bursts'. In: *Astrophys. J.* 476 (1997), pp. 232–237. DOI: 10.1086/303625. arXiv: astro-ph/9606043 [astro-ph].

[442] Soebur Razzaque, Peter Meszaros and Eli Waxman. 'Neutrino tomography of gamma-ray bursts and massive stellar collapses'. In: *Phys. Rev.* D68 (2003), p. 083001. DOI: 10.1103/PhysRevD.68.083001. arXiv: astro-ph/0303505 [astro-ph].

[443] Eli Waxman and John N. Bahcall. 'Neutrino afterglow from gamma-ray bursts: Similar to 10**18-eV'. In: *Astrophys. J.* 541 (2000), pp. 707–711. DOI: 10.1086/309462. arXiv: hep-ph/9909286 [hep-ph].

[444] J. P. Ostriker. 'The acceleration of high-energy cosmic rays by pulsars'. In: *International Cosmic Ray Conference*. Vol. 1. International Cosmic Ray Conference. Jan. 1970, p. 69.

[445] Jonathan Arons. 'Magnetars in the metagalaxy: an origin for ultrahigh-energy cosmic rays in the nearby universe'. In: *Astrophys. J.* 589 (2003), pp. 871–892. DOI: 10.1086/374776. arXiv: astro-ph/0208444 [astro-ph].

[446] G. Ghisellini et al. 'Ultra-High Energy Cosmic Rays, Spiral galaxies and Magnetars'. In: *Mon. Not. Roy. Astron. Soc.* 390 (2008), pp. L88–L92. DOI: 10.1111/j.1745-3933.2008.00547.x. arXiv: 0806.2393 [astro-ph].

[447] Kumiko Kotera. 'Ultrahigh energy cosmic ray acceleration in newly born magnetars and their associated gravitational wave signatures'. In: Phys. Rev. D 84.2, 023002 (July 2011), p. 023002. DOI: 10.1103/PhysRevD.84.023002. arXiv: 1106.3060 [astro-ph.HE].

[448] Ke Fang, Kumiko Kotera and Angela V. Olinto. 'Newly Born Pulsars as Sources of Ultrahigh Energy Cosmic Rays'. In: ApJ 750.2, 118 (May 2012), p. 118. DOI: 10.1088/0004-637X/750/2/118. arXiv: 1201.5197 [astro-ph.HE].

[449] Ke Fang, Kumiko Kotera and Angela V. Olinto. 'Ultrahigh Energy Cosmic Ray Nuclei from Extragalactic Pulsars and the effect of their Galactic counterparts'. In: *JCAP* 1303 (2013), p. 010. DOI: 10.1088/1475-7516/2013/03/010. arXiv: 1302.4482 [astro-ph.HE].

[450] Susumu Inoue. 'Origin of ultra-high energy cosmic rays in the era of Auger and Telescope Array'. In: *J. Phys. Conf. Ser.* 120 (2008), p. 062001. DOI: 10.1088/1742-6596/120/6/062001. arXiv: 0809.3205 [astro-ph].

[451] Glennys R. Farrar and Andrei Gruzinov. 'Giant AGN Flares and Cosmic Ray Bursts'. In: *Astrophys. J.* 693 (2009), pp. 329–332. DOI: 10.1088/0004-637X/693/1/329. arXiv: 0802.1074 [astro-ph].

[452] Glennys R. Farrar and Tsvi Piran. 'Tidal disruption jets as the source of Ultra-High Energy Cosmic Rays'. In: (2014). arXiv: 1411.0704 [astro-ph.HE].

[453] Daniel Biehl et al. 'Tidally disrupted stars as a possible origin of both cosmic rays and neutrinos at the highest energies'. In: (2017). arXiv: 1711.03555 [astro-ph.HE].

[454] Xavier Rodrigues et al. 'Neutrinos and Ultra-High-Energy Cosmic-Ray Nuclei from Blazars'. In: *Astrophys. J.* 854.1 (2018), p. 54. DOI: 10.3847/1538-4357/aaa7ee. arXiv: 1711.02091 [astro-ph.HE].

[455] Jorg R. Horandel. 'Cosmic rays from the knee to the second knee: 10**4 to 10**18-eV'. In: *Mod. Phys. Lett.* A22 (2007). [,63(2006)], pp. 1533–1552. DOI: 10.1142/S0217732307024139. arXiv: astro-ph/0611387 [astro-ph].

[456] S. Thoudam et al. 'Cosmic-ray energy spectrum and composition up to the ankle: the case for a second Galactic component'. In: *Astron. Astrophys.* 595 (2016), A33. DOI: 10.1051/0004-6361/201628894. arXiv: 1605.03111 [astro-ph.HE].

[457] H. Bethe and W. Heitler. 'On the Stopping of fast particles and on the creation of positive electrons'. In: *Proc. Roy. Soc. Lond.* A146 (1934), pp. 83–112. DOI: 10.1098/rspa.1934.0140.

[458] Todor Stanev et al. 'Propagation of ultrahigh-energy protons in the nearby universe'. In: *Phys. Rev.* D62 (2000), p. 093005. DOI: 10.1103/PhysRevD.62.093005. arXiv: astro-ph/0003484 [astro-ph].

[459] V. Berezinsky. 'UHECR: Signatures and Models'. In: *EPJ Web Conf.* 53 (2013), p. 01003. DOI: 10.1051/epjconf/20135301003. arXiv: 1307.4043 [astro-ph.HE].

[460] Kenneth Greisen. 'End to the cosmic ray spectrum?' In: *Phys. Rev. Lett.* 16 (1966), pp. 748–750. DOI: 10.1103/PhysRevLett.16.748.

[461] G. T. Zatsepin and V. A. Kuzmin. 'Upper limit of the spectrum of cosmic rays'. In: *JETP Lett.* 4 (1966). [Pisma Zh. Eksp. Teor. Fiz.4,114(1966)], pp. 78–80.

[462] R. Ruffini, G. V. Vereshchagin and S. S. Xue. 'Cosmic absorption of ultra high energy particles'. In: *Astrophys. Space Sci.* 361 (2016), p. 82. DOI: 10.1007/s10509-016-2668-5. arXiv: 1503.07749 [astro-ph.HE].

[463] Pedro Abreu et al. 'Interpretation of the Depths of Maximum of Extensive Air Showers Measured by the Pierre Auger Observatory'. In: *JCAP* 1302 (2013), p. 026. DOI: 10.1088/1475-7516/2013/02/026. arXiv: 1301.6637 [astro-ph.HE].

[464] Matthias Plum. 'Measurements of the Mass Composition of UHECRs with the Pierre Auger Observatory'. In: *JPS Conf. Proc.* 19 (2018), p. 011011. DOI: 10.7566/JPSCP.19.011011.

[465] R. Aloisio et al. 'Cosmogenic neutrinos and ultra-high energy cosmic ray models'. In: *JCAP* 1510.10 (2015), p. 006. DOI: 10.1088/1475-7516/2015/10/006. arXiv: 1505.04020 [astro-ph.HE].

[466] K.-H. Kampert and M. Unger. 'Measurements of the cosmic ray composition with air shower experiments'. In: *Astroparticle Physics* 35 (2012), pp. 660–678. DOI: 10.1016/j.astropartphys.2012.02.004. arXiv: 1201.0018 [astro-ph.HE].

[467] Jonas Heinze et al. 'Cosmogenic Neutrinos Challenge the Cosmic Ray Proton Dip Model'. In: *Astrophys. J.* 825.2 (2016), p. 122. DOI: 10.3847/0004-637X/825/2/122. arXiv: 1512.05988 [astro-ph.HE].

[468] M. G. Aartsen et al. 'Constraints on Ultrahigh-Energy Cosmic-Ray Sources from a Search for Neutrinos above 10 PeV with IceCube'. In: *Phys. Rev. Lett.* 117.24 (2016). [Erratum: Phys. Rev. Lett.119,no.25,259902(2017)], p. 241101. DOI: 10.1103/PhysRevLett.117.241101,10.1103/PhysRevLett.119.259902. arXiv: 1607.05886 [astro-ph.HE].

[469] Michael Unger, Glennys R. Farrar and Luis A. Anchordoqui. 'Origin of the ankle in the ultrahigh energy cosmic ray spectrum, and of the extragalactic protons below it'. In: *Phys. Rev.* D92.12 (2015), p. 123001. DOI: 10.1103/PhysRevD.92.123001. arXiv: 1505.02153 [astro-ph.HE].

[470] Noemie Globus, Denis Allard and Etienne Parizot. 'A complete model of the cosmic ray spectrum and composition across the Galactic to extragalactic transition'. In: *Phys. Rev.* D92.2 (2015), p. 021302. DOI: 10.1103/PhysRevD.92.021302. arXiv: 1505.01377 [astro-ph.HE].

[471] David Wittkowski and Karl-Heinz Kampert. 'On the anisotropy in the arrival directions of ultra-high-energy cosmic rays'. In: *Astrophys. J.* 854.1 (2018), p. L3. DOI: 10.3847/2041-8213/aaa2f9. arXiv: 1710.05617 [astro-ph.HE].

[472] Rafael Alves Batista et al. 'CRPropa 3 - a Public Astrophysical Simulation Framework for Propagating Extraterrestrial Ultra-High Energy Particles'. In: *JCAP* 1605.05 (2016), p. 038. DOI: 10.1088/1475-7516/2016/05/038. arXiv: 1603.07142 [astro-ph.IM].

[473] R. Aloisio et al. 'SimProp: a Simulation Code for Ultra High Energy Cosmic Ray Propagation'. In: *JCAP* 1210 (2012), p. 007. DOI: 10.1088/1475-7516/2012/10/007. arXiv: 1204.2970 [astro-ph.HE].

[474] M. Kachelrieß and Dmitry V. Semikoz. 'Reconciling the ultra-high energy cosmic ray spectrum with Fermi shock acceleration'. In: *Phys. Lett.* B634 (2006), pp. 143–147. DOI: 10.1016/j.physletb.2006.01.009. arXiv: astro-ph/0510188 [astro-ph].

[475] Carl Blaksley and Etienne Parizot. 'Enhancing the Relative Fe-to-Proton Abundance in Ultra-High-Energy Cosmic Rays'. In: *Astropart. Phys.* 35 (2012), pp. 342–345. DOI: 10.1016/j.astropartphys.2011.10.006. arXiv: 1111.1607 [astro-ph.HE].

[476] Alexander Aab et al. 'An Indication of anisotropy in arrival directions of ultra-high-energy cosmic rays through comparison to the flux pattern of extragalactic gamma-ray sources'. In: *Astrophys. J.* 853 (2018), p. L29. DOI: 10.3847/2041-8213/aaa66d. arXiv: 1801.06160 [astro-ph.HE].

[477] R. U. Abbasi et al. 'Testing a Reported Correlation between Arrival Directions of Ultra-high-energy Cosmic Rays and a Flux Pattern from nearby Starburst Galaxies using Telescope Array Data'. In: *Astrophys. J.* 867.2 (2018), p. L27. DOI: 10.3847/2041-8213/aaebf9. arXiv: 1809.01573 [astro-ph.HE].

[478] Eli Waxman. 'Cosmological origin for cosmic rays above 10**19-eV'. In: *Astrophys. J.* 452 (1995), pp. L1–L4. DOI: 10.1086/309715. arXiv: astro-ph/9508037 [astro-ph].

[479] Rachen, J. P. *Everything you always wanted to know about cosmic rays* (*But were afraid to ask)*. Talk at the IIHE, Brussels. June 2018.

[480] Eli Waxman and John N. Bahcall. 'High-energy neutrinos from astrophysical sources: An Upper bound'. In: *Phys. Rev.* D59 (1999), p. 023002. DOI: 10.1103/PhysRevD.59.023002. arXiv: hep-ph/9807282 [hep-ph].

[481] M. Ackermann et al. 'The Third Catalog of Active Galactic Nuclei Detected by the Fermi Large Area Telescope'. In: *Astrophys. J.* 810.1 (2015), p. 14. DOI: 10.1088/0004-637X/810/1/14. arXiv: 1501.06054 [astro-ph.HE].

[482] 'The Fourth Catalog of Active Galactic Nuclei Detected by the Fermi Large Area Telescope'. In: (2019). arXiv: 1905.10771 [astro-ph.HE].

[483] K. Mannheim and R. Schlickeiser. 'Interactions of Cosmic Ray Nuclei'. In: *Astron. Astrophys.* 286 (1994), pp. 983–996. arXiv: astro-ph/9402042 [astro-ph].

[484] B. P. Abbott et al. 'Observation of Gravitational Waves from a Binary Black Hole Merger'. In: *Phys. Rev. Lett.* 116.6 (2016), p. 061102. DOI: 10.1103/PhysRevLett. 116.061102. arXiv: 1602.03837 [gr-qc].

[485] Luis A. Anchordoqui et al. 'Galactic point sources of TeV antineutrinos'. In: *Phys. Lett.* B593 (2004), p. 42. DOI: 10.1016/j.physletb.2004.04.054. arXiv: astro-ph/0311002 [astro-ph].

[486] S. Hummer et al. 'Energy dependent neutrino flavor ratios from cosmic accelerators on the Hillas plot'. In: *Astropart. Phys.* 34 (2010), pp. 205–224. DOI: 10.1016/j.astropartphys.2010.07.003. arXiv: 1007.0006 [astro-ph.HE].

[487] Markus Ahlers and Francis Halzen. 'Pinpointing Extragalactic Neutrino Sources in Light of Recent IceCube Observations'. In: *Phys. Rev.* D90.4 (2014), p. 043005. DOI: 10.1103/PhysRevD.90.043005. arXiv: 1406.2160 [astro-ph.HE].

[488] A. Mucke et al. 'SOPHIA: Monte Carlo simulations of photohadronic processes in astrophysics'. In: *Comput. Phys. Commun.* 124 (2000), pp. 290–314. DOI: 10.1016/S0010-4655(99)00446-4. arXiv: astro-ph/9903478 [astro-ph].

[489] S. R. Kelner, Felex A. Aharonian and V. V. Bugayov. 'Energy spectra of gamma-rays, electrons and neutrinos produced by proton-proton interactions in the very high energy regime'. In: *Phys. Rev.* D74 (2006). [Erratum: Phys. Rev.D79,039901(2009)], p. 034018. DOI: 10.1103/PhysRevD.74.034018, 10.1103/PhysRevD.79.039901. arXiv: astro-ph/0606058 [astro-ph].

[490] Charles D. Dermer and Armen Atoyan. 'Ultrahigh energy cosmic rays, cascade gamma-rays, and high-energy neutrinos from gamma-ray bursts'. In: *New J. Phys.* 8 (2006), p. 122. DOI: 10.1088/1367-2630/8/7/122. arXiv: astro-ph/0606629 [astro-ph].

[491] S. Hummer et al. 'Simplified models for photohadronic interactions in cosmic accelerators'. In: *Astrophys. J.* 721 (2010), pp. 630–652. DOI: 10.1088/0004-637X/721/1/630. arXiv: 1002.1310 [astro-ph.HE].

[492] S. R. Kelner and F. A. Aharonian. 'Energy spectra of gamma-rays, electrons and neutrinos produced at interactions of relativistic protons with low energy radiation'. In: *Phys. Rev.* D78 (2008). [Erratum: Phys. Rev.D82,099901(2010)], p. 034013. DOI: 10.1103/PhysRevD.82.099901, 10.1103/PhysRevD.78.034013. arXiv: 0803.0688 [astro-ph].

[493] Daniel Biehl et al. 'Astrophysical Neutrino Production Diagnostics with the Glashow Resonance'. In: *JCAP* 1701 (2017), p. 033. DOI: 10.1088/1475-7516/2017/01/033. arXiv: 1611.07983 [astro-ph.HE].

[494] Enrico Fermi. 'High-energy nuclear events'. In: *Prog. Theor. Phys.* 5 (1950), pp. 570–583. DOI: 10.1143/PTP.5.570.

[495] J. Becker Tjus et al. 'High-energy neutrinos from radio galaxies'. In: *Phys. Rev.* D89.12 (2014), p. 123005. DOI: 10.1103/PhysRevD.89.123005. arXiv: 1406.0506 [astro-ph.HE].

[496] R. S. Fletcher et al. 'SIBYLL: An Event generator for simulation of high-energy cosmic ray cascades'. In: *Phys. Rev.* D50 (1994), pp. 5710–5731. DOI: 10.1103/PhysRevD.50.5710.

[497] Eun-Joo Ahn et al. 'Cosmic ray interaction event generator SIBYLL 2.1'. In: *Phys. Rev.* D80 (2009), p. 094003. DOI: 10.1103/PhysRevD.80.094003. arXiv: 0906.4113 [hep-ph].

[498] S. Ostapchenko. 'QGSJET-II: Towards reliable description of very high energy hadronic interactions'. In: *Nucl. Phys. Proc. Suppl.* 151 (2006), pp. 143–146. DOI: 10.1016/j.nuclphysbps.2005.07.026. arXiv: hep-ph/0412332 [hep-ph].

[499] Stefan Roesler, Ralph Engel and Johannes Ranft. 'The Monte Carlo event generator DPMJET-III'. In: *Advanced Monte Carlo for radiation physics, particle transport simulation and applications. Proceedings, Conference, MC2000, Lisbon, Portugal, October 23-26, 2000*. 2000, pp. 1033–1038. DOI: 10.1007/978-3-642-18211-2_166. arXiv: hep-ph/0012252 [hep-ph]. URL: http://www-public.slac.stanford.edu/sciDoc/docMeta.aspx?slacPubNumber=SLAC-PUB-8740.

[500] T. Pierog et al. 'EPOS LHC: Test of collective hadronization with data measured at the CERN Large Hadron Collider'. In: *Phys. Rev.* C92.3 (2015), p. 034906. DOI: 10.1103/PhysRevC.92.034906. arXiv: 1306.0121 [hep-ph].

[501] Felix Riehn. 'Charm production in SIBYLL'. In: *EPJ Web Conf.* 99 (2015), p. 12001. DOI: 10.1051/epjconf/20159912001. arXiv: 1502.06353 [hep-ph].

[502] Felix Riehn et al. 'A new version of the event generator Sibyll'. In: *PoS* ICRC2015 (2016), p. 558. DOI: 10.22323/1.236.0558. arXiv: 1510.00568 [hep-ph].

[503] Tomasz Palczewski. 'Constraints on atmospheric charmed-meson production from IceCube'. In: *EPJ Web Conf.* 130 (2016), p. 05015. DOI: 10.1051/epjconf/201613005015. arXiv: 1611.00816 [astro-ph.HE].

[504] Alexander Aab et al. 'Testing Hadronic Interactions at Ultrahigh Energies with Air Showers Measured by the Pierre Auger Observatory'. In: *Phys. Rev. Lett.* 117.19 (2016), p. 192001. DOI: 10.1103/PhysRevLett.117.192001. arXiv: 1610.08509 [hep-ex].

[505] Vasiliki Pavlidou and Theodore Tomaras. 'What do the highest-energy cosmic-ray data reveal about possible new physics around 50 TeV?' In: (2018). arXiv: 1802.04806 [astro-ph.HE].

[506] E. Waxman. 'IceCube's Neutrinos: The beginning of extra-Galactic neutrino astrophysics?' In: *Proceedings, 9th Rencontres du Vietnam: Windows on the Universe: Quy Nhon, Vietnam, August 11-17, 2013*. 2013, pp. 161–168. arXiv: 1312.0558 [astro-ph.HE]. URL: https://inspirehep.net/record/1266927/files/arXiv:1312.0558.pdf.

[507] Kohta Murase and John F. Beacom. 'Neutrino Background Flux from Sources of Ultrahigh-Energy Cosmic-Ray Nuclei'. In: *Phys. Rev.* D81 (2010), p. 123001. DOI: 10.1103/PhysRevD.81.123001. arXiv: 1003.4959 [astro-ph.HE].

[508] Claudio Giganti, Stéphane Lavignac and Marco Zito. 'Neutrino oscillations: the rise of the PMNS paradigm'. In: *Prog. Part. Nucl. Phys.* 98 (2018), pp. 1–54. DOI: 10.1016/j.ppnp.2017.10.001. arXiv: 1710.00715 [hep-ex].

[509] P. F. Harrison and W. G. Scott. 'Symmetries and generalizations of tri - bimaximal neutrino mixing'. In: *Phys. Lett.* B535 (2002), pp. 163–169. DOI: 10.1016/S0370-2693(02)01753-7. arXiv: hep-ph/0203209 [hep-ph].

[510] P. F. Harrison and W. G. Scott. 'mu - tau reflection symmetry in lepton mixing and neutrino oscillations'. In: *Phys. Lett.* B547 (2002), pp. 219–228. DOI: 10.1016/S0370-2693(02)02772-7. arXiv: hep-ph/0210197 [hep-ph].

[511] Lingjun Fu, Chiu Man Ho and Thomas J. Weiler. 'Cosmic Neutrino Flavor Ratios with Broken $\nu_\mu - \nu_\tau$ Symmetry'. In: *Phys. Lett.* B718 (2012), pp. 558–565. DOI: 10.1016/j.physletb.2012.11.011. arXiv: 1209.5382 [hep-ph].

[512] Lingjun Fu, Chiu Man Ho and Thomas J. Weiler. 'Aspects of the Flavor Triangle for Cosmic Neutrino Propagation'. In: *Phys. Rev.* D91 (2015), p. 053001. DOI: 10.1103/PhysRevD.91.053001. arXiv: 1411.1174 [hep-ph].

[513] S. N. Ahmed et al. 'Measurement of the total active B-8 solar neutrino flux at the Sudbury Neutrino Observatory with enhanced neutral current sensitivity'. In: *Phys. Rev. Lett.* 92 (2004), p. 181301. DOI: 10.1103/PhysRevLett.92.181301. arXiv: nucl-ex/0309004 [nucl-ex].

[514] M. H. Ahn et al. 'Search for electron neutrino appearance in a 250 km long baseline experiment'. In: *Phys. Rev. Lett.* 93 (2004), p. 051801. DOI: 10.1103/PhysRevLett.93.051801. arXiv: hep-ex/0402017 [hep-ex].

[515] John G. Learned and Sandip Pakvasa. 'Detecting tau-neutrino oscillations at PeV energies'. In: *Astropart. Phys.* 3 (1995), pp. 267–274. DOI: 10.1016/0927-6505(94)00043-3. arXiv: hep-ph/9405296 [hep-ph].

[516] Kohta Murase, Markus Ahlers and Brian C. Lacki. 'Testing the Hadronuclear Origin of PeV Neutrinos Observed with IceCube'. In: *Phys. Rev.* D88.12 (2013), p. 121301. DOI: 10.1103/PhysRevD.88.121301. arXiv: 1306.3417 [astro-ph.HE].

[517] H. W. M. Olbers. 'Uber die Durchsichtigkeit des Weltraums'. In: *Astron. Jahrb. für das Jahr 1826* 110 (1826).

[518] Kohta Murase, Dafne Guetta and Markus Ahlers. 'Hidden Cosmic-Ray Accelerators as an Origin of TeV-PeV Cosmic Neutrinos'. In: *Phys. Rev. Lett.* 116.7 (2016), p. 071101. DOI: 10.1103/PhysRevLett.116.071101. arXiv: 1509.00805 [astro-ph.HE].

[519] Francis Halzen. 'Cosmic Neutrinos from the Sources of Galactic and Extragalactic Cosmic Rays'. In: *Astrophys. Space Sci.* 309 (2007), pp. 407–414. DOI: 10.1007/s10509-007-9434-7. arXiv: astro-ph/0611915 [astro-ph].

[520] Karl Mannheim, R. J. Protheroe and Jorg P. Rachen. 'On the cosmic ray bound for models of extragalactic neutrino production'. In: *Phys. Rev.* D63 (2001), p. 023003. DOI: 10.1103/PhysRevD.63.023003. arXiv: astro-ph/9812398 [astro-ph].

[521] V. S. Berezinsky and A. Yu. Smirnov. 'Cosmic neutrinos of ultra-high energies and detection possibility'. In: *Astrophys. Space Sci.* 32 (1975), pp. 461–482. DOI: 10.1007/BF00643157.

[522] Kohta Murase, John F. Beacom and Hajime Takami. 'Gamma-Ray and Neutrino Backgrounds as Probes of the High-Energy Universe: Hints of Cascades, General Constraints, and Implications for TeV Searches'. In: *JCAP* 1208 (2012), p. 030. DOI: 10.1088/1475-7516/2012/08/030. arXiv: 1205.5755 [astro-ph.HE].

[523] Imre Bartos and Marek Kowalski. *Multimessenger Astronomy*. 2399-2891. IOP Publishing, 2017. ISBN: 978-0-7503-1369-8. DOI: 10.1088/978-0-7503-1369-8. URL: http://dx.doi.org/10.1088/978-0-7503-1369-8.

[524] U. F. Katz and Ch. Spiering. 'High-Energy Neutrino Astrophysics: Status and Perspectives'. In: *Prog. Part. Nucl. Phys.* 67 (2012), pp. 651–704. DOI: 10.1016/j.ppnp.2011.12.001. arXiv: 1111.0507 [astro-ph.HE].

[525] M. Honda et al. 'Atmospheric neutrino flux calculation using the NRLMSISE-00 atmospheric model'. In: *Phys. Rev.* D92.2 (2015), p. 023004. DOI: 10.1103/PhysRevD.92.023004. arXiv: 1502.03916 [astro-ph.HE].

[526] T. S. Sinegovskaya, E. V. Ogorodnikova and S. I. Sinegovsky. 'High-energy fluxes of atmospheric neutrinos'. In: *Proceedings, 33rd International Cosmic Ray Conference (ICRC2013): Rio de Janeiro, Brazil, July 2-9, 2013*. 2013, p. 0040. arXiv: 1306.5907 [astro-ph.HE]. URL: https://inspirehep.net/record/1239792/files/arXiv:1306.5907.pdf.

[527] J. A. Formaggio and G. P. Zeller. 'From eV to EeV: Neutrino Cross Sections Across Energy Scales'. In: *Rev. Mod. Phys.* 84 (2012), pp. 1307–1341. DOI: 10.1103/RevModPhys.84.1307. arXiv: 1305.7513 [hep-ex].

[528] J. Babson et al. 'Cosmic Ray Muons in the Deep Ocean'. In: *Phys. Rev.* D42 (1990), pp. 3613–3620. DOI: 10.1103/PhysRevD.42.3613.

[529] M. G. Aartsen et al. 'The IceCube Neutrino Observatory: Instrumentation and Online Systems'. In: *JINST* 12.03 (2017), P03012. DOI: 10.1088/1748-0221/12/03/P03012. arXiv: 1612.05093 [astro-ph.IM].

[530] M. Ageron et al. 'ANTARES: the first undersea neutrino telescope'. In: *Nucl. Instrum. Meth.* A656 (2011), pp. 11–38. DOI: 10.1016/j.nima.2011.06.103. arXiv: 1104.1607 [astro-ph.IM].

[531] I. A. Belolaptikov et al. 'The Baikal underwater neutrino telescope: Design, performance and first results'. In: *Astropart. Phys.* 7 (1997), pp. 263–282. DOI: 10.1016/S0927-6505(97)00022-4.

[532] M. G. Aartsen et al. 'All-sky Search for Time-integrated Neutrino Emission from Astrophysical Sources with 7 yr of IceCube Data'. In: *Astrophys. J.* 835.2 (2017), p. 151. DOI: 10.3847/1538-4357/835/2/151. arXiv: 1609.04981 [astro-ph.HE].

[533] M. G. Aartsen et al. 'The contribution of Fermi-2LAC blazars to the diffuse TeV-PeV neutrino flux'. In: *Astrophys. J.* 835.1 (2017), p. 45. DOI: 10.3847/1538-4357/835/1/45. arXiv: 1611.03874 [astro-ph.HE].

[534] M. G. Aartsen et al. 'The IceCube Neutrino Observatory - Contributions to ICRC 2017 Part II: Properties of the Atmospheric and Astrophysical Neutrino Flux'. In: (2017). arXiv: 1710.01191 [astro-ph.HE].

[535] Andrea Palladino and Walter Winter. 'A Multi-Component Model for the Observed Astrophysical Neutrinos'. In: (2018). arXiv: 1801.07277 [astro-ph.HE].

[536] M. G. Aartsen et al. 'Evidence for High-Energy Extraterrestrial Neutrinos at the IceCube Detector'. In: *Science* 342 (2013), p. 1242856. DOI: 10.1126/science.1242856. arXiv: 1311.5238 [astro-ph.HE].

[537] M. G. Aartsen et al. 'The IceCube Neutrino Observatory – Contributions to the 36th International Cosmic Ray Conference (ICRC2019)'. In: *36th International Cosmic Ray Conference (ICRC 2019) Madison, Wisconsin, USA, July 24-August 1, 2019*. 2019. arXiv: 1907.11699 [astro-ph.HE].

[538] M. G. Aartsen et al. 'A combined maximum-likelihood analysis of the high-energy astrophysical neutrino flux measured with IceCube'. In: *Astrophys. J.* 809.1 (2015), p. 98. DOI: 10.1088/0004-637X/809/1/98. arXiv: 1507.03991 [astro-ph.HE].

[539] Andrea Palladino, Maurizio Spurio and Francesco Vissani. 'On the IceCube spectral anomaly'. In: *JCAP* 1612.12 (2016), p. 045. DOI: 10.1088/1475-7516/2016/12/045. arXiv: 1610.07015 [astro-ph.HE].

[540] Chien-Yi Chen, P. S. Bhupal Dev and Amarjit Soni. 'Two-component flux explanation for the high energy neutrino events at IceCube'. In: *Phys. Rev.* D92.7 (2015), p. 073001. DOI: 10.1103/PhysRevD.92.073001. arXiv: 1411.5658 [hep-ph].

[541] Antonio Marinelli et al. 'Interpretation of astrophysical neutrinos observed by IceCube experiment by setting Galactic and extra-Galactic spectral components'. In: *EPJ Web Conf.* 116 (2016), p. 04009. DOI: 10.1051/epjconf/201611604009. arXiv: 1604.05776 [astro-ph.HE].

[542] Marco Chianese et al. 'Use of ANTARES and IceCube Data to Constrain a Single Power-law Neutrino Flux'. In: *Astrophys. J.* 851.1 (2017), p. 36. DOI: 10.3847/1538-4357/aa97e6. arXiv: 1707.05168 [hep-ph].

[543] Debasish Borah et al. 'Multi-component Fermionic Dark Matter and IceCube PeV scale Neutrinos in Left-Right Model with Gauge Unification'. In: *JHEP* 09 (2017), p. 005. DOI: 10.1007/JHEP09(2017)005. arXiv: 1704.04138 [hep-ph].

[544] Andrii Neronov and Dmitry V. Semikoz. 'Evidence the Galactic contribution to the IceCube astrophysical neutrino flux'. In: *Astropart. Phys.* 75 (2016), pp. 60–63. DOI: 10.1016/j.astropartphys.2015.11.002. arXiv: 1509.03522 [astro-ph.HE].

[545] Andrea Palladino and Francesco Vissani. 'Extragalactic plus Galactic model for IceCube neutrino events'. In: *Astrophys. J.* 826.2 (2016), p. 185. DOI: 10.3847/0004-637X/826/2/185. arXiv: 1601.06678 [astro-ph.HE].

[546] A. Neronov, M. Kachelriess and D. V. Semikoz. 'Discovery of the multi-messenger gamma-ray counterpart of the IceCube neutrino signal'. In: (2018). arXiv: 1802.09983 [astro-ph.HE].

[547] Soebur Razzaque. 'The Galactic Center Origin of a Subset of IceCube Neutrino Events'. In: *Phys. Rev.* D88 (2013), p. 081302. DOI: 10.1103/PhysRevD.88.081302. arXiv: 1309.2756 [astro-ph.HE].

[548] Markus Ahlers and Kohta Murase. 'Probing the Galactic Origin of the IceCube Excess with Gamma-Rays'. In: *Phys. Rev.* D90.2 (2014), p. 023010. DOI: 10.1103/PhysRevD.90.023010. arXiv: 1309.4077 [astro-ph.HE].

[549] W. B. Atwood et al. 'The Large Area Telescope on the Fermi Gamma-ray Space Telescope Mission'. In: *Astrophys. J.* 697 (2009), pp. 1071–1102. DOI: 10.1088/0004-637X/697/2/1071. arXiv: 0902.1089 [astro-ph.IM].

[550] W. Atwood et al. 'Pass 8: Toward the Full Realization of the Fermi-LAT Scientific Potential'. In: 2013. arXiv: 1303.3514 [astro-ph.IM]. URL: https://inspirehep.net/record/1223837/files/arXiv:1303.3514.pdf.

[551] Charles D. Dermer. 'The Extragalactic Gamma Ray Background'. In: *AIP Conf. Proc.* 921 (2007), pp. 122–126. DOI: 10.1063/1.2757282. arXiv: 0704.2888 [astro-ph].

[552] Paolo S. Coppi and Felix A. Aharonian. 'Constraints on the VHE emissivity of the universe from the diffuse GeV gamma-ray background'. In: *Astrophys. J.* 487 (1997), pp. L9–L12. DOI: 10.1086/310883. arXiv: astro-ph/9610176 [astro-ph].

[553] Piero Ullio et al. 'Cosmological dark matter annihilations into gamma-rays - a closer look'. In: *Phys. Rev.* D66 (2002), p. 123502. DOI: 10.1103/PhysRevD.66.123502. arXiv: astro-ph/0207125 [astro-ph].

[554] Lars Bergstrom, Joakim Edsjo and Piero Ullio. 'Spectral gamma-ray signatures of cosmological dark matter annihilation'. In: *Phys. Rev. Lett.* 87 (2001), p. 251301. DOI: 10.1103/PhysRevLett.87.251301. arXiv: astro-ph/0105048 [astro-ph].

[555] James E. Taylor and Joseph Silk. 'The Clumpiness of cold dark matter: Implications for the annihilation signal'. In: *Mon. Not. Roy. Astron. Soc.* 339 (2003), p. 505. DOI: 10.1046/j.1365-8711.2003.06201.x. arXiv: astro-ph/0207299 [astro-ph].

[556] M. Ajello et al. '3FHL: The Third Catalog of Hard Fermi-LAT Sources'. In: *Astrophys. J. Suppl.* 232.2 (2017), p. 18. DOI: 10.3847/1538-4365/aa8221. arXiv: 1702.00664 [astro-ph.HE].

[557] M. Ackermann et al. 'The spectrum of isotropic diffuse gamma-ray emission between 100 MeV and 820 GeV'. In: *Astrophys. J.* 799 (2015), p. 86. DOI: 10.1088/0004-637X/799/1/86. arXiv: 1410.3696 [astro-ph.HE].

[558] M. Ackermann et al. 'Resolving the Extragalactic $\gamma$-Ray Background above 50 GeV with the Fermi Large Area Telescope'. In: *Phys. Rev. Lett.* 116.15 (2016), p. 151105. DOI: 10.1103/PhysRevLett.116.151105. arXiv: 1511.00693 [astro-ph.CO].

[559] Mariangela Lisanti et al. 'Deciphering Contributions to the Extragalactic Gamma-Ray Background from 2 GeV to 2 TeV'. In: *Astrophys. J.* 832.2 (2016), p. 117. DOI: 10.3847/0004-637X/832/2/117. arXiv: 1606.04101 [astro-ph.HE].

[560] M. Ajello et al. 'The Origin of the Extragalactic Gamma-Ray Background and Implications for Dark-Matter Annihilation'. In: *Astrophys. J.* 800.2 (2015), p. L27. DOI: 10.1088/2041-8205/800/2/L27. arXiv: 1501.05301 [astro-ph.HE].

[561] M. Ackermann et al. 'Limits on Dark Matter Annihilation Signals from the Fermi LAT 4-year Measurement of the Isotropic Gamma-Ray Background'. In: *JCAP* 1509.09 (2015), p. 008. DOI: 10.1088/1475-7516/2015/09/008. arXiv: 1501.05464 [astro-ph.CO].

[562] Mattia Di Mauro et al. 'Fermi-LAT /*gamma*-ray anisotropy and intensity explained by unresolved Radio-Loud Active Galactic Nuclei'. In: *JCAP* 1411.11 (2014), p. 021. DOI: 10.1088/1475-7516/2014/11/021. arXiv: 1407.3275 [astro-ph.HE].

[563] Yoshiyuki Inoue. 'Cosmic Gamma-ray Background Radiation'. In: *5th International Fermi Symposium Nagoya, Japan, October 20-24, 2014.* 2014. arXiv: 1412.3886 [astro-ph.HE]. URL: https://inspirehep.net/record/1334100/files/arXiv:1412.3886.pdf.

[564] Luigi Costamante. 'Gamma-rays from Blazars and the Extragalactic Background Light'. In: *Int. J. Mod. Phys.* D22.13 (2013), p. 1330025. DOI: 10.1142/S0218271813300255. arXiv: 1309.0612 [astro-ph.HE].

[565] E. Lorenz. 'Status of the 17-m MAGIC telescope'. In: *New Astron. Rev.* 48 (2004), pp. 339–344. DOI: 10.1016/j.newar.2003.12.059.

[566] Stefan Funk et al. 'The Trigger system of the H.E.S.S. Telescope array'. In: *Astropart. Phys.* 22 (2004), pp. 285–296. DOI: 10.1016/j.astropartphys.2004.08.001. arXiv: astro-ph/0408375 [astro-ph].

[567] Jamie Holder et al. 'The first VERITAS telescope'. In: *Astropart. Phys.* 25 (2006), pp. 391–401. DOI: 10.1016/j.astropartphys.2006.04.002. arXiv: astro-ph/0604119 [astro-ph].

[568] J. Holder. 'TeV gamma-ray astronomy: A summary'. In: *Astroparticle Physics* 39 (2012), pp. 61–75. DOI: 10.1016/j.astropartphys.2012.02.014. arXiv: 1204.1267 [astro-ph.HE].

[569] Shin'ichiro Ando et al. 'Colloquium: Multimessenger astronomy with gravitational waves and high-energy neutrinos'. In: *Rev. Mod. Phys.* 85.4 (2013), pp. 1401–1420. DOI: 10.1103/RevModPhys.85.1401. arXiv: 1203.5192 [astro-ph.HE].

[570] XiLong Fan, Christopher Messenger and Ik Siong Heng. 'A Bayesian approach to multi-messenger astronomy: Identification of gravitational-wave host galaxies'. In: *Astrophys. J.* 795.1 (2014), p. 43. DOI: 10.1088/0004-637X/795/1/43. arXiv: 1406.1544 [astro-ph.HE].

[571] Benjamin P. Abbott et al. 'Upper Limits on the Stochastic Gravitational-Wave Background from Advanced LIGO's First Observing Run'. In: *Phys. Rev. Lett.* 118.12 (2017). [Erratum: Phys. Rev. Lett.119,no.2,029901(2017)], p. 121101. DOI: 10.1103/PhysRevLett.118.121101,10.1103/PhysRevLett.119.029901. arXiv: 1612.02029 [gr-qc].

[572] Benjamin P. Abbott et al. 'A Search for Tensor, Vector, and Scalar Polarizations in the Stochastic Gravitational-Wave Background'. In: (2018). arXiv: 1802.10194 [gr-qc].

[573] B. P. Abbott et al. 'A search for the isotropic stochastic background using data from Advanced LIGO's second observing run'. In: (2019). arXiv: 1903.02886 [gr-qc].

[574] Lukas Nellen, Karl Mannheim and Peter L. Biermann. 'Neutrino production through hadronic cascades in AGN accretion disks'. In: *Phys. Rev.* D47 (1993), pp. 5270–5274. DOI: 10.1103/PhysRevD.47.5270. arXiv: hep-ph/9211257 [hep-ph].

[575] Walter Winter. 'Photohadronic Origin of the TeV-PeV Neutrinos Observed in IceCube'. In: *Phys. Rev.* D88 (2013), p. 083007. DOI: 10.1103/PhysRevD.88.083007. arXiv: 1307.2793 [astro-ph.HE].

[576] K. Mannheim. 'High-energy neutrinos from extragalactic jets'. In: *Astropart. Phys.* 3 (1995), pp. 295–302. DOI: 10.1016/0927-6505(94)00044-4.

[577] F. W. Stecker et al. 'High-energy neutrinos from active galactic nuclei'. In: *Phys. Rev. Lett.* 66 (1991). [Erratum: Phys. Rev. Lett.69,2738(1992)], pp. 2697–2700. DOI: 10.1103/PhysRevLett.66.2697,10.1103/PhysRevLett.69.2738.

[578] Floyd W. Stecker. 'PeV neutrinos observed by IceCube from cores of active galactic nuclei'. In: *Phys. Rev.* D88.4 (2013), p. 047301. DOI: 10.1103/PhysRevD.88.047301. arXiv: 1305.7404 [astro-ph.HE].

[579] Kohta Murase, Yoshiyuki Inoue and Charles D. Dermer. 'Diffuse Neutrino Intensity from the Inner Jets of Active Galactic Nuclei: Impacts of External Photon Fields and the Blazar Sequence'. In: *Phys. Rev.* D90.2 (2014), p. 023007. DOI: 10.1103/PhysRevD.90.023007. arXiv: 1403.4089 [astro-ph.HE].

[580] Ruo-Yu Liu et al. 'Can winds driven by active galactic nuclei account for the extragalactic gamma-ray and neutrino backgrounds?' In: (2017). arXiv: 1712.10168 [astro-ph.HE].

[581] Dan Hooper. 'A Case for Radio Galaxies as the Sources of IceCube's Astrophysical Neutrino Flux'. In: *JCAP* 1609.09 (2016), p. 002. DOI: 10.1088/1475-7516/2016/09/002. arXiv: 1605.06504 [astro-ph.HE].

[582] Shigeo S. Kimura, Kohta Murase and Kenji Toma. 'Neutrino and Cosmic-Ray Emission and Cumulative Background from Radiatively Inefficient Accretion Flows in Low-Luminosity Active Galactic Nuclei'. In: *Astrophys. J.* 806 (2015), p. 159. DOI: 10.1088/0004-637X/806/2/159. arXiv: 1411.3588 [astro-ph.HE].

[583] A. Muecke et al. 'BL Lac Objects in the synchrotron proton blazar model'. In: *Astropart. Phys.* 18 (2003), pp. 593–613. DOI: 10.1016/S0927-6505(02)00185-8. arXiv: astro-ph/0206164 [astro-ph].

[584] Maria Petropoulou et al. 'Photohadronic origin of $\gamma$-ray BL Lac emission: implications for IceCube neutrinos'. In: *Mon. Not. Roy. Astron. Soc.* 448.3 (2015), pp. 2412–2429. DOI: 10.1093/mnras/stv179. arXiv: 1501.07115 [astro-ph.HE].

[585] C. Schuster, M. Pohl and R. Schlickeiser. 'Neutrinos from active galactic nuclei as a diagnostic tool'. In: *Astron. Astrophys.* 382 (2002), p. 829. DOI: 10.1051/0004-6361:20011670. arXiv: astro-ph/0111545 [astro-ph].

[586] Armen Atoyan and Charles D. Dermer. 'High-energy neutrinos from photomeson processes in blazars'. In: *Phys. Rev. Lett.* 87 (2001), p. 221102. DOI: 10.1103/PhysRevLett.87.221102. arXiv: astro-ph/0108053 [astro-ph].

[587] R. Abbasi et al. 'A Search for a Diffuse Flux of Astrophysical Muon Neutrinos with the IceCube 40-String Detector'. In: *Phys. Rev.* D84 (2011), p. 082001. DOI: 10.1103/PhysRevD.84.082001. arXiv: 1104.5187 [astro-ph.HE].

[588] M. G. Aartsen et al. 'Search for a diffuse flux of astrophysical muon neutrinos with the IceCube 59-string configuration'. In: *Phys. Rev.* D89.6 (2014), p. 062007. DOI: 10.1103/PhysRevD.89.062007. arXiv: 1311.7048 [astro-ph.HE].

[589] F. Halzen and E. Zas. 'Neutrino fluxes from active galaxies: A Model independent estimate'. In: *Astrophys. J.* 488 (1997), pp. 669–674. DOI: 10.1086/304741. arXiv: astro-ph/9702193 [astro-ph].

[590] R. J. Protheroe. 'High-energy neutrinos from blazars'. In: *ASP Conf. Ser.* 121 (1997), p. 585. arXiv: astro-ph/9607165 [astro-ph].

[591] F. Tavecchio and G. Ghisellini. 'High-energy cosmic neutrinos from spine-sheath BL Lac jets'. In: *Mon. Not. Roy. Astron. Soc.* 451.2 (2015), pp. 1502–1510. DOI: 10.1093/mnras/stv1023. arXiv: 1411.2783 [astro-ph.HE].

[592] P. Padovani et al. 'A simplified view of blazars: the neutrino background'. In: *Mon. Not. Roy. Astron. Soc.* 452.2 (2015), pp. 1877–1887. DOI: 10.1093/mnras/stv1467. arXiv: 1506.09135 [astro-ph.HE].

[593] Fabrizio Tavecchio, Gabriele Ghisellini and Dafne Guetta. 'Structured jets in BL Lac objects: efficient PeV neutrino factories?' In: *Astrophys. J.* 793 (2014), p. L18. DOI: 10.1088/2041-8205/793/1/L18. arXiv: 1407.0907 [astro-ph.HE].

[594] Julia K. Becker, Peter L. Biermann and Wolfgang Rhode. 'The Diffuse neutrino flux from FR-II radio galaxies and blazars: A Source property based estimate'. In: *Astropart. Phys.* 23 (2005), pp. 355–368. DOI: 10.1016/j.astropartphys.2005.02.003. arXiv: astro-ph/0502089 [astro-ph].

[595] Marie-Helene Ulrich, Laura Maraschi and C. Megan Urry. 'Variability of active galactic nuclei'. In: *Ann. Rev. Astron. Astrophys.* 35 (1997), pp. 445–502. DOI: 10.1146/annurev.astro.35.1.445.

[596] Henric Krawczynski et al. 'Multiwavelength observations of strong flares from the TeV - blazar 1ES 1959+650'. In: *Astrophys. J.* 601 (2004), pp. 151–164. DOI: 10.1086/380393. arXiv: astro-ph/0310158 [astro-ph].

[597] Anita Reimer, M. Bottcher and S. Postnikov. 'Neutrino emission in the hadronic synchrotron mirror model: The Orphan TeV flare from 1ES 1959+650'. In: *Astrophys. J.* 630 (2005), pp. 186–190. DOI: 10.1086/431948. arXiv: astro-ph/0505233 [astro-ph].

[598] Sarira Sahu, Andres Felipe Osorio Oliveros and Juan Carlos Sanabria. 'Hadronic-origin orphan TeV flare from 1ES 1959+650'. In: *Phys. Rev.* D87.10 (2013), p. 103015. DOI: 10.1103/PhysRevD.87.103015. arXiv: 1305.4985 [hep-ph].

[599] Eli Waxman and John N. Bahcall. 'High-energy neutrinos from cosmological gamma-ray burst fireballs'. In: *Phys. Rev. Lett.* 78 (1997), pp. 2292–2295. DOI: 10.1103/PhysRevLett.78.2292. arXiv: astro-ph/9701231 [astro-ph].

[600] Charles D. Dermer and Armen Atoyan. 'High energy neutrinos from gamma-ray bursts'. In: *Phys. Rev. Lett.* 91 (2003), p. 071102. DOI: 10.1103/PhysRevLett.91.071102. arXiv: astro-ph/0301030 [astro-ph].

[601] Svenja Hummer, Philipp Baerwald and Walter Winter. 'Neutrino Emission from Gamma-Ray Burst Fireballs, Revised'. In: *Phys. Rev. Lett.* 108 (2012), p. 231101. DOI: 10.1103/PhysRevLett.108.231101. arXiv: 1112.1076 [astro-ph.HE].

[602] M. Ahlers, M. C. Gonzalez-Garcia and F. Halzen. 'GRBs on probation: testing the UHE CR paradigm with IceCube'. In: *Astropart. Phys.* 35 (2011), pp. 87–94. DOI: 10.1016/j.astropartphys.2011.05.008. arXiv: 1103.3421 [astro-ph.HE].

[603] M. G. Aartsen et al. 'Extending the search for muon neutrinos coincident with gamma-ray bursts in IceCube data'. In: *Astrophys. J.* 843.2 (2017), p. 112. DOI: 10.3847/1538-4357/aa7569. arXiv: 1702.06868 [astro-ph.HE].

[604] Daniel Biehl et al. 'Cosmic-Ray and Neutrino Emission from Gamma-Ray Bursts with a Nuclear Cascade'. In: *Astron. Astrophys.* 611 (2018), A101. DOI: 10.1051/0004-6361/201731337. arXiv: 1705.08909 [astro-ph.HE].

[605] Mauricio Bustamante et al. 'Neutrino and cosmic-ray emission from multiple internal shocks in gamma-ray bursts'. In: (2014). [Nature Commun.6,6783(2015)]. DOI: 10.1038/ncomms7783. arXiv: 1409.2874 [astro-ph.HE].

[606] Mauricio Bustamante et al. 'Multi-messenger light curves from gamma-ray bursts in the internal shock model'. In: *Astrophys. J.* 837.1 (2017), p. 33. DOI: 10.3847/1538-4357/837/1/33. arXiv: 1606.02325 [astro-ph.HE].

[607] Kohta Murase et al. 'High Energy Neutrinos and Cosmic-Rays from Low-Luminosity Gamma-Ray Bursts?' In: *Astrophys. J.* 651 (2006), pp. L5–L8. DOI: 10.1086/509323. arXiv: astro-ph/0607104 [astro-ph].

[608] Kohta Murase et al. 'High-energy cosmic-ray nuclei from high- and low-luminosity gamma-ray bursts and implications for multi-messenger astronomy'. In: *Phys. Rev.* D78 (2008), p. 023005. DOI: 10.1103/PhysRevD.78.023005. arXiv: 0801.2861 [astro-ph].

[609] Kohta Murase and Kunihito Ioka. 'TeV–PeV Neutrinos from Low-Power Gamma-Ray Burst Jets inside Stars'. In: *Phys. Rev. Lett.* 111.12 (2013), p. 121102. DOI: 10.1103/PhysRevLett.111.121102. arXiv: 1306.2274 [astro-ph.HE].

[610] Nicholas Senno, Kohta Murase and Peter Meszaros. 'Choked Jets and Low-Luminosity Gamma-Ray Bursts as Hidden Neutrino Sources'. In: *Phys. Rev.* D93.8 (2016), p. 083003. DOI: 10.1103/PhysRevD.93.083003. arXiv: 1512.08513 [astro-ph.HE].

[611] B. Theodore Zhang et al. 'Low-luminosity gamma-ray bursts as the sources of ultrahigh-energy cosmic ray nuclei'. In: (2017). arXiv: 1712.09984 [astro-ph.HE].

[612] Enwei Liang, Bing Zhang and Z. G. Dai. 'Low Luminosity Gamma-Ray Bursts as a Unique Population: Luminosity Function, Local Rate, and Beaming Factor'. In: *Astrophys. J.* 662 (2007), pp. 1111–1118. DOI: 10.1086/517959. arXiv: astro-ph/0605200 [astro-ph].

[613] Francisco Virgili, Enwei Liang and Bing Zhang. 'Low-Luminosity Gamma-Ray Bursts as a Distinct GRB Population:A Monte Carlo Analysis'. In: *Mon. Not. Roy. Astron. Soc.* 392 (2009), p. 91. DOI: 10.1111/j.1365-2966.2008.14063.x. arXiv: 0801.4751 [astro-ph].

[614] Hui Sun, Bing Zhang and Zhuo Li. 'Extragalactic High-energy Transients: Event Rate Densities and Luminosity Functions'. In: *Astrophys. J.* 812.1 (2015), p. 33. DOI: 10.1088/0004-637X/812/1/33. arXiv: 1509.01592 [astro-ph.HE].

[615] A. W. Strong, A. W. Wolfendale and D. M. Worrall. 'Origin of the diffuse gamma ray background'. In: MNRAS 175 (1976), 23P–27P. DOI: 10.1093/mnras/175.1.23P.

[616] Andrzej M. Soltan and Jozef Juchniewicz. 'FIR galaxies and the gamma-ray background'. In: *Astrophys. Lett.* 39 (1999), p. 197. arXiv: astro-ph/9811405 [astro-ph].

[617] Vasiliki Pavlidou and Brian D. Fields. 'The Guaranteed gamma-ray background'. In: *Astrophys. J.* 575 (2002), pp. L5–L8. DOI: 10.1086/342670. arXiv: astro-ph/0207253 [astro-ph].

[618] Todd A. Thompson, Eliot Quataert and Eli Waxman. 'The Starburst Contribution to the Extra-Galactic Gamma-Ray Background'. In: *Astrophys. J.* 654 (2006), pp. 219–225. DOI: 10.1086/509068. arXiv: astro-ph/0606665 [astro-ph].

[619] Abraham Loeb and Eli Waxman. 'The Cumulative background of high energy neutrinos from starburst galaxies'. In: *JCAP* 0605 (2006), p. 003. DOI: 10.1088/1475-7516/2006/05/003. arXiv: astro-ph/0601695 [astro-ph].

[620] Irene Tamborra, Shin'ichiro Ando and Kohta Murase. 'Star-forming galaxies as the origin of diffuse high-energy backgrounds: Gamma-ray and neutrino connections, and implications for starburst history'. In: *JCAP* 1409 (2014), p. 043. DOI: 10.1088/1475-7516/2014/09/043. arXiv: 1404.1189 [astro-ph.HE].

[621] Xiao-Chuan Chang, Ruo-Yu Liu and Xiang-Yu Wang. 'Star-forming galaxies as the origin of the IceCube PeV neutrinos'. In: *Astrophys. J.* 805.2 (2015), p. 95. DOI: 10.1088/0004-637X/805/2/95. arXiv: 1412.8361 [astro-ph.HE].

[622] Kohta Murase and Eli Waxman. 'Constraining High-Energy Cosmic Neutrino Sources: Implications and Prospects'. In: *Phys. Rev.* D94.10 (2016), p. 103006. DOI: 10.1103/PhysRevD.94.103006. arXiv: 1607.01601 [astro-ph.HE].

[623] Chengchao Yuan et al. 'Cumulative Neutrino and Gamma-Ray Backgrounds from Halo and Galaxy Mergers'. In: (2017). arXiv: 1712.09754 [astro-ph.HE].

[624] F. Tavecchio et al. 'High-energy neutrinos from FR0 radio-galaxies?' In: (2017). DOI: 10.1093/mnras/sty251. arXiv: 1711.03757 [astro-ph.HE].

[625] K. Kotera et al. 'Propagation of Ultrahigh Energy Nuclei in Clusters of Galaxies: Resulting Composition and Secondary Emissions'. In: ApJ 707 (2009), pp. 370–386. DOI: 10.1088/0004-637X/707/1/370. arXiv: 0907.2433 [astro-ph.HE].

[626] Kohta Murase, Susumu Inoue and Shigehiro Nagataki. 'Cosmic Rays Above the Second Knee from Clusters of Galaxies and Associated High-Energy Neutrino Emission'. In: *Astrophys. J.* 689 (2008), p. L105. DOI: 10.1086/595882. arXiv: 0805.0104 [astro-ph].

[627] Xiang-Yu Wang et al. 'Probing the tidal disruption flares of massive black holes with high-energy neutrinos'. In: *Phys. Rev.* D84 (2011), p. 081301. DOI: 10.1103/PhysRevD.84.081301. arXiv: 1106.2426 [astro-ph.HE].

[628] Xiang-Yu Wang and Ruo-Yu Liu. 'Tidal disruption jets of supermassive black holes as hidden sources of cosmic rays: explaining the IceCube TeV-PeV neutrinos'. In: *Phys. Rev.* D93.8 (2016), p. 083005. DOI: 10.1103/PhysRevD.93.083005. arXiv: 1512.08596 [astro-ph.HE].

[629] Lixin Dai and Ke Fang. 'Can tidal disruption events produce the IceCube neutrinos?' In: *Mon. Not. Roy. Astron. Soc.* 469.2 (2017), pp. 1354–1359. DOI: 10.1093/mnras/stx863. arXiv: 1612.00011 [astro-ph.HE].

[630] Nicholas Senno, Kohta Murase and Peter Meszaros. 'High-energy Neutrino Flares from X-Ray Bright and Dark Tidal Disruption Events'. In: *Astrophys. J.* 838.1 (2017), p. 3. DOI: 10.3847/1538-4357/aa6344. arXiv: 1612.00918 [astro-ph.HE].

[631] Cecilia Lunardini and Walter Winter. 'High Energy Neutrinos from the Tidal Disruption of Stars'. In: *Phys. Rev.* D95.12 (2017), p. 123001. DOI: 10.1103/PhysRevD.95.123001. arXiv: 1612.03160 [astro-ph.HE].

[632] Claire Guépin et al. 'Ultra-High Energy Cosmic Rays and Neutrinos from Tidal Disruptions by Massive Black Holes'. In: (2017). arXiv: 1711.11274 [astro-ph.HE].

[633] M. G. Aartsen et al. 'Search for steady point-like sources in the astrophysical muon neutrino flux with 8 years of IceCube data'. In: *Eur. Phys. J.* C79.3 (2019), p. 234. DOI: 10.1140/epjc/s10052-019-6680-0. arXiv: 1811.07979 [hep-ph].

[634] Julia K. Becker et al. 'Astrophysical implications of high energy neutrino limits. 1. Overall diffuse limits'. In: *Astropart. Phys.* 28 (2008), pp. 98–118. DOI: 10.1016/j.astropartphys.2007.04.007. arXiv: astro-ph/0607427 [astro-ph].

[635] 'The Second Catalog of Active Galactic Nuclei Detected by the Fermi Large Area Telescope'. In: *Astrophys. J.* 743 (2011), p. 171. DOI: 10.1088/0004-637X/743/2/171. arXiv: 1108.1420 [astro-ph.HE].

[636] M. G. Aartsen et al. 'The IceCube Neutrino Observatory - Contributions to ICRC 2017 Part I: Searches for the Sources of Astrophysical Neutrinos'. In: (2017). arXiv: 1710.01179 [astro-ph.HE].

[637] Y. L. Chang et al. '2WHSP: A multi-frequency selected catalogue of high energy and very high energy γ-ray blazars and blazar candidates'. In: *Astron. Astrophys.* 598 (2017), A17. DOI: 10.1051/0004-6361/201629487. arXiv: 1609.05808 [astro-ph.HE].

[638] M. Ackermann et al. '2FHL: The Second Catalog of Hard Fermi-LAT Sources'. In: *Astrophys. J. Suppl.* 222.1 (2016), p. 5. DOI: 10.3847/0067-0049/222/1/5. arXiv: 1508.04449 [astro-ph.HE].

[639] M. G. Aartsen et al. 'Neutrino emission from the direction of the blazar TXS 0506+056 prior to the IceCube-170922A alert'. In: *Science* 361.6398 (2018), pp. 147–151. DOI: 10.1126/science.aat2890. arXiv: 1807.08794 [astro-ph.HE].

[640] R. Abbasi et al. 'An absence of neutrinos associated with cosmic-ray acceleration in γ-ray bursts'. In: *Nature* 484 (2012), pp. 351–353. DOI: 10.1038/nature11068. arXiv: 1204.4219 [astro-ph.HE].

[641] Nick van Eijndhoven. 'On the observability of high-energy neutrinos from gamma ray bursts'. In: *Astropart. Phys.* 28 (2008), pp. 540–546. DOI: 10.1016/j.astropartphys.2007.10.002. arXiv: astro-ph/0702029 [ASTRO-PH].

[642] M. G. Aartsen et al. 'The IceCube Realtime Alert System'. In: *Astropart. Phys.* 92 (2017), pp. 30–41. DOI: 10.1016/j.astropartphys.2017.05.002. arXiv: 1612.06028 [astro-ph.HE].

[643] M. G. Aartsen et al. 'Multimessenger observations of a flaring blazar coincident with high-energy neutrino IceCube-170922A'. In: *Science* 361.6398 (2018), eaat1378. DOI: 10.1126/science.aat1378. arXiv: 1807.08816 [astro-ph.HE].

[644] A. Albert et al. 'The Search for Neutrinos from TXS 0506+056 with the ANTARES Telescope'. In: *Astrophys. J.* 863.2 (2018), p. L30. DOI: 10.3847/2041-8213/aad8c0. arXiv: 1807.04309 [astro-ph.HE].

[645] S. Ansoldi et al. 'The blazar TXS 0506+056 associated with a high-energy neutrino: insights into extragalactic jets and cosmic ray acceleration'. In: *Astrophys. J. Lett.* (2018). [Astrophys. J.863,L10(2018)]. DOI: 10.3847/2041-8213/aad083. arXiv: 1807.04300 [astro-ph.HE].

[646] A. U. Abeysekara et al. 'VERITAS observations of the BL Lac object TXS 0506+056'. In: *Astrophys. J.* 861.2 (2018), p. L20. DOI: 10.3847/2041-8213/aad053. arXiv: 1807.04607 [astro-ph.HE].

[647] P. Padovani et al. 'Dissecting the region around IceCube-170922A: the blazar TXS 0506+056 as the first cosmic neutrino source'. In: *Mon. Not. Roy. Astron. Soc.* 480.1 (2018), pp. 192–203. DOI: 10.1093/mnras/sty1852. arXiv: 1807.04461 [astro-ph.HE].

[648] A. Keivani et al. 'A Multimessenger Picture of the Flaring Blazar TXS 0506+056: implications for High-Energy Neutrino Emission and Cosmic Ray Acceleration'. In: *Astrophys. J.* 864.1 (2018), p. 84. DOI: 10.3847/1538-4357/aad59a. arXiv: 1807.04537 [astro-ph.HE].

[649] M. Cerruti et al. 'Leptohadronic single-zone models for the electromagnetic and neutrino emission of TXS 0506+056'. In: *Mon. Not. Roy. Astron. Soc.* 483.1 (2019), pp. L12–L16. DOI: 10.1093/mnrasl/sly210. arXiv: 1807.04335 [astro-ph.HE].

[650] Shan Gao et al. 'Modelling the coincident observation of a high-energy neutrino and a bright blazar flare'. In: *Nat. Astron.* 3.1 (2019), pp. 88–92. DOI: 10.1038/s41550-018-0610-1. arXiv: 1807.04275 [astro-ph.HE].

[651] N. Sahakyan. 'Lepto-hadronic $\gamma$-ray and neutrino emission from the jet of TXS 0506+056'. In: *Astrophys. J.* 866.2 (2018), p. 109. DOI: 10.3847/1538-4357/aadade. arXiv: 1808.05651 [astro-ph.HE].

[652] Ruo-Yu Liu et al. 'A hadronuclear interpretation of a high-energy neutrino event coincident with a blazar flare'. In: (2018). arXiv: 1807.05113 [astro-ph.HE].

[653] Kai Wang et al. 'Jet-cloud/star interaction as an interpretation of neutrino outburst from the blazar TXS 0506+056'. In: (2018). arXiv: 1809.00601 [astro-ph.HE].

[654] Anita Reimer, Markus Boettcher and Sara Buson. 'Cascading Constraints from Neutrino Emitting Blazars: The case of TXS 0506+056'. In: (2018). arXiv: 1812.05654 [astro-ph.HE].

[655] Xavier Rodrigues et al. 'Leptohadronic Blazar Models Applied to the 2014–2015 Flare of TXS 0506+056'. In: *Astrophys. J.* 874.2 (2019), p. L29. DOI: 10.3847/2041-8213/ab1267. arXiv: 1812.05939 [astro-ph.HE].

[656] Kohta Murase, Foteini Oikonomou and Maria Petropoulou. 'Blazar Flares as an Origin of High-Energy Cosmic Neutrinos?' In: *Astrophys. J.* 865.2 (2018), p. 124. DOI: 10.3847/1538-4357/aada00. arXiv: 1807.04748 [astro-ph.HE].

[657] C. Righi, F. Tavecchio and S. Inoue. 'Neutrino emission from BL Lac objects: the role of radiatively inefficient accretion flows'. In: *Mon. Not. Roy. Astron. Soc.* 483.1 (2019), pp. L127–L131. DOI: 10.1093/mnrasl/sly231. arXiv: 1807.10506 [astro-ph.HE].

[658] Francis Halzen et al. 'On the Neutrino Flares from the Direction of TXS 0506+056'. In: *Astrophys. J.* 874.1 (2019), p. L9. DOI: 10.3847/2041-8213/ab0d27. arXiv: 1811.07439 [astro-ph.HE].

[659] Dan Hooper, Tim Linden and Abby Vieregg. 'Active Galactic Nuclei and the Origin of IceCube's Diffuse Neutrino Flux'. In: *JCAP* 1902 (2019), p. 012. DOI: 10.1088/1475-7516/2019/02/012. arXiv: 1810.02823 [astro-ph.HE].

[660] Foteini Oikonomou et al. 'High energy neutrino flux from individual blazar flares'. In: (2019). arXiv: 1906.05302 [astro-ph.HE].

[661] Andrea Palladino et al. 'Interpretation of the diffuse astrophysical neutrino flux in terms of the blazar sequence'. In: *Astrophys. J.* 871.1 (2019), p. 41. DOI: 10.3847/1538-4357/aaf507. arXiv: 1806.04769 [astro-ph.HE].

[662] D. Bose et al. 'Bayesian approach for counting experiment statistics applied to a neutrino point source analysis'. In: *Astropart. Phys.* 50-52 (2013), pp. 57–64. DOI: 10.1016/j.astropartphys.2013.09.009. arXiv: 1212.2008 [astro-ph.HE].

[663] Paolo Lipari. 'Proton and Neutrino Extragalactic Astronomy'. In: *Phys. Rev.* D78 (2008), p. 083011. DOI: 10.1103/PhysRevD.78.083011. arXiv: 0808.0344 [astro-ph].

[664] A. Silvestri and S. W. Barwick. 'Constraints on extragalactic point source flux from diffuse neutrino limits'. In: Phys. Rev. D 81.2, 023001 (2010), p. 023001. DOI: 10.1103/PhysRevD.81.023001. arXiv: 0908.4266 [astro-ph.HE].

[665] Marek Kowalski. 'Status of High-Energy Neutrino Astronomy'. In: *J. Phys. Conf. Ser.* 632.1 (2015), p. 012039. DOI: 10.1088/1742-6596/632/1/012039. arXiv: 1411.4385 [astro-ph.HE].

[666] Andrew M. Hopkins and John F. Beacom. 'On the normalisation of the cosmic star formation history'. In: *Astrophys. J.* 651 (2006), pp. 142–154. DOI: 10.1086/506610. arXiv: astro-ph/0601463 [astro-ph].

[667] M. Ajello et al. 'The Cosmic Evolution of Fermi BL Lacertae Objects'. In: *Astrophys. J.* 780 (2014), p. 73. DOI: 10.1088/0004-637X/780/1/73. arXiv: 1310.0006 [astro-ph.CO].

[668] Juliana STACHURSKA. *New measurement of the flavor composition of high- energy neutrino events with contained vertices in IceCube*. 2018. DOI: 10.5281/zenodo.1301122. URL: https://doi.org/10.5281/zenodo.1301122.

[669] Sheldon L. Glashow. 'Resonant Scattering of Antineutrinos'. In: *Phys. Rev.* 118 (1960), pp. 316–317. DOI: 10.1103/PhysRev.118.316.

[670] V. Barger et al. 'Glashow resonance as a window into cosmic neutrino sources'. In: *Phys. Rev.* D90 (2014), p. 121301. DOI: 10.1103/PhysRevD.90.121301. arXiv: 1407.3255 [astro-ph.HE].

[671] Andrea Palladino and Francesco Vissani. 'The natural parameterization of cosmic neutrino oscillations'. In: *Eur. Phys. J.* C75 (2015), p. 433. DOI: 10.1140/epjc/s10052-015-3664-6. arXiv: 1504.05238 [hep-ph].

[672] Andrea Palladino et al. 'Double pulses and cascades above 2 PeV in IceCube'. In: *Eur. Phys. J.* C76.2 (2016), p. 52. DOI: 10.1140/epjc/s10052-016-3893-3. arXiv: 1510.05921 [astro-ph.HE].

[673] Ignacio Taboada. *A View of the Universe with the IceCube and ANTARES Neutrino Telescopes*. 2018. DOI: 10.5281/zenodo.1286919. URL: https://doi.org/10.5281/zenodo.1286919.

[674] Peter B. Denton and Irene Tamborra. 'Invisible Neutrino Decay Resolves Ice-Cube's Track and Cascade Tension'. In: (2018). arXiv: 1805.05950 [hep-ph].

[675] Mauricio Bustamante and Markus Ahlers. 'Inferring the flavor of high-energy astrophysical neutrinos at their sources'. In: *Phys. Rev. Lett.* 122.24 (2019), p. 241101. DOI: 10.1103/PhysRevLett.122.241101. arXiv: 1901.10087 [astro-ph.HE].

[676] Andrea Palladino. 'The flavor composition of astrophysical neutrinos after 8 years of IceCube: an indication of neutron decay scenario?' In: *Eur. Phys. J.* C79.6 (2019), p. 500. DOI: 10.1140/epjc/s10052-019-7018-7. arXiv: 1902.08630 [astro-ph.HE].

[677] Kohta Murase and John F. Beacom. 'Galaxy Clusters as Reservoirs of Heavy Dark Matter and High-Energy Cosmic Rays: Constraints from Neutrino Observations'. In: *JCAP* 1302 (2013), p. 028. DOI: 10.1088/1475-7516/2013/02/028. arXiv: 1209.0225 [astro-ph.HE].

[678] B. C. Lacki and T. A. Thompson. 'Diffuse Hard X-Ray Emission in Starburst Galaxies as Synchrotron from Very High Energy Electrons'. In: ApJ 762, 29 (2013), p. 29. DOI: 10.1088/0004-637X/762/1/29. arXiv: 1010.3030 [astro-ph.HE].

[679] C. Gruppioni et al. 'The Herschel PEP/HerMES Luminosity Function. I: Probing the Evolution of PACS selected Galaxies to z 4'. In: *Mon. Not. Roy. Astron. Soc.* 432 (2013), p. 23. DOI: 10.1093/mnras/stt308. arXiv: 1302.5209 [astro-ph.CO].

[680] M. Ackermann et al. 'GeV Observations of Star-forming Galaxies with *Fermi* LAT'. In: *Astrophys. J.* 755 (2012), p. 164. DOI: 10.1088/0004-637X/755/2/164. arXiv: 1206.1346 [astro-ph.HE].

[681]    Keith Bechtol et al. 'Evidence against star-forming galaxies as the dominant source of IceCube neutrinos'. In: *Astrophys. J.* 836.1 (2017), p. 47. DOI: 10.3847/1538-4357/836/1/47. arXiv: 1511.00688 [astro-ph.HE].

[682]    Andrea Palladino et al. 'IceCube Neutrinos from Hadronically Powered Gamma-Ray Galaxies'. In: (2018). arXiv: 1812.04685 [astro-ph.HE].

[683]    M. W. E. Smith et al. 'The Astrophysical Multimessenger Observatory Network (AMON)'. In: *Astropart. Phys.* 45 (2013), pp. 56–70. DOI: 10.1016/j.astropartphys.2013.03.003. arXiv: 1211.5602 [astro-ph.HE].

[684]    M. G. Aartsen et al. 'IceCube-Gen2: A Vision for the Future of Neutrino Astronomy in Antarctica'. In: *PoS* FRAPWS2016 (2017), p. 004. arXiv: 1412.5106 [astro-ph.HE].

[685]    S. Adrian-Martinez et al. 'Letter of intent for KM3NeT 2.0'. In: *J. Phys.* G43.8 (2016), p. 084001. DOI: 10.1088/0954-3899/43/8/084001. arXiv: 1601.07459 [astro-ph.IM].

[686]    A. D. Avrorin et al. 'Status of the early construction phase of Baikal-GVD'. In: *Nucl. Part. Phys. Proc.* 273-275 (2016), pp. 314–320. DOI: 10.1016/j.nuclphysbps.2015.09.044.

[687]    G. Maggi et al. 'Obscured flat spectrum radio active galactic nuclei as sources of high-energy neutrinos'. In: *Phys. Rev.* D94.10 (2016), p. 103007. DOI: 10.1103/PhysRevD.94.103007. arXiv: 1608.00028 [astro-ph.HE].

[688]    Matias M. Reynoso, Hugo R. Christiansen and Gustavo E. Romero. 'Gamma-ray absorption in the microquasar SS433'. In: *Astropart. Phys.* 28 (2008), pp. 565–572. DOI: 10.1016/j.astropartphys.2007.10.005. arXiv: 0707.1844 [astro-ph].

[689]    Matias M. Reynoso, Gustavo E. Romero and Hugo R. Christiansen. 'Production of gamma rays and neutrinos in the dark jets of the microquasar SS433'. In: *Mon. Not. Roy. Astron. Soc.* 387 (2008), pp. 1745–1754. DOI: 10.1111/j.1365-2966.2008.13364.x. arXiv: 0801.2903 [astro-ph].

[690]    Matías M. Reynoso and Agustín M. Carulli. 'On the possibilities of high-energy neutrino production in the jets of microquasar SS433 in light of new observational data'. In: *Astropart. Phys.* 109 (2019), pp. 25–32. DOI: 10.1016/j.astropartphys.2019.02.003. arXiv: 1902.03861 [astro-ph.HE].

[691]    Wiktionary. *unionized — Wiktionary, The Free Dictionary*. [Online; accessed 15-April-2019]. 2017. URL: %5Curl%7Bhttps://en.wiktionary.org/w/index.php?title=unionized&oldid=45678883%7D.

[692]    M.J. Berger et al. *XCOM: Photon Cross Section Database (version 1.5) [online]*. National Institute of Standards and Technology, Gaithersburg, MD. 2010. URL: http://physics.nist.gov/xcom (visited on 29/09/2018).

[693]    K. A. Arnaud. 'XSPEC: The First Ten Years'. In: *Astronomical Data Analysis Software and Systems V*. Ed. by G. H. Jacoby and J. Barnes. Vol. 101. Astronomical Society of the Pacific Conference Series. 1996, p. 17.

[694] E. Anders and N. Grevesse. 'Abundances of the elements: Meteoritic and solar'. In: *Geochim. Cosmochim. Acta* 53 (1989), pp. 197–214. DOI: 10.1016/0016-7037(89)90286-X.

[695] E. Gatuzz et al. 'ISMabs: A Comprehensive X-Ray Absorption Model for the Interstellar Medium'. In: ApJ 800, 29 (2015), p. 29. DOI: 10.1088/0004-637X/800/1/29. arXiv: 1412.3813 [astro-ph.HE].

[696] M. J. Chodorowski, A. A. Zdziarski and M. Sikora. 'Reaction rate and energy-loss rate for photopair production by relativistic nuclei'. In: ApJ 400 (1992), pp. 181–185. DOI: 10.1086/171984.

[697] Mitchell C. Begelman, Bronislaw Rudak and Marek Sikora. 'Consequences of Relativistic Proton Injection in Active Galactic Nuclei'. In: ApJ 362 (Oct. 1990), p. 38. DOI: 10.1086/169241.

[698] L. D. Landau and I. Pomeranchuk. 'Limits of applicability of the theory of bremsstrahlung electrons and pair production at high-energies'. In: *Dokl. Akad. Nauk Ser. Fiz.* 92 (1953), pp. 535–536.

[699] L. D. Landau and I. Pomeranchuk. 'Electron cascade process at very high-energies'. In: *Dokl. Akad. Nauk Ser. Fiz.* 92 (1953), pp. 735–738.

[700] Arkady B. Migdal. 'Bremsstrahlung and pair production in condensed media at high-energies'. In: *Phys. Rev.* 103 (1956), pp. 1811–1820. DOI: 10.1103/PhysRev.103.1811.

[701] D.T. Haar. *Collected Papers of L.D. Landau*. Elsevier Science, 2013. ISBN: 9781483152707. URL: https://books.google.be/books?id=epc4BQAAQBAJ.

[702] Spencer Klein. 'Suppression of Bremsstrahlung and pair production due to environmental factors'. In: *Rev. Mod. Phys.* 71 (1999), pp. 1501–1538. DOI: 10.1103/RevModPhys.71.1501. arXiv: hep-ph/9802442 [hep-ph].

[703] G. B. Rybicki and A. P. Lightman. *Radiative Processes in Astrophysics*. 1986, p. 400.

[704] J. M. Pittard, T. W. Hartquist and S. A. E. G. Falle. 'The turbulent destruction of clouds - II. Mach number dependence, mass-loss rates and tail formation'. In: MNRAS 405.2 (June 2010), pp. 821–838. DOI: 10.1111/j.1365-2966.2010.16504.x. arXiv: 1002.2091 [astro-ph.GA].

[705] A. T. Araudo, V. Bosch-Ramon and G. E. Romero. 'Gamma rays from cloud penetration at the base of AGN jets'. In: A&A 522, A97 (2010), A97. DOI: 10.1051/0004-6361/201014660. arXiv: 1007.2199 [astro-ph.HE].

[706] Arnon Dar and Ari Laor. 'Hadronic production of TeV gamma-ray flares from blazars'. In: *Astrophys. J.* 478 (1997), pp. L5–L8. DOI: 10.1086/310544. arXiv: astro-ph/9610252 [astro-ph].

[707] V. Bosch-Ramon, M. Perucho and M. V. Barkov. 'Clouds and red giants interacting with the base of AGN jets.' In: A&A 539, A69 (Mar. 2012), A69. DOI: 10.1051/0004-6361/201118622. arXiv: 1201.5279 [astro-ph.HE].

[708] Maxim V. Barkov, Valentí Bosch-Ramon and Felix A. Aharonian. 'Interpretation of the Flares of M87 at TeV Energies in the Cloud-Jet Interaction Scenario'. In: ApJ 755.2, 170 (Aug. 2012), p. 170. DOI: 10.1088/0004-637X/755/2/170. arXiv: 1202.5907 [astro-ph.HE].

[709] S. del Palacio, V. Bosch-Ramon and G. E. Romero. 'Gamma rays from jets interacting with BLR clouds in blazars'. In: Astron. Astrophys. 623 (2019), A101. DOI: 10.1051/0004-6361/201834231. arXiv: 1902.07195 [astro-ph.HE].

[710] D. V. Khangulyan et al. 'Star-Jet Interactions and Gamma-Ray Outbursts from 3C454.3'. In: Astrophys. J. 774 (2013), p. 113. DOI: 10.1088/0004-637X/774/2/113. arXiv: 1305.5117 [astro-ph.HE].

[711] P. Banasiński, W. Bednarek and J. Sitarek. 'Orphan $\gamma$-ray flares from relativistic blobs encountering luminous stars'. In: Mon. Not. Roy. Astron. Soc. 463.1 (2016), pp. L26–L30. DOI: 10.1093/mnrasl/slw149. arXiv: 1607.07725 [astro-ph.HE].

[712] Stefano Bianchi, Roberto Maiolino and Guido Risaliti. 'AGN Obscuration and the Unified Model'. In: Adv. Astron. 2012 (2012), p. 782030. DOI: 10.1155/2012/782030. arXiv: 1201.2119 [astro-ph.GA].

[713] Andrea Comastri. 'Compton thick AGN: The Dark side of the x-ray background'. In: Astrophys. Space Sci. Libr. 308 (2004), p. 245. DOI: 10.1007/978-1-4020-2471-9_8. arXiv: astro-ph/0403693 [astro-ph].

[714] C. Ricci et al. 'Compton-thick Accretion in the local Universe'. In: Astrophys. J. 815 (2015), p. L13. DOI: 10.1088/2041-8205/815/1/L13. arXiv: 1603.04852 [astro-ph.HE].

[715] Martin C. Weisskopf et al. 'Chandra X-ray Observatory (CXO): overview'. In: X-Ray Optics, Instruments, and Missions III. Ed. by Joachim E. Truemper and Bernd Aschenbach. Vol. 4012. Society of Photo-Optical Instrumentation Engineers (SPIE) Conference Series. July 2000, pp. 2–16. DOI: 10.1117/12.391545. arXiv: astro-ph/0004127 [astro-ph].

[716] Junyao Li et al. 'Piercing Through Highly Obscured and Compton-thick AGNs in the Chandra Deep Fields: I. X-ray Spectral and Long-term Variability Analyses'. In: (2019). arXiv: 1904.03827 [astro-ph.GA].

[717] G. Risaliti, R. Maiolino and M. Salvati. 'The Distribution of absorbing column densities among seyfert 2 galaxies'. In: Astrophys. J. 522 (1999), pp. 157–164. DOI: 10.1086/307623. arXiv: astro-ph/9902377 [astro-ph].

[718] Murray Brightman et al. 'Compton thick active galactic nuclei in Chandra surveys'. In: Mon. Not. Roy. Astron. Soc. 443.3 (2014), pp. 1999–2017. DOI: 10.1093/mnras/stu1175. arXiv: 1406.4502 [astro-ph.HE].

[719] I. García-Bernete et al. 'Torus model properties of an ultra-hard X-ray selected sample of Seyfert galaxies'. In: arXiv e-prints, arXiv:1904.03694 (Apr. 2019), arXiv:1904.03694. arXiv: 1904.03694 [astro-ph.GA].

[720] Hans A. Krimm et al. 'The Swift/BAT Hard X-ray Transient Monitor'. In: *Astrophys. J. Suppl.* 209 (2013), p. 14. DOI: 10.1088/0067-0049/209/1/14. arXiv: 1309.0755 [astro-ph.HE].

[721] Michael J. Koss et al. 'A New Population of Compton-thick AGNs Identified Using the Spectral Curvature above 10 keV'. In: ApJ 825.2, 85 (July 2016), p. 85. DOI: 10.3847/0004-637X/825/2/85. arXiv: 1604.07825 [astro-ph.HE].

[722] S. Mateos et al. 'Survival of the obscuring torus in the most powerful active galactic nuclei'. In: *Astrophys. J.* 841.2 (2017), p. L18. DOI: 10.3847/2041-8213/aa7268. arXiv: 1705.04323 [astro-ph.HE].

[723] P. F. Roche, A. Alonso-Herrero and O. Gonzalez-Martin. 'The silicate absorption profile in the interstellar medium towards the heavily obscured nucleus of NGC 4418'. In: MNRAS 449.3 (May 2015), pp. 2598–2603. DOI: 10.1093/mnras/stv495. arXiv: 1503.02964 [astro-ph.GA].

[724] Kazushi Sakamoto et al. 'Submillimeter Interferometry of the Luminous Infrared Galaxy NGC 4418: A Hidden Hot Nucleus with an Inflow and an Outflow'. In: *Astrophys. J.* 764 (2013), p. 42. DOI: 10.1088/0004-637X/764/1/42. arXiv: 1301.1878 [astro-ph.CO].

[725] F. Costagliola et al. 'A high-resolution mm and cm study of the obscured LIRG NGC 4418 - A compact obscured nucleus fed by in-falling gas?' In: *Astron. Astrophys.* 556 (2013), A66. DOI: 10.1051/0004-6361/201220634. arXiv: 1306.2211 [astro-ph.GA].

[726] A. Lawrence and M. Elvis. 'Misaligned Discs as Obscurers in Active Galaxies'. In: *ArXiv e-prints* (2010). arXiv: 1002.1759 [astro-ph.GA].

[727] Charles D. Dermer, Kohta Murase and Hajime Takami. 'Variable Gamma-Ray Emission Induced by Ultra-high Energy Neutral Beams: Application to 4C +21.35'. In: ApJ 755.2, 147 (Aug. 2012), p. 147. DOI: 10.1088/0004-637X/755/2/147. arXiv: 1203.6544 [astro-ph.HE].

[728] S. P. O'Sullivan and D. C. Gabuzda. 'Magnetic field strength and spectral distribution of six parsec-scale active galactic nuclei jets'. In: MNRAS 400.1 (Nov. 2009), pp. 26–42. DOI: 10.1111/j.1365-2966.2009.15428.x. arXiv: 0907.5211 [astro-ph.CO].

[729] Lukas Merten et al. 'On the non-thermal electron-to-proton ratio at cosmic ray acceleration sites'. In: *Astropart. Phys.* 90 (2017), pp. 75–84. DOI: 10.1016/j.astropartphys.2017.02.007. arXiv: 1702.07523 [astro-ph.HE].

[730] N.N. Kalmykov, S.S. Ostapchenko and A.I. Pavlov. 'Quark-gluon-string model and EAS simulation problems at ultra-high energies'. In: *Nuclear Physics B - Proceedings Supplements* 52.3 (1997), pp. 17–28. ISSN: 0920-5632. DOI: https://doi.org/10.1016/S0920-5632(96)00846-8. URL: http://www.sciencedirect.com/science/article/pii/S0920563296008468.

[731] H. Netzer. 'AGN emission lines.' In: *Active Galactic Nuclei*. Ed. by R. D. Blandford et al. 1990, pp. 57–160.

[732] Sjoert van Velzen et al. 'Radio galaxies of the local universe: all-sky catalog, luminosity functions, and clustering'. In: *Astron. Astrophys.* 544 (2012), A18. DOI: 10.1051/0004-6361/201219389. arXiv: 1206.0031 [astro-ph.CO].

[733] G. Helou et al. 'The NASA/IPAC extragalactic database.' In: *Databases and On-line Data in Astronomy*. Ed. by M. A. Albrecht and D. Egret. Vol. 171. Astrophysics and Space Science Library. 1991, pp. 89–106. DOI: 10.1007/978-94-011-3250-3_10.

[734] M. Massi. 'Steady jets and transient jets: observational characteristics and models'. In: *Mem. Soc. Ast. It.* 82 (2011), p. 24. arXiv: 1010.3861 [astro-ph.HE].

[735] Andrea Merloni, Sebastian Heinz and Tiziana Di Matteo. 'A Fundamental plane of black hole activity'. In: *Mon. Not. Roy. Astron. Soc.* 345 (2003), p. 1057. DOI: 10.1046/j.1365-2966.2003.07017.x. arXiv: astro-ph/0305261 [astro-ph].

[736] Heino Falcke, Elmar Koerding and Sera Markoff. 'A Scheme to unify low - power accreting black holes: Jet - dominated accretion flows and the radio / x-ray correlation'. In: *Astron. Astrophys.* 414 (2004), pp. 895–903. DOI: 10.1051/0004-6361:20031683. arXiv: astro-ph/0305335 [astro-ph].

[737] G. Ghisellini et al. 'General physical properties of bright Fermi blazars'. In: MNRAS 402 (2010), pp. 497–518. DOI: 10.1111/j.1365-2966.2009.15898.x. arXiv: 0909.0932.

[738] R. Schlickeiser. *Cosmic Ray Astrophysics*. Astronomy and Astrophysics Library. Springer Berlin Heidelberg, 2014. ISBN: 9783662048153. URL: https://books.google.be/books?id=1K8PswEACAAJ.

[739] P. Giommi. 'A Catalog of 157 X-ray Spectra and 84 Spectral Energy Distributions of Blazars Observed with BeppoSAX'. In: *Blazar Astrophysics with BeppoSAX and Other Observatories*. Ed. by P. Giommi, E. Massaro and G. Palumbo. 2002, p. 63. eprint: astro-ph/0209596.

[740] S. T. Myers et al. 'The Cosmic Lens All-Sky Survey - I. Source selection and observations'. In: MNRAS 341 (2003), pp. 1–12. DOI: 10.1046/j.1365-8711.2003.06256.x. eprint: astro-ph/0211073.

[741] S. E. Healey et al. 'CRATES: An All-Sky Survey of Flat-Spectrum Radio Sources'. In: ApJS 171 (2007), pp. 61–71. DOI: 10.1086/513742. eprint: astro-ph/0702346.

[742] R. S. Dixon. 'A Master List of Radio Sources'. In: ApJS 20 (1970), pp. 1–503. DOI: 10.1086/190216.

[743] H. Kuehr et al. 'A catalogue of extragalactic radio sources having flux densities greater than 1 Jy at 5 GHz'. In: A&AS 45 (1981), pp. 367–430.

[744] E. Nieppola et al. '37 GHz Observations of a Large Sample of BL Lacertae Objects'. In: AJ 133 (2007), pp. 1947–1953. DOI: 10.1086/512609. arXiv: 0705.0887.

[745]  J. J. Condon et al. 'The NRAO VLA Sky Survey'. In: AJ 115 (1998), pp. 1693–1716.
       DOI: 10.1086/300337.

[746]  M. Moshir and et al. 'IRAS Faint Source Catalogue, version 2.0.' In: *IRAS Faint
       Source Catalogue, version 2.0 (1990)*. 1990, p. 0.

[747]  W. G. Joint Iras Science. 'VizieR Online Data Catalog: IRAS catalogue of Point
       Sources, Version 2.0 (IPAC 1986)'. In: *VizieR Online Data Catalog* 2125 (1994), p. 0.

[748]  Planck Collaboration et al. 'Planck early results. VII. The Early Release Compact
       Source Catalogue'. In: A&A 536, A7 (2011), A7. DOI: 10.1051/0004-6361/
       201116474. arXiv: 1101.2041.

[749]  P. C. Gregory et al. 'The GB6 Catalog of Radio Sources'. In: ApJS 103 (1996),
       p. 427. DOI: 10.1086/192282.

[750]  R. L. White and R. H. Becker. 'A new catalog of 30,239 1.4 GHz sources'. In: ApJS
       79 (1992), pp. 331–467. DOI: 10.1086/191656.

[751]  Planck Collaboration et al. 'Planck 2013 results. XXVIII. The Planck Catalogue
       of Compact Sources'. In: A&A 571, A28 (2014), A28. DOI: 10.1051/0004-
       6361/201321524. arXiv: 1303.5088.

[752]  Planck Collaboration et al. 'Planck 2015 results. XXVI. The Second Planck Cata-
       logue of Compact Sources'. In: *ArXiv e-prints* (2015). arXiv: 1507.02058.

[753]  E. L. Wright et al. 'Five-Year Wilkinson Microwave Anisotropy Probe Observa-
       tions: Source Catalog'. In: ApJS 180 (2009), pp. 283–295. DOI: 10.1088/0067-
       0049/180/2/283. arXiv: 0803.0577.

[754]  E. L. Wright et al. 'The Wide-field Infrared Survey Explorer (WISE): Mission
       Description and Initial On-orbit Performance'. In: AJ 140, 1868 (2010), pp. 1868–
       1881. DOI: 10.1088/0004-6256/140/6/1868. arXiv: 1008.0031 [astro-ph.IM].

[755]  L. Bianchi et al. 'Catalogues of hot white dwarfs in the Milky Way from GALEX's
       ultraviolet sky surveys: constraining stellar evolution'. In: MNRAS 411 (2011),
       pp. 2770–2791. DOI: 10.1111/j.1365-2966.2010.17890.x. arXiv: 1011.1733
       [astro-ph.SR].

[756]  V. D'Elia et al. 'The seven year Swift-XRT point source catalog (1SWXRT)'. In:
       A&A 551, A142 (2013), A142. DOI: 10.1051/0004-6361/201220863. arXiv:
       1302.7113 [astro-ph.IM].

[757]  P. A. Evans et al. '1SXPS: A Deep Swift X-Ray Telescope Point Source Catalog
       with Light Curves and Spectra'. In: *The Astrophysical Journal Supplement Series*
       210.1 (2014), p. 8. URL: http://stacks.iop.org/0067-0049/210/i=1/a=8.

[758]  M. Elvis et al. 'The Einstein Slew Survey'. In: ApJS 80 (1992), pp. 257–303. DOI:
       10.1086/191665.

[759]  W. Voges et al. 'The ROSAT all-sky survey bright source catalogue'. In: A&A 349
       (1999), pp. 389–405. eprint: astro-ph/9909315.

[760] T. Boller et al. 'Second ROSAT all-sky survey (2RXS) source catalogue'. In: *aap* 588, A103 (2016), A103. DOI: 10.1051/0004-6361/201525648.

[761] R. D. Saxton et al. 'The first XMM-Newton slew survey catalogue: XMMSL1'. In: A&A 480 (2008), pp. 611–622. DOI: 10.1051/0004-6361:20079193. arXiv: 0801.3732.

[762] A. A. Abdo et al. 'Fermi Large Area Telescope First Source Catalog'. In: ApJS 188 (2010), pp. 405–436. DOI: 10.1088/0067-0049/188/2/405. arXiv: 1002.2280 [astro-ph.HE].

[763] P. L. Nolan et al. 'Fermi Large Area Telescope Second Source Catalog'. In: ApJS 199, 31 (2012), p. 31. DOI: 10.1088/0067-0049/199/2/31. arXiv: 1108.1435 [astro-ph.HE].

[764] F. Acero et al. 'Fermi Large Area Telescope Third Source Catalog'. In: ApJS 218, 23 (2015), p. 23. DOI: 10.1088/0067-0049/218/2/23. arXiv: 1501.02003 [astro-ph.HE].

[765] B. Bartoli et al. 'TeV Gamma-Ray Survey of the Northern Sky Using the ARGO-YBJ Detector'. In: *The Astrophysical Journal* 779.1 (2013), p. 27. URL: http://stacks.iop.org/0004-637X/779/i=1/a=27.

[766] *SSDC SED Builder*. https://tools.ssdc.asi.it/. Version 1.21.

[767] Hasan Yuksel et al. 'Revealing the High-Redshift Star Formation Rate with Gamma-Ray Bursts'. In: *Astrophys. J.* 683 (2008), pp. L5–L8. DOI: 10.1086/591449. arXiv: 0804.4008 [astro-ph].

[768] C. D. Wilson et al. 'Extreme Dust Disks in Arp 220 as Revealed by ALMA'. In: ApJ 789.2, L36 (July 2014), p. L36. DOI: 10.1088/2041-8205/789/2/L36. arXiv: 1406.4530 [astro-ph.GA].

[769] Carol Lonsdale, Duncan Farrah and Harding Smith. 'Ultraluminous infrared galaxies'. In: (2006). DOI: 10.1007/3-540-30313-8_9. arXiv: astro-ph/0603031 [astro-ph].

[770] D. L. Clements, L. Dunne and S. Eales. 'The submillimetre properties of ultraluminous infrared galaxies'. In: MNRAS 403 (2010), pp. 274–286. DOI: 10.1111/j.1365-2966.2009.16064.x. arXiv: 0911.3593.

[771] Neil M. Nagar et al. 'The AGN content of ultraluminous IR galaxies: High resolution VLA imaging of the IRAS 1Jy ULIRG sample'. In: *Astron. Astrophys.* 409 (2003), pp. 115–122. DOI: 10.1051/0004-6361:20031069. arXiv: astro-ph/0309298 [astro-ph].

[772] D. Farrah et al. 'Starburst and agn activity in ultraluminous infrared galaxies'. In: *Mon. Not. Roy. Astron. Soc.* 343 (2003), p. 585. DOI: 10.1046/j.1365-8711.2003.06696.x. arXiv: astro-ph/0304154 [astro-ph].

[773] L. G. Hou, X.-B. Wu and J. L. Han. 'Ultra-luminous Infrared Galaxies in Sloan Digital Sky Survey Data Release 6'. In: ApJ 704 (2009), pp. 789–802. DOI: 10.1088/0004-637X/704/1/789. arXiv: 0908.2297 [astro-ph.CO].

[774] D. Fadda et al. 'Ultra-deep Mid-infrared Spectroscopy of Luminous Infrared Galaxies at z~1 and z~2'. In: ApJ 719 (2010), pp. 425–450. DOI: 10.1088/0004-637X/719/1/425. arXiv: 1006.2873.

[775] E. Nardini et al. 'Exploring the active galactic nucleus and starburst content of local ultraluminous infrared galaxies through 5–8 $\mu$m spectroscopy'. In: *Monthly Notices of the Royal Astronomical Society* 399.3 (2009), pp. 1373–1402. DOI: 10.1111/j.1365-2966.2009.15357.x. eprint: /oup/backfile/content_public/journal/mnras/399/3/10.1111_j.1365-2966.2009.15357.x/1/mnras0399-1373.pdf. URL: http://dx.doi.org/10.1111/j.1365-2966.2009.15357.x.

[776] F. G. Saturni et al. '"Zombie" or active? An alternative explanation to the properties of star-forming galaxies at high redshift'. In: A&A 617, A131 (2018), A131. DOI: 10.1051/0004-6361/201833261. arXiv: 1806.07423.

[777] G. Kauffmann and M. G. Haehnelt. 'The clustering of galaxies around quasars'. In: MNRAS 332 (2002), pp. 529–535. DOI: 10.1046/j.1365-8711.2002.05278.x. eprint: astro-ph/0108275.

[778] K. M. Dasyra et al. 'Dynamical Properties of Ultraluminous Infrared Galaxies. I. Mass Ratio Conditions for ULIRG Activity in Interacting Pairs'. In: *The Astrophysical Journal* 638.2 (2006), p. 745. URL: http://stacks.iop.org/0004-637X/638/i=2/a=745.

[779] K. M. Dasyra et al. 'Dynamical Properties of Ultraluminous Infrared Galaxies. II. Traces of Dynamical Evolution and End Products of Local Ultraluminous Mergers'. In: *The Astrophysical Journal* 651.2 (2006), p. 835. URL: http://stacks.iop.org/0004-637X/651/i=2/a=835.

[780] Philip F. Hopkins et al. 'A Physical Model for the Origin of Quasar Lifetimes'. In: *The Astrophysical Journal Letters* 625.2 (2005), p. L71. URL: http://stacks.iop.org/1538-4357/625/i=2/a=L71.

[781] Philip F. Hopkins et al. 'A Unified, merger-driven model for the origin of starbursts, quasars, the cosmic x-ray background, supermassive black holes and galaxy spheroids'. In: *Astrophys. J. Suppl.* 163 (2006), pp. 1–49. DOI: 10.1086/499298. arXiv: astro-ph/0506398 [astro-ph].

[782] Scott C. Chapman et al. 'A Redshift survey of the submillimeter galaxy population'. In: *Astrophys. J.* 622 (2005), pp. 772–796. DOI: 10.1086/428082. arXiv: astro-ph/0412573 [astro-ph].

[783] L. L. Cowie et al. 'The Evolution of the Ultraluminous Infrared Galaxy Population from Redshift 0 to 1.5'. In: ApJ 603 (2004), pp. L69–L72. DOI: 10.1086/383198. eprint: astro-ph/0402235.

[784] Hao-Ning He et al. 'Diffuse PeV neutrino emission from ultraluminous infrared galaxies'. In: *Phys. Rev.* D87.6 (2013), p. 063011. DOI: 10.1103/PhysRevD.87.063011. arXiv: 1303.1253 [astro-ph.HE].

[785] M. Schmidt. 'The Rate of Star Formation.' In: ApJ 129 (1959), p. 243. DOI: 10.1086/146614.

[786] R. C. Kennicutt Jr. 'The star formation law in galactic disks'. In: ApJ 344 (1989), pp. 685–703. DOI: 10.1086/167834.

[787] R. C. Kennicutt Jr. 'The Global Schmidt Law in Star-forming Galaxies'. In: ApJ 498 (1998), pp. 541–552. DOI: 10.1086/305588. eprint: astro-ph/9712213.

[788] M. T. Sargent et al. 'No Evolution in the IR-Radio Relation for IR-luminous Galaxies at $z < 2$ in the COSMOS Field'. In: ApJ 714 (2010), pp. L190–L195. DOI: 10.1088/2041-8205/714/2/L190. arXiv: 1003.4271.

[789] David W. Hogg. 'Distance measures in cosmology'. In: (1999). arXiv: astro-ph/9905116 [astro-ph].

[790] S. Weinberg. *Cosmology*. OUP Oxford, 2008. ISBN: 9780198526827. URL: https://books.google.be/books?id=rswTDAAAQBAJ.

[791] Yoshiyuki Inoue et al. 'Extragalactic Background Light from Hierarchical Galaxy Formation: Gamma-ray Attenuation up to the Epoch of Cosmic Reionization and the First Stars'. In: *Astrophys. J.* 768 (2013), p. 197. DOI: 10.1088/0004-637X/768/2/197. arXiv: 1212.1683 [astro-ph.CO].

[792] Floyd W. Stecker, M. A. Malkan and S. T. Scully. 'Intergalactic photon spectra from the far ir to the uv lyman limit for $0 < Z < 6$ and the optical depth of the universe to high energy gamma-rays'. In: *Astrophys. J.* 648 (2006), pp. 774–783. DOI: 10.1086/506188. arXiv: astro-ph/0510449 [astro-ph].

[793] Floyd W. Stecker, M. A. Malkan and S. T. Scully. 'Corrected Table for the Parametric Coefficients for the Optical Depth of the Universe to Gamma-rays at Various Redshifts'. In: *Astrophys. J.* 658 (2007), p. 1392. DOI: 10.1086/511738. arXiv: astro-ph/0612048 [astro-ph].

[794] Tanja M. Kneiske et al. 'Implications of cosmological gamma-ray absorption. 2. Modification of gamma-ray spectra'. In: *Astron. Astrophys.* 413 (2004), pp. 807–815. DOI: 10.1051/0004-6361:20031542. arXiv: astro-ph/0309141 [astro-ph].

[795] Justin D. Finke, Soebur Razzaque and Charles D. Dermer. 'Modeling the Extragalactic Background Light from Stars and Dust'. In: *Astrophys. J.* 712 (2010), pp. 238–249. DOI: 10.1088/0004-637X/712/1/238. arXiv: 0905.1115 [astro-ph.HE].

[796] R. C. Gilmore et al. 'Semi-analytic modeling of the EBL and consequences for extragalactic gamma-ray spectra'. In: *Mon. Not. Roy. Astron. Soc.* 422 (2012), p. 3189. DOI: 10.1111/j.1365-2966.2012.20841.x. arXiv: 1104.0671 [astro-ph.CO].

[797] A. Dominguez et al. 'Extragalactic Background Light Inferred from AEGIS Galaxy SED-type Fractions'. In: *Mon. Not. Roy. Astron. Soc.* 410 (2011), p. 2556. DOI: 10.1111/j.1365-2966.2010.17631.x. arXiv: 1007.1459 [astro-ph.CO].

[798] Alberto Franceschini, Giulia Rodighiero and Mattia Vaccari. 'The extragalactic optical-infrared background radiations, their time evolution and the cosmic photon-photon opacity'. In: *Astron. Astrophys.* 487 (2008), p. 837. DOI: 10.1051/0004-6361:200809691. arXiv: 0805.1841 [astro-ph].

[799] Kohta Murase. 'High-Energy Emission Induced by Ultra-high-Energy Photons as a Probe of Ultra-high-Energy Cosmic-Ray Accelerators Embedded in the Cosmic Web'. In: *Astrophys. J.* 745 (2012), p. L16. DOI: 10.1088/2041-8205/745/2/L16. arXiv: 1111.0936 [astro-ph.HE].

[800] K. Murase et al. 'Blazars as Ultra-high-energy Cosmic-ray Sources: Implications for TeV Gamma-Ray Observations'. In: ApJ 749, 63 (2012), p. 63. DOI: 10.1088/0004-637X/749/1/63. arXiv: 1107.5576 [astro-ph.HE].

[801] Kohta Murase and John F. Beacom. 'Constraining Very Heavy Dark Matter Using Diffuse Backgrounds of Neutrinos and Cascaded Gamma Rays'. In: *JCAP* 1210 (2012), p. 043. DOI: 10.1088/1475-7516/2012/10/043. arXiv: 1206.2595 [hep-ph].

[802] Sangjin Lee. 'On the propagation of extragalactic high-energy cosmic and gamma-rays'. In: *Phys. Rev.* D58 (1998), p. 043004. DOI: 10.1103/PhysRevD.58.043004. arXiv: astro-ph/9604098 [astro-ph].

[803] Krijn D. de Vries et al. 'Constraints and prospects on gravitational-wave and neutrino emissions using GW150914'. In: *Phys. Rev.* D96.8 (2017), p. 083003. DOI: 10.1103/PhysRevD.96.083003. arXiv: 1612.02648 [astro-ph.HE].

[804] Charles W. Misner, K. S. Thorne and J. A. Wheeler. *Gravitation*. San Francisco: W. H. Freeman, 1973. ISBN: 9780716703440, 9780691177793.

[805] B. Schutz. *A First Course in General Relativity*. Cambridge University Press, 2009. ISBN: 9780521887052. URL: https://books.google.be/books?id=V1CGLi58W7wC.

[806] Benjamin P. Abbott et al. 'The basic physics of the binary black hole merger GW150914'. In: *Annalen Phys.* 529.1-2 (2017), p. 1600209. DOI: 10.1002/andp.201600209. arXiv: 1608.01940 [gr-qc].

[807] Lior M. Burko. 'Gravitational Wave Detection in the Introductory Lab'. In: *Phys. Teacher* 55 (2017), pp. 288–292. DOI: 10.1119/1.4981036. arXiv: 1602.04666 [physics.ed-ph].

[808] Bernard F. Schutz. 'GRAVITATIONAL WAVES ON THE BACK OF AN ENVELOPE'. In: *Am. J. Phys.* 52 (1984), pp. 412–419. DOI: 10.1119/1.13627.

[809] W. E. Couch and E. T. Newman. 'Generalized lienard-wiechert fields'. In: *J. Math. Phys.* 13 (1972), pp. 929–931. DOI: 10.1063/1.1666088.

[810] Albert Einstein. 'Über Gravitationswellen'. In: *Sitzungsber. Preuss. Akad. Wiss. Berlin (Math. Phys.)* 1918 (1918), pp. 154–167.

[811] Curt Cutler and Éanna E. Flanagan. 'Gravitational waves from merging compact binaries: How accurately can one extract the binary's parameters from the inspiral wave form?' In: *Phys. Rev.* D49 (1994), pp. 2658–2697. DOI: 10.1103/PhysRevD.49.2658. arXiv: gr-qc/9402014 [gr-qc].

[812] Samaya Nissanke et al. 'Exploring short gamma-ray bursts as gravitational-wave standard sirens'. In: *Astrophys. J.* 725 (2010), pp. 496–514. DOI: 10.1088/0004-637X/725/1/496. arXiv: 0904.1017 [astro-ph.CO].

[813] Ben Farr et al. 'Parameter estimation on gravitational waves from neutron-star binaries with spinning components'. In: *Astrophys. J.* 825.2 (2016), p. 116. DOI: 10.3847/0004-637X/825/2/116. arXiv: 1508.05336 [astro-ph.HE].

[814] K. Schwarzschild. 'Über das Gravitationsfeld eines Massenpunktes nach der Einsteinschen Theorie'. In: *Sitzungsberichte der Königlich Preußischen Akademie der Wissenschaften (Berlin), 1916, Seite 189-196* (1916).

[815] K. Schwarzschild. 'Über das Gravitationsfeld einer Kugel aus inkompressibler Flüssigkeit nach der Einsteinschen Theorie'. In: *Sitzungsberichte der Königlich Preussischen Akademie der Wissenschaften zu Berlin, Phys.-Math. Klasse, 424-434 (1916).* 1916.

[816] G. Hobbs et al. 'The international pulsar timing array project: using pulsars as a gravitational wave detector'. In: *Class. Quant. Grav.* 27 (2010), p. 084013. DOI: 10.1088/0264-9381/27/8/084013. arXiv: 0911.5206 [astro-ph.SR].

[817] B. P. Abbott et al. 'GW150914: The Advanced LIGO Detectors in the Era of First Discoveries'. In: *Phys. Rev. Lett.* 116.13 (2016), p. 131103. DOI: 10.1103/PhysRevLett.116.131103. arXiv: 1602.03838 [gr-qc].

[818] J. Aasi et al. 'Advanced LIGO'. In: *Class. Quant. Grav.* 32 (2015), p. 074001. DOI: 10.1088/0264-9381/32/7/074001. arXiv: 1411.4547 [gr-qc].

[819] F. Acernese et al. 'Advanced Virgo: a second-generation interferometric gravitational wave detector'. In: *Class. Quant. Grav.* 32.2 (2015), p. 024001. DOI: 10.1088/0264-9381/32/2/024001. arXiv: 1408.3978 [gr-qc].

[820] Kentaro Somiya. 'Detector configuration of KAGRA: The Japanese cryogenic gravitational-wave detector'. In: *Class. Quant. Grav.* 29 (2012), p. 124007. DOI: 10.1088/0264-9381/29/12/124007. arXiv: 1111.7185 [gr-qc].

[821] C. S. Unnikrishnan. 'IndIGO and LIGO-India: Scope and plans for gravitational wave research and precision metrology in India'. In: *Int. J. Mod. Phys.* D22 (2013), p. 1341010. DOI: 10.1142/S0218271813410101. arXiv: 1510.06059 [physics.ins-det].

[822] B. P. Abbott et al. 'GW170814: A Three-Detector Observation of Gravitational Waves from a Binary Black Hole Coalescence'. In: *Phys. Rev. Lett.* 119.14 (2017), p. 141101. DOI: 10.1103/PhysRevLett.119.141101. arXiv: 1709.09660 [gr-qc].

[823] B. P. Abbott et al. 'Binary Black Hole Mergers in the first Advanced LIGO Observing Run'. In: *Phys. Rev.* X6.4 (2016), p. 041015. DOI: 10.1103/PhysRevX.6.041015. arXiv: 1606.04856 [gr-qc].

[824] Peter S. Shawhan. 'Rapid alerts for following up gravitational wave event candidates'. In: *Proc. SPIE Int. Soc. Opt. Eng.* 8448 (2012), p. 844825. DOI: 10.1117/12.926372. arXiv: 1206.6163 [astro-ph.IM].

[825] Pau Amaro-Seoane et al. 'eLISA/NGO: Astrophysics and cosmology in the gravitational-wave millihertz regime'. In: *GW Notes* 6 (2013), pp. 4–110. arXiv: 1201.3621 [astro-ph.CO].

[826] C. J. Moore, R. H. Cole and C. P. L. Berry. 'Gravitational-wave sensitivity curves'. In: *Class. Quant. Grav.* 32.1 (2015), p. 015014. DOI: 10.1088/0264-9381/32/1/015014. arXiv: 1408.0740 [gr-qc].

[827] Benjamin P. Abbott et al. 'Sensitivity of the Advanced LIGO detectors at the beginning of gravitational wave astronomy'. In: *Phys. Rev.* D93.11 (2016). [Addendum: Phys. Rev.D97,no.5,059901(2018)], p. 112004. DOI: 10.1103/PhysRevD.93.112004,10.1103/PhysRevD.97.059901. arXiv: 1604.00439 [astro-ph.IM].

[828] B. P. Abbott et al. 'Properties of the Binary Black Hole Merger GW150914'. In: *Phys. Rev. Lett.* 116.24 (2016), p. 241102. DOI: 10.1103/PhysRevLett.116.241102. arXiv: 1602.03840 [gr-qc].

[829] K. S. Thorne. 'NONSPHERICAL GRAVITATIONAL COLLAPSE: A SHORT REVIEW'. In: (1972).

[830] B. P. Abbott et al. 'Prospects for Observing and Localizing Gravitational-Wave Transients with Advanced LIGO, Advanced Virgo and KAGRA'. In: *Living Rev. Rel.* 21.1 (2018), p. 3. DOI: 10.1007/s41114-018-0012-9,10.1007/lrr-2016-1. arXiv: 1304.0670 [gr-qc].

[831] B. P. Abbott et al. 'Astrophysical Implications of the Binary Black-Hole Merger GW150914'. In: *Astrophys. J.* 818.2 (2016), p. L22. DOI: 10.3847/2041-8205/818/2/L22. arXiv: 1602.03846 [astro-ph.HE].

[832] Clifford M. Will. 'The Confrontation between general relativity and experiment'. In: *Living Rev. Rel.* 9 (2006), p. 3. DOI: 10.12942/lrr-2006-3. arXiv: gr-qc/0510072 [gr-qc].

[833] Event Horizon Telescope Collaboration et al. 'First M87 Event Horizon Telescope Results. I. The Shadow of the Supermassive Black Hole'. In: ApJ 875, L1 (2019), p. L1. DOI: 10.3847/2041-8213/ab0ec7.

[834] B. P. Abbott et al. 'Tests of general relativity with GW150914'. In: *Phys. Rev. Lett.* 116.22 (2016). [Erratum: Phys. Rev. Lett.121,no.12,129902(2018)], p. 221101. DOI: 10.1103/PhysRevLett.116.221101,10.1103/PhysRevLett.121.129902. arXiv: 1602.03841 [gr-qc].

[835] Bernard Carr, Florian Kuhnel and Marit Sandstad. 'Primordial Black Holes as Dark Matter'. In: *Phys. Rev.* D94.8 (2016), p. 083504. DOI: 10.1103/PhysRevD.94.083504. arXiv: 1607.06077 [astro-ph.CO].

[836] Misao Sasaki et al. 'Primordial Black Hole Scenario for the Gravitational-Wave Event GW150914'. In: *Phys. Rev. Lett.* 117.6 (2016). [erratum: Phys. Rev. Lett.121,no.5,059901(2018)], p. 061101. DOI: 10.1103/PhysRevLett.121.059901,10.1103/PhysRevLett.117.061101. arXiv: 1603.08338 [astro-ph.CO].

[837] Bernard Carr et al. 'Primordial black hole constraints for extended mass functions'. In: *Phys. Rev.* D96.2 (2017), p. 023514. DOI: 10.1103/PhysRevD.96.023514. arXiv: 1705.05567 [astro-ph.CO].

[838] B. P. Abbott et al. 'Localization and broadband follow-up of the gravitational-wave transient GW150914'. In: *Astrophys. J.* 826.1 (2016), p. L13. DOI: 10.3847/2041-8205/826/1/L13. arXiv: 1602.08492 [astro-ph.HE].

[839] V. Connaughton et al. 'Fermi GBM Observations of LIGO Gravitational Wave event GW150914'. In: *Astrophys. J.* 826.1 (2016), p. L6. DOI: 10.3847/2041-8205/826/1/L6. arXiv: 1602.03920 [astro-ph.HE].

[840] M. Ackermann et al. 'Fermi-LAT Observations of the LIGO event GW150914'. In: *Astrophys. J.* 823.1 (2016), p. L2. DOI: 10.3847/2041-8205/823/1/L2. arXiv: 1602.04488 [astro-ph.HE].

[841] V. Savchenko et al. 'INTEGRAL upper limits on gamma-ray emission associated with the gravitational wave event GW150914'. In: *Astrophys. J.* 820.2 (2016), p. L36. DOI: 10.3847/2041-8205/820/2/L36. arXiv: 1602.04180 [astro-ph.HE].

[842] M. Tavani et al. 'AGILE Observations of the Gravitational Wave Event GW150914'. In: *Astrophys. J.* 825.1 (2016), p. L4. DOI: 10.3847/2041-8205/825/1/L4. arXiv: 1604.00955 [astro-ph.HE].

[843] P. A. Evans et al. 'Swift follow-up of the Gravitational Wave source GW150914'. In: *Mon. Not. Roy. Astron. Soc.* 460.1 (2016), pp. L40–L44. DOI: 10.1093/mnrasl/slw065. arXiv: 1602.03868 [astro-ph.HE].

[844] J. Greiner et al. 'On the FERMI-GBM event seen 0.4 s after GW150914'. In: *Astrophys. J.* 827.2 (2016), p. L38. DOI: 10.3847/2041-8205/827/2/L38. arXiv: 1606.00314 [astro-ph.HE].

[845] V. Connaughton et al. 'On the Interpretation of the Fermi-GBM Transient Observed in Coincidence with LIGO Gravitational-wave Event GW150914'. In: *Astrophys. J.* 853.1 (2018), p. L9. DOI: 10.3847/2041-8213/aaa4f2. arXiv: 1801.02305 [astro-ph.HE].

[846] A. Gando et al. 'Search for electron antineutrinos associated with gravitational wave events GW150914 and GW151226 using KamLAND'. In: *Astrophys. J.* 829.2 (2016), p. L34. DOI: 10.3847/2041-8205/829/2/L34. arXiv: 1606.07155 [astro-ph.HE].

[847] K. Abe et al. 'Search for Neutrinos in Super-Kamiokande associated with Gravitational Wave Events GW150914 and GW151226'. In: *Astrophys. J.* 830.1 (2016), p. L11. DOI: 10.3847/2041-8205/830/1/L11. arXiv: 1608.08745 [astro-ph.HE].

[848] Alexander Aab et al. 'Ultrahigh-energy neutrino follow-up of Gravitational Wave events GW150914 and GW151226 with the Pierre Auger Observatory'. In: *Submitted to: Phys. Rev. D* (2016). arXiv: 1608.07378 [astro-ph.HE].

[849] S. Adrian-Martinez et al. 'High-energy Neutrino follow-up search of Gravitational Wave Event GW150914 with ANTARES and IceCube'. In: *Phys. Rev.* D93.12 (2016), p. 122010. DOI: 10.1103/PhysRevD.93.122010. arXiv: 1602.05411 [astro-ph.HE].

[850] M. G. Aartsen et al. 'Searches for Extended and Point-like Neutrino Sources with Four Years of IceCube Data'. In: *Astrophys. J.* 796.2 (2014), p. 109. DOI: 10.1088/0004-637X/796/2/109. arXiv: 1406.6757 [astro-ph.HE].

[851] M. G. Aartsen et al. 'Search for Time-independent Neutrino Emission from Astrophysical Sources with 3 yr of IceCube Data'. In: *Astrophys. J.* 779 (2013), p. 132. DOI: 10.1088/0004-637X/779/2/132. arXiv: 1307.6669 [astro-ph.HE].

[852] R. Abbasi et al. 'IceCube Sensitivity for Low-Energy Neutrinos from Nearby Supernovae'. In: *Astron. Astrophys.* 535 (2011). [Erratum: Astron. Astrophys.563,C1(2014)], A109. DOI: 10.1051/0004-6361/201117810e, 10.1051/0004-6361/201117810. arXiv: 1108.0171 [astro-ph.HE].

[853] Stephen Boughn. 'Electromagnetic radiation induced by a gravitational wave'. In: *Phys. Rev.* D11 (1975), pp. 248–252. DOI: 10.1103/PhysRevD.11.248.

[854] M. Bustamante. Private communication. 16th Dec. 2016.

[855] Alexander Dolgov and Konstantin Postnov. 'Electromagnetic Radiation Accompanying Gravitational Waves from Black Hole Binaries'. In: *JCAP* 1709.09 (2017), p. 018. DOI: 10.1088/1475-7516/2017/09/018. arXiv: 1706.05519 [astro-ph.HE].

[856] Justine Tarrant, Geoff Beck and Sergio Colafrancesco. 'Making light of gravitational-waves'. In: (2019). arXiv: 1904.12678 [astro-ph.HE].

[857] P. Meszaros. 'Gamma-ray bursts: The fireball shock model and its implications'. In: *eConf* C9808031 (1998). [,539(1998)], p. 32.

[858] T. Piran. 'Gamma-ray bursts and the fireball model'. In: *Phys. Rept.* 314 (1999), pp. 575–667. DOI: 10.1016/S0370-1573(98)00127-6. arXiv: astro-ph/9810256 [astro-ph].

[859] R. D. Blandford and R. L. Znajek. 'Electromagnetic extractions of energy from Kerr black holes'. In: *Mon. Not. Roy. Astron. Soc.* 179 (1977), pp. 433–456. DOI: 10.1093/mnras/179.3.433.

[860] Rosalba Perna, Davide Lazzati and Bruno Giacomazzo. 'Short Gamma-Ray Bursts from the Merger of Two Black Holes'. In: *Astrophys. J.* 821.1 (2016), p. L18. DOI: 10.3847/2041-8205/821/1/L18. arXiv: 1602.05140 [astro-ph.HE].

[861] Shigeo S. Kimura, Sanemichi Z. Takahashi and Kenji Toma. 'Evolution of an Accretion Disk in Binary Black Hole Systems'. In: *Mon. Not. Roy. Astron. Soc.* 465 (2016), p. 4406. DOI: 10.1093/mnras/stw3036. arXiv: 1607.01964 [astro-ph.HE].

[862] Agnieszka Janiuk et al. 'On the possible gamma-ray burst–gravitational wave association in GW150914'. In: *New Astron.* 51 (2017), pp. 7–14. DOI: 10.1016/j.newast.2016.08.002. arXiv: 1604.07132 [astro-ph.HE].

[863] Abraham Loeb. 'Electromagnetic Counterparts to Black Hole Mergers Detected by LIGO'. In: *Astrophys. J.* 819.2 (2016), p. L21. DOI: 10.3847/2041-8205/819/2/L21. arXiv: 1602.04735 [astro-ph.HE].

[864] Imre Bartos et al. 'Rapid and Bright Stellar-mass Binary Black Hole Mergers in Active Galactic Nuclei'. In: *Astrophys. J.* 835.2 (2017), p. 165. DOI: 10.3847/1538-4357/835/2/165. arXiv: 1602.03831 [astro-ph.HE].

[865] F. Fraschetti. 'Possible role of magnetic reconnection in the electromagnetic counterpart of binary black hole merger'. In: *JCAP* 1804.04 (2018), p. 054. DOI: 10.1088/1475-7516/2018/04/054. arXiv: 1603.01950 [astro-ph.HE].

[866] Kohta Murase et al. 'Ultrafast Outflows from Black Hole Mergers with a Minidisk'. In: *Astrophys. J.* 822.1 (2016), p. L9. DOI: 10.3847/2041-8205/822/1/L9. arXiv: 1602.06938 [astro-ph.HE].

[867] P. Veres et al. 'Gravitational-wave Observations may Constrain Gamma-ray Burst Models: the Case of Gw150914–GBM'. In: *Astrophys. J.* 827.2 (2016), p. L34. DOI: 10.3847/2041-8205/827/2/L34. arXiv: 1607.02616 [astro-ph.HE].

[868] Kumiko Kotera and Joseph Silk. 'Ultrahigh Energy Cosmic Rays and Black Hole Mergers'. In: *Astrophys. J.* 823.2 (2016), p. L29. DOI: 10.3847/2041-8205/823/2/L29. arXiv: 1602.06961 [astro-ph.HE].

[869] Reetanjali Moharana et al. 'High Energy Neutrinos from the Gravitational Wave event GW150914 possibly associated with a short Gamma-Ray Burst'. In: *Phys. Rev.* D93.12 (2016), p. 123011. DOI: 10.1103/PhysRevD.93.123011. arXiv: 1602.08436 [astro-ph.HE].

[870] R. Abbasi et al. 'The Design and Performance of IceCube DeepCore'. In: *Astropart. Phys.* 35 (2012), pp. 615–624. DOI: 10.1016/j.astropartphys.2012.01.004. arXiv: 1109.6096 [astro-ph.IM].

[871] B. P. Abbott et al. 'Binary Black Hole Population Properties Inferred from the First and Second Observing Runs of Advanced LIGO and Advanced Virgo'. In: (2018). arXiv: 1811.12940 [astro-ph.HE].

[872] M. G. Aartsen et al. 'The IceCube Neutrino Observatory - Contributions to ICRC 2015 Part II: Atmospheric and Astrophysical Diffuse Neutrino Searches of All Flavors'. In: *Proceedings, 34th International Cosmic Ray Conference (ICRC 2015): The Hague, The Netherlands, July 30-August 6, 2015*. 2015. arXiv: 1510.05223 [astro-ph.HE]. URL: https://inspirehep.net/record/1398539/files/arXiv:1510.05223.pdf.

[873] M. G. Aartsen et al. 'Observation and Characterization of a Cosmic Muon Neutrino Flux from the Northern Hemisphere using six years of IceCube data'. In: *Astrophys. J.* 833.1 (2016), p. 3. DOI: 10.3847/0004-637X/833/1/3. arXiv: 1607.08006 [astro-ph.HE].

[874] M. Spurio. 'Results from the ANTARES neutrino telescope'. In: *EPJ Web Conf.* 116 (2016), p. 11006. DOI: 10.1051/epjconf/201611611006.

[875] A. Karle and for the IceCube Collaboration. 'IceCube'. In: *arXiv e-prints*, arXiv:1003.5715 (Mar. 2010), arXiv:1003.5715. arXiv: 1003.5715 [astro-ph.HE].

[876] S. Adrian-Martinez et al. 'First Search for Point Sources of High Energy Cosmic Neutrinos with the ANTARES Neutrino Telescope'. In: *Astrophys. J.* 743 (2011), p. L14. DOI: 10.1088/2041-8205/743/1/L14. arXiv: 1108.0292 [astro-ph.HE].

[877] Neil Gehrels et al. 'Galaxy Strategy for LIGO-Virgo Gravitational Wave Counterpart Searches'. In: *Astrophys. J.* 820.2 (2016), p. 136. DOI: 10.3847/0004-637X/820/2/136. arXiv: 1508.03608 [astro-ph.HE].

[878] Gary C. Hill and Katherine Rawlins. 'Unbiased cut selection for optimal upper limits in neutrino detectors: The Model rejection potential technique'. In: *Astropart. Phys.* 19 (2003), pp. 393–402. DOI: 10.1016/S0927-6505(02)00240-2. arXiv: astro-ph/0209350 [astro-ph].

[879] B. P. Abbott et al. 'The Rate of Binary Black Hole Mergers Inferred from Advanced LIGO Observations Surrounding GW150914'. In: *Astrophys. J.* 833 (2016), p. 1. DOI: 10.3847/2041-8205/833/1/L1. arXiv: 1602.03842 [astro-ph.HE].

[880] Emanuele Berti et al. 'Inspiral, merger and ringdown of unequal mass black hole binaries: A Multipolar analysis'. In: *Phys. Rev.* D76 (2007), p. 064034. DOI: 10.1103/PhysRevD.76.064034. arXiv: gr-qc/0703053 [GR-QC].

[881] John G. Baker et al. 'Mergers of non-spinning black-hole binaries: Gravitational radiation characteristics'. In: *Phys. Rev.* D78 (2008), p. 044046. DOI: 10.1103/PhysRevD.78.044046. arXiv: 0805.1428 [gr-qc].

[882] B. P. Abbott et al. 'GWTC-1: A Gravitational-Wave Transient Catalog of Compact Binary Mergers Observed by LIGO and Virgo during the First and Second Observing Runs'. In: (2018). arXiv: 1811.12907 [astro-ph.HE].

[883] B. P. Abbott et al. 'GW151226: Observation of Gravitational Waves from a 22-Solar-Mass Binary Black Hole Coalescence'. In: *Phys. Rev. Lett.* 116.24 (2016), p. 241103. DOI: 10.1103/PhysRevLett.116.241103. arXiv: 1606.04855 [gr-qc].

[884] Benjamin P. Abbott et al. 'GW170104: Observation of a 50-Solar-Mass Binary Black Hole Coalescence at Redshift 0.2'. In: *Phys. Rev. Lett.* 118.22 (2017). [Erratum: Phys. Rev. Lett.121,no.12,129901(2018)], p. 221101. DOI: 10.1103/PhysRevLett.118.221101,10.1103/PhysRevLett.121.129901. arXiv: 1706.01812 [gr-qc].

[885] B.. P.. Abbott et al. 'GW170608: Observation of a 19-solar-mass Binary Black Hole Coalescence'. In: *Astrophys. J.* 851.2 (2017), p. L35. DOI: 10.3847/2041-8213/aa9f0c. arXiv: 1711.05578 [astro-ph.HE].

[886] S. E. Woosley. 'Pulsational Pair-Instability Supernovae'. In: *Astrophys. J.* 836.2 (2017), p. 244. DOI: 10.3847/1538-4357/836/2/244. arXiv: 1608.08939 [astro-ph.HE].

[887] Cody Messick et al. 'Analysis framework for the prompt discovery of compact binary mergers in gravitational-wave data'. In: Phys. Rev. D 95.4, 042001 (Feb. 2017), p. 042001. DOI: 10.1103/PhysRevD.95.042001. arXiv: 1604.04324 [astro-ph.IM].

[888] Surabhi Sachdev et al. 'The GstLAL Search Analysis Methods for Compact Binary Mergers in Advanced LIGO's Second and Advanced Virgo's First Observing Runs'. In: (2019). arXiv: 1901.08580 [gr-qc].

[889] J. L. Racusin et al. 'Searching the Gamma-ray sky for Counterparts to Gravitational Wave Sources: Fermi Gamma-ray Burst Monitor and Large Area Telescope Observations of LVT151012 and GW151226'. In: *Astrophys. J.* 835.1 (2017), p. 82. DOI: 10.3847/1538-4357/835/1/82. arXiv: 1606.04901 [astro-ph.HE].

[890] A. Albert et al. 'Search for High-energy Neutrinos from Gravitational Wave Event GW151226 and Candidate LVT151012 with ANTARES and IceCube'. In: *Phys. Rev.* D96.2 (2017), p. 022005. DOI: 10.1103/PhysRevD.96.022005. arXiv: 1703.06298 [astro-ph.HE].

[891] A. Albert et al. 'Search for Multimessenger Sources of Gravitational Waves and High-energy Neutrinos with Advanced LIGO during Its First Observing Run, ANTARES, and IceCube'. In: *Astrophys. J.* 870.2 (2019), p. 134. DOI: 10.3847/1538-4357/aaf21d. arXiv: 1810.10693 [astro-ph.HE].

[892] A. Albert et al. 'Search for High-energy Neutrinos from Binary Neutron Star Merger GW170817 with ANTARES, IceCube, and the Pierre Auger Observatory'. In: *Astrophys. J.* 850.2 (2017), p. L35. DOI: 10.3847/2041-8213/aa9aed. arXiv: 1710.05839 [astro-ph.HE].

[893] Neil Cornish, Diego Blas and Germano Nardini. 'Bounding the speed of gravity with gravitational wave observations'. In: *Phys. Rev. Lett.* 119.16 (2017), p. 161102. DOI: 10.1103/PhysRevLett.119.161102. arXiv: 1707.06101 [gr-qc].

[894] V. Alan Kostelecky and Neil Russell. 'Data Tables for Lorentz and CPT Violation'. In: *Rev. Mod. Phys.* 83 (2011), pp. 11–31. DOI: 10.1103/RevModPhys.83.11. arXiv: 0801.0287 [hep-ph].

[895] M. Fishbach et al. 'A Standard Siren Measurement of the Hubble Constant from GW170817 without the Electromagnetic Counterpart'. In: *Astrophys. J.* 871.1 (2019), p. L13. DOI: 10.3847/2041-8213/aaf96e. arXiv: 1807.05667 [astro-ph.CO].

[896] B. P. Abbott et al. 'A gravitational-wave standard siren measurement of the Hubble constant'. In: *Nature* 551.7678 (2017), pp. 85–88. DOI: 10.1038/nature24471. arXiv: 1710.05835 [astro-ph.CO].

[897] S. Adrián-Martínez et al. 'Optical and X-ray early follow-up of ANTARES neutrino alerts'. In: *JCAP* 1602.02 (2016), p. 062. DOI: 10.1088/1475-7516/2016/02/062. arXiv: 1508.01180 [astro-ph.HE].

[898] M. G. Aartsen et al. 'The IceCube Neutrino Observatory - Contributions to ICRC 2015 Part I: Point Source Searches'. In: (2015). arXiv: 1510.05222 [astro-ph.HE].

[899] Abdelhak Djouadi. 'The Anatomy of electro-weak symmetry breaking. II. The Higgs bosons in the minimal supersymmetric model'. In: *Phys. Rept.* 459 (2008), pp. 1–241. DOI: 10.1016/j.physrep.2007.10.005. arXiv: hep-ph/0503173 [hep-ph].

[900] ATLAS Collaboration. *ATLAS EXOTICS Summary plots*. https://atlas.web.cern.ch/Atlas/GROUPS/PHYSICS/CombinedSummaryPlots/EXOTICS/. Accessed on 04-08-2019.

[901] Michele Papucci, Joshua T. Ruderman and Andreas Weiler. 'Natural SUSY Endures'. In: *JHEP* 1209 (2012), p. 035. DOI: 10.1007/JHEP09(2012)035. arXiv: 1110.6926 [hep-ph].

[902] Michele Papucci et al. 'Fastlim: a fast LHC limit calculator'. In: *Eur.Phys.J.* C74.11 (2014), p. 3163. DOI: 10.1140/epjc/s10052-014-3163-1. arXiv: 1402.0492 [hep-ph].

[903] Jong Soo Kim et al. ''Stop' that ambulance! New physics at the LHC?' In: *JHEP* 1412 (2014), p. 010. DOI: 10.1007/JHEP12(2014)010. arXiv: 1406.0858 [hep-ph].

[904] Philipp Grothaus, Seng Pei Liew and Kazuki Sakurai. 'A closer look at a hint of SUSY at the 8 TeV LHC'. In: *JHEP* 1505 (2015), p. 133. DOI: 10.1007/JHEP05(2015)133. arXiv: 1502.05712 [hep-ph].

[905] Thomas Jacques and Karl Nordström. 'Mapping monojet constraints onto Simplified Dark Matter Models'. In: *JHEP* 06 (2015), p. 142. DOI: 10.1007/JHEP06(2015)142. arXiv: 1502.05721 [hep-ph].

[906] N. Jackson et al. 'A survey of polarization in the JVAS/CLASS flat-spectrum radio source surveys - I. The data and catalogue production'. In: MNRAS 376 (2007), pp. 371–377. DOI: 10.1111/j.1365-2966.2007.11442.x. eprint: astro-ph/0703273.

[907] A. Wright and R. Otrupcek. 'Parkes Catalog, 1990, Australia telescope national facility.' In: *PKS Catalog (1990)*. 1990, p. 0.

[908] R. L. White et al. 'A Catalog of 1.4 GHz Radio Sources from the FIRST Survey'. In: ApJ 475 (1997), pp. 479–493.

[909] S. R. Rosen et al. 'The XMM-Newton serendipitous survey. VII. The third XMM-Newton serendipitous source catalogue'. In: *ArXiv e-prints* (2015). arXiv: 1504.07051 [astro-ph.HE].

[910] W. Forman et al. 'The fourth Uhuru catalog of X-ray sources.' In: ApJS 38 (1978), pp. 357–412. DOI: 10.1086/190561.

[911]  F. Verrecchia et al. 'The BeppoSAX WFC X-ray source catalogue'. In: A&A 472 (2007), pp. 705–713. DOI: 10.1051/0004-6361:20067040.

[912]  Eanna E. Flanagan and Scott A. Hughes. 'The Basics of gravitational wave theory'. In: *New J. Phys.* 7 (2005), p. 204. DOI: 10.1088/1367-2630/7/1/204. arXiv: gr-qc/0501041 [gr-qc].

[913]  Richard A. Isaacson. 'Gravitational Radiation in the Limit of High Frequency. I. The Linear Approximation and Geometrical Optics'. In: *Phys. Rev.* 166 (1968), pp. 1263–1271. DOI: 10.1103/PhysRev.166.1263.

[914]  Richard A. Isaacson. 'Gravitational Radiation in the Limit of High Frequency. II. Nonlinear Terms and the Ef fective Stress Tensor'. In: *Phys. Rev.* 166 (1968), pp. 1272–1279. DOI: 10.1103/PhysRev.166.1272.